\documentclass[twoside,12pt,openright]{report}

\usepackage{axodraw,color}
\usepackage{latexsym,amssymb,epsf}
\usepackage{setspace} %double spacing for text, single for captions,
\usepackage{cite}
\usepackage{subfloat}
\usepackage{amsmath}
\usepackage{fancyhdr,graphicx}
\usepackage{makeidx,subfigure}

\makeindex
\newcommand{\newc}{\newcommand}
\newc{\gev}{\,GeV}
\newc{\mev}{\,MeV}
\newc{\ra}{\rightarrow}
\newc{\rpv}{$\mathrm{\not\!R_p}$}
\newc{\rp}{$\mathrm{R_p}$}
\newc{\real}{\mathcal{R}e}
\newc{\alsm}{{\displaystyle \sum_{\alpha=1,2}}}
\newc{\besm}{{\displaystyle \sum_{\beta=1,2}}}
\newc{\al}{\alpha}
\newc{\sgn}{\mr{sgn}\,}
\newc{\be}{\beta}
\newc{\ga}{\gamma}
\newc{\de}{\delta}
\newc{\sla}{\!\!\!\!\!\not\:\:\!}
\newc{\slab}{\!\!\!\!\!\not\,\,\,}
\newc{\slac}{\!\!\!\!\!\!\!\not\,\,\,\,}
\newc{\met}{$\not\!\!E_T$}
\newc{\cw}{\cos\theta_W}
\newc{\sw}{\sin\theta_W}
\newc{\ssw}{\sin^2\theta_W}
\newc{\ccw}{\cos^2\theta_W}
\newc{\cbe}{\cos\beta}
\newc{\sbe}{\sin\beta}
\newc{\ort}{\frac1{\sqrt{2}}}
\newc{\sh}{\hat{s}}
\newc{\uh}{\hat{u}}
\newc{\tha}{\hat{t}}
\newc{\sa}{\sin\al}
\newc{\ca}{\cos\al}
\newc{\mz}{M_{\mr{Z}}}
\newc{\mw}{M_{\mr{W}}}
\newc{\bv}{$\mathrm{\not\!B}$}
\newc{\lv}{$\mathrm{\not\!L}$}
\newc{\beq}{\begin{equation}}
\newc{\eeq}{\end{equation}}
\newc{\ie}{{\it i.e.\/}\ }
\newc{\lam}{\lambda}
\newc{\cht}{\tilde{\chi}}
\newc{\glt}{\tilde{g}}
\newc{\upt}{\tilde{u}}
\newc{\qkt}{\tilde{q}}
\newc{\elt}{\tilde{\ell}}
\newc{\hgt}{\tilde{H}}
\newc{\nut}{\tilde{\nu}}
\newc{\dnt}{\tilde{d}}
\newc{\ftl}{\mr{\tilde{f}}}
\newc{\psb}{\bar{\psi}}
\newc{\rtt}{\sqrt{2}}
\newc{\mut}{\tilde{\mu}}
\newc{\mr}{\mathrm}
\newc{\bath}{\bar{\theta}}
\newc{\tht}{\theta}
\newc{\JC}{{\bf J}}
\newc{\lra}{\longrightarrow}
\newc{\eg}{{\it e.g.\  }}
\newc{\barr}{\begin{eqnarray}}
\newc{\earr}{\end{eqnarray}}
\newc{\me}{\mathcal{M}}
\newc{\dbm}{\partial_\mu}
\newc{\dbmu}{\stackrel{\leftrightarrow\  }{\partial^\mu}}
\newc{\sgm}{\sigma_\mu}
\newc{\captionB}[2]{\caption[{#1}]{{\small {#2}}}}
\begin{document}
%
%  Hack for subfigure package
%
\renewcommand{\thesubfigure}{\alph{subfigure}}
\newcommand{\thesubfigureold}{(\thesubfigure)}
\makeatletter
\renewcommand{\@thesubfigure}{\thesubfigureold\space}

\newpage

\pagenumbering{arabic}
\setcounter{page}{1} \pagestyle{fancy} %\pagestyle{plain}
\renewcommand{\chaptermark}[1]{\markboth{\chaptername%
\ \thechapter:\,\ #1}{}}
\renewcommand{\sectionmark}[1]{\markright{\thesection\,\ #1}}

%%%%%%%%%%%%%%%%%%%%%%%%%%%%%%%%%%%%%%%%%%%%%%%%%%%%%%%%%%%%%%%%%%%%%%%%%%%
% This is a sample header for a sample dissertation. Fill in the name,
% and the other information. LaTeX will work out the table of
% content, the list of figures and of tables for you. 
%%%%%%%%%%%%%%%%%%%%%%%%%%%%%%%%%%%%%%%%%%%%%%%%%%%%%%%%%%%%%%%%%%%%%%%%%%%
\newpage
\thispagestyle{empty}

\vspace*{1cm}
\begin{center}
    {\Huge{\bf Simulations of R-parity Violating}}\\
\vspace{5mm}
{\Huge{\bf  SUSY Models}} \vspace{12pt} \\ 
\vspace{2cm}
{\large 
Peter Richardson\\
Balliol College\\
\vspace{0.8cm}
Department of Physics $\cdot$
Theoretical Physics\\
University of Oxford}
\vspace{1cm}
\end{center}

\begin{center}
{\Large\font\oxcrest=oxcrest40
\oxcrest\char'01}
\hspace{2cm}
\includegraphics[width=11.3mm]{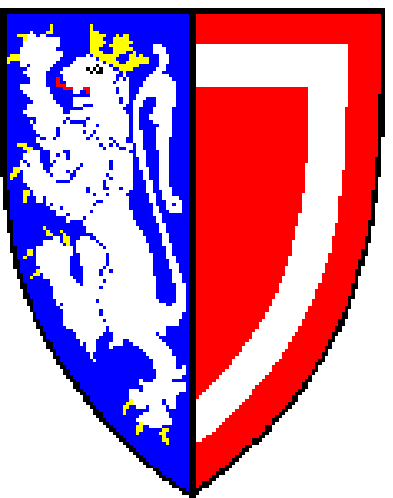}
\end{center}
\vspace{5.5cm}

\begin{center}
{\large Thesis submitted in partial fulfillment of the requirements
for\\ the Degree of Doctor of Philosophy at the\\ University of
Oxford\\
\vspace{0.9cm}
$\cdot$ August 2000 $\cdot$}
\end{center}
\newpage
\thispagestyle{empty}
\vspace*{20cm}
\phantom{please at a blank page}
\newpage

\singlespacing

%ABSTRACT
\begin{abstract}

  In recent years there has been a great deal of interest in R-parity
  violating supersymmetric models. We explain the motivation for studying
  these models and explore the various phenomenological consequences
  of R-parity violation.
  In particular, we argue that if we are to explore all channels
  for the discovery of supersymmetry then these models must be investigated.

  It has become essential for the experimental study of any new model to have a
  Monte Carlo event generator which includes the processes predicted by that
  model. We review the techniques used in the construction of these
  simulations and show how we have extended the HERWIG event generator to
  include R-parity violating processes. We discuss how to treat the emission of
  QCD radiation in these processes including colour coherence effects
  via the angular-ordered parton shower. 

  We then make use of this simulation to investigate the discovery potential for
  resonant slepton production, via either supersymmetric
  gauge or R-parity violating decay modes, in
  hadron--hadron collisions. In particular, we show that although
  the colour coherence properties of the R-parity violating decay modes can be used to 
  improve the extraction of a signal
  above the QCD background these processes
  will only be visible for large values of the R-parity
  violating Yukawa couplings. However a signal, \ie like-sign dilepton production,
  from the supersymmetric gauge decay modes is
  visible above the background for much smaller values of the R-parity violating Yukawa
  couplings.

  Finally, we look at the possibility that the KARMEN time anomaly can be explained by
  the existence of a light neutralino which is produced in the decay of charged
  pions via R-parity violation. This neutralino then decays inside the KARMEN 
  detector, into three leptons via R-parity violation, explaining the excess of 
  events observed by the KARMEN experiment.

\vspace*{5cm}

\begin{center}
\begin{quote}
\it 
   The road goes ever on and on,\\
   down from the door where it began,\\
   now far ahead the road has gone\\
   and I must follow if I can \ldots
\end{quote}
\end{center}
\hfill{\small J.R.R. Tolkien}

\end{abstract}
\newpage
\thispagestyle{empty}
\vspace*{20cm}
\phantom{please at a blank page}
\newpage
\singlespacing
\thispagestyle{empty}

\vspace*{7cm}
\begin{center}
\it \LARGE
To my parents,\\
\vspace{1cm}
this wouldn't have been possible\\
\vspace{0.2cm}
without their support and encouragement.

\hspace*{5cm}
\end{center}
\newpage
\thispagestyle{empty}
\vspace*{20cm}
\phantom{please at a blank page}
%
% Set double spacing for the rest of the thesis
%
%\doublespacing
%\onehalfspacing

%\pagestyle{empty}
\pagenumbering{roman}
\setcounter{page}{0} \pagestyle{plain}

\tableofcontents

\listoffigures
\listoftables

% ********************************** 
% ******** Acknowledgments  ******** 
% ********************************** 
\chapter*{Acknowledgments} 
\addcontentsline{toc}{chapter} 
		 {\protect\numberline{Acknowledgments\hspace{-96pt}}} 
  
   Although, in principle, this is all my own work many people have helped along the way.
   During my time in Cambridge, particularly during my final-year project, I was
   very lucky to have been taught by Bryan Webber to whom I owe a great debt for
   awakening my interest in particle physics.
   Here in Oxford I have increasingly come to realize how fortunate
   I was in my choice of supervisors. It has been a pleasure to work with both
   Herbi Dreiner and  Mike Seymour. I have benefited from their understanding of
   their respective fields and while I have learnt a great deal from both of them
   I hope that they have learnt something as well.

   A number of people read this thesis at various stages of the writing process
   and I am indebted to them for correcting my occasionally poor grammar and
   making me explain what I meant in what I hope is a clear way. So thank you
   very much Herbi, Jenny and Mike.

   During my time in Oxford I have been able to travel widely due to the
   support of various funding bodies, in particular PPARC. During my second
   year I was able to attend, thanks to PPARC, the Tevatron Run II Workshop
   at Fermilab were I presented some of the results in Chapter~\ref{chap:slepton}.
   I was also able to spend an interesting fortnight at CERN, again thanks to PPARC.

   I would 
   like to thank Torbj\"{o}rn Sj\"{o}strand and Paula Eerola for an invitation
   to speak at the Nordic LHC physics workshop, and for paying for my trip.
   At a time when my work was going slowly it was a wonderful experience to
   meet experimentalists who were running my simulation code and producing
   interesting results which encouraged me to actually get something finished when
   I returned, thanks a lot to Christophe Clement and 
   Linda Megner.

   In the final year of my D. Phil. I was also able to attend the SUSY Tools
   workshop in Colmar, thanks to the generosity of PPARC, where again
   I was lucky to meet a number of experimentalists who had been using my results.

   In the middle of writing this thesis I went to the CTEQ summer school at
   Lake Geneva, in Wisconsin, where I spent an enjoyable ten days. I was also
   able to visit Fermilab and Argonne after this. So thanks to the Fermilab theory
   group for their hospitality and to both Fermilab and PPARC for paying for my trip.

   Shortly after this trip to Fermilab I attended the SUSY2000 conference, at CERN,
   where I gave a talk on the results in Chapters~\ref{chap:monte} and
   \ref{chap:slepton}. This trip was funded by the physics department and Balliol
   college.

   A large amount of my time has been spent working on the HERWIG Monte Carlo
   event generator. It has been a pleasure to work with the many authors
   of HERWIG  and they have provided
   a lot of support and encouragement. 

   The Cambridge experimental group, in particular Andy Parker, Lee Drage,
   Debbie Morgan, and Alan Barr,
   have produced a lot of interesting results on the search for baryon number
   violating SUSY and I am grateful for their interest in my work and the 
   occasional bugs they found. I look forward to spending the next two years
   working with them in Cambridge.

   The theory group at RAL have also been very helpfully during my time in Oxford
   and during my occasional visits to the lab. 

   I have spent an enjoyable three years in Oxford for which I must
   thank various members of both the physics department and Balliol College.
   Here in the in the physics department Jenny, Peter, Alex, Ed, Marc,
   Paul, Ankush, Joern, Todd, Armin, and in college Kyle, Chris, Michaels (all of them),
   Zoe, Mark, Craig and Joao.

   Finally none of this would have been possible without the support of my family.
   They put up with me when I was very stressed and various deadlines were getting
   ever closer. So thanks a lot Mam, Dad and Angus.

\vspace{60mm}

\subsection*{Publications}

  Some of the work contained in this thesis has been previously published.
  The work in Chapter~\ref{chap:monte} was published in \cite{Dreiner:2000qz}, 
  as were the results in Section~\ref{sect:sleptonrpv}.
  The implementation of the work in Chapter~\ref{chap:monte} into the HERWIG
  event generator is described in the release note for the latest version, and the
  new manual, \cite{HERWIG61}.

  The work in Section~\ref{sect:sleptongauge} is available in \cite{Dreiner:2000vf}
  and has been accepted for publication in Phys. Rev. D. 
  This work was also reported in various conference
  proceedings. Preliminary results of the Tevatron studies were presented at the 
  Tevatron Run II workshop on supersymmetry and Higgs physics and are described in 
  \cite{Dreiner:1998gz} and in the
  Beyond the MSSM working group report \cite{Allanach:1999bf}.
  The results of the LHC studies are available in the summary of the
  SUSY working group of the 1999 Les Houches workshop \cite{Abdullin:1999zp}
  and in \cite{Dreiner:2000qf}.

  The work on the KARMEN anomaly presented in Chapter~\ref{chap:karmen} was
  published as \cite{Choudhury:1999tn}.

\pagestyle{fancy}

\newpage

%{\setlength{\baselineskip}{24pt} 
%\pagestyle{fancy}
\addtolength{\headheight}{3pt}    % more space for rule under running head
\fancyhead{}
%\fancyhead[LE]{\sl\leftmark}
%\fancyhead[LO,RE]{\rm\thepage}
%\fancyhead[RO]{\sl\rightmark}
\fancyhead[LE]{\sl\leftmark}
\fancyhead[LO,RE]{\rm\thepage}
\fancyhead[RO]{\sl\rightmark}
\fancyfoot[C,L,E]{}

\pagenumbering{arabic}

%\chapter{Introduction}
%%%%%%%%%%%%%%%%%%%%%%%%%%%%%%%%%%%%%%%%%%%%%%%%%%%%%%%%%%%%%%%%%%%%%%%%%%%%%
%									    %
%    Chapter 1 of my thesis					            %
%									    %
%    This chapter will contain					            %
%									    %
%   1. A brief introduction to the Standard Model concentrating on how the  %
%      SM can be obtained from the particle content and symmetries          %
%									    %
%   2. Introduce the problems and present SUSY as a solution to these and   %
%      a new symmetry. Then explain the SUSY theory                         %
%                                                                           %
%%%%%%%%%%%%%%%%%%%%%%%%%%%%%%%%%%%%%%%%%%%%%%%%%%%%%%%%%%%%%%%%%%%%%%%%%%%%%

\chapter{Introduction} \label{chap:intro}

  For more than twenty years the Standard Model (SM)
  \cite{Glashow:1961tr:Weinberg:1967tq:Salam:1968} of  particle physics
  has provided predictions of the results of high energy particle
  physics experiments. Despite the excellent agreement between
  the predictions of this theory and the experimental results it is
  widely believed that the Standard Model cannot be a complete
  theory of everything. 
  This thesis will be concerned with one possible extension of the Standard
  Model, R-parity violating (\rpv) supersymmetry (SUSY). 

  In this chapter we will briefly describe the Standard Model, emphasizing
  the r\^{o}le
  that symmetries play in its construction. This will lead to a discussion
  of a
  possible new symmetry of nature, supersymmetry, and the problems in the 
  Standard Model that it solves. In the most na\"{\i}ve supersymmetric
  extension of the Standard Model 
  lepton and baryon number are violated. This leads to fast proton decay,
  in conflict with the experimental lower limit on the proton lifetime. This
  has led to the imposition of a new symmetry called R-parity (\rp),
  to prevent the decay of
  the proton. However, we will present other symmetries that can prevent
  proton decay
  without the imposition of \mbox{R-parity}. These symmetries lead to the
  violation
  of either lepton
  {\it or} baryon number but not both simultaneously.
  These models have been much less studied than those which conserve
  R-parity. We will therefore briefly discuss the experimental
  consequences of R-parity violation.

  The next chapter discusses the idea of Monte Carlo simulations and
  presents the necessary calculations to include R-parity violating
  processes in these simulations. The simulation program produced using
  these results
  is then used in Chapter~\ref{chap:slepton} to study resonant 
  slepton production in hadron--hadron collisions. Chapter~\ref{chap:karmen} 
  explores the possibility that the anomaly seen by the KARMEN
  experiment\footnote{The KARMEN collaboration has recently announced new
			 results which are discussed in the Addendum.} 
  can be explained by R-parity violation. The conclusions are presented in
  Chapter~\ref{chap:conclude}.
%
%  Section on the Standard Model
%
\section{Standard Model} \label{sect:introSM}

  In the present view of particle physics the symmetries of the theory play a
  central r\^{o}le. In fact, the conventional approach for obtaining the
  Standard Model
  Lagrangian is
  to specify the particle content of the theory and the symmetries it obeys.
  We can then write the most general Lagrangian for the particles given the
  symmetries and the requirement that the theory be renormalizable, \ie  
  that the ultraviolet divergences involved in calculations beyond tree 
  level can be absorbed into the bare parameters of the theory.

  In general, we will be considering two types of symmetries. The first is
  the requirement that the theory be Lorentz invariant, \ie obey the laws
  of special relativity. All the Lagrangians we will consider will be
  Lorentz invariant.\footnote{In general by invariant we mean that the
		 Lagrangian is invariant up to the addition of 
		 a total derivative under the symmetry transformation.
		 As we are really interested in the action, 
		 $S=\int\,d^4x\, \mathcal{L}$, the total derivative
	 	 will lead to a surface term which will vanish provided
		 the fields go to zero as $x\ra\infty$.}
  The other symmetries of the Standard Model are gauge symmetries which
  lead to gauge field theories. In these theories the forces
  between the fundamental particles are mediated by the exchange of spin-1
  gauge bosons.
  In the Standard Model the electromagnetic force between charged particles
  is  carried by the photon, the weak force by the W and Z bosons, and the
  strong force between coloured particles by the gluon. 

  All of these theories are based on the simplest example, \ie 
  Quantum Electrodynamics~(QED). 
  The Dirac Lagrangian for $n$ fermions is given by
\begin{equation}
\mathcal{L}_{\mr{Dirac}} = \sum_{i=1}^n\psb_i(i\dbm\gamma^\mu-m_i)\psi_i,
\label{eqn:dirac}
\end{equation}
  where $\psi_i$ are the fermionic fields,
  $\gamma^\mu$ are the Dirac matrices and $m_i$ is
  the mass of the fermion~$i$.
  This Lagrangian is invariant under a global phase change
\begin{equation}
 \psi_i \ra e^{iq_i\al} \psi_i,
\end{equation}
  where $\al$ is the phase change and $q_i$ is an arbitrary 
  flavour-dependent parameter.
  If instead we consider a local change of phase, $\al\ra\al(x)$, 
  then the Dirac Lagrangian is no longer invariant under this 
  transformation. 
  The Lagrangian can be made invariant under a local change of phase by
  introducing a new vector field, $A^\mu$, which has the kinetic term
\begin{equation}
\mathcal{L}_{\mr{kinetic}} = -\frac1{4}F^{\mu\nu}F_{\mu\nu},
\end{equation}
 where $F_{\mu\nu} = \dbm A_\nu-\partial_\nu A_\mu$. We can introduce an 
 interaction of the vector field with the fermion via the substitution
\begin{equation}
 \dbm \ra D_\mu = \dbm+i QA_\mu,
\label{eqn:covderab}
\end{equation}
  where $D_\mu$ is the covariant derivative and $Q$ is the charge operator
  defined by $Q\psi_i=q_i\psi_i$. The arbitrary constants  $q_i$ which 
  we introduced for
  the case of the global transformation are the couplings of the fermions
  to the gauge field. If the vector field transforms as
\begin{equation}
  A_\mu \ra A_\mu -\dbm\al,
\end{equation}
 under a local change of phase then the Lagrangian, 
\begin{equation}
 \mathcal{L} =  \sum_{i=1}^n\psb_i(i\dbm\gamma^\mu-m_i)\psi_i
		 +\mathcal{L}_{\mr{kinetic}},
\end{equation}
 is invariant under a local change of phase. This gives the interactions
 of the fermions,
 \linebreak 
 \eg the electron, with the electromagnetic field. This is the simplest
  example of
 a gauge transformation, namely a $U(1)$ transformation. Here the couplings
 $q_i$ are
 the electric charges of the fermions. The gauge theories which form the
 Standard Model are generalizations of this principle to non-Abelian gauge
 groups.
 The non-Abelian gauge transformation is
\begin{equation}
 \psi_a(x) \ra {\psi'}_a(x) = 
		\left[e^{i\tht_A(x){\bf t}^A}\right]_{ab} \psi_b(x) 
		\equiv \Omega_{ab}(x)\psi_b(x),
\end{equation}
  where $\tht^A(x)$ are the parameters of the transformation and  ${\bf t}^A$
  are the generators of the non-Abelian group in the same representation as 
  the fermions. The generalized covariant derivative has the form
\begin{equation}
  D^\mu_{ab} = \delta_{ab}\partial^\mu+i g {\bf t}^A_{ab} A_A^\mu,
\label{eqn:covder}
\end{equation}
  where $A^\mu_A$ are the gauge bosons in the adjoint representation of the
  gauge group and $g$ is the coupling of the fermion to the gauge field.
  The non-Abelian gauge transformation for the gauge field is given by
\begin{equation}
 {\bf t}^A A^A_\mu\ra {{\bf t}}^A {A'}^A_\mu = 
		 \Omega(x){\bf t}^A A^A_\mu\Omega^{-1}(x)
		    +\frac{i}{g}\left(\dbm\Omega(x)\right)\Omega^{-1}(x).
\end{equation}
  The Lagrangian for the non-Abelian gauge theory can then be constructed
\begin{equation}
  \mathcal{L} = \mathcal{L}_{\mr{gauge}} +\mathcal{L}_{\mr{fermion}},
\end{equation}
\begin{subequations}
 where
\begin{eqnarray}
 \mathcal{L}_{\mr{gauge}}  &=& -\frac1{4}F^A_{\mu\nu}F_A^{\mu\nu},  \\
 \mathcal{L}_{\mr{fermion}}&=& 
	\psb_a\left(iD_\mu\gamma^\mu-m\right)_{ab}\psi_b.
\end{eqnarray}
\end{subequations}
  The generalization to a non-Abelian symmetry of the $U(1)$ field strength
  tensor is given by
\begin{equation}
 F^A_{\mu\nu} =
	 \partial_\mu A^A_\nu-\partial_\nu A^A_\mu-gf^{ABC}  A^B_\mu A^C_\nu.
\label{eqn:QCDF}
\end{equation}
  The additional terms due to the non-Abelian structure of the field
  strength tensor lead to self-interactions of the gauge bosons.
  These non-Abelian gauge theories form the basis of the Standard Model.
  There is a problem because a mass term for the gauge boson, \ie
\begin{equation}
 \mathcal{L}^{\mr{boson}}_{\mr{mass}} = \frac12m^2A_A^\mu A^A_\mu,
\end{equation}
  is not gauge invariant and therefore cannot be included in the Lagrangian. 
  However in the Standard Model the bosons which mediate
  the weak force, \ie the W and Z bosons, are massive.
  There is a way of including a mass for the gauge bosons in a
  gauge-invariant manner. This is called the Higgs mechanism 
  \cite{Higgs:1964ia:Higgs:1964pj:Higgs:1966ev:Englert:1964et:Kibble:1967sv}.
  It is easiest to illustrate this mechanism by considering the simplest
  example, \ie a $U(1)$ gauge field coupled
  to a scalar field with the following Lagrangian
\begin{equation}
\mathcal{L} = \mathcal{L}_{\mr{derivative}}-V(\phi) = 
	(D^\mu\phi)^*(D_\mu\phi) -V(\phi),
\end{equation}
 where $\phi$ is a complex scalar field and $D^\mu$ is given by
 Eqn.\,\ref{eqn:covderab}.
 Here the charge operator acts on the scalar field, \ie $Q\phi=q\phi$,
 where $q$ is the coupling of the gauge field to the scalar field.
 The most general renormalizable, gauge-invariant potential has the form 
\begin{equation}
 V(\phi) = \mu^2\phi^*\phi+\lam(\phi^*\phi)^2.
\label{eqn:Higgspot}
\end{equation}
  The shape of the potential depends on the sign of $\mu^2$:
\begin{enumerate}
\item if $\mu^2>0$ then the minimum of the potential is at $|\phi|=0$;
\item if $\mu^2<0$ then the minimum of the potential is at 
	$|\phi|^2=-\frac{\mu^2}{2\lam}\equiv \frac{v^2}{2}$.
\end{enumerate}
  For the case $\mu^2<0$ we must consider fluctuations of the
  field about its ground state, which rather than being $\phi=0$ is a point
  on the circle $|\phi|=v/\sqrt{2}$. This breaks the symmetry because while
  any point on the circle $|\phi|=v/\sqrt{2}$ is equally likely only one of
  these points gets chosen.
  We can consider fluctuations by rewriting the field as
\begin{equation}
  \phi = 
  	\frac{1}{\sqrt{2}}\left(v + \rho \right)e^{i\left(\xi/v+\tht\right)},
\end{equation}
  where $\rho$ and $\xi$ are real scalar fields and $ve^{i\tht}/\sqrt{2}$ 
  is the point on the
  circle $|\phi|=v/\sqrt{2}$ about which we are expanding the field.
  The covariant derivative part of the Lagrangian can be expanded in terms
  of these fields,
\begin{equation}
\mathcal{L}_{\mr{derivative}} = \frac12\left[ i\left(v+\rho\right)
				\left\{qA_\mu+\frac1{v}\dbm\xi\right\}
				+\dbm\rho \right]
\times			 \left[ -i\left(v+\rho\right)
      	\left\{qA^\mu+\frac1{v}\partial^\mu\xi\right\}
		+\partial^\mu\rho \right]\!.
\end{equation}
  We can then perform a gauge transformation on this Lagrangian,
\begin{equation}
	A^\mu \ra A^\mu-\frac1{qv}\partial^\mu\xi,
\end{equation}
  to eliminate all dependence on the field $\xi$. This choice
  of gauge is called the unitary gauge. In the unitary gauge the
  Lagrangian is
\begin{eqnarray}
\mathcal{L} &=& -\frac1{4}F_{\mu\nu}F^{\mu\nu}+\frac{1}{2}v^2q^2A^\mu A_\mu
	 +\frac12\dbm\rho\partial^\mu\rho \nonumber\\
	&&     +\frac12(2v\rho+\rho^2)q^2A^\mu A_\mu
	      -\frac1{4}\mu^2v^2+\mu^2\rho^2
		-\lam v\rho^3-\frac{\lam}{4}\rho^4.
\end{eqnarray}
  This Lagrangian represents a massive gauge boson with mass $M^2_A=q^2v^2$,
  a massive real scalar field with mass $m^2_\rho=-2\mu^2$, the 
  self-interactions of the scalar field and its interactions with the gauge
  boson.
  This mechanism allows us to introduce a gauge boson mass term in 
  a manifestly gauge-invariant way. In this
  process the initial complex scalar field $\phi$ possesses two degrees of
  freedom, whereas the real scalar field $\rho$ has only one degree of
  freedom. The second degree of freedom has been `eaten' to provide the
  longitudinal polarization of the massive gauge boson, which has three
  degrees of freedom rather than the two degrees of freedom of a massless
  gauge boson.

  This theory describes a massive $U(1)$ gauge boson, \ie a massive photon. 
  Although in the Standard Model the photon is massless the gauge bosons
  which mediate the weak force, the W and Z, are massive and their masses are
  generated by a generalization of this mechanism to non-Abelian symmetries.

  We can
  now discuss the two theories which form the Standard Model in slightly
  more detail, \ie Quantum Chromodynamics which describes the strong force
  and the Glashow-Weinberg-Salam model 
  \cite{Glashow:1961tr:Weinberg:1967tq:Salam:1968}
  which describes the electroweak force.

\subsection{Quantum Chromodynamics}
\label{subsect:QCD}
  Quantum Chromodynamics (QCD)
  describes the colour force between the quarks which is carried by the
  exchange of gluons. The Lagrangian is
\begin{equation}
  \mathcal{L}_{\mr{QCD}} =  -\frac1{4}F^A_{\mu\nu}F_A^{\mu\nu} 
			+\sum_i\bar{q}^i_a
	\left(iD_\mu\gamma^\mu-m_i\right)_{ab}q^i_b,\label{eqn:LQCD}
\end{equation}
  where $F^A_{\mu\nu}$ is the non-Abelian field strength tensor given by
  Eqn.\,\ref{eqn:QCDF}, $q^i$ are the quark fields with mass $m_i$ and
  the covariant derivative $D^\mu_{ab}$ is given by Eqn.\,\ref{eqn:covder}.
  This is an $SU(3)$ gauge theory based on the general non-Abelian theory we
  have already considered. The fermions are in the fundamental
  representation of $SU(3)$ and the gauge bosons, \linebreak \ie the gluons,
  in the adjoint representation. There are some subtleties 
  involved in quantizing non-Abelian gauge theories which are discussed in,
  for example, \cite{Peskin:1995ev,Ellis:1991qj}. The major difference
  between this non-Abelian theory and, for example, QED is the presence of
  self-interactions of the gauge bosons due to the different structure of
  the field strength tensor, Eqn.\,\ref{eqn:QCDF}.

  This theory possesses two important features which occur due to the running
  of the coupling, \ie the fact that after we renormalize the theory the
  coupling depends on the energy scale at which it is evaluated. This is
  described in terms of the $\be$-function
\begin{equation}\
   Q^2\frac{\partial\al_S(Q^2)}{\partial Q^2} = \beta(\al_S),
\end{equation}
  where $\al_s=g_s^2/(4\pi)$, $g_s$ is the strong coupling
  and $Q$ is the energy scale at which the coupling is calculated.
  To leading order in $\al_s$ the $\be$-function is given by 
\begin{equation}
    \be(\al_s) = -\al^2_s\frac{(11C_A-2n_f)}{12\pi},
\end{equation}
  where $C_A$ is the Casimir in the adjoint representation and $n_f$ is the
  number of massless quark flavours. The quarks can be considered as
  effectively massless for
  energy scales \mbox{$Q\gg m_i$}.  The important features of QCD 
  are due to the sign of this
  $\be$-function. Given that $C_A=3$ for $SU(3)$, the $\be$-function is
  negative for less than 17 quark flavours. Six quark flavours have currently
  been discovered,
  so even at energy scales above the top quark mass the $\be$-function is
  negative. This leads to the following properties of QCD:
\begin{enumerate}
\item Asymptotic freedom. The coupling decreases as the energy scale at
	which it is evaluated increases.
        Hence if we are dealing with high energy processes we can use
 	perturbation theory in the small coupling constant to calculate
  	physical
	quantities. In general, we can only make use of this fact if the
 	 quantity 
	we are calculating is infra-red safe, \ie does not have large
	corrections due to long-range physics. 
\item Confinement. At low energies the coupling becomes large and we can no
	longer use perturbation theory to perform calculations. No free
	quarks or gluons have been observed experimentally. All the quarks
	and gluons are bound together to give the hadrons which are observed.
	These bound states either contain a valence quark--antiquark pair,
 	a meson, or three valence quarks, a baryon. These bound states also
	contain gluons which bind the quarks together and a
	sea of virtual quark--antiquark pairs produced by gluon splitting.
\end{enumerate}

\subsection{Electroweak Theory}
\label{subsect:electroweak}
  A major success of quantum field theory in general, and gauge theories in
  particular, is the unification of the electromagnetic and weak forces into
  a single gauge theory \cite{Glashow:1961tr:Weinberg:1967tq:Salam:1968}.
  This theory has the symmetry
  group $SU(2)_L\times U(1)$. The
  Lagrangian is given by
\begin{equation}
  \mathcal{L}_{\mr{electroweak}} =   \mathcal{L}_{\mr{Higgs}}
			+\mathcal{L}_{\mr{gauge}}
			+\mathcal{L}_{\mr{fermions}},
\label{eqn:Lelectroweak}
\end{equation}
  where $\mathcal{L}_{\mr{Higgs}}$, $\mathcal{L}_{\mr{gauge}}$ and 
  $\mathcal{L}_{\mr{fermions}}$ are the Lagrangians for the Higgs field,
  the gauge fields and the fermions respectively,
\begin{subequations}
\begin{eqnarray}
 \mathcal{L}_{\mr{Higgs}}   &=& (D_\mu\phi)^\dagger(D^\mu\phi)
				-\mu^2\phi^\dagger\phi
				-\lam(\phi^\dagger\phi)^2,\\
 \mathcal{L}_{\mr{gauge}}   &=& 	-\frac1{4}F^A_{\mu\nu}F_A^{\mu\nu}
				-\frac1{4}B_{\mu\nu}B^{\mu\nu},\\
 \mathcal{L}_{\mr{fermions}}&=& \sum_{\mr{fermions}}
	\psb_{f,L}i\ga^\mu D_\mu\psi_{f,L}+\psb_{f,R}i\ga^\mu D_\mu\psi_{f,R}
		-\psb_{f,L}g_{ff'}\phi\psi_{f',R}
		-\psb_{f,R}g_{ff'}^*\phi^\dagger\psi_{f',L}.\nonumber\\
&&
\label{eqn:Lfermions}
\end{eqnarray}
\end{subequations}
  We have two gauge fields. The first is the $SU(2)_L$ field, $W^A_\mu$,
  with field strength tensor,~$F^A_{\mu\nu}$. This field is in the 
  adjoint of $SU(2)_L$, \ie an $SU(2)_L$ triplet with coupling $g$. 
  The associated charge is called weak isospin. The second gauge field
  is the $U(1)$ field, $B_\mu$, with field strength tensor $B_{\mu\nu}$
  and coupling $g'$. The associated $U(1)$ charge is called
  hypercharge, by analogy with the electric charge of QED.
  The fields $\psi_{f,L}=\frac12(1-\ga_5)\psi$ are the left-hand components
  of the fermion fields and are $SU(2)_L$ doublets, as is the Higgs field
  $\phi$. This is because experimentally the W bosons only couple to
  left-handed fermions. The fields $\psi_{f,R}=\frac12(1+\ga_5)\psi$ are
  the right-hand components of the fermion fields and are $SU(2)_L$
  singlets, because they do not couple to the W bosons. 
  The coupling, $g_{ff'}$, of the Higgs fields to the fermions $f$ and $f'$
  is only non-zero for those combinations of fermions which give a 
  gauge-invariant term in the Lagrangian.

  We will first consider the Higgs part of the Lagrangian,
  $\mathcal{L}_{\mr{Higgs}}$. The structure of the Higgs potential is the
  same as that which we considered for the $U(1)$ case, 
  Eqn.\,\ref{eqn:Higgspot}. Here we take the vacuum state to be
\begin{equation}
    \phi = \left(\begin{array}{c} 0 \\ v \end{array} \right)\!,
\end{equation}
  and expand about this point in terms of one real field and three angles
  in $SU(2)_L$ space to give the mass terms for the gauge bosons,
\begin{equation}
 \mathcal{L}_{\mr{gauge}}^{\mr{mass}} = 
	\frac14g^2v^2W^+_\mu W^{-\mu} +\frac14g^2v^2W^0_\mu W^{0\mu}
	-\frac12gg'v^2W^0_\mu B^\mu +\frac14{g'}^2v^2B_\mu B^\mu.
\end{equation}
  After the breaking of the $SU(2)_L$ symmetry there are two neutral gauge
  bosons, $\mr{W^0}$ and B, which
  mix to give mass eigenstates A and $\mr{Z^0}$,
\begin{subequations}
\label{eqn:bosonmix}
\begin{eqnarray}
 A_\mu &=& \phantom{-}\cw B_\mu+\sw W^0_\mu, \\
 Z_\mu &=& -\sw B_\mu +\cw W^0_\mu,
\end{eqnarray}
\end{subequations}
  where the weak mixing angle is given by $\tan\tht_W=\frac{g'}{g}$. This
  gives the mass terms for the gauge bosons in terms of the physical fields,
\begin{equation}
 \mathcal{L}_{\mr{gauge}}^{\mr{mass}} = 
	\frac14g^2v^2W^+_\mu W^{-\mu} +\frac{g^2v^2}{4\ccw} Z^0_\mu Z^{0\mu}.
\end{equation}
  Hence the gauge boson masses are
\begin{subequations}
\begin{eqnarray}
    M^2_\mr{W} &=& \frac12g^2v^2, \\
    M^2_\mr{Z} &=& \frac{g^2v^2}{2\ccw}, \\
    M^2_\mr{A} &=& 0.
\end{eqnarray}
\end{subequations}
  So by using the Higgs mechanism masses have been generated for the gauge
  bosons carrying the weak force, \ie the W and Z, while leaving the photon
  massless.

  As with QCD the second term in Eqn.\,\ref{eqn:Lelectroweak}, 
  $\mathcal{L}_{\mr{gauge}}$, contains the kinetic energy terms for the
  gauge fields and terms which lead to interactions between the gauge 
  bosons, for example the interaction of the photon with
  the $\mr{W^{\pm}}$ bosons. These gauge boson interaction terms 
  will not be important for the 
  rest of the discussion of the Standard Model and of supersymmetry.

  The final term in the Lagrangian, Eqn.\,\ref{eqn:Lelectroweak}, gives the 
  interactions of the gauge and Higgs bosons with the Standard Model
  fermions. Using Eqn.\,\ref{eqn:bosonmix} we can express the couplings
  of the fermions to the photon and the Z in terms of their hypercharge and
  isospin:
\begin{subequations}
\begin{eqnarray}
 \!\!\!\!\!\!\!\!\!\!\! \!\!\!\!\!\!\!\!\!\!\!\!\!\!\!\!\!\!\!\!\!\!\!
 \!\!\!\!\!\!\!\!\!
 \!\!\!\!\!\!\!\!\!\!\!\!\mr{photon\  couplings} &\!\!=
	\!\!& (t^3+\frac12 Y)g\sw\equiv Qe, \\
 \mr{Z\  couplings} &\!\!=\!\!& (t^3\ccw-\frac12 Y\ssw) \frac{g}{\cw}
	 = (t^3-Q\ssw)\frac{g}{\cw}, \,\,\,\,\,\,
\end{eqnarray}
\end{subequations}
  where $t^3$ is the third component of the weak isospin, $Y$ is the
  hypercharge and $e$ is the magnitude of the electron's electric charge. 
  The hypercharge is assigned given the weak isospin and 
  electric charges, $Q$, of the fermions, \ie $Y=2(Q-t^3)$.
  This gives the charges in Table~\ref{tab:SMcharges}.

\begin{table}
\renewcommand{\arraystretch}{1.5}
\begin{center}
\begin{tabular}{|c|c|c|c|c|}
\hline
 Fermion & Hypercharge, $Y$ & Isospin & $t^3$ &  Charge, $Q$\\
\hline
   $d_R$ & $-\frac23$ & $0$ & $\phantom{+}0$ & $-\frac13$ \\ 
\hline
   $u_R$ & $+\frac43$ & $0$ & $\phantom{+}0$ & $\phantom{+}\frac23$ \\ 
\hline
   {\renewcommand{\arraystretch}{1.0}
\rule{0mm}{8mm}
$\left(\begin{array}[c]{c}\!\!\! u\!\!\!\\\!\!\! d\!\!\!\end{array}\right)_L
  $} &
  $+\frac13$& $\frac12$ &
  $\begin{array}{c} +\frac12\\-\frac12\end{array}$ &
  $\begin{array}{c}+\frac23 \\ -\frac13\end{array}$  \\
\hline 
   $e^-_R$ & $-2$ & $0$ & $\phantom{+}0$ & $-1$ \\
\hline
   {\renewcommand{\arraystretch}{1.0}
\rule{0mm}{8mm}
$\left(\begin{array}{c}\!\!\!\nu_e\!\!\!\\\!\!\!e^-\!\!\!\end{array}
		\right)_L$} &
 $-1$& $\frac12$& $\begin{array}{c} +\frac12\\-\frac12\end{array}$ &
  $\begin{array}{c}\phantom{+} 0\\-1\end{array}$\\
\hline 
\end{tabular}
\end{center}
\captionB{Gauge quantum numbers of the Standard Model fermions.}
	{Gauge quantum numbers of the Standard Model fermions.}
\label{tab:SMcharges}
\end{table}

  The Feynman rules for the interactions of the Standard Model fermions with
  the gauge bosons are given in Appendix~\ref{chap:Feynman}. The relevant
  Lagrangian is given in Eqn.\,\ref{eqn:LfermionZ} with the Feynman rules
  given in Fig.\,\ref{fig:Zquarks}.

  This only leaves the problem of generating a mass for the fermions. Simply
  adding a Dirac mass term, as in Eqn.\,\ref{eqn:dirac}, violates the gauge
  invariance of the theory. We can reexpress the Dirac mass term in terms of
  the left- and right-handed fields,
\begin{equation}
	\mathcal{L}_{\mr{Dirac}}^{\mr{mass}} = -m\psb\psi 
		\equiv -m\left( \psb_L\psi_R+\psb_R\psi_L \right)\!. 
\end{equation}
  As the $SU(2)_L$ gauge transformation only acts upon the left-handed fields
  this term is not gauge invariant. This is the origin of the last two terms
  in Eqn.\,\ref{eqn:Lfermions} which couple the Higgs field and the fermions
  in a gauge-invariant manner.  After the breaking of
  the electroweak symmetry this gives fermion masses,
\begin{equation}
\mathcal{L}_{\mr{fermions}}^{\mr{mass}} = 
	 -g_{ee} v \left(\bar{e}_Le_R+\bar{e}_Re_L\right)
	-g_{dd} v \left(\bar{d}_Ld_R+\bar{d}_Rd_L\right)
	-g_{uu} v \left(\bar{u}_Lu_R+\bar{u}_Ru_L\right)\!,
\end{equation}
  by defining $m_f = g_{ff}v$ the fermions obtain a mass in a gauge-invariant
  way. Here we have only considered one generation of fermions. It should be
  noted that to give mass to both the up- and down-type quarks we have used
  both the Higgs field $\phi$ and its hermitian conjugate $\phi^\dagger$.

  In general, with the three generations of fermions which have been
  discovered experimentally, the mass eigenstates of the fermions and their
  weak interaction states can be different. We must write down all the
  possible terms in the Lagrangian which are invariant under the symmetries
  and are renormalizable. We obtain the following Lagrangian
\begin{eqnarray}
\mathcal{L}_{\mr{fermions}}^{\mr{mass}} &=& 
 -v\left(\begin{array}{ccc}\bar{d}_L&\bar{s}_L&\bar{b}_L\end{array}\right)
  \left(\begin{array}{ccc} g_{dd} & g_{ds} & g_{db} \\ 
			   g_{sd} & g_{ss} & g_{sb} \\
			   g_{bd} & g_{bs} & g_{bb} \\
   \end{array}\right)
	\left(\begin{array}{c} d_R\\s_R\\b_R\end{array}\right)\nonumber\\
 &&-v\left(\begin{array}{ccc}\bar{u}_L&\bar{c}_L&\bar{t}_L\end{array}\right)
  \left(\begin{array}{ccc} g_{uu} & g_{uc} & g_{ut} \\ 
			   g_{cu} & g_{cc} & g_{ct} \\
			   g_{tu} & g_{tc} & g_{tt} \\
   \end{array}\right)
	\left(\begin{array}{c} u_R\\c_R\\t_R\end{array}\right)+\mr{h.c.}\\
 &=& -\bar{D}_LM_dD_R-\bar{U}_LM_uU_R+\mr{h.c.},\label{eqn:fermionmassnomix}
\end{eqnarray}
  where $D_{L,R}$ is a vector containing the three (left/right) down-type
  quark fields and $U_{L,R}$ is a vector containing the three (left/right)
  up-type quark fields. $M_d$ and $M_u$ are the mass matrices for the up-
  and down-type quarks, respectively. This leads to mixing between the quark 
  generations.\footnote{If the neutrino is massless and the right-handed
		        neutrino does not exist there can be no mixing
		  	between the lepton generations. However, as we now
			believe the neutrinos are massive, it is possible
			that the leptons also undergo mixing.}
  The fields in Eqn.\,\ref{eqn:fermionmassnomix} are the weak interaction
  eigenstates. We can express Eqn.\,\ref{eqn:fermionmassnomix} in terms of
  the mass eigenstates by applying separate rotations to the left- and
  right-handed components of the quark fields. If we rotate the left-hand
  components of the field by the unitary matrix $L$ and the right-hand
  components by the unitary matrix $R$ we can obtain a diagonal mass matrix,
  $M'=LMR^\dagger$. We will need different rotation matrices for the 
  up- and down-type quarks which we shall denote with the subscripts $u$ and
  $d$, respectively. We must also rotate the quark fields in the other terms
  in the Standard Model Lagrangian. The neutral-current part of the
  electroweak and the QCD Lagrangians are flavour diagonal in terms of either
  the mass or weak interaction eigenstates. The only parts of the Lagrangian
  which are not are those which involve the coupling of the W boson to the
  quark fields. These terms involve the combination of fields 
\begin{equation}
  W\bar{U}_LD_L = W \bar{U'}_LL_uL^\dagger_dD'_L\equiv W\bar{U'}_L VD'_L,
\end{equation}
  where $V = L_uL^\dagger_d$ is called the Cabibbo-Kobayashi-Maskawa (CKM)
  matrix. We can therefore work in a physical, \ie mass, basis for the quarks
  and simply include the relevant element of the CKM matrix at the vertex
  coupling the W boson to the quarks. This is the only physical combination
  of the rotation matrices and can be associated with the down-type quarks.

\subsection{Symmetries of the Standard Model}

  In addition to the gauge and Lorentz symmetries of the Standard Model
  there are a number of other symmetries. These symmetries are consequences
  of the gauge and Lorentz symmetries and the requirement that the Standard
  Model be renormalizable.
  
  In particular there are two
   symmetries\footnote{These symmetries are violated in
			 the Standard Model by non-perturbative
   	  		effects \cite{'tHooft:1976fv:'tHooft:1976up}.}
 which will be important in the rest of this thesis:
\begin{description}
\item[Lepton Number.] We assign a  quantum number, called
	lepton number, such that the lepton fields have lepton number $+1$, 
	the antileptons have lepton number $-1$ and all the other fields
	have lepton number zero. The electroweak Lagrangian,
 	Eqn.\,\ref{eqn:Lelectroweak}, conserves this quantum number. 
\item[Baryon Number.] We assign the baryon number 
	such that the quarks have baryon number $+1/3$, the antiquarks
	baryon number $-1/3$ and all the other fields have baryon number
        zero. This is done so that
	the baryons, \eg the proton, have baryon number $+1$.
	Both the QCD Lagrangian, Eqn.\,\ref{eqn:LQCD}, and the electroweak
	Lagrangian, Eqn.\,\ref{eqn:Lelectroweak}, respect this symmetry.
\end{description}
  The important point to note for the rest of the thesis is that we did not
  construct the Lagrangian to have these symmetries. It is impossible to
  write down a term in the Lagrangian which is renormalizable, Lorentz and
  gauge invariant, but violates these discrete symmetries given the particle
  content of the Standard Model. This is important as in a supersymmetric
  theory it is possible to have terms in the Lagrangian which are
  renormalizable, Lorentz and gauge invariant, but violate either lepton or
  baryon number.

\section{Supersymmetry} \label{sect:introSUSY}

  Despite the great success of the Standard Model in explaining all the
  current experimental results it is widely believed that it cannot
  be a complete theory, if for no other reason than that it does not
  include gravity. The Standard Model is viewed as some
  low-energy effective theory of some larger theory which may be:
\begin{itemize}
  \item a grand unified theory (GUT) in which the gauge group of the
  Standard Model is unified as a part of a larger gauge group,
  \eg $SU(5)$;
  \item a string theory which would also include gravity.
\end{itemize} 
  This theory would hope to explain, for example, why there are three
  generations of fermions and predict
  some of the free parameters of the Standard Model, \eg the particle masses.
  While supersymmetry does not provide a solution to any of these questions
  it does provide a solution to other problems in the Standard Model such as
  the hierarchy problem, which we will discuss in 
  Section~\ref{subsect:SUSYmot}. 

  We will first discuss the idea of supersymmetry, followed by the problems
  it solves and other reasons for favouring it as a possible extension of
  the Standard Model. The simplest supersymmetric extension of the Standard
  Model is then discussed in some detail, concentrating on those parts of
  the theory which will be important in the rest of this thesis.

\subsection{Introduction to Supersymmetry}

  The basic idea of supersymmetry is that for every bosonic degree of
  freedom there is a corresponding fermionic one. Therefore the operator
  $Q$, which generates the transformation from a boson to a fermion must
  be a spinor and fermionic, \ie
\begin{subequations}
\begin{eqnarray}
 Q | \mr{Fermion} \rangle  & = & | \mr{Boson}\rangle, \\
 Q | \mr{Boson}   \rangle  & = & | \mr{Fermion} \rangle.
\end{eqnarray}
\end{subequations}
  These fermionic generators form part of an 
  extended Poincar\'{e} group. In addition
  to the standard algebra for the generators of the Poincar\'{e} group
\begin{subequations}
\begin{eqnarray}
[P^\mu,P^\nu] &=& 0, \\
\left[M^{\mu\nu},P^\rho\right] &=&  i(g^{\nu\rho}P^\mu-g^{\mu\rho}P^\nu), \\
 \left[M^{\mu\nu},M^{\rho\lam}\right] &=&  i(g^{\nu\rho}M^{\mu\lam}
					+g^{\mu\lam}M^{\nu\rho}
					-g^{\mu\rho}M^{\nu\lam}
					-g^{\nu\lam}M^{\mu\rho}),
\end{eqnarray}
\end{subequations}
  we have the following relations for the anti-commuting generator $Q$,
\begin{subequations}
\begin{eqnarray}
[Q_\al,P_\mu] &=& [\bar{Q}_{\dot{\al}},P_\mu] = 0, \\
 \{ Q_{\al},Q_{\be} \} &=& \{ \bar{Q}_{\dot{\al}},\bar{Q}_{\dot{\be}}\} 
				= 0, \\
\left[M^{\mu\nu},Q_\al \right] &=&  -i {(\sigma^{\mu\nu})_\al}^\be Q_\be, \\
\left[M^{\mu\nu},\bar{Q}^{\dot{\al}}\right] & =&
   -i {(\bar{\sigma}^{\mu\nu})^{\dot{\al}}}_{\dot{\be}} \bar{Q}^{\dot{\be}},
					 \\
\{ Q_{\al},\bar{Q}_{\dot{\be}}\} & =&2{\sigma^\mu}_{\al\dot{\be}}P_\mu.
\end{eqnarray}
\end{subequations}
  The generators $Q$ and $\bar{Q}$ are two-component left- and right-handed 
  Weyl spinors, respectively. We use the index $\al$ (or $\be$) to denote the
  component of the left-handed spinors and the index $\dot{\al}$ 
  (or $\dot{\be}$)
  to denoted the component of the right-handed spinors, since the two types
  of spinors transform differently under Lorentz transformations.

  This gives a closed algebra for the extended Poincar\'{e} group 
  with the addition of the fermionic generator $Q$. As with
  the Poincar\'{e} group we can construct Lagrangians which are invariant
  under this symmetry. For example the simplest possible supersymmetric
  Lagrangian contains a complex field $\phi$, with two degrees of freedom,
  and a Weyl fermion $\psi$, also with two degrees of freedom.
  The Lagrangian \cite{Wess:1974kz}
\begin{equation}
 \mathcal{L} =\dbm \phi^*\partial^\mu \phi
		+i\psb\bar{\sigma}^\mu\partial_\mu\psi \label{eqn:SUSYleq},
\end{equation}
 is invariant under the supersymmetry transformation 
\begin{subequations}
\begin{eqnarray}
 \delta \phi & = & \rtt\xi\psi, \\
 \delta \psi & = & i\rtt\dbm\phi\sigma^\mu\bar{\xi},
\end{eqnarray}
\end{subequations}
  where we have a spinor parameter $\xi$ for the transformation.
  While it is easier to construct the Lagrangian for a supersymmetric theory
  in two-component notation, using Weyl spinors, to actually derive the
  Feynman rules we will use the standard four-component notation with Dirac
  spinors. In this case for every Standard Model Dirac fermion with four
  degrees of freedom there will be two corresponding complex scalar fields,
  each with two degrees of freedom. Similarly the massless gauge bosons of
  the Standard Model which each have two degrees of freedom, \ie two
  polarizations, will have as partners a Majorana fermion with two degrees
  of freedom. The particles and their superpartners will have the same mass, 
  and the masses of the massive gauge bosons and gauginos are generated via
  the Higgs mechanism.

\subsection{Motivations for Supersymmetry}
\label{subsect:SUSYmot}
  As none of the superpartners of the Standard Model fields have been
  observed supersymmetry cannot be
  an exact symmetry, \ie it must be broken in such a way that most of the
  superpartners are more massive than their Standard Model
  partners.\footnote{Given the current experimental limits apart from the
  supersymmetric partners of the top quark all the sfermions must be heavier
  than their Standard Model partners.}
  Despite this there are a number of theoretical reasons for favouring a
  supersymmetric extension of the Standard Model:

\begin{enumerate}

\item{Coleman-Mandula Theorem.}

  Perhaps the most persuasive argument in favour of supersymmetry is based
  on the Coleman-Mandula theorem \cite{Coleman:1967ad}. As we saw in 
  Section~\ref{sect:introSM} the Standard Model can be constructed by
  imposing the symmetries of the theory, \ie gauge and Lorentz invariance,
  and by the requirement that the theory is renormalizable.  It is therefore 
  interesting to consider possible extensions of these symmetries, \eg by
  unifying the gauge groups of the Standard Model into one group. However if
  we consider possible extensions of the Poincar\'{e} group, Coleman and
  Mandula showed that the addition of any new generators which transform as
  bosons leads to a trivial S-matrix, 
  \linebreak[2]
  \ie~in particle scattering experiments the particles could only scatter
  through certain discrete angles, which is not observed. While extending
  the Poincar\'{e} group with additional bosonic generators is forbidden by
  this theorem we can extend the group with generators which transform as
  fermions. It was later shown \cite{Haag:1975qh} that supersymmetry is the
  only possible extension of the Poincar\'{e} group which does not lead to a
  trivial S-matrix.

\item{Hierarchy Problem.}
\nopagebreak

  A second argument in favour of supersymmetry can be seen by considering the
  one-loop corrections to the Higgs mass, given by the diagrams in 
  Fig.\,\ref{fig:higgs1}. In particular, if we only consider the fermion
  loops we get a contribution
\begin{equation}
\delta {M_H^2}_f = \Pi_{H}(M^2_H),
\end{equation}
  where 
\begin{equation}
  i\Pi_{H}(p^2) =  -\frac{|g_f|^2}{4}\int \frac{d^4k}{(2\pi)^4}
		\frac{	\mr{tr}\,
	\left[(k\!\sla\,+p\sla+m_f)(k\!\sla\,+m_f)\right]}
	{\left[(k+p)^2-m_f^2\right]\left[k^2-m_f^2\right]}
\end{equation}
  and $g_f=m_f/v$ is the coupling of the fermion to the Higgs field.
  Na\"{\i}vely, from power counting, this diagram is quadratically
  divergent. This  divergence can be regulated by imposing a cut-off,
  $\Lambda$, giving the following result
\begin{equation}
\delta {M_H^2}_f = \frac{|g_f|^2}{16\pi^2}\left[ -2\Lambda^2
			+6m_f^2\ln\left(\Lambda/m_f\right) \right]\!,
\end{equation}
  where we have neglected terms which are finite in the limit
  $\Lambda\ra\infty$.

\begin{figure}[t]
\begin{center}
% Fermion Loop
\subfigure[Fermion loops.]{
\begin{picture}(440,80)
\SetOffset(120,-10)
\DashArrowLine(40,50)(80,50){5}
\ArrowArcn(100,50)(20,0,180)
\ArrowArcn(100,50)(20,180,360)
\DashArrowLine(120,50)(160,50){5}
\Text(30,50)[c]{$\mr{H_0}$}
\Text(170,50)[c]{$\mr{H_0}$}
\Text(100,17)[c]{$\mr{\bar{f}}$}
\Text(100,83)[c]{$\mr{f}$}
\end{picture}}
% W and Z loops
\subfigure[W and Z boson loops.]{
\begin{picture}(440,80)
\SetOffset(20,-10)
\DashArrowLine(40,50)(80,50){5}
\PhotonArc(100,50)(20,0,180){3}{4}
\PhotonArc(100,50)(20,180,360){3}{4}
\DashArrowLine(120,50)(160,50){5}
\Text(30,50)[c]{$\mr{H_0}$}
\Text(170,50)[c]{$\mr{H_0}$}
\Text(100,17)[c]{$\mr{W^+}$}
\Text(100,83)[c]{$\mr{W^-}$}
\SetOffset(220,-10)
\DashArrowLine(40,50)(80,50){5}
\PhotonArc(100,50)(20,0,180){3}{4}
\PhotonArc(100,50)(20,180,360){3}{4}
\DashArrowLine(120,50)(160,50){5}
\Text(30,50)[c]{$\mr{H_0}$}
\Text(170,50)[c]{$\mr{H_0}$}
\Text(100,17)[c]{$\mr{Z^0}$}
\Text(100,83)[c]{$\mr{Z^0}$}
\end{picture}}
% Higgs loops
\subfigure[Higgs loops.]{
\begin{picture}(440,80)
\SetOffset(20,-10)
\DashArrowLine(40,50)(80,50){5}
\DashArrowArcn(100,50)(20,0,180){5}
\DashArrowArcn(100,50)(20,180,360){5}
\DashArrowLine(120,50)(160,50){5}
\Text(30,50)[c]{$\mr{H_0}$}
\Text(170,50)[c]{$\mr{H_0}$}
\Text(100,17)[c]{$\mr{H_0}$}
\Text(100,83)[c]{$\mr{H_0}$}
\SetOffset(220,-30)
\DashArrowLine(40,50)(100,50){5}
\DashCArc(100,70)(20,0,180){5}
\DashCArc(100,70)(20,180,360){5}
\DashArrowLine(100,50)(160,50){5}
\Text(30,50)[c]{$\mr{H_0}$}
\Text(170,50)[c]{$\mr{H_0}$}
\Text(100,103)[c]{$\mr{H_0}$}
\end{picture}}
\captionB{One-loop contributions to the Higgs mass.}
	{One-loop contributions to the Higgs mass.}
\label{fig:higgs1}
\end{center}
\end{figure}

  The Standard Model Higgs mass depends quadratically on the cut-off scale,
  $\Lambda$. This is not a problem provided that we renormalize the theory
  and absorb the divergent terms into a redefinition of the Higgs mass.
  However if we take the modern
  view that the Standard Model is only a low-energy effective theory we would
  expect the cut-off to be the scale of new physics, \eg the GUT or Planck
  scale. This means that the natural value of the Higgs mass is
  $10^{14}-10^{17} \,\mr{\gev}$ rather than the upper limit of around
  $300\, \mr{\gev}$ suggested by the precision electroweak data.
  This is the hierarchy problem, \ie the natural scale for the Higgs mass
  is the scale of new physics. We would therefore have to tune the bare
  Higgs mass in the Standard Model,
\begin{equation}
  M^2_H = {M^2_H}_{\mr{bare}}+\delta M^2_H,
\end{equation}
  in order to obtain a Higgs mass at the electroweak scale.
  If we require the Higgs mass to be around the electroweak scale the
  cancellation between
  $ M_H^{\mr{bare}}$ and $\delta M_H$ must be $\sim 1$ part in $10^{12}$
  which requires an 
  enormous fine-tuning to the parameters of the bare Lagrangian.

  We can see how this problem is solved by supersymmetry by considering an 
  additional scalar field \cite{Martin:1997ns}
  which interacts with the Higgs boson via the Lagrangian 
 \linebreak
 \mbox{$\mathcal{L} = -\lam_s\phi^2_sH^2S^2$}. This gives an extra one-loop
 contribution, Fig.\,\ref{fig:higgs2}, to the
 Higgs mass
\begin{equation}
\delta {M_H^2}_S = \frac{\lam_s}{16\pi^2}
		   \left(\Lambda^2-2M_S^2\ln\left(\Lambda/M_S\right)
  \right)\!,
\end{equation}
  where we have again neglected terms which are finite in the limit
  $\Lambda\ra\infty$. In supersymmetry, as we have two complex scalar
  fields for each Dirac fermion, this
  contribution will cancel the contribution from the fermion loops 
  provided that  \mbox{$\lam_S=|g_f|^2$}, which is the case if
  supersymmetry is unbroken. 
\begin{figure}
\begin{center}
\begin{picture}(440,70)
% The new scalar loop
\SetOffset(120,-50)
\DashArrowLine(40,50)(100,50){5}
\DashCArc(100,70)(20,0,180){5}
\DashCArc(100,70)(20,180,360){5}
\DashArrowLine(100,50)(160,50){5}
\Text(30,50)[c]{$\mr{H_0}$}
\Text(170,50)[c]{$\mr{H_0}$}
\Text(100,103)[c]{$\mr{S}$}
\end{picture}
\end{center}
\captionB{Contribution to the Higgs mass from an extra scalar field.}
	{Contribution to the Higgs mass from an extra scalar field.}
\label{fig:higgs2}
\end{figure}

\item{Supersymmetric Grand Unified Theories.}

  The idea in grand unified theories is that at some high scale the strong
  and electroweak forces can be unified into a single gauge theory, the
  simplest example being $SU(5)$. This is supported by the running of the
  couplings measured by the LEP 
  experiments. If we evolve the couplings from their values at current day 
  collider energies the strong, electromagnetic and weak couplings
  seem to unify at about $10^{16}\, \mr{\gev}$. 
  However, in the Standard Model these couplings do
  not quite unify. If we introduce new particles into the spectrum, as in
  supersymmetry, this will change the evolution of the couplings with
  energy and in a supersymmetric theory the couplings do indeed unify. 
  While this may seem to favour SUSY as an extension of the Standard Model,
  the unification of the couplings can also be achieved
  by the addition of other kinds of new particles into the spectrum.

\end{enumerate}

  While there are many theoretical reasons,
 \ie the Coleman-Mandula theorem and the hierarchy
  problem, and perhaps even the suggestion that we need supersymmetry, or
  some other new physics if we want the couplings to unify, there is no
  direct experimental evidence for supersymmetry. Supersymmetry is a very
  elegant theory but this does not mean that it is realized in nature.
  However, as one possible extension of the Standard Model, it  has a number
  of attractive theoretical features and we must investigate whether it is
  indeed a symmetry of nature.

\subsection{Construction of Supersymmetric Lagrangians}

  In principle we can start from the particle content of the theory 
  and write the most general Lagrangian consistent with the symmetries and 
  renormalizablity. However for supersymmetric theories it is easier to
  construct the supersymmetrically-invariant Lagrangian using the superfield
  formalism rather than the basic bosonic and fermionic fields.
  A more detailed description of this formalism can be found in, for
  example, \cite{Haber:1985rc,Bailin:1994qt,Wess}.

  The momentum operator and the generators, $Q$ and $\bar{Q}$, of the
  SUSY transformations form a sub-group of the extended Poincar\'{e} group. 
  This allows us to construct a function
  $S(x^\mu,\tht,\bath)$, the superfield,
  which is a linear representation of this sub-group. The change in this
  function induced by the action of a member of the sub-group,
\begin{equation}
 G(a^\mu,\xi,\bar{\xi}) = \exp i(\xi Q+\bar{\xi}\bar{Q}-a^\mu P_\mu),
\end{equation} 
  is generated by 
\begin{subequations}
\begin{eqnarray}
    P_\mu  & = & i\dbm, \\
    Q_\al  & = & -i \frac{\partial}{\partial\tht^\al}
                 -\sigma^\mu_{\al\dot{\al}}\bath^{\dot{\al}}\dbm, \\
\bar{Q}_{\dot{\al}}  & = &
	\phantom{-} i \frac{\partial}{\partial\bath^{\dot{\al}}}
                 +\tht^{\al}\sigma^\mu_{\al\dot{\al}}\dbm,
\end{eqnarray}
\end{subequations}
  where $\xi$, $\bar{\xi}$, $\tht$ and $\bath$ are anti-commuting Grassmann
  variables which transform as Weyl spinors. Further derivatives which
  anti-commute with the generators can also be defined
\begin{subequations}
\begin{eqnarray}
 D_\al			 & = & \phantom{-}\frac{\partial}{\partial\tht^\al}
                 +i\sigma^\mu_{\al\dot{\al}}\bath^{\dot{\al}}\dbm, \\
 \bar{D}_{\dot{\al}} 	 & = & - \frac{\partial}{\partial\bath^{\dot{\al}}}
                 -i\tht^{\al}\sigma^\mu_{\al\dot{\al}}\dbm.
\end{eqnarray}
\end{subequations}
  The general superfield,  $S(x^\mu,\tht,\bath)$, is a reducible 
  representation of the supersymmetric algebra. However, we can obtain
  irreducible representations by imposing further conditions:
\begin{subequations}
\begin{eqnarray}
 \bar{D}_{\dot{\al}}S  & = & 0 
	 \,\,\,\,\,\,\,\,\,\mr{chiral \  superfield};\\
  S^\dagger 	       & = & S 
	 \,\,\,\,\,\,\,\,\mr{vector \  superfield}. 
\end{eqnarray}
\end{subequations}
  It should be noted that because $\tht$ and $\bath$ are two-component
  Grassmann variables
  the expansion of the superfield as a power series in $\tht$ and $\bath$
  cannot involve terms with more than two powers of $\tht$ or $\bath$.  
  The chiral superfield can be written as an expansion in terms of the
  Grassmann variables $\tht$ and $\bath$ giving
\begin{eqnarray}
   \Phi(x^\mu,\tht,\bath) &=& \phi(x) + \rtt\tht\psi(x)+\tht\tht F(x)
				+i\dbm\phi(x)\tht\sigma^\mu\bath\nonumber\\
			&&-\frac{i}{\rtt}\tht\tht\dbm\psi(x)\sigma^\mu\bath
			-\frac1{4}\partial^\mu\dbm\phi(x)\tht\tht\bath\bath.
\end{eqnarray}
 This superfield includes a Weyl spinor, $\psi$, and a complex scalar field,
  $\phi$. The field $F$ is an auxiliary field which can be eliminated using
  the equations of motion. The component fields transform in the following
  way under the SUSY transformation:
\begin{subequations}
\begin{eqnarray}
\delta \phi &=& \rtt \xi \psi; \\
\delta \psi &=& \rtt \xi F +i\rtt\dbm\phi\sigma^\mu\bar{\xi};\\
\delta F    &=& i\rtt\dbm\psi\sigma^\mu\bar{\xi}. \label{eqn:fterm}
\end{eqnarray}
\end{subequations}
  This is the left chiral superfield, the right chiral superfield can be
  obtained by taking the hermitian conjugate. We refer to the coefficient
  of the $\tht\tht$ term  as the $F$-term. Eqn.\,\ref{eqn:fterm} shows that
  the change, under the  SUSY transformations, of the $F$-term
  is a total derivative. Hence the $F$-term is suitable for use
  as a supersymmetrically-invariant Lagrangian.

  Similarly, the vector superfield can be expanded in powers of $\tht$
  and $\bath$,
\begin{eqnarray}
 S(x^\mu,\tht,\bath) 	& = & C(x)+i\tht\chi(x)-i\bath\bar{\chi}(x)
				+\frac{i}{2}\tht\tht\left[M(x)+iN(x)\right]
				 \nonumber \\
			&   & -\frac{i}{2}\bath\bath\left[M(x)-iN(x)\right]
				+\tht\sigma^\mu\bath V_\mu(x) \nonumber \\
			&   & +i\tht\tht\bath\left[\bar{\lam}(x)
			+\frac{i}{2}\bar{\sigma}^\mu\dbm\chi(x)\right]
			        -i\bath\bath\tht\left[\lam(x)
			+\frac{i}{2}\sigma^\mu\dbm\bar{\chi}(x)\right] 
				\nonumber \\
			&   & +\frac1{2}\tht\tht\bath\bath\left[D(x)
				-\frac1{2}\dbm\partial^\mu C(x)\right]\!,
\end{eqnarray}
  where the real 
  scalar fields $C$, $M$, $N$ and the Weyl fermion $\chi$
  can be eliminated by a SUSY gauge transformation leaving the physical
  degrees of freedom, \ie~the gauge field $V_\mu$ and its superpartner
  gaugino field $\lam$, and the auxiliary field $D$. The component fields
  transform in the following way under the SUSY transformations:
\begin{subequations}
\begin{eqnarray}
 \delta C 	& = & i\left(\xi\chi-\bar{\xi}\bar{\chi}\right)\!; \\
 \delta\lam_\al & = & -iD\xi_\al
			-\frac12{(\sigma^\mu\bar{\sigma}^\nu)_\al}^\be\xi_\be
			\left(\dbm V_\nu-\partial_\nu V_\mu\right)\!;\\
 \delta V^\mu   & = & i\left(\xi\sigma^\mu\bar{\lam}
			-\lam\sigma^\mu\bar{\xi}\right)
			-\partial^\mu\left(\xi\chi+
			\bar{\xi}\bar{\chi}\right)\!; \\
 \delta D	& = & \dbm\left(-\xi\sigma^\mu\bar{\lam}
				+\lam\sigma^\mu\bar{\xi}\right)\!.
\end{eqnarray}
\end{subequations}
  Here the variation of the coefficient of the $\tht\tht\bath\bath$ term,
  the $D$-term, is a total derivative and hence this can also be used as a
  supersymmetrically-invariant Lagrangian.

\textheight 24.0cm 
  Using these superfields we can construct supersymmetric Lagrangians. In
  the  supersymmetric extension of the Standard Model we will use left 
  chiral superfields to represent the left-hand components of the
  Standard Model fermions, together with their superpartners,
  and right chiral superfields to represent the right-hand components,
  and their superpartners. The gauge 
  bosons are represented by vector superfields. 

  Here we will only consider  how to construct the Lagrangian for the chiral
  superfields as these are the terms in the supersymmetric Lagrangian we are
  interested in. This can be done by taking products of chiral superfields. 
  In particular the product of two left chiral superfields is also a left
  chiral superfield. Hence the $F$-term of a product of left chiral
  superfields can be used to give a suitable
  term in the Lagrangian. The product of a left and a right chiral superfield
  gives a vector superfield. The $D$-term of the 
  product of a left and a right chiral superfield can therefore
  also be used to give a term in the Lagrangian.
  The simplest example of this is a
  single  left chiral superfield. We can form the product of this field 
  with its hermitian conjugate and take the $D$-term. This gives
\begin{equation}
 [\Phi\Phi^\dagger]_{\tht\tht\bath\bath} = 
	FF^\dagger + \dbm\phi^*\partial^\mu\phi
	  +i\psb\bar{\sigma}^\mu\partial_\mu\psi,
\end{equation} 
  which, after eliminating the auxiliary field $F$ using the equations of
  motion, is the Lagrangian given in Eqn\,\ref{eqn:SUSYleq}. Therefore using
  the $D$-term of the product of the superfield and its hermitian 
  conjugate we can form the kinetic term for
  the fermionic field and its superpartner.
  We can add interaction and mass terms to this theory by taking products of
  the left chiral superfields. This can be done by forming the superpotential
  for the theory. For example, in a theory with only one chiral superfield,
\begin{equation}
  {\bf W}(\Phi) = \frac{m}{2} \Phi\Phi + \frac{\lam}{3} \Phi\Phi\Phi.
\end{equation}
  In general we can only include terms which are at most cubic in the
  superfields in order for the theory to be renormalizable. This gives
  the interaction Lagrangian
\begin{eqnarray}
\mathcal{L} & = & [W(\Phi)]_{\tht\tht} + \mr{h.c.}, \\
	    & = & m( \phi F -\frac{1}{2}\psi\psi) + 
		  \lam(\phi^2 F-\phi\psi\psi) +\mr{h.c.}.
\end{eqnarray}
  We can then write the full Lagrangian for this theory and use the
  equations of motion to eliminate the auxiliary field F,
\begin{equation}
  F^\dagger = -m\phi -\lam\phi^2,
\end{equation}
  giving the result
\begin{eqnarray}
\mathcal{L} & = &	 [\Phi\Phi^\dagger]_{\tht\tht\bath\bath}+ 
			\left([W(\Phi)]_{\tht\tht} + \mr{h.c.}\right)\!, \\
 	    & = & \dbm\phi^*\partial^\mu\phi 
		 +i\psb\bar{\sigma}^\mu\partial_\mu\psi
		  -|\lam\phi^2+m\phi|^2
	-\left(\frac{m}{2}\psi\psi+\lam\psi\psi\phi +\mr{h.c.}\right)\!.
\end{eqnarray}{\nopagebreak
  By taking the relevant combinations of the chiral superfields, the kinetic 
  terms and the interactions of the chiral fields with each other can be
  constructed. In general the superpotential gives the Yukawa-type
  interactions and part of the scalar potential of the theory.
  Similarly the kinetic energy terms of the gauge fields and their
  supersymmetric
  partners  can  be constructed,  as well as the interactions of the vector
  and chiral superfields. In the  next section we will construct a
  supersymmetric extension of the
  Standard Model using these superfield techniques.}

\textheight 23.0cm 
\subsection{The Minimal Supersymmetric Standard Model}

  We can construct the minimal supersymmetric extension of the Standard Model
  by including the superpartners of the Standard Model particles and then
  using the
  superfield formalism to construct the most general renormalizable
  supersymmetrically- and gauge-invariant
  Lagrangian.
  In the supersymmetric version of the Standard Model we cannot use
  the same Higgs boson to give mass to both the up- and down-type quarks.
  Hence we need two $SU(2)_L$ doublet Higgs fields to give all the particles
  in the theory
  mass. This gives the particle content in Table~\ref{tab:superfield}.

\begin{table}
\begin{center}
\begin{tabular}{|cc|cc|c|c|c|}
\hline
\multicolumn{2}{|c|}{Superfields} &
 Bosonic Fields & Fermionic Fields & $SU(3)_C$ & $SU(2)_L$ & $Y$ \\
Field & Type & & & & & \\
\hline
\multicolumn{2}{|c|}{Gauge Multiplets} & & & & & \\\cline{1-2}
 $G^a$ & Vector       & Gluons & Gluinos & Octet & Singlet &
						 $\phantom{-}0$ \\
 $W^a$ & Vector       & $W$ & Winos & Singlet & Triplet    &
						 $\phantom{-}0$ \\
 $B^a$ & Vector       & $B$ & Bino & Singlet & Singlet     &
						 $\phantom{-}0$ \\
 \hline
\multicolumn{2}{|c|}{Matter Multiplets} & & & & & \\ \cline{1-2}
 $L_i$ & Left Chiral  & $(\nut_L,\elt^-_L)$ & $(\nu_L,\ell_L)$ &
				 Singlet & Doublet & $-1$\\
 $E_i$ & Right Chiral & $\elt^-_R$        & $\ell_R$           &
				 Singlet & Singlet & $-2$ \\
 $Q_i$ & Left Chiral  & $(\upt_L,\dnt_L)$ & $(u_L,d_L)$    &
				 Triplet & Doublet & $\phantom{-}1/3$\\
 $U_i$ & Right Chiral & $\upt_R$ & $u_R$ & 
				 Triplet & Singlet & $\phantom{-}4/3$ \\
 $D_i$ & Right Chiral & $\dnt_R$ & $d_R$ &
				 Triplet & Singlet & $-2/3$  \\
\hline
\multicolumn{2}{|c|}{Higgs Multiplets} & & & & & \\\cline{1-2}
 $H_1$ & Left Chiral & $(H^1_1,H^2_1)$ & $(\hgt^0_1,\hgt^-_1)_L$ &
 					 Singlet & Doublet & $-1$\\
 $H_2$ & Left Chiral & $(H^1_2,H^2_2)$ & $(\hgt^+_2,\hgt^0_2)_L$ &
 					 Singlet & Doublet & $\phantom{-}1$\\
\hline
\end{tabular}
\end{center}
\captionB{Superfields in the MSSM.}
{Superfields in the Minimal Supersymmetric Standard Model (MSSM). The
 subscript $i=1,2,3$ gives the generation of the matter fields.}
\label{tab:superfield}
\end{table}

  The Standard Model is extended by including the superpartners of all the 
  Standard Model particles. 
  As with the Standard Model we should write down the most general Lagrangian
  consistent with the symmetries of the theory and renormalizablity.
  This means we should write down the most general renormalizable 
  supersymmetrically-invariant theory
  consistent with the gauge symmetries. The most general renormalizable
  gauge-invariant superpotential is given by
\begin{equation}
  {\bf W} = {\bf W_{\mr{MSSM}}} + {\bf W_{\not R_p}},
\end{equation}
 where
\begin{eqnarray}
 {\bf W_{\mr{MSSM}}} &=&-h^E_{ij}\varepsilon_{ab}L^a_iH_1^b\overline{E}_j
			-h^D_{ij}\varepsilon_{ab}Q^a_iH_1^b\overline{D}_j
			+h^U_{ij}\varepsilon_{ab}Q^a_iH_2^b\overline{U}_j
			+\mu\varepsilon_{ab} H_1^aH_2^b,
\label{eqn:MSSMsuper}\\
{\bf W_{\not R_p}} &=&
\frac{1}{2}\lam_{ijk}\varepsilon^{ab}L_{a}^{i}L_{b}^{j}\overline{E}^{k}
+ \lam_{ijk}'\varepsilon^{ab}L_{a}^{i}Q_{b}^{j}\overline{D}^{k} +
\frac{1}{2}\lam_{ijk}''\varepsilon^{c_1c_2c_3}\overline{U}_{c_1}^{i}
\overline{D}_{c_2}^{j}\overline{D}_{c_3}^{k} +
 \kappa_i\varepsilon^{ab}L_a^{i}H_b^2, \nonumber \\\label{eqn:Rsuper1} 
\end{eqnarray}
  using the superfields given in Table~\ref{tab:superfield}. Here
  $a,b=1,2$ are the $SU(2)_L$  indices,\linebreak $c_1,c_2,c_3=1,2,3$
  are the $SU(3)_C$ indices and $i,j,k=1,2,3$ are the
  generations of the matter fields. As we can only
  include left chiral superfields in the superpotential we must take the
  hermitian conjugate, denoted with a bar, of the right chiral superfields.
  Another consequence of the requirement that we can only use left chiral
  superfields in the superpotential is that we cannot use the conjugate of
  the Higgs field to give
  mass to both the up- and down-type quarks, as in the Standard Model, and 
  we therefore need two
  Higgs doublets to give mass to both types of quark.
  The first term in this superpotential, ${\bf W_{\mr{MSSM}}}$, gives the
  Yukawa terms of the Standard Model and the additional terms required by
  supersymmetry. However
  the second term\footnote{It should be noted that some authors choose to
  define this superpotential without the factors of one half in the LLE
  and UDD terms. This will lead to differences in the Feynman rules, but
  the results with this second convention can always be obtained by
  taking $\lam$ or $\lam''$ to be twice their value in our convention.}
% End of footnote 
 ${\bf W_{\not R_p}}$ gives additional interactions.
  A recent summary of
  the bounds on the couplings in Eqn.\,\ref{eqn:Rsuper1} can be found in 
  \cite{Allanach:1999ic}. This superpotential gives, for example from the
  third term, interactions of a squark and two quarks or of four squarks.
  When combined with the MSSM superpotential, ${\bf W_{\mr{MSSM}}}$, there
  are also terms involving the interactions of three squarks/sleptons with
  a Higgs boson. The presence of this second term in the superpotential, 
  ${\bf W_{\not R_p}}$,
  leads to a problem with the supersymmetric extension of the
  Standard Model. If we consider the second
  two terms of Eqn.\,\ref{eqn:Rsuper1} the proton can decay by the
  process shown in Fig.\,\ref{fig:protondecay}.
%
% Proton Decay Figure
%
\begin{figure}
\begin{center}
\begin{picture}(180,100)(0,40)
\SetScale{1.0}
\SetOffset(40,25)
\Line(-20,90)(5,90)
\ArrowLine(5,90)(60,90)
\Curve{(-20,15)(5,26)}
\ArrowLine(5,26)(60,52)
\Curve{(-20,85)(5,78)}
\GOval(-20,52.5)(37.5,4)(0){0.7}
\ArrowLine(-100,52.5)(-20,52.5)
\ArrowLine(5,78)(60,52)
\Curve{(60,90)(90,91)(170,110)}
\DashArrowLine(100,52)(60,52){5}
\ArrowLine(160,26)(100,52)
\ArrowLine(170,95)(100,52)
\Text(135,67)[]{$\mr{\bar{u}}$}
\Text(30,73)[]{$\mr{d}$}
\Text(170,25)[]{$\mr{e^{+}}$}
\Text(30,30)[]{$\mr{u}$}
\Text(30,97)[]{$\mr{u}$}
\Text(140,107)[]{$\mr{u}$}
\Text(185,105)[]{$\mr{\pi^{0}}$}
\Text(80,60)[]{$\mr{\tilde{s_{R}}^{*}}$}
\Text(-60,45)[c]{Proton}
\Vertex(60,52){1}
\Vertex(100,52){1}
\GOval(170,103)(8,4)(0){0.7}
\end{picture}
\end{center}
\captionB{\rpv\  proton decay.}
	{\rpv\  proton decay, $\mr{p\ra\pi^0e^+}$.}
\label{fig:protondecay}
\end{figure}
% End of proton decay figure

  We can use the experimental limit on the
  proton lifetime \cite{Caso:1998tx}
\begin{equation}
\tau\left(\mr{Proton}\longrightarrow e^+\pi^0\right) > 10^{32} \,\mr{yr},
\label{eqn:proton}
\end{equation}
  to obtain a bound on the product of the couplings at  the two vertices in
  Fig.\,\ref{fig:protondecay}, as a function of the exchanged squark mass,
\begin{equation}
  {\lam'}_{11k}{\lam''}_{11k}\lesssim 2\times 10^{-27}
  \left(\frac{M_{\mr{\dnt}_{kR}}}{100\, \mr{\gev}}\right)\!.
\end{equation}
  Hence as we require that $M_{\mr{\dnt}_{kR}} \lesssim 1 \, \mr{TeV}$ to
  solve the hierarchy problem the only natural way to satisfy this bound
  is to have one of these couplings set to zero. The standard way
  to achieve this is to introduce a new multiplicatively conserved
  quantum number called R-parity which is defined as \cite{Farrar:1978xj}
\begin{equation}
\mr{R_p} = (-1)^{3\mr{B}+\mr{L}+2\mr{S}}, 
\end{equation}
  where L is the lepton number, B is the baryon number and S is the spin of
  the particle. This new quantum number is $+1$ for the Standard Model
  particles and $-1$ for their SUSY partners. This prevents
  proton decay by forbidding all the terms in
  Eqn.\,\ref{eqn:Rsuper1} but not the terms in Eqn.\,\ref{eqn:MSSMsuper}.
  The conservation of R-parity in addition to
  the symmetries, \linebreak[3] 
  \ie~supersymmetry and the Standard Model gauge symmetries,
  and particle content defines the Minimal Supersymmetric
  Standard Model (MSSM). 
  Another symmetry which has the same effect, but is easier to see at the
  superfield level, is called matter parity. Here we change the sign of the
  matter, \ie quark and lepton, superfields but not the Higgs or gauge
  superfields
\begin{eqnarray}
 (Q_i,\bar{U}_i,\bar{D}_i,L_i,\bar{E}_i) \rightarrow
   - (Q_i,\bar{U}_i,\bar{D}_i,L_i,\bar{E}_i),\ \ \ \ \ &
 (H_1,H_2) \rightarrow (H_1,H_2).
\end{eqnarray}
  This also forbids all the terms in Eqn.\,\ref{eqn:Rsuper1} but none of
  the terms in Eqn.\,\ref{eqn:MSSMsuper}.

  The Lagrangian of the theory can then be specified by taking the relevant
  combinations of the superfields and extracting the terms which are
  invariant under supersymmetry. In general these Lagrangians are given in 
  \cite{Haber:1985rc,Gunion:1986yn}.
  The Feynman rules for this new theory can be
  found in \cite{Haber:1985rc,Gunion:1986yn} and the Feynman rules for
  those interactions that will be important in this thesis are given
  in a more general form in Appendix~\ref{chap:Feynman}. 

  The problem is that in a theory with unbroken supersymmetry
  the Standard Model particles and their superpartners would have the same
  mass. However the superpartners of the known fundamental particles have
  not been detected experimentally. Thus if supersymmetry
  is realized in nature it must be broken.

  There are mechanisms, based on either the $F$- or $D$-terms developing a
  non-zero
  vacuum expectation value, which spontaneously break supersymmetry. However
  in these models there are generally mass sum rules, for example, the
  supertrace in models with \linebreak 
  $F$-term supersymmetry breaking \cite{Ferrara:1979wa} ,
\begin{equation}
     \mr{Str}\,M^2\equiv\sum_{J}(-1)^{2J}(2J+1)m^2_J=0,
\end{equation}
  where the supertrace, $\mr{Str}\,M^2$, denotes the trace of the
  mass-squared
  matrix over the real fields, of spin J. This formula can be modified by 
  radiative corrections.
  However at tree level it  
  implies that while one of the superpartners of the Standard Model
  fermions would be heavier than the fermion the other would be lighter. This
  is not observed. Given this tree-level result, realistic models of broken
  supersymmetry based on either $F$- or $D$-term supersymmetry breaking
  are difficult to construct.

  The approach which is taken in the MSSM is to add by hand all terms that 
  violate supersymmetry in such a way that the quadratic divergences, \eg in
  the Higgs mass, are not reintroduced. This is called soft supersymmetry
  breaking. However soft SUSY breaking is ``ad-hoc'' and leads to over one
  hundred additional
  parameters \cite{Haber:1997if}. Given the number of free
  parameters which must be specified it is also 
  of limited use in experimental searches.

  Another approach is to break supersymmetry in some
  ``hidden-sector'' which only couples to the MSSM fields via either
  non-renormalizable operators \cite{Nilles:1984ge,Bailin:1994qt} or loop
  diagrams
  \cite{Dine:1982gu}.

  In general the supersymmetry breaking in the ``hidden-sector'' can be due
  to either $F$-term supersymmetry breaking or the production of gaugino
  condensates \cite{Nilles:1982ik:Ferrara:1983qs}.
  The two types of models differ in how this supersymmetry breaking is
  transmitted
  from the ``hidden-sector'' to the ``visible-sector'', \ie the MSSM fields.

  In supergravity models the supersymmetry breaking is transmitted to the
  ``visible-sector'' through gravitational interactions, represented by
  non-renormalizable terms suppressed by inverse powers of the Planck mass. 
  In gauge-mediated models
  of supersymmetry breaking new vector fields are introduced which transmit 
  the supersymmetry breaking between the ``hidden-'' and ``visible-sectors''.

  In this thesis we will not be concerned with the details of the
  supersymmetry-breaking mechanism. While in principle we could perform all
  our analyses within the framework of the MSSM, \ie by specifying all the
  soft SUSY-breaking parameters, it is much easier to work with a smaller
  set of parameters. We will therefore 
  use the standard supergravity (SUGRA) scenario
  where the soft SUSY-breaking masses for the
  gauginos ($M_{1/2}$) and scalars ($M_0$), and the trilinear SUSY-breaking
  terms ($A_0$) are universal at the GUT scale. In addition we will require
  that electroweak symmetry is radiatively broken \cite{Ibanez:1982fr},
  \ie the Higgs mass squared 
  becomes negative due to its renormalization group evolution.
  This gives five parameters: the soft masses and trilinear SUSY-breaking
  terms $M_{1/2}$, $M_0$, $A_0$, the ratio of the Higgs vacuum expectation
  values of
  the two Higgs doublets $\tan\beta$ and $\sgn\mu$. The absolute value of the
  $\mu$ term in Eqn.\,\ref{eqn:MSSMsuper} is fixed by the requirement of
  radiative electroweak symmetry breaking. 

\section{R-parity Violating Supersymmetry} \label{sect:intoRPV}
 
  In the Standard Model the
  Lagrangian is constructed by writing down all the terms consistent with
  the gauge symmetries, renormalizablity and the particle content of the
  theory. The discrete symmetries, \ie lepton and baryon number, then emerge
  as consequences of the other symmetries of the theory. However in the
  construction of the MSSM we were forced to impose a new discrete symmetry,
  \rp,  to prevent the decay of the proton.

  There is no reason  to impose
  R-parity as a symmetry, all that is required is a symmetry such that either
  the second or third terms in Eqn.\,\ref{eqn:Rsuper1} are forbidden. We
  can achieve this for example by imposing baryon parity,
\begin{eqnarray}
 (Q_i,\bar{U}_i,\bar{D}_i) \rightarrow - (Q_i,\bar{U}_i,\bar{D}_i),\ \ &
 (L_i,\bar{E}_i,H_1,H_2) \rightarrow (L_i,\bar{E}_i,H_1,H_2),
\end{eqnarray}
  which prevents the third term in Eqn.\,\ref{eqn:Rsuper1}. This
  leads to lepton number violation but no baryon number violation and hence
  prevents the decay of the
  proton. Similarly there are symmetries, \eg lepton parity
\begin{eqnarray}
 (L_i,\bar{E}_i) \rightarrow - (L_i,\bar{E}_i),\ \ &
 (Q_i,\bar{U}_i,\bar{D}_i,H_1,H_2) \rightarrow
	 (Q_i,\bar{U}_i,\bar{D}_i,H_1,H_2),
\end{eqnarray}
  which forbids the second term
  in Eqn.\,\ref{eqn:Rsuper1} giving baryon number violation with no lepton
  number violation and preventing the decay of the proton.

  As we are imposing, by hand, a new multiplicative symmetry on the
  theory there is no reason to favour imposing either R-parity or,
  for example, baryon parity. As there is no reason for favouring either
  of these models, if we are to search for supersymmetry both must be
  studied. This is particularly important as there are major differences
  in the experimental signatures of these processes.

  The conservation of \rp\  in the MSSM gives a number of effects:
\begin{itemize}
\item as the initial state in any collider experiment contains only
   Standard Model particles SUSY particles must be produced in
   pairs;
\item the lightest supersymmetric particle (LSP) is stable;
\item cosmological bounds on electric- or colour-charged stable
 relics imply that a stable LSP must be a neutral colour singlet
 \cite{Ellis:1984ew}.
\end{itemize}
  This leads to a classical signature for supersymmetry in all collider
  experiments. As the lightest SUSY particle is stable, any SUSY
  particle produced will tend to cascade decay to the LSP. The LSP,
  which is weakly interacting, will then escape from the
  detector without interacting giving missing energy and
  momentum. This type of signature is part of most search channels
  which have been used by experiments to look for \rp-conserving SUSY.

  However when \rp\  is violated the following can happen:
\begin{itemize}
\item single sparticles can be produced;
\item the LSP can decay to Standard Model particles;
\item the LSP can be any SUSY particle and need not be neutral;
\item either lepton or baryon number is violated.
\end{itemize}

  This leads to different approaches for searching for the experimental
  signatures of these 
  models. In particular as the LSP can decay, depending on the \rpv\ 
  couplings, inside the detector the standard missing energy and
  momentum signatures of the MSSM no longer exist. In general there have been
  two types of studies of these models: the first has studied the production
  of sparticle pairs by \rp-conserving processes followed by \rpv\  decays, 
  usually of the LSP; the second has studied the possibility of single
  sparticle production via \rpv\  processes.

  There has been a great deal of interest in these models in recent years
  motivated by
  the possible explanations of various experimental discrepancies, \eg
  \cite{Grant:1996bc,Dreiner:1996dd,Carena:1997xu,Chankowski:1996mx,
	Choudhury:1996gd,
	Dreiner:1997cd,Altarelli:1997ce,Choudhury:1997dt,Kalinowski:1997fk}.
  It has become clear that if we are to explore all possible channels
  for the discovery of supersymmetry then \rpv\    models must be
  investigated. Recent reviews of R-parity violating models can be found in
  \cite{Dreiner:1997uz:Bhattacharyya:1997vv}.
  We will now briefly discuss the phenomenology of these two different
  scenarios and review the studies which have been made.

\subsection{Sparticle Pair Production}

  In this case the sparticles are produced by MSSM processes, hence the
  production cross sections only depend on the parameters of the MSSM.
  There are a number of possible scenarios
  depending on the size of the \rpv\  couplings and the lifetime of the
  lightest supersymmetric particle:
\begin{itemize}
\item 	For large values of the \rpv\  couplings it is possible that
	particles other than the LSP will have significant branching ratios
        for \rpv\  decay modes.
\item 	For small values of the \rpv\  couplings the main effect is
  	the decay of the LSP. The experimental situation will  depend on
        the lifetime of the LSP:
	\begin{enumerate}
	\item If the lifetime of the LSP is such that it is stable on
	      collider time scales it can escape the detector before it
	      decays. The experimental search strategy depends on the nature
	      of the LSP. If it is the lightest neutralino the searches are
	      identical to those for the MSSM.
	      However if the LSP is charged it can be also be detected. 
	      In this case, it would look
	      like a muon in the detector apart from losing energy due to
	      ionization
	      at a different rate, due to the different mass. There have
	      been a 
	      number of searches for these heavy stable charged particles 
              \cite{Abe:1989es}.
	\item The LSP lifetime can be such that while it decays inside the
	      detector it can travel a significant distance away from the
	      primary interaction point before decaying. There has been
	      little study of this case.
	\item If the LSP lifetime is sufficiently short it will decay at the
 	      primary interaction point. This is the case which has been
	      most studied. 
	\end{enumerate}
\end{itemize}

  The signatures in the case where the LSP decays inside the detector depend
  on whether lepton or baryon number is violated and we will briefly
  consider both of these cases:
\begin{itemize}

% First the lepton number violating scenario
  \item If lepton number is violated the typical experimental signature now
  	involves large numbers of high transverse momentum leptons in the
  	final state which can be easily detected by collider experiments. In
  	some ways more information can potentially be
  	extracted in these models experimentally. As the decay products of
        the LSP can be detected it is possible that the LSP mass can be
  	reconstructed. This is impossible in the MSSM because the LSP
        escapes the detector.

	There have been a number of searches for these processes by both the
	LEP \cite{Acton:1993xj,Abbiendi:1998ff,Abbiendi:1999is,
	       Keranen:1997br:Abreu:1999qz,Buskulic:1995az,Buskulic:1996uj,
	       Barate:1997ra,Barate:1998gy,Barate:1999fc,Acciarri:1999yk}
	 and Tevatron 
	\cite{Wu:1997yk:Chertok:1998ri:Abe:1998gu,Abbott:1999nh} 
	experiments, and studies of the range of
	parameters which can be discovered by the LHC \cite{Mirea:1999fs}.
% Second the baryon number violating scenario
  \item If, however, baryon number is violated the situation is much worse
	and this is considered a potential ``worst case'' scenario for the
  	discovery of SUSY. In this case instead of the clean missing
 	transverse momentum signatures of the MSSM the LSP will decay giving
	jets of hadrons. Extracting this signal of SUSY from the QCD
	backgrounds in hadron colliders will prove to be a challenging
	experimental problem.
	These models have been much less studied experimentally.
\end{itemize}

\subsection{Single Sparticle Production}

  There have been many theoretical and experimental studies of single
  sparticle
  production. The cross sections for these processes depend on the \rpv\  
  couplings, in addition to the parameters of the MSSM. The kinematic reach
  of these processes is typically twice that of sparticle pair production.
  While there have been
  some studies of non-resonant sparticle production \cite{Allanach:1997sa}
  most of the studies have been of resonant sparticle production.

\begin{figure}[b!]\begin{center}
\vskip 3mm
\renewcommand{\subfigcapmargin}{-10mm}
\subfigure[Resonant sneutrino production in ${\rm{e^+e^-}}$ collisions.]{
\begin{picture}(200,60)(-50,0)
\SetOffset(20,-20)
\ArrowLine(60,52)(5,26)
\ArrowLine(5,78)(60,52)
\DashArrowLine(60,52)(90,52){5}
\Text(0,26)[r]{\large$\mr{e^+}$}
\Text(0,78)[r]{\large $\mr{e^-}$}
\Text(100,52)[l]{\large $\mr{\tilde{\nu}_{L}}$}
\Vertex(60,52){1}
\end{picture}
\label{fig:resonantneut}}\\
\vskip 3mm
\subfigure[Resonant squark production in electron-proton collisions.]{
\begin{picture}(200,60)(-50,0)
\SetOffset(-80,-20)
\ArrowLine(60,52)(5,26)
\ArrowLine(5,78)(60,52)
\DashArrowLine(90,52)(60,52){5}
\Text(0,26)[r]{\large $\mr{\bar{d}}$}
\Text(0,78)[r]{\large $\mr{e^-}$}
\Text(100,52)[l]{\large $\mr{\tilde{u}^*_{L}}$}
\Vertex(60,52){1}
\SetOffset(120,-20)
\ArrowLine(5,26)(60,52)
\ArrowLine(5,78)(60,52)
\DashArrowLine(60,52)(90,52){5}
\Text(0,26)[r]{\large $\mr{u}$}
\Text(0,78)[r]{\large $\mr{e^-}$}
\Text(100,52)[l]{\large $\mr{\tilde{d}_{R}}$}
\Vertex(60,52){1}
\end{picture}
\label{fig:resonantsqu1}}\\
\vskip 3mm
\subfigure[Resonant slepton production in hadron--hadron collisions.]{
\begin{picture}(200,60)(-50,0)
\SetOffset(-80,-20)
\ArrowLine(60,52)(5,26)
\ArrowLine(5,78)(60,52)
\DashArrowLine(60,52)(90,52){5}
\Text(0,26)[r]{\large $\mr{\bar{d}}$}
\Text(0,78)[r]{\large $\mr{d}$}
\Text(100,52)[l]{\large $\mr{\tilde{\nu}_{L}}$}
\Vertex(60,52){1}
\SetOffset(120,-20)
\ArrowLine(5,26)(60,52)
\ArrowLine(60,52)(5,78)
\DashArrowLine(60,52)(90,52){5}
\Text(0,26)[r]{\large $\mr{\bar{u}}$}
\Text(0,78)[r]{\large $\mr{d}$}
\Text(100,52)[l]{\large $\mr{\tilde{e}_{L}}$}
\Vertex(60,52){1}
\end{picture}
\label{fig:resonantlep}}\\
\vskip 3mm
\subfigure[Resonant squark production in hadron--hadron collisions.]{
\begin{picture}(200,60)(-50,0)
\SetOffset(-80,-20)
\ArrowLine(5,26)(60,52)
\ArrowLine(5,78)(60,52)
\DashArrowLine(90,52)(60,52){5}
\Text(0,26)[r]{\large $\mr{d}$}
\Text(0,78)[r]{\large $\mr{u}$}
\Text(100,52)[l]{\large $\mr{\tilde{d}^*_{R}}$}
\Vertex(60,52){1}
\SetOffset(120,-20)
\ArrowLine(5,26)(60,52)
\ArrowLine(5,78)(60,52)
\DashArrowLine(90,52)(60,52){5}
\Text(0,26)[r]{\large $\mr{d}$}
\Text(0,78)[r]{\large $\mr{d}$}
\Text(100,52)[l]{\large $\mr{\tilde{u}^*_{R}}$}
\Vertex(60,52){1} 
\end{picture}
\label{fig:resonantsqu2}}
\vskip 3mm
\end{center}
\captionB{Different possible resonant sparticle production mechanisms.}
	{Different possible resonant sparticle production mechanisms.
	 While the produced sparticles can be of any generation the incoming
	 particles will usually be first generation.}
\label{fig:resonant}
\end{figure}

  The terms in Eqn.\,\ref{eqn:Rsuper1} lead to different resonant
  production mechanisms in various collider experiments, which are
  shown in Fig.\,\ref{fig:resonant}. The first term leads
  to resonant sneutrino production in $\mr{e^+e^-}$ collisions 
  \cite{Dimopoulos:1988jw,Barger:1989rk,Giudice:1996dm,Accomando:1997wt,
        Erler:1997ww,Kalinowski:1997bc}, Fig.\,\ref{fig:resonantneut}.
  The second term gives both resonant 
  squark production in $\mr{ep}$ collisions
  \cite{Butterworth:1993tc,Dreiner:1994tj,Dreiner:1997cd,Altarelli:1997ce,
	Kalinowski:1997fk}, Fig.\,\ref{fig:resonantsqu1},
  and 
  resonant slepton production in hadron--hadron collisions 
  \cite{Dimopoulos:1990fr,Kalinowski:1997zt,Hewett:1998fu,Allanach:1999bf,
	Moreau:1999bt:Moreau:2000ps:Moreau:2000bs,Abdullin:1999zp},
   Fig.\,\ref{fig:resonantlep}.
  The third term gives resonant
  squark production in hadron--hadron collisions 
  \cite{Dimopoulos:1990fr,Allanach:1999bf,Datta:1997us,
	Yang:1997uw,Oakes:1998zg,Berger:1999zt}, 
  Fig.\,\ref{fig:resonantsqu2}.

  In general the search strategies depend on the decay modes of the
  sparticle. Most of the studies have concentrated on the \rpv\  decay 
  modes of the resonant sparticles, although there have also been a number
  of studies of the gauge decay modes in various collider experiments.

\section{Summary}

  In this chapter we have introduced the concept of supersymmetry and the 
  theoretical reasons for favouring it as a possible extension of the
  Standard
  Model. We then argued that models in which \rp\  is conserved are no
  better motivated than those in which R-parity is violated and therefore
  both should be studied if we are to discover supersymmetry.

  Despite the interest in \rpv\  SUSY models,
  and the potential experimental problems,
  there have been few experimental studies particularly at hadron colliders.
  The
  first systematic study of \rpv\   signatures at hadron colliders was
  presented in \cite{Dreiner:1991pe}. More recent overviews of the
  search potential at the LHC and Run II of the Tevatron have been
  presented in \cite{Barbier:1998fe,Allanach:1999bf}. These studies have
  been limited by the fact that few simulations have been available. In
  hadron--hadron collisions the only available Monte Carlo event
  generator is ISAJET \cite{Baer:1999sp} where the \rpv\  decays can be
  implemented using the \texttt{FORCE} command, \ie the decay mode of a given
  particle, \eg the LSP, can be specified by hand.  However there has
  been no simulation which includes all the decay modes and the single
  sparticle production processes.

  We shall therefore present in this thesis the calculations required
  to produce a simulation of the \rpv\  processes and decays together
  with some results from these simulations looking at the possibility
  of detecting \rpv\  SUSY processes. We also look at the latest
  experimental anomaly which can be explained by \rpv\  SUSY.

  We present in Chapter~\ref{chap:monte} an introduction to
  the techniques of parton-shower Monte Carlo simulations followed by
  the calculations required to apply these techniques to \rpv\  SUSY.
  
  Chapter~\ref{chap:slepton} then uses these simulations to look at the
  production of resonant
  sleptons via \rpv\  SUSY in hadron--hadron colliders and some ways
  of detecting these processes.

  Chapter~\ref{chap:karmen} considers a possible \rpv\  SUSY
  explanation of the results of the KARMEN experiment.

%\chapter{Monte Carlo sims}
%%%%%%%%%%%%%%%%%%%%%%%%%%%%%%%%%%%%%%%%%%%%%%%%%%%%%%%%%%%%%%%%%%%%%%%%%%%%%%%
%							       	              %
%  Chapter on Monte-Carlo event generators.				      %
%									      %
%  This is an expansion of the introduction to monte Carlo given in the RPV   %
%  montecarlo paper followed by the angular ordering results from that        %
%  paper.                                                                     %
%  The hadronization results are also included as are some of the results on  %
%  angular ordering, basically the results showing the differences between the%
%  resonant slepton production and the background, and the resonant squark    %
%  production processes.						      %
%									      %
%  1. Start with basic idea of the monte Carlo				      %
%									      %
%  2. Explain the Monte-Carlo in detail					      %
%	a. parton-shower in particular angular ordering etc		      %
%	b. hadronization						      %
%	c. Brief summary of the monte-Carlo s				      %
%									      %
%  3. Extension of R-parity violation					      %
%	a. Explain angular ordering in R-parity				      %
%	b. Explain hadronization in R-parity				      %
%									      %
%  4. Conclude and explain results in next chapter			      %
%									      %
%%%%%%%%%%%%%%%%%%%%%%%%%%%%%%%%%%%%%%%%%%%%%%%%%%%%%%%%%%%%%%%%%%%%%%%%%%%%%%%

\chapter{Monte Carlo Simulations} \label{chap:monte}
%
%  Introduction on monte Carlo
%
\section{Introduction} \label{sect:monteintro}

  If we wish to examine the experimental signatures of any model of
  beyond the Standard Model physics
  we need a Monte Carlo event generator which includes the processes predicted
  by that model. This is particularly important in experimental studies so 
  that the
  effects of the cuts applied and the resolution of the
  detector can be included.
  There is currently only one Monte Carlo event generator,
  ISAJET \cite{Baer:1999sp}, which can simulate
  \rpv\  processes in hadron--hadron collisions. However these processes
  are only included in an ad-hoc way via the \texttt{FORCE} command
  which allows the decay modes of a given particle, for example the LSP,
  to be specified by hand. It does
  not contain any hard resonant \rpv\  production processes or a 
  calculation of the \rpv\  decay rates.
  If we wish to study these processes, in particular
  resonant sparticle production, we must add these processes to a Monte Carlo
  event generator. This chapter will start by discussing the physics used in
  Monte Carlo simulations and then go on to show how \rpv\  processes
  can be added.

  In principle it should be possible to calculate all the observables
  measured 
  experimentally using  perturbative QCD, or the electroweak theory if they
  do not involve particles which interact via the strong force.
  There are however  two problems with this approach:
\begin{enumerate}
  \item In practice most observables have been calculated to at most
  next-to-leading order in perturbative QCD. A few inclusive
  observables have been calculated to higher orders, \eg the cross section
  for $\mr{e^+}\mr{e^-}\ra\mr{hadrons}$. Given the
  complexity of QCD calculations beyond leading order it is unlikely
  that many higher orders will be calculated in the near future. Even
  the calculation of final states including more partons is very
  difficult, \eg the first general next-to-leading order calculation of
  three-jet
  observables in $\mr{e^+e^-}$ collisions \cite{Ellis:1980nc} was performed
  in 1980, while it is only recently that the four-jet calculations
  \cite{Glover:1997eh}
  were completed.
 
  \item An experiment observes hadrons not the quarks and gluons of
  a perturbative QCD calculation. It is currently impossible to
  calculate the hadronization process in QCD and we must use
  various phenomenological models with
  adjustable parameters which are fitted to data.
\end{enumerate}

  The idea of the Monte Carlo procedure is to provide a full
  description of the events which are seen in modern particle physics
  experiments. To do this we need some way of starting with
  a given hard process and obtaining the hadrons which are observed by
  the detector as a result of that process.
  This is done by considering the regions of phase space
  where the emission of QCD radiation is enhanced and taking these
  into account to all orders in perturbation theory. This leads to the
  idea of a parton shower where we take a parton at some high scale
  and then evolve this down to a lower scale with the emission of QCD
  radiation. Finally at some low scale typical of the hadronization
  process we resort to a non-perturbative model for the hadronization
  process to give the observed hadrons. 

  In general the Monte Carlo event generation process can be divided
  into main three phases:
\begin{enumerate}

  \item The hard process where the particles in the hard collision and
 	their momenta are generated, usually according to the leading-order
 	matrix element. This can be of either the incoming fundamental
   	particles in lepton collisions or of a parton extracted from a hadron
 	in hadron-initiated processes. In the example event shown in
	Fig.\,\ref{fig:monteeg} the hard process is $\mr{e^+e^-\ra q\bar{q}}$.

  \item The parton-shower phase where the coloured particles in the event are
  	perturbatively evolved from the hard scale of the collision to the
  	infrared cut-off. This is done for both the particles produced in
  	the collision, the final-state shower, and the initial partons
        involved in the collision for processes with incoming hadrons,
	the initial-state shower. This is shown by the gluon
	radiation in Fig.\,\ref{fig:monteeg}.
	The emission of electromagnetic radiation from charged
  	particles can be handled in the same way.
 
  \item A hadronization phase in which the partons left after the perturbative
 	evolution are formed into the observed hadrons. For
 	processes with hadrons in the initial state after the removal of the
 	partons in the hard process we are left with a hadron remnant.
        This remnant is also formed into hadrons by the hadronization model.
        In the example shown in Fig.\,\ref{fig:monteeg} the cluster model,
	which is used in HERWIG, is shown.
\end{enumerate}
%
% Monte Carlo Event generator picture
% 
%  This needs to be in black and white and probably be a lot better
%
\begin{figure}
\begin{center} \begin{picture}(360,230)(0,40)
\SetOffset(70,0)
%Hard Process
\ArrowLine(0,180)(30,150)
\ArrowLine(30,150)(0,120)
\Photon(30,150)(70,150){5}{5}
\ArrowLine(70,150)(130,210)
\ArrowLine(130,90)(70,150)
%PartonShower
\Gluon(130,210)(150,190){-5}{3}
\Gluon(130,90)(150,110){5}{3}
\Gluon(150,190)(170,210){-5}{3}
\Gluon(150,190)(170,170){-5}{2.5}
\ArrowLine(170,130)(150,110)
\ArrowLine(150,110)(170,90)
\ArrowLine(130,210)(170,250)
\ArrowLine(170,50)(130,90)
\ArrowLine(170,250)(185,265)
\ArrowLine(205,245)(170,210)
\ArrowLine(170,210)(185,195)
\ArrowLine(185,35)(170,50)
\ArrowLine(170,90)(205,55)
\ArrowLine(185,145)(170,130)
\ArrowLine(185,185)(170,170)
\ArrowLine(170,170)(185,155)
\GOval(195,255)(20,5)(45){0.7}
\GOval(195,45)(20,5)(-45){0.7}
\GOval(185,150)(10,3)(0){0.7}
\GOval(185,190)(10,3)(0){0.7}
%Labels
\Text(20,175)[]{$\mr{e^-}$}
\Text(20,130)[]{$\mr{e^+}$}
\Text(50,165)[]{$\mr{Z_0/\gamma}$}
\Text(100,190)[]{$\mr{q}$}
\Text(100,108)[]{$\mr{\bar{q}}$}
% Now some vertical lines to split up the diagram
%\DashLine(110,275)(110,25){5}
%\DashLine(165,275)(165,25){5}
%\Text(50,260)[]{Hard Process}
%\Text(132.5,260)[]{Parton}
%\Text(132.5,240)[]{Shower}
\end{picture}
\end{center}
\captionB{Example of a Monte Carlo event.}
	{Example of a Monte Carlo event.}
\label{fig:monteeg}
\end{figure}
%
%  End of the figure
%
  There are usually two additional stages, one between the parton-shower and
  hadronization phases and the other after the hadronization phase, which are
  conceptually less important but are necessary for a full simulation of
  a hard collision process. In these phases those particles
  which are produced, but
  are unstable, decay. These secondary decays are handled in different ways
  depending on whether the particle decays before or after 
  hadronization:\footnote{The stage at which the various secondary decays
  			  occur is different in the different event generators
			  and the procedure we discuss here is that adopted
			  in the HERWIG Monte Carlo event generator.}
\begin{enumerate}
\item Those particles which decay before hadronizing, \eg the top quark,
      are decayed before the hadronization phase.
      Any coloured particles produced in
      these decays are then evolved by the parton-shower algorithm. 
      The hadronization
      phase occurs after all such particles have been decayed. These decays 
      are usually performed according to a calculated branching ratio 
      and often, \eg in top
      decay, use a matrix element to give the momenta of the decay products.  
\item Those unstable hadrons which are produced in the hadronization phase
      must also
      be decayed. These decays are usually performed using the experimentally
      measured branching ratios and a phase-space
      distribution for the momenta of the decay products. It is at this stage
      that those unstable fundamental particles which are not coloured,
      and hence
      do not hadronize, are decayed. For example the W and Z boson
      decays occur at this
      stage of the event generation process, however the decays of the
      colourless
      SUSY particles are handled by the previous secondary decay stage because
      unstable coloured sparticles are often produced in these decays.
      Any coloured
      particles produced in these decays are then evolved according to the 
      parton-shower algorithm and hadronized. This procedure
      is repeated until
      all the unstable particles have been decayed.
\end{enumerate}

  We will describe the three main phases in some detail,
  concentrating on those used in the HERWIG event generator 
  \cite{HERWIG61} but mentioning the other available approaches.
  We then study the extension of these simulations to include  \rpv\  
  hard processes and decays. 
 
\section{Hard Processes}

  The first stage of the Monte Carlo event generator is to generate the
  momenta of the
  particles involved in the hard process. Usually these momenta are
  generated according
  to the leading-order cross section. As we will mainly be dealing with
  hadron--hadron 
  collisions we will look at the procedure here in more detail.

  The cross section, for example, for a two-to-two process in hadron--hadron
  collisions is given by
\begin{equation}
  \sigma =\int^1_0 dx_1 \int^1_0 dx_2 \int^1_{-1} 
	  d\!\cos\theta \int^{2\pi}_0 d\phi
  \sum_{ij} \frac{d\hat{\sigma}_{ij}}{d\Omega}
	\!\stackrel{{\displaystyle(\sh,\theta,\phi,\mu^2)}}{\ }
                       f_i(x_1,\mu^2)f_j(x_2,\mu^2),
\end{equation}
  where $\frac{d\hat{\sigma}_{ij}}{d\Omega}
	\!\stackrel{{\scriptstyle(\sh,\theta,\phi,\mu^2)}}{\ }$ is the
  differential cross section for
  the partons $i$ and $j$ to go to whatever final state we are
  interested in and $f_i(x,\mu^2)$ is
  the parton distribution function,
  \ie the probability of finding a parton $i$ with a fraction $x$ of
  the incoming 
  hadron's momentum. The parton distributions are also dependent on the
  factorization scale, $\mu$. Partons which are emitted from those partons
  involved in the hard collision and have transverse momenta below this scale
  are treated
  as part of the hadron structure whereas those with momenta above this scale
  are part of the hard collision process. In principle the cross section
  should not
  depend on this scale. If we only perform the calculation to leading order
  however
  there can be a sizable factorization-scale dependence. This dependence
  is usually
  reduced if higher-order corrections are included. 
  The cross-section integral can then be performed using the Monte
  Carlo method.

  The Monte Carlo procedure is based on the following result.
  For a simple one-dimensional integral,
\begin{equation}
  \int^{x_2}_{x_1} f(x) dx = (x_2-x_1) \langle f(x) \rangle.
\end{equation}
  The average, $\langle f(x) \rangle$, can be approximated by
  calculating $f(x)$ at $N$ 
  randomly chosen points, in the interval $(x_1,x_2)$, \ie
\begin{equation}
 \langle f(x) \rangle \simeq \frac1{N}\sum_{i=1}^N f(x_i) = \overline{f_N},
\end{equation}
  giving an estimate, $ \overline{f_N}$, of the average. This method
  is particularly
  useful as we can also calculate an error on the estimate by computing
  the standard
  deviation and applying the central limit theorem
\begin{equation}
  \langle f(x) \rangle =  \overline{f_N} \pm \frac{\sigma_N}{\sqrt{N}},
\end{equation}
  where $\sigma_N=\sqrt{  \overline{f^2_N}-\overline{f_N}^2}$ and
  $\overline{f^2_N}= \frac1{N}\displaystyle{\sum_{i=1}^N} f^2(x_i)$.

  The convergence of this method for  numerically evaluating the integral
  goes as 
  $1/\sqrt{N}$ with the number of function evaluations, $N$.
  This is slower than other commonly used
  techniques for numerical integration, \eg the trapezium rule converges as
  $1/N^2$ and Simpson's  rule as $1/N^4$. 
  While the convergence of these other methods becomes far slower for higher 
  dimensional integrals, \eg  the trapezium rule converges as
  $1/N^{2/d}$ and Simpson's  rule as $1/N^{4/d}$ where $d$ is the dimension of
  the integral, the Monte Carlo method will always
  converge as $1/\sqrt{N}$.

  Hence for the performance of high-dimensional integrals the Monte Carlo
  technique is
  more efficient. This is particularly important in particle physics where
  we need to
  perform high dimensional phase-space integrals. The Monte Carlo procedure
  is also well
  suited to integrating over complex regions, which are difficult with 	
  other methods, and
  often occur in particle physics, \eg due to experimental cuts.
  Another advantage of 
  this method is that we can evaluate a number of different quantities, \eg  
  differential distributions, at the same time whereas with other methods
  each distribution
  would have to be calculated separately.

  The convergence of the Monte Carlo technique can be improved by reducing 
  the standard deviation, $\sigma_N$. In principle if the integral can
  be performed
  analytically a Jacobian transform can be used to reduce the standard
  deviation to
  zero. In practice we can use a simple function which approximates
  the shape of the
  function we are integrating to improve the convergence of the integral, thus
  considerably reducing the time required for the numerical computation of the
  integral. 

  In the cross sections for the \rpv\  processes we are studying there are 
  Breit-Wigner resonances. These lead to a large variance in the calculation
  of the 
  cross section, which can be improved by applying a Jacobian transformation.
  First
  we consider the case of only one resonance. The cross section can be
  rewritten as
\begin{equation}
  \sigma = \int^1_{\tau_0} d\tau \int^1_\tau \frac{dx_1}{x_1} 
                  \int^1_{-1} d\!\cos\theta \int^{2\pi}_0 d\phi  \sum_{ij}
	 \frac{d\hat{\sigma}_{ij}}{d\Omega}
	\!\stackrel{\displaystyle(\sh,\theta,\phi,\mu^2)}{\ }
                       f_i(x_1,\mu^2)f_j\left(\frac{\tau}{x_1},\mu^2\right)\!,
\end{equation}
  where $\tau=x_1 x_2$, $\tau_0=(m_3+m_4)^2/s$, $s$ is the hadron--hadron
  centre-of-mass energy squared, and $m_3$ and $m_4$ are
  the masses of the final-state particles.
  The parton-level centre-of-mass energy squared is given by $\sh=\tau s$.
  The form of the Breit-Wigner
  peak in the cross section is
\begin{equation}
   \frac{d\hat{\sigma}_{ij}}{d\Omega}
	\!\stackrel{\displaystyle(\sh,\theta,\phi,\mu^2)}{\ } \sim 
	\frac{1}{\left(\sh-M^2\right)^2 + \Gamma^2 M^2},
\end{equation}
  where $M$ is the mass of the resonant particle and $\Gamma$ is its
  width. We can perform a change of variables 
\begin{equation}
  \tau = \tau_M
       +\sqrt{\tau_M\tau_\Gamma}
	\tan\left(\rho \sqrt{\tau_M\tau_\Gamma}\right)\!,
\end{equation}
  where $\tau_M = M^2/s$ and $\tau_\Gamma = \Gamma^2/s$. 
  This gives an integral over $\rho$,
\begin{eqnarray}
  \sigma & = & \int^{\rho_1}_{\rho_0} d\rho \int^1_\tau \frac{dx_1}{x_1} 
                  \int^1_{-1} d\!\cos\theta \int^{2\pi}_0 d\phi \nonumber 
		\\\nopagebreak
	 &   &
		\left[\tau_M\tau_\Gamma+(\tau-\tau_M)^2\right] 
 \sum_{ij} \frac{d\hat{\sigma}_{ij}}{d\Omega}
	\!\stackrel{\displaystyle(\sh,\theta,\phi,\mu^2)}{\ }
                       f_i(x_1,\mu^2)f_j\left(\frac{\tau}{x_1},\mu^2\right)\!,
\end{eqnarray}
  where
\begin{subequations}
\begin{eqnarray} 
\rho_0 &=&\frac1{\sqrt{\tau_M\tau_\Gamma}}\tan^{-1}\left(\frac{\tau_0-\tau_M}
	  {\sqrt{\tau_M\tau_\Gamma}}\right)\!,\\
\rho_1 &=&\frac1{\sqrt{\tau_M\tau_\Gamma}}\tan^{-1}\left(\frac{1-\tau_M}
	  {\sqrt{\tau_M\tau_\Gamma}}\right)\!.
\end{eqnarray}
  If the only dependence of the cross section on $\sh$ was the
  Breit-Wigner resonance
  the $\rho$ integral would be the integral of a constant,
  \ie the variance would
  be zero. In practice there is some remaining $\sh$ dependence in 
  the integrand but
  this is far smoother than the Breit-Wigner resonance and hence the
  variance is
  dramatically reduced.  
\end{subequations}

  In the \rpv\  cross sections there can be more than one accessible resonance
  depending on the number of non-zero \rpv\  couplings. In this case we can
  use a multi-channel Monte Carlo integration technique. The cross section
  can be rewritten in the following way
\begin{eqnarray}
  \sigma & = & \int^1_{\tau_0} d\tau \int^1_\tau \frac{dx_1}{x_1} 
              		\int^1_{-1} d\!\cos\theta \int^{2\pi}_0 d\phi
	\sum^N_k \frac{W_k}
			{F(\tau)\left[ \left(\sh-M^2_k\right)^2
					+\Gamma^2_k M^2_k\right]}
	\nonumber \\\nopagebreak
         &   &	 
		\sum_{ij}
		 \frac{d\hat{\sigma}_{ij}}{d\Omega}
	\!\stackrel{\displaystyle(\sh,\theta,\phi,\mu^2)}{\ }
                       f_i(x_1,\mu^2)f_j\left(\frac{\tau}{x_1},\mu^2\right)\!,
\end{eqnarray}
  where
\begin{equation}
   F(\tau) = \sum_i^N \frac{W_i}{\left(\sh-M^2_i\right)^2+\Gamma^2_i M^2_i},
\end{equation}
  and $M_i$ and $\Gamma_i$ are the mass and width of the $i$th resonance. 
  The weight, $W_i$, is chosen to approximate the contribution of
  the $i$th resonance
  to the total cross section.
  We can perform a Jacobian transform for these integrals treating each of the
  terms in the sum as we  did before for the single resonance.
  This allows us to perform
  a change of variables for each of the integrals in the sum 
\begin{equation}
  \tau = \tau^k_M
          +\sqrt{\tau^k_M\tau^k_\Gamma}
   		\tan\left(\rho_k \sqrt{\tau^k_M\tau^k_\Gamma}\right)\!,
\end{equation}
  where $\tau^k_M = M_k^2/s$ and $\tau^k_\Gamma = \Gamma_k^2/s$. 
  This gives a set of integrals over the new variables
  $\rho_k$,
\begin{eqnarray}
  \sigma & = &  
              		\int^1_{-1} d\!\cos\theta \int^{2\pi}_0 d\phi 
			\sum^N_k \int^{\rho^k_1}_{\rho^k_0} d\rho_k \frac{W_k}
			{F(\tau)}\,\frac1{s^2} \int^1_\tau \frac{dx_1}{x_1}
	\sum_{ij} \frac{d\hat{\sigma}_{ij}(\sh)}{d\Omega}
          f_i(x_1,\mu^2)f_j\left(\frac{\tau}{x_1},\mu^2\right)\!,\nonumber\\
&&
\end{eqnarray}
  where
\begin{subequations}
\begin{eqnarray} 
\rho^k_0 &=&\frac1{\sqrt{\tau^k_M\tau^k_\Gamma}}
            \tan^{-1}\left(\frac{\tau_0-\tau^k_M}
	  {\sqrt{\tau^k_M\tau^k_\Gamma}}\right)\!,\\
\rho^k_1 &=&\frac1{\sqrt{\tau^k_M\tau^k_\Gamma}}
           \tan^{-1}\left(\frac{1-\tau^k_M}
	  {\sqrt{\tau^k_M\tau^k_\Gamma}}\right)\!.
\end{eqnarray}
\end{subequations}
  The use of this technique significantly increases the efficiency of 
  the Monte Carlo simulation.
  Each different hard collision process must be studied separately
  and a Jacobian
  transformation applied to reduce the variance. In, for example, QCD jet and
  heavy quark production the cross section falls like $p_T^{-4}$,
  where $p_T$ is the
  transverse momentum of the partons produced in the hard collision, and
  we must therefore use a Jacobian transform to smooth this fall-off.
  In other processes, for example Drell-Yan, a multi-channel approach
  must be used because in addition to a power law fall-off, like
  $\sh^{-2}$, of the cross section due to photon exchange there is
  a Breit-Wigner resonance due to Z exchange.

  We can then use the parton-level 
  centre-of-mass energy and angles
  which we randomly generate while performing
  the cross-section integral to construct the four-momenta
  of the particles involved in the hard collision.
  In the next section we will
  describe how these particles can be evolved from the high scale of the
  parton--parton collision to some low scale typical of the
  hadronization process.

%
%  Now the section on the Parton Shower Phase 
%
\section{Parton Showers}
\label{subsect:monteshower}

  In the previous section we described how to generate the momenta
  of the particles involved in the hard scattering process. This
  scattering process usually involves
  coloured particles in either the initial or final state. After the
  hard collision
  these coloured particles must be evolved from the high-energy scale
  of the collision
  to some lower scale with the emission of QCD radiation. The exact
  calculation of
  the matrix elements for processes with large numbers of final-state
  partons is not
  possible and we must therefore treat the regions of phase space
  for which the
  emission of QCD radiation is enhanced and take these into account
  to all orders
  in perturbation theory. 
  There are two regions of phase space where the emission of QCD radiation
  is enhanced:
\begin{enumerate}
 \item Collinear Emission;
 \item Soft Emission.
\end{enumerate}

  This can be seen by considering the process
  $\mr{e^+ e^- \ra q(p_1) \bar{q}(p_2) g(p_3)}$,
   shown in Fig.\,\ref{fig:eenextreal}.
  The leading-order cross section for this process is given by
  \cite{Ellis:1976uc}
\begin{equation}
   \sigma_{q\bar{q}g}= N_c\sigma_0  \sum_q Q^2_q \int dx_1 dx_2 C_F
                      \frac{\al_s}{2\pi}
                      \frac{x_1^2+x^2_2}{(1-x_1)(1-x_2)},
 \label{eqn:qqgcross}
\end{equation}
  where $\sigma_0=4\pi\al^2/(3s)$ is the leading-order cross section
  for $\mr{e^+e^-\ra\mu^+\mu^-}$. The energy of the final-state partons in the
  laboratory frame is $E_i$, and
  $x_i=2E_i/\sqrt{s}$, where $\sqrt{s}$ is the centre-of-mass energy of the
  $\mr{e^+e^-}$ collision.
  Momentum conservation therefore leads to
  $x_1+x_2+x_3=2$. $C_F=(N_c^2-1)/(2N_c)$
  is the Casimir in the fundamental representation, where $N_c$ is the
  number of colours. If we look at Eqn.\,\ref{eqn:qqgcross} the cross section
  diverges when either $x_1$, $x_2$, or both tend to one.
  We can now consider the physical origins of these divergences:
\begin{enumerate}
 \item $(1-x_1)= \frac{E_2E_3}{Q^2}\left(1-\cos\theta_{2g}\right)$, where 
 $\theta_{2g}$ is the angle between the antiquark and the gluon.
 The singularity as $x_1\ra 1$ therefore occurs as the antiquark and the gluon
 become collinear. Similarly the singularity as $x_2\ra1$ occurs as the
 quark and
 the gluon become collinear. This is the collinear singularity.
 \item As $x_1\ra 1$ and $x_2\ra 1$ energy conservation implies that
 $x_3\ra 0$, \ie
 the energy of the gluon tends to zero. This is the soft singularity. 
\end{enumerate}

%
% Feynman diagrams for e+e- ---> hadrons at NLO
%
\begin{figure}
\begin{center}
\subfigure[Leading-order diagram.]{
\begin{picture}(360,45)
% first the leading order diagram
\SetScale{0.5}
\SetOffset(140,0)
\ArrowLine(0,0)(50,50)
\ArrowLine(50,50)(0,100)
\Photon(50,50)(100,50){10}{5}
\ArrowLine(100,50)(150,0)
\ArrowLine(150,100)(100,50)
\end{picture}
\label{fig:eenextlo}}
% then the real emission diagrams
\subfigure[Real emission diagrams.]{
\begin{picture}(360,45)
% First Real Emission Diagram
\SetScale{0.5}
\SetOffset(70,0)
\ArrowLine(0,0)(50,50)
\ArrowLine(50,50)(0,100)
\Photon(50,50)(100,50){10}{5}
\ArrowLine(100,50)(150,0)
\ArrowLine(150,100)(125,75)
\ArrowLine(125,75)(100,50)
\Gluon(125,75)(150,75){-5}{3}
% Second Real Emission Diagram
\SetOffset(210,0)
\ArrowLine(0,0)(50,50)
\ArrowLine(50,50)(0,100)
\Photon(50,50)(100,50){10}{5}
\ArrowLine(100,50)(125,25)
\ArrowLine(125,25)(150,0)
\ArrowLine(150,100)(100,50)
\Gluon(125,25)(150,25){5}{3}
\end{picture}
\label{fig:eenextreal}}
% then the loop diagrams
\subfigure[Loop diagrams.]{
\begin{picture}(360,45)
\SetScale{0.5}
% First loop diagram
\SetOffset(20,0)
\ArrowLine(0,0)(50,50)
\ArrowLine(50,50)(0,100)
\Photon(50,50)(100,50){10}{5}
\ArrowLine(100,50)(125,25)
\ArrowLine(125,25)(150,0)
\ArrowLine(150,100)(125,75)
\ArrowLine(125,75)(100,50)
\GlueArc(100,50)(35.36,-45,45){5}{3}
% Second loop diagram
\SetOffset(140,0)
\ArrowLine(0,0)(50,50)
\ArrowLine(50,50)(0,100)
\Photon(50,50)(100,50){10}{5}
\ArrowLine(100,50)(150,0)
\ArrowLine(150,100)(100,50)
\GlueArc(125,75)(17.67,-135,45){5}{3}
% Third loop diagram
\SetOffset(260,0)
\ArrowLine(0,0)(50,50)
\ArrowLine(50,50)(0,100)
\Photon(50,50)(100,50){10}{5}
\ArrowLine(100,50)(150,0)
\ArrowLine(150,100)(100,50)
\GlueArc(125,25)(17.67,-45,135){5}{3}
\end{picture}
\label{fig:eenextnlo}}
\end{center}
\captionB{Feynman diagrams for $\mr{e^+e^-\ra hadrons}$ at next-to-leading
	 order.}
	{Feynman diagrams for $\mr{e^+e^-\ra hadrons}$ at next-to-leading
	 order in $\al_s$.}
\label{fig:eenext}
\end{figure}
% End of the figure

  If we were to perform the full next-to-leading-order calculation of
  the cross section for $\mr{e^+e^-\ra hadrons}$ these singularities
  would cancel with the
  singularities that occur in the loop diagrams. The real emission diagrams in
  Fig.\,\ref{fig:eenextreal} give a contribution to the cross section which is
  of order $\al_s$, Eqn.\,\ref{eqn:qqgcross}. The product of the loop
  diagrams, Fig.\,\ref{fig:eenextnlo}, and the leading-order diagram, 
  Fig.\,\ref{fig:eenextlo}, also
  gives a contribution of the same order in $\al_s$, and the singularities
  in this term exactly
  cancel those in the real emission diagrams.

  However for less inclusive observables, \eg the thrust
  in $\mr{e^+e^-}$ events, it can be the case
  that after this cancellation
  large logarithms remain. The aim of the parton-shower phase is
  to resum these logarithms to all orders in perturbation theory.
  In general the simplest algorithm for the parton shower only resums
  the leading collinear logarithms \cite{Webber:1986mc} whereas with a 
  slight modification 
  \cite{Marchesini:1984bm,Marchesini:1988cf} both the leading soft
  and collinear logarithms are resummed. We will first describe how the
  collinear logarithms can be resummed and then generalize this to include
  the soft logarithms as well. Good reviews of the parton-shower
  algorithm can be found in 
  \cite{Ellis:1991qj,Webber:1986mc}.

%
%  First the collinear parton showers
%
\subsection{Collinear Parton Showers}
\label{subsubsect:montecollinear}

\begin{figure}
\begin{picture}(360,160)
\SetOffset(80,-70)
\SetScale{1.5}
\GCirc(58.33,100){16.67}{0.6}
\ArrowLine(75,100)(125,100)
\ArrowLine(125,100)(175,50)
\DashLine(125,100)(195.7,100){5}
\Gluon(125,100)(175,150){4}{7}
\CArc(125,100)(30,-45,52.5)
\Text(90,110)[]{{\LARGE $\me_n$}}
\Text(150,135)[]{{\LARGE $i$}}
\Text(275,227)[]{{\LARGE $k$}}
\Text(275,75)[]{{\LARGE $j$}}
\Text(242,168)[]{{\large$\tht_k$}}
\Text(242,130)[]{{\large$\tht_j$}}
\end{picture}
\captionB{Radiation of a collinear gluon.}
  	{Radiation of a collinear gluon $k$ by a
         quark $i$ giving the quark $j$.}
\label{fig:colgluon}
\end{figure}
  We will start by considering the simplest type of process, \ie the emission
  of a collinear gluon by a final-state quark.
  The matrix element for the radiation of an additional gluon from a
  final-state quark, Fig.\,\ref{fig:colgluon}, can be written as
\begin{equation}
\me_{n+1} = \frac{g_s}{t}{\bf t}^a \bar{u}(p_j)\gamma^\mu p\sla_i
			\me'_n \varepsilon^*_\mu,
\end{equation}
  where ${\bf t}^a$ is the colour generator of $SU(3)$ in the fundamental
	representation,
        $p_i$ is the momentum of the quark before the gluon is radiated,
        $p_j$ is the momentum of the quark after the gluon radiation,
	$p_k$ is the momentum of the radiated gluon,
        $\varepsilon_\mu$ is the polarization vector of the gluon 
        and $\me'_n$ represents
	the matrix element, without the spinor for the quark $i$, before the
	gluon radiation. The initial quark $i$ is off mass-shell, 
	$t=p_i^2$, while both the gluon and the quark after the branching
        are on mass-shell. We will assume that the quarks are massless.

        We can  square this matrix element and sum over
	the final-state spins giving,
\begin{equation}
\sum_{\mr{spins}}|\me_{n+1}| = 
	\frac{g^2_sC_F}{2t^2}
	\mr{tr}\left({\me'_n}^\dagger p\sla_i \gamma_\nu p\sla_j \gamma_\mu 
		p\sla_i	\me'_n \right)
	\left(-g^{\mu\nu}+\frac{p^\mu_k l^\nu_{\phantom{k}}
	+p^\nu_k l^\mu_{\phantom{k}}}{p_k\cdot l}\right)\!,
\end{equation}
  where we have used an axial gauge for the gluon propagator. The four-vector
  $l^\mu_{\phantom{k}}$ is an arbitrary light-like four-vector
  which must not be collinear with any of the momenta.

  We can  evaluate this expression in the collinear limit.
  The traces in this
  expression can be treated individually. The first term gives
\begin{eqnarray}
  -\mr{tr}\left({\me'_n}^\dagger p\sla_i \gamma_\nu p\sla_j \gamma_\mu p\sla_i
			\me'_n \right)g^{\mu\nu}&=&
   2\, \mr{tr}\left({\me'_n}^\dagger p\sla_i p\sla_j p\sla_i\me'_n\right)\!,
   \nonumber\\
   & = & 2\, \mr{tr}\left({\me'_n}^\dagger p\sla_k p\sla_j 
   p\sla_k\me'_n\right)\!,
\end{eqnarray}
  where the first line comes from the properties of the Dirac matrices and
  the second line from momentum conservation, \ie $p_i=p_j+p_k$, and
  $p_j^2=0$. As the gluon is on
  mass-shell, \ie $p_k^2=0$, we can make use of the following identity,
\begin{equation}
         p\sla_k = \sum_{\mr{spins}} u_s(p_k)\bar{u}_s(p_k),
\end{equation}
 giving
\begin{equation}
     -\mr{tr}\left({\me'_n}^\dagger p\sla_i \gamma_\nu p\sla_j \gamma_\mu
	 p\sla_i \me'_n \right)g^{\mu\nu}=
	8p_j\cdot p_k\,\mr{tr}({\me'_n}^\dagger p\sla_k \me'_n).
\end{equation}
  Finally $p_k$ can be replaced by $p_k=(1-z)p_i$ in the collinear limit,
  where $z=E_j/E_i$ is the fraction of the initial quark $i$'s energy carried
  by the quark $j$ after the gluon emission, giving
\begin{equation} -\mr{tr}\left(\me^\dagger_n p\sla_i \gamma_\nu p\sla_j
	 \gamma_\mu p\sla_i \me_n \right)g^{\mu\nu}=
        4t(1-z)\sum_{\mr{spins}}|\me_n|^2,
\end{equation}
  where $\me_n$ is the full matrix element for the process before the
  emission of the collinear gluon. The remaining terms can be calculated in
  the same way giving
\begin{equation}
 \sum_{\mr{spins}}|\me_{n+1}|^2 =  \frac{2g^2_s}{t}\, C_F\frac{1+z^2}{1-z}\,
				   \sum_{\mr{spins}}  |\me_n|^2.
\end{equation}
   The matrix element squared for the emission of a collinear gluon 
   {\it factorizes} into the matrix element squared for the process without
   the collinear gluon and a universal, \ie process independent,
   Altarelli-Parisi splitting function 
   \cite{Gribov:1972rt:Gribov:1972ri:Dokshitzer:1977sg:Altarelli:1977zs} for 
   the radiation of a gluon from a quark, \ie
\begin{equation}
 P_{qq} =   C_F\frac{1+z^2}{1-z}.
\label{eqn:splitqq}
\end{equation}
   In general, if we consider a process involving $n$ partons 
   the amplitude squared for a process with the emission of an extra collinear
   parton can be written in terms of the matrix element for the $n$
   parton process and a universal splitting function
\begin{equation}
 \sum_{\mr{spins}}|\me_{n+1}|^2 =   \frac{2g^2_s}{t} P_{ji}(z) 
	\sum_{\mr{spins}}|\me_n|^2,
\label{eqn:colme}
\end{equation}  
  where $P_{ji}$ are the unregularized
  Altarelli-Parisi splitting functions. $P_{qq}$ is given
  above for gluon radiation from a quark and
\begin{eqnarray}
 P_{gg} = &  \displaystyle{C_A \left[\frac{1-z}{z}
			+\frac{z}{1-z}+z(1-z)\right]\!,} \\
 P_{qg} = &  \displaystyle{T_R\left[z^2+(1-z)^2\right]\!,}  
\end{eqnarray}
  for gluon radiation from a gluon and for a gluon splitting into a
  quark--antiquark pair, respectively. $T_R=\frac{1}{2}$ and $C_A=N_c$ is
  the Casimir in the adjoint representation of $SU(3)$. There is an
  additional splitting function, $P_{gq}(z)$, which describes the splitting
  of a quark to give a gluon with a fraction $z$ of its energy. This
  splitting function can be obtained by making the replacement $z\ra1-z$ in
  Eqn.\,\ref{eqn:splitqq}.

  The phase space for this splitting also factorizes. The phase space for the
  $n$ body process before the gluon radiation is given by
\begin{equation}
d{\bf\Phi}_{n} = d{\bf \Gamma}\, \frac{d^3p_i}{2E_i(2\pi)^3},
\label{eqn:lophase}
\end{equation}
  where $d{\bf \Gamma}$ represents the phase-space integrals over all the
  particles apart from the quark~$i$ which radiates the gluon. The phase space
  for the process after the gluon radiation can be written as
\begin{equation}
d{\bf\Phi}_{n+1} = d{\bf \Gamma}\, \frac{d^3p_j}{2E_j(2\pi)^3}\,
			 \frac{d^3p_k}{2E_k(2\pi)^3}.
\label{eqn:nlophase}
\end{equation}
  As momentum is conserved, \ie $p_i=p_j+p_k$, at fixed $p_j$ this implies
  $d^3p_k=d^3p_i$. We therefore obtain
\begin{equation}
  d{\bf\Phi}_{n+1} = 
	d{\bf\Phi}_{n}\frac1{(1-z)} \, \frac{d^3p_j}{2E_j(2\pi)^3},
\end{equation}
  using $E_k=(1-z)E_i$ in the collinear limit. The phase-space integral over
  the momentum of the quark $j$ can  be expanded in the small angle 
  limit giving
\begin{eqnarray}
  d{\bf\Phi}_{n+1} &=& d{\bf\Phi}_{n}\frac1{2(2\pi)^3}
			\frac1{(1-z)}E_jdE_j\tht_jd\tht_jd\phi,\nonumber\\
      &=&d{\bf\Phi}_{n}\frac1{2(2\pi)^3}\, E_jdE_j\, \tht_jd\tht_j\, d\phi 
		   dt\delta\left(t-\frac{E_jE_k\tht_j^2}{(1-z)^2}\right)\, 
	\frac{dz}{(1-z)}
	\delta\left(z-\frac{E_j}{E_i}\right)\!,\,\,\,\,\,\,\,\,\,\,\,
\label{eqn:coldelta}
\end{eqnarray}
  where we have inserted the definitions of $t$ and $z$,
  and $\tht_j$ is angle between the directions of the partons $i$ and $j$ as
  shown in Fig.\,\ref{fig:colgluon}. In the small angle limit
\begin{equation}
     t= (p_j+p_k)^2 = 2E_jE_k(1-\cos\tht)\simeq E_jE_k\tht^2,
\end{equation}
  where $\tht$ is the angle between $j$ and $k$. Using transverse momentum
  conservation gives $z\tht_j=(1-z)\tht_k$. As $\tht=\tht_j+\tht_k$ we obtain
  $\tht=\tht_j/(1-z)$. Hence the virtual mass of the initial quark
  is
\begin{equation}
 t \simeq \frac{E_jE_k\tht_j^2}{(1-z)^2}.
\end{equation}
  We can now use the $\delta$-functions in Eqn.\,\ref{eqn:coldelta} to 
  integrate out the dependence on the energy of the quark, $E_j$, and its
  angle, $\tht_j$, giving
\begin{equation}
   d{\bf\Phi}_{n+1} = d{\bf\Phi}_{n} \frac1{4(2\pi)^3}\,d\phi\,dt\,dz.
\label{eqn:colphase}
\end{equation}
 
  Therefore, using the equations for the collinear factorization of the matrix
  element, Eqn.\,\ref{eqn:colme}, and for the phase space,
  Eqn.\,\ref{eqn:colphase}, we obtain  
\begin{equation}
 d\sigma_{n+1} = d\sigma_n\frac{dt}{t}\,dz\frac{\al_s}{2\pi}P_{ji}(z),
\label{eqn:collinearfact}
\end{equation}
  where we have averaged over the azimuthal angle of the emitted parton. 
  The same equation also holds for the radiation of a gluon from an
  incoming parton in processes involving hadrons in the initial state 
  \cite{Ellis:1991qj}.

  So for the emission of QCD radiation in the collinear limit,
  after azimuthal averaging, the cross section obeys a
  factorization theorem. The cross section for a process in which one
  parton pair is much more collinear than any other pair can be
  written as the convolution of a universal
  splitting function and the cross section for the same process where
  the collinear pair is replaced by a single parton of the
  corresponding flavour. This functional form allows us to apply the procedure
  to the next most collinear pair in the final state, and so on. We thus have
  an iterative rule which leads to a description of multi-parton final states
  as a Markov chain. 

  This can be reexpressed as an evolution in some energy-like scale, such as
  the virtuality, where a parton at a high scale is evolved by successive
   branchings to a lower scale. The normal approach
  \cite{Ellis:1991qj} is to consider the evolution of the parton density, 
  $f_i(x,t)$, \linebreak \ie the probability of finding a parton of type $i$
  with a given fraction $x$ of the momentum of an incoming hadron at a given
  scale $t$. As we will mainly be dealing with partons in the final state we
  will instead consider the evolution of the fragmentation function,
  $d^h_i(x,t)$, \ie the probability of a parton of type $i$ giving a hadron
  of type $h$ with a fraction $x$ of the parton's momentum at a given scale
  $t$. 

  We can use Eqn.\,\ref{eqn:collinearfact} to write the evolution of the
  fragmentation function at a given virtual mass-squared and momentum
  fraction, $d(x,t)$. For simplicity we will only consider one type of
  branching, \ie the radiation of a gluon by a quark, and one type of hadron,
  but the results generalize easily.

  The change in the fragmentation function consists of two parts: firstly
  there is an increase due to partons branching to give a parton with a
  momentum fraction greater than $x$,
\begin{eqnarray}
  \delta d(x,t)_{\mr{in}} & = & \frac{\delta t}{t} 
				\int^1_xdx'dz\frac{\al_s}{2\pi}
				P(z)d(x',t)\delta(x-zx'), \nonumber \\
 		   & = &   \frac{\delta t}{t}  \int^1_0 \frac{dz}{z}
				\frac{\al_s}{2\pi}P(z)d(x/z,t);
\label{eqn:evolveder1}
\end{eqnarray}
  secondly there is a decrease due to partons radiating to give partons with
  momentum fractions smaller than $x$,
\begin{eqnarray}
  \delta d(x,t)_{\mr{out}} & = & 
	 \frac{\delta t}{t}d(x,t)\int^x_0dx'dz \frac{\al_s}{2\pi}
				P(z)\delta(x'-zx), \nonumber \\
 		    & = &   \frac{\delta t}{t} d(x,t) \int^1_0 dz
			    	\frac{\al_s}{2\pi}P(z).
\end{eqnarray}
  The fragmentation function, $d(x,t)$, is zero for $x>1$ and therefore the
  integral in Eqn.\,\ref{eqn:evolveder1} is zero for $z<x$.
  Hence the evolution of the fragmentation function with
  the virtual mass-squared is given by
\begin{equation}
   \frac{\partial}{\partial t}d(x,t)
	= \frac1{t}  \int^1_0 dz \frac{\al_s}{2\pi}
			P(z) \left[ \frac1{z}d(x/z,t)-d(x,t) \right]\!.
\label{eqn:evolve1}
\end{equation}
  In the standard treatment of the evolution of the fragmentation function, or
  the parton distribution function, we would now regulate the Altarelli-Parisi
  splitting functions to obtain the DGLAP equation 
  \cite{Gribov:1972rt:Gribov:1972ri:Dokshitzer:1977sg:Altarelli:1977zs}
  for the evolution of the
  fragmentation function. However, it is more convenient for us to continue
  to work with the unregulated splitting functions and to reexpress the
  evolution equation in a more useful way using the Sudakov form factor
  \cite{Sudakov:1956sw} which we will regularize below.
  The Sudakov form factor is given by
\begin{equation}
 \Delta(t) = \exp\left[-\int^t_{t_0} \frac{dt'}{t'}\int^1_0 dz
			\frac{\al_s}{2\pi} P(z) \right]\!,
\label{eqn:sudakov1}
\end{equation}
  where $t_0$ is the low scale at which we stop the evolution of the
  fragmentation function.
  Using the Sudakov form factor we can rewrite Eqn.\,\ref{eqn:evolve1}
  in the following way
\begin{equation}
  t  \frac{\partial}{\partial t} \left(\frac{d}{\Delta}\right) = 
	\frac1{\Delta}\int^1_0\frac{dz}{z}\frac{\al_s}{2\pi} P(z)d(x/z,t).
\end{equation}
  This differential equation can be solved to give an integral equation for
  the fragmentation function, $d(x,t)$, in terms of its initial value,
  $d(x,t_0)$, at some low scale $t_0$,
\begin{equation}
     d(x,t)  =  \Delta(t)d(x,t_0)
	        + \int^t_{t_0}\frac{dt'}{t'}
					\frac{ \Delta(t)}{ \Delta(t')}
					\int^1_0 \frac{dz}{z}
					\frac{\al_s}{2\pi} P(z) d(x/z,t').
\label{eqn:evolve2}
\end{equation}

  This equation can be interpreted in the following way. The first term is
  the contribution from those partons which do not branch between the scales
  $t$ and $t_0$. The second term is the contribution from partons which last
  branched at the scale $t'$. This suggests that we can interpret the ratio
  of  Sudakov form factors, $\Delta(t)/\Delta(t')$, as the probability of
  evolving down from the scale $t$ to $t'$ without any branching. This is
  consistent with the first term in Eqn.\,\ref{eqn:evolve2} as 
  $\Delta(t_0)=1$. We can therefore interpret $\Delta(t)$
  as the probability of no emission between the scales $t$ and $t_0$.

  The Sudakov form factor can be generalized to include more than one
  type of parton, by defining the Sudakov form factor for parton type $i$
\begin{equation}
 \Delta_i(t) = \exp\left[-\sum_j\int^t_{t_0} \frac{dt'}{t'}\int^1_0 dz
			\frac{\al_s}{2\pi} P_{ji}(z) \right]\!.
\end{equation}
  The evolution equation for more than one type of parton can be written as
\begin{equation}
  t  \frac{\partial}{\partial t} \left(\frac{d_i}{\Delta_i}\right) = 
		\frac1{\Delta_i}\sum_j
		\int^1_0\frac{dz}{z}\frac{\al_s}{2\pi} P_{ij}(z)d_j(x/z,t).
\end{equation}
  There is however still a problem because the splitting functions,
  $ P_{ij}(z)$, contain the collinear singularity as $z \ra 1$ which needs
  to be regularized.
  We can do this by introducing an infrared cut-off, $z < 1-\epsilon$. We then
  classify branchings above this cut-off as unresolvable.
  This means that the regularized Sudakov form factor,
\begin{equation}
 \Delta_i(t) = \exp\left[-\sum_j\int^t_{t_0} \frac{dt'}{t'}
			\int^{1-\epsilon}_\epsilon dz
			\frac{\al_s}{2\pi} P_{ji}(z) \right]\!,
\end{equation}
  gives the probability of evolving between two scales without
  resolvable emission. This implicitly resums the virtual contributions as
  well as the real emissions we have been considering to all orders. This is
  because the virtual contributions will affect the probability of no
  branching and are hence included by unitarity, \ie because we have imposed
\begin{equation}
\mr{   P(no\  resolvable\  emission) + P(resolvable\  emission) = 1}.
\end{equation}
  This correctly  resums the leading collinear singularities to all orders in 
  perturbation theory \cite{Webber:1986mc}. The infrared cut-off can be
  chosen as a cut-off, $t_0$, on the virtuality of the parton $i$, by defining
\begin{equation}
    \epsilon = \frac{t_0}{t}.
\end{equation}
  The value of $t_0$ used in simulations is typically of order
  $1\, \mr{\gev}^2$.

  In this section we have seen how we can resum the collinear singularities
  to all orders in perturbation theory using the Sudakov form factor. In the
  next section we will discuss how the soft singularities can be resummed as
  well.

%
%  Now the subsection on the soft Parton showers
%
\subsection{Angular-ordered Parton Showers}
\label{subsubsect:montesoft}

    We can consider the emission of QCD radiation in the soft limit in much
    the same way as we did for the collinear emission 
    in Section~\ref{subsubsect:montecollinear}. The matrix
    element for the emission of a gluon from an outgoing quark is given by
\begin{equation}
  \me_{n+1} = \frac{g_s}{(p_i^2-m^2)} {\bf t}^a_{\be\al}C_\al\bar{u}_\be(p_j)
	 \gamma^\mu	\left(p\sla_i +m\right)  \me_n \varepsilon_\mu^*,
\end{equation} 
    where we are considering the splitting $i\ra jk$ as in
    Fig.\,\ref{fig:colgluon}. $C_\al$ represents the colour structure of the
    leading-order process without the soft gluon and $m$ is the mass of the
    quark. The quark has colour $\al$ before the emission of the gluon and
    colour $\be$ afterwards. As the gluon and the quark $j$, after the
    splitting, are on mass-shell we can rewrite the denominator as
    $p_i^2-m^2=2p_j\cdot p_k$. We can then use the anti-commutation
    relations for the Dirac matrices to obtain
\begin{equation}
   \me_{n+1} = \frac{g_s}{2p_j\cdot p_k}
		 {\bf t}^a_{\be\al}C_\al\bar{u}_\be(p_j) 
		\left[ 2p_i^\mu + (m-p\sla_i) \gamma^\mu \right]\me_n
		 \varepsilon_\mu^*.
\end{equation}
    In the soft limit $p_i\approx p_j$ and we can therefore use the Dirac
    equation $\bar{u}(p_j)(p\sla_j-m)=0$ to give
\begin{equation}
   \me_{n+1} = \frac{g_sp_j^\mu}{p_j\cdot p_k} {\bf t}^a_{\be\al}
	C_\al\bar{u}_\be(p_j) \me_n \varepsilon_\mu^*.
\end{equation}
    In general, a factorization theorem exists for the
    emission of QCD radiation in the soft limit. However, this
    theorem applies to the amplitude for the process, rather than
    the cross section. The amplitude for a process in which one gluon is much
    softer than the other energy scales in the process can be written
    as a product of a universal eikonal current and the amplitude for
    the same process without the soft gluon. The matrix element
    including the emission of an extra soft gluon is given by
\begin{equation}
    {\bf {\cal M}}  = g_s {\bf m} \cdot {\bf J}(q),
\end{equation}
  where ${\bf {\cal M}}$ is the matrix element for the process including
	the emission of an extra soft gluon,
	${\bf m}$ is the tree-level amplitude for the
        underlying process and $\JC(q)$ is the non-Abelian semi-classical
        current for the emission of the soft gluon with momentum $q$
	from the hard partons. In general the eikonal current $\JC(q)$ is
	given by
\begin{equation}
    \JC(q) = \varepsilon^*_\mu
	\sum_\mr{{\stackrel{\scriptstyle external}{partons}}} 
	C^b_\al P^{ab}_\al
	\left(\frac{p_{\mr{parton}}}{p_{\mr{parton}}\cdot q}\right)^\mu
\end{equation}
    for the emission of a soft gluon with momentum $q$. $C^b_\al$ represents
    the colour structure of the leading-order process without the soft gluon
    and $P^{ab}_\al$ the colour matrix for the emission of a gluon with
    colour $a$. These colour matrices are given in 
    Table~\ref{tab:softcolour} for the various possible radiating partons.

\begin{table}
\renewcommand{\arraystretch}{1.5}
\renewcommand{\tabcolsep}{30pt}
\begin{center}
\begin{tabular}{|c|c|c|}
\hline
Radiating & \multicolumn{2}{c|}{Colour Matrix} \\ \cline{2-3}
Parton    & incoming & outgoing \\
\hline 
 quark     & $ -{\bf t}^c_{\be\al}$ & $ \phantom{-}{\bf t}^c_{\al\be}$\\
\hline
 antiquark & $\phantom{-}{\bf t}^c_{\al\be}$   & $-{\bf t}^c_{\be\al}$\\
\hline
 gluon     & $-if^{abc}$ 	   & $-if^{abc}$ \\
\hline
\end{tabular}
\captionB{Eikonal current factors.}
	{Colour matrices for the eikonal current. As the eikonal current is
	independent of the spin of the emitting parton the colour 
	factors for squarks are the same as those for quarks and those for
	gluinos the same as those for gluons. In the case of radiation from
	quarks the colour of the external parton is $\al$ and the colour of
	the parton in the hard process is $\be$. Similarly for gluons the
	colour of the external parton is $a$ and the colour of the gluon
	participating in the hard process is $b$. In both cases the colour
	of the radiated gluon is $c$. $f^{abc}$ is the $SU(3)$ colour
	generator in the adjoint representation.}
\label{tab:softcolour}
\end{center}
\end{table}

    After we square the
    amplitude and sum over the spins of the external partons we obtain
    a result which depends on the momenta of all the external
    partons. It therefore seems unlikely that we can recover a factorization
    theorem for the cross section as in the previous section. The
    surprising result  \cite{Marchesini:1984bm,Marchesini:1988cf}
    is that, after azimuthal averaging, these effects
    can be incorporated into a collinear algorithm by simply using a different
    choice for the evolution scale, \ie the opening angle.

    We can illustrate this with a simple example, \ie the process 
    \mbox{$\mr{e^+e^- \ra q\, {\bar q}\, g_1}$}, shown in Fig.\,\ref{fig:qqg}.
    The semi-classical eikonal current can be used to study the emission
    of an extra soft gluon in this process, \ie the process
    \mbox{$\mr{e^+e^-\ra q \, \bar{q} \, g_1 \, g_2}$} where the second gluon 
    is much softer than the other partons. The matrix element
    including the emission of the extra soft gluon is given by
\begin{equation}
    {\bf {\cal M}} (k_1,k_2,p_1,p_2,p_3;q) = g_s {\bf m}
   (k_1,k_2,p_1,p_2,p_3) \cdot {\bf J}(q),
\end{equation}
where
\begin{itemize}

\item ${\bf m}(k_1,k_2,p_1,p_2,p_3)$ is the tree-level amplitude for the
        underlying process, \linebreak  
        $\mr{e^+}(k_1)\,\mr{e^-}(k_2)\, \ra  \mr{q}(p_1)\, 
        \mr{\bar{q}}(p_2)\, \mr{g_1}(p_3)$.

  \item ${\bf {\cal M}} (k_1,k_2,p_1,p_2,p_3;q)$ is the matrix element for the
        process \linebreak $\mr{e^+}(k_1)\, \mr{e^-}(k_2)\, \ra 
             \mr{q}(p_1)\, \mr{\bar{q}}(p_2)\, \mr{g_1}(p_3)\,  \mr{g_2}(q)$,
        \ie including the emission of an extra soft
        gluon, $\mr{g_2}$, with momentum $q$.

\item $\JC(q)$ is the non-Abelian semi-classical current for the
        emission of the soft gluon with momentum $q$ from the hard
        partons.
\end{itemize}

Explicitly in our example the current, ${\bf J}(q)$, is given by
\begin{equation} 
    \JC(q) = \sum_{s=1,2}  \JC^{b,\mu}(q) \varepsilon^*_{\mu,s},
\end{equation}
where 
\begin{equation}  
   \JC^{b,\mu}(q) = 
                {\bf t}^{b,\mr{q}}_{c_1c_1'}{\bf t}^a_{c_1'c_2}
                                      \left(\frac{p^\mu_1}{p_1 \cdot q}\right)
                -{\bf t}^{a}_{c_1c_2'}{\bf t}^{b,\mr{\bar{q}}}_{c_2'c_2}
 				      \left(\frac{p^\mu_2}{p_2 \cdot q}\right)
         	-i f^{aa'b} {\bf t}^{a'}_{c_1c_2} 
			      \left(\frac{p^\mu_3}{p_3 \cdot q}\right)\!.
\label{eqn:eegcurrent}
\end{equation}
  If we now define the radiation functions, as was done in 
  \cite{Marchesini:1990yk}, we can express the square of this current in a
  useful way. We can define the dipole radiation function
\begin{equation}
\frac{2}{\omega^2}W_{ij}(q)=
- \left( \frac{p_{i}}{p_i
                   \cdot q}-\frac{p_{j}}{p_j\cdot q} \right)^2 =
                   \frac{2}{\omega^2} \left( \frac{\xi_{ij}}{\xi_{i}\xi_{j}}
                   -\frac{1}{2\gamma_{i}^2\xi_{i}^2}
   -\frac{1}{2\gamma_{j}^2\xi_{j}^2} \right)\!,
\end{equation}
 where $\omega$ is the energy of the soft gluon,
       ${\xi_{ij}} = \frac{p_i \cdot p_j}{E_{i}E_{j}} = 1-v_{i}v_{j}
                  \cos\theta_{ij}$,
       ${\xi_{i}} = 1-v_{i} \cos\theta_{i}$, 
       $\gamma_{i} = E_{i}/m_{i} = 1/\sqrt{(1-v_{i}^2)}$,
       $v_{i}$ is the velocity of parton $i$, 
       $\theta_{i}$ is the angle between the direction of motion of
         the soft gluon and the parton $i$, and
       $\theta_{ij}$ is the angle between the partons $i$ and $j$.

%
% Feynman diagram and colour flow for e^+e^-\raq\bar{q}g
%
\begin{figure}
\begin{center} \begin{picture}(360,130)(50,0)
\SetScale{1.0}
% Feynman diagram
\SetOffset(50,0)
\ArrowLine(5,30)(50,60)
\ArrowLine(50,60)(5,90)
\Photon(50,60)(100,60){5}{5}
\ArrowLine(100,60)(150,100)
\ArrowLine(150,20)(100,60)
\Gluon(125,80)(150,80){-3}{3}
\Text(75,-30)[]{{\small (a) Feynman Diagram}}
% Colour Flow
\ArrowLine(250,65)(300,120)
\ArrowLine(300,65)(250,65)
\DashArrowLine(300,0)(250,55){5}
\DashArrowLine(250,55)(300,55){5}
\Text(280,-30)[]{{\small (b) Colour Flow}}
% Labels
\Text(30,85)[]{$\mr{e^+}$}
\Text(30,35)[]{$\mr{e^-}$}
\Text(75,75)[]{$\mr{Z_0/\gamma}$}
\Text(130,95)[]{$\mr{q}$}
\Text(130,25)[]{$\mr{\bar{q}}$}
\Text(310,-5)[]{$\mr{\bar{q}}$}
\Text(310,60)[]{g}
\Text(310,125)[]{q}
\end{picture}
\end{center}\
\captionB{Feynman diagram and colour flow for $\mr{e^+e^- \ra q \bar{q} g}$.}
	{Feynman diagram and colour flow for $\mr{e^+e^- \ra q \bar{q} g}$.}
\label{fig:qqg}
\end{figure}
% End of the Figure
  In general this dipole radiation function can be used to express the current
  squared for a process in the form
\begin{equation}
   \JC^2(q) = \frac{ C_{\mathbf{m}}}{\omega^2}W(\Omega_q),
\label{eqn:current}
\end{equation}
  where  $C_{\mathbf{m}}$ is the colour factor for the tree-level
  process, $\Omega_q$ is the direction of the gluon and $W(\Omega_q)$ is the
  soft gluon radiation pattern.
  For the example we are considering $C_{\mathbf{m}}=C_FN_c$ and the radiation
  pattern is given by
\begin{equation}
  W_{q\bar{q}g}(\Omega_q) =   C_A\left[W_{qg}(\Omega_q)
				+W_{\bar{q}g}(\Omega_q)\right]
              			-\frac1{N_c}W_{q\bar{q}}(\Omega_q).
\end{equation}
  This corresponds to emission of a soft gluon from a colour dipole,
\ie  $W_{qg}$ is emission from the dipole formed by the quark and the
anticolour line of the gluon, $W_{\bar{q}g}$ is emission from the colour
line of the gluon and the antiquark, and $W_{q\bar{q}}$ is emission from
the quark and antiquark. This then shows that the $\mr{q\bar{q}}$ dipole is
negative, which is a problem if we wish to use a probabilistic approach
  to treat the soft gluon radiation.

 These dipole radiation functions can 
 be split into two parts as was done in \cite{Marchesini:1990yk}, 
\ie
\begin{equation}
   W_{ij}(\Omega_q) = W_{ij}^{i}(\Omega_q) +W_{ij}^{j}(\Omega_q),
\end{equation}  
  where 
\begin{equation}
   W_{ij}^{i} = \frac{1}{2\xi_{i}} \left(
  	      1-\frac{1}{\gamma_{i}^2\xi_{i}}+\frac{\xi_{ij}-\xi_{i}}{\xi_{j}}
       		 \right)\!.
\label{eqn:radfunction}
\end{equation}
  This allows us to rewrite the square of the current for 
  $\mr{e^+e^-\ra q\bar{q}g}$, Eqn.\,\ref{eqn:eegcurrent},
  using these radiation functions, in the following form, 
\begin{eqnarray}
 W_{q\bar{q}g}(\Omega_q) = & 2 C_F \left( W^q_{qg} +  W^{\bar{q}}_{\bar{q}g}
				 \right) 
                 +C_A \left( W^g_{g\bar{q}} +  W^{g}_{gq} \right)\nonumber\\
                &  +\frac1{N_c} \left( W^{q}_{qg} -W^q_{q\bar{q}} 
               + W^{\bar{q}}_{\bar{q}g} -W^{\bar{q}}_{\bar{q}q}
                 \right)\!.  
 \label{eqn:radqqg}
\end{eqnarray}
  The last term in Eqn.\,\ref{eqn:radqqg}, and other terms of this type,
  can be neglected for two reasons: firstly it is  
  suppressed by $1/N^2_c$ with respect to the leading-order term; and secondly
  it is dynamically suppressed because it does not contain a 
  collinear singularity in the massless limit. Typically the size of these
  suppressed terms is at most a few percent of the size of the non-$N_c$
  suppressed terms.

  The function $W^i_{ij}$ has a number of important properties which we will
  now consider:
\begin{enumerate}
\item In the massless limit it  contains the collinear singularity as
        $\theta_{i} \rightarrow 0 $,\cite{Marchesini:1990yk}. This can be seen
	by taking the massless limit of Eqn.\,\ref{eqn:radfunction},
\begin{equation}
   W_{ij}^{i} = \frac{1}{2\left(1-\cos\tht_i\right)} \left(
  	      1+\frac{\cos\tht_i-\cos\tht_{ij}}{1-\cos\tht_j}
       		 \right)\!.
\end{equation}
  This shows that the radiation function is singular in the collinear limit,
  \ie as $\tht_i\ra 0$, and not in the other collinear limit $\tht_j\ra0$
  since $\tht_i\ra\tht_{ij}$ as $\tht_j\ra0$.

\item   After averaging over the azimuthal angle of the soft gluon about the
        parton~$i$ the function $W^i_{ij}$
        corresponds to emission in a cone about the direction
        of the parton~$i$ up to the direction of  $j$
        \cite{Marchesini:1984bm,Marchesini:1988cf}. This can be seen
  by writing the angular integral in terms of the polar and azimuthal angles
  of the gluon with respect to the parton~$i$. We can then consider the 
  integral of the radiation
  functions over the azimuthal angle
\begin{equation}
\left\langle W^i_{ij} \right\rangle =\int^{2\pi}_0 
	\frac{d\phi_i}{2\pi}W^i_{ij} = 
	\frac{v_i}{2\xi_i}\int^{2\pi}_0 \frac{d\phi_i}{2\pi}
	\left[\frac{v_i-\cos\tht_i}{1-v_i\cos\tht_i}+
	\frac{\cos\tht_i-v_j\cos\tht_{ij}}{1-v_j\cos\tht_j}\right]\!.
\end{equation}
  The only dependence of the integrand on the azimuthal angle is contained in
  the term $1/(1-v_j\cos\tht_j)$. We can rewrite $\tht_j$ in terms of the
  other angles in the problem. It is easiest to take the $z$-axis along the
  direction of the parton $i$ and define the $xz$ plane to be the plane
  containing $j$. In this co-ordinate system unit
  vectors along the direction of the parton $j$ and the gluon are given by
\begin{subequations}
\begin{eqnarray}
   \underline{\hat{\jmath}} & = & \left( \sin\tht_{ij},\  0,\  \cos\tht_{ij}
				 \right)\!, \\
   \underline{\hat{g}} & = & \left( \cos\phi_i\sin\tht_i,\
			  \sin\phi_i\sin\tht_i,\    \cos\tht_{i} \right)\!.
\end{eqnarray}
\end{subequations}
  This allows us to express $\tht_j$ in terms of the other angles by taking
  the scalar product,
\begin{equation}
    \cos\tht_j=\underline{\hat{\jmath}} \cdot \underline{\hat{g}} = 
	\cos\phi_i\sin\tht_i\sin\tht_{ij}+\cos\tht_{i}\cos\tht_{ij}.
\end{equation}
  The only $\phi_i$ dependent part of the integral can  be written as
\begin{equation}
   I = \int^{2\pi}_0 \frac{d\phi_i}{2\pi} \frac1{1-v_j\cos\tht_j} = 
       \int^{2\pi}_0 \frac{d\phi_i}{2\pi} \frac1{a-b\cos\phi_i},
\end{equation}
  where $a=1-v_j\cos\tht_{i}\cos\tht_{ij}$ and $b=v_j\sin\tht_i\sin\tht_{ij}$,
  as was done in \cite{Ellis:1991qj} for the massless case.
  This integral can be performed via contour
  integration \cite{Ellis:1991qj} giving
\begin{equation}
   I = \frac1{\sqrt{a^2-b^2}}=\frac1{\sqrt{
	\left(\cos\tht_i-v_j\cos\tht_{ij}\right)^2+\left(\sin\tht_i/\gamma_j
	\right)^2}}.
\end{equation}
  This allows us to perform the azimuthal average for the radiation function
  giving
\begin{equation}
\left\langle W^i_{ij} \right\rangle  = 	
	\frac{v_i}{2\left(1-v_i\cos\tht_i\right)}
		\left[\frac{A_i}{\left(v_iA_i+\gamma_i^{-2}\right)}
     			+\frac{B_i}{\sqrt{B_i^2+(\sin\tht_i/\gamma_j)^2}}
\right]\!,
\label{eqn:fullangles}
\end{equation}
   where $A_i=v_i-\cos\tht_i$ and $B_i=\cos\tht_i-v_j\cos\tht_{ij}$. This
   result was first derived in \cite{Marchesini:1990yk} for massive partons.
   If we now take the massless limit of this function we obtain
\begin{equation}
\left\langle W^i_{ij} \right\rangle  = 
	\frac1{2\left(1-\cos\tht_i\right)}
	\left[1+\frac{\cos\tht_i-\cos\tht_{ij}}{|\cos\tht_i-
	\cos\tht_{ij}|}\right]\!,
\end{equation}
   which can be rewritten as
\begin{eqnarray}
\left\langle W^i_{ij} \right\rangle =& 
{\displaystyle\frac1{1-\cos\tht_i}} &
	 \ \ \ \ \ \ \ \ \ \ \ \ \mr{if}\  \tht_i<\tht_{ij}, \nonumber \\
 \ \ \ \ \ \ \ \ \ \ \ \ \ \ \ \ \ \ \ =&   0 & \ \ \ \ \ \ \ \ \ \ \ \  
\mr{otherwise}.
\end{eqnarray}
  We have therefore shown that, after azimuthally averaging, the emission of
  a soft gluon from the parton $i$ can only occur in a cone about the
  direction of
  $i$, with the opening angle of the cone given by the direction of $j$.

  \item If the parton $i$ is massive we should use the
	full azimuthally averaged radiation function, given in
	Eqn.\,\ref{eqn:fullangles}, rather than massless result.
	 This gives two main effects:
	firstly the step function at $\cos\tht_i=\cos\tht_{ij}$ moves to
        $\cos\tht_i=v_j\cos\tht_{ij}$ and the fall-off of the radiation is
	smoothed, \ie rather than being a step-function the fall-off occurs
	over a region in $\cos\tht_i$ of order $\gamma_j^{-1}$.
	Secondly soft radiation in the direction of the parton is reduced, 
        \ie emission within an angle of order $\theta \sim m_i/E_i$
        vanishes \cite{Marchesini:1990yk}. Again rather than a step function
	the soft gluon radiation distribution goes to zero at $\cos\tht_i=v_i$
	with a width, in $\cos\tht_i$, of order $\gamma_i^{-2}$. This full
        radiation
	function cannot be implemented numerically and therefore in practice
	we use the massless form of the function together with the `dead-cone'
        prescription \cite{Marchesini:1990yk} in which there is no emission
	of soft gluons for angles $\theta < m_i/E_i$.

\item	While $W_{ij}/\omega^2$ is Lorentz invariant the individual functions 
	$W^i_{ij}/\omega^2$ and $W^j_{ij}/\omega^2$ are not.
\end{enumerate}

% Angular Ordering Figure
\begin{figure}
\vskip -5mm
\begin{center} 
\begin{picture}(360,110)(0,0)
\SetOffset(100,-10)
\LongArrow(0,60)(150,79.75)
\LongArrow(0,60)(150,40.25)
\Line(0,60)(96.41,99.94)
%\Line(0,60)(96.41,20.06)
\Oval(100,73.17)(27.01,5)(7.5)
%\Oval(100,46.83)(27.01,5)(-7.5)
\Text(223.5,80)[]{1. Direction of the parton}
\Text(205,40)[]{2. Direction of the}
\Text(208,25)[]{ colour partner}
\end{picture}
\vskip -8mm
\end{center}
\captionB{Emission in angular-ordered cones.}
	{Emission in angular-ordered cones.}
\label{fig:cones}
\end{figure}

  If we first define the concept of a {\it colour connected} parton
  we can then use the properties of the radiation functions to look at
  the radiation pattern from $\mr{e^+e^-\ra q\bar{q}g}$.
  Two partons
  are considered to be colour connected if they share the same colour
  line. The colour flow, in the large $N_c$ limit,
  for the process $\mr{e^+e^- \ra q {\bar q} g}$ is
  shown in Fig.\,\ref{fig:qqg}b. 
  The $\mr{{\bar q}}$ and $\mr{g}$ are colour connected and
  the $\mr{q}$ and $\mr{g}$
  are colour connected, while the $\mr{{\bar q}}$ and $\mr{q}$ are not colour
  connected. Each quark only has one colour-connected partner in a
  given Feynman diagram and each gluon has two. Colour-connected
  partners are defined at each stage of the iterative parton-shower
  procedure. If the final-state $\mr{q}$ were to emit another gluon,
  $\mr{g_2}$,
  the new final-state $\mr{q}$ would be colour connected to $\mr{g_2}$ and no
  longer to $\mr{g}$. The gluons
  $\mr{g}$ and $\mr{g_2}$ would then also be colour connected.

  We see from Eqn.\,\ref{eqn:radqqg} that after neglecting the
  final term, using the properties of the function $W^i_{ij}$, and
  averaging over the azimuthal angle of the gluon about a parton,
  the radiation can only occur in a cone about the
  direction of the parton up to the direction of its colour
  partner. This is shown in Fig.\,\ref{fig:cones}. We can draw a
  cone around parton one with half-angle given by the angle between
  the momenta of partons one and two. The emission from parton one
  within the cone defined by its colour-connected partner, parton two,
  is called angular-ordered emission.

\begin{figure}[t]
\includegraphics[angle=90,width=0.9\textwidth]{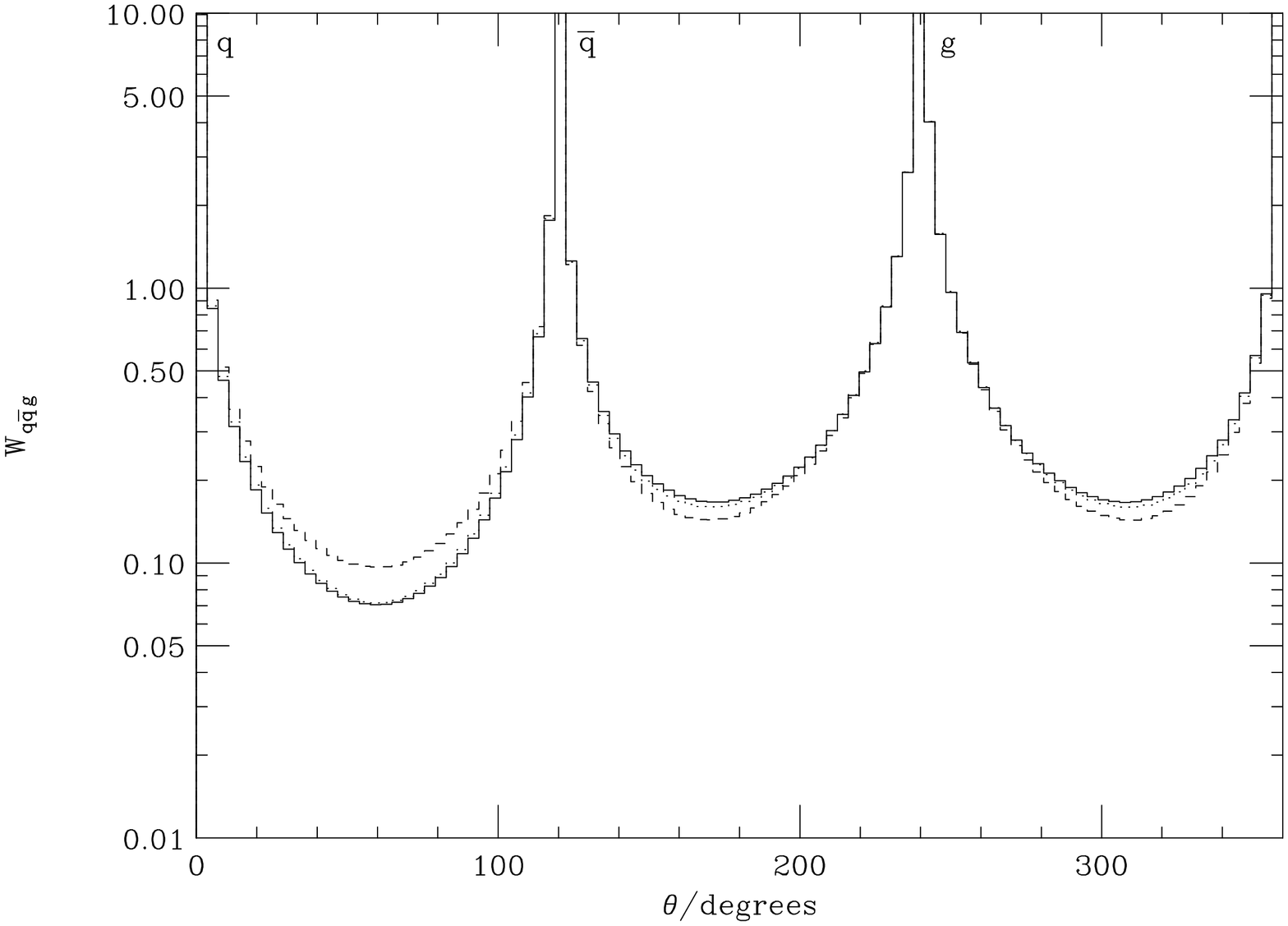}
\captionB{Radiation pattern from the process $\mr{e^+e^-\ra q \bar{q}g}$.}
	{Soft gluon radiation pattern from the process 
	 $\mr{e^+e^-\ra q \bar{q}g}$. The hard quark, antiquark and gluon
	 directions were fixed at $\tht=0^0$, $\tht=120^0$ and $\tht=240^0$
	 respectively. The solid line gives the full radiation pattern and the
	 dashed line gives the angular-ordered radiation pattern after the
	 $1/N_c^2$-suppressed terms have been neglected. The dotted line
 	 gives the improved angular-ordered approximation.}
\label{fig:eegpatt}
\end{figure}

  The radiation pattern for the process $\mr{e^+e^-\ra q \bar{q}g}$ is shown
  in Fig.\,\ref{fig:eegpatt}. The solid histogram gives the full radiation
  pattern for the soft gluon given in Eqn.\,\ref{eqn:radqqg} using the
  non-azimuthally averaged radiation functions given in
  Eqn.\,\ref{eqn:radfunction}.
  The dashed line gives the radiation pattern in the angular-ordered
  approximation where we have dropped the $1/N_c^2$-suppressed terms in
  Eqn.\,\ref{eqn:radqqg} and used the azimuthal-averaged radiation 
  function. \linebreak
  Azimuthal correlations can be included inside the angular-ordered cones
  after the full parton shower has been generated 
  \cite{Knowles:1988hu:Knowles:1988vs:Knowles:1988cu}. This leads to the
  improved angular-ordered approximation where we use the full radiation
  function, Eqn.\,\ref{eqn:radfunction}, inside the angular ordered cone and
  the azimuthal average, \ie zero, outside the cone.
  This gives the replacement \cite{Marchesini:1988cf}
\begin{equation}
 	W^i_{ij} \longrightarrow W^i_{ij}\Theta\left(\tht_i
				-\tht_{ij}\right)\!.
\end{equation}
  This is shown by a dotted line in Fig.\,\ref{fig:eegpatt}.

  The angular-ordering procedure is one way of implementing the phenomenon of 
  {\it colour coherence}.
  The idea of colour coherence is that if we consider a pair of partons and a
  soft gluon at a large angle to them, with respect to the angle between the
  parton pair,  then this gluon can only resolve the total colour charge of
  the pair of smaller angle partons. It is therefore as
  if the larger-angle soft gluon was emitted before the smaller angle
  branchings. There have been a number of experimental studies of colour 
  coherence
  effects. In particular the ``string effect'' in $\mr{e^+e^-}$ collisions
  \cite{Bartel:1984ij:Bartel:1985zx:Aihara:1985du:Akrawy:1991ag:Akers:1995xs},
  where there is a suppression of soft QCD radiation between the two
  quark jets in three jet events, has been studied.
  Fig.\,\ref{fig:eegpatt} shows that both angular-ordering 
  approximations reproduce the ``string effect'', \ie the large dip in the
  soft gluon radiation about $\tht=60^0$.
 There have also been studies of 
  colour coherence effects between the initial and final
  states in hadron--hadron collisions \cite{Abbott:1997bk,Abe:1994nj}.

  In general, in the soft limit the matrix element can be written as
\begin{equation}
  \sum_{\mr{spins}}|\me_{n+1}|^2 = 
	\frac{g^2_s}{\omega^2}\sum_{\mr{spins}}|\me_{n}|^2
	                         \sum_{i,j\neq i}^nC^i_{ij}W^{i}_{ij},
\label{eqn:softme}
\end{equation}
  where $C^i_{ij}$ are the colour factors and $W^{i}_{ij}$ the dipole
  radiation functions. After neglecting the $1/N^2_c$-suppressed terms the
  colour factors are $C^q_{qj}=2C_F$ for a quark and its colour partner, 
  $C^g_{gj}=C_A$ for a gluon and its colour partner and zero otherwise.
  The tree-level colour factor $C_{\mathbf{m}}$ has been absorbed
  into the tree-level matrix element squared $|\me_{n}|^2$. We have now
  obtained a factorized form for the matrix element squared for the
  emission of a soft gluon by neglecting the $1/N_c^2$-suppressed terms.

  As in the collinear limit, we also need to consider the factorization of the
  phase space. As before the phase space for the leading-order
  process is given by Eqn.\,\ref{eqn:lophase} and the phase space for the
  process after the radiation of a soft gluon by Eqn.\,\ref{eqn:nlophase}.
  In the soft limit the momentum of the quark before and after the radiation
  of the gluon are the same and therefore
\begin{equation} 
   \frac{d^3p_i}{2E_i(2\pi)^3} \stackrel{w\ra0}{=}
	 \frac{d^3p_j}{2E_j(2\pi)^3}.
\end{equation}
  The momentum of the radiated soft gluon can be written in terms of its
  energy giving
\begin{equation}
\frac{d^3p_k}{2E_k(2\pi)^3} = \omega d\omega \frac{d\Omega}{16\pi^3}.
\end{equation}
  Hence, as in the collinear limit, the phase space factorizes giving
\begin{equation}
   d\Phi_{n+1} = d\Phi_n \omega d\omega \frac{d\Omega}{16\pi^3}.
\label{eqn:softphase}
\end{equation}
  This allows us to write the cross section for the emission of a soft gluon
  in the following factorized form using Eqns.\,\ref{eqn:softme} and
  \ref{eqn:softphase},
\begin{equation}
 d\sigma_{n+1} =d\sigma_n  \frac{\al_s}{2\pi}\frac{d\omega}{\omega}
		\sum_{i,j\neq i}\frac{d\Omega_i}{2\pi}C^i_{ij}W^{i}_{ij}.
\end{equation}
  After averaging over the
  azimuthal direction
\begin{equation}
  d\sigma_{n+1}= d\sigma_n  
		\frac{\al_s}{2\pi}\frac{d\omega}{\omega}d\!\cos\theta
		\sum_{i,j \ne i} C^i_{ij}\left\langle W^i_{ij}\right\rangle\!.
\label{eqn:softfact}
\end{equation}
  Now if we use the angle for the evolution, or as is done in practice 
\begin{equation}
 \zeta = \frac{p_j \cdot p_k}{E_jE_k} \simeq 1 -\cos\theta, 
\end{equation}
   for the branching $i \ra j k$, in the collinear limit
   we can replace $dt/t$ with $d\zeta/\zeta$ in Eqn.\,\ref{eqn:collinearfact}.
   If we take the soft limits
   of the splitting functions we obtain
\begin{subequations}
\begin{eqnarray}
      \lim_{z\ra1} P_{qq} &=& \frac{2C_F}{1-z}, \\
      \lim_{z\ra1} P_{gg} &=& \frac{C_A}{1-z}.
\end{eqnarray}
  Hence both of these splitting functions are singular in the soft limit,
  \ie as $z\ra1$. However
  the splitting function for $\mr{g\ra q\bar{q}}$ is non-singular in the soft
  limit.
\end{subequations}
  The energy of the soft gluon is given by $\omega=(1-z)E_i$ where $E_i$ is
  the momentum of the particle before the gluon radiation. Hence 
  $d\omega/\omega=dz/(1-z)$. This means we can replace 
  $C^i_{ij}d\omega/\omega$ in Eqn.\,\ref{eqn:softfact} with $P(z)dz$ in the
  soft limit. We can therefore combine Eqn.\,\ref{eqn:collinearfact} and
  Eqn.\,\ref{eqn:softfact} to give the following equation which correctly
  includes both the soft and collinear singularities:
\begin{equation}
 d\sigma_{n+1} = d\sigma_n\frac{d\zeta}{\zeta}\frac{\al_s}{2\pi}P_{ji}(z)dz.
\end{equation}

  The evolution equation, Eqn.\,\ref{eqn:evolve2}, can be written in terms of
  these angular variables instead of the virtualities we used before, 
\begin{equation}
     d(x,\zeta)  =  \Delta(\zeta) d(x_0,\zeta_{0})
			+ \int^\zeta_{\zeta_0}\frac{d\zeta'}{\zeta'}
					\frac{ \Delta(\zeta)}{ \Delta(\zeta')}
			\int^1_0 dz\frac{\al_s}{2\pi} P(z) d(x/z,\zeta').
\end{equation}
  The new Sudakov-like form factor defined in terms of angles is given by
\begin{equation}
 \Delta(\zeta) = \exp\left[-\int^\zeta_{\zeta_0} \frac{d\zeta'}{\zeta'}
	\int^1_0 dz \frac{\al_s}{2\pi} P(z) \right]\!.
\end{equation}
  As before, we need to choose an infra-red cut-off to regularize the
  integral, \ie the Sudakov form factor becomes
\begin{equation}
 \Delta(\zeta) = \exp\left[-\int^\zeta_{\zeta_0} \frac{d\zeta'}{\zeta'}
			    \int^{1-\epsilon}_\epsilon dz
			\frac{\al_s}{2\pi} P(z) \right]\!.
\end{equation}
  Whereas in the collinear case we imposed a cut-off on the virtuality of the
  parton, here we impose a cut-off on the minimum angle of the emission. It
  is easier however to impose a cut-off on some energy-like variable and we
  therefore choose the variable
\begin{equation}
 \tilde{t}=E^2\zeta,
\end{equation}
  as the evolution variable. It was shown in \cite{Marchesini:1990yk} that
  for particles with a mass $m_i$ and energy $E_i$
  there is no soft radiation for angles $\theta\lesssim m_i/E_i$.
  This corresponds to
\begin{equation}
  \tilde{t}\lesssim m^2_i.
\end{equation}
  Hence if we wish to impose a cut-off, $t_0$, on the virtuality of the
  particle we can impose
\begin{equation}
  \tilde{t}\geq t_0.
\end{equation}
  This is the simplest choice of the cut-off, however by using a different
  choice we can include the next-to-leading-log
  terms (at least for large $x$) as well \cite{NLO}.
  To do this we must impose a cut-off on the transverse momentum 
  \cite{Marchesini:1984bm}
\begin{equation}
    p^2_T=\frac{{\bf q}_j^2 {\bf q}_k^2\sin^2\theta}{{\bf q}_i^2}=
	\frac{\left[{\bf q}_j^2 {\bf q}_k^2 -E^2_jE^2_k(1-\zeta)^2\right]}
	{{\bf q}_i^2}\ge t_0,
\end{equation}  
  where ${\bf q}_i$, ${\bf q}_j$ and  ${\bf q}_k$ are the three-momenta of
  the partons $i$, $j$ and $k$, respectively. If we now assume that the
  particles produced in the branching are massless, this implies
\begin{equation}
  z,(1-z) > \sqrt{\frac{t_0}{2\tilde{t}}}.
\end{equation}
  To next-to-leading-log accuracy we can neglect the factor of $1/\sqrt{2}$ 
  which gives the following cut-off condition for $z$,
\begin{equation}
   \sqrt{t_0/\tilde{t}}<z<1-\sqrt{t_0/\tilde{t}}.
\end{equation}
  These two limits only allow some phase space for the branching if
  $\tilde{t}\ge4t_0$.
  This gives the Sudakov form factor in terms of this new variable
\begin{equation}
 \Delta_i(\tilde{t}) = \exp\left[-\sum_j\int^{\tilde{t}}_{4t_0}
	 \frac{d\tilde{t}'}{\tilde{t}'}
		    \int^{1-\sqrt{t_0/\tilde{t}'}}_{\sqrt{t_0/\tilde{t}'}}
			\frac {dz}{2\pi}
		\al_s\!\left(z^2(1-z)^2\tilde{t}'\right) P_{ji}(z) \right]\!.
\end{equation}

  Although we have averaged over the azimuthal angle for the emission
  of the gluon in both the soft and collinear cases, azimuthal effects,
  \eg due to spin correlations, can be included
  \cite{Knowles:1988hu:Knowles:1988vs:Knowles:1988cu} after the
  full parton shower has been generated.

  In processes where there is more than one Feynman diagram it is
  possible for the colour flows in the diagrams to be different. This
  leads to so called ``non-planar'' terms from the interference
  between diagrams with different colour flows.
  These are not positive definite and hence cannot be
  interpreted in a probabilistic way for implementation in the Monte Carlo
  procedure. The ``non-planar'' terms are always suppressed by inverse
  powers of $N_c$. A procedure must be adopted to split up the
  ``non-planar'' parts of the tree-level matrix element to give
  redefined planar terms with positive-definite coefficients that can
  be used in the Monte Carlo procedure. Such a procedure was proposed
  in \cite{Marchesini:1988cf} and shown to work correctly for all QCD
  processes. However,  as shown in \cite{Odagiri:1998ep}, this is
  inadequate for MSSM processes and hence
  a new procedure was proposed, which we  adopt here. In this
  procedure the ``non-planar'' parts of the matrix element are split up
  according to 
\begin{equation}
|\overline{M}|^2_{\mr{full},i} =
\frac{|\overline{M}|^2_i}{|\overline{M}|^2_{\mr{planar}}}
 | \overline{M}|^2_{\mr{tot}},
\label{eqn:kosuke}
\end{equation}
  where $|\overline{M}|^2_i$ is the matrix element squared for the $i$th
  colour flow, ${|\overline{M}|^2_{\mr{planar}}}$ is the sum of the matrix
  elements squared for the planar colour flows, and
  $|\overline{M}|^2_{\mr{tot}}$ is the total matrix element squared.
  This ensures that the redefined
  planar terms, $|\overline{M}|^2_{\mr{full},i}$, are positive definite and 
  have the correct pole structure. This can be
  implemented numerically.  

  In this section we have explained how by using a Markov branching
 procedure we can resum both the soft and collinear singularities in QCD. 
 In the next section we will
 discuss how this procedure can be implemented numerically.

%
%  Sub section on the Monte Carlo procedure
%
\subsection{Monte Carlo Procedure}

  The ratio of Sudakov form factors $\Delta(t_1)/\Delta(t_2)$ gives the
  probability of evolving downwards from the scale $t_1$ to the scale $t_2$
  without resolvable emission. We can therefore implement the parton-shower
  algorithm numerically by solving
\begin{equation}
 \frac{\Delta(t_1)}{\Delta(t_2)} = \mathcal{R},
\label{eqn:branching1}
\end{equation}
  where $\mathcal{R}$ is a random number uniformly distributed between 0 and
  1.\footnote{This interval includes both of the end points,
		    \ie $0\le \mathcal{R}\le1$.}
  Using Eqn.\,\ref{eqn:branching1}, given the initial scale
  $t_1$, we can generate the scale of 
  the next branching. If the value of $\mathcal{R}$ is smaller than
  $\Delta(t_1)$  there is no solution of Eqn.\,\ref{eqn:branching1} for
  $t_2>t_0$. This is because $\Delta(t)$ increases as the scale decreases and
  $\Delta(t_0)=1$. This procedure correctly gives the
  probability that there is no resolvable branching.

 If a solution  $t_2>t_0$ exists we need to generate the momentum fraction 
 $z=x_2/x_1$ of the parton after the branching. This is done by solving
\begin{equation}
  \int^z_{\sqrt{t_0/\tilde{t}'}} dz'\frac{\al_s}{2\pi}P(z') =
  \mathcal{R}' \int^{1-{\sqrt{t_0/\tilde{t}'}}}_{\sqrt{t_0/\tilde{t}'}} dz'
 \frac{\al_s}{2\pi}P(z'),
\label{eqn:momentafrac1}
\end{equation} 
  where $\mathcal{R}'$ is a second uniformly distributed random number in the
  interval $[0,1]$.

  We have discussed the parton-shower procedure for the final state, \eg in 
  $\mr{e^+e^-\ra q\bar{q}}$. However in processes involving hadrons in the
  initial state the QCD radiation from the incoming partons must also be
  generated, \eg in  Drell-Yan $\mr{q\bar{q}\ra\ell^+\ell^-}$. There are two
  ways in which this can be achieved:
\begin{description}
  \item[Forward Evolution.]

     Starting at the cut-off scale with the parton distribution function,
  $f(x_0,t_0)$, which is evolved to the scale of the hard collision. This
  starts with the initial momentum fraction $x_0$ and generates the momentum
  fraction $x_n$ of the partons in the hard collisions after $n$ branchings.

  \item[Backward Evolution.]

    Starting with the momentum fractions of the partons involved in the hard
  collision the partons are evolved backwards to give the parton branchings
  from which they came.
      
\end{description}	
 
  The problem is that forward evolution will often generate momentum
  fractions, $x_n$, which give a small contribution to the total cross
  section and are therefore rejected,\linebreak \ie~forward evolution is
  inefficient. However with backward evolution we can generate those momentum
  fractions which give large
  contributions to the cross section, hence this is more efficient.

  The backward evolution is set up in the following way. First we can write
  an  evolution equation for the parton distribution functions in the same 
  way as for the
  fragmentation functions in Eqn.\,\ref{eqn:evolve2},
\begin{eqnarray}
  t  \frac{\partial}{\partial t} \left(\frac{f}{\Delta}\right) & = & 
		\frac1{\Delta}\int^1_0\frac{dz}{z}\frac{\al_s}{2\pi} 
			P(z)f(x/z,t), \\
     f(x,t) & = &  \Delta(t) f(x,t_0)+ \int^t_{t_0}\frac{dt'}{t'}
					\frac{ \Delta(t)}{ \Delta(t')}
					\int^1_0 \frac{dz}{z}
				\frac{\al_s}{2\pi} P(z) f(x/z,t').
\label{eqn:evolve3}
\end{eqnarray}
  We can now 
  define $df(x,t_2)$ as the fraction of partons with momentum fraction $x$ and
  scale $t_2$ which came from branchings between the scale $t$ and 
  $t+\delta t$. This
  gives the probability of no branching between the scales $t_1$ and $t_2$,
\begin{equation}
  \Pi = 1 - \int^{t_2}_{t_1}df.
\label{eqn:backevolve}
\end{equation}
  Now from the evolution equation, Eqn.\,\ref{eqn:evolve3}, 
\begin{eqnarray}
 f(x,t_2)df &=& \frac{\delta t}{t}\frac{ \Delta(t_2)}{ \Delta(t)}
		\int^1_0 \frac{dz}{z}\frac{\al_s}{2\pi} P(z) f(x/z,t), \\
            &=&\delta t\frac{\partial}{\partial t}
	\left[\frac{ \Delta(t_2)}{ \Delta(t)}f(x,t)\right]\!.
\end{eqnarray}
  Hence we can substitute into Eqn.\,\ref{eqn:backevolve} for $df$ and perform
  the integral to give the probability of no branching between the scales
  $t_1$ and $t_2$, $\Pi$, in a more useful form
\begin{equation} 
  \Pi = \frac{f(x,t_1)\Delta(t_2)}{f(x,t_2)\Delta(t_1)}.
\end{equation}
  Therefore instead of the Sudakov form factor for the backwards evolution we
  should use $\Delta(t)/f(x,t)$. We can then generate the correct scale for
  the branching by solving Eqn.\,\ref{eqn:branching1} with this modified
  Sudakov form factor, \ie by solving $\Pi=\mathcal{R}$ where $\mathcal{R}$
  is a random number uniformly distributed between 0 and 1.

  As in the forward evolution case after we have generated the scale of the
  next branching we need to find the momentum fractions of the partons
  produced in the branching. In the forward evolution case this is done by
  solving Eqn.\,\ref{eqn:momentafrac1}. This equation is modified for the
  backwards evolution algorithm to give
\begin{equation}
  \int^z_{\sqrt{t_0/\tilde{t}'}} dz'\frac{\al_s}{2\pi}\frac{P(z')}{z'}
 f(x_2/z',t_1) =
  \mathcal{R}' \int^{1-{\sqrt{t_0/\tilde{t}'}}}_{\sqrt{t_0/\tilde{t}'}} dz'
 \frac{\al_s}{2\pi}\frac{P(z')}{z'}f(x_2/z',t_1),
\label{eqn:momentafrac2}
\end{equation} 
  where again $\mathcal{R}'$ is a random number uniformly distributed
  between 0 and 1.

\subsection{Summary}

  We have explained how the cross section for $n+1$ partons factorizes in
  both the
  collinear and soft limits into a universal splitting term and the
  cross section for $n$ partons. Both of these limits can be implemented
  by using angles as the evolution variable in a Markov branching procedure.
  We start at the hard
  cross section, normally with a two-to-two process. The maximum angle
  of emission from a parton is set by the direction of the colour partner.
  We then generate some smaller angle parton, \eg a gluon from a quark. Then
  we repeat the procedure, \eg the gluon's colour partner is now the
  colour partner of the original quark, and its anticolour partner the
  quark, and the colour partner of the quark is the gluon. One of the
  partons will now radiate with the maximum angle given by the
  direction of the new colour
  partner and so on until the cut-off below which emission does not occur is
  reached. This procedure  resums both the leading soft and collinear
  singularities. If the initial state contains partons the radiation from
  these incoming partons can be generated by using a backward evolution
  algorithm which is more efficient than forward evolution in this case.
 
  After the parton-shower phase we are left with partons with a low
  virtuality which must then form the observed hadrons. In the next section
  we will discuss the various phenomenological models for the hadronization
  process which are currently used in Monte Carlo event generators.

%
%   Section on the hadronization models
%
\section{Hadronization}
\label{sect:montehadron}
  There are a number of different phenomenological models of the
  hadronization process which are used in different Monte Carlo event
  generators:\footnote{A more detailed review of the various hadronization
			models
		        can be found in \cite{Ellis:1991qj,Webber:1986mc}.}
\begin{description}
% Independent fragmentation
\item[Independent Fragmentation.] This was the first proposed
  hadronization model \cite{Field:1977ve} and is the simplest scheme. In this
  method, for example for the hadronization of a quark,
   a quark--antiquark pair is created from the vacuum, the original quark and
  the antiquark then form a meson. The procedure is then repeated for the
  quark which was created and so on until the energy of the remaining quark
  falls below some cut-off. This model leads to violations of energy and
  momentum conservation which must be corrected after the hadronization phase
  is finished. The colours and flavours of the left-over partons must also be
  neutralized at this stage.
% String
\item[String Model.] In this model \cite{Andersson:1983jt} the quark
  and  antiquark produced in
  $\mr{e^+e^-}$ collisions are assumed to be joined by a relativistic
  string. As the quark and antiquark move apart the string breaks via
  the production of a $\mr{q\bar{q}}$ pair in the colour field of the string.
  The original quark is now connected by a string to the produced antiquark
  and the original antiquark to the produced quark. This
  procedure is repeated until there is insufficient energy to break the colour
  strings any further. The $\mr{q\bar{q}}$ pairs connected by the strings
  then give the observed hadrons.
% Cluster Model
\item[Cluster Model.] This model is based on the idea of colour
  preconfinement \cite{Bassetto:1979vy:Marchesini:1981cr}.
  This suggests that if we consider the pairs
  of colour-connected partons left after the parton-shower phase they
  have a mass spectrum which falls rapidly at high masses, is
  independent of $Q^2$ and universal (see Fig.\,\ref{fig:cluster}a).
  The model then decays these
  clusters into the observed hadrons. As this is the model used in the
  HERWIG event generator we will now discuss it in more detail.  
\end{description} 

  The cluster model proceeds in the following way. The gluons left
  after the end of the parton-shower phase are non-perturbatively split
  into $\mr{q\bar{q}}$ pairs. The colour-connected quarks and antiquarks are
  then formed into colour-singlet clusters.  The cluster mass
  spectrum is shown in Fig.\,\ref{fig:cluster}a. The
  exact form of this spectrum will depend on the QCD scale $\Lambda$,
  the cut-off scale $t_0$ and the mechanism used to split the
  gluons into $\mr{q\bar{q}}$ pairs. Fig.\,\ref{fig:cluster}a shows the 
  spectrum for a low value of the cut-off $t_0$. Most clusters have
  masses of a few \gev\  and it is therefore reasonable to assume they are
  superpositions of the known hadrons. In general, as the clusters are too
  massive to be any of the observed hadrons they are  
  assumed to decay
  into a pair of hadrons, either two mesons or a baryon and an antibaryon,
  with the type of hadron determined by the available density of states,
  \ie phase space times the spin degeneracy.

  A simple extension of this model is used for hadron remnants. If we
  consider the example of a collision in which a valence quark in the proton
  participates in a hard process, the two remaining valence quarks are left
  in the final state. These valence quarks are paired up into a ``diquark''
  which, in the planar approximation, carries an anticolour index and can be
  treated like an antiquark. The resulting
  cluster has baryonic quantum numbers and decays into a baryon and a meson.

  The procedure for selecting the hadrons produced in these cluster decays
  works as follows \cite{Webber:1984if}. We will consider the procedure for a
  cluster containing a quark--antiquark pair  $\mr{q_1\bar{q}_2}$, where
  $1,2$ are any of the quark flavours d, u, s, c and b
  which hadronize before decaying. If the cluster is too light to decay into
  two hadrons  it is taken to represent the lightest single hadron of its
  flavour. Its mass is shifted to the appropriate value by an exchange of
  momentum with a neighbouring cluster in the jet.
  Those clusters massive enough to decay into two hadrons decay
  into pairs of hadrons selected in the following way.
  Another flavour $\mr{q_3}$ or $\mr{d_3}$ is randomly selected where 
  $\mr{q_3=\ u,\ d,\ s}$ is one of the three light quark
  flavours and $\mr{d_3}$ is one of the six corresponding diquarks, \ie 
  $\mr{dd}$, $\mr{du}$, $\mr{ds}$, $\mr{uu}$, $\mr{us}$, $\mr{ss}$. The 
  flavours
  of the decay products are  taken to be either $\mr{q_1\bar{q}_3}$ and 
  $\mr{q_3\bar{q}_2}$, a two-meson decay, or
  $\mr{q_1d_3}$ and $\mr{\bar{d}_3\bar{q}_2}$, a baryon--antibaryon decay.
  Each decay product is then randomly
  selected from a list of resonances with the correct
  flavours. A weight, $W$, for a given pair of resonances is  calculated by
  taking the available phase space weighted with the spin degeneracy. This is
  compared with a random number, \ie the pair of hadrons is accepted if
\begin{equation}
    W \geq \mathcal{R}\times W_{\mr{max}},
\end{equation}
  where $W_{\mr{max}}$ is the maximum possible weight and $\mathcal{R}$ is a
  uniformly distributed random number in the range $[0,1]$. If this weight is
  rejected the whole procedure is repeated.  
  In the original model \cite{Webber:1984if,Field:1983dg} each
  cluster was assumed to decay isotropically in the rest frame of
  the cluster into a pair of
  hadrons. However, in the current implementation of the model \cite{HERWIG61}
  hadrons containing quarks from the perturbative stage of the event continue
  in the same direction, in the cluster rest frame, as the original quark. 

  While it is reasonable to assume that low mass clusters are superpositions
  of hadron resonances there is a small fraction of high mass clusters for
  which this is not a reasonable approximation. These clusters must first be
  split using a  string-like mechanism \cite{Webber:1984if}, into lighter
  clusters, before they are  decayed into hadrons.   

%
%  Section on the different event generators
%
\section{Monte Carlo Event Generators}

  There are a number of Monte Carlo event generators currently
  available which implement different hadronization models and treat
  the parton-shower phase in different ways. In general there are
  three main types of Monte Carlo event generator depending on which
  hadronization model is used.

  The ISAJET event generator \cite{Baer:1999sp} uses the original independent
  fragmentation model of \cite{Field:1977ve}. However the treatment of
  the parton-shower phase does not include colour coherence
  effects. This model was successful in explaining moderate energy
  $\mr{e^+e^-}$ data but gives poor agreement with the current LEP and
  Tevatron data. 
  ISAJET is still commonly used, particularly for studies of supersymmetric
  processes, due to the large number of SUSY production processes and decays 
  which are implemented.

  The JETSET event generator \cite{Sjostrand:1994yb} 
  uses the string hadronization model and a
  final-state parton shower which includes colour coherence
  effects. However, colour coherence effects are only partially included in
  the initial-state parton shower via a veto algorithm. This simulation
  is in impressive agreement with the experimental measurements
  of hadronic final states, particularly in $\mr{e^+e^-}$ collisions, up to
  the highest energies currently studied.
  JETSET is  used to
  perform the hadronization for a range of event generators, PYTHIA 
  \cite{Sjostrand:1994yb}, LEPTO \cite{Ingelman:1997mq} and 
  ARIADNE \cite{Lonnblad:1992tz}, which uses a different formalism for the 
  parton-shower phase.
  There is also a supersymmetric extension, SPYTHIA \cite{Mrenna:1997hu},
  and a simulation of SUSY in $\mr{e^+e^-}$ collisions,
  SUSYGEN \cite{Katsanevas:1997fb}, which
  use JETSET to perform the parton-shower and hadronization phases.
 
  The HERWIG event generator uses a cluster hadronization model with
  full treatment of colour coherence effects in both the initial-
  and final-state parton showers. The agreement with the $\mr{e^+e^-}$ data
  is not as good as with the string hadronization model used in
  JETSET, however there are fewer adjustable parameters in the model.
  HERWIG provides better agreement, in hadron--hadron 
  collisions, for observables which are sensitive to colour coherence effects.

  In the next two sections we will discuss how to implement \rpv\  SUSY
  processes into the HERWIG event generator. 

%
%  Angular Ordering in R-parity violation
%
\section[Angular Ordering in \rpv]{Angular Ordering in \boldmath{\rpv}}
\label{subsect:monteRPVangles}

  In Standard Model and MSSM processes, apart from
complications involving processes where there are ``non-planar'' terms 
\cite{Odagiri:1998ep}, the angular-ordering procedure is relatively
straightforward to implement. However in \rpv\   SUSY there are additional
complications. 

  The lepton number violating processes, which come from
the first two terms in the superpotential, Eqn.\,\ref{eqn:Rsuper1},
 have colour flows which are
the same as those which occur in the MSSM. 
On the other hand the baryon number violating interactions,
which come from the third  term in Eqn.\,\ref{eqn:Rsuper1}, have a very
different colour structure involving the totally antisymmetric 
tensor, $\epsilon^{c_1c_2c_3}$. We look first at the colour structure of
the various baryon number violating decays which we include
in the Monte Carlo simulation and then at the structure of the various hard
scattering processes. 

\subsection{Decays}
 From the point of view of the colour structure 
 there are three types of baryon number violating decays which we
 include in the Monte Carlo simulation:
\begin{enumerate} 
 \item two-body \bv\  decay of an antisquark to two quarks or a
 squark to two antiquarks;
 \item three-body \bv\  decay of a colourless sparticle, \ie 
        neutralino and chargino, to three quarks or antiquarks;
 \item three-body \bv\  decay of the gluino to three quarks or antiquarks.
\end{enumerate}

  In general it is possible to consider, for example, the decay
  of a neutralino to three quarks as either a three-body decay or two
  sequential two-body decays, of the neutralino to an antisquark and a quark,
  and then of the antisquark to two quarks. If either of the two sequential
  two-body decays are kinematically forbidden, \ie they can only proceed if
  the internal
  particle in the three-body decay is off-shell, we consider the decay to be
  three-body, otherwise we
  treat the decay as two sequential two-body decays.

  The problem is how to implement the angular-ordering procedure
  for these processes. We shall consider these processes using the
  eikonal current with an arbitrary number of colours as was done in
  Section~\ref{subsubsect:montesoft} for the process
  $\mr{e^+e^-\ra q\bar{q}g}$.
  In these \rpv\  processes this means we need to consider the decay of an
  antisquark to $(N_c-1)$ quarks and of the neutralino, chargino and
  gluino to $ N_c$ quarks. We also have to use the generalization
  to $N_c$ colours of the antisymmetric tensor, \ie 
  $\epsilon^{c_1 \ldots c_{N_c}}$.

\subsubsection{Squark Decays}

  For the decay of an antisquark to $(N_c-1)$ quarks the leading
  infrared contribution to the soft gluon distribution 
  has the factorized form
\begin{equation}
    {\bf {\cal M}} (p_0,p_1,p_2,\ldots\!\,,p_{N_c-1};q) = g_s {\bf m}
   (p_0,p_1,p_2,\ldots\!\,,p_{N_c-1}) \cdot \JC(q),
\end{equation}
where
\begin{itemize}
  \item $ {\bf m} (p_0,p_1,p_2,\ldots\!\,,p_{N_c-1})$ is the tree-level
         matrix element for an antisquark, with momentum $p_0$, to
         decay to $N_c-1$ quarks, with momenta $p_1,\ldots\!\,,p_{N_c-1}$.
  \item ${\bf {\cal M}} (p_0,p_1,p_2,\ldots\!\,,p_{N_c-1};q)$ is the
	tree-level matrix element for the decay of an antisquark to
	$N_c-1$ quarks including
        the emission of an extra soft gluon with momentum~$q$. 
  \item $ \JC(q)$ is the non-Abelian semi-classical current for
        the emission of the soft gluon, with momentum $q$, from the hard
	partons.
  \item $c_0$ is the colour of the decaying antisquark and $c_1,
        \ldots\!\,, c_{N_c-1}$ are the colours of the quarks.
\end{itemize}

  Again the current, $ \JC(q)$, is given by,
  ${\displaystyle \JC(q) = \sum_{s=1,2}  \JC^{b,\mu}(q) 
\varepsilon^*_{\mu,s}}$ where
\begin{equation}  
   \JC^{b,\mu}(q) = \left(\frac{p_{0}^{\mu}}{p_0 \cdot q}\right) {\bf
                                 t}_{c_0c_{0}'}^{b,\mr{\tilde{q}^{*}}}
                                 \epsilon^{c_{0}'c_1 \ldots c_{N_c-1}}
                          +\sum_{i=1}^{N_c-1} \left(\frac{p_{i}^{\mu}}{p_i
   			 \cdot q}\right) {\bf t}_{c_ic_{i}'}^{b,\mr{q}_i}
			 \epsilon^{c_{0} \ldots c_{i}' \ldots c_{N_c-1}},
\end{equation}
  $b$ and $\mu$ are the colour and Lorentz indices of the emitted
  gluon, and
  ${\bf t}^{b,\mr{\tilde{q}^{*}}}$ and ${\bf t}^{b,\mr{q}_{i}}$ are the colour
  matrices of the antisquark and quarks, respectively.

  We can now obtain the soft gluon distribution simply by squaring the
current,
\begin{equation}  
   \JC^{2}(q) =  -C_{F}N_{c}(N_{c}-2)!  \left[
                      \displaystyle{\sum_{i=1}^{N_c-1}}
                      \left( \frac{p_{0}}{p_0
                   \cdot q}-\frac{p_{i}}{p_i\cdot q} \right)^2
                   + \displaystyle{\sum_{i=1}^{N_c-2}}
                     \displaystyle{\sum_{j>i}^{N_c-1}} \left(
                   \frac{p_{i}}{p_i
                   \cdot q}-\frac{p_{j}}{p_j\cdot q} \right)^2\right]\!.
\end{equation}

  This can be expressed in terms of the radiation functions,
  as in Eqn.\,\ref{eqn:current}. The tree-level colour
  factor is now
  $C_{\mathbf{m}} = \epsilon^{c_{0} \ldots c_{N_c-1}} \epsilon^{c_{0}
\ldots c_{N_c-1}} = N_{c}!$, where we have not averaged over the initial
  colours, and the radiation pattern is given by
\begin{equation}  
   W(\Omega_q) = \frac{-\omega^2 C_{F}}{(N_c-1)} \left[ \sum_{i=1}^{N_c-1}
                   \left( \frac{p_{0}}{p_0
                   \cdot q}-\frac{p_{i}}{p_i\cdot q} \right)^2
                   + \sum_{i=1}^{N_c-2}\sum_{j>i}^{N_c-1} \left(
                   \frac{p_{i}}{p_i
                   \cdot q}-\frac{p_{j}}{p_j\cdot q} \right)^2\right]\!.
\end{equation}
   We can reexpress this result in terms of the functions given in 
\cite{Marchesini:1990yk},
\begin{equation}  
    W(\Omega_q) = \frac{2 C_{F}}{(N_c-1)}  \sum_{i=0}^{N_c-1}
                   \sum_{j \neq i}^{N_c-1} W_{ij}^{i}.
\label{eqn:squarkradpat}
\end{equation}

  This is exactly the same result as was obtained in \cite{Gibbs:1995cw}, in
the context of baryon number violation in the Standard Model, except that
 the massless radiation functions of \cite{Gibbs:1995cw} are replaced by
the massive functions here.

\begin{figure}[t]
\includegraphics[angle=90,width=0.9\textwidth]{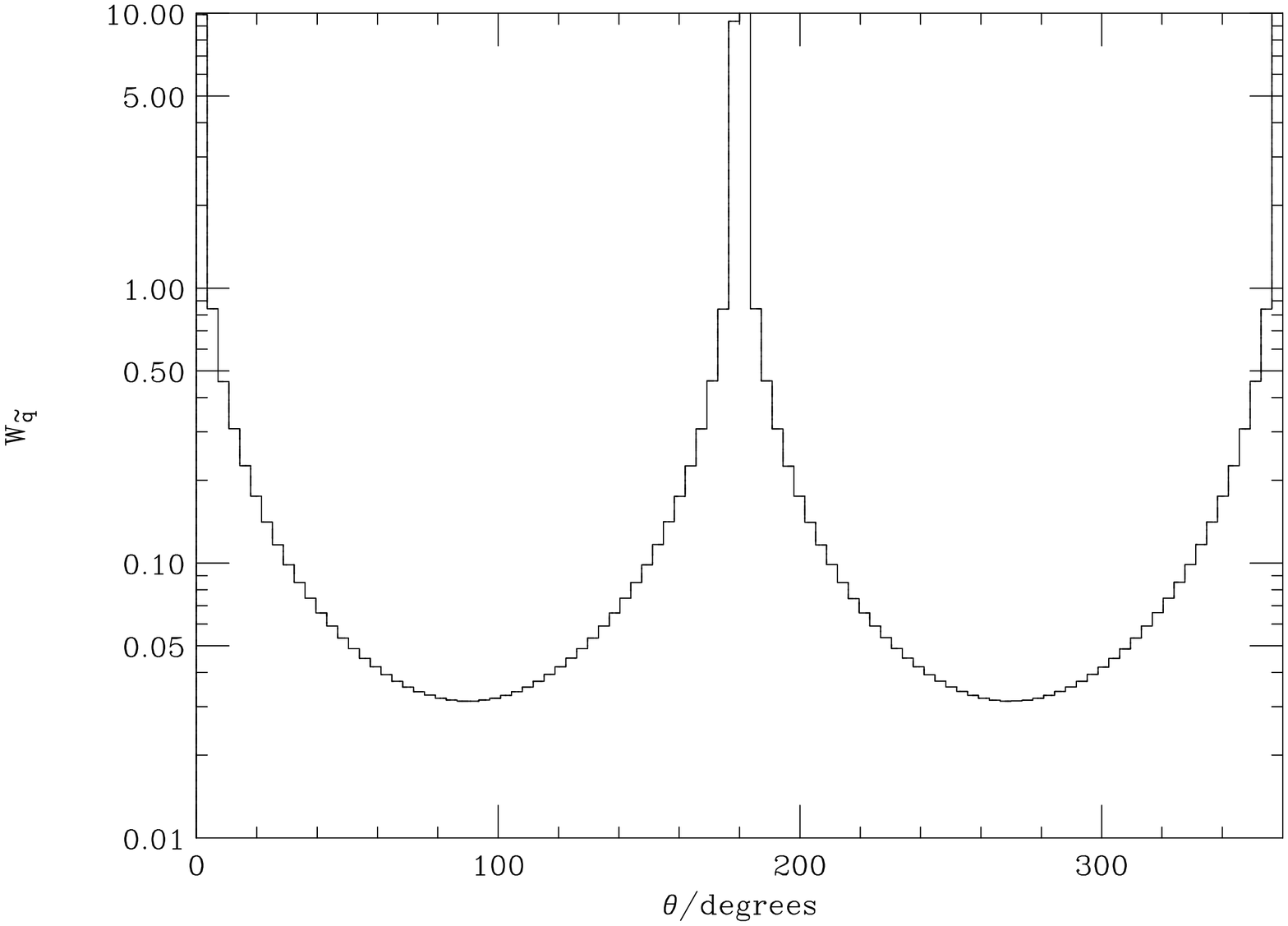}
\captionB{Radiation pattern for the decay $\mr{\qkt^*\ra q q}$.}
	{Radiation pattern for the decay $\mr{\qkt^*\ra q q}$. The quarks
         produced in the decay are in the directions $\tht=0^0$ and 
	 $\tht=180^0$. In this case the lines for the two different
	 angular-ordering approximations
	are indistinguishable from the full result.}
\label{fig:squarkrad}
\end{figure}

  We can now consider the radiation pattern given by this result. If we
  average over the azimuthal angle for the emission of the soft gluon
  about one of the final-state quarks we have two contributions
  to the radiation from this quark: the first is from radiation in
  a cone about the direction of the quark up to the direction of the decaying
  squark; the second is from radiation in a cone up to the direction of the
  other quark produced in the decay, if we only consider three colours.
  The coefficient of the radiation functions describing the radiation into
  these cones is suppressed by a factor
  of $1/(N_c-1)=1/2$ with respect to the Standard Model case where the quark
  can only radiate up to the direction of its colour partner. If we wish
  to include these decays in the parton shower we need to have one maximum
  angle for the emission of QCD radiation from the quark, to give the
  initial conditions for the parton-shower algorithm. 
  This leads to the following approach for treating the soft gluon
  radiation from this process: the quarks from the decay
  are randomly colour connected to either the decaying antisquark or
  the other quark. The direction of this colour partner gives the
  maximum angle for the QCD radiation from
  the quark.
  On average, this  correctly treats the soft gluon  
  radiation from the decay products. We cannot treat this radiation
  pattern correctly on an event-by-event basis, a parton can only have one
  colour partner whose direction gives the maximum angle for the emission of
  QCD radiation. However if we consider all the events the radiation is
  treated correctly because we have picked the colour partners with the
  correct probabilities, \ie the coefficients in Eqn.\,\ref{eqn:squarkradpat},
  $1/(N_c-1)=1/2$. A similar
  situation also occurs in Standard Model events where the gluon has both a 
  colour and an anticolour partner. When we perform the evolution of the gluon
  we must pick one of these, with equal probability, to give the maximum
  angle for the radiation from the gluon.

  It should be noted that this is only the colour connection for the 
  angular-ordering procedure. The colour connections for the hadronization 
  of these events will be
  discussed in Section~\ref{subsect:monteRPVhadron}.
  The radiation pattern for this process is shown in Fig.\,\ref{fig:squarkrad}
  in the rest frame of the decaying antisquark. This shows radiation between 
  the two quarks produced in the antisquark decay as there is no QCD
  radiation from the antisquark in its rest frame.

\begin{figure}
\begin{center} 
\begin{picture}(360,80)(0,0)
\SetScale{0.7}
\SetOffset(-50,0)
\ArrowLine(185,78)(240,78)
\ArrowLine(240,78)(285,105)
\ArrowLine(264,53)(309,26)
\ArrowLine(264,53)(309,80)
\DashArrowLine(264,53)(240,78){5}
\Text(150,63)[]{$\mr{\cht}^+_l$}
\Text(200,54)[]{$\mr{u}_j$}
\Text(200,20)[]{$\mr{d}_k$}
\Text(180,70)[]{$\mr{u}_i$}
\Text(170,40)[]{$\mr{\dnt}_{i\al}$}
\Vertex(240,78){1}
\Vertex(264,53){1}
\ArrowLine(365,78)(420,78)
\ArrowLine(420,78)(465,105)
\ArrowLine(444,53)(489,26)
\ArrowLine(444,53)(489,80)
\DashArrowLine(444,53)(420,78){5}
\Text(277,63)[]{$\mr{\cht}^+_l$}
\Text(330,20)[]{$\mr{d}_k$}
\Text(310,73)[]{$\mr{u}_j$}
\Text(330,55)[]{$\mr{u}_i$}
\Text(300,40)[]{$\mr{\dnt}_{j\al}$}
\Vertex(420,78){1}
\Vertex(444,53){1}
\end{picture}
\begin{picture}(360,80)(0,0)
\SetScale{0.7}
\ArrowLine(5,78)(60,78)
\ArrowLine(105,105)(60,78)
\ArrowLine(129,26)(84,53)
\ArrowLine(129,80)(84,53)
\DashArrowLine(60,78)(84,53){5}
\Text(25,63)[]{$\mr{\cht}^+_l$}
\Text(55,73)[]{$\mr{\bar{d}}_i$}
\Text(75,18)[]{$\mr{\bar{d}}_k$}
\Text(75,57)[]{$\mr{\bar{d}}_j$}
\Text(45,40)[]{$\mr{\upt}_{i\al}$}
\Vertex(60,78){1}
\Vertex(84,53){1}
\ArrowLine(185,78)(240,78)
\ArrowLine(285,105)(240,78)
\ArrowLine(309,26)(264,53)
\ArrowLine(309,80)(264,53)
\DashArrowLine(240,78)(264,53){5}
\Text(150,63)[]{$\mr{\cht}^+_l$}
\Text(200,56)[]{$\mr{\bar{d}}_i$}
\Text(200,20)[]{$\mr{\bar{d}}_k$}
\Text(180,72)[]{$\mr{\bar{d}}_j$}
\Text(170,40)[]{$\mr{\upt}_{j\al}$}
\Vertex(240,78){1}
\Vertex(264,53){1}
\ArrowLine(365,78)(420,78)
\ArrowLine(465,105)(420,78)
\ArrowLine(489,26)(444,53)
\ArrowLine(489,80)(444,53)
\DashArrowLine(420,78)(444,53){5}
\Text(277,63)[]{$\mr{\cht}^+_l$}
\Text(330,18)[]{$\mr{\bar{d}}_i$}
\Text(310,75)[]{$\mr{\bar{d}}_k$}
\Text(330,59)[]{$\mr{\bar{d}}_j$}
\Text(300,40)[]{$\mr{\upt}_{k\al}$}
\Vertex(420,78){1}
\Vertex(444,53){1}
\end{picture}
\vspace{-10mm}
\end{center}
\captionB{UDD decays of the $\mr{{\tilde\chi}^+}$.}
	{UDD decays of the $\mr{{\tilde\chi}^+}$. The index $l=1,2$ gives
	 the mass eigenstate of the chargino, the indices $i,j,k=1,2,3$ give
         the
	generation of the fermions and sfermions, and the index $\al=1,2$
	gives the mass eigenstate of the sfermion. The conventions for
	the mixings of the sfermions and electroweak gauginos are discussed
        in Appendix~\ref{chap:Feynman}.}
\label{fig:UDDchar}
\end{figure}

  In general the QCD radiation from
  sparticles, which are in the initial state here,
  is neglected in HERWIG. We would expect this
  approximation to be valid for two reasons: firstly the sparticles will
  usually have a short lifetime and secondly, due to their heavy masses,
  the QCD radiation will also be suppressed unless they have momenta much
  greater than their masses. However for the decays we are considering, we
  can include the effects of radiation from the decaying sparticles. This is
  done  by treating the radiation in the rest frame of the decaying squark
  where there is no radiation from the decaying sparticle, which HERWIG would
  not generate anyway. However, as stated in
  Section~\ref{subsubsect:montesoft}, while the radiation
  from individual partons, \ie $W^i_{ij}$, is not Lorentz invariant
  the dipole radiation functions are. Hence the total radiation pattern
  is Lorentz invariant and
  therefore, by treating the decay in the rest frame of the decaying
  particle, we correctly include the QCD radiation from the decaying
  particle when we boost back to the laboratory frame.

\subsubsection{Neutralino and Chargino Decays}

  The charginos decay via the processes shown in
Fig.\,\ref{fig:UDDchar} and the neutralinos via the processes in
Fig.\,\ref{fig:UDDneut}. If we consider the QCD radiation from the
decay of a colour-neutral
object which decays, for an arbitrary number of colours $N_c$, to $N_c$
quarks, then we see that there is only one possible colour flow for
this process. The squarks appearing in these processes, $\mr{\qkt}_{i\al}$,
can be either of the states $\al=1,2$ resulting from the mixing of 
$\mr{\qkt}_{iL}$ and  $\mr{\qkt}_{iR}$. This is discussed in more detail in
Appendix~\ref{chap:Feynman}.

In fact, the colour structure of this process is very
similar to that of the squark decay and the matrix element in the soft
limit can be written in the same factorized form as before. Again we
can express the current as in Eqn.\,\ref{eqn:current}, where here the
tree-level colour factor 
$C_{\mathbf{m}}=\epsilon^{c_{0} \ldots c_{N_c-1}} \epsilon^{c_{0}
\ldots c_{N_c-1}}=N_c!$, and the radiation
function is given by
\begin{equation}  
    W(\Omega_q) = \frac{2 C_{F}}{(N_c-1)}  \sum_{i=1}^{N_c}
                   \sum_{j \neq i}^{N_c} W_{ij}^{i}.
\end{equation}

  This result can be interpreted in the same way as for the squark decay
  considered in the previous section. If we now only consider three colours,
  this radiation pattern gives two
  contributions to the radiation from a given final-state quark, after
  averaging over the azimuthal angle of the radiated gluon about that quark.
  For this process both of these
  contributions are from radiation in cones up to the
  directions of either of the other quarks produced in the decay, for three
  colours. The contribution from the radiation into these cones in suppressed
  by a factor of $1/(N_c-1)=1/2$ relative to the case where the quark has a
  unique colour partner. Hence we can interpret this radiation pattern as
  saying that a quark in the final state should be randomly connected to any
  of the other quarks from the neutralino or chargino decay. Again this is
  only the colour connection for the angular-ordering procedure, the colour
  connections for the hadronization phase
  are discussion in Section~\ref{subsect:monteRPVhadron}.

\begin{figure}
\begin{center} 
\begin{picture}(360,80)(0,0)
\SetScale{0.7}
\ArrowLine(5,78)(60,78)
\ArrowLine(60,78)(105,105)
\ArrowLine(84,53)(129,26)
\ArrowLine(84,53)(129,80)
\DashArrowLine(84,53)(60,78){5}
\Text(25,63)[]{$\mr{\cht}^{0}_l$}
\Text(55,70)[]{$\mr{u}_i$}
\Text(75,20)[]{$\mr{d}_k$}
\Text(75,56)[]{$\mr{d}_j$}
\Text(45,40)[]{$\mr{\upt}_{i\al}$}
\Vertex(60,78){1}
\Vertex(84,53){1}
\ArrowLine(185,78)(240,78)
\ArrowLine(240,78)(285,105)
\ArrowLine(264,53)(309,26)
\ArrowLine(264,53)(309,80)
\DashArrowLine(264,53)(240,78){5}
\Text(150,63)[]{$\mr{\cht}^{0}_l$}
\Text(200,54)[]{$\mr{u}_i$}
\Text(200,20)[]{$\mr{d}_k$}
\Text(180,72)[]{$\mr{d}_j$}
\Text(170,40)[]{$\mr{\dnt}_{j\al}$}
\Vertex(240,78){1}
\Vertex(264,53){1}
\ArrowLine(365,78)(420,78)
\ArrowLine(420,78)(465,105)
\ArrowLine(444,53)(489,26)
\ArrowLine(444,53)(489,80)
\DashArrowLine(444,53)(420,78){5}
\Text(277,63)[]{$\mr{\cht}^{0}_l$}
\Text(330,20)[]{$\mr{u}_i$}
\Text(310,74)[]{$\mr{d}_k$}
\Text(330,57)[]{$\mr{d}_j$}
\Text(300,40)[]{$\mr{\dnt}_{k\al}$}
\Vertex(420,78){1}
\Vertex(444,53){1}
\end{picture}
\end{center}
\vspace{-10mm}
\captionB{UDD decays of the $\mr{{\tilde\chi}^0}$.}
	{UDD decays of the $\mr{{\tilde\chi}^0}$. The indices $i$, $j$, $k$
 	 and
	 $\al$ are defined in the caption of Fig.\,\ref{fig:UDDchar}. Here the
	index $l=1,\ldots\,\!,4$ gives the mass eigenstate of the neutralino.}
\label{fig:UDDneut}
\end{figure}
\begin{figure}[t]
\includegraphics[angle=90,width=0.9\textwidth]{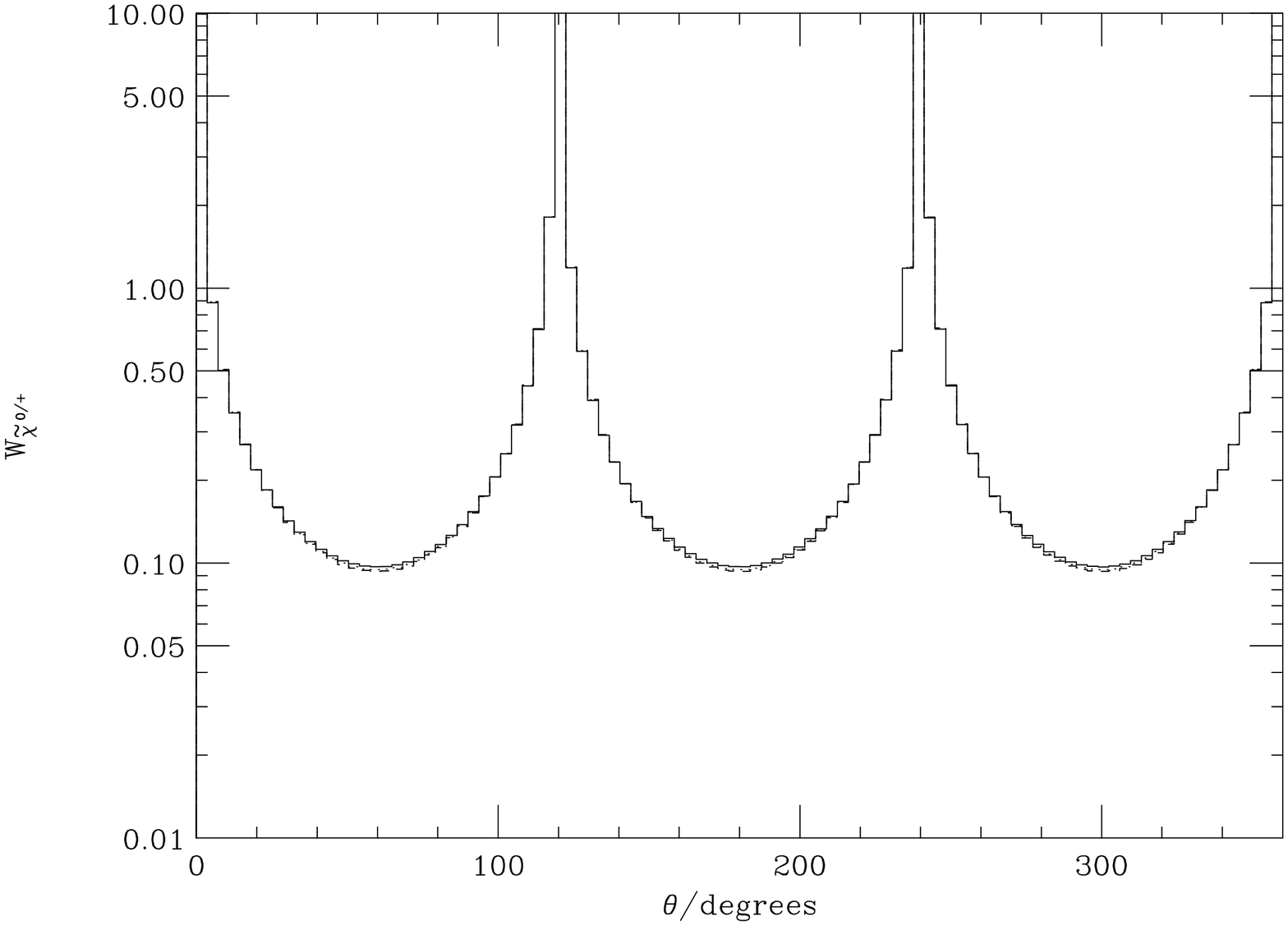}
\captionB{Radiation pattern for the decay $\mr{\cht^{0/+}\ra q q q}$.}
	{Radiation pattern for the radiation of a soft gluon in 
	 the decay $\mr{\cht^{0/+}\ra q q q}$.
	 The decay is in the rest frame of the gaugino and the quarks
	 are along the directions $\tht=0^0$, $\tht=120^0$ and $\tht=240^0$.
	 The solid line gives the full radiation pattern, the dashed
	 line the angular-ordered approximation and the dotted line
	 the improved angular-ordered approximation.}
\label{fig:neutpattern}
\end{figure}

  The radiation pattern for this process is shown in
  Fig.\,\ref{fig:neutpattern}. As can been seen 
  this pattern is more symmetric than the radiation from the
  process $\mr{e^+e^-\ra q \bar{q} g}$, Fig.\,\ref{fig:eegpatt}.
  In particular there is an equal amount
  of soft gluon radiation between all the quarks, due to the random colour
  connection structure at the \bv\  vertex, rather than the reduced radiation
  between the
  quark and antiquark in $\mr{e^+e^-\ra q \bar{q} g}$ which occurs because the
  quark and the antiquark are not colour connected.

\subsubsection{Gluino Decays}

  The colour structure of the gluino decay is very different from that
of the colourless objects or the squarks which we have already
considered, the diagrams for this process are shown in 
Fig.\,\ref{fig:UDDgluino}. Again if we consider an arbitrary number of
colours, $N_c$, the gluino will decay to $N_c$  quarks. In this case there
will be
$N_c$ possible colour flows, corresponding to the Feynman diagrams and
colour flows shown in Fig.\,\ref{fig:gluino}. These different colour
flows will lead to ``non-planar'' terms which must be dealt with.

The leading
infrared contribution to the soft gluon distribution can be written in the 
factorized form,
\begin{equation}
    {\bf {\cal M}} (p_0,p_1,p_2, \ldots\!\,,p_{N_c};q) = 
	g_s \sum_{i=1}^{N_c}{\bf m}_i
   (p_0,p_1,p_2,  \ldots\!\, ,p_{N_c}) \cdot \JC_i(q),
\end{equation}
   where 
\begin{itemize}
 \item $ {\bf m}_i (p_0,p_1,p_2, \ldots\!\,,p_{N_c})$ is the tree-level
   matrix element for the three-body gluino decay for the $i$th possible
   colour flow.
  \item$ {\bf {\cal M}} (p_0,p_1,p_2, \ldots\!\, ,p_{N_c};q)$ is the 
	tree-level matrix element for the three-body gluino decay including
  	the emission of an extra soft gluon of momentum q.
 \item $ \JC_i(q)$ is the non-Abelian semi-classical
  current for the emission of a soft gluon, with momentum q, from the
  hard partons for the $i$th possible colour flow.
\end{itemize}

  Again the current has the form $\JC_i(q) = \displaystyle{\sum_{s=1,2}}
  \JC_i^{b,\mu}(q) \varepsilon^*_{\mu,s}$, where in this case 
\begin{eqnarray}  {\displaystyle
   \JC_i^{b,\mu}(q) }&=&
	 \displaystyle{i\left(\frac{p_0^\mu}{p_0 \cdot q}\right) 
                        {\bf f}^{ba'a}{\bf t}_{c_ic'_i}^{a'}
                        \epsilon^{c_1 \ldots c'_i \ldots c_{N_c}}
                         + \left(\frac{p_i^\mu}{p_i \cdot q}\right)
                        {\bf t}_{c_ic'_i}^b{\bf t}_{c'_ic''_i}^a
                   \epsilon^{c_1 \ldots c''_i \ldots c_{N_c}} }\nonumber \\
&&  {\displaystyle+\sum_{j=1,j \neq i}^{N_c}
	 \left(\frac{p_j^\mu}{p_j \cdot q}\right)
                         {\bf t}_{c_jc'_j}^b{\bf t}_{c_ic'_i}^a
                        \epsilon^{c_1 \ldots c'_i  \ldots c'_j
                         \ldots c_{N_c}}.}
\end{eqnarray}
  The first term describes the radiation of a gluon by the decaying gluino
  and the second two terms describe radiation from the quarks produced in 
  the gluino decay.

  We can write the matrix element squared for this process as
\begin{eqnarray}
   |{\bf {\cal M}} (p_0,p_1,p_2, \ldots\!\,,p_{N_c};q)|^2&= & g^2_s
 \displaystyle    \sum_{i=1}^{N_c} |{\bf m_i}
 (p_0,p_1,p_2,  \ldots\!\,,p_{N_c})|^2 \cdot |\JC_i(q)|^2 \nonumber \\
    & & + g^2_s  
   \displaystyle \sum_{i=1}^{N_c} \displaystyle \sum_{j=1,j \neq i}^{N_c}
 {\bf m_im_j^*} \cdot \JC_i(q)\cdot \JC_j^*(q). 
\end{eqnarray}

\begin{figure}
\begin{center} 
\begin{picture}(360,80)(0,0)
\SetScale{0.7}
\ArrowLine(5,78)(60,78)
\ArrowLine(60,78)(105,105)
\ArrowLine(84,53)(129,26)
\ArrowLine(84,53)(129,80)
\DashArrowLine(84,53)(60,78){5}
\Text(25,63)[]{$\mr{\glt}$}
\Text(55,70)[]{$\mr{u}_i$}
\Text(75,20)[]{$\mr{d}_k$}
\Text(75,56)[]{$\mr{d}_j$}
\Text(45,40)[]{$\mr{\upt}_{i\al}$}
\Vertex(60,78){1}
\Vertex(84,53){1}
\ArrowLine(185,78)(240,78)
\ArrowLine(240,78)(285,105)
\ArrowLine(264,53)(309,26)
\ArrowLine(264,53)(309,80)
\DashArrowLine(264,53)(240,78){5}
\Text(150,63)[]{$\mr{\glt}$}
\Text(200,54)[]{$\mr{u}_i$}
\Text(200,20)[]{$\mr{d}_k$}
\Text(180,72)[]{$\mr{d}_j$}
\Text(170,40)[]{$\mr{\dnt}_{j\al}$}
\Vertex(240,78){1}
\Vertex(264,53){1}
\ArrowLine(365,78)(420,78)
\ArrowLine(420,78)(465,105)
\ArrowLine(444,53)(489,26)
\ArrowLine(444,53)(489,80)
\DashArrowLine(444,53)(420,78){5}
\Text(277,63)[]{$\mr{\glt}$}
\Text(330,18)[]{$\mr{u}_i$}
\Text(310,75)[]{$\mr{d}_k$}
\Text(330,57)[]{$\mr{d}_j$}
\Text(300,40)[]{$\mr{\dnt}_{k\al}$}
\Vertex(420,78){1}
\Vertex(444,53){1}
\end{picture}
\end{center}
\vspace{-10mm}
\captionB{UDD decays of the $\mr{{\glt}}$.}
	{UDD decays of the $\mr{{\glt}}$. As before, the indices $i,j,k=1,2,3$
	 are the generations of the fermions and sfermions.}
\label{fig:UDDgluino}
\end{figure}
% Gluino Decay
\begin{figure}
\begin{center} \begin{picture}(360,130)(0,40)
\SetScale{1.0}
\ArrowLine(5,128)(60,128)
\ArrowLine(60,128)(105,155)
\ArrowLine(84,103)(129,76)
\ArrowLine(84,103)(129,130)
\DashArrowLine(84,103)(60,128){5}
\Text(30,140)[]{$\mr{\glt}$}
\Text(73,145)[]{$\mr{q}_i$}
\Text(108,80)[]{$\mr{q}_{N_c}$}
\Text(108,127)[]{$\mr{q_1}$}
\Text(60,115)[]{$\mr{\tilde{q}}_{i\al}$}
\Vertex(60,128){1}
\Vertex(84,103){1}
\SetOffset(20,0)
\ArrowLine(185,128)(240,128)
\ArrowLine(240,123)(185,123)
\ArrowLine(240,128)(285,155)
\ArrowLine(264,98)(309,71)
\ArrowLine(264,98)(309,125)
\ArrowLine(264,98)(240,123)
\Text(215,140)[]{$\mr{\glt}$}
\Text(284,122)[]{$\mr{q_1}$}
\Text(284,77)[]{$\mr{q}_{N_c}$}
\Text(257,146)[]{$\mr{q}_i$}
\Text(242,111)[]{$\mr{\tilde{q}}_{i\al}$}
\Vertex(240,128){1}
\Vertex(240,123){1}
\Vertex(264,98){1}
% Dots for missing particles
\Vertex(90,112){1}
\Vertex(90,102){1}
\Vertex(90,92){1}
\Vertex(290,107){1}
\Vertex(290,97){1}
\Vertex(290,87){1}
% Diagram labels
\Text(55,50)[]{{\small (a) Feynman Diagram}}
\Text(248,50)[]{{\small (b) Colour Flow}}
\end{picture}
\captionB{Baryon number violating decay of the $\mr{\glt}$.}
	{Baryon number violating decay of the $\mr{\glt}$. There are $N_c-1$
	 quarks coupling to the squark at the lower vertex, \ie all the
	 quarks produced in the gluino decay apart from the quark~$i$.}
\label{fig:gluino}
\end{center}
\end{figure}

  The procedure of \cite{Odagiri:1998ep}, which was described in
 Section~\ref{subsubsect:montesoft}, can be used 
 with the matrix elements for this process, given
in Appendix~\ref{chap:decay}, to deal with the ``non-planar'' terms.
 We will now consider the radiation pattern of the planar terms.
 The current can be written as in Eqn.\,\ref{eqn:current} where
 here the tree-level colour factor 
\begin{equation}
  C_{\mathbf{m}} = {\bf t}_{c_ic'_i}^b\epsilon^{c_1 \ldots c'_i
                  \ldots c_{N_c}}{\bf t}_{c_i''c_i}^b \epsilon^{c_{0}
                   \ldots c''_i \ldots c_{N_c}} = C_F N_{c}!.
\end{equation}
 We have not averaged over the initial colours and the
 radiation function is given by
\begin{eqnarray}  
   W(\Omega_q) &= & C_A W_{0i}^{0} +2 C_F W_{i0}^{i} 
                   +\frac{2 C_{F}}{(N_c-1)}\displaystyle
                         \sum_{j \neq i, k \neq
                        j}^{N_c} W^{j} _{jk}  +\frac1{(N_c-1)}\displaystyle 
			\sum_{j \neq i}^{N_c} 
                        \left( C_{A}W_{0j}^{0} + 2C_{F}W_{j0}^{j}
                        \right) \nonumber \\
                 &   & + \frac1{N_c} W_{i0}^{i} 
                        +\frac1{N_c(N_c-1)}\displaystyle 
                        \sum_{j \neq i}^{N_c}
                         \left(W_{j0}^{j}-W_{ij}^{i}-W_{ji}^{j}
                     \right)\!. \label{eqn:planar}
\end{eqnarray}
  This planar part of the soft radiation pattern gives us the result
  we would na\"{\i}vely expect. The radiation pattern,
  Eqn.\,\ref{eqn:planar}, contains terms,
  the second line of Eqn.\,\ref{eqn:planar}, which are of order
  $1/N^2_c$ with respect to the leading-order terms which we shall
  neglect as in Section~\ref{subsubsect:montesoft}. The
  first term, in Eqn.\,\ref{eqn:planar}, describes radiation from the gluino
  up to the direction of the $i$th quark.
  The second term describes radiation from the direction of the $i$th quark
  up to the direction of the gluino. Hence the colour line of the gluino
  and the $i$th quark should be colour connected as for MSSM processes. 

\begin{figure}[t]
\includegraphics[angle=90,width=0.9\textwidth]{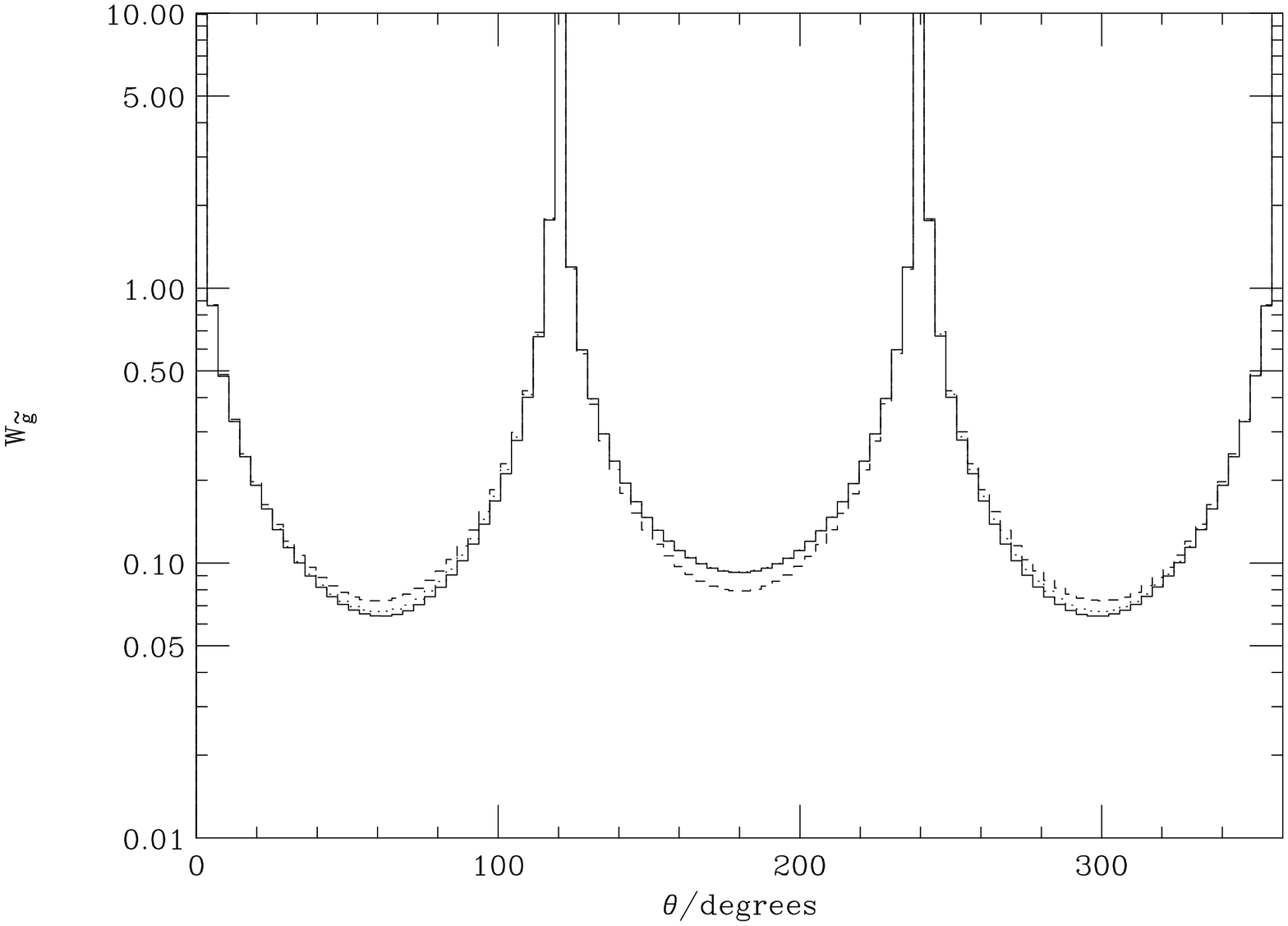}
\captionB{Radiation pattern for $\mr{\glt\ra qqq}$.}
	{Soft gluon radiation pattern for $\mr{\glt\ra qqq}$.
	 The decay is in the rest frame of the gluino and the quarks
	 are along the directions $\tht=0^0$, $\tht=120^0$ and $\tht=240^0$.
	 The quark at $\tht=0^0$ is colour connected to the gluino in the
	 standard MSSM way while the other two quarks are colour connected to
	 the gluino via the \bv\  vertex. As before, the solid line gives the
	 full radiation pattern, the dashed
	 line the angular-ordered radiation pattern and the dotted line
	 the improved angular-ordered radiation pattern.} 
\label{fig:gluinorad}
\end{figure}

  As for both the squark and electroweak gaugino decays we have already
  considered we must now assign colour partners, with the correct
  probabilities, for the remaining quarks
  and the anticolour line of the gluino.
  The third term, in Eqn.\,\ref{eqn:planar}, after azimuthal averaging,
  describes radiation from the direction of one of the final-state quarks,
  which is not connected to the colour
  line of the gluino, up to the directions of all the
  other final-state quarks, apart from the quark which is connected to the
  colour line of the gluino. The next term then describes radiation from the
  anticolour line of the gluino in cones up to the directions of all the
  final-state quarks, apart from the quark which is connected to the colour
  line of the gluino. The final, non-suppressed, term describes radiation
  from one of the final-state quarks, not connected to the colour line of
  the gluino, up to the direction of the
  gluino. We can therefore treated the anticolour line of the gluino and
  the remaining quarks in the same way as a decaying antisquark. The
  anticolour line of the gluino should be randomly connected to one of the
  quarks
  which is not connected to the colour line of the gluino and the final-state
  quarks which are not connected to the colour line of the gluino should be
  connected at random to either the anticolour line of the gluino or the other
  final-state quark which is not connected to the colour line of the gluino. 
  This assigns the colour partners with the correct probabilities.
  This is only the procedure for one of the $N_c$ possible planar diagrams.

  If we now consider all the possible planar colour flows the correct
  procedure is to
  connect the $i$th quark to the colour line of the gluino in the
  standard MSSM way with probability given by      
  $\frac{|\overline{M}|^2_{\mr{full},i}}{|\overline{M}|^2_{\mr{tot}}}$,
  where $|\overline{M}|^2_{\mr{full},i}$ is given by Eqn.\,\ref{eqn:kosuke}.
  We can then treat the anticolour line of the gluino and the remaining
  quarks as an antisquark decaying to quarks.

  The soft gluon radiation pattern for one of the possible planar diagrams 
  is shown in Fig.\,\ref{fig:gluinorad}. There is slightly less radiation
  between the quark at $\tht=0^0$ and the other 
  two quarks as the quark at $\tht=0^0$ is not colour
  connected to either of the other two quarks. There is however more radiation
  between the other two quarks which are colour connected.
 
\subsection{Hard Processes}

  In addition to the decays which we have already discussed there are a
number of baryon number violating hard subprocesses which we 
include in the simulation. All of the colour structures of the
hard processes which actually violate baryon number have already been
discussed, as these processes are merely crossed versions of the various
decays discussed above. However, in addition to these processes there are
some hard processes which occur via the third term in the
superpotential but involve no net baryon number violation, \linebreak \eg
Fig.\,\ref{fig:bump}.

  In this section we will only discuss this type of process which cannot be
  obtained by crossing the previous results.

\subsubsection[Resonant Squark production followed by \bv\  Decay]
	{Resonant Squark production followed by \boldmath{\bv}\  Decay}

\begin{figure}
\begin{center} \begin{picture}(180,70)(0,40)
\SetScale{1.3}
% Feynman Diagram
\ArrowLine(5,26)(60,52)
\ArrowLine(5,78)(60,52)
\DashArrowLine(90,52)(60,52){5}
\ArrowLine(90,52)(145,26)
\ArrowLine(90,52)(145,78)
% Labels
\Text(150,40)[]{$\mr{q}_{N_c-1}$}
\Text(150,95)[]{$\mr{q}_l$}
\Text(43,95)[]{$\mr{q}_i$}
\Text(55,40)[]{$\mr{q}_{N_c-1}$}
\Text(100,80)[]{$\mr{\tilde{q}}_{R_{N_c}}$}
% Vertex and Dots
\Vertex(60,52){1}
\Vertex(90,52){1}
\Vertex(25,43){1}
\Vertex(25,53){1}
\Vertex(25,63){1}
\Vertex(125,43){1}
\Vertex(125,53){1}
\Vertex(125,63){1}
\end{picture}
\end{center}
\captionB{Resonant squark production followed by \bv\  decay.}
	{Resonant squark production followed by \bv\  decay for an
	arbitrary number of colours $N_c$.\label{fig:bump}}
\end{figure}

  As before, we will consider the process in Fig.\,\ref{fig:bump}
for an arbitrary number of colours, $N_c$. 
  We can write the matrix element for the emission of an
extra soft gluon in the form
\begin{equation}
    {\bf {\cal M}} (p_1,\ldots\!\,,p_{N_c-1}: k_1,\ldots\!\,,k_{N_c-1} ;q) = 
    g_s {\bf m}(p_1,\ldots\!\,,p_{N_c-1}:k_1,\ldots\!\,,k_{N_c-1}),
   \cdot \JC(q),
\end{equation}
where
\begin{itemize}
   \item ${\bf m}(p_1,\ldots\!\,,p_{N_c-1}:k_1,\ldots\!\,,k_{N_c-1}) $
         is the tree-level matrix element for the $(N_c-1)$ quarks to 
	 $(N_c-1)$ quarks scattering.
   \item $ {\bf {\cal M}} (p_1,\ldots\!\,,p_{N_c-1}:k_1,\ldots\!\,,k_{N_c-1}
	  ;q)$
         is the tree-level matrix element for the $(N_c-1)$ quarks to 
	 $(N_c-1)$ quarks scattering with the
         emission of an extra soft gluon with momentum q.
   \item $\JC(q)$ is the non-Abelian semi-classical current for 
         the emission of the soft gluon, with momentum q, from the hard
	 partons.
   \item $p_1,\ldots\!\,,p_{N_c-1}$ are the momenta of the partons in the
         initial state.
   \item $k_1,\ldots\!\,,k_{N_c-1}$ are the momenta of the partons in the
         final state.
\end{itemize}

  Again the current has the form $\JC(q) = \displaystyle{\sum_{s=1,2}}
  \JC^{b,\mu}(q) \varepsilon^*_{\mu,s}$, where in this case 
\begin{eqnarray}  
   \JC^{b,\mu}(q) &=& -\displaystyle \sum_{i=1}^{N_c-1} 
\left(\frac{p_{i}^{\mu}}{p_i
    \cdot q}\right) {\bf t}_{c'_ic_i}^b \epsilon^{c_1 \ldots c'_i 
    \ldots c_{N_c}}  \epsilon^{d_1 \ldots d_{N_c-1} c_{N_c}}\nonumber \\ 
   &&    +\displaystyle\sum_{i=1}^{N_c-1} \left(\frac{k_{i}^{\mu}}{k_i
    \cdot q}\right) {\bf t}_{d_id_{i}'}^b \epsilon^{c_1 \ldots c_{N_c-1} 
    d_{N_c}}  \epsilon^{d_1 \ldots d'_i \ldots d_{N_c}}, 
\end{eqnarray}
  where $b$ and $\mu$ are the colour and Lorentz indices of the emitted
gluon, respectively.

\begin{figure}[t]
\includegraphics[angle=90,width=0.9\textwidth]{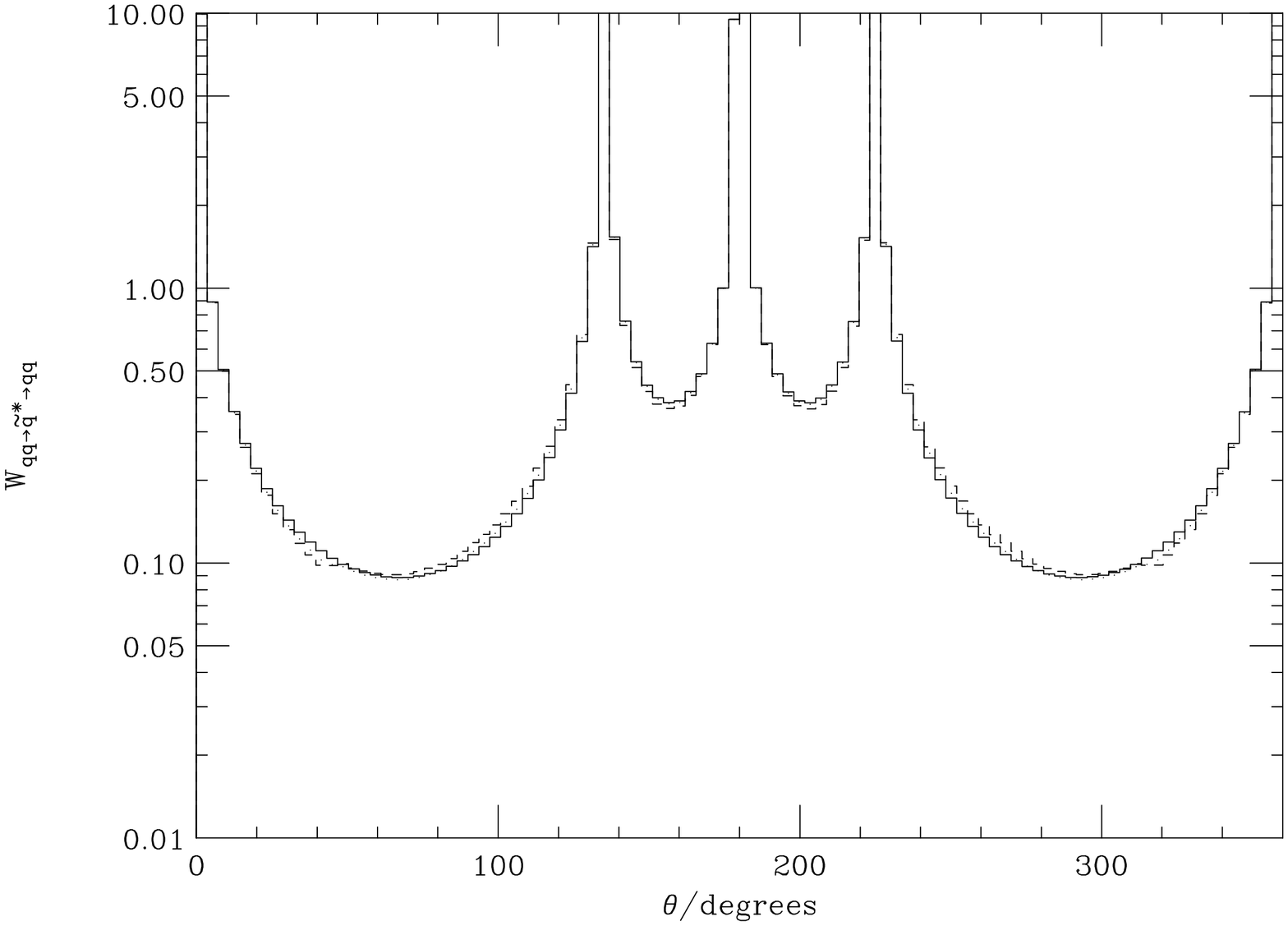}
\captionB{Radiation pattern for resonant squark production.}
	{Soft gluon radiation pattern for $\mr{qq\ra\qkt^*\ra qq}$. The
	 incoming quark directions are $\tht=0^0$ and $\tht=180^0$. The
	 outgoing quark directions are $\tht=135^0$ and $\tht=225^0$.
	 This process is shown in a frame which is boosted along the beam
	 direction as will usually occur in hadron--hadron collisions.
	 The solid line gives the full radiation pattern, the dashed line
	 the angular-ordered approximation and the dotted line the
	 improved angular-ordered approximation.}
\label{fig:hardpat}
\end{figure}

  We can obtain the soft gluon distribution by squaring the
  current. The result can be rewritten using Eqn.\,\ref{eqn:current} where
  the tree-level colour factor is given by\linebreak
  \mbox{$C_{\mathbf{m}}=N_c!(N_c-1)!$}. We have not averaged over the initial
  colours, and the radiation
  function is given by
\begin{eqnarray}  
  { \displaystyle  W(\Omega_q) }&=&{  \displaystyle \frac{2 C_{F}}{(N_c-1)}
                    \sum_{i=1}^{N_c-1}
                   \sum_{j \neq i} W_{ij}^{i}
                   +\frac{2 C_{F}}{(N_c-1)} 
                     \sum_{l=1}^{N_c-1}
                   \sum_{m \neq l} W_{lm}^{l} \nonumber} \\
 &&          {  \displaystyle       + \frac{2 C_{F}}{(N_c-1)^2}
                     \sum_{i=1}^{N_c-1}
       \sum_{l=1}^{N_c-1}\left( W_{li}^{i}+ W_{il}^{l} \right)\!,}
\end{eqnarray}
  where the partons $i$ and $j$ are in the initial state and the partons $l$
  and $m$ are in the final state.

  If we now only consider three colours, the
  first term in this radiation pattern describes radiation from one of the
  initial-state quarks in cones up to the direction of the other initial-state
  quark, after azimuthal averaging. Similarly, the second term describes
  radiation from the direction of one of the final-state quarks up to the
  direction of the other final-state quark. Both of these have probability
  $1/(N_c-1)=1/2$ relative to the case of a quark and its Standard Model
  colour partner. The remaining terms describe radiation from  one of the
  final-state quarks up to the direction of either of the initial-state quarks
  and radiation from one of the initial-state quarks up to the direction of
  either of the final-state quarks. The probability for radiation from 
  one of the final-state quarks up to the direction of one of the
  incoming quarks is $1/(N_c-1)^2=1/4$ relative to the Standard Model case.

  This radiation pattern gives quite an unusual angular-ordering
  procedure. If we consider one of the quarks in the initial state, this
  quark should be randomly connected to any of the other quarks in the
  initial state or to the intermediate squark. If the quark is connected to
  the intermediate squark it should then be randomly connected to any of the
  final-state quarks. Similarly the final-state quarks are connected at
  random to any of the other final-state quarks or the intermediate squark,
  and again quarks connected to the intermediate squark are then randomly
  connected to any of the initial-state quarks. This correctly assigns the
  probabilities of the colour partners described above.

  The radiation pattern for this process is shown in Fig.\,\ref{fig:hardpat}.
  There is less radiation between the two outgoing quarks and the quark at 
  $\tht=0^0$ than between the outgoing quarks and the quark at  $\tht=180^0$.
  This is because, while there are some colour connections between the
  incoming
  and outgoing quarks, due to the random nature of the colour connection at
  the \bv\  vertex,  the main colour connection is between the two quarks in
  the initial state and the two quarks in the final state. As the 
  angular distance between the two outgoing quarks is smaller about
  $\tht=180^0$ there is more radiation in this direction.
\section[Hadronization of \rpv\  Processes]
	{Hadronization of \boldmath{\rpv}\  Processes}
\label{subsect:monteRPVhadron}

% Showering and hadronization of a neutralino decay
\begin{figure}
\begin{center} \begin{picture}(360,200)(0,0)
\SetScale{0.7}
\SetOffset(50,0)
% Feynman Diagram
% Incoming Line
\DashArrowLine(-50,150)(0,150){5}
% Decay
\ArrowLine(0,150)(40,190)
\ArrowLine(0,150)(115,150)
\ArrowLine(0,150)(40,110)
% Gluon Radiation
\Gluon(40,190)(80,190){3}{4}
\Gluon(40,110)(80,110){3}{4}
\Gluon(80,190)(100,210){3}{3}
\Gluon(80,190)(100,170){3}{3}
% Gluon Splitting
\ArrowLine(40,190)(115,265)
\ArrowLine(40,110)(115,35)
\ArrowLine(80,110)(115,145)
\ArrowLine(115,75)(80,110)
\ArrowLine(115,225)(100,210)
\ArrowLine(100,210)(115,195)
\ArrowLine(115,185)(100,170)
\ArrowLine(100,170)(115,155)
% Ovals for clusters
\GOval(115,245)(25,5)(0){0.7}
\GOval(115,150)(15,5)(0){0.2}
\GOval(115,190)(15,5)(0){0.7}
\GOval(115,55)(25,5)(0){0.7}
% Labels
\Text(30,0)[]{{\small (a) Feynman diagram}}
\Text(-35,90)[]{neutralino}
\Text(110,170)[]{meson}
\Text(110,105)[]{baryon}
%%%%%%%%%%%%% Colour Flows
\SetOffset(125,0)
% Incoming Line
\DashArrowLine(150,150)(200,150){5}
% Decay
\ArrowLine(200,150)(235,185)
\ArrowLine(200,150)(315,150)
\ArrowLine(200,150)(235,115)
% Gluon Radiation
\ArrowLine(235,185)(280,185)
\ArrowLine(280,195)(245,195)
\ArrowLine(235,115)(280,115)
\ArrowLine(280,105)(245,105)
\ArrowLine(280,185)(300,165)
\ArrowLine(300,175)(285,190)
\ArrowLine(285,190)(300,205)
\ArrowLine(300,215)(280,195)
% Gluon Splitting
\ArrowLine(245,195)(315,265)
\ArrowLine(245,105)(315,35)
\ArrowLine(315,190)(300,175)
\ArrowLine(300,165)(315,150)
\ArrowLine(300,205)(315,190)
\ArrowLine(315,230)(300,215)
\ArrowLine(315,70)(280,105)
\ArrowLine(280,115)(315,150)
% Ovals for clusters
\GOval(315,245)(25,5)(0){0.7}
\GOval(315,150)(15,5)(0){0.2}
\GOval(315,190)(15,5)(0){0.7}
\GOval(315,55)(25,5)(0){0.7}
% Labels
\Text(160,0)[]{{\small (b) Colour flow}}
\Text(110,90)[]{neutralino}
\Text(255,170)[]{meson}
\Text(255,105)[]{baryon}
\end{picture}\end{center}
\captionB{Hadronization of a \bv\  neutralino decay.}
	{The Feynman diagram and colour flow for the hadronization of
	 a \bv\  neutralino decay. No QCD radiation from the central quark 
	 has been shown for simplicity, in general this quark can also
	 radiate.}
\label{fig:neuthad}
\end{figure}
% End of the Figure

  As  we saw in  Section~\ref{subsect:monteRPVangles}, it is possible to
  angular order  the baryon number violating decays and hard  processes.
  It is then necessary to decide how to hadronize these events using the
  cluster hadronization model \cite{Webber:1984if} in order to include a full
  simulation of these processes in the HERWIG event generator. The procedure
  described in Section~\ref{sect:montehadron} also works for the MSSM provided
  that the lifetime of
  the coloured sparticles does not exceed the hadronization time-scale.
  However, some modifications to this model  are required 
  for \rpv\  processes.
  
  In the Standard Model and MSSM cases the colour partner for the
  colour coherence effects and for the hadronization phase are always
  the same. However in the \bv\   decays and hard processes  we see
  for the first time cases where the colour connection for the angular
  ordering and for the hadronization can be different.  This is
  because while the colour connection for the angular-ordering procedure
  is determined by the eikonal current, the colour connection for the
  hadronization phase is defined by the colour flow in the leading-order
  diagram. When baryon number is conserved these are identical, however
  when baryon number is violated, there are cases where
  the two are different. 

  First we consider the simplest  type  of decay,  \ie a
  neutralino or chargino decaying to three quarks. The method described in
  Section~\ref{subsect:monteRPVangles} correctly implements the
  angular-ordering procedure. After the parton-shower phase (and the
  splitting of the remaining gluons into quark--antiquark pairs) we will be 
  left with pairs of colour-connected partons forming colour singlets as well
  as three further quarks. An example of this is
  shown in Fig.\,\ref{fig:neuthad}. These three remaining quarks form a colour
  singlet with baryonic quantum numbers, a baryonic cluster.
  To handle baryonic clusters HERWIG needs the
  constituents to be labelled as one quark and one diquark rather than
  three quarks, so we randomly pair two of them into a diquark.
  In our example in Fig.\,\ref{fig:neuthad} the three quarks in the middle
  together form a colour singlet. Two of these quarks are paired into a
  diquark which combines with the remaining quark to form a baryonic cluster.
  The mesonic clusters will decay to give either two mesons, or a
  baryon--antibaryon pair, while the baryonic cluster will decay to give a
  meson and a baryon. 

% Hadronization if one BV decay of a squark
\begin{figure}[htp]
\begin{center} \begin{picture}(180,100)(0,20)
\SetScale{1.0}
\SetOffset(40,-10)
\DashArrowLine(60,100)(5,70){5}
\DashArrowLine(5,70)(60,40){5}
\ArrowLine(60,40)(100,60)
\ArrowLine(60,40)(100,20)
\ArrowLine(60,100)(100,120)
\ArrowLine(60,100)(100,80)
\Text(80,80)[]{$\mr{d}_j$}
\Text(80,120)[]{$\mr{d}_k$}
\Text(80,57)[]{$\mr{u}_i$}
\Text(80,20)[]{$\mr{\cht^{0}}$}
\Text(35,98)[]{$\mr{\upt^{*}}_{i\al}$}
\Text(35,45)[]{$\mr{\upt}_{i\al}$}
\Vertex(5,70){1}
\Vertex(60,100){1}
\Vertex(60,40){1}
\end{picture}\end{center}
\captionB{Hadronization with one \bv\  decay.}
	{Hadronization with one \bv\  decay.}
\label{fig:had1}
\end{figure}
% End of the figure

  This procedure is relatively easy to implement in the case of
  electroweak gaugino decays. However it becomes more difficult in the case of
  the \bv\  decay of an antisquark to two quarks. If
  the anticolour partner of the decaying antisquark is a particle which decays
  via a baryon number conserving  process then the two quarks and the
  particle which gets the colour of the second decaying particle can be
  clustered as in the neutralino case, \linebreak[4] \eg~in 
  Fig.\,\ref{fig:had1} the $\mr{u}_i$, $\mr{d}_j$, and $\mr{d}_k$ should be
  formed into a baryonic cluster. In general, we do not form these quarks 
  into the colour-singlet baryonic cluster but the parton produced in the
  parton shower of these partons which inherits the colour of the showering
  parton, as in Fig.\,\ref{fig:neuthad}.

% Hadronization with two BV squarks decays
\begin{figure}[htp]
\begin{center} \begin{picture}(180,100)(0,30)
\SetScale{1.0}
\SetOffset(40,0)
\DashArrowLine(60,100)(5,70){5}
\DashArrowLine(5,70)(60,40){5}
\ArrowLine(100,60)(60,40)
\ArrowLine(100,20)(60,40)
\ArrowLine(60,100)(100,120)
\ArrowLine(60,100)(100,80)
\Text(80,80)[]{$\mr{d}_j$}
\Text(80,120)[]{$\mr{d}_k$}
\Text(80,60)[]{$\mr{\bar{d}}_l$}
\Text(80,20)[]{$\mr{\bar{d}}_m$}
\Text(35,98)[]{$\mr{\upt^{*}}_{i\al}$}
\Text(35,45)[]{$\mr{\upt}_{i\al}$}
\Vertex(5,70){1}
\Vertex(60,100){1}
\Vertex(60,40){1}
\end{picture}\end{center}
\captionB{Hadronization with two \bv\  decays.}
	{Hadronization with two \bv\  decays.}
\label{fig:had2}
\end{figure}
% End of the figure

   However, if this second particle
   decays via \bv\  then the procedure must be different, as shown in
   Fig.\,\ref{fig:had2}. Here, instead of forming one baryonic cluster, we
   form two mesonic clusters. This is done by pairing the $\mr{d}_k$ randomly
   with either the $\mr{\bar{d}}_l$ or $\mr{\bar{d}}_m$ into a standard
   colour-singlet cluster, the remaining quark and antiquark  are also paired
   into a colour-singlet cluster. This is not the colour
   connection for the angular ordering procedure but the colour
   connection for the hadronization phase, which is different in this
   case and determined by the colour flow in the tree-level diagram.

   This leaves the case of the gluino decay which looks more
   complicated but can be considered by treating the colour line as
   normal and the anticolour line like a decaying antisquark. If the
   anticolour partner of the gluino is a Standard Model particle or
   decays via a baryon number conserving MSSM decay mode we form the
   three quarks into a baryonic cluster. However if the anticolour
   partner decays via a \bv\  mode we then form two mesonic clusters.

   There is one further type of colour flow to be considered, which is
   the production of a resonant squark via baryon number violation
   which then also decays
   via \bv. The correct hadronization procedure in this case is
   similar to that adopted for the case of two colour
   connected \bv\  decays. We randomly connect one of the final-state
   quarks to the colour partners of either of the initial-state
   quarks to form a colour-singlet cluster. The remaining final-state quark
   can then be paired with the colour partner of the other initial-state
   quark. This gives two colour-singlet clusters. Again
   the colour partner for hadronization  is determined by the  colour
   flow in the tree-level diagram.

   Using the procedures we have outlined above it is possible to
   hadronize any of the \bv\  decays or hard processes. There is however one 
   potential problem. 
   The cluster model is based on the idea of colour preconfinement.
   In \bv\   processes we see a very different
   spectrum for the baryonic clusters formed from the baryon number
   violation to that seen for clusters in Standard Model events.

% Cluster Mass Spectrum
\begin{figure}
\centering
\includegraphics[angle=90,width=0.45\textwidth]{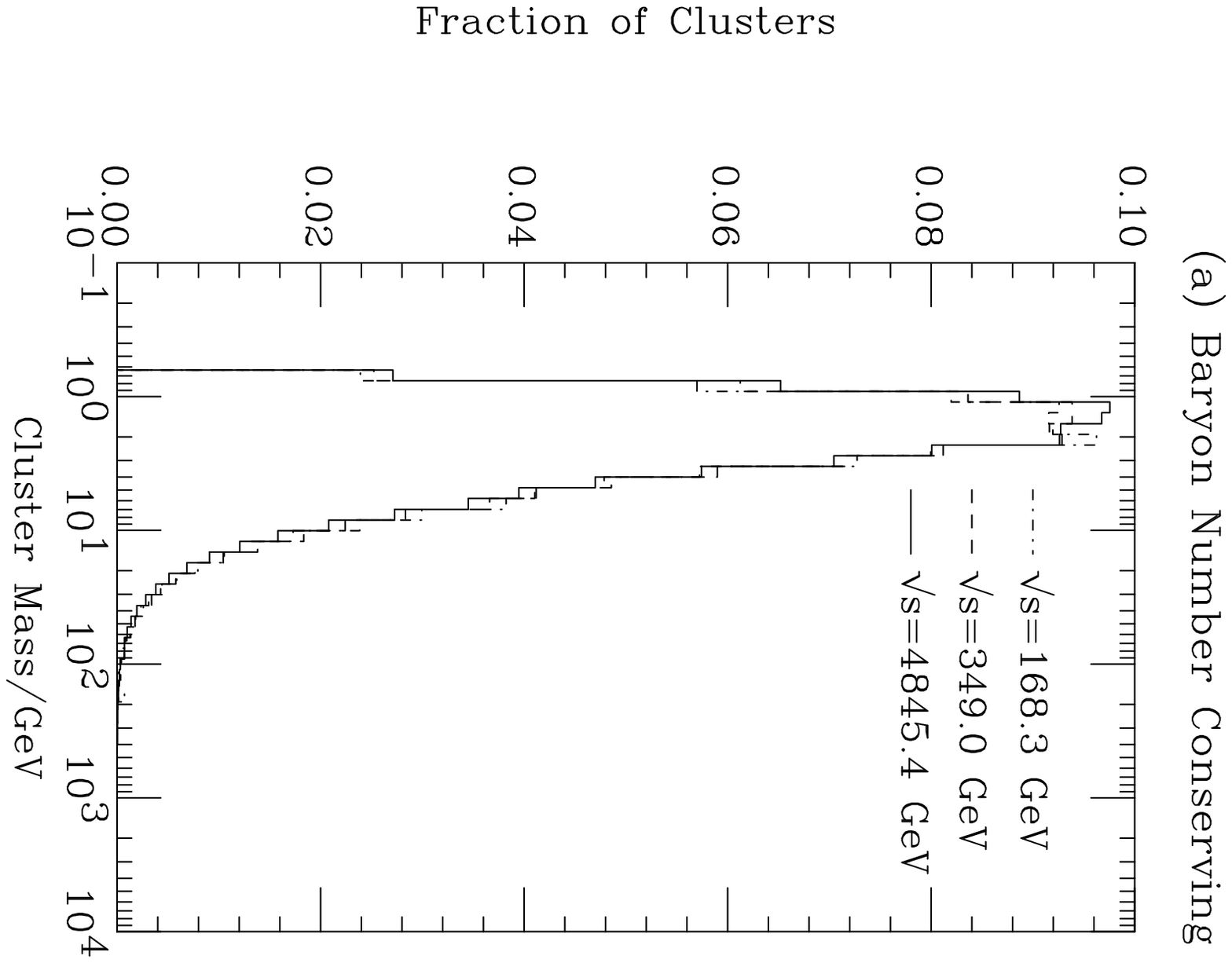}
\hfill
\includegraphics[angle=90,width=0.45\textwidth]{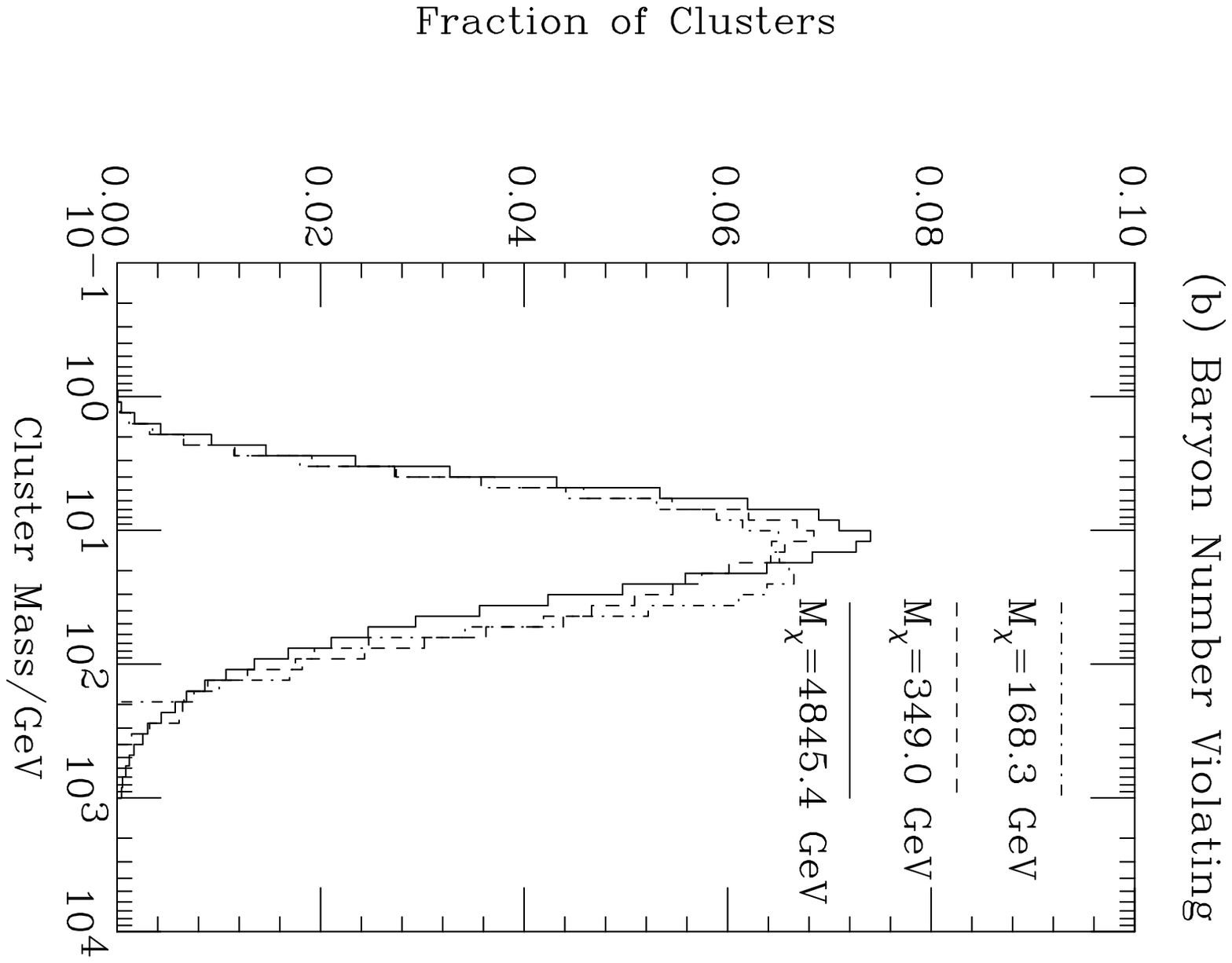}\\
\captionB{Distributions of the colour-singlet cluster masses.}
	{Distributions of the colour-singlet cluster masses.
 	 The baryon number conserving clusters come from $\mr{e^+e^-}$ events
         at the given centre-of-mass energy, $\sqrt{s}$, whereas the baryon
	 number violating clusters come from decays of neutralinos at the
	 given masses.}
\label{fig:cluster}
\end{figure}
% End of the Figure

   Fig.\,\ref{fig:cluster} shows the spectra for both normal
   and \bv\   clusters. The spectrum for the baryon number
   violating clusters peaks at a
   much higher mass than the baryon number conserving clusters and has a
   large tail at high masses. This therefore means that before these clusters
   are decayed to hadrons most of them must be split into lighter clusters. 
   The baryon number conserving clusters in these events have the
   same spectrum as in Standard Model events.
   Fig.\,\ref{fig:cluster} contains the mass spectrum of pairs of
   colour-connected partons after the parton-shower phase and the
   non-perturbative splitting of the gluons into quark--antiquark pairs.
   The baryon number conserving clusters, Fig.\,\ref{fig:cluster}a, contains
   all the clusters in $\mr{e^+e^-}$ events
   at the given centre-of-mass energies, whereas the baryon number violating
   clusters, Fig.\,\ref{fig:cluster}b, only contains those clusters which
   contain the three quarks left after all the other quarks are paired into
   colour singlets from neutralino decays at the given mass.

\begin{figure}
\centering
%\vskip 10mm
\includegraphics[angle=90,width=0.45\textwidth]{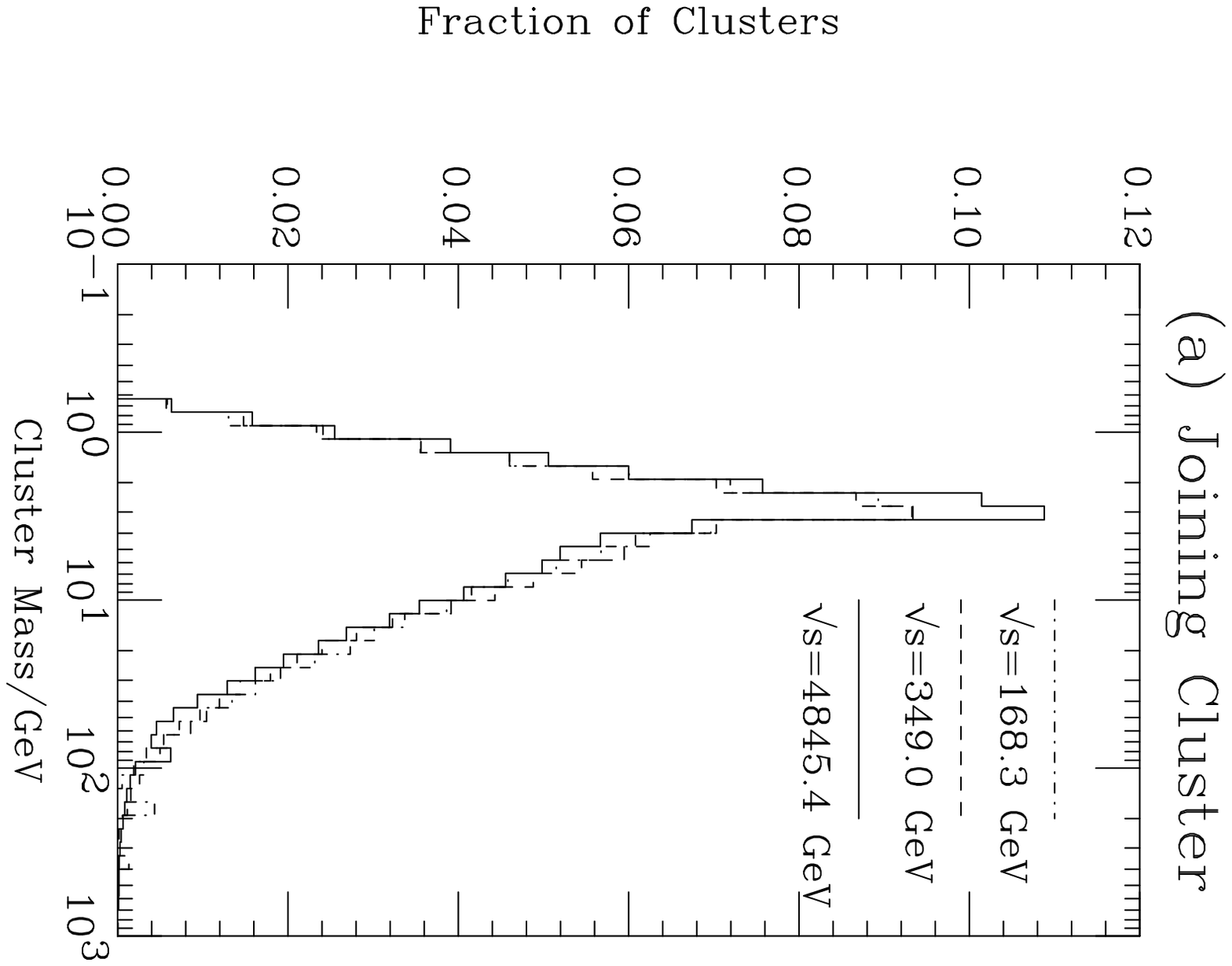}
\hfill
\includegraphics[angle=90,width=0.45\textwidth]{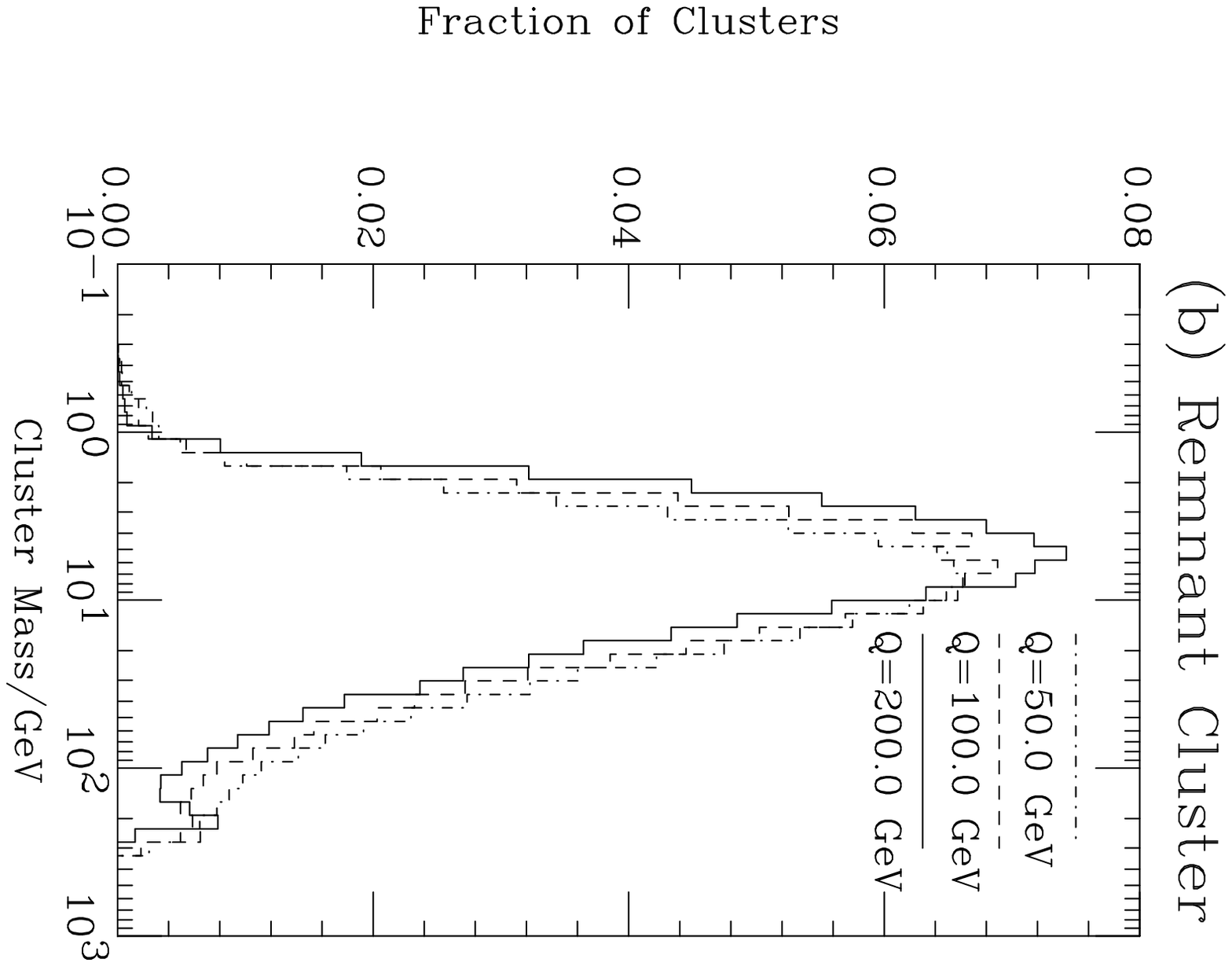}\\
\captionB{Masses of the joining and remnant clusters.}
	{Masses of the joining and remnant clusters. The joining clusters are
         from $\mr{e^+e^-}$ events at the given centre-of-mass energy,
         $\sqrt{s}$. The remnant  clusters were generated
         in $\mr{e^-p}$ events with $30\, \mr{\gev}$
	 electrons and $820\, \mr{\gev}$ protons with the given
         scale as the minimum value of the momentum transfer, Q.}
\label{fig:cluster2}
\end{figure}

   Fig.\,\ref{fig:cluster2}a shows the joining clusters, \ie the clusters in
   $\mr{e^+e^-}$ events with a quark from the parton shower of the quark
   and an antiquark from the parton shower of the antiquark. The remnant
   clusters, Fig.\,\ref{fig:cluster2}b, come from the cluster in
   deep-inelastic scattering events which contains
   the diquark, formed from two of the valence quarks.

% Cluster Mass Spectrum for joining and DIS clusters
% End of the Figure

  We would expect the baryon number violating clusters to be
  heavier than the standard
  baryon number conserving clusters because:
\begin{enumerate}
  \item The baryonic cluster is formed from three quarks originating from
        three different jets, as shown in Fig.\,\ref{fig:cluster} from
	the neutralino decay. In normal, \ie baryon number conserving,
  	 $\mr{e^+e^-\ra hadrons}$ events
        the clusters joining partons from different jets,
  	\ie containing a parton from each jet, are heavier than the clusters
	which come entirely from partons from one jet,
	Fig.\,\ref{fig:cluster2}a.
   \item The \bv\  
 	 cluster contains a diquark and in general the clusters containing
	 diquarks, in for example the hadron remnant in deep-inelastic
	 scattering, are heavier than the quark--antiquark clusters,
         Fig.\,\ref{fig:cluster2}b. 
\end{enumerate}
  As these clusters are heavier they will be more sensitive to the fine
  details of the hadronization model. In particular, these clusters are
  sensitive to the maximum cluster mass before the clusters are split into
  lighter clusters and the details of
  this splitting mechanism. This is also true for the clusters which join jets
  in $\mr{e^+e^-\ra 3\ jet}$ events and it is these clusters which contribute
  to the ``string effect'', which is well described by HERWIG. The mass
  distribution of the remnant in deep-inelastic scattering events at HERA is
  also reasonably well described by the cluster model.

\section{Results}
\label{sect:monteresults}

  We have implemented R-parity violating decays and hard processes
  into the HERWIG Monte Carlo event generator according to the
  algorithms given in Sections~\ref{subsect:monteRPVangles} and
  \ref{subsect:monteRPVhadron}. They are available in the current version
  HERWIG\,6.1 \cite{HERWIG61}. Having taken care to implement colour coherence
  effects, it is of immediate interest to see whether they have a
  significant influence on observable final-state distributions. To
  this end we have studied some jet production processes and
  compared the final-state distributions with those from standard QCD
  di-jet events. 
  It was observed in \cite{Abe:1994nj} that  certain
  variables can be constructed which are particularly sensitive to
  colour coherence effects. In particular these variables are sensitive
  to the presence of colour connections that link the initial and final
  states.  To investigate these effects for the different colour-connection
  structures of the
  \rpv\  models, we will study these variables for jet production via resonant
  sparticle production in hadron--hadron collisions.
  We essentially follow the details of the analysis of \cite{Abe:1994nj}.

  As examples, we study the processes:
\begin{itemize}
 \item $\mr{\bar{u} d \ra  \bar{u} d}$ via a resonant
 stau, and $\mr{\bar{d}d \ra  \bar{d} d}$ via a
 resonant tau sneutrino. These occur via the coupling ${\lam'}_{311}$.  The
 Feynman diagrams for these processes are shown in
 Fig.\,\ref{fig:LQDbump}. This process involves lepton number violating
 couplings but no baryon number violating vertices.

 \item Resonant squark production via the coupling ${\lam''}_{212}$. This
       leads to resonant down, strange and charm squark production. The
       resonant diagram for this process is shown in Fig.\,\ref{fig:bump}. 
\end{itemize}

  These processes were chosen to try to maximize the cross section at the
  Tevatron given the current low energy limits on the couplings. The coupling
  ${\lam'}_{311}$ has an upper bound, at the $2\sigma$ level, given by
  \cite{Allanach:1999bf}
\begin{equation}
   {\lam'}_{311} < 
	0.11\left(\frac{M_{\mr{\dnt}_R}}{100\, \mathrm{\gev}}\right)\!.
\label{eqn:limit311}
\end{equation}
  While the bounds on other LQD couplings are weaker, they involve higher
  generation quarks and hence the cross sections will be suppressed by the
  parton luminosities. As we will see in Chapter~\ref{chap:slepton}, for the
  values of the couplings which are still allowed by the current experimental
  limits \cite{Allanach:1999bf}, these processes may not be visible above the
  large QCD background.

%
% Feynman diagrams of the processes and the colour flow
%
\begin{figure}
\begin{center} \begin{picture}(360,80)(20,0)
\SetScale{0.7}
\SetOffset(30,0)
\ArrowLine(5,26)(60,52)
\ArrowLine(60,52)(5,78)
\DashArrowLine(90,52)(60,52){5}
\ArrowLine(90,52)(145,26)
\ArrowLine(145,78)(90,52)
\Text(85,20)[]{ $\mr{u}$}
\Text(85,55)[]{ $\mr{\bar{d}}$}
\Text(23,55)[]{ $\mr{\bar{d}}$}
\Text(23,20)[]{ $\mr{u}$}
\Text(55,44)[]{ $\mr{\tilde{\tau}_{\al}}$}
\Vertex(60,52){1}
\Vertex(90,52){1}
\SetOffset(155,0)
\ArrowLine(5,26)(60,52)
\ArrowLine(60,52)(5,78)
\DashArrowLine(90,52)(60,52){5}
\ArrowLine(90,52)(145,26)
\ArrowLine(145,78)(90,52)
\Text(85,18)[]{ $\mr{d}$}
\Text(85,55)[]{ $\mr{\bar{d}}$}
\Text(23,55)[]{ $\mr{\bar{d}}$}
\Text(23,20)[]{ $\mr{d}$}
\Text(55,44)[]{ $\mr{\nut_\tau}$}
\Vertex(60,52){1}
\Vertex(90,52){1}
\SetOffset(280,-35)
\ArrowLine(60,128)(5,128)
\ArrowLine(115,128)(60,128)
\ArrowLine(5,76)(60,76)
\ArrowLine(60,76)(115,76)
\DashArrowLine(60,76)(60,128){5}
\Text(65,45)[]{ $\mr{d}$}
\Text(65,98)[]{ $\mr{\bar{d}}$}
\Text(23,98)[]{ $\mr{\bar{d}}$}
\Text(23,45)[]{ $\mr{d}$}
\Text(50,71)[]{$\mr{\nut_\tau}$}
\Vertex(60,128){1}
\Vertex(60,76){1}
\end{picture}
\end{center}
\captionB{Feynman diagrams for $\mr{\bar{u} d \rightarrow \bar{u} d}$ and
 $\mr{\bar{d} d \ra  \bar{d} d}$ via slepton exchange.}
	{Feynman diagrams for $\mr{\bar{u} d \rightarrow \bar{u} d}$ via
 	 resonant charged slepton production and 
	 $\mr{\bar{d} d \ra  \bar{d} d}$ via resonant sneutrino production
	 and $t$-channel sneutrino exchange.}
\label{fig:LQDbump}
\end{figure}
% end of the figure
  Similarly for resonant squark production, the couplings which couple to two
  first generation quarks would in principle give the highest cross sections.
  However, the
  limits on these couplings are so strict that we used the
  ${\lam''}_{212}$ coupling
  which is only limited by perturbativity \cite{Allanach:1999bf}.
  At the Tevatron these processes have a very low cross section due to
  the requirement for
  two quarks to be 
  carrying a high fraction of the incoming proton or anti-proton's momentum in
  the initial state. This means that resonant squark production followed by
  a \bv\  decay will never by visible above the QCD background.
  Here we will simply look at the difference in the observables used in 
  \cite{Abe:1994nj} for the two different processes and the QCD background.

  We would expect very different results for the variables which are
  sensitive to the initial--final state colour connections for these two
  processes. In particular the first process only has this type of
  connection for the $t$-channel sneutrino diagram, which gives
  a small contribution close to the resonance, whereas the second process does
  have  initial--final state colour connections, due to the random colour
  structure at the \bv\  vertex, for the resonant production mechanism.

  We looked at both processes and QCD di-jet production using events
  generated with the program described in
  \cite{HERWIG61}. The cone algorithm, with a cone-size in ($\eta,\phi$)
  of 0.7~radians, was used to define the jets for this study. The actual
  algorithm used is very similar to that used by CDF apart from taking the
  midpoint between two particles as a seed for the algorithm in addition to
  the particles themselves. The inclusion of the midpoints as seeds improves
  the infra-red safety of the cone algorithm. The only cut was to require the
  presence of at least one jet with transverse energy, $E_T$, 
  greater than $200\, \mr{\gev}$ in
  the event. A parton-level cut requiring the transverse momentum, $p_T$,
  of the two final-state partons to
  be greater than $150\, \mr{\gev}$ was used to reduce the number of events we
  needed to simulate, however this should not affect the results. The
  signal points were generated using the following SUGRA parameters:

\begin{description}
  \item[Resonant Slepton] 
     $M_0=600\, \mr{\gev}$, $M_{1/2}=200\, \mr{\gev}$, $A_0=0\, \mr{\gev}$,
               $\tan\beta=10$, \mbox{$\sgn\mu=+$} and  ${\lam'}_{311}=0.8$;
  \item[Resonant Squark]
     $M_0=430\, \mr{\gev}$, $M_{1/2}=200\, \mr{\gev}$, $A_0=0\, \mr{\gev}$, 
               $\tan\beta=10$,\linebreak \mbox{$\sgn\mu=+$}  and
	 ${\lam''}_{212}=1.0$.
\end{description}

  These points were chosen so that the resonant sleptons and squarks have 
  approximately the same mass. At the first point the slepton masses are 
  $M_{\tilde{\tau}_1}=599\, \mr{\gev}$, $M_{\tilde{\tau}_2}=617\, \mr{\gev}$
  and $M_{\tilde{\nu}_\tau}=610\, \mr{\gev}$, while at the second point the
  squark masses are $M_{\dnt_R,\tilde{s}_R,\tilde{c}_R}=602\, \mr{\gev}$.

  We can now study the
  events about the resonance by imposing a cut\linebreak 
  \mbox{$580\, \mr{\gev}  \leq M_{jj} \leq 640\, \mr{\gev} $} and plotting
  the variables that are sensitive to angular ordering for these events.
  These variables depend on the distribution of a third jet in the events
  which is generated in the simulation by the parton-shower algorithm.
   Three variables, $\eta_3$, $R$ and $\al$ were considered.
   They are defined in the following way \cite{Abe:1994nj}:
\begin{itemize}
\item If we define the
   jets in the event in order of their $E_T$, with jet 1 being the
   hardest jet in the event, then $\eta_3$ is the pseudo-rapidity of
   the third jet.
\item Defining $\Delta\eta = \eta_3-\eta_2$ and the difference in polar angles
   $\Delta\phi = \phi_3-\phi_2$, then  the variable $R$ is the distance
   between the second and third jets in $(\eta,\phi)$ space, \ie $R=
  \sqrt{\Delta\eta^2+\Delta\phi^2}$.
\item If we define $\Delta H =
   \mr{sgn}(\eta_2)\Delta\eta$, we can consider
   the polar angle in $(|\Delta\phi|,\Delta H)$ space,
    $\al= \tan^{-1}(\Delta H/|\Delta\phi|)$.
\end{itemize}

   In the analysis of \cite{Abe:1994nj} additional cuts had to be imposed,
   which we also use here:
\begin{enumerate}
	\item a pseudo-rapidity cut on the highest two $p_T$ jets in the
		event, $|\eta_1|,|\eta_2|<0.7$;

	\item requiring the two leading jets to be back-to-back 
		$||\phi_1-\phi_2|-180^0|<20^0$;

	\item third jet transverse energy $E_{T_3}>10\, \mr{\gev}$  to avoid
	       background from the underlying event;

	\item a final cut used only for the study of $\al$ is that
	$1.1<R<\pi$ to avoid problems with the jet clustering algorithm. 
\end{enumerate}

   We can now study these variables for resonant slepton and squark
   production, and for QCD di-jet events. The results in all the graphs
   correspond to the number of events at the Run II Tevatron centre-of-mass
   energy of 2~TeV and integrated luminosity of 2~$\mr{fb^{-1}}$. 

\begin{figure}
\vskip -8mm
\subfigure[Resonant slepton]{
\includegraphics[angle=90,width=0.3\textwidth]{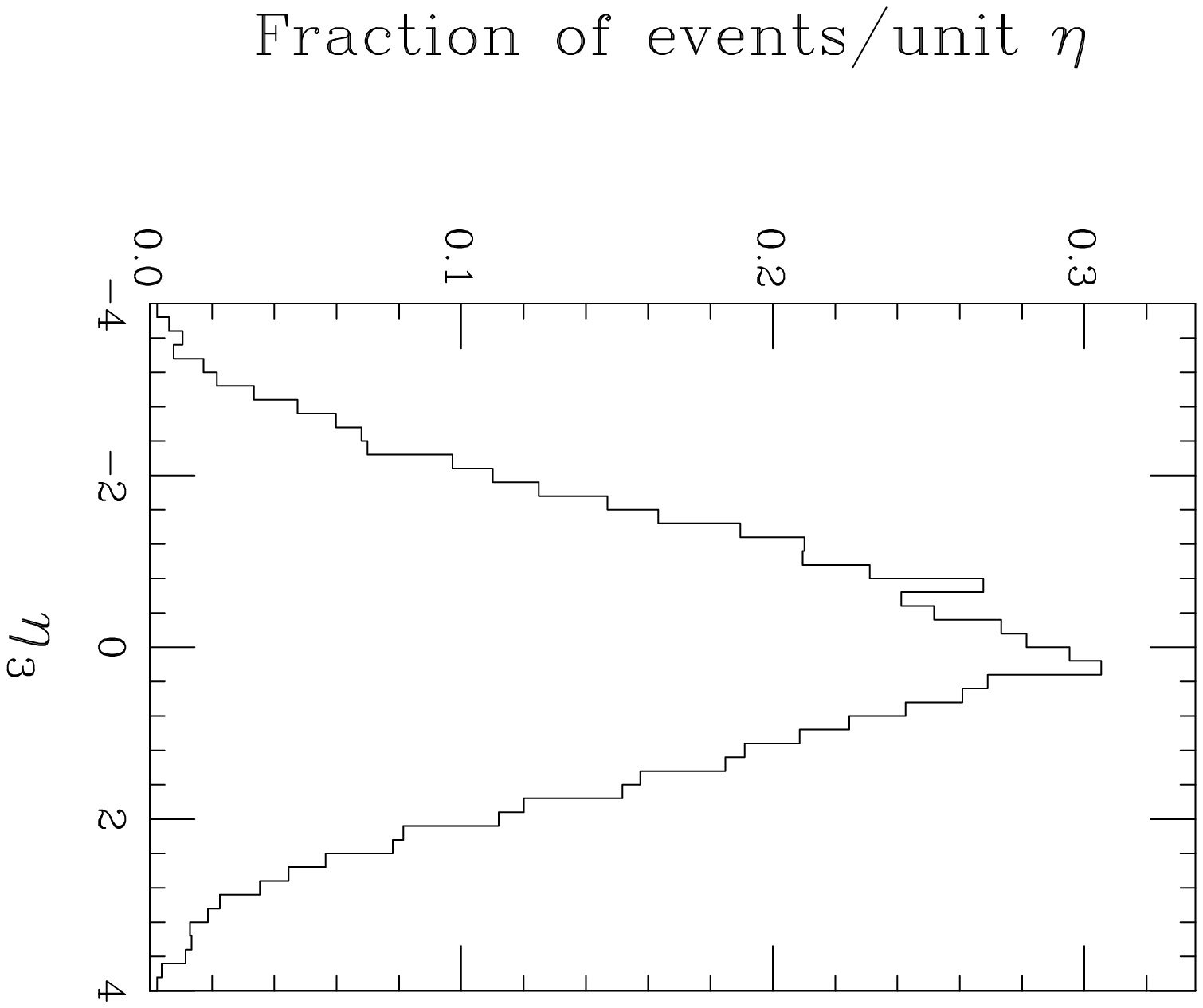}
\label{fig:etaLQD}}\hfill
\subfigure[Resonant squark]{
\includegraphics[angle=90,width=0.3\textwidth]{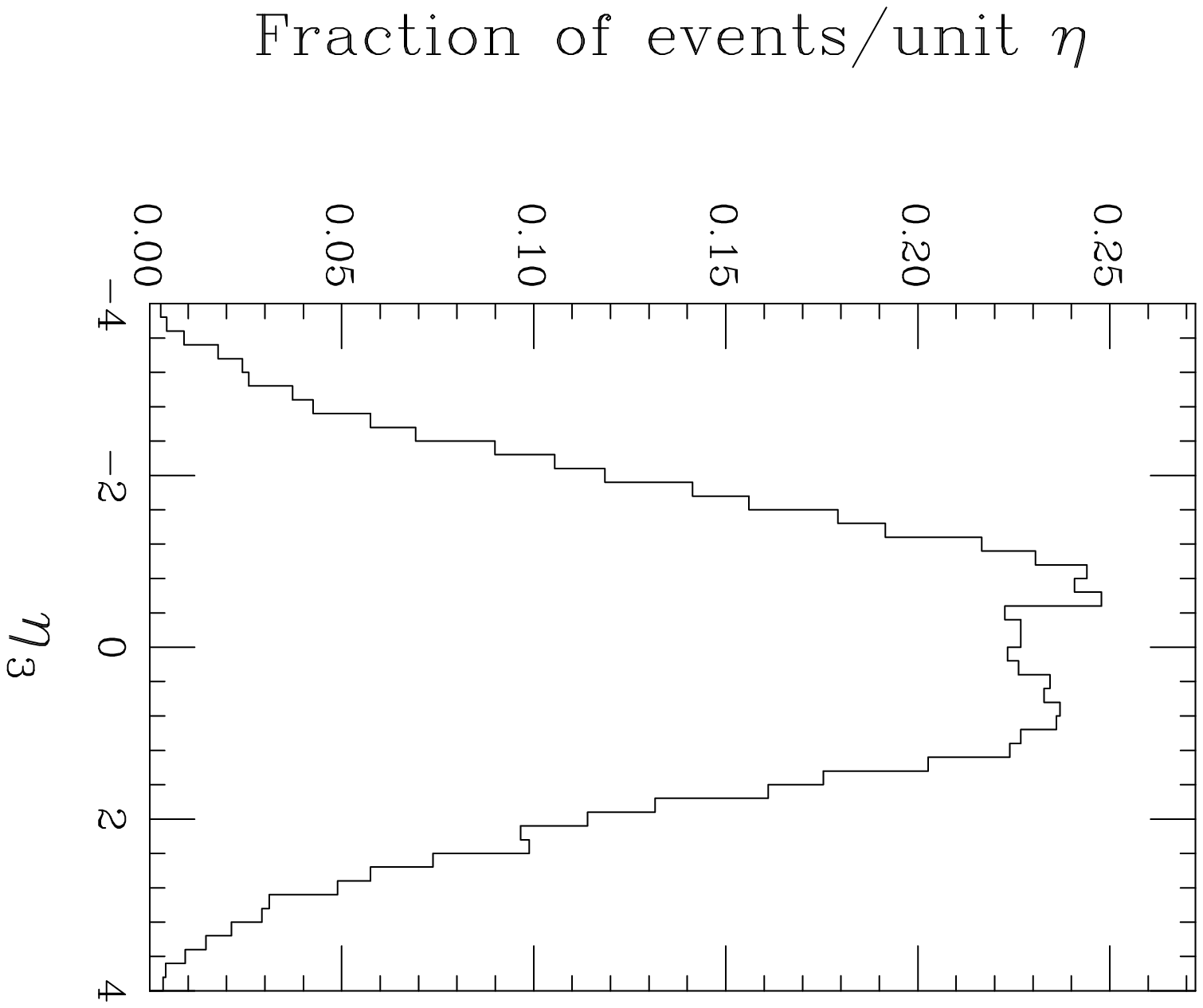}
\label{fig:etaUDD}}\hfill
\subfigure[QCD di-jet]{
\includegraphics[angle=90,width=0.3\textwidth]{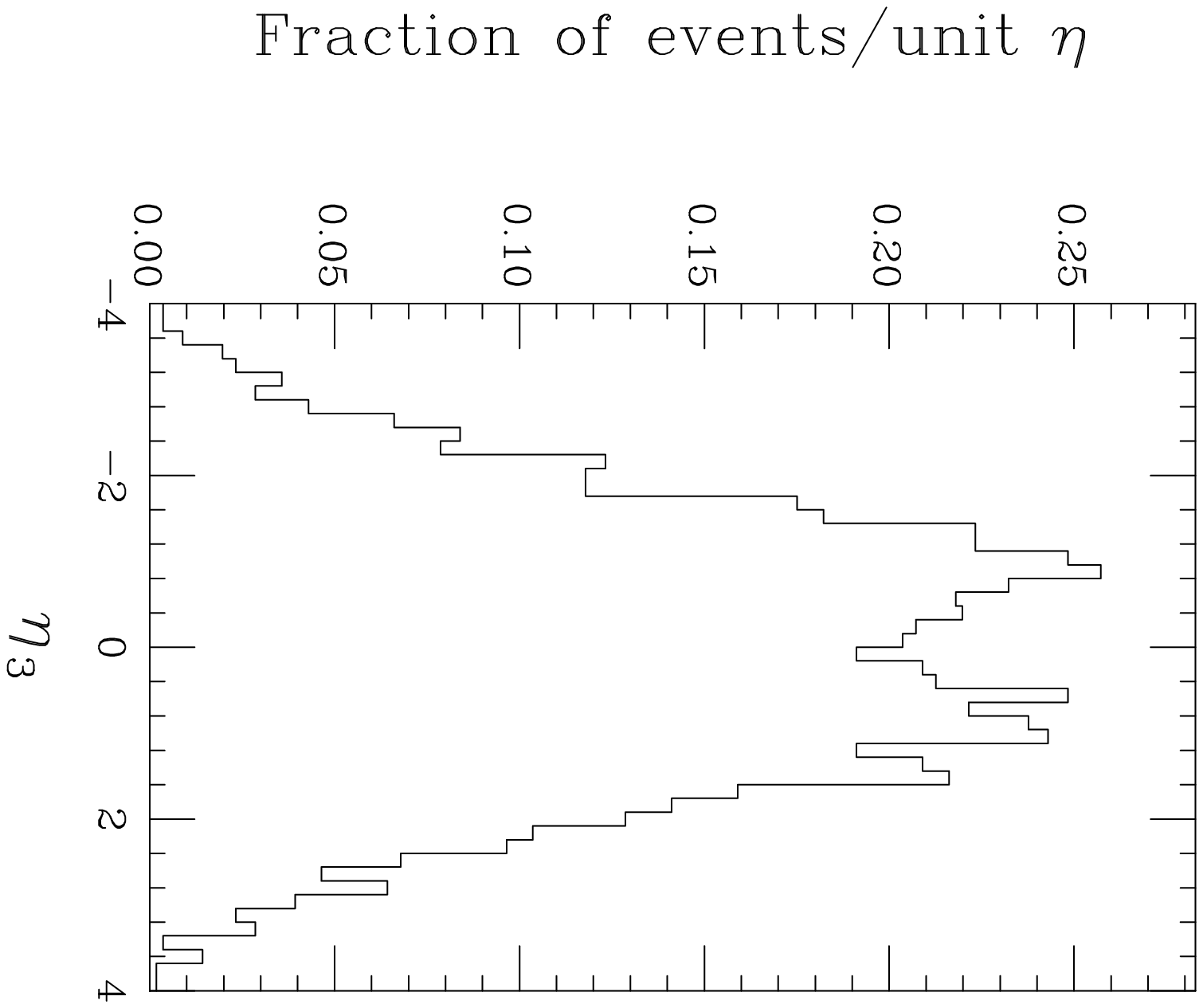}
\label{fig:etaQCD}} 
\\
\vskip -4mm
\captionB{Distribution of $\eta_3$ for resonant slepton, squark and QCD
	 jet production.}
	{Distribution of $\eta_3$ for resonant slepton, resonant squark
	 and QCD jet production.}
\label{fig:etatotal}
%\end{figure}
% end of the figure
%
%  Distribution of R
%
%\begin{figure}
\subfigure[Resonant slepton]{
\includegraphics[angle=90,width=0.3\textwidth]{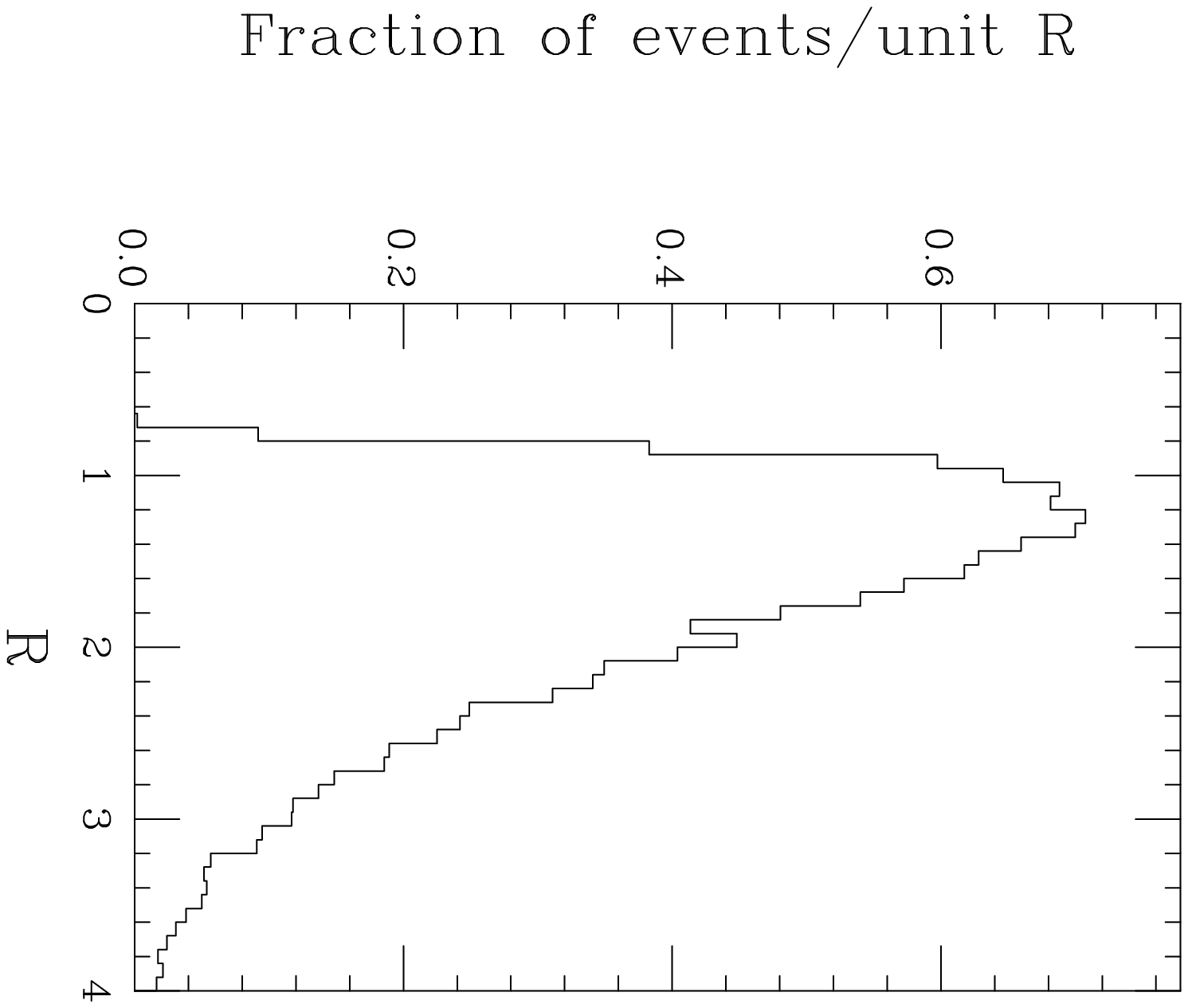}
\label{fig:RLQD}}\hfill
\subfigure[Resonant squark]{
\includegraphics[angle=90,width=0.3\textwidth]{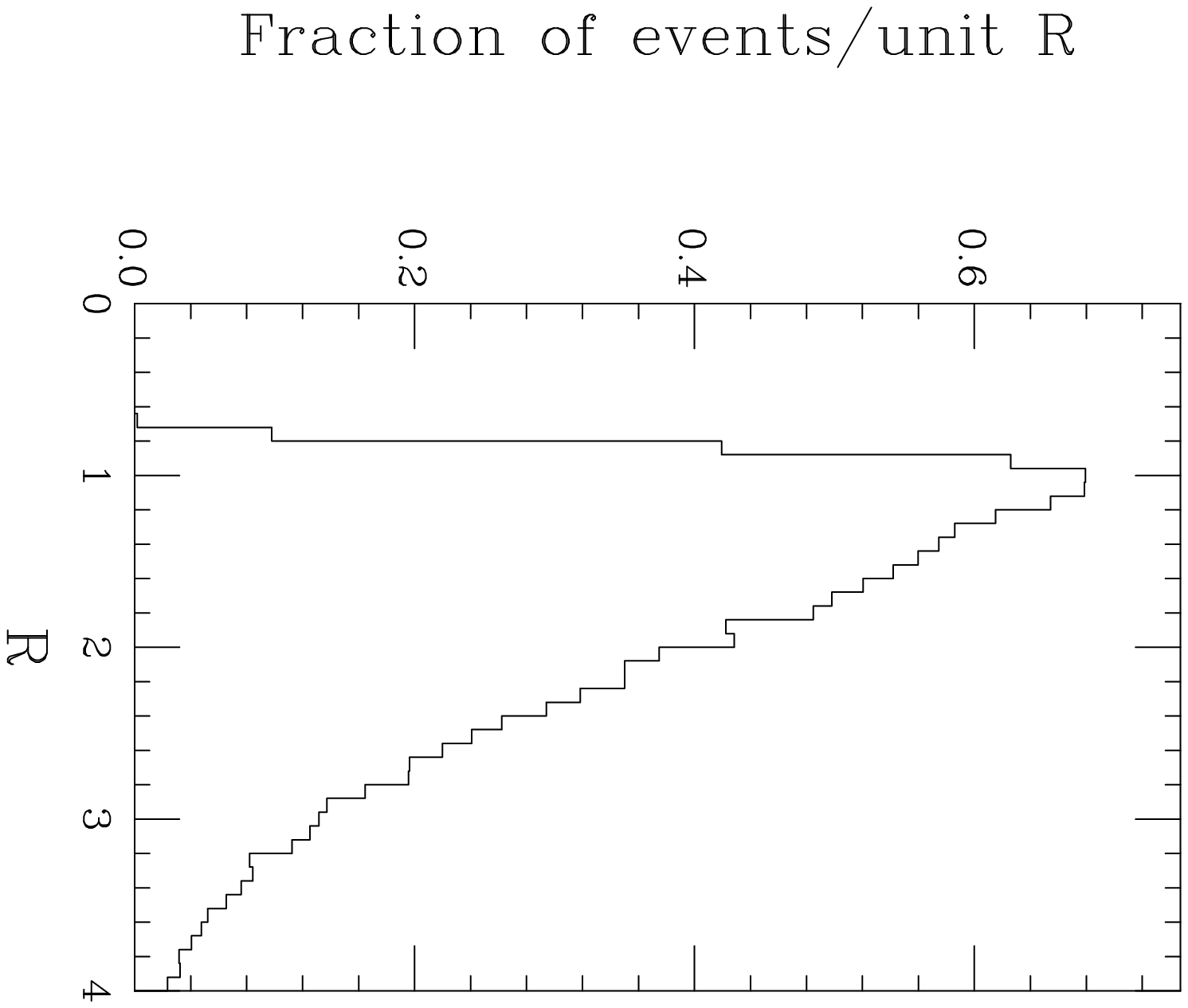}
\label{fig:RUDD}}\hfill
\subfigure[QCD di-jet]{
\includegraphics[angle=90,width=0.3\textwidth]{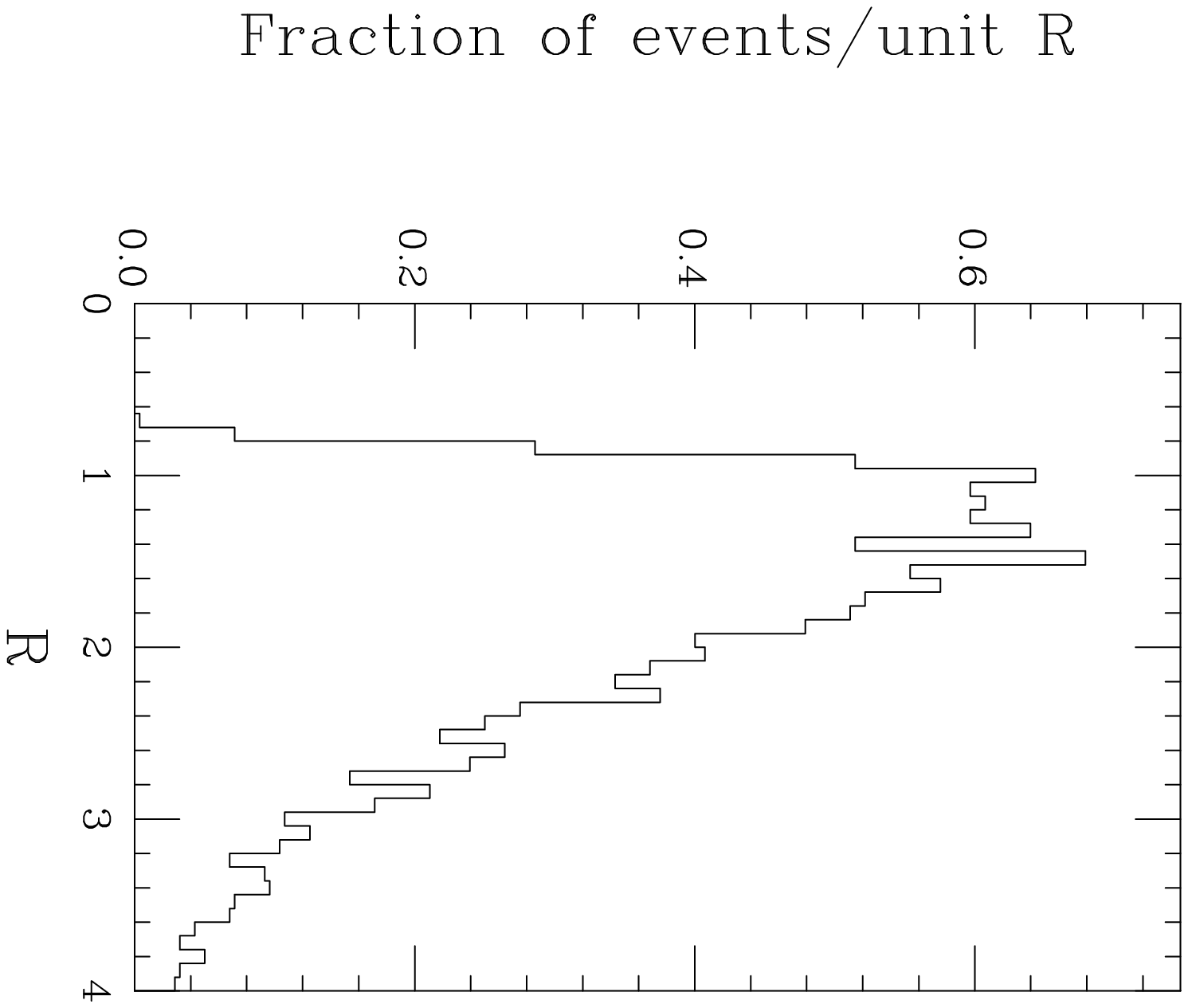}\label{fig:RQCD}}
\\
\vskip -4mm
\captionB{Distribution of $R$ for resonant slepton, squark and QCD
	 jet production.}
	{Distribution of $R$ for resonant slepton, resonant squark and
         QCD jet  production.}
\label{fig:Rtotal}
%\end{figure}
% end of the figure
%
%  Distribution of alpha 
%
%\begin{figure}
\subfigure[Resonant slepton]{\hfill
\includegraphics[angle=90,width=0.3\textwidth]{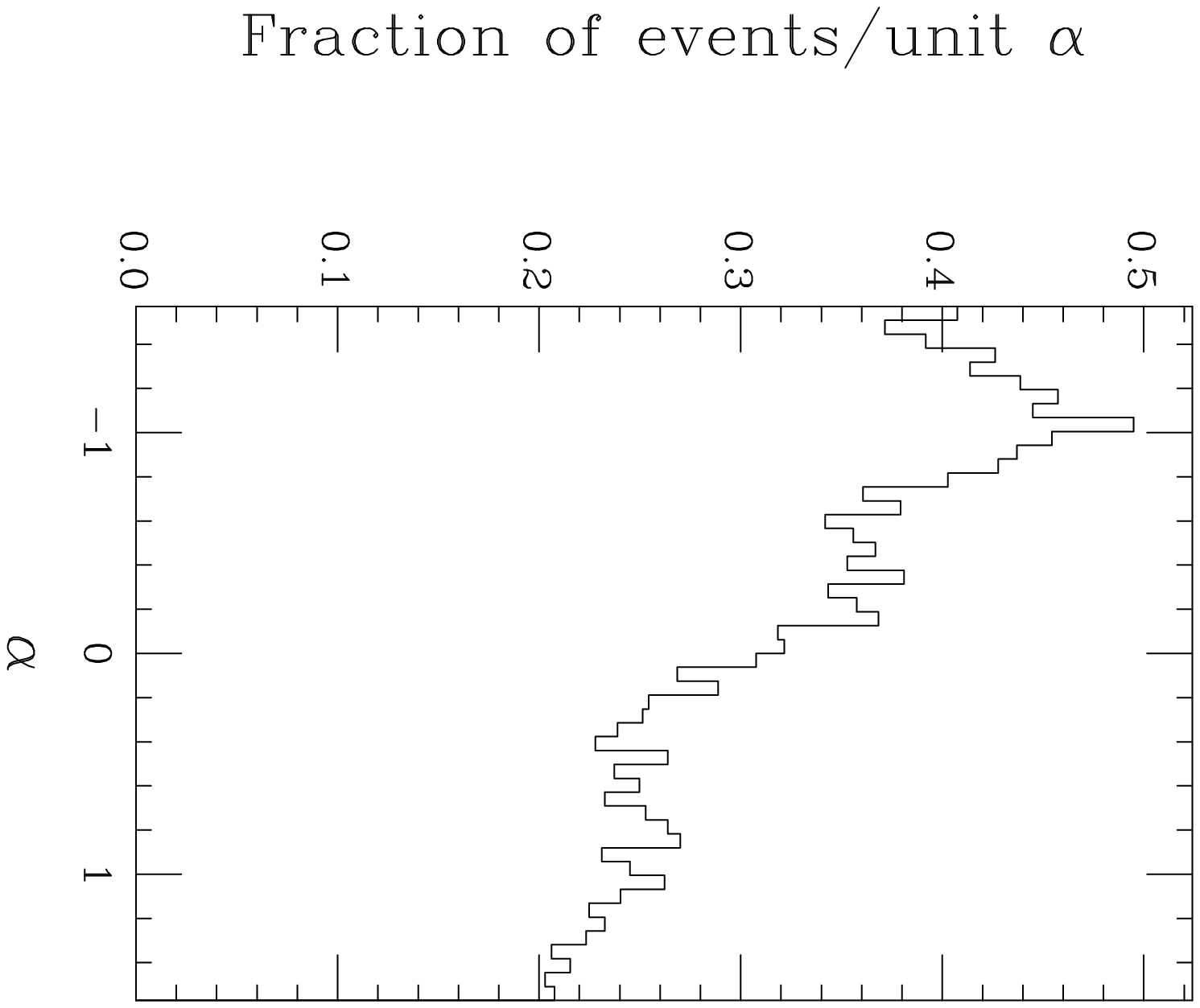}
\label{fig:alphaLQD}}\hfill
\subfigure[Resonant squark]{\hfill
\includegraphics[angle=90,width=0.3\textwidth]{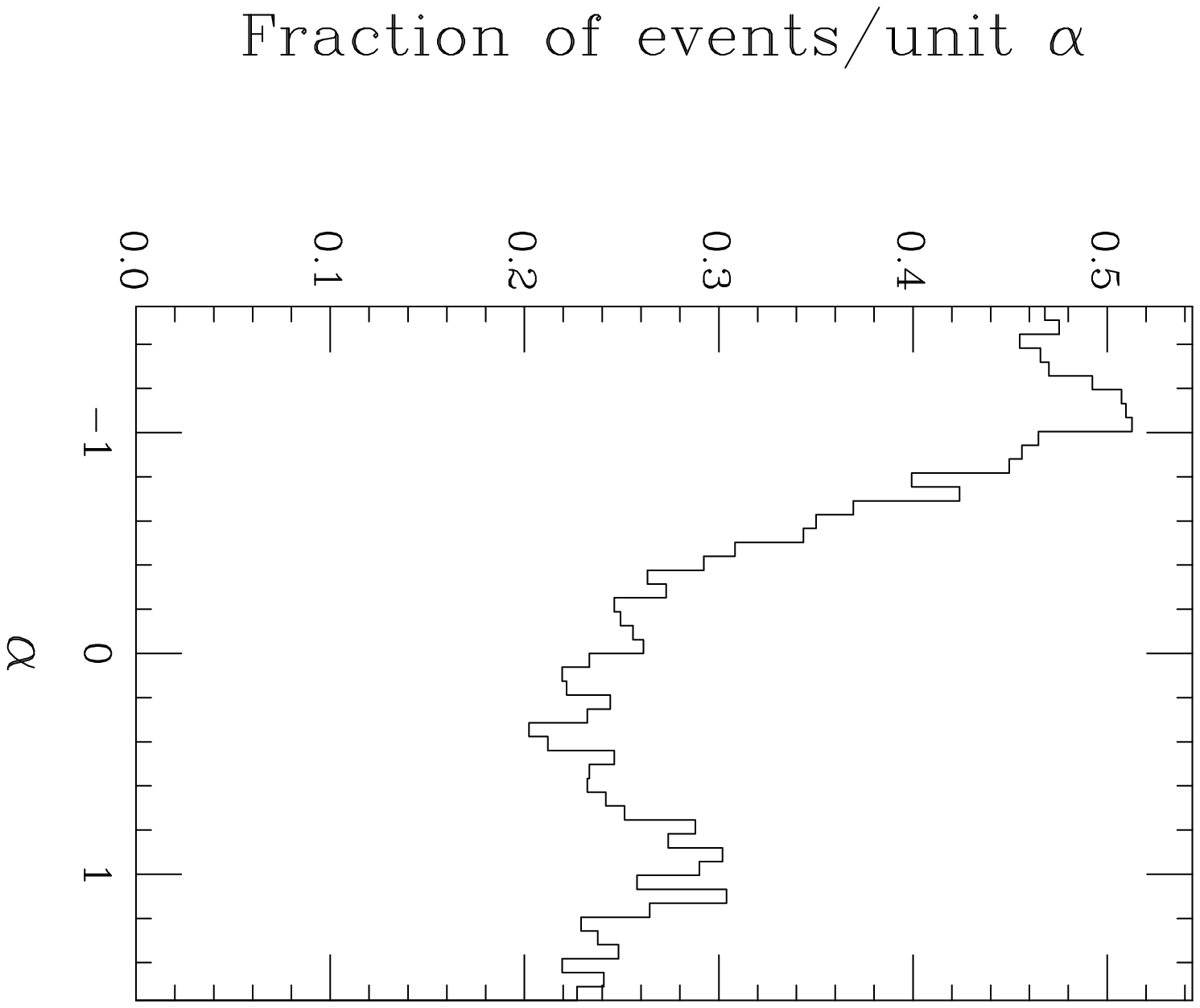}
\label{fig:alphaUDD}}\hfill
\subfigure[QCD di-jet]{\hfill
\includegraphics[angle=90,width=0.3\textwidth]{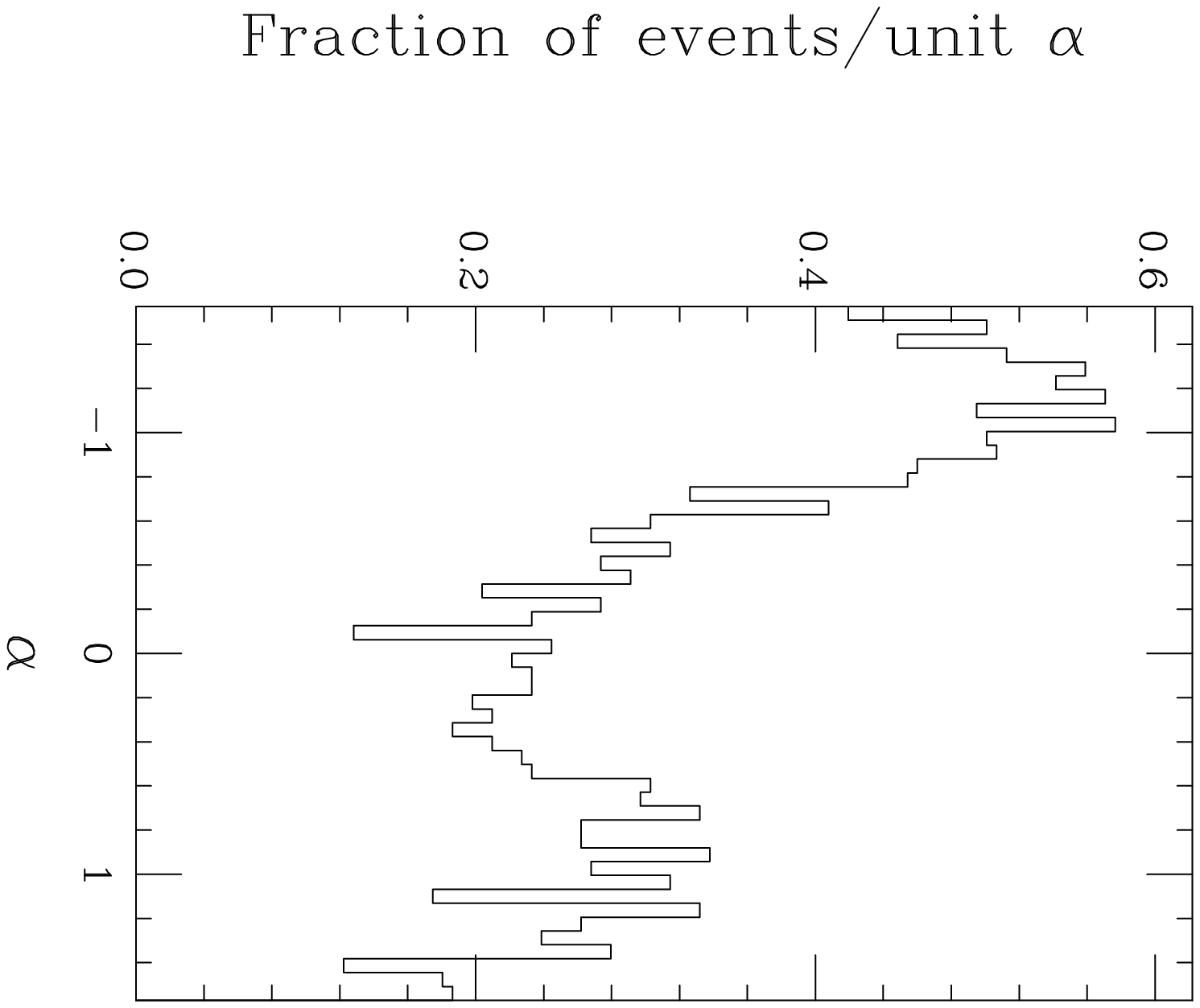}
\label{fig:alphaQCD}}
\\
\vskip -4mm
\captionB{Distribution of $\al$ for resonant slepton, squark and QCD
	 jet production.}
        {Distribution of $\al$ for resonant slepton, resonant squark and
         QCD jet production.}
\label{fig:alphatotal}
\end{figure}

   There are a number of differences in the observables $\eta_3$, $R$ and
   $\al$ for resonant slepton production and QCD jet production.
   In particular, in the $\eta_3$ distributions
   instead of a dip in the distribution at $\eta_3=0$ for the QCD jet
   production, Fig.\,\ref{fig:etaQCD},
   there is a bump for the resonant slepton production, 
   Fig.\,\ref{fig:etaLQD}. This dip in the QCD jet production was
   observed in \cite{Abe:1994nj}, and is a feature of the initial--final
   state colour connections which are present in QCD jet production, but not
   in resonant slepton production.

%
%  Distribution of eta_3 
%

   The distribution of the distance in $(\Delta\eta,\Delta\phi)$ space, $R$,
   is very similar for both resonant slepton production, 
   Fig\,\ref{fig:RLQD}, and QCD jet production, Fig\,\ref{fig:RQCD} .
   In the study of \cite{Abe:1994nj} all the event generators, even
   those which do not include angular
   ordering, gave good agreement with the data for this observable.
   The distribution of
   the polar angle  $\al$
   also shows a
   difference between resonant slepton production,
   Fig.\,\ref{fig:alphaLQD}, and QCD jet production,
   Fig.\,\ref{fig:alphaQCD}, with the resonant slepton production
   not showing the dip in the middle which is again an effect of the
   initial--final state colour connections which are present in
   QCD jet production but not in resonant slepton production.

   There is less difference for
  the variables $\eta_3$, $R$ and $\al$ between resonant squark production 
  and QCD jet production shown in
  Figs.\,\ref{fig:etatotal}, \ref{fig:Rtotal} and \ref{fig:alphatotal}. The
  distributions for the two resonant production processes are different. The
  distribution for resonant
  squark production shows the dip at $\eta_3=0$ and the rise as
  $\al\ra\frac{\pi}{2}$ which is due to the colour connections in
  these processes between the initial and final states. The effect of
  the colour flow between the initial and final states is less
  for this process than for QCD jet production as there are combinatorially 
  fewer such
  connections for resonant squark production.

  The fact that the final-state distributions of the resonant slepton
  and resonant squark production processes are so different, despite
  the identities and kinematics of the jets themselves being so
  similar, clearly shows that colour coherence plays an important r\^{o}le
  in determining the properties of R-parity violating processes. Even
  if this is not used as a tool to enhance the signal, it is likely
  that it will effect the efficiency of any cuts that are applied, so
  it is essential that any experiments looking for R-parity violating
  processes take into account colour coherence in their simulations of
  the signal. 

  Even if R-parity violating hard processes were added to ISAJET
  \cite{Baer:1999sp}, this event generator would not be expected
  to describe the final
  state well, as it is based on the incoherent parton shower and
  independent fragmentation models. Thus, in our case for example, 
  resonant slepton and resonant squark processes would have very
  similar properties. It is worth noting that ISAJET gives a 
  poor description of the CDF data  \cite{Abe:1994nj} on $\eta_3$
  and $\al$ in standard QCD di-jet events.

%
%  Conclusions
%
\section{Summary}

    We have presented a procedure for implementing colour coherence
  effects via the angular-ordering procedure in \rpv\   SUSY models. We
  find that the \bv\  processes have a random colour connection structure
  for angular ordering. In these processes we see, for the first time,
  differences in the colour partners for the colour coherence effects
  and those used, with the idea of colour preconfinement, for
  hadronization in the cluster model. 

  A full set of decays and hadron--hadron cross sections have now been 
  implemented in the HERWIG Monte Carlo event generator 
  \cite{HERWIG61,SUSYimplement},
  using the matrix elements given in Appendices~\ref{chap:decay} and 
  \ref{chap:cross}. The first preliminary
  results for these processes show that the inclusion of colour
  coherence is important.
  In the next chapter we will look at the possibility of using
  these colour coherence properties to improve the extraction of a resonant
  slepton signature at the Tevatron.

  These simulations have allowed the first experimental studies of the
  possibility of discovering \rpv\  supersymmetry via the baryon number
  violating decay of the neutralino at the LHC \cite{Drage:1999th}.

%\chapter{Resonant slepton}
%%%%%%%%%%%%%%%%%%%%%%%%%%%%%%%%%%%%%%%%%%%%%%%%%%%%%%%%%%%%%%%%%%%%%%%%%%%%%%%%%
%										%	
%  Chapter on the production of resonant sleptons in hadron--hadron collisions	%
%										%
%  This is basically an expansion of my Tevatron Run II workshop talk,		%
%  the LHC results and the results section of the RPV montecarlo paper		%
%										%
%   The structure will be as follows						%
%										%
%   1. Introduction  -  List the various possible decays modes and the ones we	%
%			we will be looking at in more detail + references for 	%
%			the other.						%
%										%
%   2. RPV Decays    -  Resonant slepton followed by gauge decay at the 	%
%			TEVATRON. Basically the results section from the RPV	%
%			Montecarlo papers					%
%										%
%   3. Gauge Decays  -	The decay to lepton neutralino followed by the decay of	%
%			the neutralino giving like sign dileptons. Basically 	%
%			TEVATRON and LHC workshop stuff				%
%										%
%   4. Conclusion    -  Some conclusions highlighting what can now be done as 	%
%			we have a full simulation				%
%										%
%%%%%%%%%%%%%%%%%%%%%%%%%%%%%%%%%%%%%%%%%%%%%%%%%%%%%%%%%%%%%%%%%%%%%%%%%%%%%%%%%

\chapter{Resonant Slepton Production in Hadron--Hadron Collisions}
\label{chap:slepton}
%
%  Introduction giving slepton decay modes etc..
%
\section{Introduction}

  In the previous chapter we discussed how to include R-parity violating
  processes into a Monte Carlo simulation. We then briefly looked at the
  differences in the final states for resonant squark and slepton production,
  which were solely due to the different colour structures of the two
  different
  processes. In this chapter we will look at resonant slepton production in
  hadron colliders and ways in which it can be detected using the simulations
  described in the previous chapter. In hadron colliders, as we saw in the
  previous chapter, sleptons can be produced on resonance via the \rpv\  
  $L_i Q_j{\overline D}_k$ term in the superpotential,
  Eqn.\,\ref{eqn:Rsuper1}.

%%%%%%%%%%%%%%%%%%%%%%%%%%%%%%%%%%%%%%%%%%%%%%%%%%%%%%%%%%%%%%%%%%%%%%%%%%%%%%
%
% charged slepton and sneutrino decays
%
\begin{table}[!b]
\renewcommand{\arraystretch}{1.2}
\begin{center}
\begin{tabular}{|l|c|c|}
\hline
 & Charged Sleptons   & Sneutrinos  \\
\hline
%\vspace{5mm}
 Supersymmetric  & $\mr{\elt}_{i\al} 	\ra
		\mr{\ell}^{-}_{i} 	\mr{\cht^0}$ 
                 & $\mr{\nut}_i 	\ra
	 	\mr{\nu}_i 		\mr{\cht^0	}$ \\
 Gauge Decays	  & $\mr{\elt}_{i\al} 	\ra
	 		\mr{\nu}_i	 	\mr{\cht^-	}$
 		 & $\mr{\nut}_i 	\ra
	 		\mr{\ell}^-_i 		\mr{\cht^+	}$\\
\hline
 \rpv\  Decays	 & $\mr{\elt}_{i\al} 	\ra
	 		\mr{\bar{u}}_j 		\mr{d}_k$ 
		 & $\mr{\nut}_i 	\ra
	 		\mr{\bar{d}}_j 		\mr{d}_k$ \\
		 & $\mr{\elt}_{i\al} 	\ra
	 		\mr{\bar{\nu}}_j 	\mr{\ell}^-_k$ 
 		 & $\mr{\nut}_i 	\ra
	 		\mr{\ell}^+_j 		\mr{\ell}^-_k$ \\
\hline
 Weak Decays	 & $\mr{\elt}_{i\al} 	\ra
		\mr{\nut}_i		\mr{W^-	}$
 		 & $\mr{\nut}_i		\ra
		\mr{\elt}_{i\al}	\mr{W^+	}$ \\
		 & $\mr{\elt}_{i2}	\ra
		\mr{\elt}_{i1}		\mr{Z_0	}$
		 & \\
\hline
 Higgs Decays	 & $\mr{\elt}_{i\al}	\ra
		\mr{\nut}_i		\mr{H^-	}$
 		 & $\mr{\nut}_i		\ra
		\mr{\elt}_{i\al}	\mr{H^+	}$ \\
		 & $\mr{\elt}_{i2}	\ra
		\mr{\elt}_{i1}	\mr{h_0,H_0,A_0}$
 		 & \\
\hline	
\end{tabular}
\captionB{Decay modes of charged sleptons and sneutrinos.}
	{Decay modes of charged sleptons and sneutrinos. The index 
	$i,j,k=1,2,3$ gives
	 the generation of the fermion or sfermion and
	 the index $\al=1,2$ the mass eigenstate of the sfermion as
	 described in Appendix~\ref{chap:Feynman}.}
\label{tab:decaymodes}
\end{center}
\end{table}
%%%%%%%%%%%%%%%%%%%%%%%%%%%%%%%%%%%%%%%%%%%%%%%%%%%%%%%%%%%%%%%%%%%%%%%%%%%%%%

  Resonant slepton production in hadron--hadron collisions has previously been
  considered in \cite{Dimopoulos:1990fr,Kalinowski:1997zt,Hewett:1998fu,
  Allanach:1999bf,Moreau:1999bt:Moreau:2000ps:Moreau:2000bs,Abdullin:1999zp}.
  The signature of this process depends
  on the decay mode of the resonant slepton. The various possible decay
  channels are given in Table\,\ref{tab:decaymodes}.
  The cross sections for resonant slepton production followed by these decay
  modes are given in Appendix~\ref{chap:cross}.  Most of the previous 
  studies have only considered the \rpv\  decays of the resonant slepton to
  either leptons via the first term in Eqn.\,\ref{eqn:Rsuper1} 
  \cite{Dimopoulos:1990fr,Kalinowski:1997zt,Hewett:1998fu,Allanach:1999bf},
  or to quarks via the second term in Eqn.\,\ref{eqn:Rsuper1} 
  \cite{Dimopoulos:1990fr,Hewett:1998fu,Allanach:1999bf}.

  We shall first consider the \rpv\  decay modes and look at the use of 
  the angular-ordering properties of these processes to improve the
  extraction of
  a signal over the QCD background in Section~\ref{sect:sleptonrpv}.

  There has been little study  of the supersymmetric gauge decays of the
  resonant sleptons. The cross sections for these processes were first
  presented in \cite{Dimopoulos:1990fr}\footnote{The cross sections for the
  resonant processes were only presented in the narrow-width approximation
  and the interference between the resonant and $t$-channel diagrams was
  neglected. Neutralino and chargino mixing was also neglected, and there
  were errors in the calculations of the non-resonant diagrams.}
  where there was a discussion of the
  possible experimental signatures, however the signal
  we will consider
  was not discussed and there was no calculation of the Standard Model
  background.
  The supersymmetric gauge decays of the sneutrinos have been studied
  \cite{Moreau:1999bt:Moreau:2000ps:Moreau:2000bs,Abdullin:1999zp}. 
  These studies were performed using a detector-level
  Monte Carlo simulation for both the signal and background processes.
  These analyses looked at the trilepton
  signature for resonant sneutrino production, which comes from the decay
  chain $\nut\ra\ell^-\cht^+$, followed by a decay of the chargino, 
  $\mr{\cht^+\ra\ell^+\nu_\ell\cht^0_1}$, and the decay of the neutralino to
  a lepton and a quark--antiquark pair.
  A study of the supersymmetric  gauge decays of the sleptons is presented in
  Section~\ref{sect:sleptongauge},
  where we look at the like-sign dilepton signature for these processes.

  There has been no study of either the weak or Higgs decay modes for which,
  in general, the resonance is not kinematically accessible,
  although for completeness these have
  been included in the HERWIG event generator \cite{HERWIG61} and the
  cross sections are presented in Appendix~\ref{chap:cross}.

%
% Section on the rpv decays from the rpv montecarlo papers
%
\section[\rpv\  Decays of the Resonant Slepton]
	{\boldmath{\rpv}\  Decays of the Resonant Slepton}
\label{sect:sleptonrpv}

  There are two types of \rpv\  decay modes of the resonant sleptons:
\begin{enumerate}
% Drell-Yan production
\item 	The first requires that in addition to the \rpv\  Yukawa coupling
 	which allows the resonant slepton production there is a second 
  	non-zero coupling, $\lam_{ijk}$, allowing the decays
	$\mr{\elt}_{i\al}\ra\mr{\bar{\nu}}_j \mr{\ell}^-_k$
	and $\mr{\nut}_i \ra\mr{\ell}^+_j \mr{\ell}^-_k$.
	This has been extensively studied
  \cite{Dimopoulos:1990fr,Kalinowski:1997zt,Hewett:1998fu,Allanach:1999bf}.
	As there is only initial-state QCD radiation, from the quarks
 	involved in
	the hard collision, in these processes the parton-level results of
	\cite{Dimopoulos:1990fr,Kalinowski:1997zt,Hewett:1998fu,
		Allanach:1999bf}
	should not be significantly affected
	by the addition of a full simulation. We therefore
	have not considered these processes.
% other RPV
\item   It is also possible that the resonant slepton decays back to the
	initial state via the same coupling required for the resonant slepton
	production. We studied the colour coherence properties of these
	 processes
	in the previous chapter and we will discuss them in more detail here.
\end{enumerate}

  We considered resonant slepton production followed by an \rpv\  decay
  using the same cuts which were applied in Section~\ref{sect:monteresults}.
  We also used the same SUGRA point \linebreak \mbox{$M_0=600\, \mr{\gev}$},
	 $M_{1/2}=200\, \mr{\gev}$, $A_0=0\, \mr{\gev}$,
               $\tan\beta=10$ and $\sgn\mu=+$.
The results in all the graphs correspond to the
number of events at Run II of the Tevatron, with centre-of-mass energy
of $2\, \mr{TeV}$
 and integrated luminosity of 2~$\rm{fb}^{-1}$. Again we used a cone algorithm
with radius parameter of 0.7 radians to define the jets.

\begin{figure}[t]
\includegraphics[angle=90,width=0.47\textwidth]{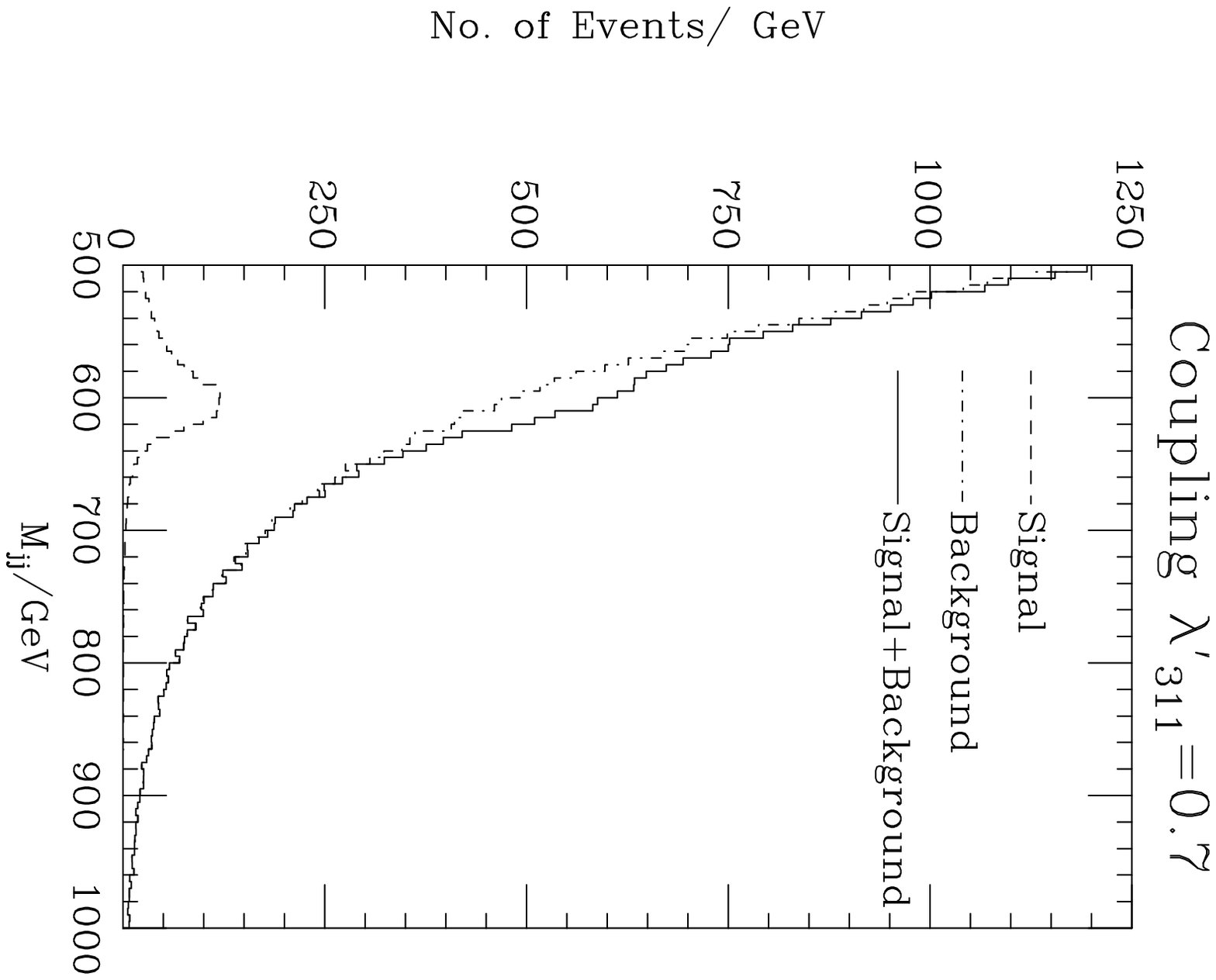}
\hfill
\includegraphics[angle=90,width=0.47\textwidth]{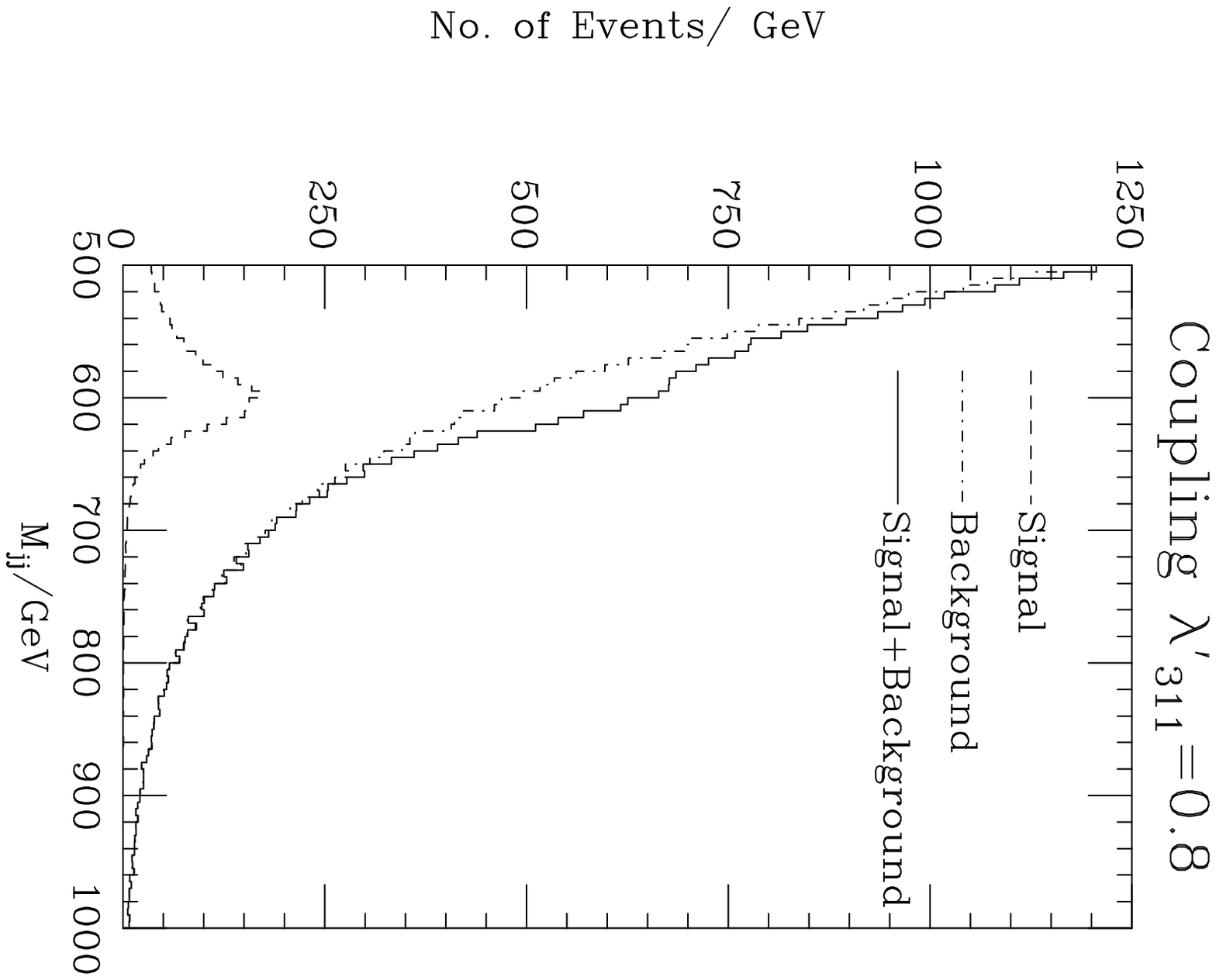}
\captionB{Di-jet invariant-mass distribution for ${\lam'}_{311}=
0.7$ and ${\lam'}_{311}=0.8$.}
	{Di-jet invariant-mass distribution for ${\lam'}_{311}=
0.7$ and ${\lam'}_{311}=0.8$.}
\label{fig:invm}
\end{figure}

At this SUGRA point the right sdown mass is $728\, \mr{\gev}$  
which corresponds to a limit on the
coupling of ${\lam'}_{311}<0.80$, from Eqn.\,\ref{eqn:limit311}.
As can been seen in Fig.\,\ref{fig:invm}, the results for two
different values of the coupling show that there is a bump in the
di-jet invariant-mass distribution, $M_{jj}$, from the resonant
slepton production for large values of the coupling.
%
% Invariant mass distribution of slepton
%
% End of the Figure

We can now consider the events around the bump, $580\, \mr{\gev}\leq M_{jj}
\leq 640\, \mr{\gev}$, in the distribution and plot the variables that
are sensitive to angular ordering for these events.
These variables depend on the distribution of a third jet in the
events which is generated in the simulation by the parton-shower
algorithm. The definitions of the variables and the cuts
used are the same as in Section~\ref{sect:monteresults}.

%
%  Distribution of eta_3 for the LQD processes
%
\begin{figure}
\subfigure[Signal]
{\includegraphics[angle=90,width=0.3\textwidth]{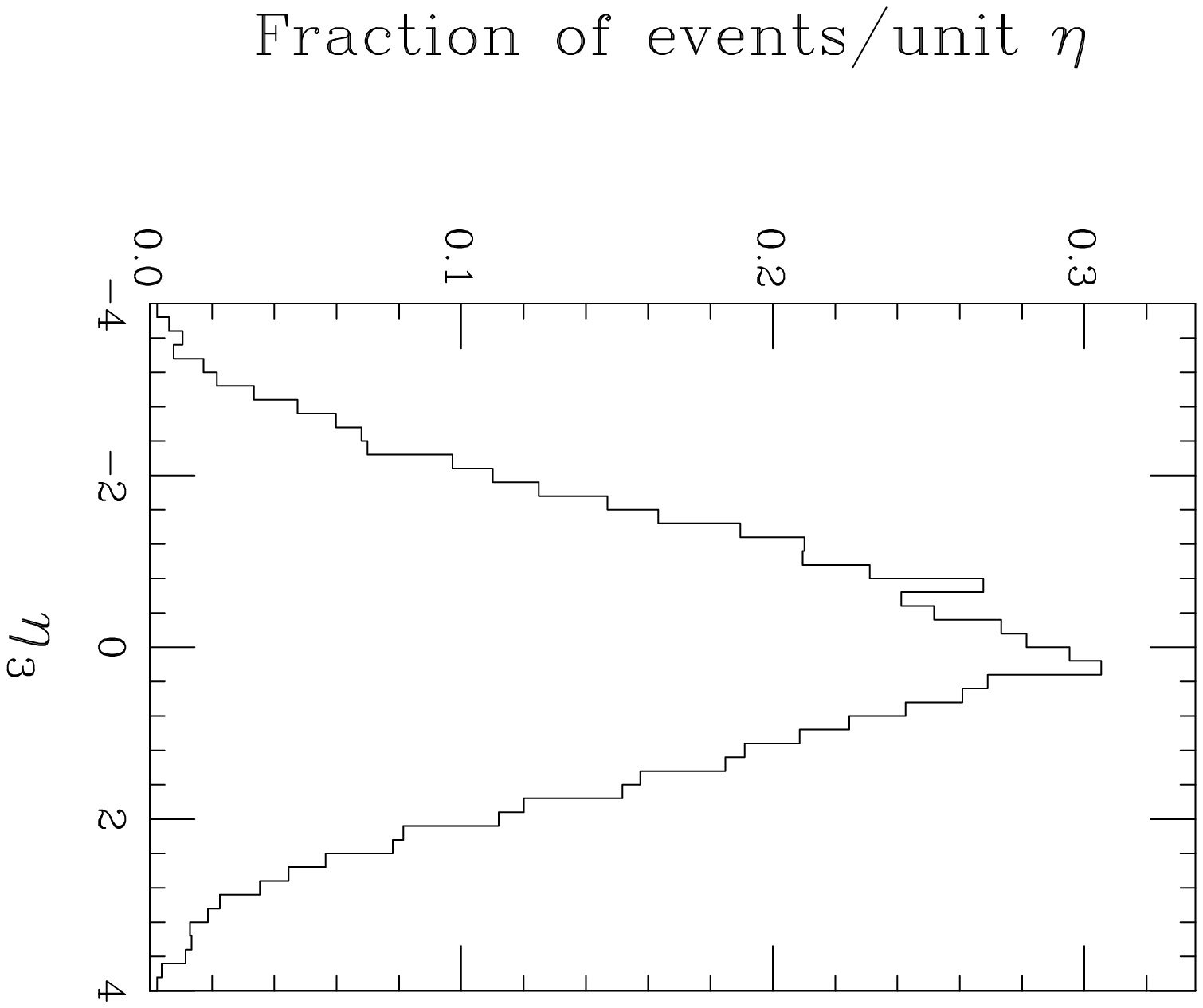}
\hfill}
\subfigure[QCD Background]
{\includegraphics[angle=90,width=0.3\textwidth]{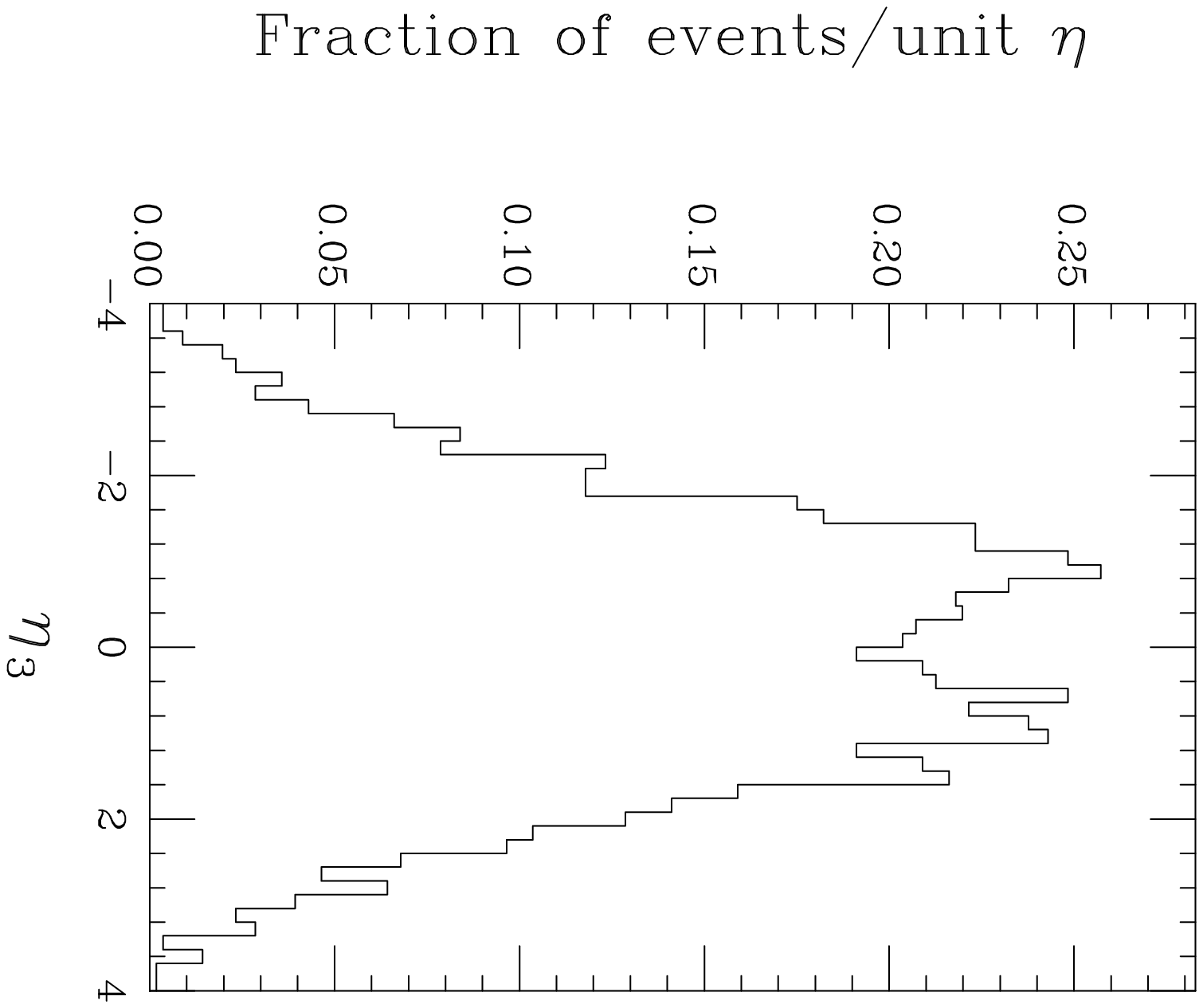}
\hfill}
\subfigure[Signal+Background]
{\includegraphics[angle=90,width=0.3\textwidth]{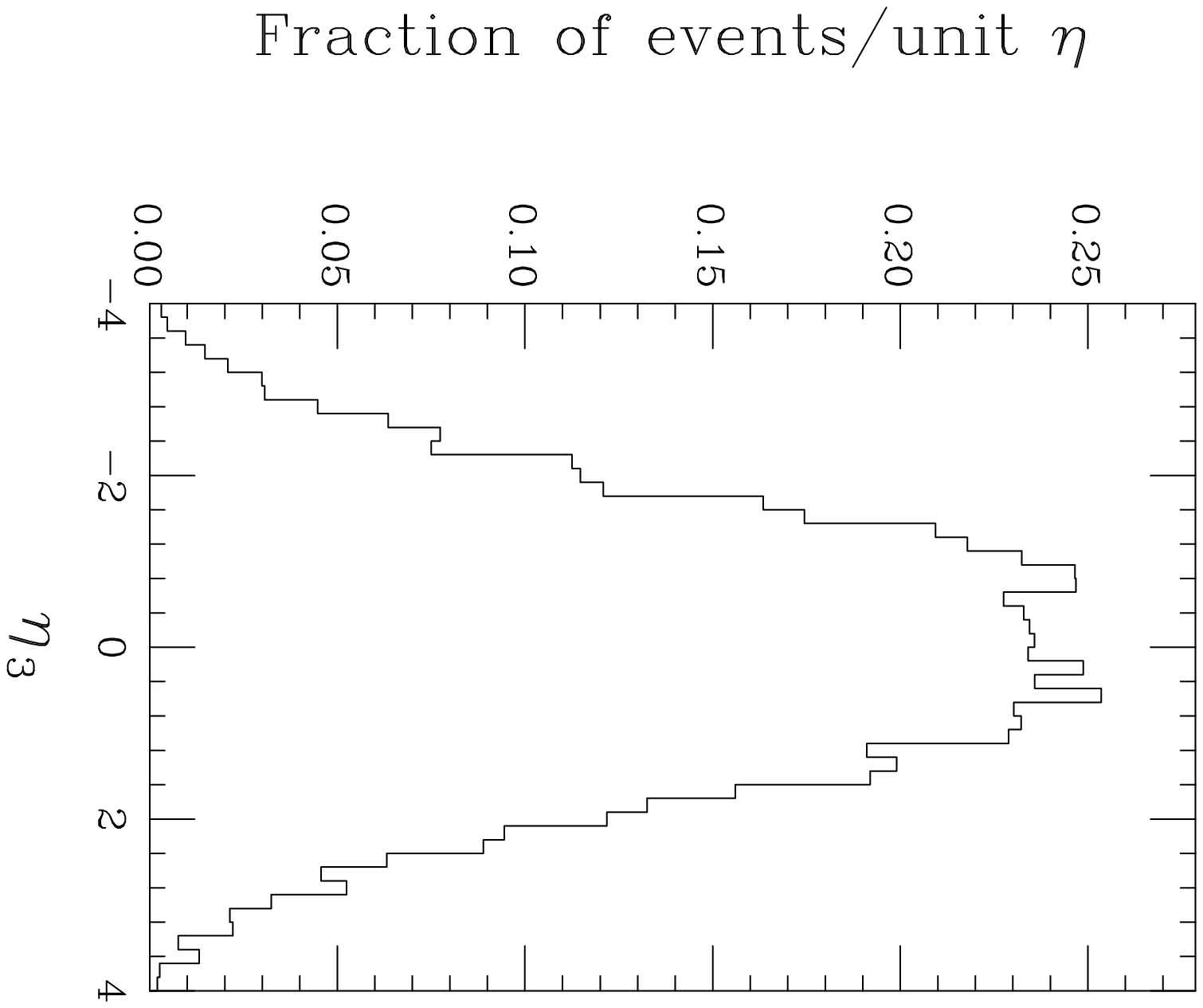}} 
\\
\vskip -4mm
\captionB{Distribution of $\eta_3$
	 for resonant slepton production and the QCD background.}
	{Distribution of events in $\eta_3$ for resonant slepton
	 production and the QCD background.}
\label{fig:LQDeta}
%\vskip 20mm
%\end{figure}
% end of the figure
%
%  Distribution of R for the LQD processes
%
%\begin{figure}[htp]
\subfigure[Signal]
{\includegraphics[angle=90,width=0.3\textwidth]{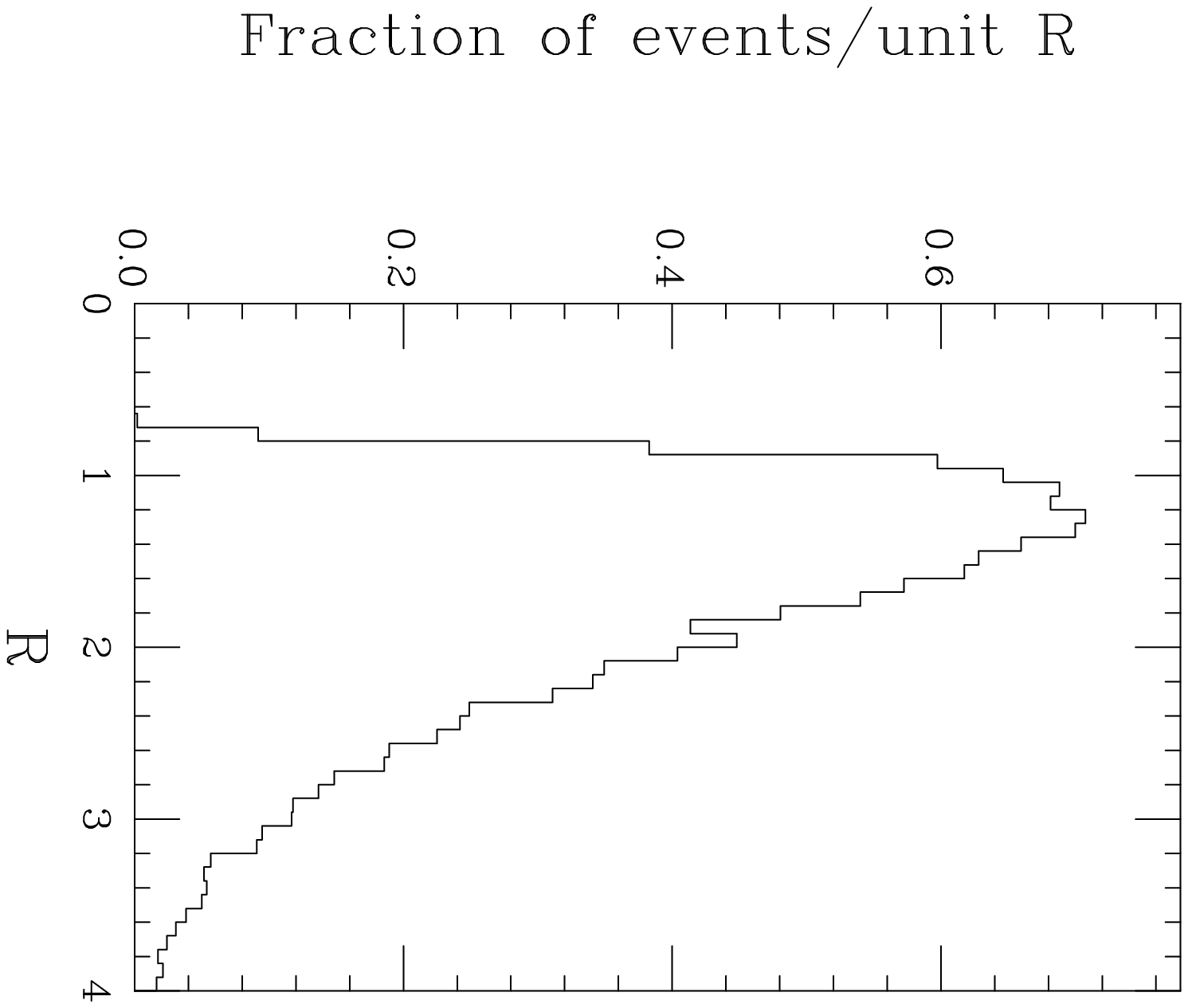}
\hfill}
\subfigure[QCD Background]
{\includegraphics[angle=90,width=0.3\textwidth]{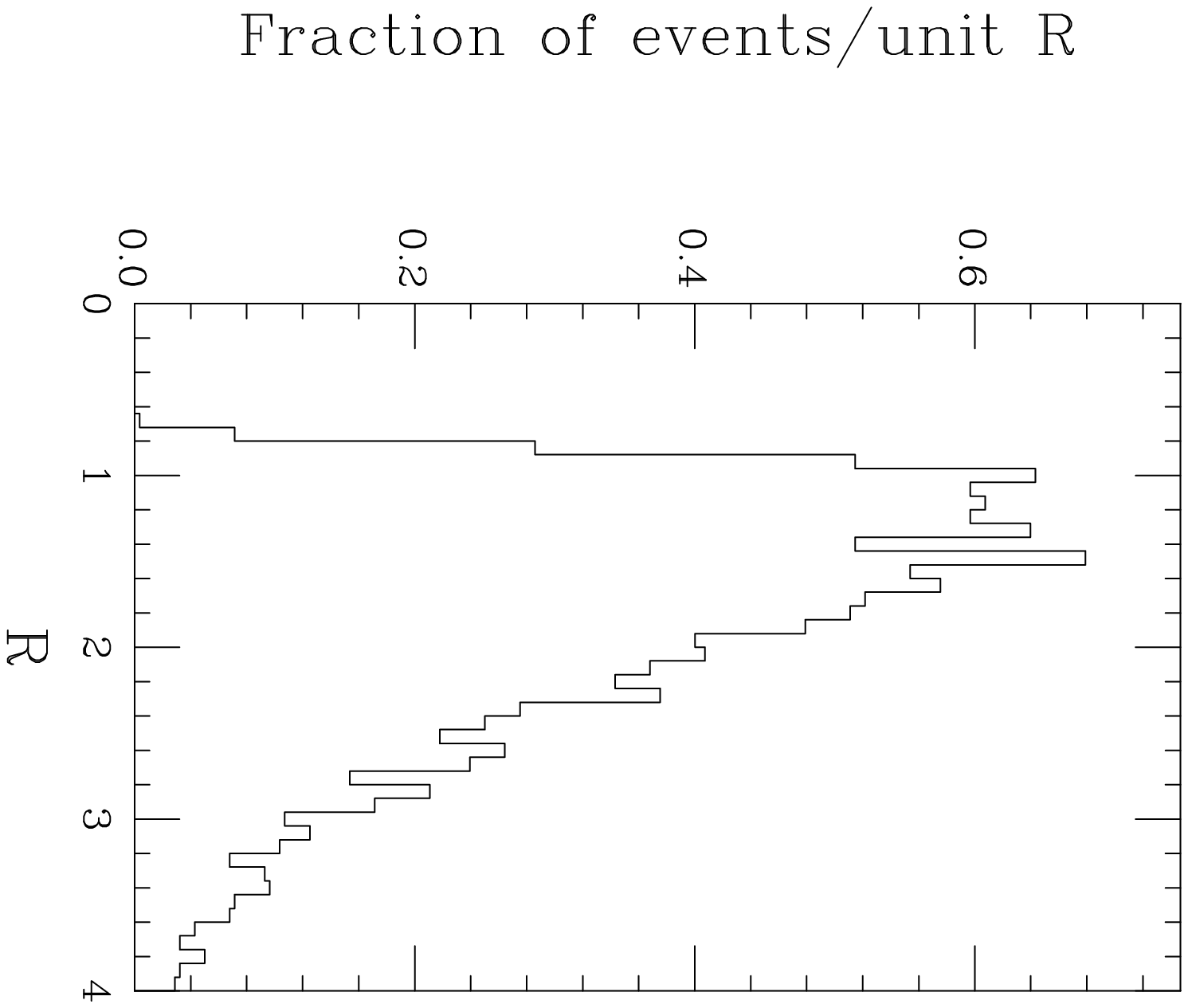}
\hfill}
\subfigure[Signal+Background]
{\includegraphics[angle=90,width=0.3\textwidth]{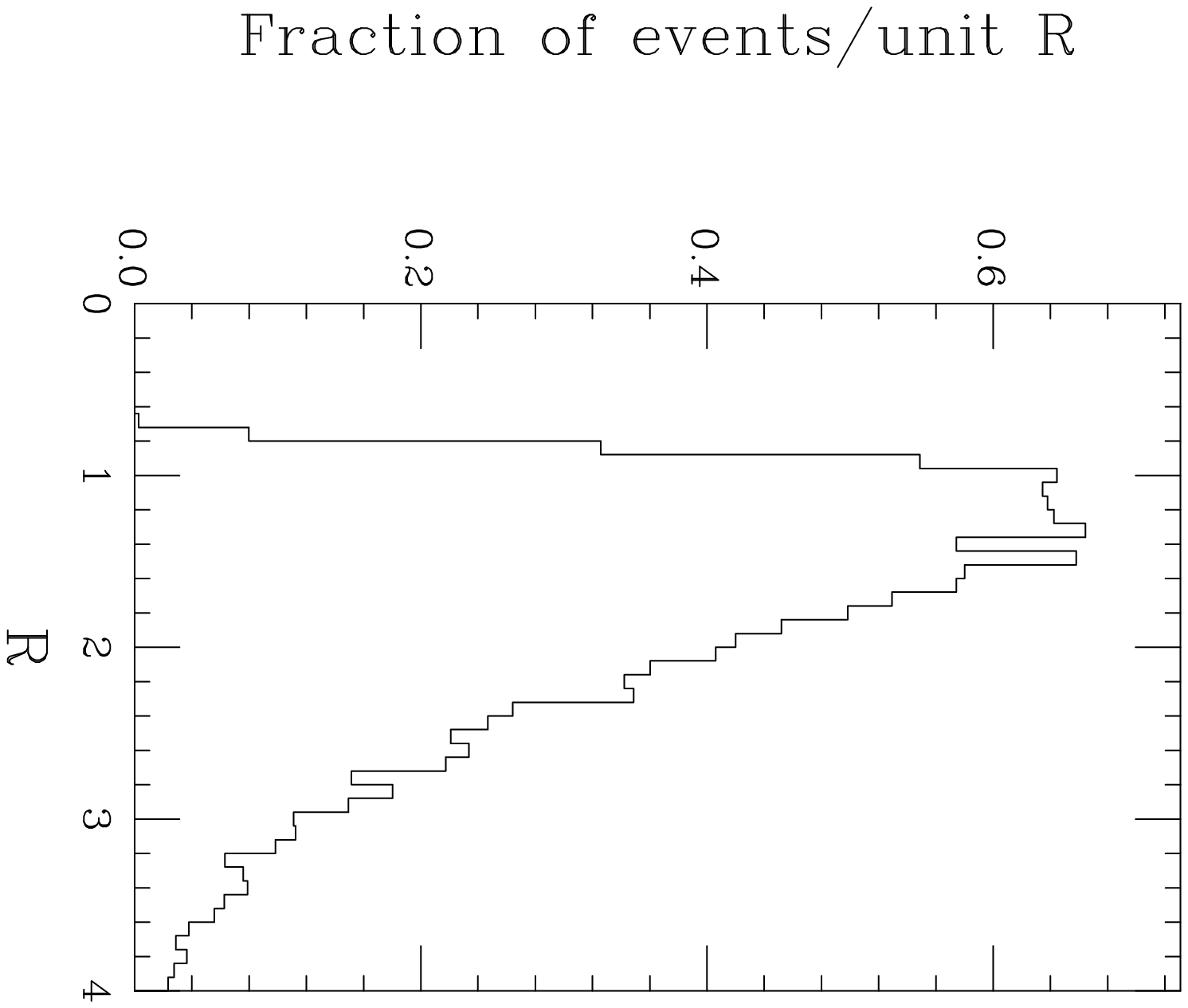}}
\\
\vskip -4mm
\captionB{Distribution of $R$
	 for resonant slepton production and the QCD  background.}
	{Distribution of events in $R$ for resonant slepton production
 	and the QCD background.}
\label{fig:LQDR}
%\end{figure}
% end of the figure
%
%  Distribution of alpha for the LQD processes
%
%\begin{figure}
\subfigure[Signal]
{\includegraphics[angle=90,width=0.3\textwidth]{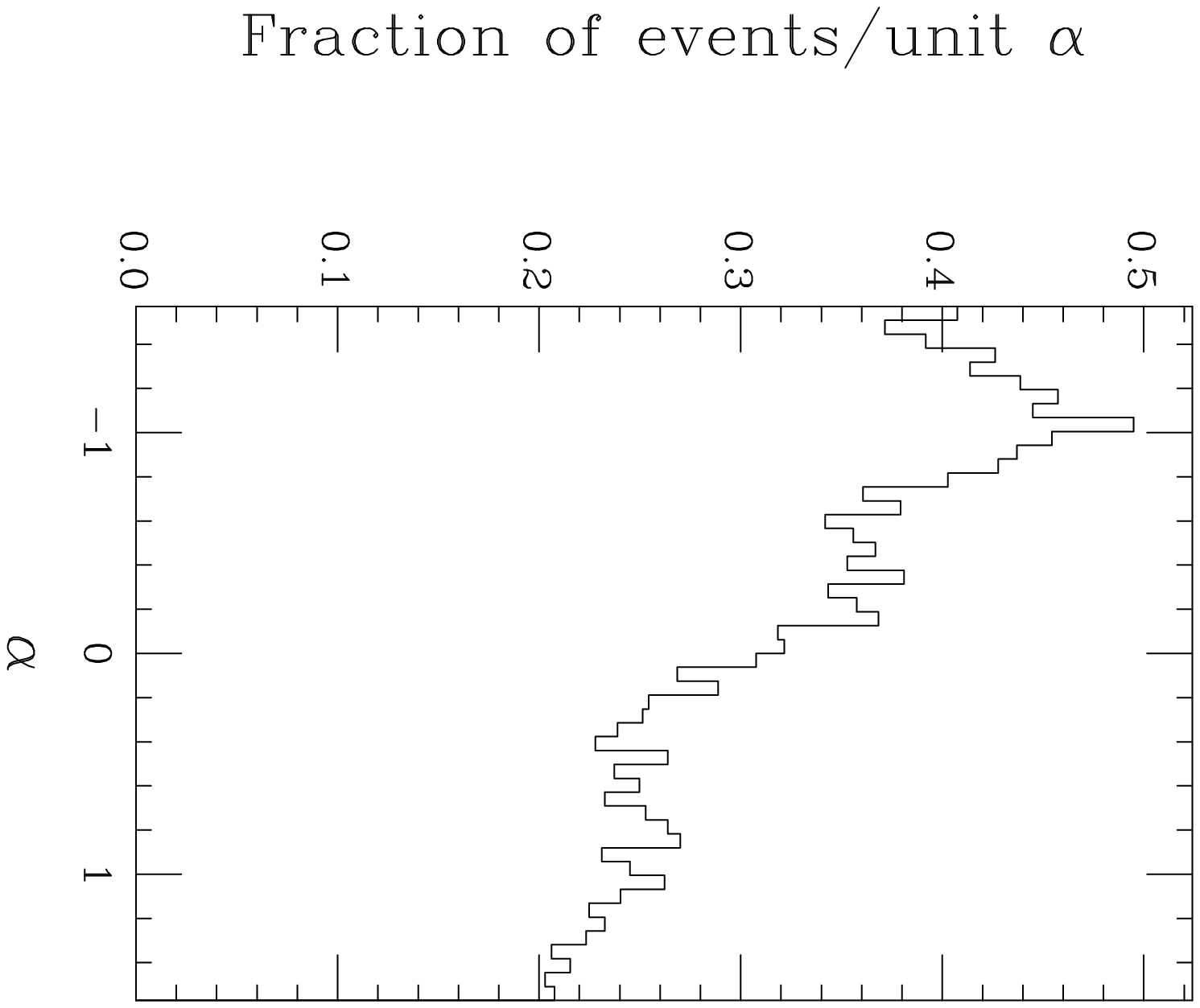}
\hfill}
\subfigure[QCD Background]
{\includegraphics[angle=90,width=0.3\textwidth]{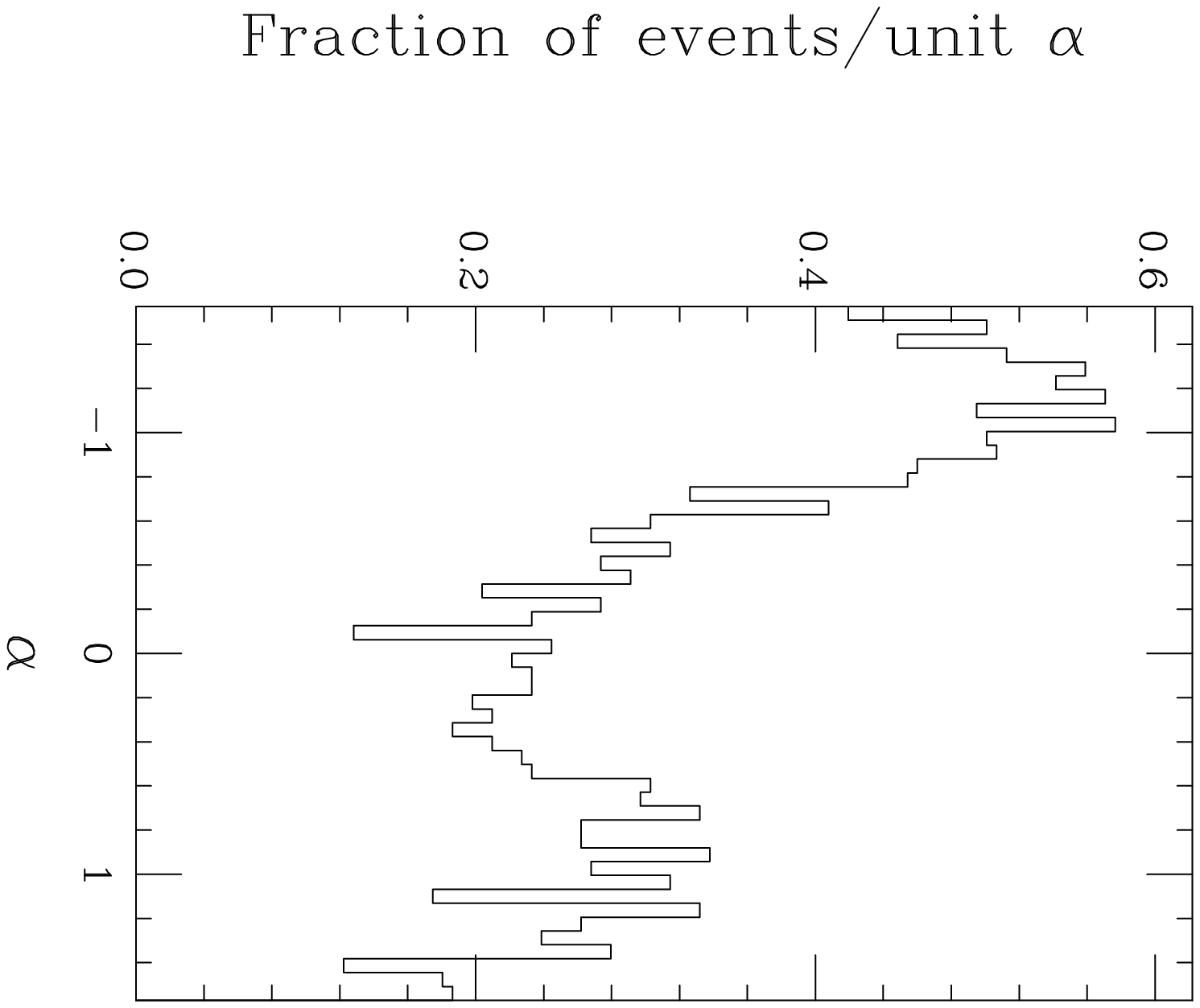}
\hfill}
\subfigure[Signal+Background]
{\includegraphics[angle=90,width=0.3\textwidth]{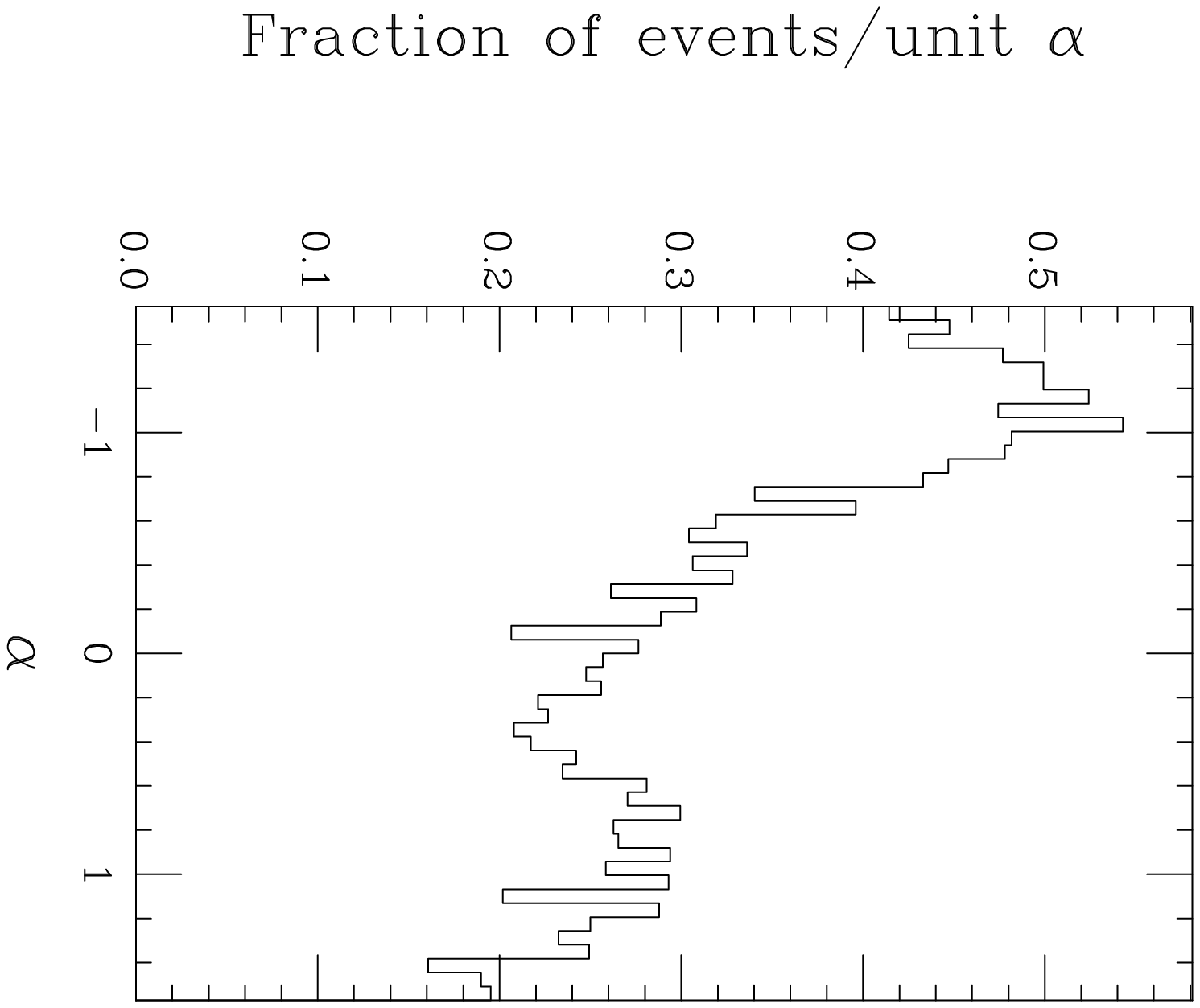}}
\\
\vskip -4mm
\captionB{Distribution of $\al$
	 for resonant slepton production and the QCD  background.}
	{Distribution of events in $\al$ for resonant slepton
	production and the QCD background.}
\label{fig:LQDalpha}
\end{figure}
We can now study the distributions for the signal, background, and
signal plus background for resonant slepton production with
coupling ${\lam'}_{311}=0.8$. As we saw in the previous chapter, 
there are significant differences
between the signal and the background for this process which we will
briefly recall here. In the $\eta
_3$ distributions, Fig.\,\ref{fig:LQDeta}, instead of a dip in at $
\eta_3=0$ there is a bump in the signal. This dip in the QCD
background was observed in \cite{Abe:1994nj}, and is a feature of the
initial-final state colour connection. In our study it is present in
the background, but not the signal.
The distribution
of events in $\al$, Fig.\,\ref{fig:LQDalpha}, also shows a difference
between the signal and the background, with the signal not showing the
dip in the middle.  This is again an effect of the initial-final state
colour connection which is present in the background but not in the
signal.

\begin{figure}
\begin{center}
\includegraphics[angle=90,width=0.47\textwidth]{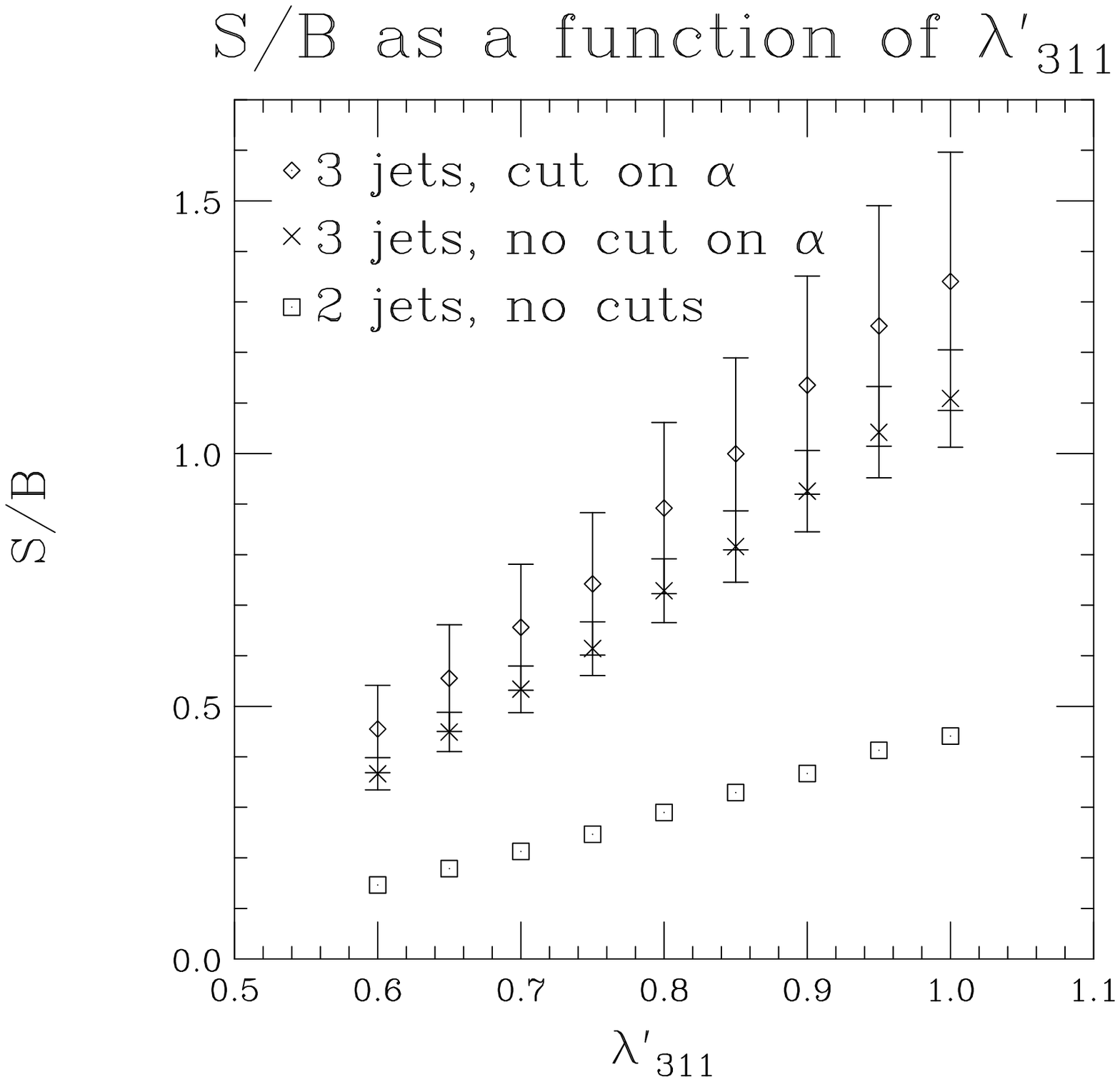}
\hfill
\includegraphics[angle=90,width=0.47\textwidth]{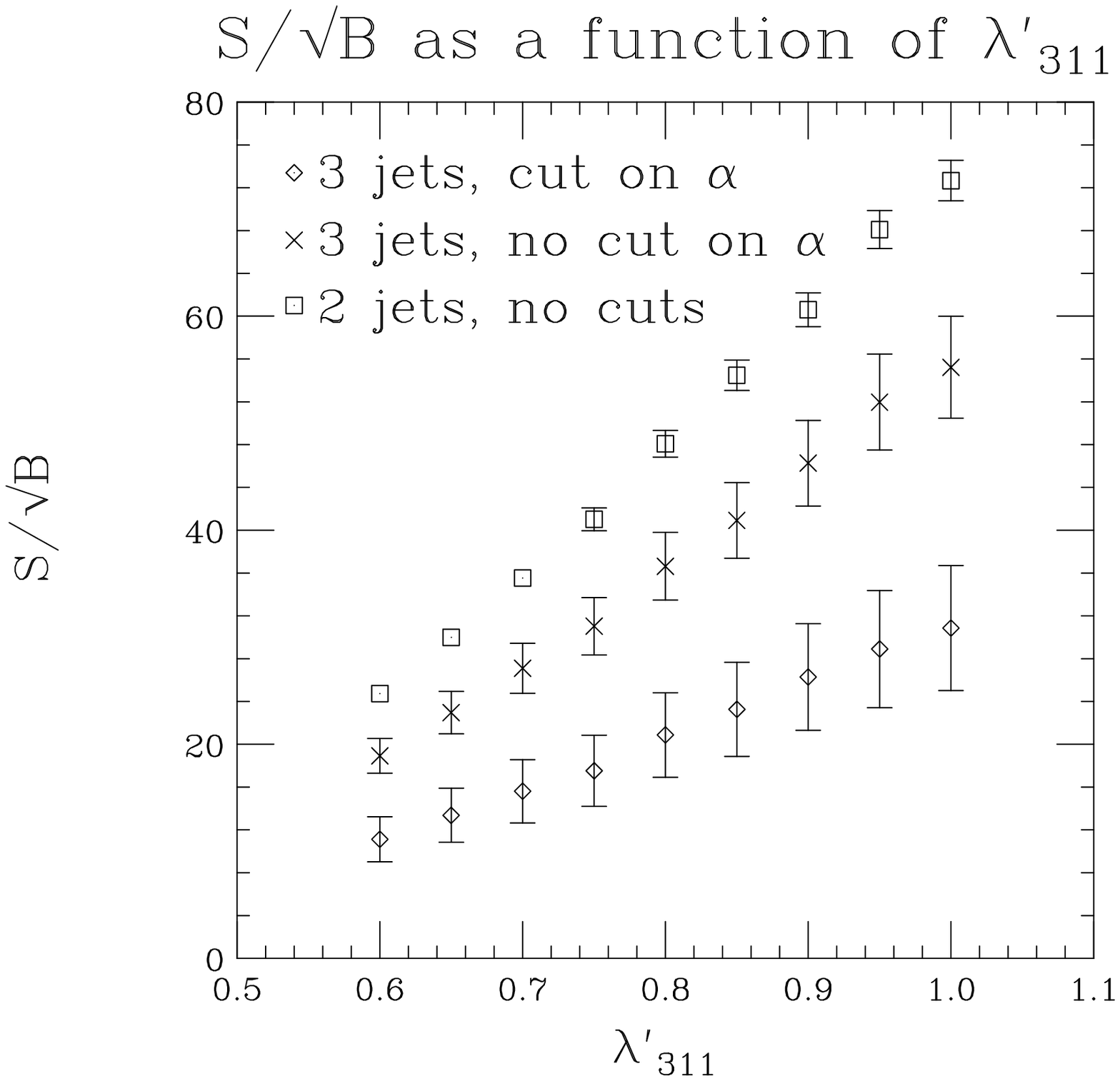}
\captionB{Effect of the cuts on the angular-ordering variables as a
	function of ${\lam'}_{311}$.}
	{Effect of the cuts on the angular-ordering variables as a
	function of ${\lam'}_{311}$.}
\label{fig:cuts}
\end{center}
\end{figure}

\begin{figure}
\includegraphics[angle=90,width=0.47\textwidth]{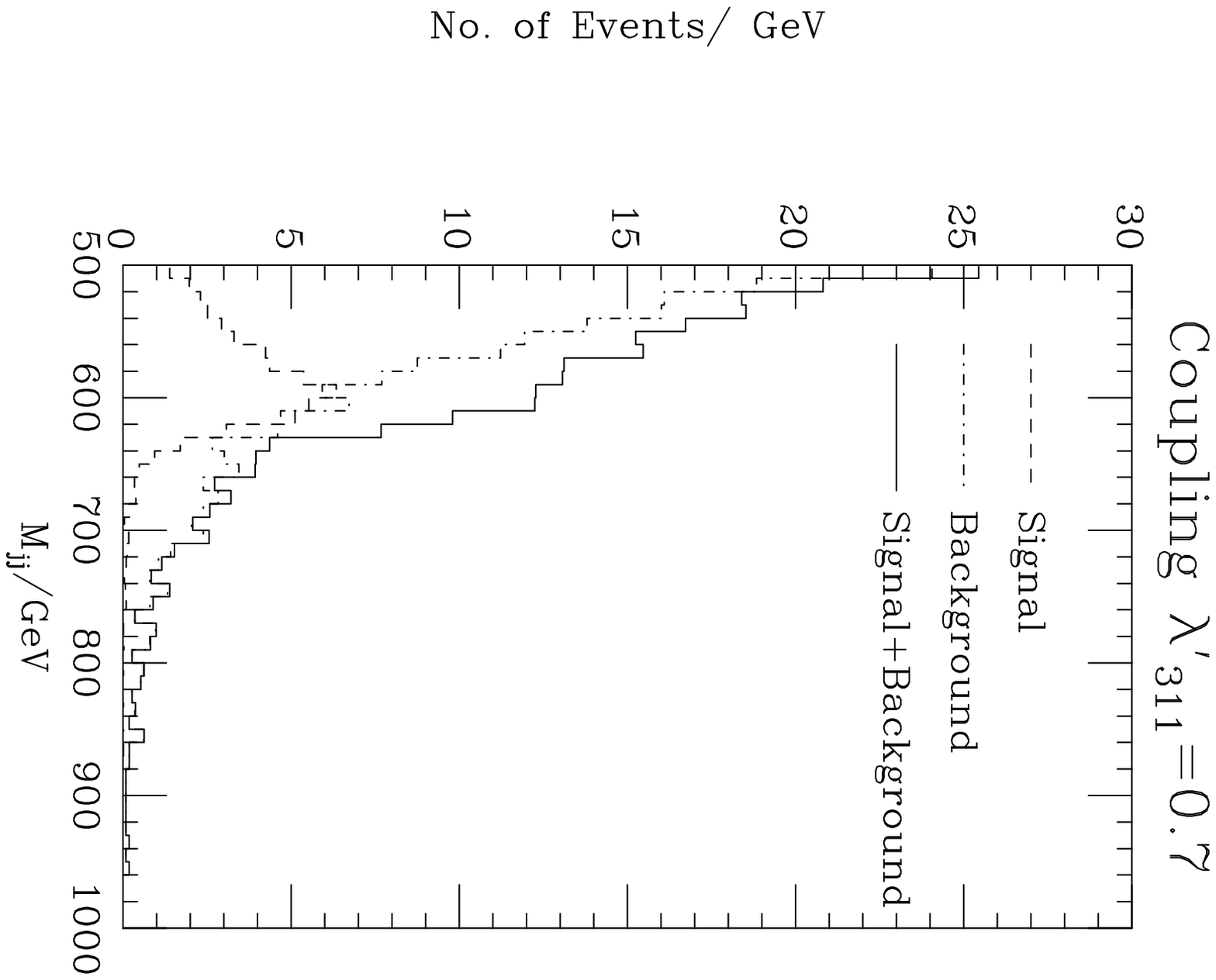}
\hfill
\includegraphics[angle=90,width=0.47\textwidth]{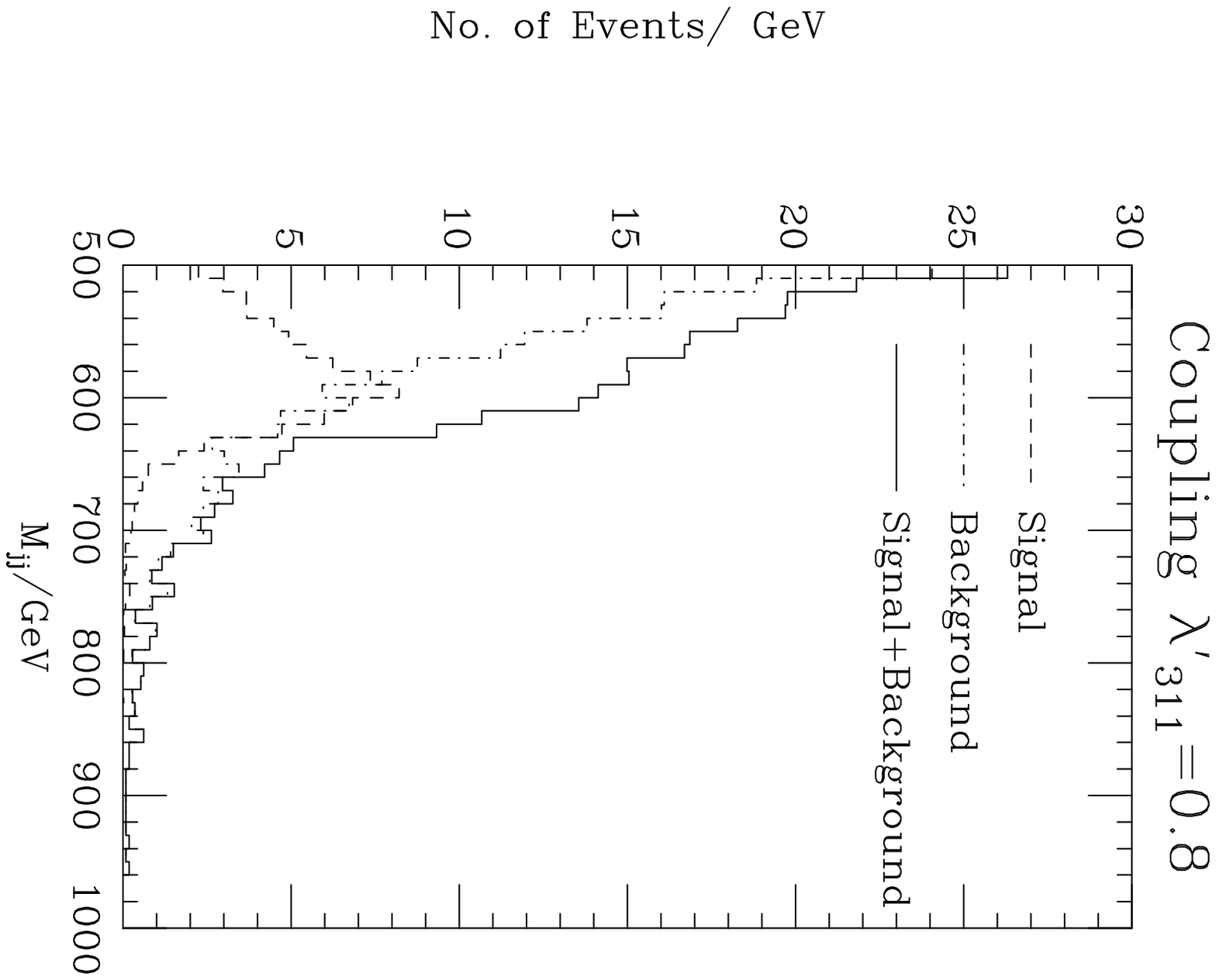}
\captionB{Invariant-mass distributions after cuts on the angular-ordering
	  variables.}
	{Invariant-mass distribution for ${\lam'}_{311}=0.7$ and
	${\lam'}_{311}=0.8$ after cuts on the angular-ordering variables.}
\label{fig:invm2}
\end{figure}

As can be seen in all the distributions, apart from the disappearance of
the dip at $\eta_3=0$, once the signal and
background are added the effect of the signal is minimal.
While there are differences between the signal and background it is hard to
see how cuts can be applied on these variables to improve the extraction of a
signal over the QCD background. The only major difference which can be cut on
is the difference in the distribution of $\al$.
  We consider two approaches to increase the 
ratio of signal to background, $S/B$:
\begin{enumerate}
 \item Accept all the events with at least three jets, provided they
       pass the cuts described in Section~\ref{sect:monteresults}
       from the analysis of \cite{Abe:1994nj}.

 \item Reject all the two-jet events and only accept the events with
 more than two jets provided that $|\al|\le\al_{\mr{cut}}$. We apply a cut
 of $\al_{\mr{cut}}=0.4$ for these jet events.
\end{enumerate}
These cuts were chosen to maximize $S/B$ while not reducing $S/\sqrt
{B}$ below five. As can been seen in Fig.\,\ref{fig:cuts} both of these cuts 
significantly increase the $S/B$. The effects of the second cut on the
invariant-mass distribution is shown in Fig.\,\ref{fig:invm2}.
In the invariant-mass distribution 
the signal is now more visible over the background.

%
%  Effect of the cuts for varying couplings
%
% end of figure

This shows that by using the colour coherence effects we can improve
the extraction of a signal. Obtaining a large $S/B$ is
important for this process because we do not have an accurate
prediction for the QCD background. However given the limits on the coupling, 
${\lam'}_{311}$, 
the signal will only be visible above the background at the highest values
of the coupling 
currently allowed by low-energy experiments. In 
\cite{Hewett:1998fu,Allanach:1999bf}
it was suggested that by using the sidebands to normalize the background,
resonant slepton production could be probed to much smaller values of the
\rpv\  coupling.
Indeed the $S/\sqrt{B}$ numbers in Fig.\,\ref{fig:cuts} suggest that without 
any of our additional cuts the signal is visible at a much lower coupling. 
The results in \cite{Hewett:1998fu,Allanach:1999bf} were obtained using the
narrow-width limit for the production cross section and did not included the
effects of QCD radiation. Our results suggest that after including these
effects the signal will only be visible for large values of the coupling.
It may be possible  to use the
sidebands which we have removed with our cuts to normalize this
background, as in \cite{Hewett:1998fu,Allanach:1999bf}, to improve the
extraction of the signal. However this may not be possible due to the
increased width of the resonance, Fig.\,\ref{fig:invm}, due to QCD radiation.
The situation will hopefully improve with the availability of
a next-to-leading order calculation for the QCD background.
At present, if we require a $S/B$ ratio of 25\% in addition to $S/\sqrt{B}>5$,
then looking at the di-jet invariant masses will only allow a coupling of 
${\lam'}_{311}>0.75$ to be probed. However, by using the cuts we described,
based on the colour structure, couplings
as low as ${\lam'}_{311}>0.55$ can be probed.

%
%  Effect of the cuts on the LQD mass distribution
%
%  end of figure
  
%
%  Section on the gauge decays
%
\section[Gauge Decays of the Resonant Slepton]
	{Gauge Decays of the Resonant Slepton}
\label{sect:sleptongauge}

  As we saw in the previous section, due to the large QCD background, the 
  \rpv\  decay modes of the resonant sleptons can only be observed above the
  QCD background for large values of the \rpv\  Yukawa couplings. If we
  are to observe resonant slepton production we must therefore examine other
  possible
  decay modes of the resonant sleptons which have a lower background from
  Standard Model processes. One possibility is the decay of the slepton
  via the \rpv\  operator $L_iL_j\overline{E}_k$.
  Another which we shall consider here is a 
  supersymmetric gauge decay of the resonant slepton.

  We will consider a specific signature, \ie like-sign dilepton production,
  for these processes rather than any
  one given resonant production mechanism.
  We would expect like-sign dilepton production
  to have a low background from 
  Standard Model processes.

  In Section~\ref{sec:signal} we will consider the signal processes in more 
  detail, followed by a discussion of the various  processes which
  contribute to the background in 
  Section~\ref{sec:backgrounds}. We will also discuss the
  different cuts which can be used to reduce the background. In
  Section~\ref{sec:results} we will then consider the discovery potential
  at both Run II of the Tevatron and the LHC. We also
  consider the possibility of reconstructing the neutralino and slepton masses
  using their decay products.
%
%  Section on the Signal process
%
\subsection{Signal}
\label{sec:signal}

  There are a number of different possible production mechanisms for a
  like-sign
  dilepton pair via resonant slepton production. The dominant production
  mechanism is the production of a charged slepton followed by a 
  supersymmetric gauge decay
  of the charged slepton to a neutralino and
  a charged lepton. The neutralino can then decay via the crossed process
  to give a second charged lepton which, due to the Majorana nature
  of the neutralino,
  can have the same charge as the lepton produced in the slepton decay.
  The production of a charged lepton and a neutralino via the LQD term in the
  \rpv\  superpotential, Eqn.\,\ref{eqn:Rsuper1}, occurs at tree-level 
  via the Feynman
  diagrams given in Fig.\,\ref{fig:cross}. 
  The decay of the neutralino occurs at tree-level via the diagrams
  given in 
  Fig.\,\ref{fig:decay}.

%
%  Signal production processes
%
\begin{figure}
\begin{center}
\begin{picture}(360,60)(0,30)
\SetScale{0.7}
\SetOffset(0,-25)
\ArrowLine(5,76)(60,102)
\ArrowLine(60,102)(5,128)
\DashArrowLine(90,102)(60,102){5}
\ArrowLine(90,102)(145,76)
\ArrowLine(145,128)(90,102)
\Text(80,52)[]{$\mathrm{\tilde{\chi}^{0}}$}
\Text(80,88)[]{$\mathrm{\ell^{+}}$}
\Text(25,89)[]{$\mathrm{\bar{d}}$}
\Text(25,57)[]{$\mathrm{u}$}
\Text(53,82)[]{$\mathrm{\tilde{\ell}}_{L}$}
\Vertex(60,102){1}
\Vertex(90,102){1}
\SetScale{0.7}
\ArrowLine(240,128)(185,128)
\ArrowLine(240,128)(295,128)
\ArrowLine(185,76)(240,76)
\ArrowLine(295,76)(240,76)
\DashArrowLine(240,76)(240,128){5}
\Text(190,98)[]{$\mathrm{\tilde{\chi}^{0}}$}
\Text(190,45)[]{$\mathrm{\ell^{+}}$}
\Text(150,45)[]{$\mathrm{u}$}
\Text(150,98)[]{$\mathrm{\bar{d}}$}
\Text(160,70)[]{$\mathrm{\tilde{d}}_{R}$}
\Vertex(240,128){1}
\Vertex(240,76){1}
\ArrowLine(420,128)(365,128)
\ArrowLine(475,128)(420,128)
\ArrowLine(365,76)(420,76)
\ArrowLine(420,76)(475,76)
\DashArrowLine(420,76)(420,128){5}
\Text(315,45)[]{$\mathrm{\tilde{\chi}^{0}}$}
\Text(315,98)[]{$\mathrm{\ell^{+}}$}
\Text(277,98)[]{$\mathrm{\bar{d}}$}
\Text(277,45)[]{$\mathrm{u}$}
\Text(287,70)[]{$\mathrm{\tilde{u}}_{L}$}
\Vertex(420,128){1}
\Vertex(420,76){1}
\end{picture}
\end{center}
\captionB{Production of $\tilde{\chi}^{0}\ell^{+}\!$.}
	{Production of $\tilde{\chi}^{0}\ell^{+}$.}
\label{fig:cross}
\end{figure}
% End of the Figure %%%%%%%%%%%%%%%%%%%%%%%%%%%%%%%%%%%%%%%%%%%%%%%%%%%%%%%%%%

  Like-sign dileptons can also be produced in resonant charged slepton
  production
  with a supersymmetric gauge decay of the slepton to a chargino and neutrino,
  $\mr{\elt}^+\ra\mr{\cht}^+_1\mr{\nu_\ell}$. The chargino
  can then decay $\mr{\cht}^+_1\ra\mr{\ell}^+\mr{\nu_\ell}\mr{\cht}^0_1$.
  Again given the
  Majorana nature of the neutralino it can decay to give a like-sign dilepton
  pair.

  The production of like-sign dileptons is also possible in resonant sneutrino
  production  followed by a supersymmetric 
  gauge decay to a chargino and a charged lepton,
  $\nut\ra\ell^-\cht^+_1$. This can be followed by 
  $\mr{\cht}^+_1\ra \mr{q}\mr{\bar{q}'}\mr{\cht}^0_1$,
  the neutralino can then decay as in Fig.\,\ref{fig:decay} to give 
  a like-sign dilepton pair. 

  All the resonant \rpv\  production mechanisms and the decays of the SUSY
  particles
  have been included in the HERWIG event generator \cite{HERWIG61}.
  The implementation of both 
  R-parity conserving and R-parity violating SUSY is described in
  \cite{SUSYimplement}.
  The matrix elements used for the various \rpv\  processes are given in
  Appendices~\ref{chap:decay} and \ref{chap:cross}. 
 
  We will only consider one of the \rpv\  Yukawa couplings
  to be non-zero at a
  time, either ${\lam'}_{111}$ or ${\lam'}_{211}$, which lead to resonant
  selectron and smuon production, respectively.
  The cross section depends quadratically on the
  \rpv\  Yukawa coupling.
  These couplings have upper bounds from low
  energy experiments. The bound on the coupling ${\lam'}_{111}$
  from neutrino-less double beta decay is very
  strict \cite{Allanach:1999ic,Hirsch:1995zi:Hirsch:1996ek:Babu:1995vh}.
  We therefore consider smuon production via the coupling
  ${\lam'}_{211}$, which has a much weaker bound,
\begin{equation}
{\lam'}_{211} < 0.059 \times \left(\frac{M_{\dnt_R}}{100\,
 \mr{\gev}}\right)\!,
\end{equation}
  from the ratio
  $R_\pi=\Gamma(\mr{\pi\ra e\nu_e})/\Gamma(\mr{\pi\ra \mu\nu_\mu})$
  \cite{Allanach:1999ic,Barger:1989rk}.
  Our results will however apply for resonant selectron production if the
  coupling ${\lam'}_{111}$ is large enough 
  to give an observable signal
  while still satisfying the bound from neutrino-less double beta decay.

  As we are considering a dominant ${\lam}'_{211}$
  coupling the leptons produced in the
  neutralino decays and the hard processes will be muons. We will
  therefore require throughout that both leptons are muons
  because this reduces the background,
  where electrons
  and muons are produced with equal probability, with respect to the signal.
  This typically reduces the Standard Model background by a factor of four
  while leaving the dominant signal process almost unaffected.
  It will lead to some reduction of the signal from
  channels where some of
  the leptons are produced in cascade decays from the decay of a W or
  Z boson.

  The signal has a number of features, in addition to the presence of a
  like-sign dilepton pair, which will enable us to extract it above the
  background: 
\begin{itemize}
% Isolation and pt
\item   Provided that the difference between the slepton and
	the neutralino/chargino masses is large enough both 
	the leptons will have a high transverse momentum, $p_T$,
 	and be well isolated.
% Missing et
\item 	As the neutralino decays inside the detector, for this signature,
	there
	will be little missing transverse energy, \met, in the event. Any
	\met\  will come from semi-leptonic hadron decays or from cascade
	decays following the production of a chargino or 
        one of the heavier neutralinos.
% lack of other leptons
\item 	The presence of a third lepton can only come from semi-leptonic
	hadron decays, or in SUSY cascade decays if a chargino or
	one of the heavier
	neutralinos is produced.
% jets?
\item 	The presence of two hard jets from the decay of the neutralino.
\end{itemize}

%%%%%%%%%%%%%%%%%%%%%%%%%%%%%%%%%%%%%%%%%%%%%%%%%%%%%%%%%%%%%%%%%%%%%%%%%%%%%%
%
%  Neutralino Decay Figure
%
\begin{figure}
\vskip 5mm
\begin{center}
\begin{picture}(360,35)(0,0)
\SetScale{0.7}
\SetOffset(0,-30)
\ArrowLine(5,78)(60,78)
\ArrowLine(105,105)(60,78)
\ArrowLine(84,53)(129,26)
\ArrowLine(129,80)(84,53)
\DashArrowLine(60,78)(84,53){5}
\Text(23,64)[]{$\mathrm{\tilde{\chi^{0}}}$}
\Text(55,70)[]{$\mathrm{\ell^{+}}$}
\Text(75,20)[]{$\mathrm{d}$}
\Text(75,54)[]{$\mathrm{\bar{u}}$}
\Text(45,40)[]{$\mathrm{\tilde{\ell}}_{L}$}
\Vertex(60,78){1}
\Vertex(84,53){1}
\ArrowLine(185,78)(240,78)
\ArrowLine(285,105)(240,78)
\ArrowLine(264,53)(309,26)
\ArrowLine(309,80)(264,53)
\DashArrowLine(240,78)(264,53){5}
\Text(150,63)[]{$\mathrm{\tilde{\chi}^{0}}$}
\Text(200,54)[]{$\mathrm{\ell^{+}}$}
\Text(200,20)[]{$\mathrm{d}$}
\Text(180,70)[]{$\mathrm{\bar{u}}$}
\Text(170,40)[]{$\mathrm{\tilde{u}}_{L}$}
\Vertex(240,78){1}
\Vertex(264,53){1}
\ArrowLine(365,78)(420,78)
\ArrowLine(420,78)(465,105)
\ArrowLine(489,26)(444,53)
\ArrowLine(489,80)(444,53)
\DashArrowLine(444,53)(420,78){5}
\Text(277,63)[]{$\mathrm{\tilde{\chi}^{0}}$}
\Text(330,20)[]{$\mathrm{\ell^{+}}$}
\Text(310,70)[]{$\mathrm{d}$}
\Text(330,54)[]{$\mathrm{\bar{u}}$}
\Text(300,40)[]{$\mathrm{\tilde{d}}_{R}$}
\Vertex(420,78){1}
\Vertex(444,53){1}
\end{picture}
\end{center}
\captionB{Feynman diagrams for the decay 
	$\mathrm{{\tilde\chi}^0}\ra\ell^+\mr{d}\mr{\bar{u}}$.}
        {Feynman diagrams for the decay 
	$\mathrm{{\tilde\chi}^0}\ra\ell^+\mr{d}\mr{\bar{u}}$.
	 The neutralino is a Majorana
	fermion and decays to the charge conjugate final state as well.
	There is a further decay mode $\mr{\cht^0\ra\nu d\bar{d}}$.}
\label{fig:decay}
\end{figure}
% End of the Figure %%%%%%%%%%%%%%%%%%%%%%%%%%%%%%%%%%%%%%%%%%%%%%%%%%%%%%%%%%

%%%%%%%%%%%%%%%%%%%%%%%%%%%%%%%%%%%%%%%%%%%%%%%%%%%%%%%%%%%%%%%%%%%%%%%%%%%%%%
%
%  Figure containing the masses at the various SUSY points
%
\begin{figure}
\begin{center}
\includegraphics[angle=90,width=0.48\textwidth]{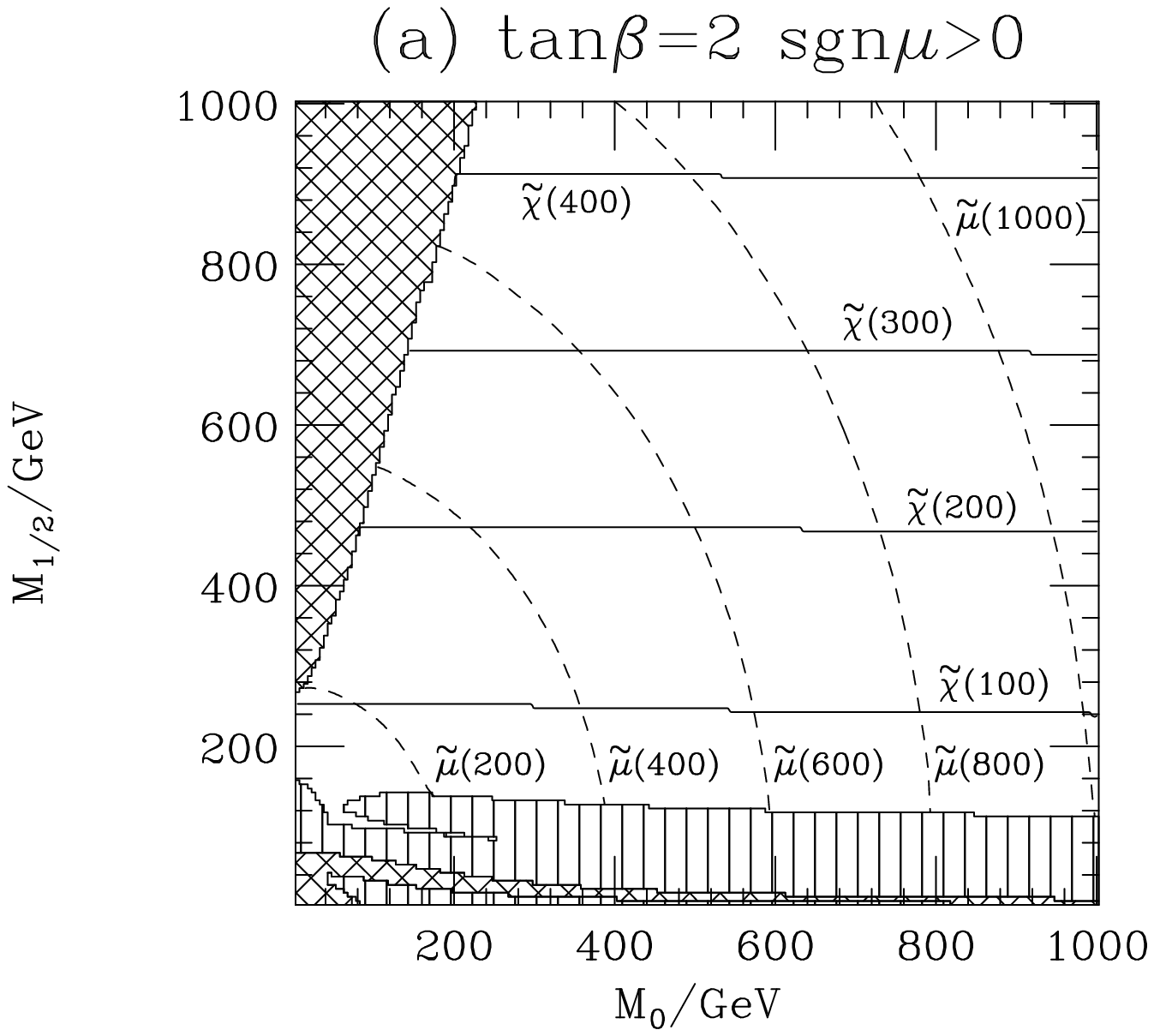}
\hfill
\includegraphics[angle=90,width=0.48\textwidth]{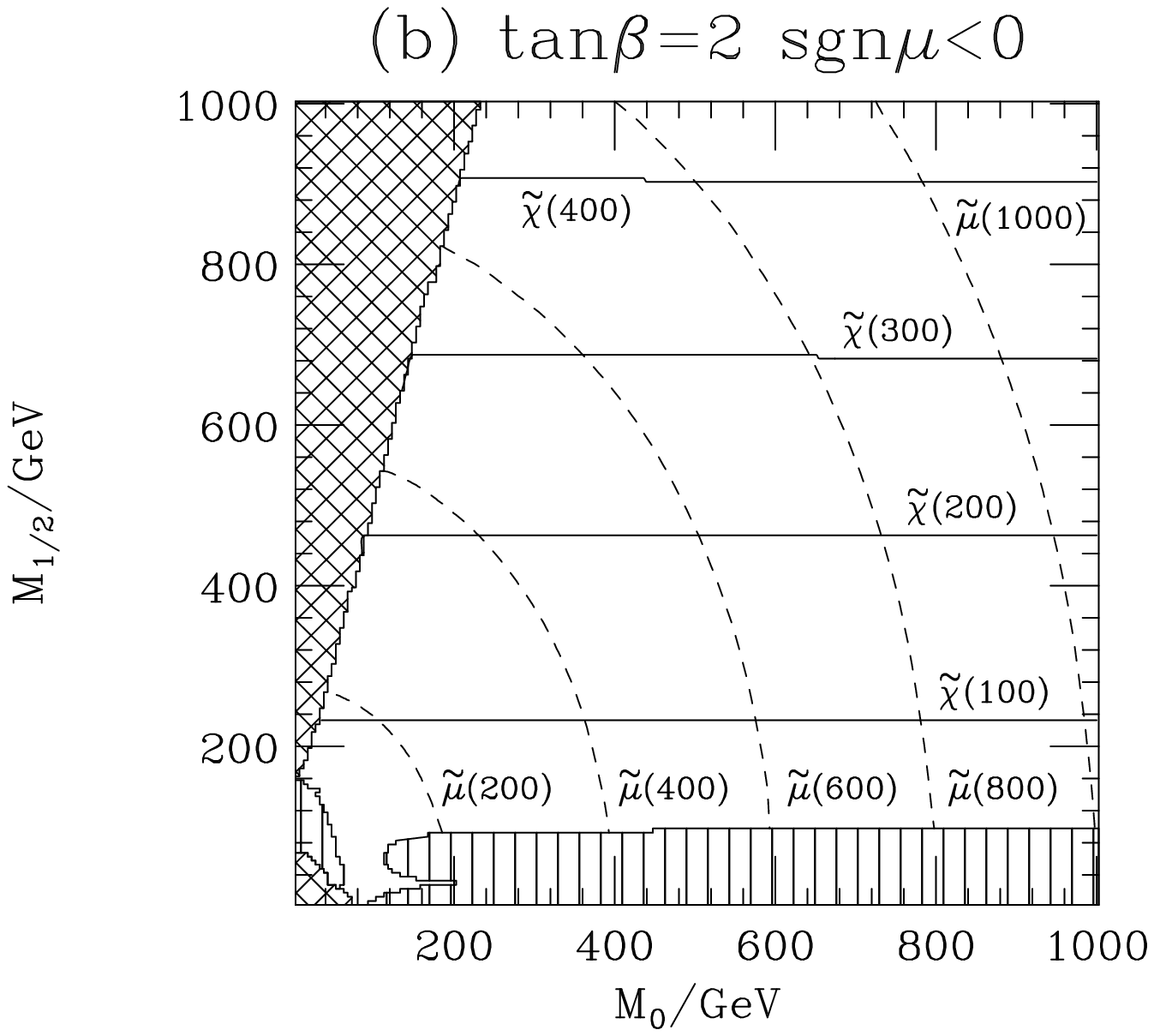}\\
\vskip 15mm
\includegraphics[angle=90,width=0.48\textwidth]{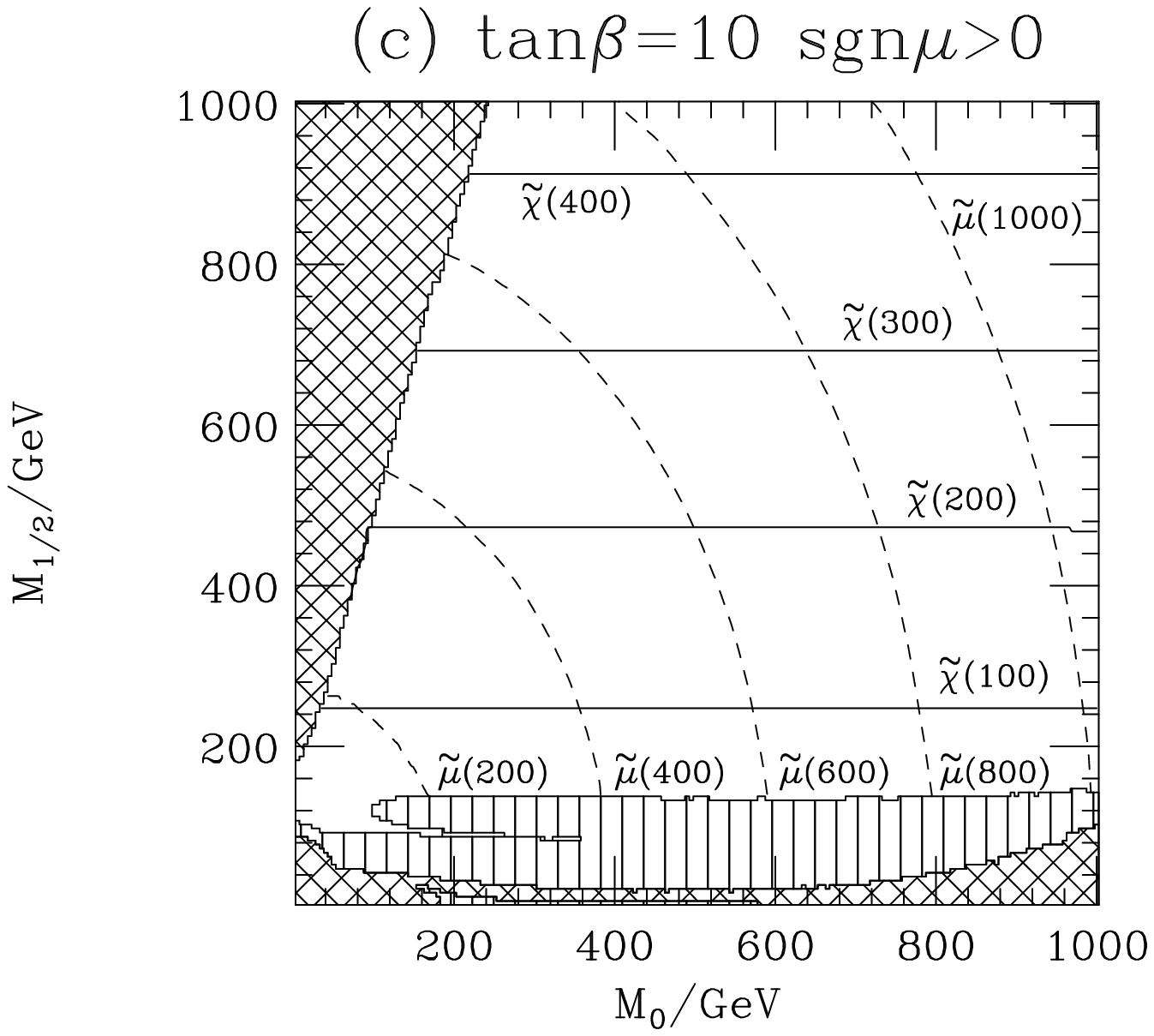}
\hfill
\includegraphics[angle=90,width=0.48\textwidth]{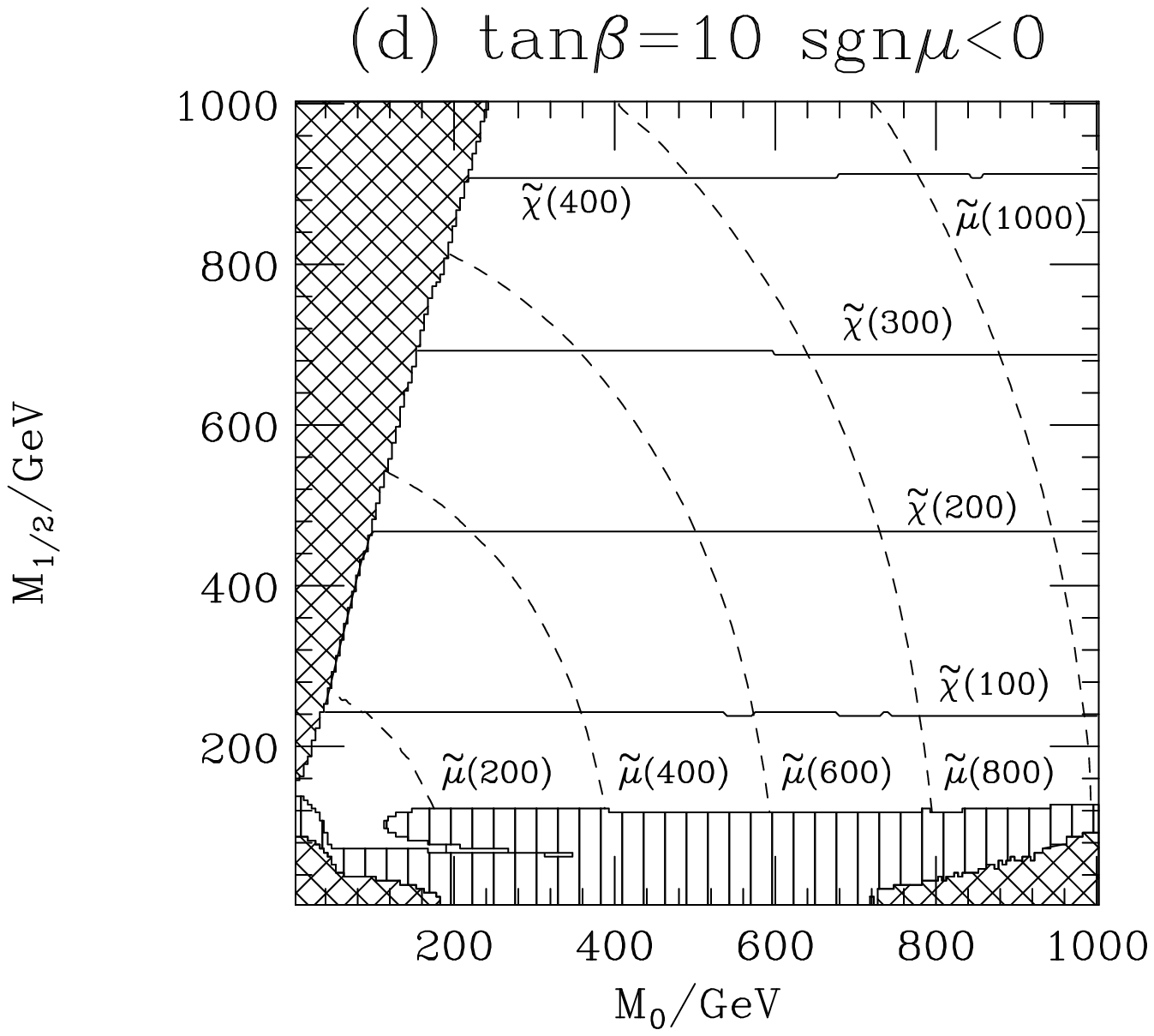}\\
\captionB{The $\mr{\cht^0_1}$ and $\mr{\tilde{\mu}}_L$ masses in the $M_0$,
	 $M_{1/2}$ plane.}
  	{Contours showing the lightest neutralino mass, solid lines, and the
	$\mr{\tilde{\mu}}_L$ mass, dashed lines, in the $M_0$, $M_{1/2}$
	plane with $A_0=0\, \mr{\gev}$  for different values of $\tan\beta$
	 and $\sgn\mu$.
	The hatched regions at small $M_0$ are excluded by the requirement
	that the $\mr{\cht^0_1}$ be the LSP.
	The region at large $M_0$ and $\tan\beta$
  	is excluded because there is no radiative 
	electroweak symmetry breaking. 
  	The vertically-striped region is excluded by the LEP experiments.
	This region was obtained
	using the limits on the chargino \cite{Abbiendi:1998ff} and smuon 
	\cite{Abbiendi:1999is} production cross sections,
	and the chargino mass \cite{Barate:1998gy}. 
        This analysis was performed using ISAJET 7.48 \cite{Baer:1999sp}.} 
\label{fig:SUSYmass}
\end{center}
\end{figure}
% End of the Figure %%%%%%%%%%%%%%%%%%%%%%%%%%%%%%%%%%%%%%%%%%%%%%%%%%%%%%%%%%

  The cross section for the signal processes and the 
  acceptance\footnote{We define the acceptance to be the
	fraction of signal events which
  pass the cuts.}
  will depend upon the
  various SUSY parameters.
  We have performed a scan in $M_0$ and $M_{1/2}$ with $A_0=0\, \mr{\gev}$  
  for two different values of $\tan\beta$ and both values of
  $\sgn\mu$. The masses of the left-handed smuon and the lightest neutralino
  are shown in Fig.\,\ref{fig:SUSYmass}.  There are
  regions in these plots which we have not considered either due to the lack
  of radiative electroweak symmetry breaking, or because the lightest
  neutralino is not the lightest supersymmetric particle~(LSP). In the
  MSSM, the LSP must be a neutral colour singlet \cite{Ellis:1984ew}, from
  cosmological bounds on electric- or colour-charged stable
  relics. However if R-parity is violated the LSP can decay and these
  bounds no longer apply. We should therefore consider cases where one
  of the other SUSY
  particles is the LSP. We have only considered the case where the neutralino
  is the LSP for two reasons:
\begin{enumerate}
\item Given the unification of the SUSY-breaking parameters at the GUT scale
	it is hard to find points in parameter space where the lightest 
	neutralino 
	is not the LSP without the lightest neutralino becoming heavier
	than the 
	sleptons, which tend to be the lightest sfermions in these models.
	If the neutralino is heavier than the sleptons the resonance will not
 	be accessible for
 	the supersymmetric 
	gauge decay modes we are considering and the slepton will decay
	via \rpv\  modes.

\item The ISAJET code for the running of the couplings and the calculation of 
	the MSSM decay modes only works when the neutralino is the LSP.
\end{enumerate}

  The plots in Fig.\,\ref{fig:SUSYmass}
  also include the current experimental limits on the SUSY parameters 
  from LEP. This experimentally excluded region comes from two sources: the
  region at large $M_0$ is excluded by the limit on the cross section for
  chargino pair production \cite{Abbiendi:1998ff} and the limit on the
  chargino mass \cite{Barate:1998gy}; the region at small $M_0$ is
  excluded by the limit on the production of smuons 
  \cite{Abbiendi:1999is}.
  There is also a limit on the neutralino production cross section 
  \cite{Abbiendi:1998ff}. However, for most of the SUGRA parameter space this
  is weaker than the limit on chargino pair production. The gap in the
  excluded region between $M_0$ of about $50\, \mr{\gev}$  and
  $100\, \mr{\gev}$  is due to the presence
  of a destructive interference between the $t$-channel sneutrino exchange and
  the $s$-channel photon and Z exchanges in the chargino production cross
  section in $\mr{e^+e^-}$ collisions.

%%%%%%%%%%%%%%%%%%%%%%%%%%%%%%%%%%%%%%%%%%%%%%%%%%%%%%%%%%%%%%%%%%%%%%%%%%%%%%
%
%  Figure containing the limits at the various SUSY points
%
\begin{figure}
\begin{center}
\includegraphics[angle=90,width=0.48\textwidth]{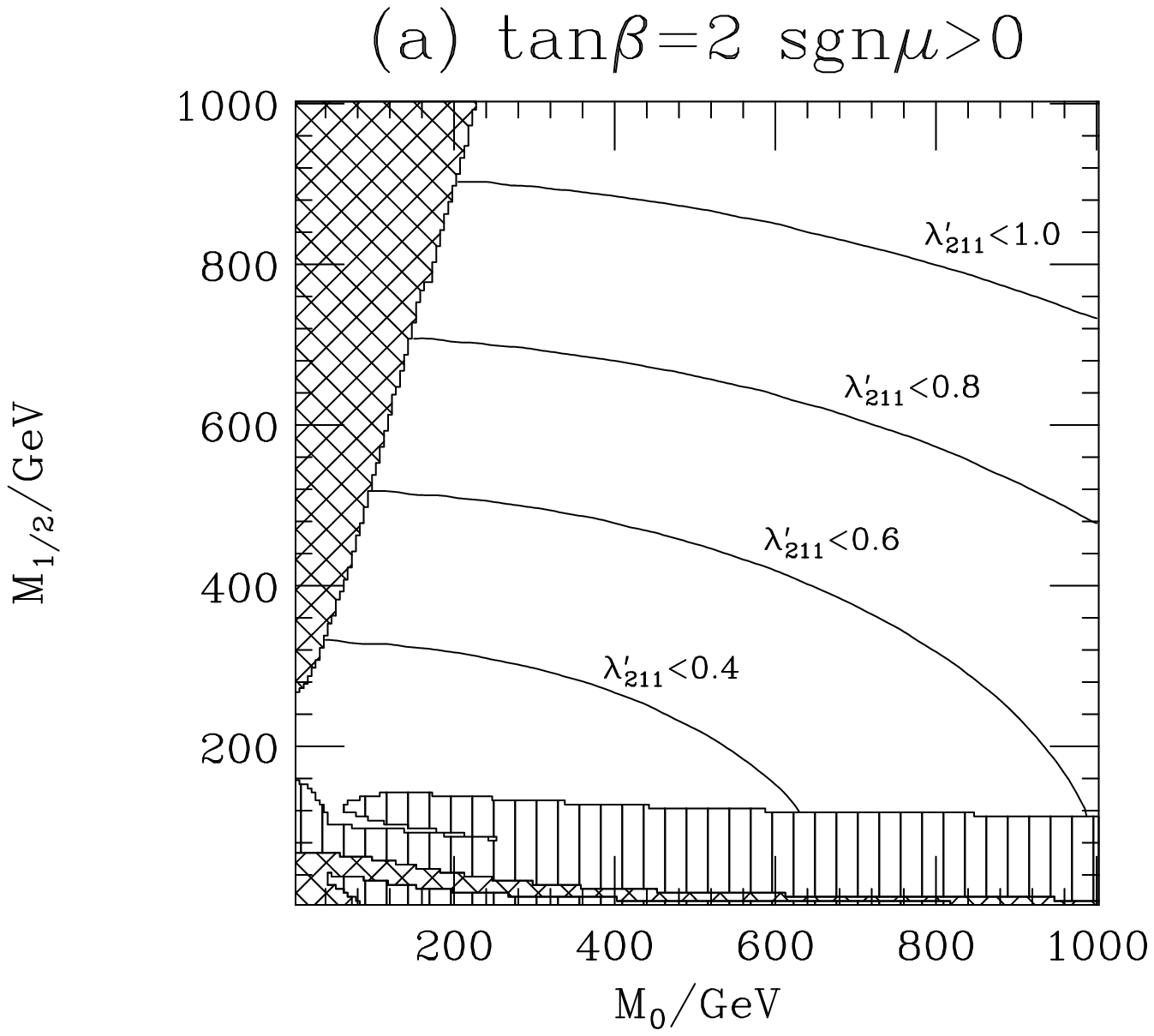}
\hfill
\includegraphics[angle=90,width=0.48\textwidth]{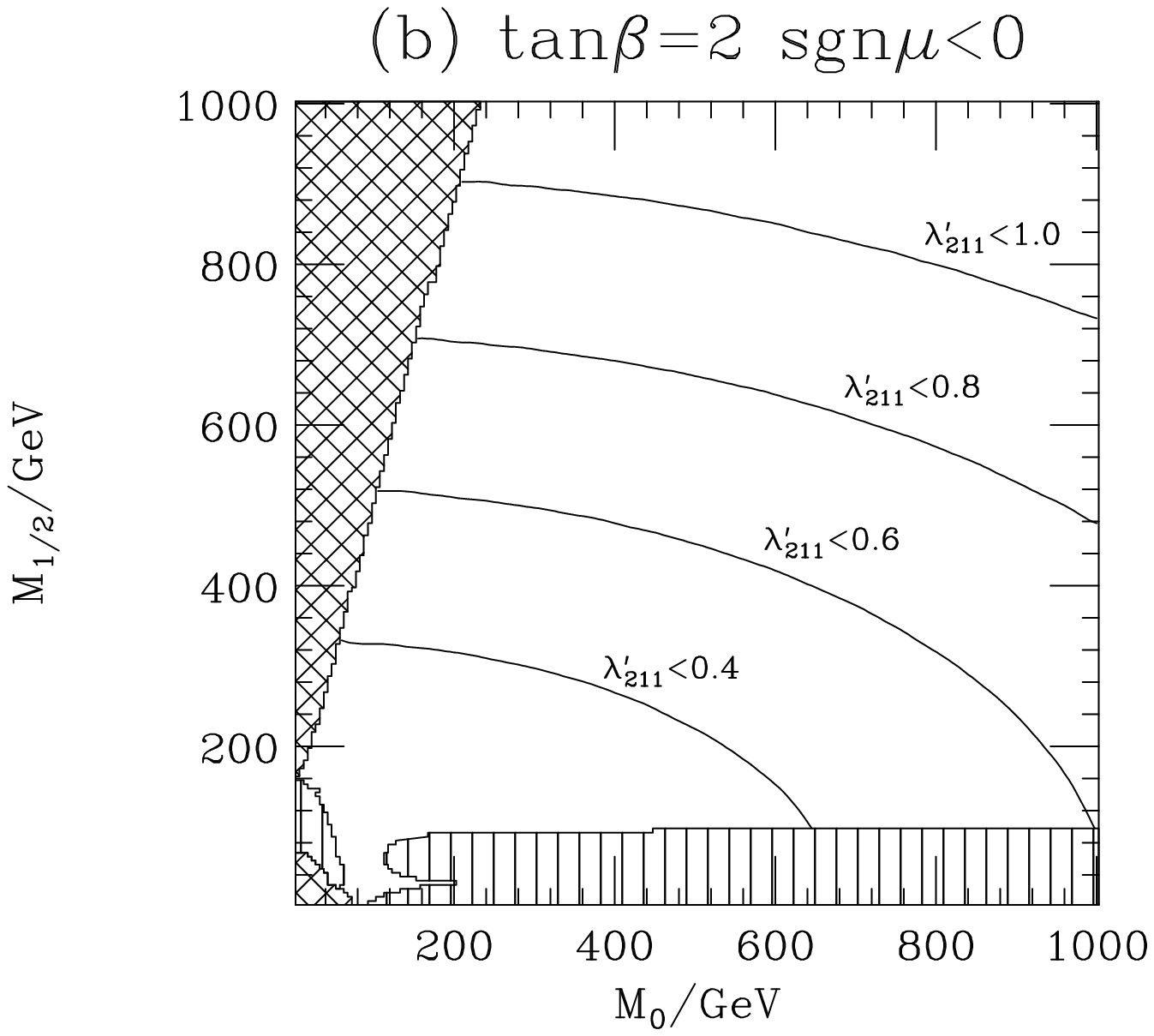}\\
\vskip 15mm
\includegraphics[angle=90,width=0.48\textwidth]{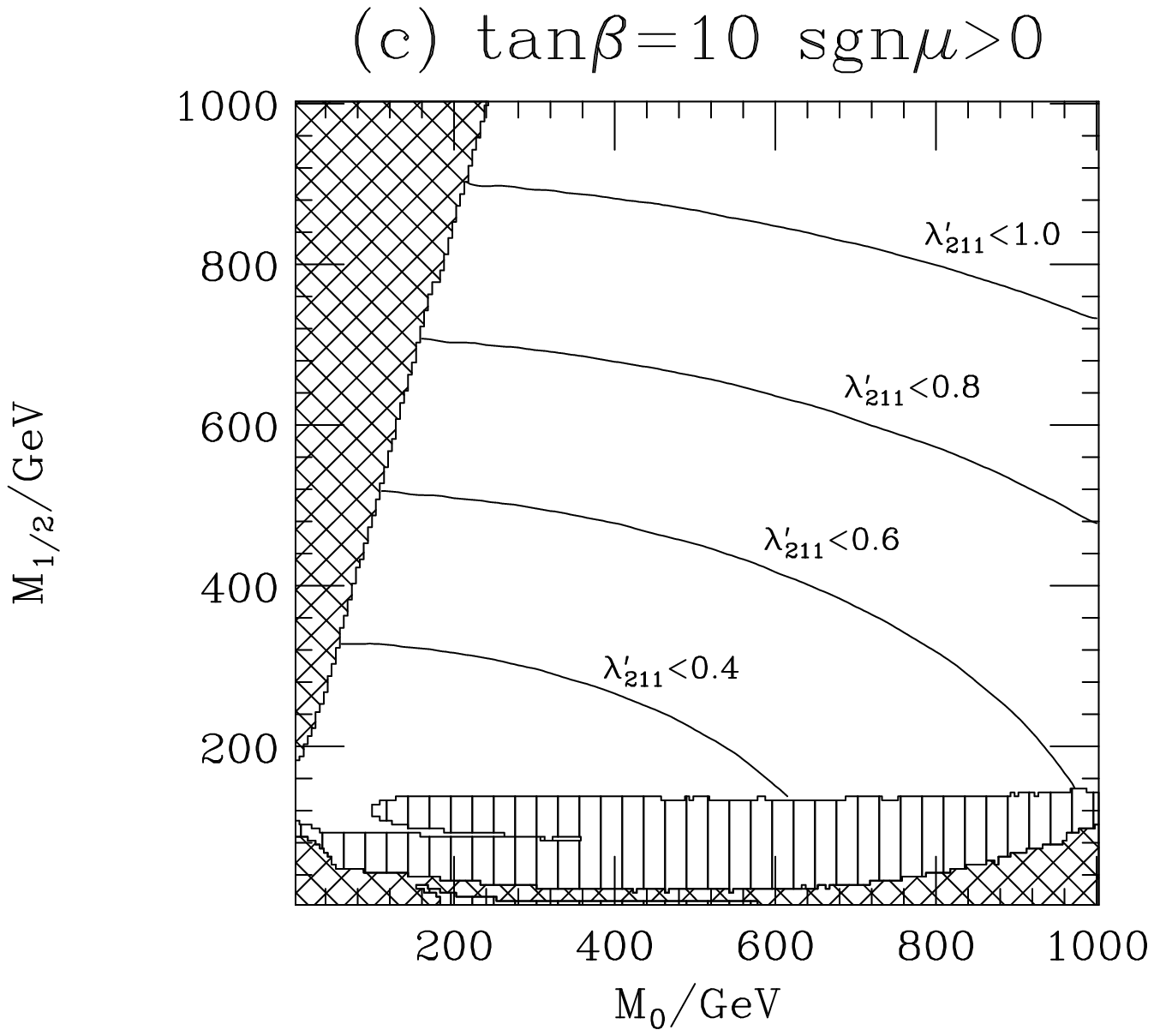}
\hfill
\includegraphics[angle=90,width=0.48\textwidth]{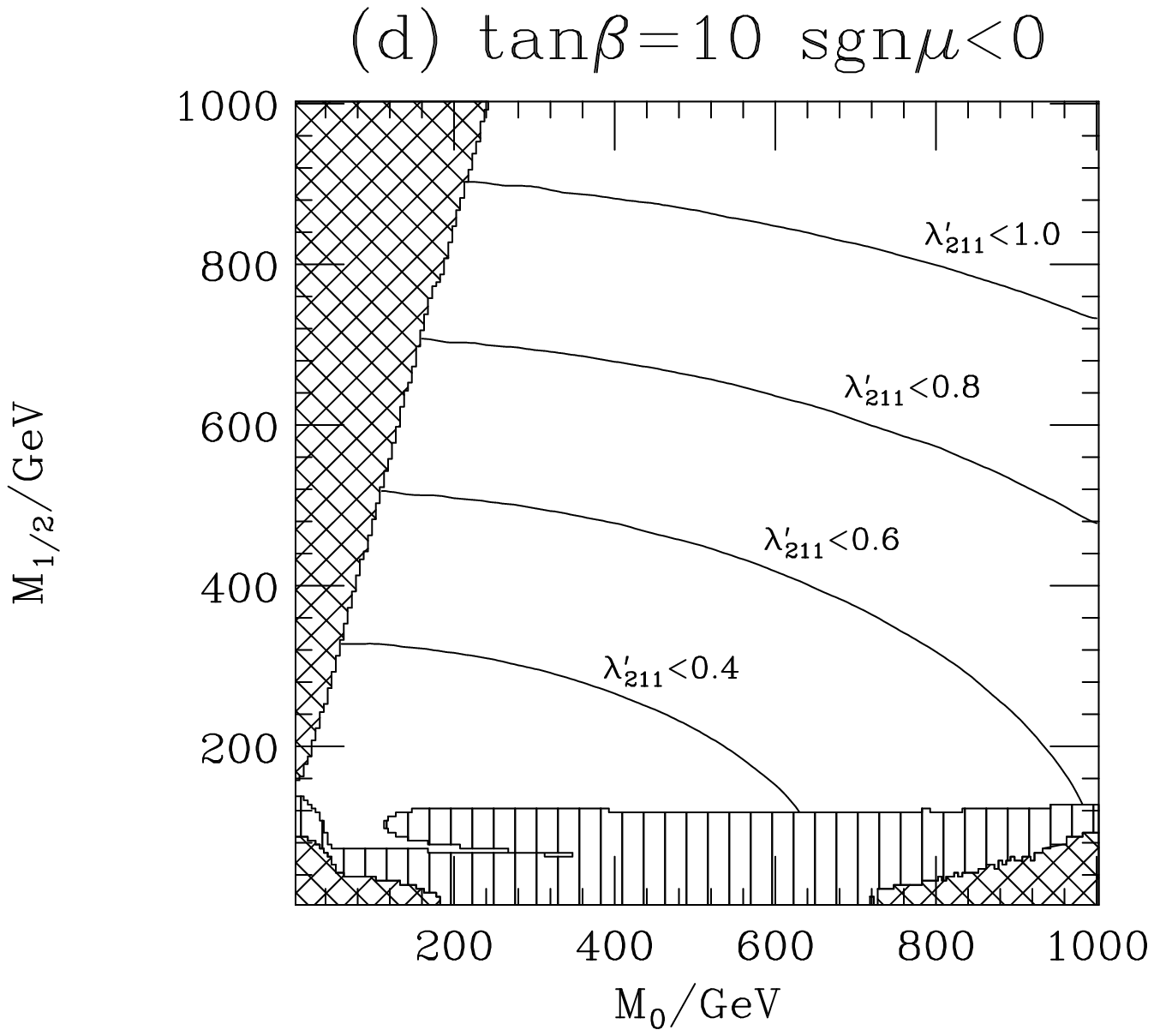}\\
\captionB{Limits on the coupling ${\lam'}_{211}$ in the  $M_0$, $M_{1/2}$
	 plane.}
	{Contours showing the limit on the \rpv\  Yukawa coupling 
	 ${\lam'}_{211}$ in the 
	 $M_0$, $M_{1/2}$ plane for $A_0=0\, \mr{\gev}$  
	 and different values of $\tan\beta$ and  $\sgn\mu$. The
	 striped and hatched regions are described in the caption of
	 Fig.\,\ref{fig:SUSYmass}.} 
\label{fig:SUSYlimit}
\end{center}
\end{figure}
% End of the Figure %%%%%%%%%%%%%%%%%%%%%%%%%%%%%%%%%%%%%%%%%%%%%%%%%%%%%%%%%%

%%%%%%%%%%%%%%%%%%%%%%%%%%%%%%%%%%%%%%%%%%%%%%%%%%%%%%%%%%%%%%%%%%%%%%%%%%%%%%
%
%  Figure containing the decay length at the various SUSY points
%
\begin{figure}
\begin{center}
\includegraphics[angle=90,width=0.48\textwidth]{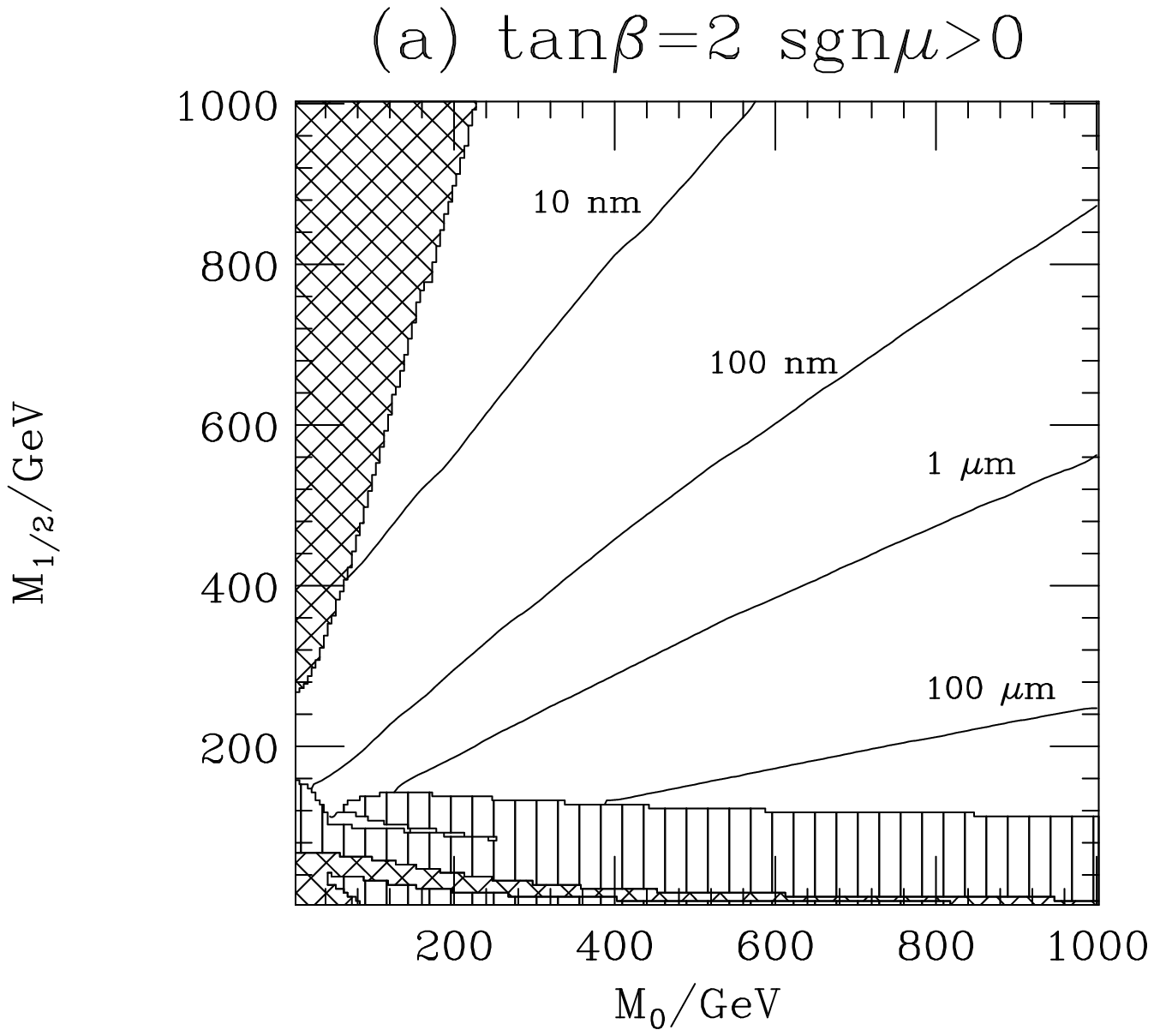}
\hfill
\includegraphics[angle=90,width=0.48\textwidth]{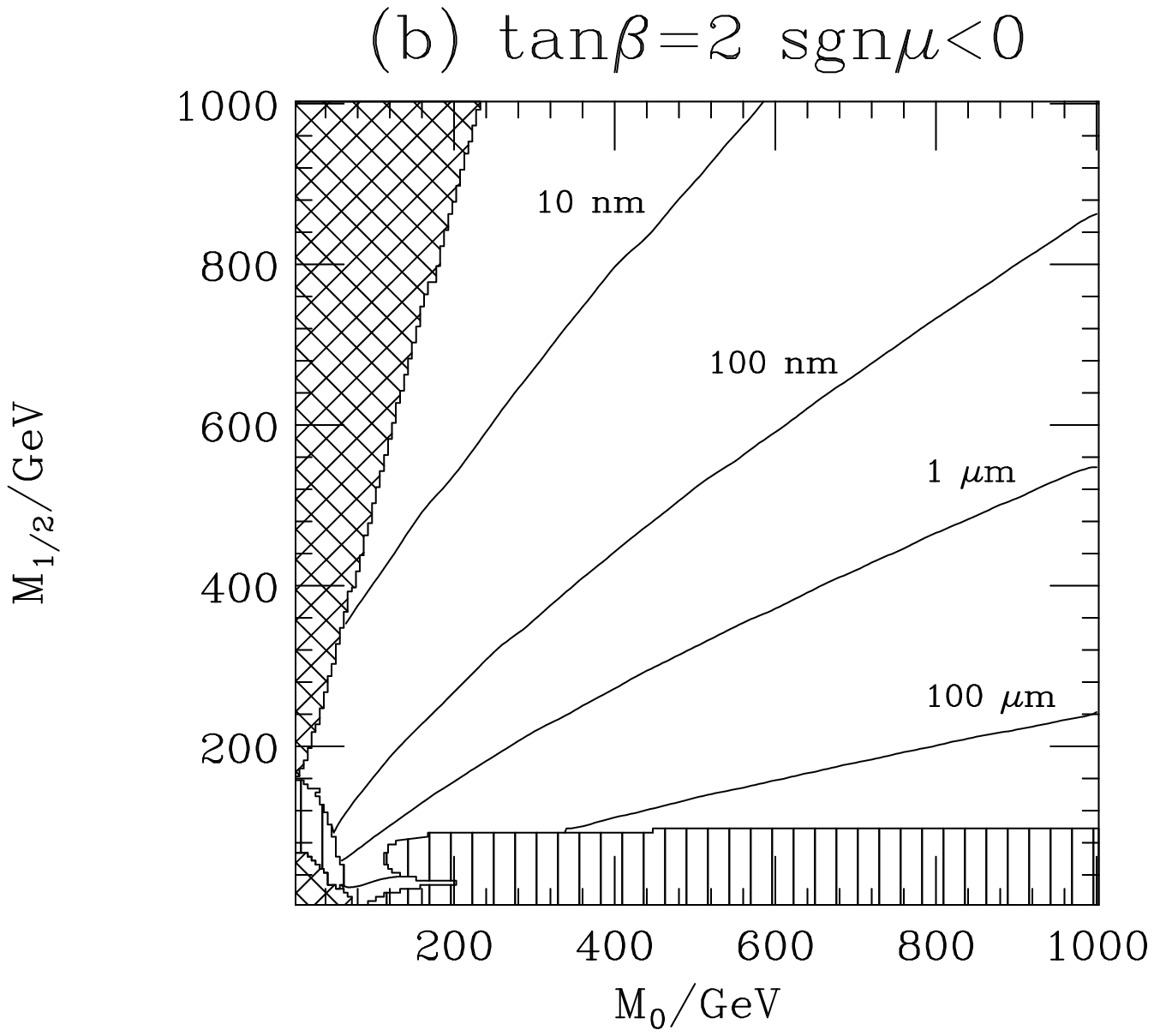}\\
\vskip 15mm
\includegraphics[angle=90,width=0.48\textwidth]{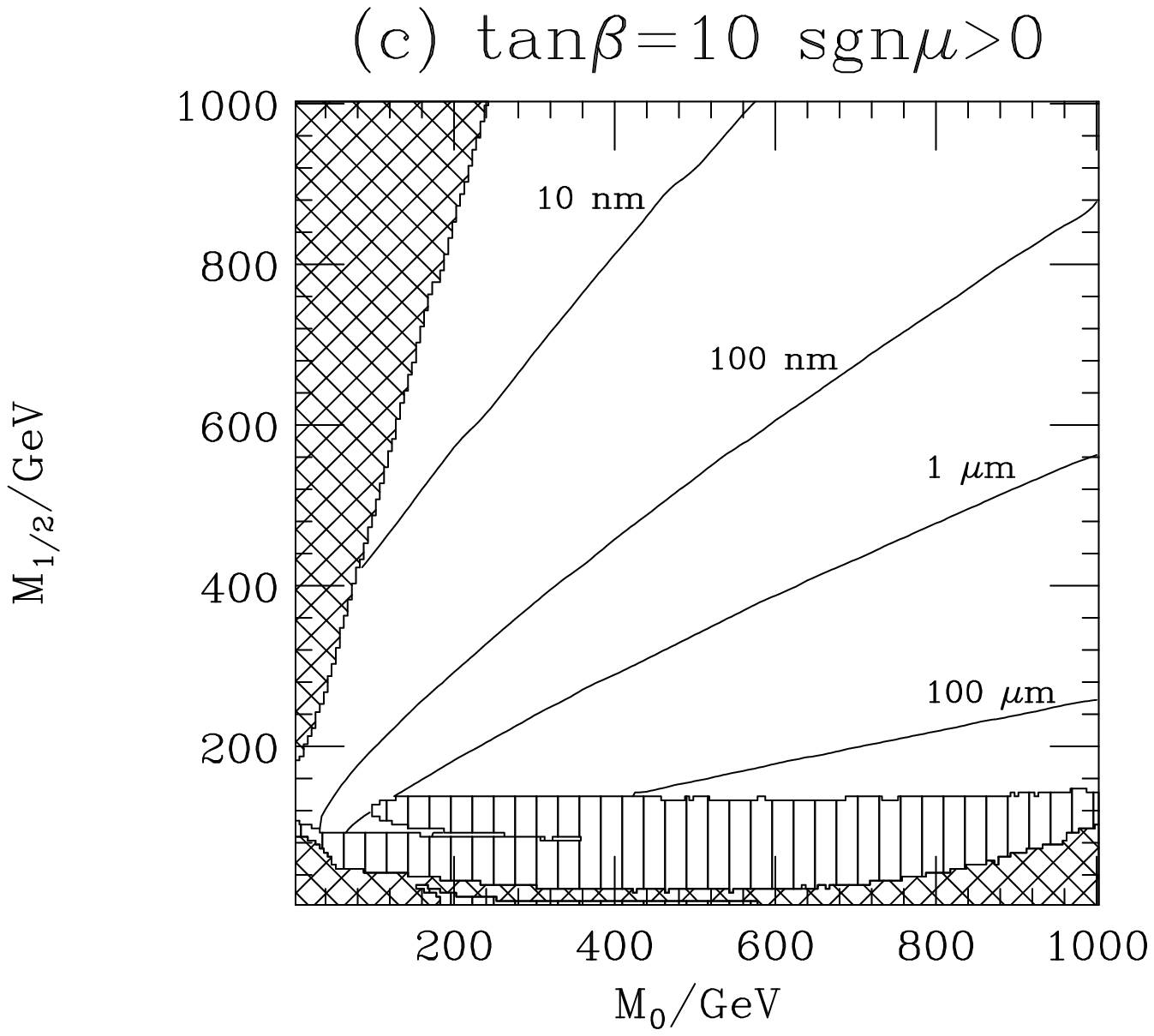}
\hfill
\includegraphics[angle=90,width=0.48\textwidth]{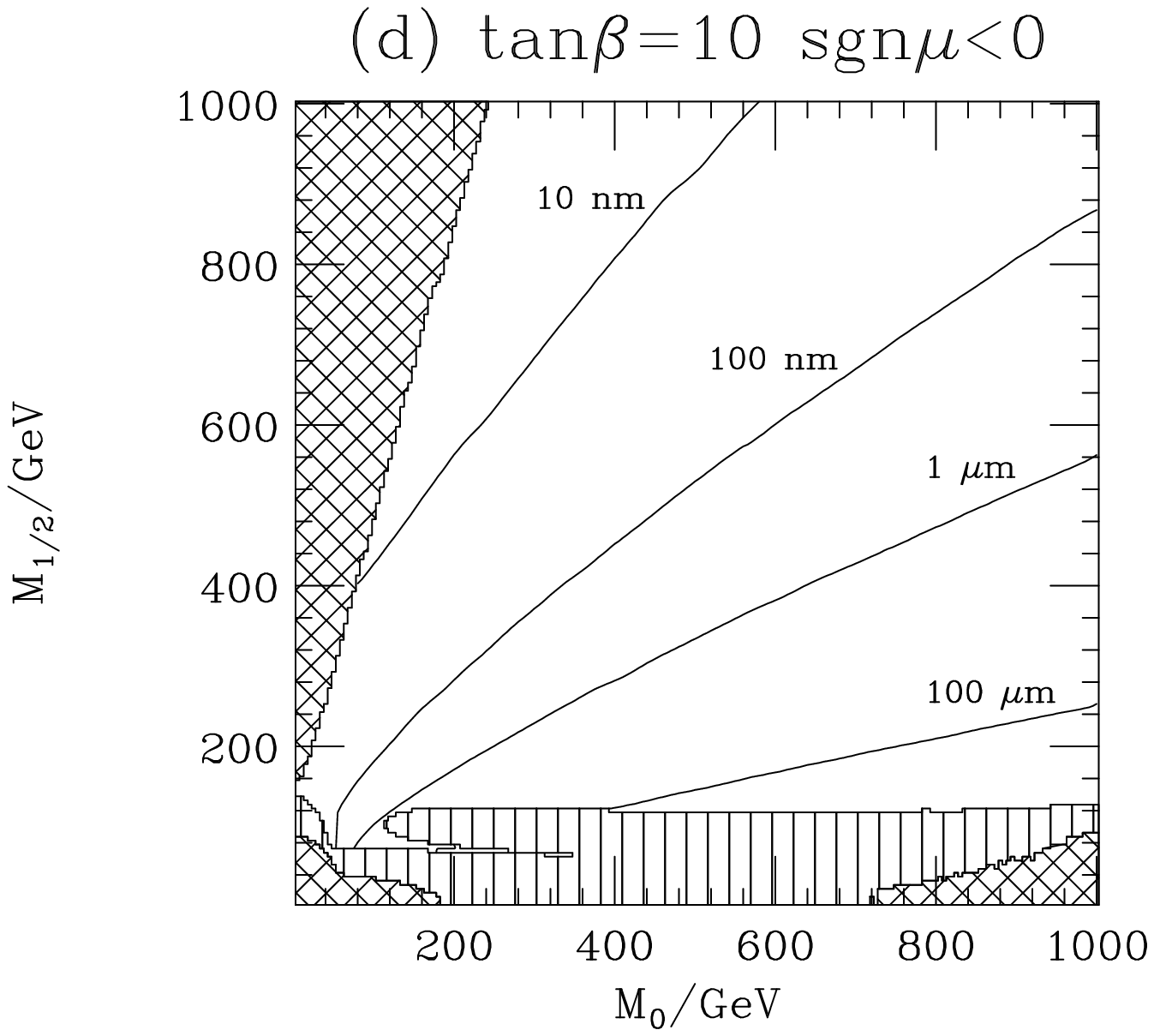}\\
\captionB{Lightest neutralino decay length in the $M_0$, $M_{1/2}$ plane.}
	{Contours showing the decay length of a neutralino produced in the 
	 decay of an on-mass-shell slepton  in the $M_0$, $M_{1/2}$ plane for
	 $A_0=0\, \mr{\gev}$, ${\lam'}_{211}=10^{-2}$
	 and different values of $\tan\beta$ and  $\sgn\mu$. The
	 striped and hatched regions are described in the caption of 
	 Fig.\,\ref{fig:SUSYmass}.} 
\label{fig:SUSYlength}
\end{center}
\end{figure}
% End of the Figure %%%%%%%%%%%%%%%%%%%%%%%%%%%%%%%%%%%%%%%%%%%%%%%%%%%%%%%%%%

  The limit on the coupling ${\lam'}_{211}$ is shown in
  Fig.\,\ref{fig:SUSYlimit}. As can be seen from Figs.\,\ref{fig:SUSYmass}
  and 
  \ref{fig:SUSYlimit} the limit on the coupling is fairly weak for large
  regions of parameter space, even when the smuon is relatively light. This is
  due to the squark masses, upon which the limit depends, being larger than
  the slepton masses in the SUGRA models.

  The signature we are considering requires the neutralino to decay
  inside
  the detector. In practice, if the neutralino decays more than a few
  centimeters from the primary interaction point a different analysis
  including displaced vertices would be necessary. The neutralino decay
  length is shown in Fig.\,\ref{fig:SUSYlength} and is small for all the
  currently allowed values of the SUGRA parameters. There will, however, be a
  lower limit on the \rpv\  
  couplings which can be probed using this process as the decay 
  length~$\sim1/{\lam'}^2_{211}$ \cite{Dawson:1985vr}.

%
% Section on the various backgrounds
%
\subsection{Backgrounds}
\label{sec:backgrounds}
\subsubsection{Standard Model Backgrounds}

  The dominant Standard Model backgrounds to like-sign dilepton production
come from:
\begin{itemize}
\item 	Gauge boson pair production, \ie production of WZ or ZZ 
followed by leptonic decays  with some of the
leptons not being detected.

\item 	$\mr{t\bar{t}}$ production. Either the t or $\mr{\bar{t}}$ 
        decays semi-leptonically, giving one lepton. The second top
        decays hadronically. A second lepton with the same charge can
        be produced in a semi-leptonic decay of the bottom hadron
        formed in the decay of the second top, \ie
\begin{eqnarray}
 \mr{t}	&\ra&  \mr{W^+ b} \ra \mr{\mu^{+}\bar{\nu}_{\mu} b},\nonumber \\
 \mr{\bar{t}}	&\ra& \mr{W^{-}\bar{b}}	\ra \mr{q\bar{q}\bar{b}},\quad 
 \mr{\bar{b}} \ra \mr{\mu^{+}\bar{\nu}_{\mu}\bar{c}}.
\end{eqnarray}

\item 	$\mathrm{b\bar{b}}$ production. If either of these quarks 
        hadronizes to form a $\mr{B^0_{d,s}}$ meson this can mix to
        give a $\mr{\bar{B}^0_{d,s}}$. This means that both the
 	bottom hadrons in the event will contain a $\mr{b}$ quark, if 
	a $\mr{B^0_{d,s}}$ undergoes mixing, or both bottom hadrons will
	contain a $\mr{\bar{b}}$, if a  $\mr{\bar{B}^0_{d,s}}$ mixes.
	Thus if both the
        bottom hadrons decay semi-leptonically the leptons will have
        the same charge as they are both coming from either b or
        $\mr{\bar{b}}$ decays.

\item 	Single top production. A single top quark can be produced together 
        with a $\mr{\bar{b}}$ quark by either an $s$-  or $t$-channel W
        exchange. This can give one charged lepton from the top
        decay, and a second lepton with the same charge from the decay
        of the meson formed after the b quark hadronizes.

 \item Non-physics backgrounds. There are two major sources: (i) from
 misidentifying the charge of a lepton, \eg in Drell-Yan production, and 
 (ii) from incorrectly identifying an isolated hadron as a lepton. This
 means that there is a major source of background from W production
 with an additional jet faking a lepton.
\end{itemize}

  These processes have been extensively studied 
  \cite{Baer:1992ef:Baer:1992xs:Jarlskog:1900dv:Barger:1985qa:Barnett:1993ea,
	Dreiner:1994ba:Guchait:1995zk,Armstrong:1994it:Baer:1994zt,
	Baer:1996va:Abdullin:1998nv,
       Matchev:1999nb,Matchev:1999yn,Nachtman:1999ua,Baer:1999bq} as they are
  also the major backgrounds to the production of like-sign dileptons 
  in the MSSM. The first studies of like-sign dilepton production at the LHC 
  \cite{Dreiner:1994ba:Guchait:1995zk} only considered the background from
  heavy quark production, \ie $\mr{t\bar{t}}$ and  $\mr{b\bar{b}}$
  production. More recent studies for both the LHC 
  \cite{Armstrong:1994it:Baer:1994zt,Baer:1996va:Abdullin:1998nv}
  and Run II of the Tevatron 
  \cite{Matchev:1999nb,Matchev:1999yn,Nachtman:1999ua,Baer:1999bq} 
  have also considered the background from gauge boson
  pair production. In addition, the Tevatron studies 
  \cite{Matchev:1999nb,Matchev:1999yn,Nachtman:1999ua,Baer:1999bq}
  have included the non-physics backgrounds.
  We have considered all the physics backgrounds, from both heavy quark
  production and gauge boson pair production, but have neglected the
  non-physics backgrounds which would require a full simulation of the
  detector.

  In these studies a number of different cuts have been used to suppress the
  backgrounds. These cuts can be split into two groups. The first of these
  sets of cuts
  is  designed to reduce the background from
  heavy quark production:
\begin{itemize}
% p_t cut
\item A cut on the $p_T$ of the leptons requiring 
	\begin{equation}
		p_T^{\mr{lepton}}>p^{\mr{CUT}}_T.
	\end{equation}
	The values of $p^{\mr{CUT}}_T$ for Tevatron studies have been between
	5 and $20\, \mr{\gev}$. Higher values, between 20 and
 	$50\, \mr{\gev}$, have usually
 	been used in LHC simulations.
% lepton isolation cut
\item A cut requiring that the leptons are isolated, \ie imposing a cut on the
      transverse energy, $E^{IC}_T$, of the particles other than the lepton 
      in a cone about the direction of the lepton such that
	\begin{equation}
		E^{IC}_T < E_0.
		\label{eqn:etcut}
	\end{equation}
	$E_0$ has been taken to be less than $5\, \mr{\gev}$  for Tevatron
	simulations and between 5 and $10\, \mr{\gev}$ for LHC studies.
      The radius of the cone is usually taken to be 
	\begin{equation}
  		\Delta R = \sqrt{\Delta\phi^2+\Delta\eta^2} < 0.4,
	\end{equation}
      where $\Delta\phi$ is the azimuthal angle and $\Delta\eta$ 
      the pseudo-rapidity of the particles with respect to the lepton.
      
\end{itemize}
  It was shown in \cite{Dreiner:1994ba:Guchait:1995zk} that these cuts
  can reduce the background from heavy quark production by several orders
  of magnitude. Any high-$p_T$ lepton from a bottom hadron decay must come
  from a high-$p_T$ hadron. This is due
  to the small mass of the bottom hadron relative
  to the lepton $p_T$ which means the lepton will be travelling in
  the same direction as the other decay products 
  \cite{Mondal:1994vi:Godbole:1983yb:Barger:1983qr:Roy:1987mg}. Hence the
  isolation and $p_T$ cuts remove the majority of these events.

  The analyses of 
  \cite{Baer:1996va:Abdullin:1998nv,Matchev:1999nb,Matchev:1999yn,
        Baer:1999bq,Nachtman:1999ua}
  then imposed further cuts to reduce the backgrounds from gauge boson pair
  production, which is the major contribution to the SM background after 
  the imposition of the isolation and $p_T$ cuts:
\begin{itemize}
% Z mass rejection
\item A cut on the invariant mass, $m_{\ell^+\ell^-} $,
       of any pair of opposite sign same 
      flavour (OSSF)
      leptons to remove those leptons coming from Z decays, \ie
	\begin{equation}
		|M_{\mr{Z}}-m_{\ell^+\ell^-}| < m^{\mr{CUT}}_{\ell^+\ell^-},
	\end{equation}	
      was used in 
	\cite{Matchev:1999nb,Nachtman:1999ua,Baer:1999bq}.
% OSSF veto
\item Instead of a cut on the mass of OSSF lepton pairs, some analyses
      considered a veto on the presence of an OSSF lepton pair in the event.
% transverse mass cut
\item In \cite{Baer:1999bq,Matchev:1999yn} a cut on the transverse mass was
      imposed to reject leptons which come from the decays of W bosons. 
      The transverse mass, $M_T$, of a lepton--neutrino pair is given by
	\begin{equation}
	M^2_T = 2|p_{T_{\ell}}||p_{T_\nu}|(1-\cos\Delta\phi_{\ell\nu}),
	\label{eqn:MTdef}
	\end{equation}
	where   $p_{T_{\ell}}$ is the transverse momentum of the charged
  	          	 lepton,
		$p_{T_\nu}$ is the transverse momentum of the neutrino 
			 (assumed to be the total missing transverse momentum
			 in the event) and
	        $\Delta\phi_{\ell\nu}$ is the azimuthal angle between
			 the lepton and the neutrino 
			(the direction of the neutrino is taken to be
			 the direction of the missing momentum in the event). 
			
        This cut is applied to both of the like-sign leptons in the event
	to reject events in which either of them came from the decay of a
	W boson. A cut removing events with
	\mbox{$60\, \mr{\gev}<M_T<85\, \mr{\gev}$} was
	used in \cite{Baer:1999bq}
	to reduce the background from WW and WZ production.
% cuts of the missing transverse energy
\item For the MSSM signatures considered in 
      \cite{Matchev:1999nb,Matchev:1999yn,Baer:1999bq,Nachtman:1999ua}
      there is missing transverse energy, \met, due to the LSP escaping
      from the detector. This allowed them to impose a cut on the
      \met, $\not\!\!E_T > E_T^{\mr{CUT}}$, to reduce the background.
\end{itemize}

  There are, however, differences between the MSSM signatures which were
  considered in 
  \cite{Matchev:1999nb,Matchev:1999yn,Baer:1999bq,Nachtman:1999ua}
  and the \rpv\  processes we are considering here. In particular as the
  LSP decays, there will be little missing transverse energy in the \rpv\  
  events. 
  This means that instead of a cut requiring the \met\  to be above
  some value we will consider a cut requiring the \met\  to be less than
  some value, \ie
	\begin{equation}
		\not\!\!E_T < E_T^{\mr{CUT}}.
	\end{equation}
  This cut will remove events from some of the possible resonant production
  mechanisms, \ie those channels where a neutrino is produced in either
  the slepton decay or the cascade decay of a chargino, or one of the
  heavier neutralinos, to the lightest neutralino.
  However it will not affect the decay of a charged slepton to the lightest
  neutralino
  which is the dominant production mechanism over most of the SUSY parameter
  space.

  Similarly, the signal we are considering in general will not contain more
  than two leptons. Further leptons can only come from cascade decays
  following the production of either a chargino or one of the heavier
  neutralinos, or from semi-leptonic
  hadron decays. This means that instead of the cut on the invariant mass of
  OSSF lepton pairs we will only consider the effect of a veto on the presence
  of OSSF pairs. This veto was considered in 
   \cite{Matchev:1999nb,Matchev:1999yn} but for the MSSM signal  
  considered there it removed more signal than background.

%
%  SUSY backgrounds bit
%
\subsubsection{SUSY Backgrounds}

  So far we have neglected what may be the major source of background to this
  process, \ie supersymmetric particle pair production. If we only consider 
  small \rpv\  couplings the dominant effect in sparticle pair production is 
  that the LSP produced at the end of the cascade decays of the other SUSY
  particles will decay. For large \rpv\ 
  couplings the cascade decay chains can also
  be affected by the heavier SUSY particles decaying via \rpv\  modes.
  We will not consider this effect here.\footnote{These additional decays are
  included in HERWIG~6.1 and the matrix elements are given in
  Appendix~\ref{chap:decay}.} The LSP will decay giving a quark--antiquark
  pair and either a charged lepton or a
  neutrino. There will usually be two LSPs in each event, one from the 
  decay chain of each of the sparticles produced in the hard collision. This
  means that they can both decay to give leptons with the same charge. Leptons
  can also be produced in the cascade decays. These processes will therefore
  be a major background to like-sign dilepton production via resonant slepton
  production.

  The cuts which were intended to reduce the Standard Model background will
  also significantly reduce the background from sparticle pair production.
  However we will need to impose additional cuts to suppress this background.
  In the signal events there will be at least two high-$p_T$ jets from the
  neutralino decay and there may be more jets from either initial-state QCD
  radiation or radiation from quarks produced in the neutralino decay.
  In the dominant production mechanism, \ie $\mr{\mut\ra\mu^-\cht^0_1}$,
  this will be the only source  of jets, however additional jets can be
  produced
  in the cascade SUSY decays if a chargino or one of the heavier neutralinos
  is produced.
  In the SUSY background there will be at least four high-$p_T$ jets
  from the neutralino decays, plus other jets formed in the decays of the
  coloured sparticles which are predominantly formed in hadron--hadron
  collisions. This suggests two possible strategies for reducing
  the sparticle 
  pair production background:
\begin{enumerate}
\item	A cut such that there are at most 2 or 3 jets (allowing for some QCD
 	radiation) above a given $p_T$. This will reduce the SUSY background
	which typically has more than four high-$p_T$ jets.

\item	A cut such that there are exactly two jets, or only two or three
	jets above a given $p_T$. This will reduce the gauge boson pair
	 production
	background where typically the only jets come from initial-state
 	radiation, as well as the background from sparticle pair production.

\end{enumerate}

  In practice we would use a much higher momentum cut in the first case, as
  we only need to ensure that the cut is sufficiently high that most of the 
  sparticle pair production events give more than 2 or 3 jets above the cut.
  However with the second cut we need to ensure that the jets in the signal
  have sufficiently high-$p_T$ to pass the cut as well. In practice we found
  that the first cut significantly reduced the sparticle pair production 
  background while
  having little effect on the signal, while the second cut dramatically
  reduced the signal as well. In the next section we will
  consider the effects of these cuts on both the signal and background
  at Run II of the Tevatron and the LHC.

%
%  Section describing the cuts we used and what signal reaches we can achieve
%
\subsection{Simulations}
\label{sec:results}
  
  HERWIG~6.1 \cite{HERWIG61} was used to simulate the
  signal and the backgrounds from sparticle pair, $\mr{t\bar{t}}$,
  $\mr{b\bar{b}}$ and single top production. HERWIG does not include gauge
  boson pair production in hadron--hadron collisions and we therefore used  
  PYTHIA~6.1 \cite{Sjostrand:1994yb} to simulate this background. The
  simulation of the signal includes all the R-parity conserving  decay modes
  given in Table~\ref{tab:decaymodes}. We used the cone algorithm described in
  the previous chapter for all the jet reconstructions, although for this
  study we took the radius of the cone to be 0.4 radians rather than the value
  of 0.7 radians used in both the previous section and the last chapter.
  
  Due to the large cross sections for some of the Standard Model backgrounds
  before any cuts, we imposed parton-level cuts and forced certain decay
  modes, \ie we required that certain particles decay via a particular
  channel, in order to simulate a sufficient number of events with the
  resources available. We designed these cuts in such a way that hopefully
  they are weaker than any final cut we apply, so that we do not lose any of
  the events that would pass the final cuts. We imposed the following cuts
  for the various backgrounds:
\begin{itemize}
\item \underline{$\mr{b\bar{b}}$ production.}
	We forced the B hadrons produced by the
	hadronization to decay semi-leptonically. This neglects the production
        of leptons in charm decays which has a higher cross section but
        which we would expect to have a lower $p_T$
	and be less well isolated than those leptons produced in
	bottom decays. If there was only one 
  	$\mr{B^0_{d,s}}$ meson in the event this was forced to mix.
	When there was more than one $\mr{B^0_{d,s}}$ 
	meson then one of them was
	forced to mix and the others were forced not to mix. Similarly we
	imposed a parton-level cut on the transverse momentum of the initial
	b and $\mr{\bar{b}}$, 
	$p_T^{\mr{b},\mr{\bar{b}}}\geq p_T^{\mr{parton}}$.
 	This parton level cut should not affect the
	background provided that we impose a
        cut on the transverse momentum of
  	the leptons produced in the decay, 
	$p_T^{\mr{lepton}}\geq p_T^{\mr{parton}}$.

\item 	\underline{$\mr{t\bar{t}}$ production.}
        While not as large as the $\mr{b\bar{b}}$
	production cross section the cross section for $\mr{t\bar{t}}$ is
	large, particularly at the LHC. We improved the efficiency
	by forcing one of the top quarks in each
	event to decay semi-leptonically, again this neglects events in which
	there are leptons from charm decay. However we did not impose a cut
	on the $p_T$ of the top quarks as, due to the large top quark mass,
        even relatively low $p_T$ top quarks can give high-$p_T$ leptons.

\item 	\underline{Single top production.}
        While the cross section for this process is
      	relatively small compared to the heavy quark pair production cross
      	sections, we forced the tops to decay semi-leptonically to 
      	reduce the number of events we needed to simulate.

\item 	\underline{Gauge boson pair production.}
	The cross sections for these processes are relatively small and it
	 was not necessary to impose any parton level
	cuts, or force particular decay modes.
\end{itemize}

  Where possible, the results of the Monte Carlo simulations have been
  normalized by using next-to-leading-order cross sections for the various
  background processes. We used the next-to-leading-order calculation of 
  \cite{Campbell:1999ah} for gauge boson pair production.
  The $\mr{t\bar{t}}$ simulations were normalized using 
  the next-to-leading-order, with next-to-leading-log resummation, calculation
  from \cite{Bonciani:1998vc}.

  The calculation of a next-to-leading-order cross section for $\mr{b\bar{b}}$
  production
  is more problematic due to the parton-level cuts we imposed on the simulated
  events. There are a range of possible options for applying the $p_T$ cut we
  imposed on the bottom quark at next-to-leading order. At leading order the
  transverse momenta of the quarks are identical and therefore the cut
  requires them both to have transverse momentum 
  \mbox{$p_T>p_T^{\mr{CUT}}$}. However at 
  next-to-leading order, due to gluon radiation, the transverse momenta of 
  the quarks are no longer equal. Therefore a cut on, for example, the
  $p_T$ of 
  the hardest quark, \mbox{$p_{T_1}>p_T^{\mr{CUT}}$}, together with a
  cut on the lower $p_T$ quark, \mbox{$p_{T_2}>p_T^{\mr{CUT}}-\delta$},
  with any positive value of \mbox{$\delta<p_T^{\mr{CUT}}$}, is the same
  as the leading-order cut we applied. Given that we need a high
  transverse momentum bottom hadron to give a high-$p_T$ lepton and 
  only events with two such high-$p_T$ leptons will contribute to the
  background, a cut requiring both bottom quarks to have 
  \mbox{$p_{T}>p_T^{\mr{CUT}}$}, \ie $\delta=0$, is most appropriate.
  However at this point perturbation theory is
  unreliable \cite{Frixione:1997ks} and for the cuts we applied the
  next-to-leading-order cross section is smaller than the leading-order
  result. We therefore applied the cut 
  \mbox{$p_{T_1}>p_T^{\mr{CUT}}$} with no cut on the softer bottom 
  as this avoids the point 
  at which the perturbative expansion is unreliable, 
  \ie $\delta=p_T^{\mr{CUT}}$. We used the program 
  of \cite{Mangano:1992jk} to calculate the next-to-leading-order
  cross section with these cuts.

  All of the simulations and SUSY cross section calculations 
  used the latest MRS parton distribution set 
  \cite{Martin:1999ww:Martin:1998ab}, as did the calculation of the
  single top production cross section. The parton distribution sets used
  in the various next-to-leading-order cross sections are described in the
  relevant papers.

  We can now study the signal and background in more detail for both the
  Tevatron and the LHC. This is followed by a discussion of the methods used
  to 
  reconstruct the masses of both the lightest neutralino and the resonant
  slepton.

%
%  a subsection on the Tevatron
%
\subsubsection{Tevatron}
\label{sub:tevatron}

%%%%%%%%%%%%%%%%%%%%%%%%%%%%%%%%%%%%%%%%%%%%%%%%%%%%%%%%%%%%%%%%%%%%%%%%%%%%%%
%
%  Figure containing the signal cross sections at the Tevatron
%
\begin{figure}
\begin{center}
\includegraphics[angle=90,width=0.48\textwidth]{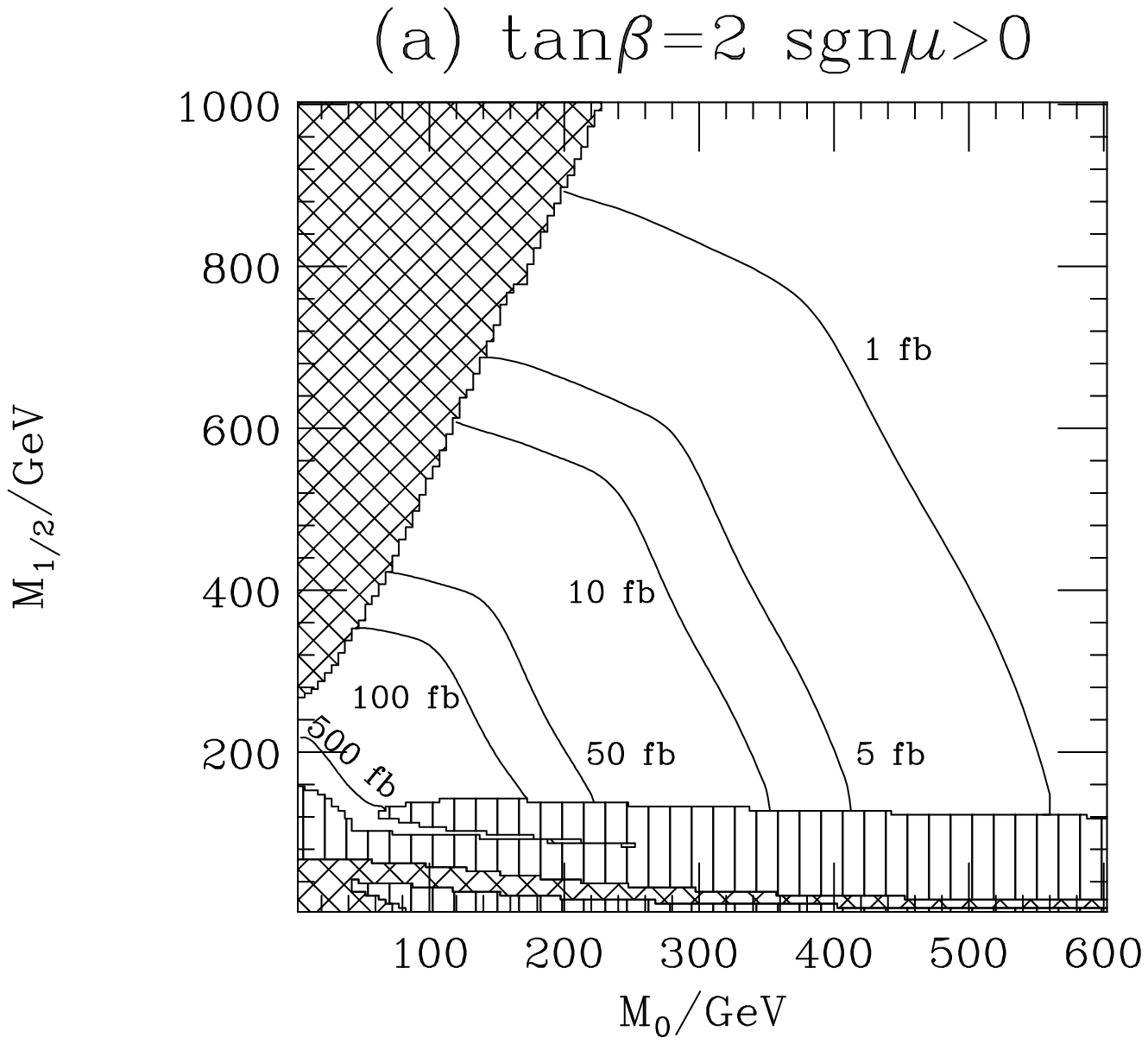}
\hfill
\includegraphics[angle=90,width=0.48\textwidth]{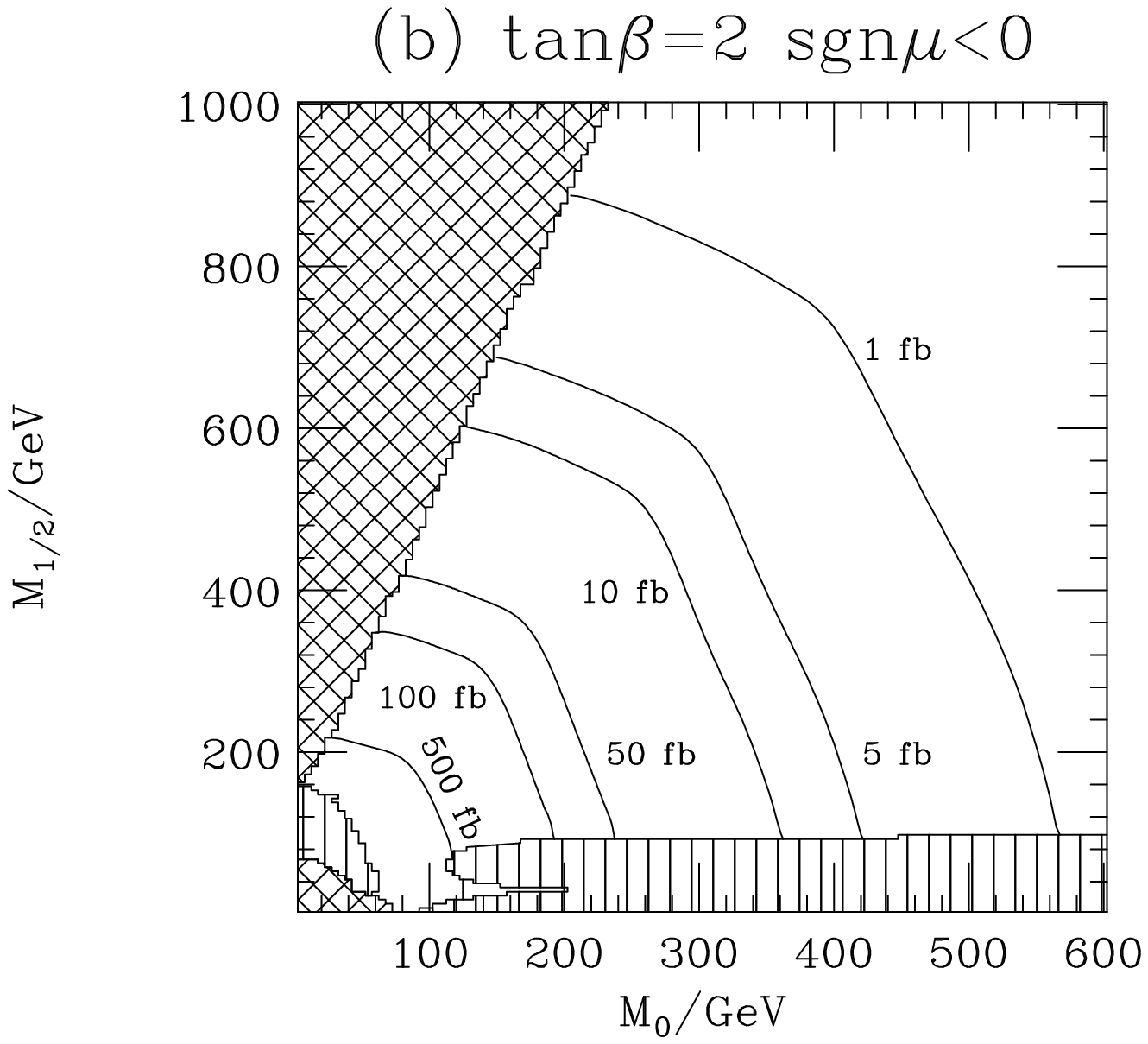}\\
\vskip 15mm
\includegraphics[angle=90,width=0.48\textwidth]{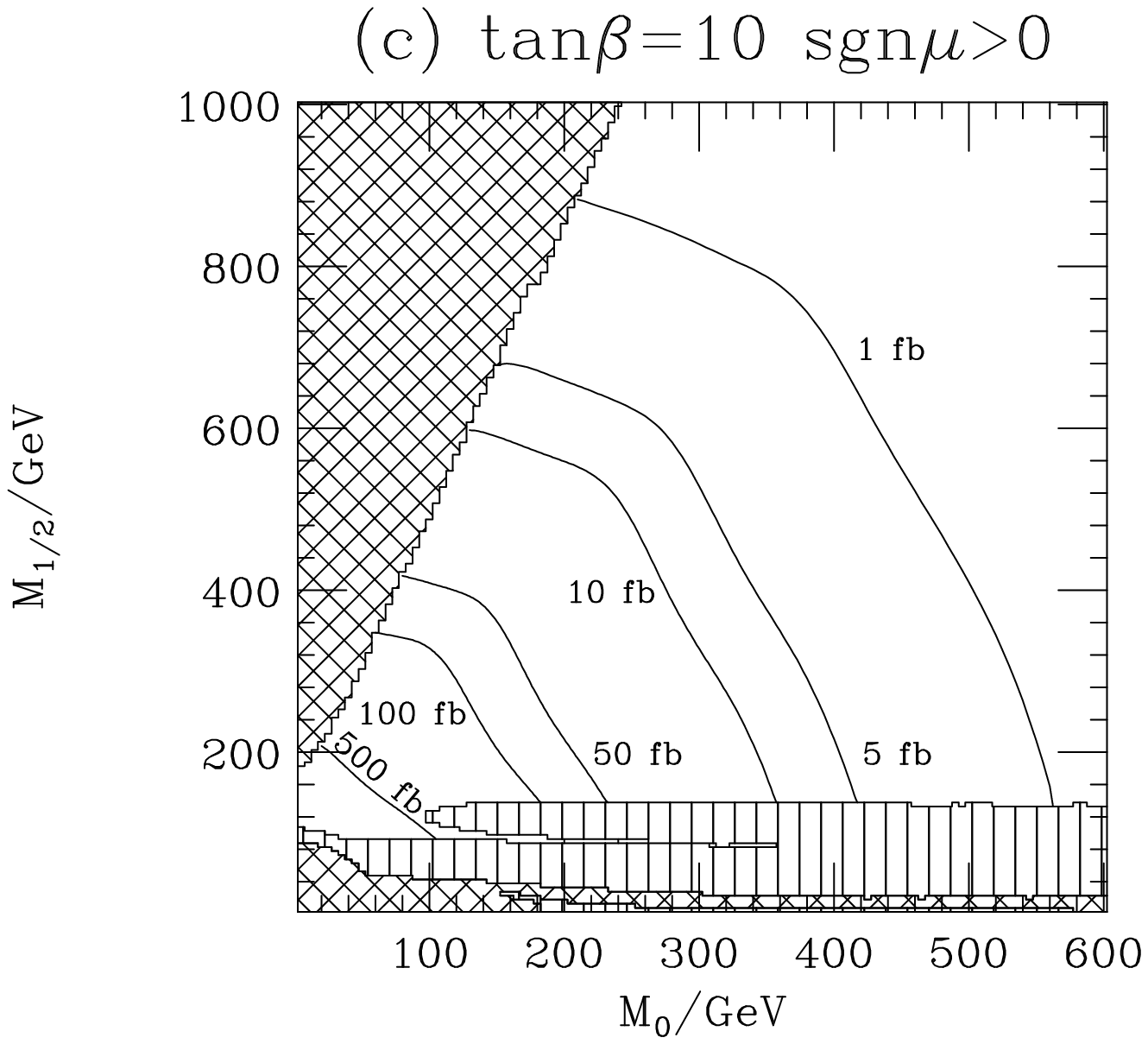}
\includegraphics[angle=90,width=0.48\textwidth]{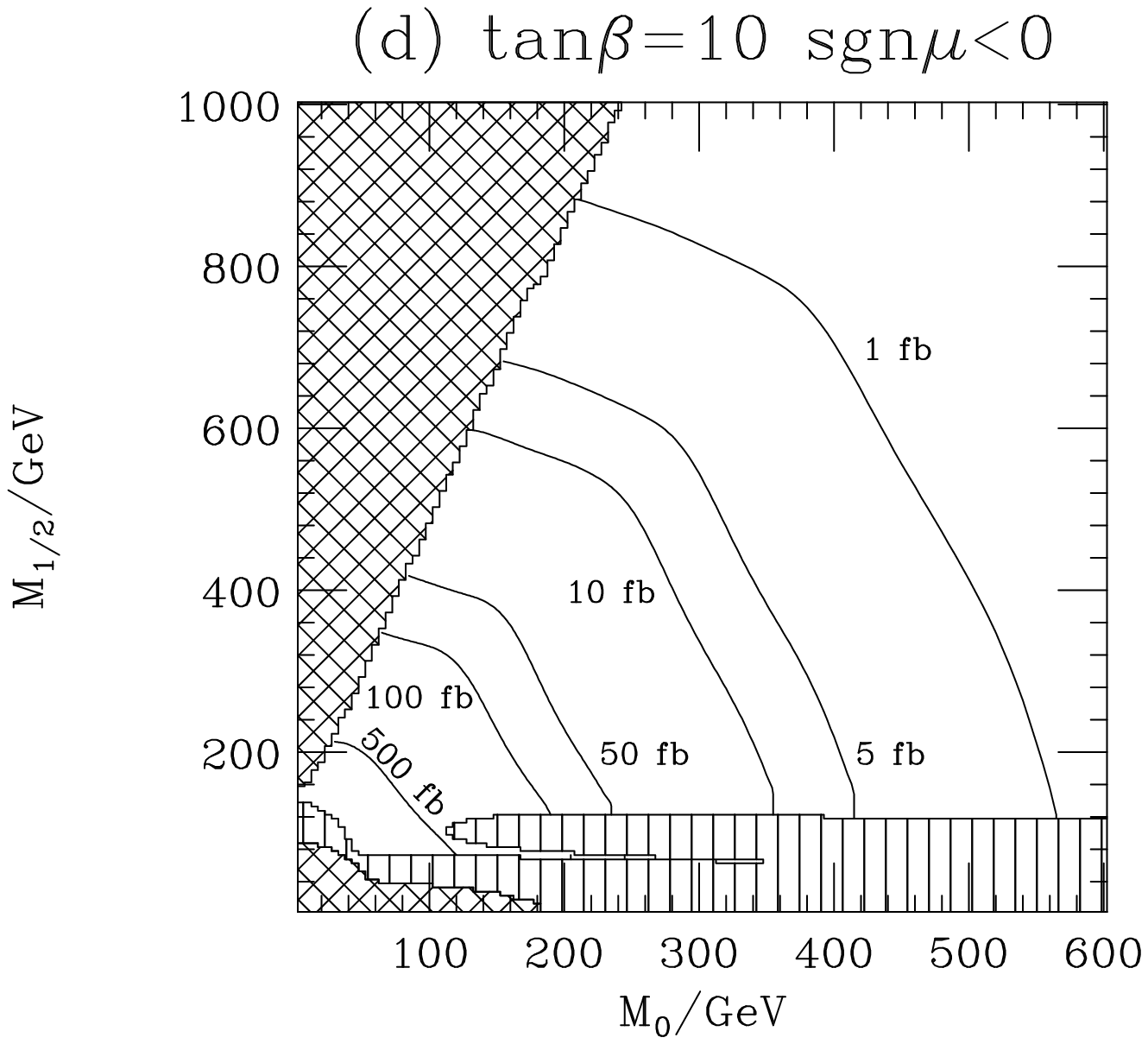}\\
\captionB{Production cross section for $\cht^0\ell$ at the Tevatron in $M_0$,
	 $M_{1/2}$ plane.}
	{Contours showing the cross section for the production of a neutralino
	 and a charged lepton at Run II of the Tevatron
	 in the $M_0$, $M_{1/2}$ plane for $A_0=0\, \mr{\gev}$ 
         and ${\lam'}_{211}=10^{-2}$ with different values
	 of $\tan\beta$ and $\sgn\mu$. The striped and hatched regions are
	 described in the caption of Fig.\,\ref{fig:SUSYmass}.} 
\label{fig:tevcross}
\end{center}
\end{figure}
% End of the Figure %%%%%%%%%%%%%%%%%%%%%%%%%%%%%%%%%%%%%%%%%%%%%%%%%%%%%%%%%%

%%%%%%%%%%%%%%%%%%%%%%%%%%%%%%%%%%%%%%%%%%%%%%%%%%%%%%%%%%%%%%%%%%%%%%%%%%%%%%
%
%  Figure containing the signal cross sections at the Tevatron
%
\begin{figure}
\begin{center}
\includegraphics[angle=90,width=0.48\textwidth]{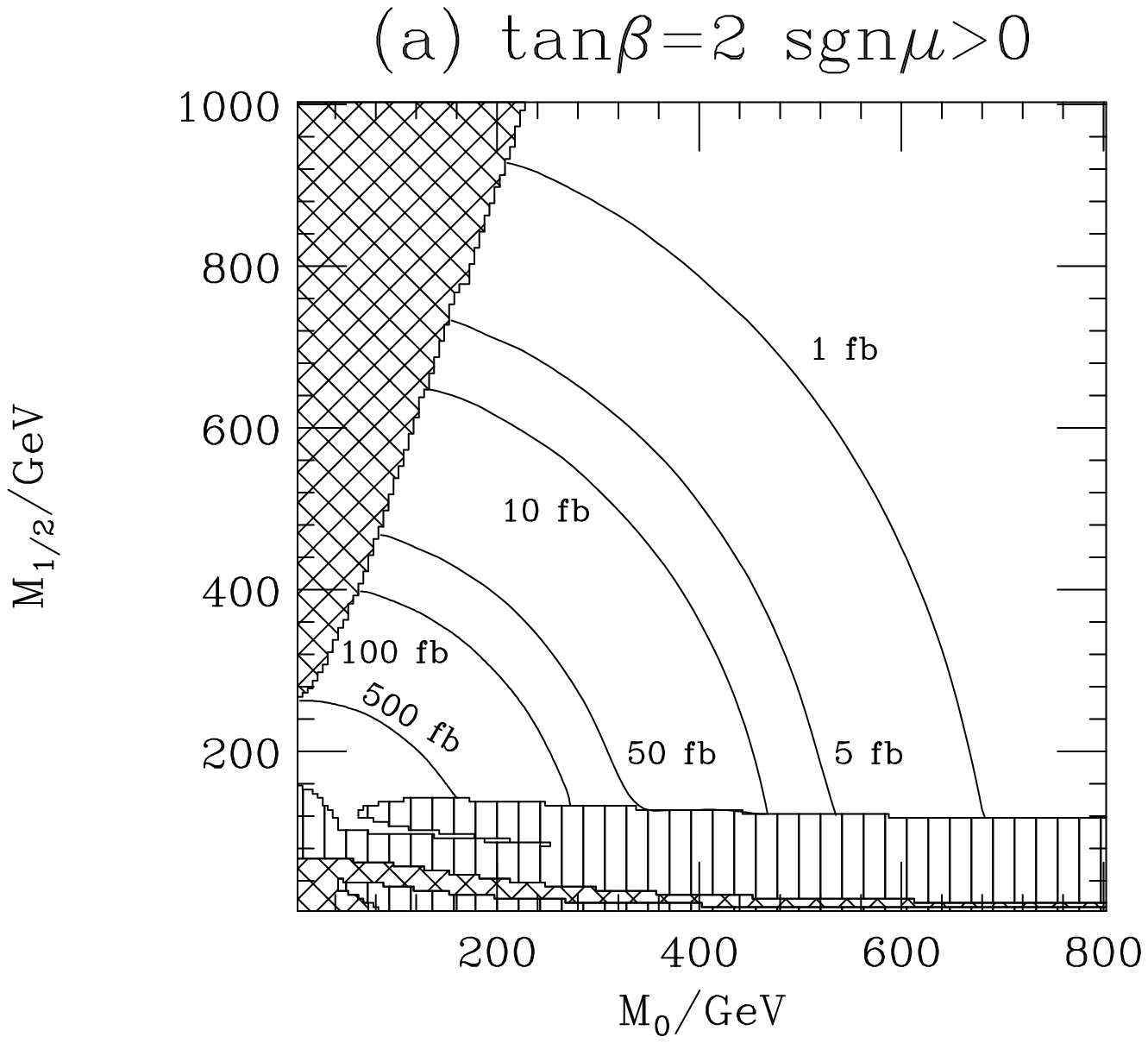}
\hfill
\includegraphics[angle=90,width=0.48\textwidth]{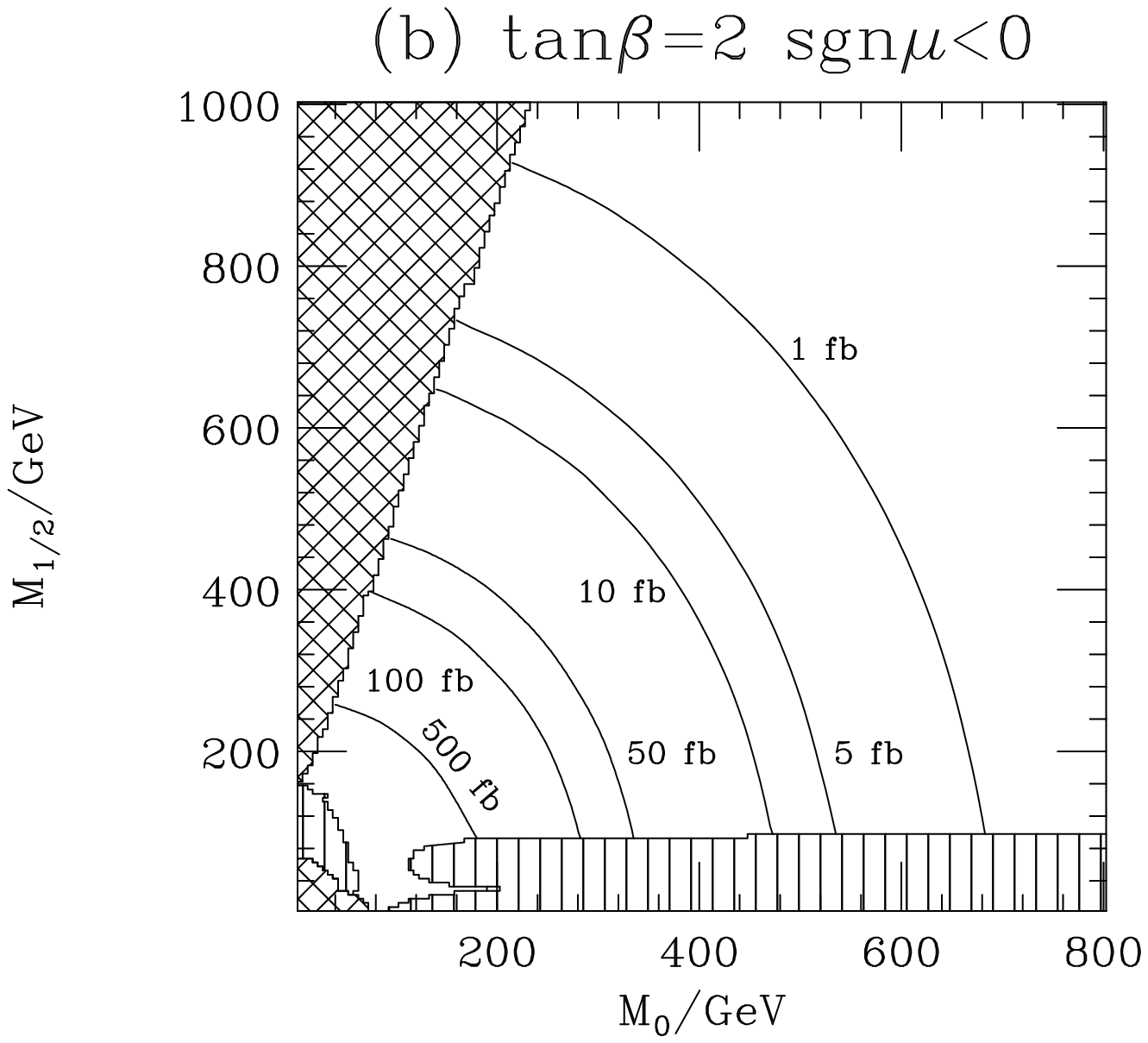}\\
\vskip 15mm
\includegraphics[angle=90,width=0.48\textwidth]{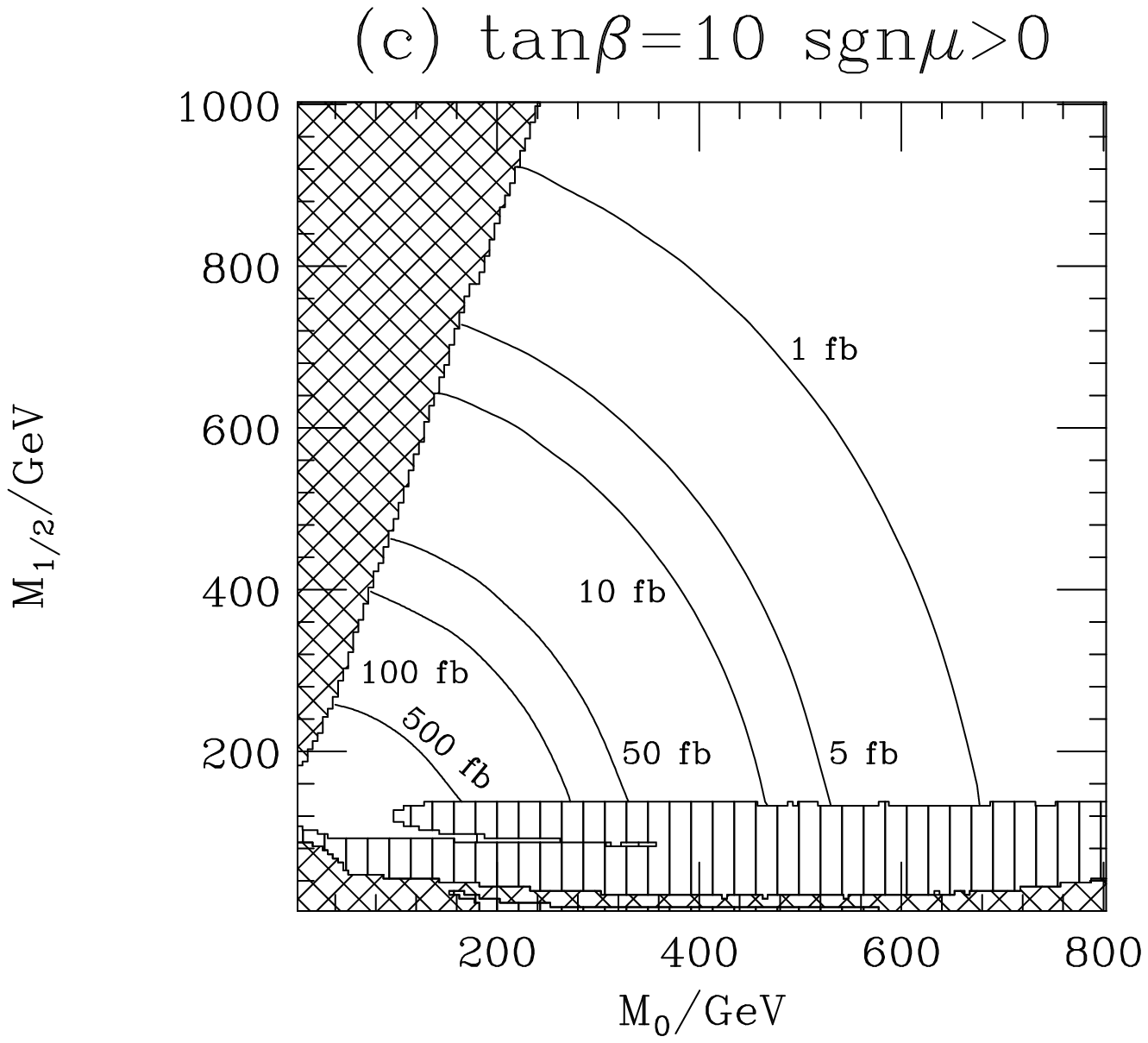}
\includegraphics[angle=90,width=0.48\textwidth]{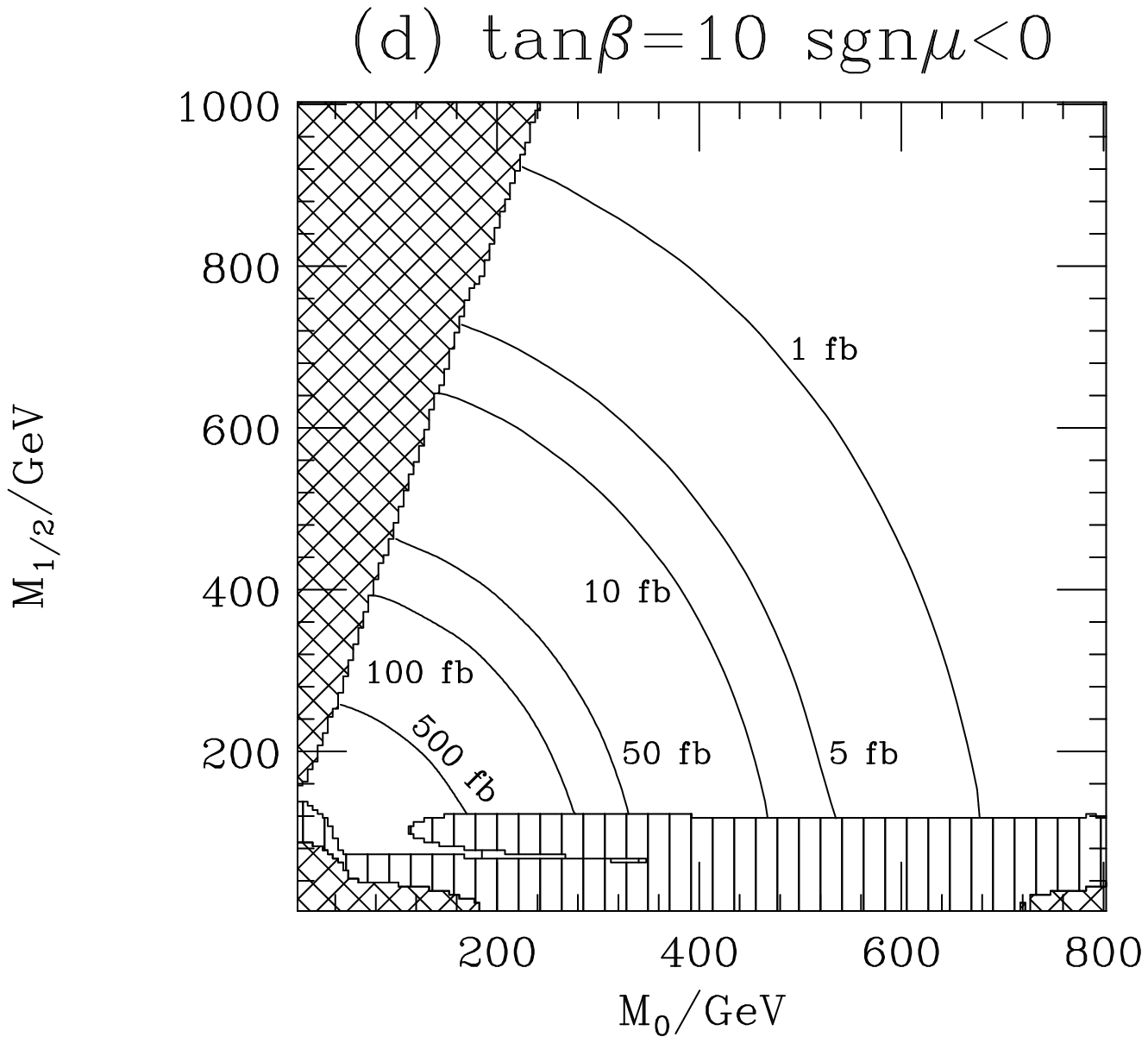}\\
\captionB{Cross section for the production of a resonant slepton followed by a
	supersymmetric gauge decay
 	 at the Tevatron in the $M_0$, $M_{1/2}$ plane.}
	{Contours showing the cross section for the production of a slepton
	 followed by a supersymmetric gauge decay at
	 Run II of the Tevatron
	 in the $M_0$, $M_{1/2}$ plane for $A_0=0\, \mr{\gev}$ 
	 and ${\lam'}_{211}=10^{-2}$ with different values
	 of $\tan\beta$ and $\sgn\mu$. The striped and hatched regions are
	 described in the caption of Fig.\,\ref{fig:SUSYmass}.} 
\label{fig:tevcross2}
\end{center}
\end{figure}
% End of the Figure %%%%%%%%%%%%%%%%%%%%%%%%%%%%%%%%%%%%%%%%%%%%%%%%%%%%%%%%%%

  The cross section for the production of a neutralino and a charged lepton,
  which is the dominant like-sign dilepton production mechanism,
  is shown in Fig.\,\ref{fig:tevcross} in the $M_0$, $M_{1/2}$ plane 
  with \mbox{$A_0=0\, \mr{\gev}$} and
  ${\lam'}_{211}=10^{-2}$ for two different values of $\tan\beta$ and both
  values of $\sgn\mu$. The total cross section for resonant slepton
  production followed by supersymmetric
  gauge decays is shown in Fig\,\ref{fig:tevcross2}. As can be seen from these
  figures, the total cross section closely follows the slepton 
  mass contours shown in Fig.\,\ref{fig:SUSYmass}, whereas the 
  neutralino--lepton cross section falls off more quickly at small
  $M_{1/2}$ where the charginos and heavier neutralinos can be produced.
  This cross section must be multiplied by the 
  acceptance, \ie the fraction of signal events which
  pass the cuts, to give the number of events detected in the experiment.

  We will first discuss the cuts applied to reduce the various Standard Model
  backgrounds and then present the discovery potential at the Tevatron if we
  only consider these backgrounds. This is followed by a discussion of the
  additional cuts needed to reduce the background from sparticle pair
  production.

\vskip 5mm
\noindent{\underline{Standard Model Backgrounds}}
\nopagebreak 
\vskip 5mm
\nopagebreak
  We have applied the following cuts to reduce the Standard Model backgrounds:
\begin{enumerate}

\item A cut requiring all the leptons to be in the central region of the
      detector, $|\eta|<2.0$.

\item	A cut on the transverse momentum of each of the like-sign leptons,
	$p_T^{\mr{lepton}} \geq 20\, \mr{\gev}$.
	This is the lowest cut we could apply given our parton level cut of
	$p_T^{\mr{parton}}=20\, \mr{\gev}$ for the $\mr{b\bar{b}}$ background.
	
\item 	An isolation cut on the like-sign leptons so that the
	transverse energy in a cone of radius
	$\Delta R = \sqrt{\Delta\phi^2+\Delta\eta^2} = 0.4$ about the
        direction of the lepton is less than $5\, \mr{\gev}$.

\item   We reject events with \mbox{$60\, \mr{\gev} < M_T < 85\, \mr{\gev}$}
	($c.f.$ Eqn.\,\ref{eqn:MTdef}).
	This cut is applied to both of the like-sign leptons.

\item   A veto on the presence of a lepton in the event with the same flavour
        but opposite charge as either of the leptons in the like-sign
        pair if the lepton has  $p_T>10\, \mr{\gev}$  
	and passes the same isolation cut as the like-sign leptons.

\item   A cut on the missing transverse energy, 
	$\not\!\!\!E_T<20\, \mr{\gev}$.
	In our analysis we have assumed that the missing transverse energy 
	is solely the due to the momenta of the neutrinos produced. 
\end{enumerate}

\begin{figure}
\begin{center}
\includegraphics[angle=90,width=0.48\textwidth]{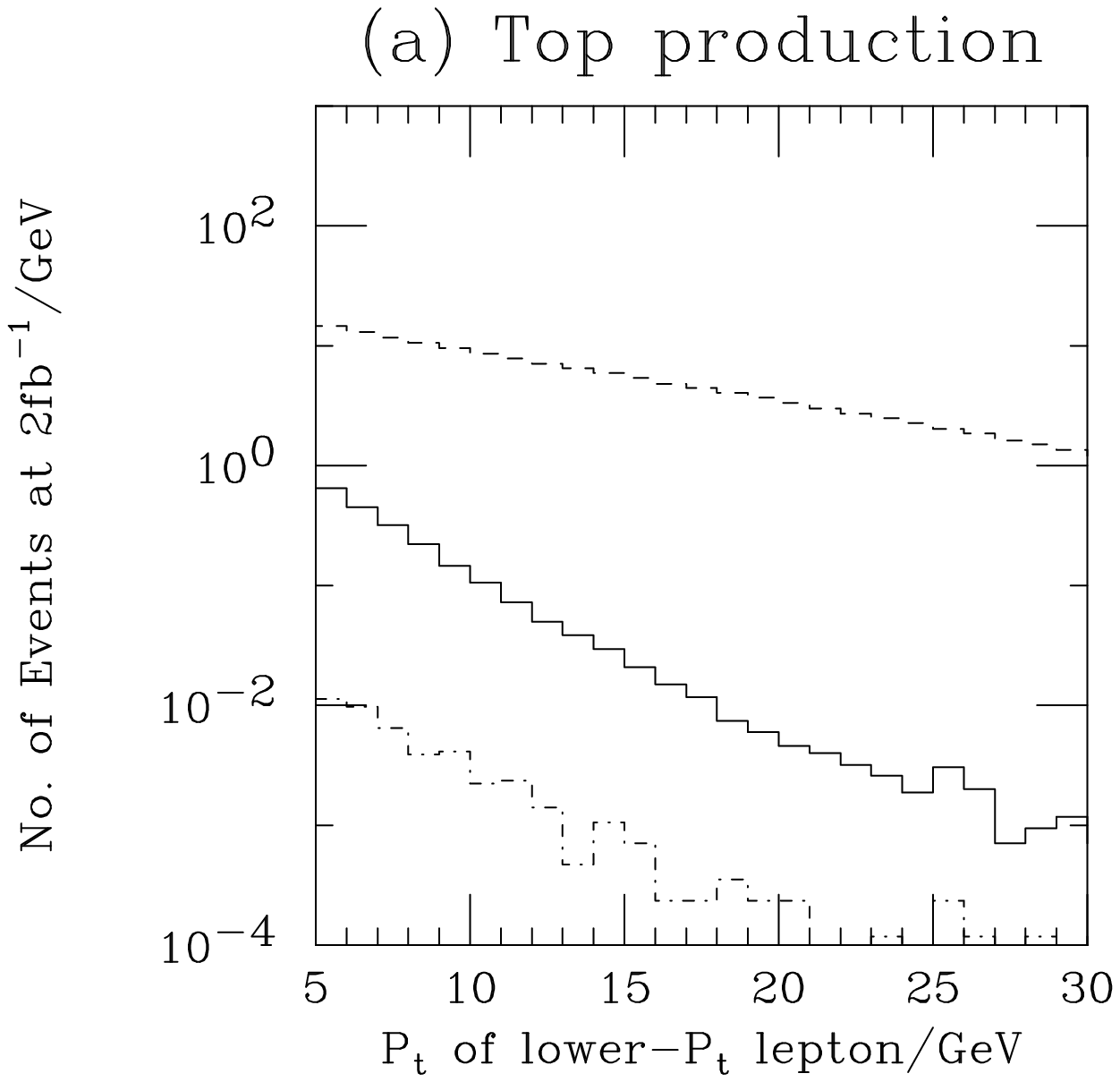}
\hfill
\includegraphics[angle=90,width=0.48\textwidth]{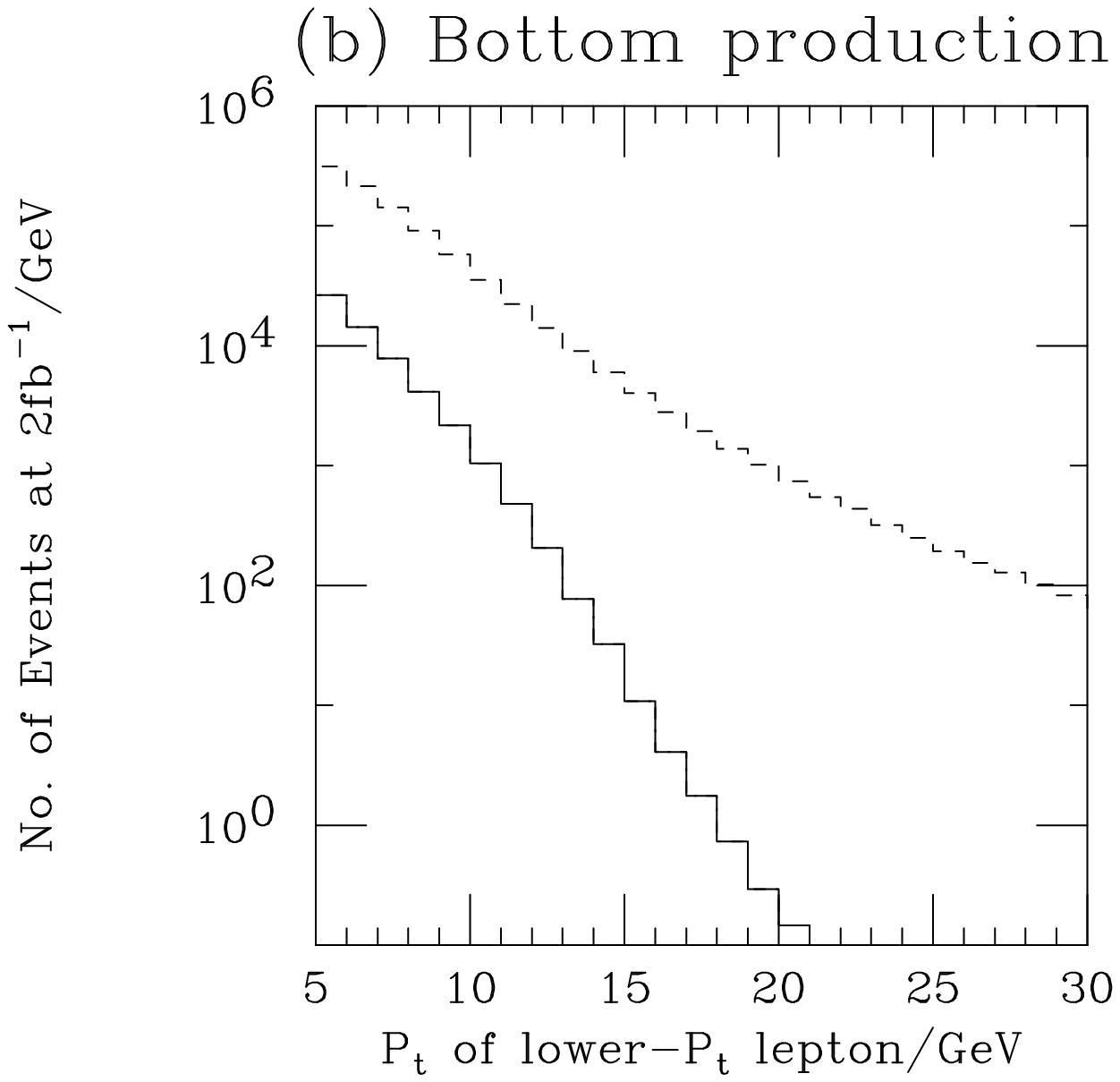}\\
\captionB{Effect of the isolation cuts on the $\mr{t\bar{t}}$ and 
	 $\mr{b\bar{b}}$ backgrounds at the Tevatron.}
	{Effect of the isolation cuts on the $\mr{t\bar{t}}$ and 
	 $\mr{b\bar{b}}$ backgrounds at Run II of
	the Tevatron. The dashed line gives the background before any cuts and
	the solid line shows the effect of the isolation cut described in the
	text. The dot-dash line gives the effect of all the cuts, including
	the cut on the number of jets (for the $\mr{b\bar{b}}$ background 
 	this is  indistinguishable from the solid line). As a parton-level cut
	of $20\, \mr{\gev}$  was used in simulating the $\mr{b\bar{b}}$
        background the results below $20\, \mr{\gev}$  for the lepton $p_T$
	do not correspond to the full number of background events.
	The distributions have been normalized
 	to an  integrated luminosity of $2\  \mr{fb}^{-1}$.} 
\label{fig:tevheavyiso}
\end{center}
\end{figure}

  The first two cuts are designed to reduce the background from heavy
  quark production, which is the major source of background before any
  cuts. Fig.\,\ref{fig:tevheavyiso} shows that the cut on the transverse
  momentum, $p_T>20\, \mr{\gev}$,  reduces the background by several
  orders of magnitude and the addition of the isolation cut reduces this
  background to less than one event at Run II of the Tevatron.

\begin{figure}
\begin{center}
\includegraphics[angle=90,width=0.48\textwidth]{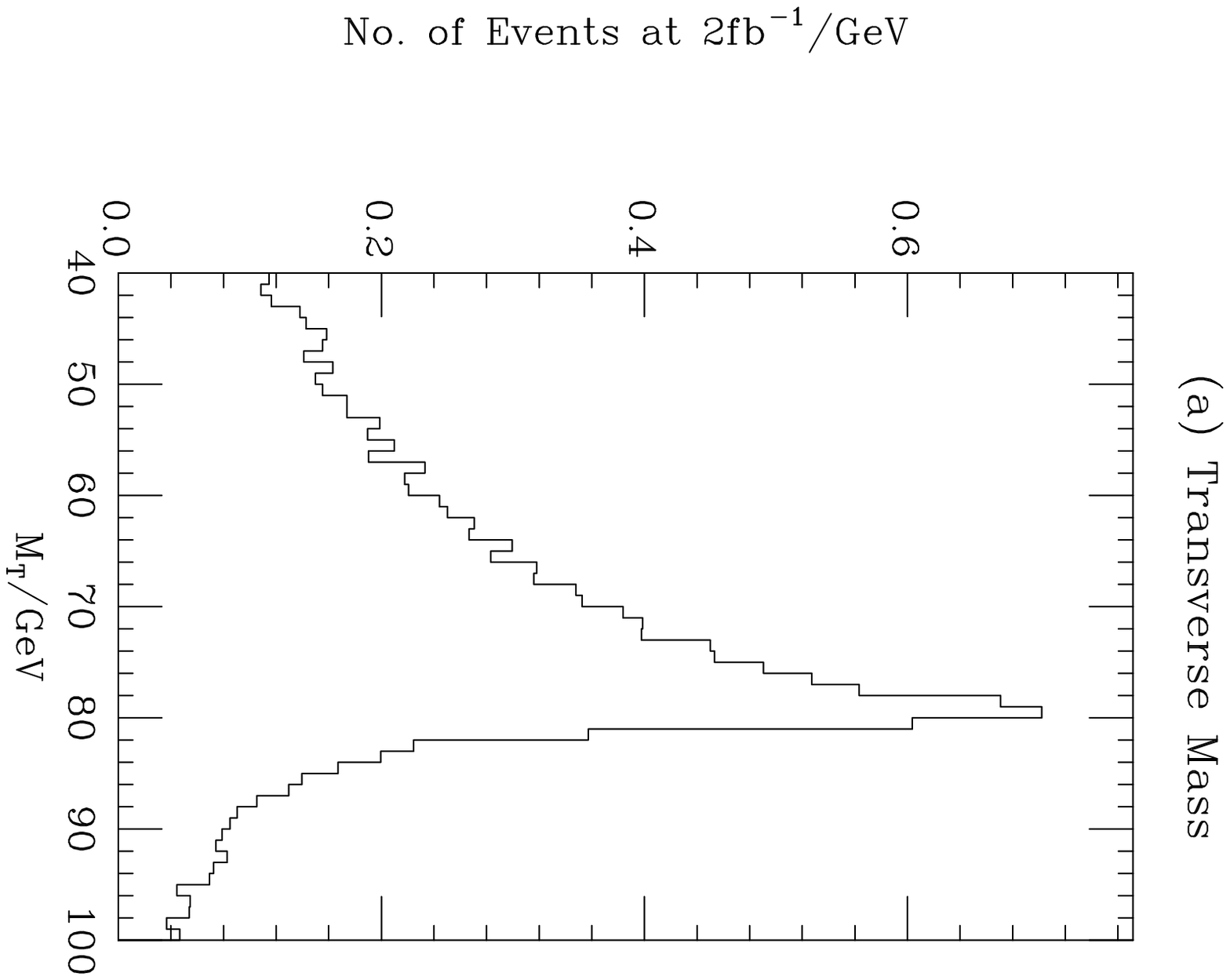}
\hfill
\includegraphics[angle=90,width=0.48\textwidth]{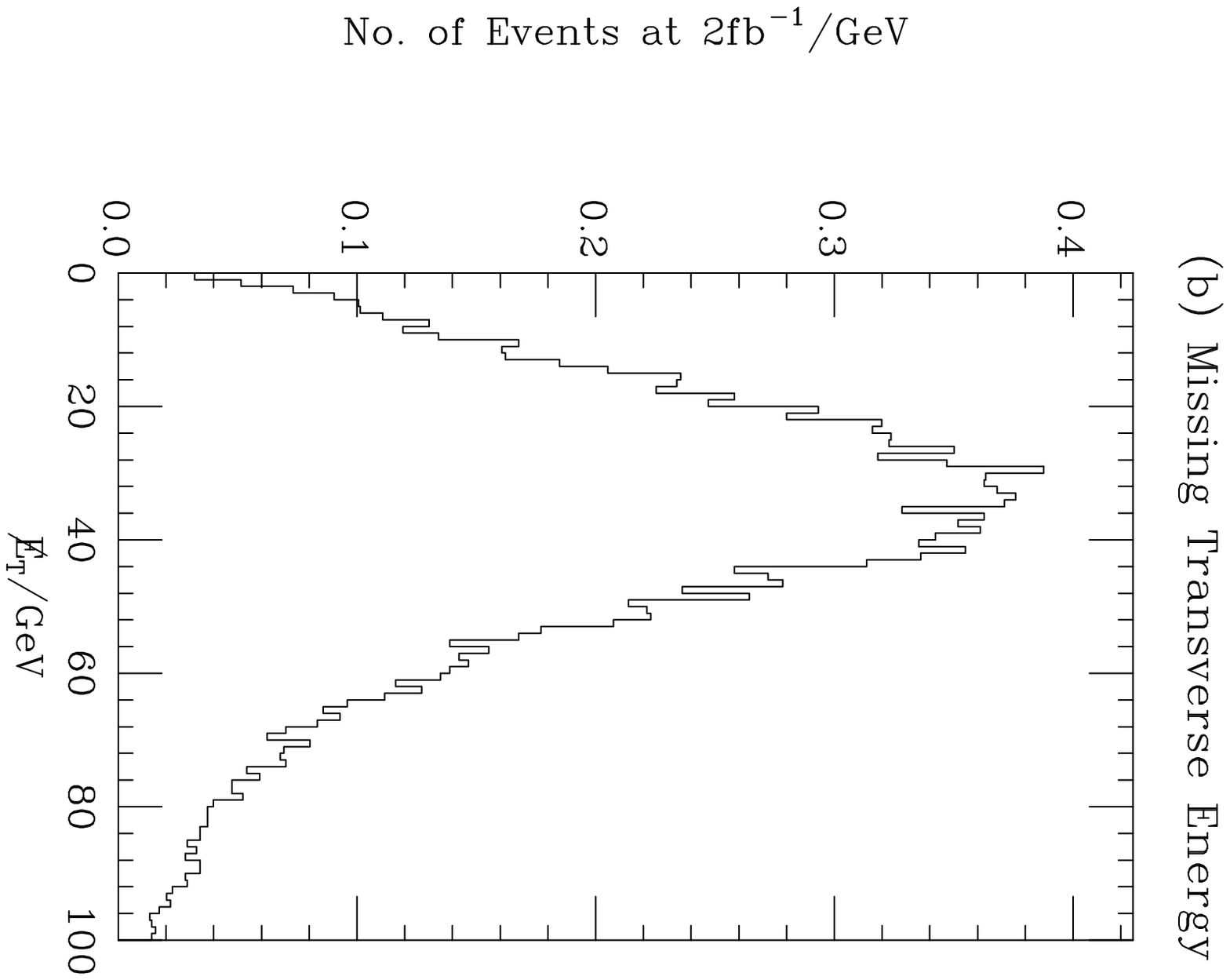}\\
\captionB{$M_T$ and \met\  in WZ events at the Tevatron.}
	{The transverse mass and missing transverse energy in WZ events
	 at Run II of the Tevatron. 
	The distributions are normalized to
	an integrated luminosity of $2\  \mr{fb}^{-1}$.} 
\label{fig:tevemiss}
\end{center}
%\end{figure}
%% End of the Figure %%%%%%%%%%%%%%%%%%%%%%%%%%%%%%%%%%%%%%%%%%%%%%%%%%%%%%%%%
%%%%%%%%%%%%%%%%%%%%%%%%%%%%%%%%%%%%%%%%%%%%%%%%%%%%%%%%%%%%%%%%%%%%%%%%%%%%%%
%%
%%  Figure containing the effect of the cuts on WZ and ZZ at the Tevatron
%%
%\begin{figure}
\begin{center}
\includegraphics[angle=90,width=0.48\textwidth]{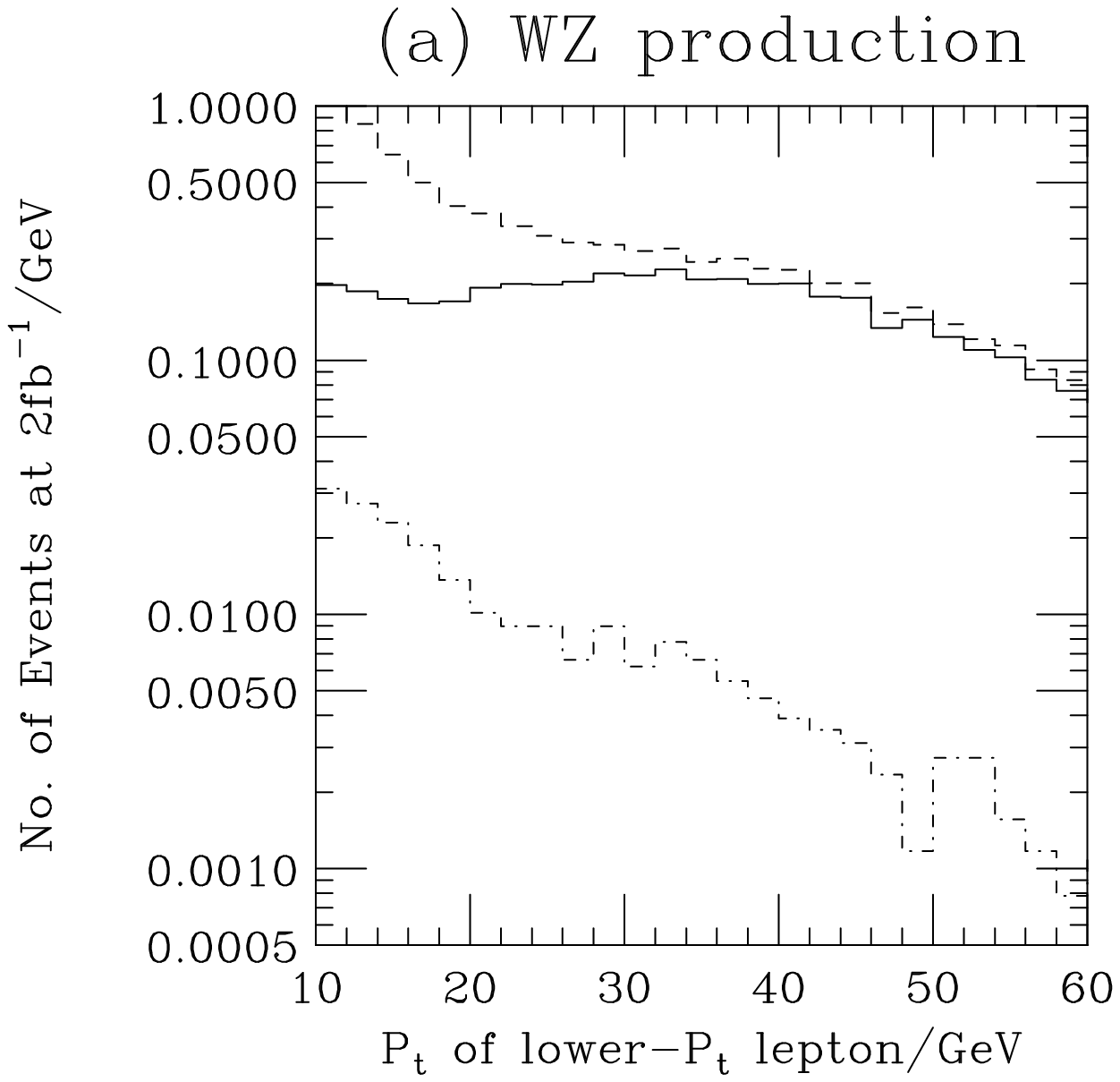}
\hfill
\includegraphics[angle=90,width=0.48\textwidth]{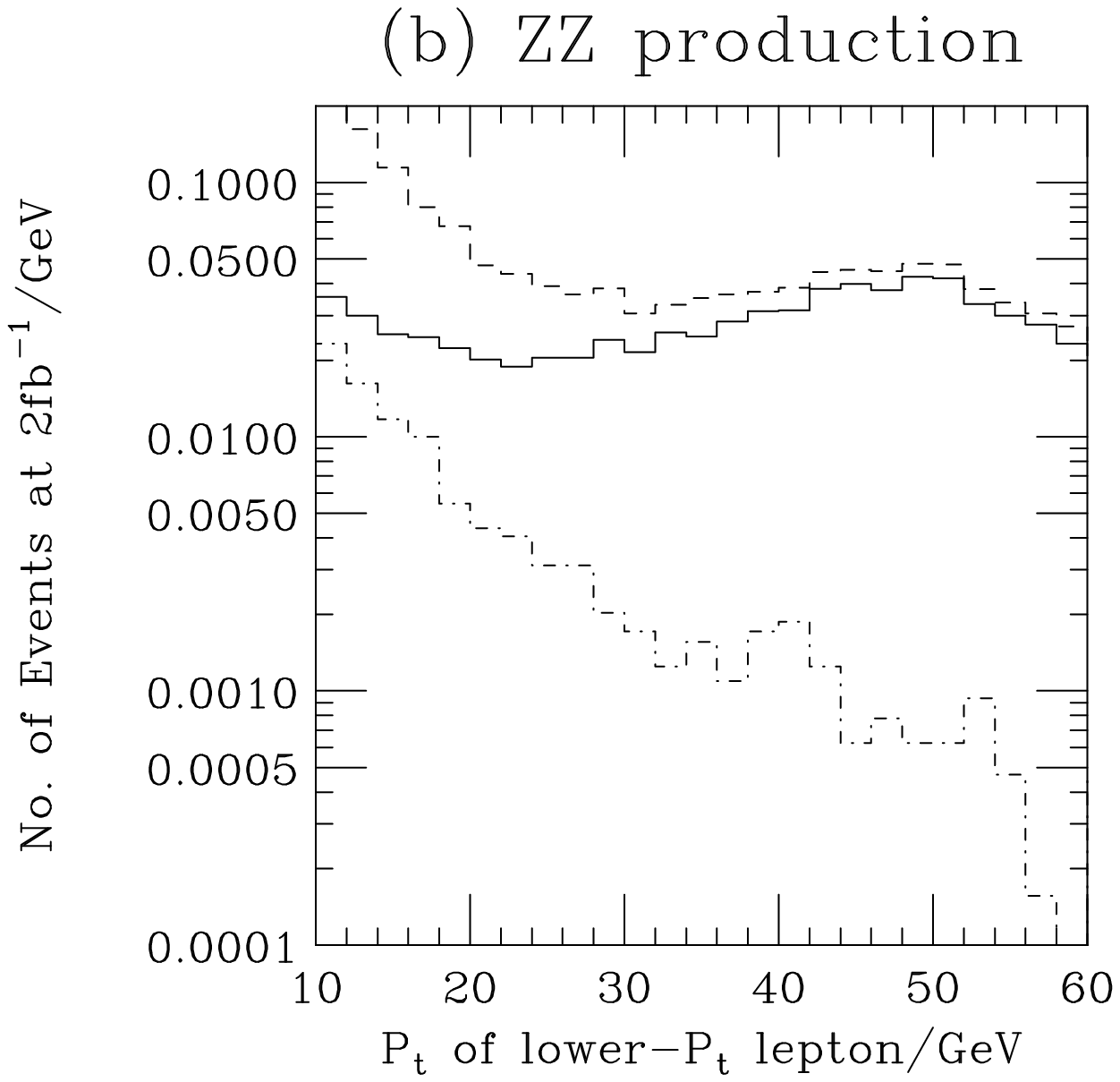}\\
\captionB{Effect of the isolation cuts on the WZ and ZZ backgrounds at
	the Tevatron.}
	{Effect of the isolation cuts on the WZ and ZZ backgrounds at Run II
	of
	the Tevatron. The dashed line gives the background before any cuts and
	the solid line shows the effect of the isolation cut described in the
	text. The dot-dash line gives the effect of all the cuts, including
	the cut on the number of jets. The distributions are
	 normalized to an integrated luminosity of 
	$2\  \mr{fb}^{-1}$.} 
\label{fig:tevgaugeiso}
\end{center}
\end{figure}

  The remaining cuts reduce the background from gauge boson pair
  production which dominates the Standard Model background after the
  imposition of the isolation and $p_T$ cuts. Fig.\,\ref{fig:tevemiss}a
  shows that the cut on the transverse mass, \ie removing the region
  $60\, \mr{\gev} < M_T < 85\, \mr{\gev}$, for each of the like-sign leptons
  will reduce the background
  from WZ production, which is the largest of the gauge boson pair
  production backgrounds. Similarly the cut on the missing transverse
  energy, $\not\!\!E_T<20\, \mr{\gev}$, 
  will significantly reduce the background from WZ production.
  This is shown in Fig.\,\ref{fig:tevemiss}b. The effect of these cuts is
  shown in Fig.\,\ref{fig:tevgaugeiso}.  Our simulations do not
  include $\mr{W\gamma}$ production which was recently found to be
  a major source of background to like-sign
  dilepton production in the MSSM \cite{Matchev:1999yn}. However, we
  would expect this to be less important here due to the different cuts
  we have applied. In particular, in the analysis of \cite{Matchev:1999yn} a
  cut on the invariant mass of OSSF lepton pairs was imposed to reduce the
  background from Z production, rather than the veto on the presence of OSSF
  leptons which we have used. The veto and missing transverse energy cut
  will reduce the number of events from
  $\mr{W\gamma}$ production while the cut on the invariant mass will not
  suppress this background.

%%%%%%%%%%%%%%%%%%%%%%%%%%%%%%%%%%%%%%%%%%%%%%%%%%%%%%%%%%%%%%%%%%%%%%%%%%%%%%
%
%  Figure containing the effect of the cuts on top and bottom at the Tevatron
%
% End of the Figure %%%%%%%%%%%%%%%%%%%%%%%%%%%%%%%%%%%%%%%%%%%%%%%%%%%%%%%%%%

%%%%%%%%%%%%%%%%%%%%%%%%%%%%%%%%%%%%%%%%%%%%%%%%%%%%%%%%%%%%%%%%%%%%%%%%%%%%%%
%
%  Figure containing the transverse mass and missing ET at the Tevatron
%
% End of the Figure %%%%%%%%%%%%%%%%%%%%%%%%%%%%%%%%%%%%%%%%%%%%%%%%%%%%%%%%%%

  The effect of all these cuts on the background is given in 
  Table~\ref{tab:tevback}.
  While the dominant background is from WZ production, the dominant
  contribution to the error comes from $\mr{b\bar{b}}$ production. 
  This can only be reduced with a significantly more elaborate simulation.

  We also need to calculate the acceptance of these cuts for the signal.
  To estimate the acceptance of the cuts we simulated  twenty thousand events
  at each of one hundred points in the $M_0$, $M_{1/2}$ plane.
  The acceptance was then interpolated between the points and multiplied by
  the cross section to give the number of
  signal events passing the cuts.
  This can be used to find the discovery potential
  by comparing the number of signal events with a $5\sigma$ 
  statistical fluctuation of the background.

\begin{figure}
\begin{center}
\includegraphics[angle=90,width=0.48\textwidth]{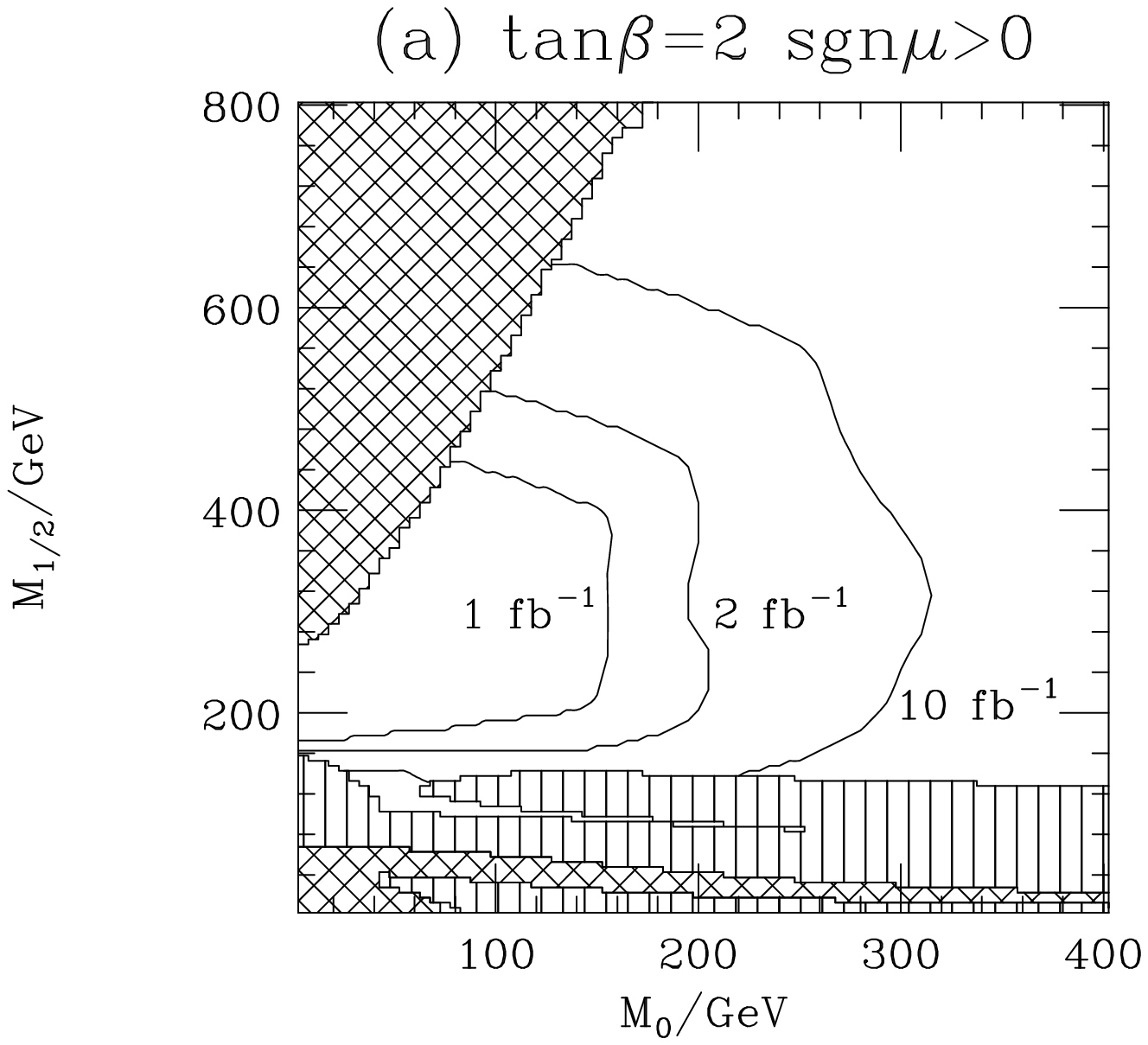}
\hfill
\includegraphics[angle=90,width=0.48\textwidth]{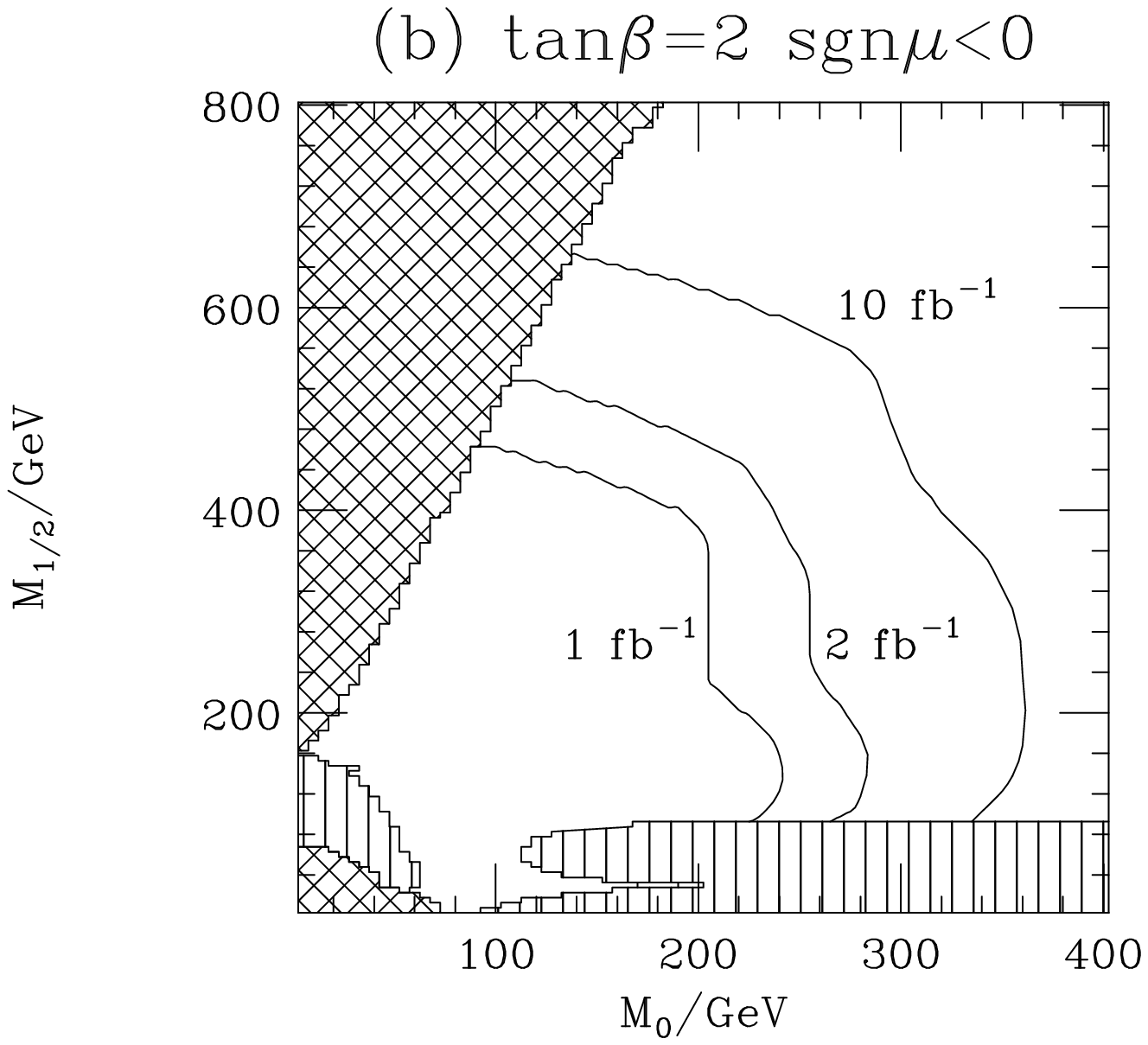}\\
\vskip 15mm
\includegraphics[angle=90,width=0.48\textwidth]{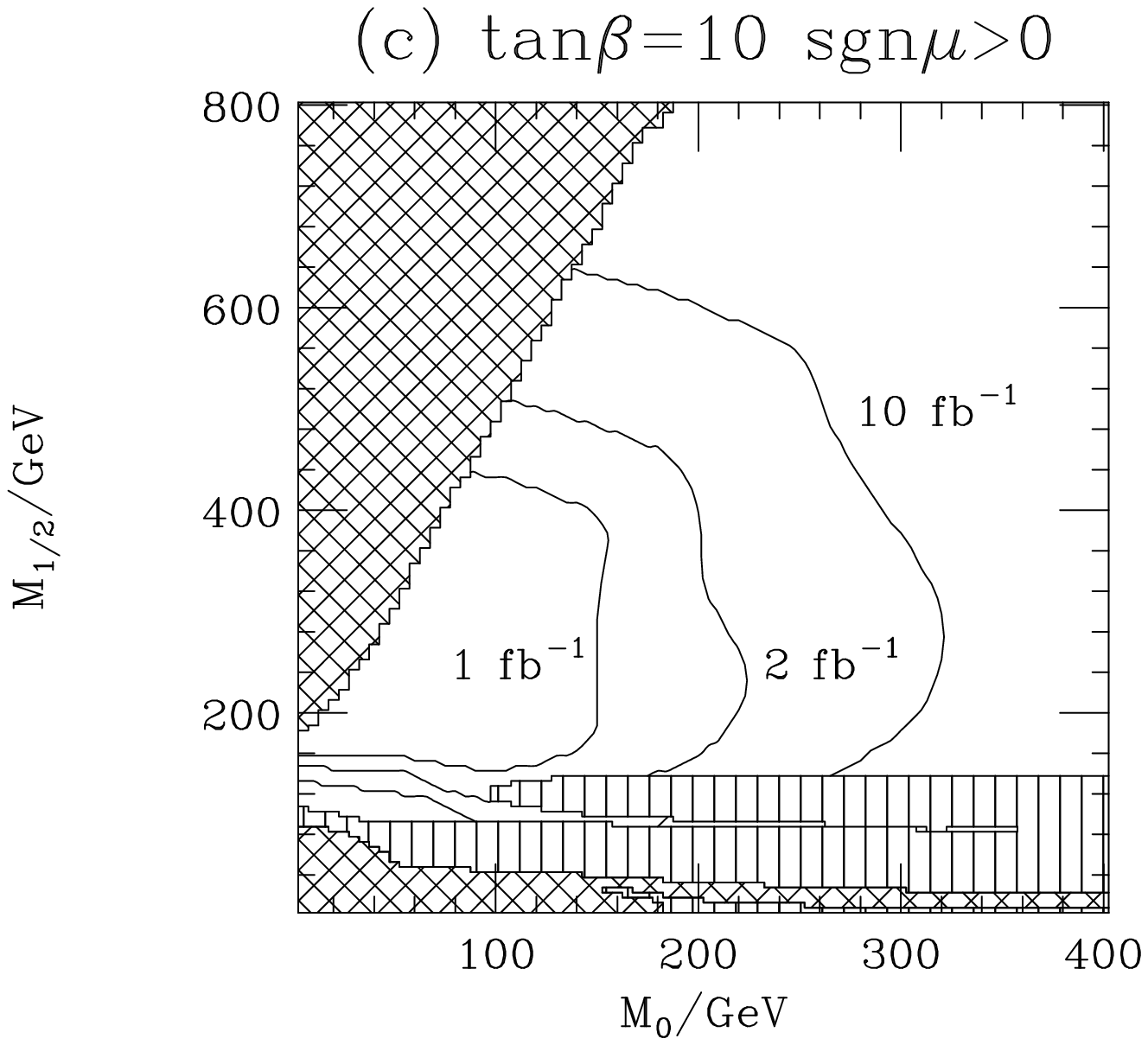}
\hfill
\includegraphics[angle=90,width=0.48\textwidth]{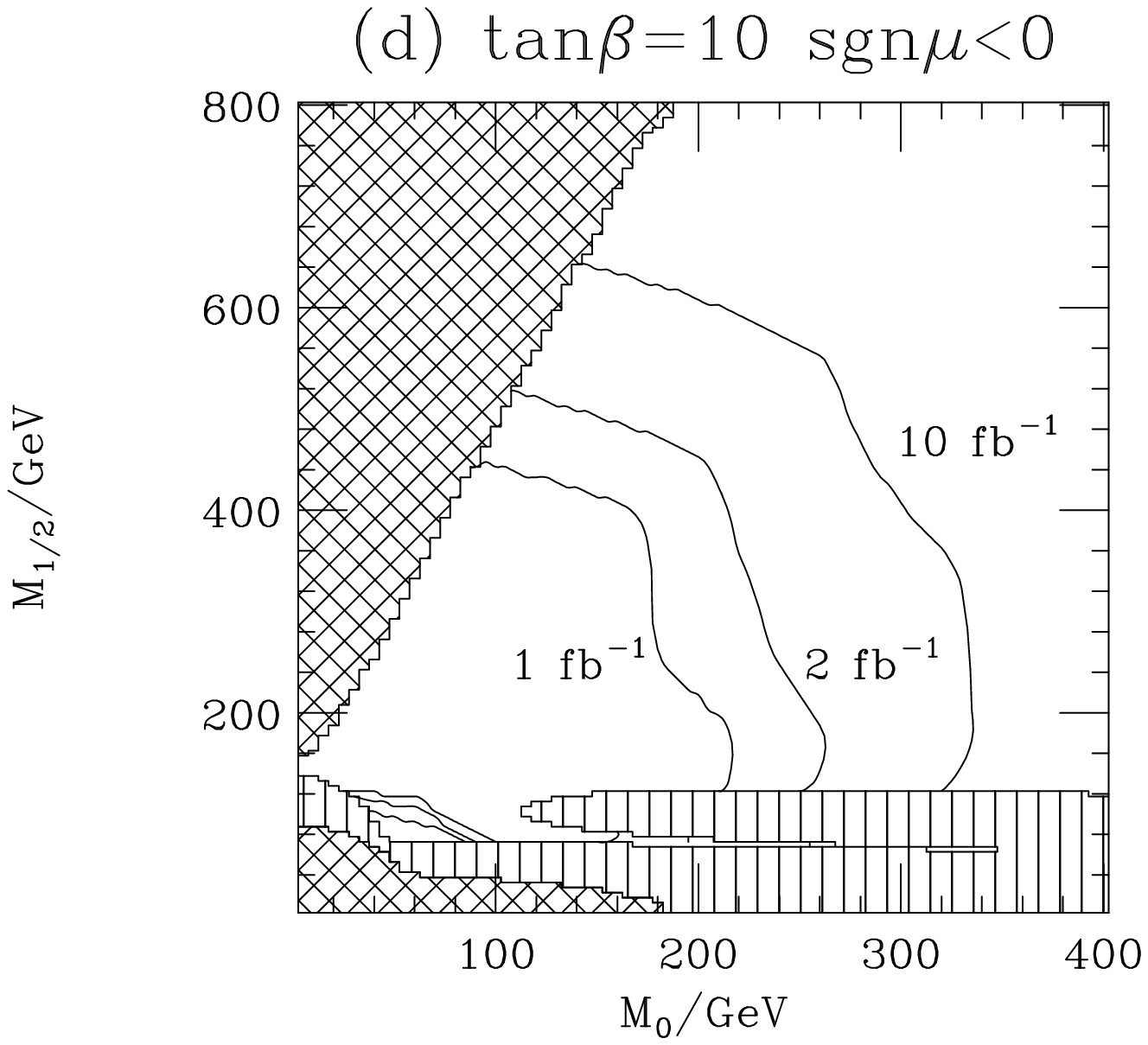}\\
\captionB{Discovery potential at the Tevatron for the Standard Model
          backgrounds
	in the $M_0$, $M_{1/2}$ plane with different integrated luminosities.}
	{Contours showing the discovery potential of the Tevatron in the
	 $M_0$,
	 $M_{1/2}$ plane for ${\lam'}_{211}=10^{-2}$ and $A_0=0\, \mr{\gev}$.
	 These contours are a $5\sigma$ excess of the signal above the
	 background. Here we have imposed the cuts on the isolation and $p_T$
  	 of the leptons, the transverse mass and the missing transverse energy
	 described in the text, and a veto on the presence of OSSF leptons.
	 We have only considered the Standard Model background. 
	The striped and hatched regions are
	 described in the caption of Fig.\,\ref{fig:SUSYmass}.} 
\label{fig:tevSMnojet}
\end{center}
\end{figure}

\begin{figure}
\begin{center}
\includegraphics[angle=90,width=0.48\textwidth]{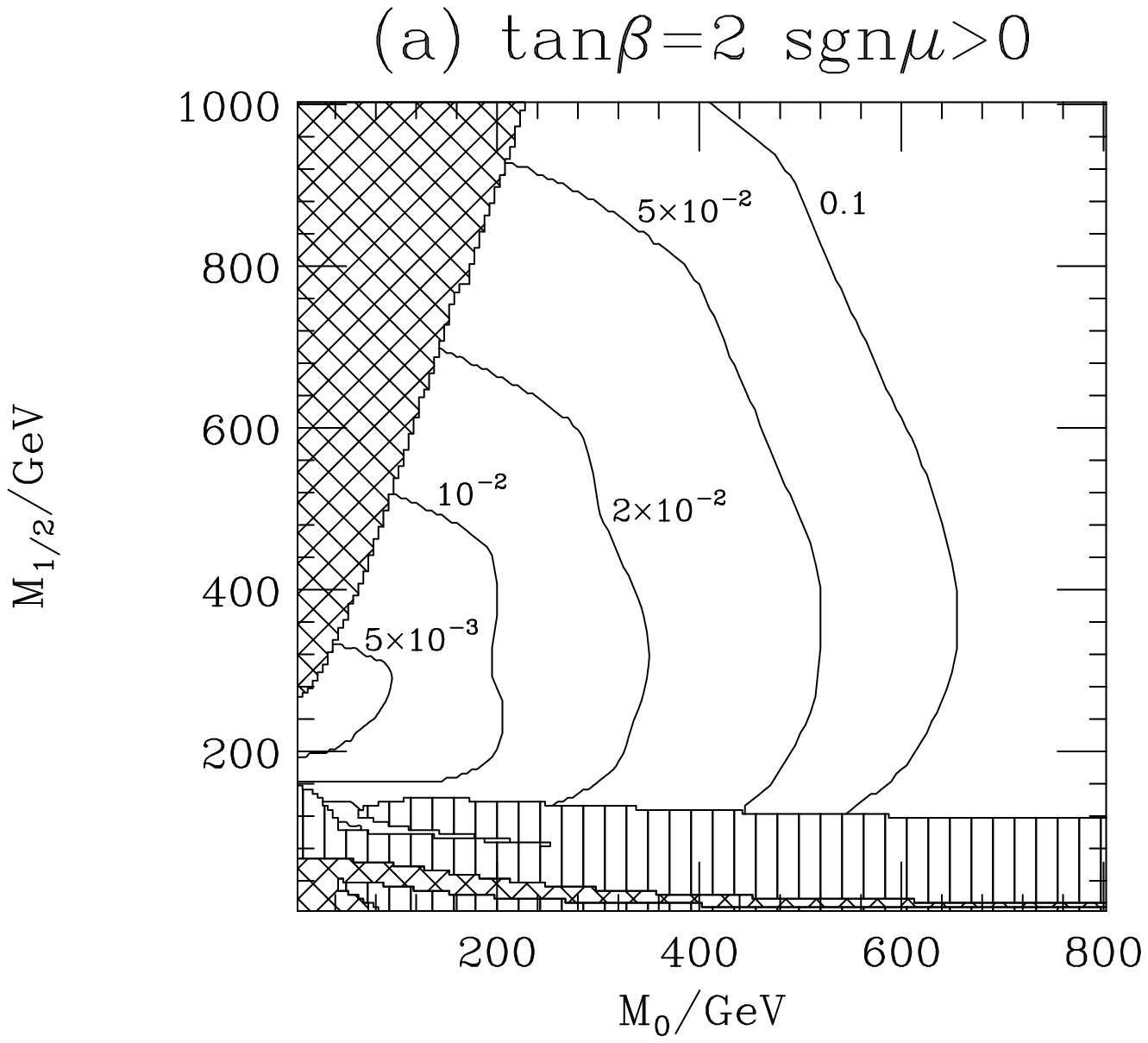}
\hfill
\includegraphics[angle=90,width=0.48\textwidth]{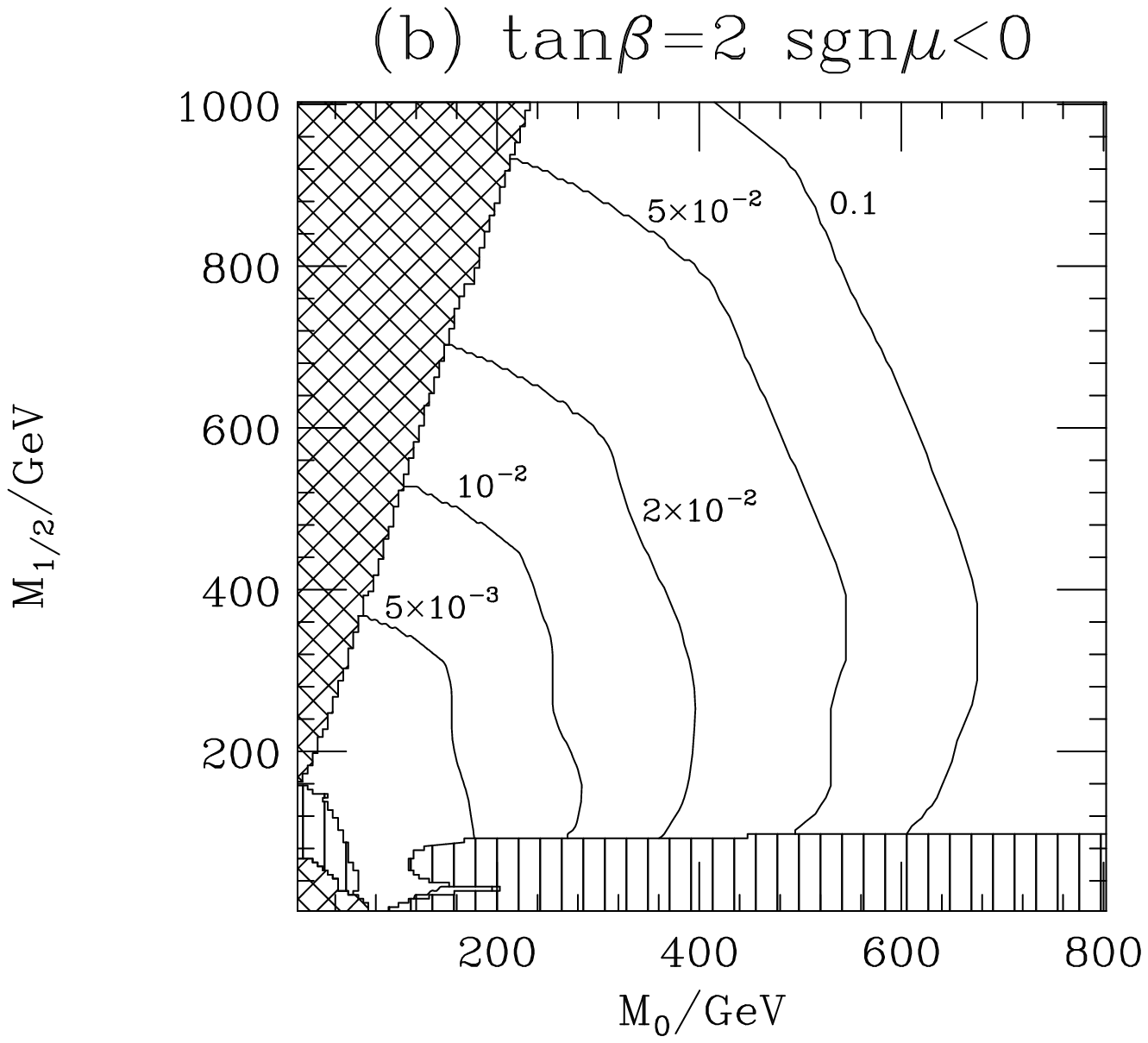}\\
\vskip 15mm
\includegraphics[angle=90,width=0.48\textwidth]{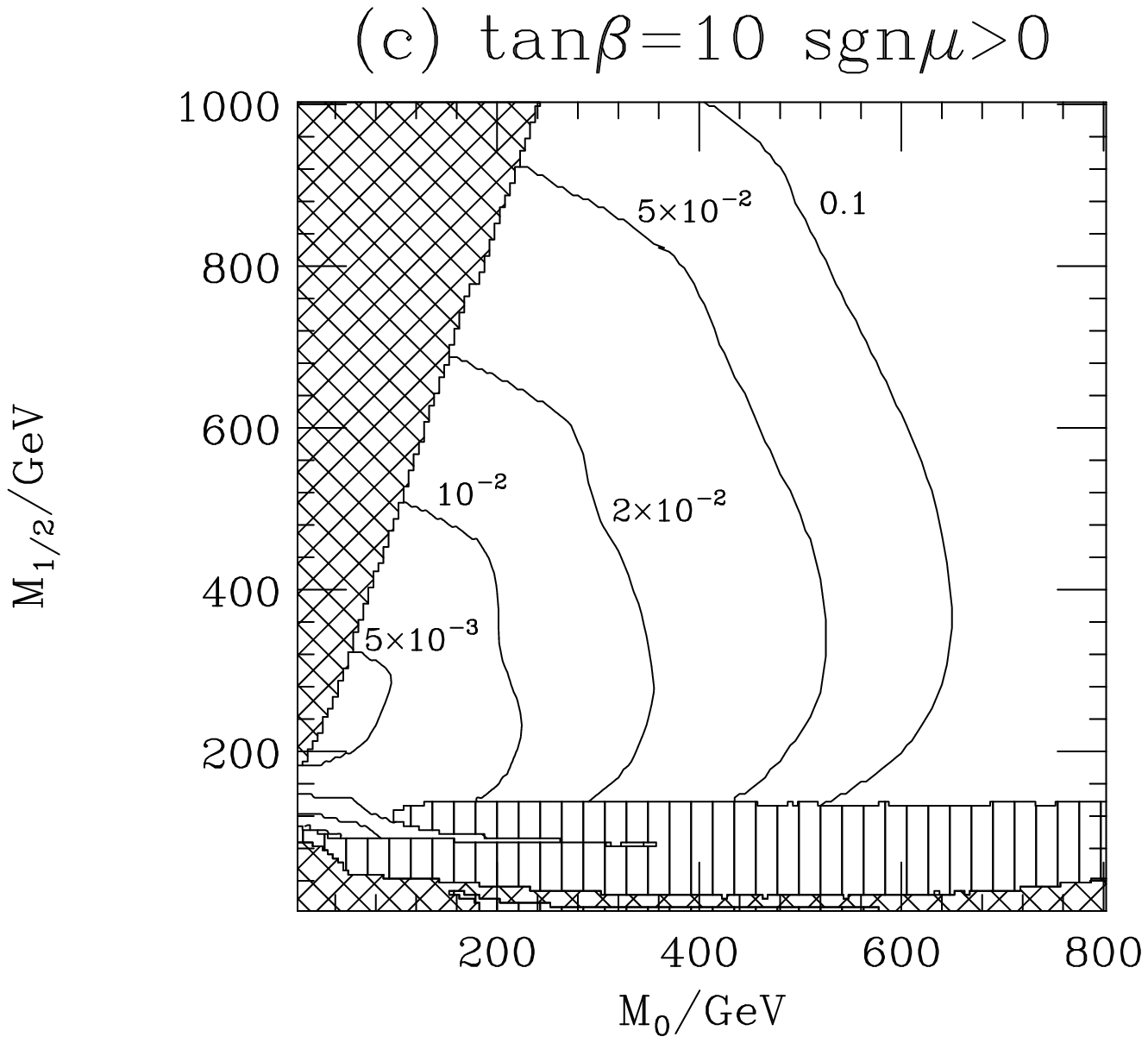}
\hfill
\includegraphics[angle=90,width=0.48\textwidth]{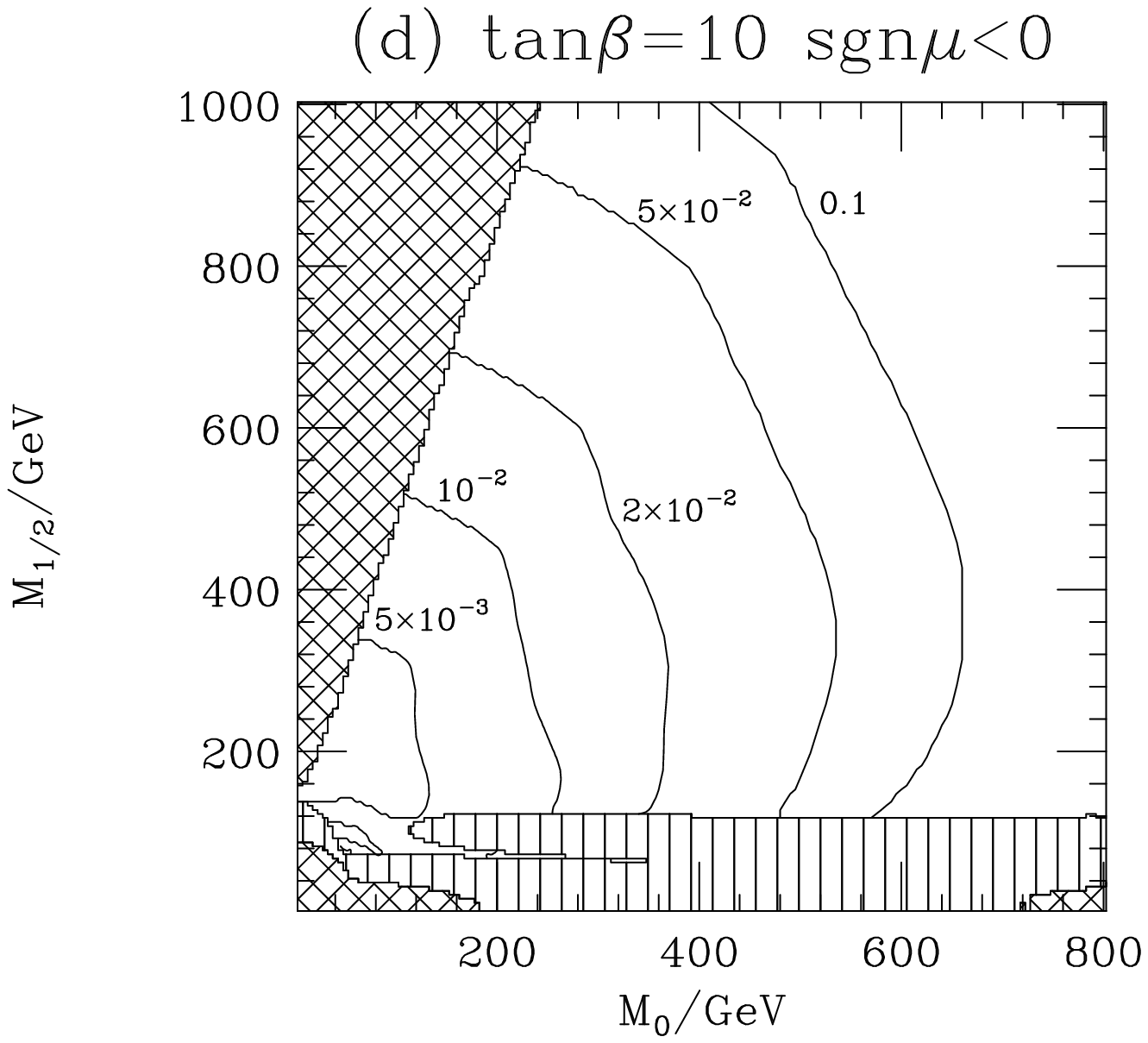}\\
\captionB{Discovery potential at the Tevatron for the Standard Model
	 backgrounds in the $M_0$, $M_{1/2}$ plane with different values of
	 ${\lam'}_{211}$.}
	{Contours showing the discovery potential of the Tevatron in the
	 $M_0$,
	 $M_{1/2}$ plane for $A_0=0\, \mr{\gev}$   and 
         $2\ \mr{fb}^{-1}$ integrated luminosity for different values
	 of ${\lam'}_{211}$.
	 These contours are a $5\sigma$ excess of the signal above the
	 background. Here we have imposed the cuts on the isolation and $p_T$
  	 of the leptons, the transverse mass and the missing transverse energy
	 described in the text, and a veto on the presence of OSSF leptons.
	 We have only considered the Standard Model background.
	 The striped and hatched regions are
	 described in the caption of Fig.\,\ref{fig:SUSYmass}.} 
\label{fig:tevSMnojetb}
\end{center}
\end{figure}

  Fig.\,\ref{fig:tevSMnojet} shows the discovery potential,
  for different integrated luminosities and a fixed
  value of the coupling ${\lam'}_{211}=10^{-2}$, if we only consider
  the Standard Model backgrounds and apply the cuts we described to suppress
  these backgrounds. Fig.\,\ref{fig:tevSMnojetb} shows the effect of varying
  the \rpv\  coupling for $2\  \mr{fb}^{-1}$ integrated luminosity with the
  same assumptions.

  We have taken a conservative approach where the background
  is taken to be one standard deviation above the central value,
  \ie we take the background to be 0.59 events. Due to the 
  small number of events we must use Poisson statistics. For
  the Standard Model background given in Table~\ref{tab:tevback}, 7 events
  corresponds to the same probability as a  $5 \sigma$ statistical
  fluctuation for a
  Gaussian distribution.

%%%%%%%%%%%%%%%%%%%%%%%%%%%%%%%%%%%%%%%%%%%%%%%%%%%%%%%%%%%%%%%%%%%%%%%%%%%%%%
%
%  Table giving the backgrounds for the Tevatron
%
\begin{table}
\begin{center}
\begin{tabular}{|c|c|c|c|c|}
\hline
	& \multicolumn{4}{c|}{Number of Events} \\
\cline{2-5}
 		& 	     	  &  		    & After isolation,&\\
Background      & After $p_T$ cut & After isolation & $p_T$, $M_T$, \met\  
cuts & 
		 After all cuts \\
 process        &                 & and $p_T$ cuts  & and OSSF lepton & \\
	        & 		  &		    & veto.  & \\
\hline
WW 		& $0.23\pm0.02$			& $0.0\pm0.003$ 
	        & $0.0\pm0.003$	 		& $0.0\pm0.003$ \\
\hline		                
WZ 		& $9.96\pm0.09$			& $7.93\pm0.08$
		& $0.21\pm0.01$ 		& $0.21\pm0.01$ \\
\hline				                
ZZ 		& $2.05\pm0.03$			& $1.61\pm0.02$	
		& $0.069\pm0.005$ 		& $0.069\pm0.005$ \\
\hline		                
$\mr{t\bar{t}}$ & $34.1\pm1.6$ 			& $0.028\pm0.002$
		& $0.0032\pm0.0006$ 		& $0.0016\pm0.0004$ \\
\hline		                
$\mr{b\bar{b}}$ & $(3.4\pm1.3)\times10^3$	& $0.15\pm0.16$
		& $0.15\pm0.16$ 		& $0.15\pm0.16$ \\
\hline	                
Single top 	& $1.77\pm0.01$ 		& $0.0014\pm0.0003$	
		& $0.0001\pm0.0001$ 		& $0.0001\pm0.0001$ \\
\hline\hline	                
Total 	 	& $(3.4\pm1.3)\times10^3$	& $9.72\pm0.18$ 
		& $0.43\pm0.16$ 		& $0.43\pm0.16$ \\
\hline
\end{tabular}
\end{center}
\captionB{Backgrounds to like-sign dilepton production at the Tevatron.}
	{Backgrounds to like-sign dilepton production at Run II of the
         Tevatron. The numbers of events
	are based on an integrated luminosity of $2\  \mr{fb}^{-1}$.
	We have calculated an error on the cross section by varying the
	scale between half and twice the hard scale, apart from the
	gauge boson pair production cross section where we do not have this
	information and the effect of varying the scale is expected
	to be small
	anyway. The error on the number of events is the error on
	the cross section and the statistical error from the Monte
	Carlo simulation added in quadrature. If no events passed
 	the cut the statistical error was taken to be the same
 	as if one event had passed the cuts.}
\label{tab:tevback}
\end{table}
% End of the Table %%%%%%%%%%%%%%%%%%%%%%%%%%%%%%%%%%%%%%%%%%%%%%%%%%%%%%%%%%%

  For small couplings there are regions, for low $M_{1/2}$, which cannot
  be observed even for small smuon masses. For larger couplings, however,
  we can probe masses of up to $430\,(500)\, \mr{\gev}$  for a coupling
  ${\lam'}_{211}=0.05$ with 2\,(10)~$\mr{fb}^{-1}$ integrated luminosity.
  Masses of up to $520\,(600)\, \mr{\gev}$  can be observed for a coupling of
  ${\lam'}_{211}=0.1$ with 2\,(10)~$\mr{fb}^{-1}$ integrated luminosity.

%%%%%%%%%%%%%%%%%%%%%%%%%%%%%%%%%%%%%%%%%%%%%%%%%%%%%%%%%%%%%%%%%%%%%%%%%%%%%%
%
%  Figure containing the signal SM background, no jet cut at the Tevatron
%
% End of the Figure %%%%%%%%%%%%%%%%%%%%%%%%%%%%%%%%%%%%%%%%%%%%%%%%%%%%%%%%%%
%%%%%%%%%%%%%%%%%%%%%%%%%%%%%%%%%%%%%%%%%%%%%%%%%%%%%%%%%%%%%%%%%%%%%%%%%%%%%%

%
%  Figure containing the signal SM background, no jet cut at the Tevatron
%
% End of the Figure %%%%%%%%%%%%%%%%%%%%%%%%%%%%%%%%%%%%%%%%%%%%%%%%%%%%%%%%%%

  We have neglected the non-physics background. This mainly comes from fake 
  leptons in W production. The cuts we have applied to reduce the gauge boson
  pair production backgrounds, in particular the cuts on the missing
  transverse
  energy and the transverse mass, will significantly reduce this
  background. It was noted in \cite{Matchev:1999nb} that the cross
  section
  falls extremely quickly with the $p_T$ of the fake lepton, and hence the
  large $p_T$ cut we have imposed will suppress this background. 
  A proper treatment of the non-physics background requires a simulation of 
  the detector and this is beyond the scope of this work.

  In Figs.\,\ref{fig:tevSMnojet} and \ref{fig:tevSMnojetb} the background from
  sparticle pair production is neglected. This is
  reasonable in an initial search where presumably an experiment would be
  looking for an excess of like-sign dilepton pairs, rather than worrying
  about
  precisely which model was giving the excess. If such an
  excess were observed,
  it would then be necessary to establish which physical processes were
  producing
  the excess. In the \rpv\  MSSM there are two possible mechanisms which
  could produce such an excess: either resonant sparticle production; or
  sparticle pair production followed by the decay of the LSP.
  We will now consider additional cuts which will suppress the
  background to resonant slepton production 
 from sparticle pair production and hopefully allow these two 
  scenarios to be distinguished.

\vskip 5mm  
\noindent{\underline{SUSY Backgrounds}}
\nopagebreak
\vskip 5mm
\nopagebreak
  We have seen that by imposing cuts on the transverse momentum and isolation
  of the like-sign dileptons, the missing transverse energy, the transverse 
 mass and the presence of OSSF leptons the Standard Model backgrounds can be 
  significantly reduced. However a significant
  background from sparticle pair production still remains.
  We therefore imposed the following additional cut to 
  reduce this background:
\begin{itemize}
\item 	Vetoing all events when there are more than two jets each with
	$p_T>20\, \mr{\gev}$. 
\end{itemize}
  While this cut only slightly reduces the signal it dramatically reduces the
  background from sparticle pair production. We performed a scan of the
  SUGRA parameter space at the four values of $\tan\beta$ and $\sgn\mu$
  considered in Section~\ref{sec:signal}. We generated fifty thousand
  events at each of one hundred points in the $M_0$, $M_{1/2}$ plane at
  each value of $\tan\beta$ and $\sgn\mu$, and then interpolated between
  these points as for the signal process. This allowed us to estimate an
  acceptance for the cuts which we multiplied by the sparticle pair
  production cross section to give a number of
  background events.

%%%%%%%%%%%%%%%%%%%%%%%%%%%%%%%%%%%%%%%%%%%%%%%%%%%%%%%%%%%%%%%%%%%%%%%%%%%%%%
%
%  Figure containing the signal all backgrounds+jet cut at the Tevatron
%
\begin{figure}
\begin{center}
\includegraphics[angle=90,width=0.48\textwidth]{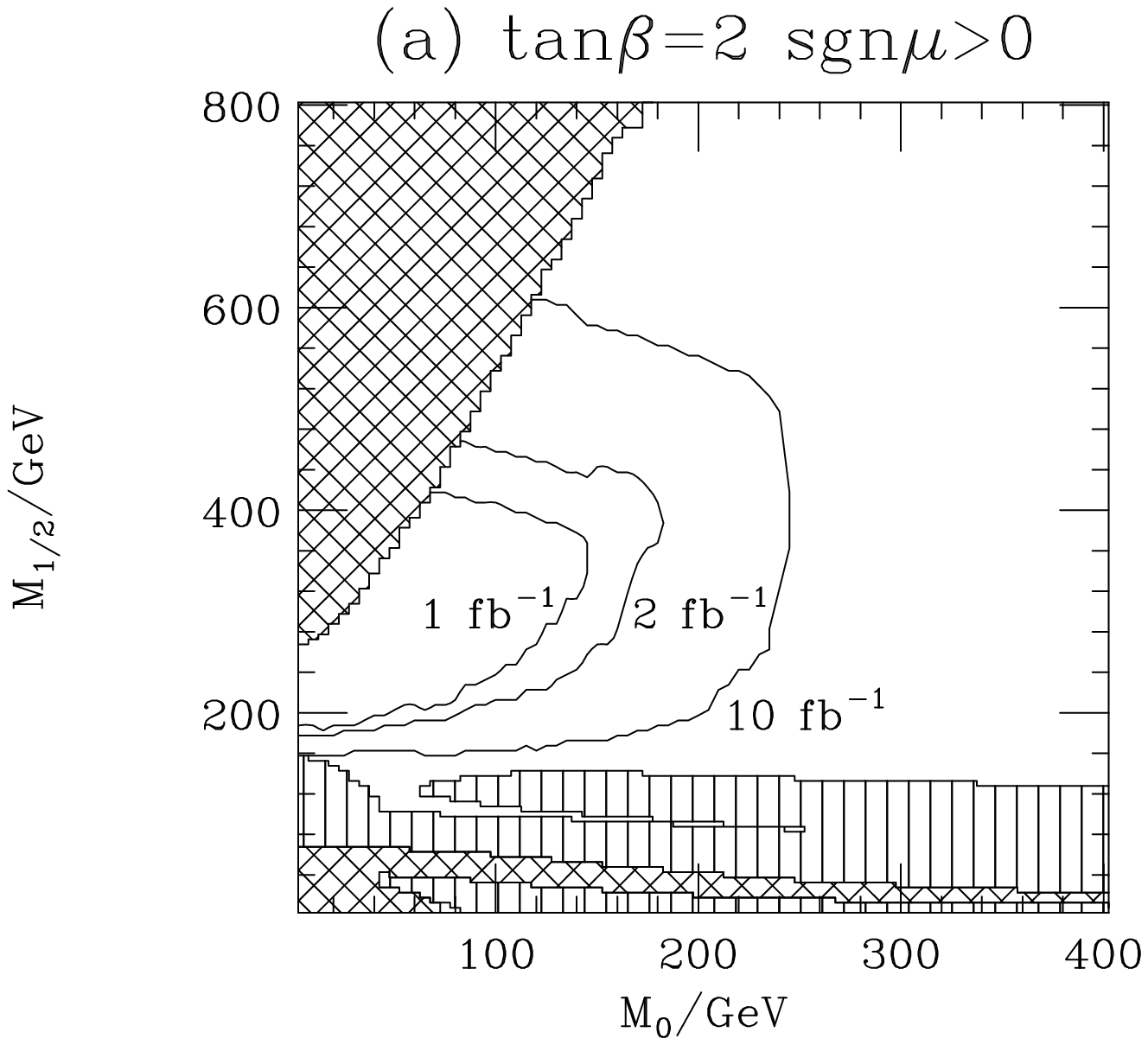}
\hfill
\includegraphics[angle=90,width=0.48\textwidth]{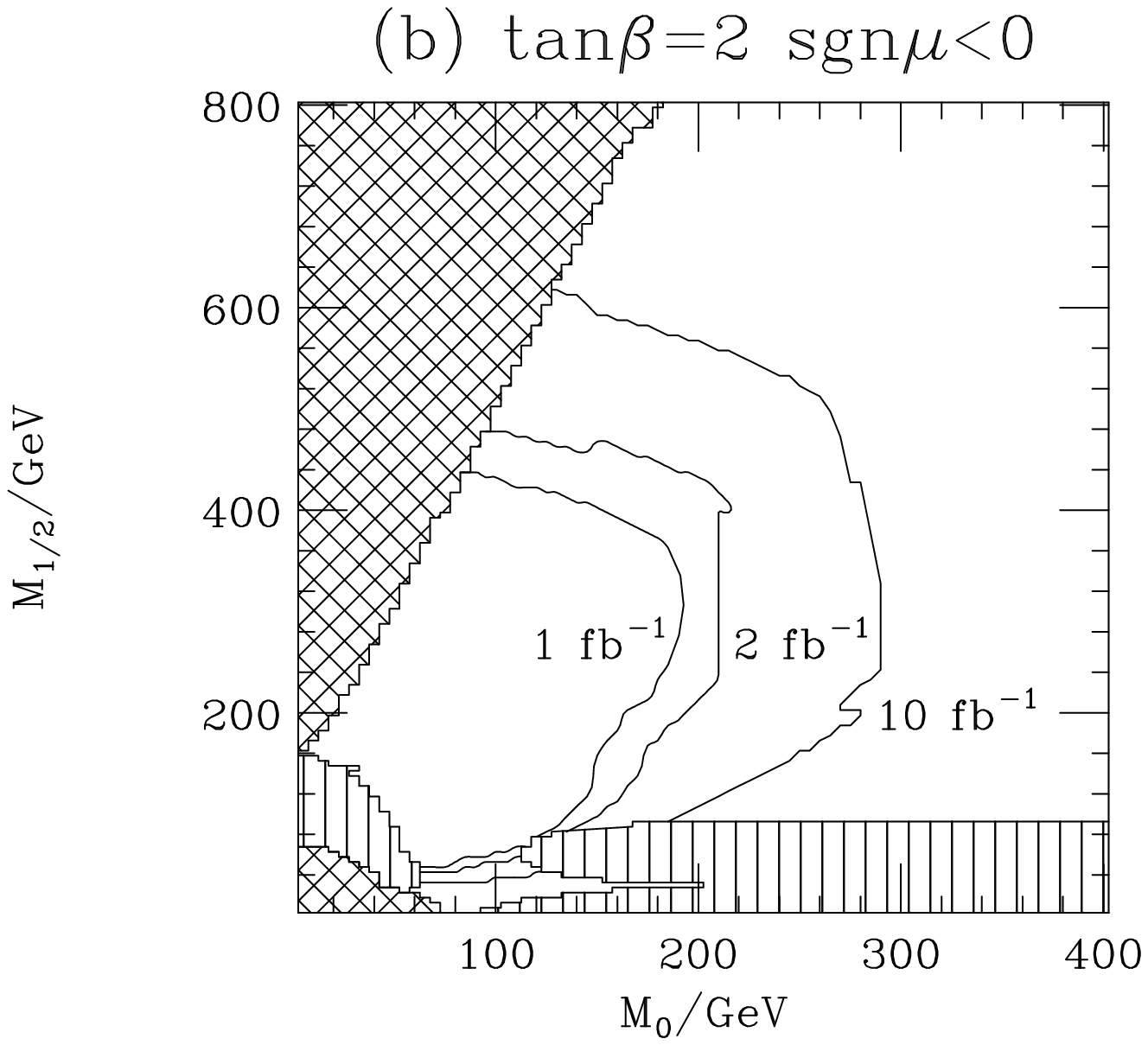}\\
\vskip 15mm
\includegraphics[angle=90,width=0.48\textwidth]{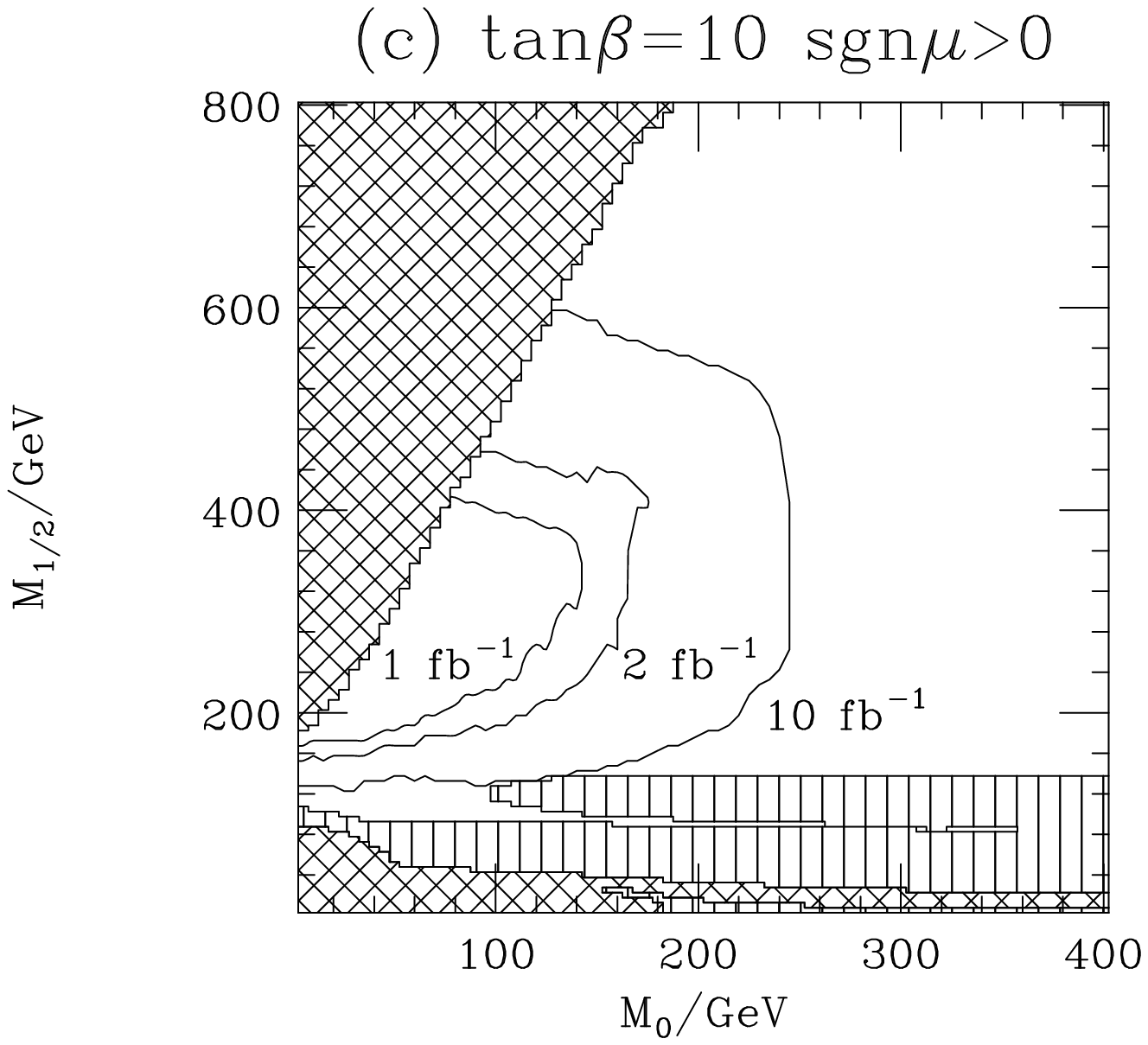}
\hfill
\includegraphics[angle=90,width=0.48\textwidth]{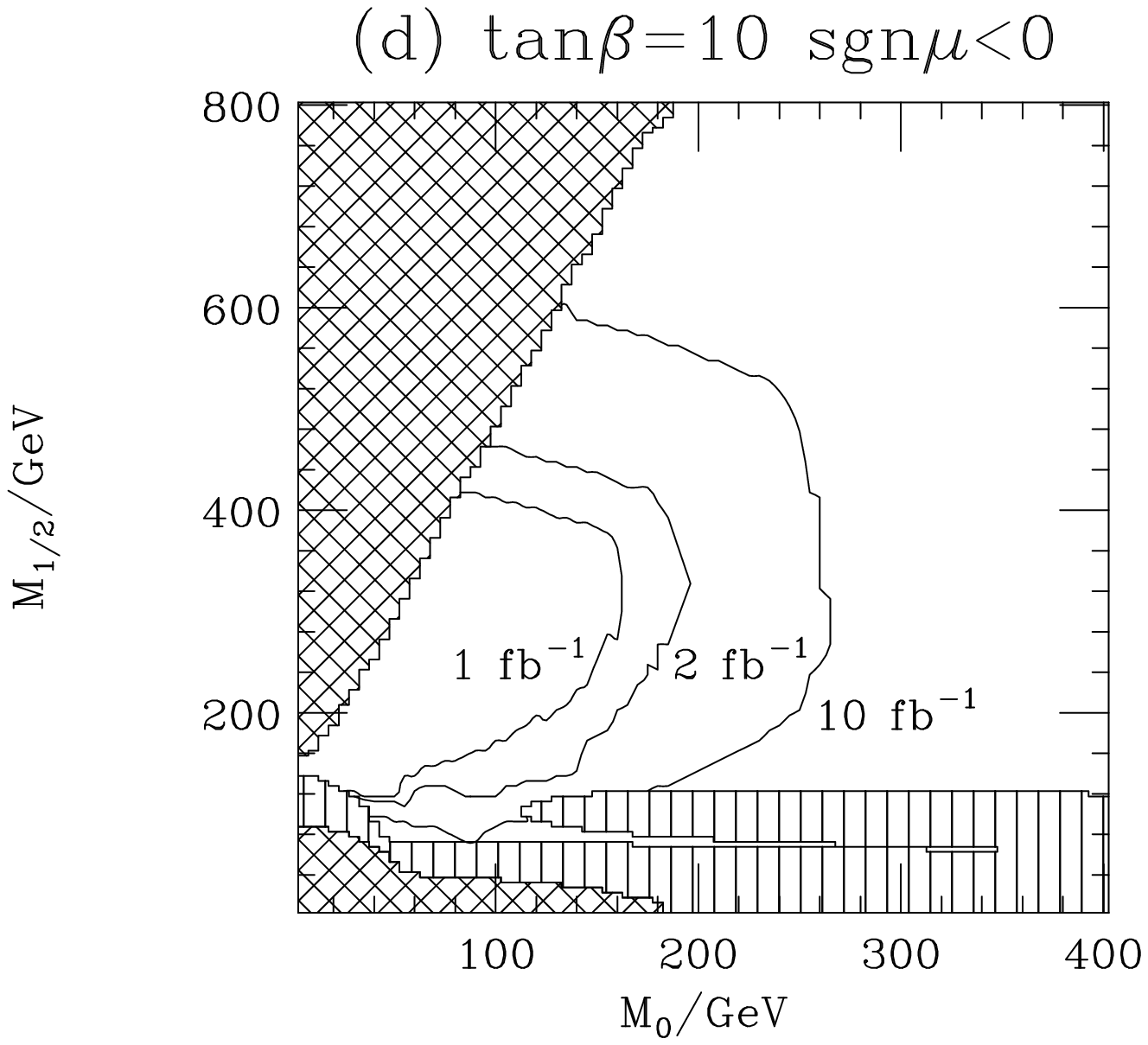}\\
\captionB{Discovery potential for the Tevatron in the $M_0$,
	 $M_{1/2}$ plane, including all the backgrounds,
	 with different integrated luminosities.}
	{Contours showing the discovery potential of the Tevatron in the
	 $M_0$,
	 $M_{1/2}$ plane for ${\lam'}_{211}=10^{-2}$ and $A_0=0\, \mr{\gev}$.
	 These contours are a $5\sigma$ excess of the signal above the
	 background. Here, in addition to the cuts on the isolation and $p_T$
  	 of the leptons, the transverse mass and the missing transverse energy
	 described in the text, and a veto on the presence of OSSF leptons
	 we have imposed a cut on the presence of more than two jets. This
	 includes the sparticle pair production background as well as the
	 Standard Model backgrounds. The striped and hatched regions
	 are as described in the caption of Fig.\,\ref{fig:SUSYmass}.}
\label{fig:tevSUSYjet}
\end{center}
\end{figure}
% End of the Figure %%%%%%%%%%%%%%%%%%%%%%%%%%%%%%%%%%%%%%%%%%%%%%%%%%%%%%%%%%

%%%%%%%%%%%%%%%%%%%%%%%%%%%%%%%%%%%%%%%%%%%%%%%%%%%%%%%%%%%%%%%%%%%%%%%%%%%%%%
%
%  Figure containing the signal all backgrounds+jet cut at the Tevatron
%
\begin{figure}
\begin{center}
\includegraphics[angle=90,width=0.48\textwidth]{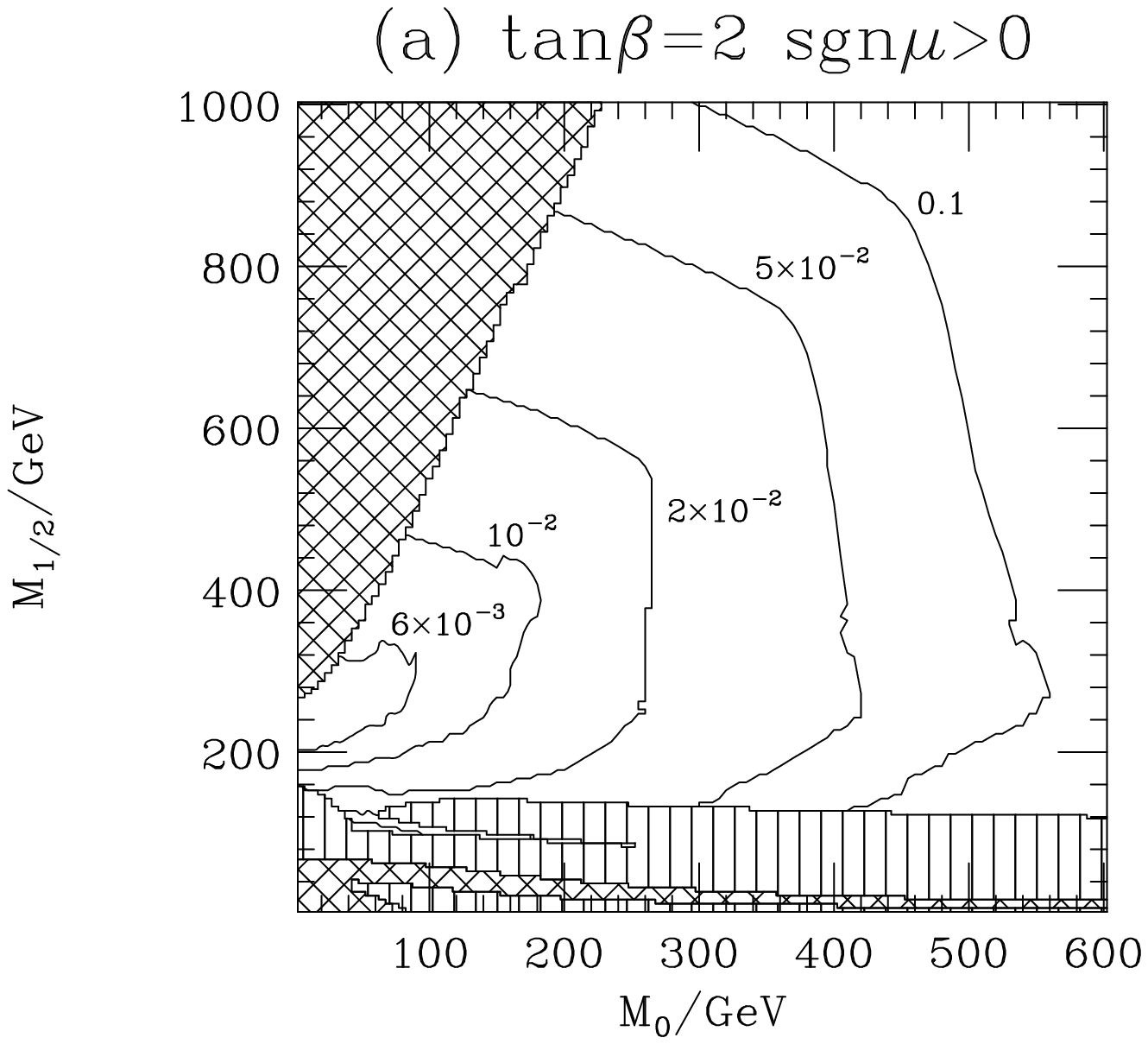}
\hfill						
\includegraphics[angle=90,width=0.48\textwidth]{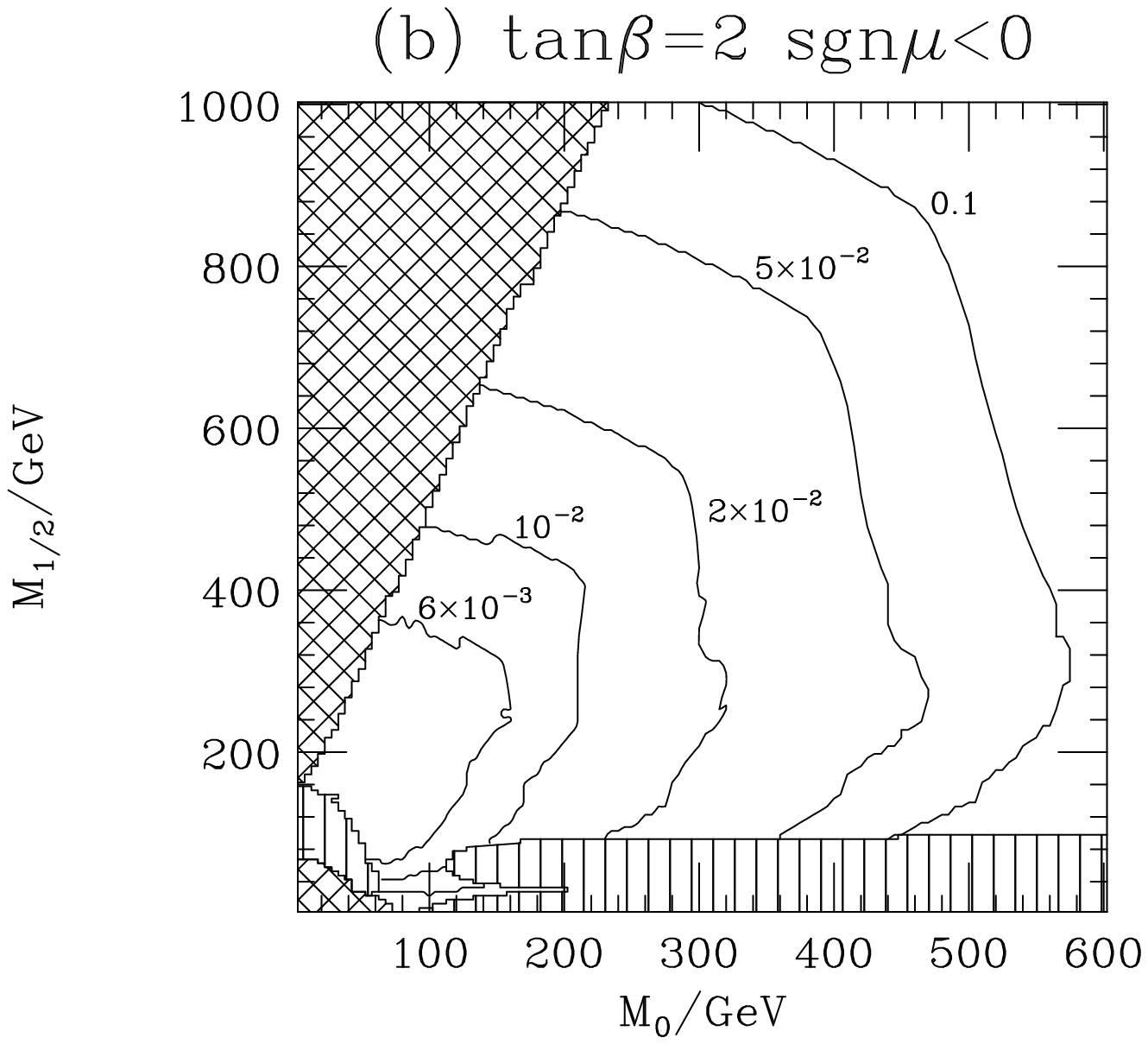}\\
\vskip 15mm					
\includegraphics[angle=90,width=0.48\textwidth]{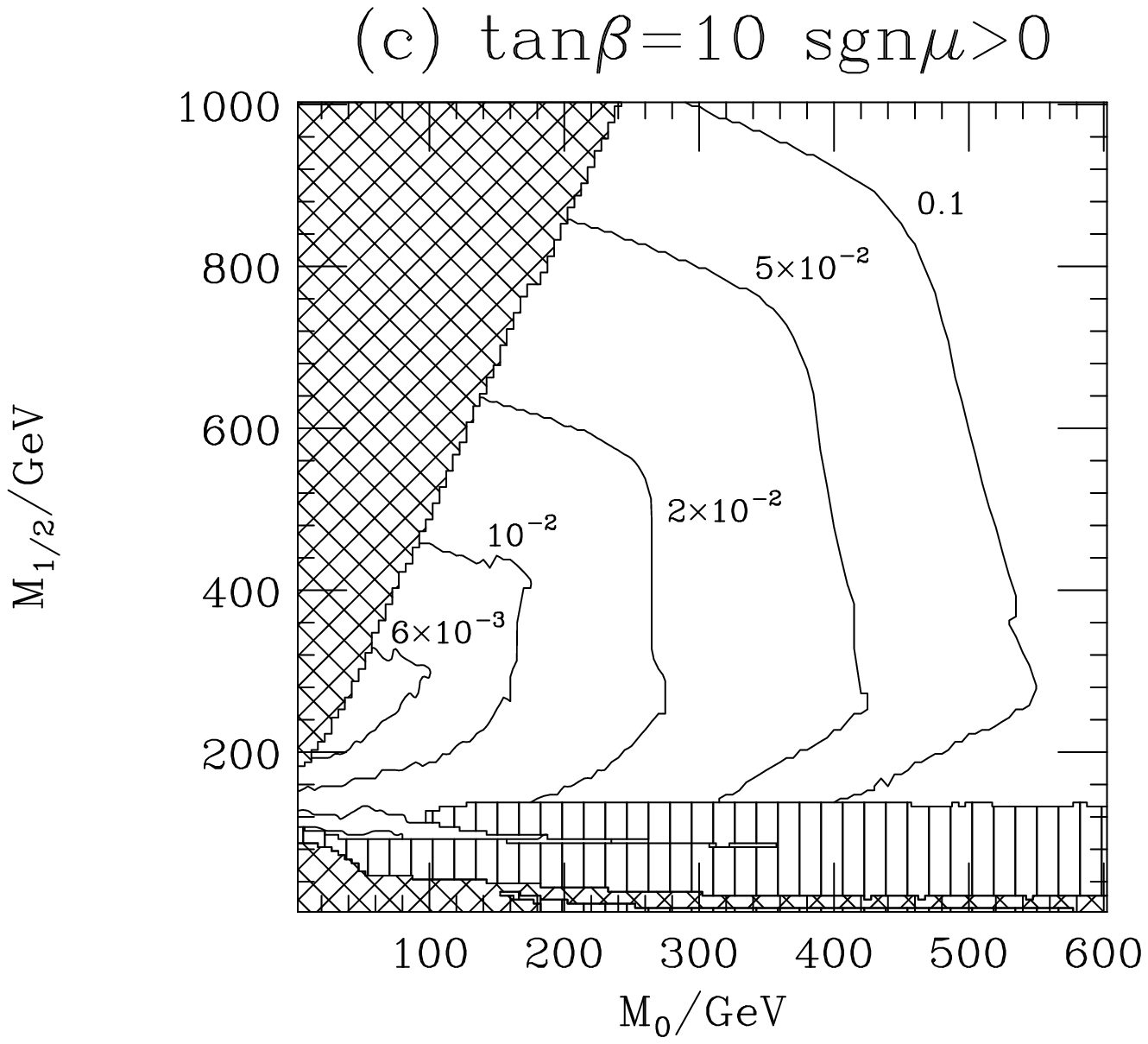}
\hfill						
\includegraphics[angle=90,width=0.48\textwidth]{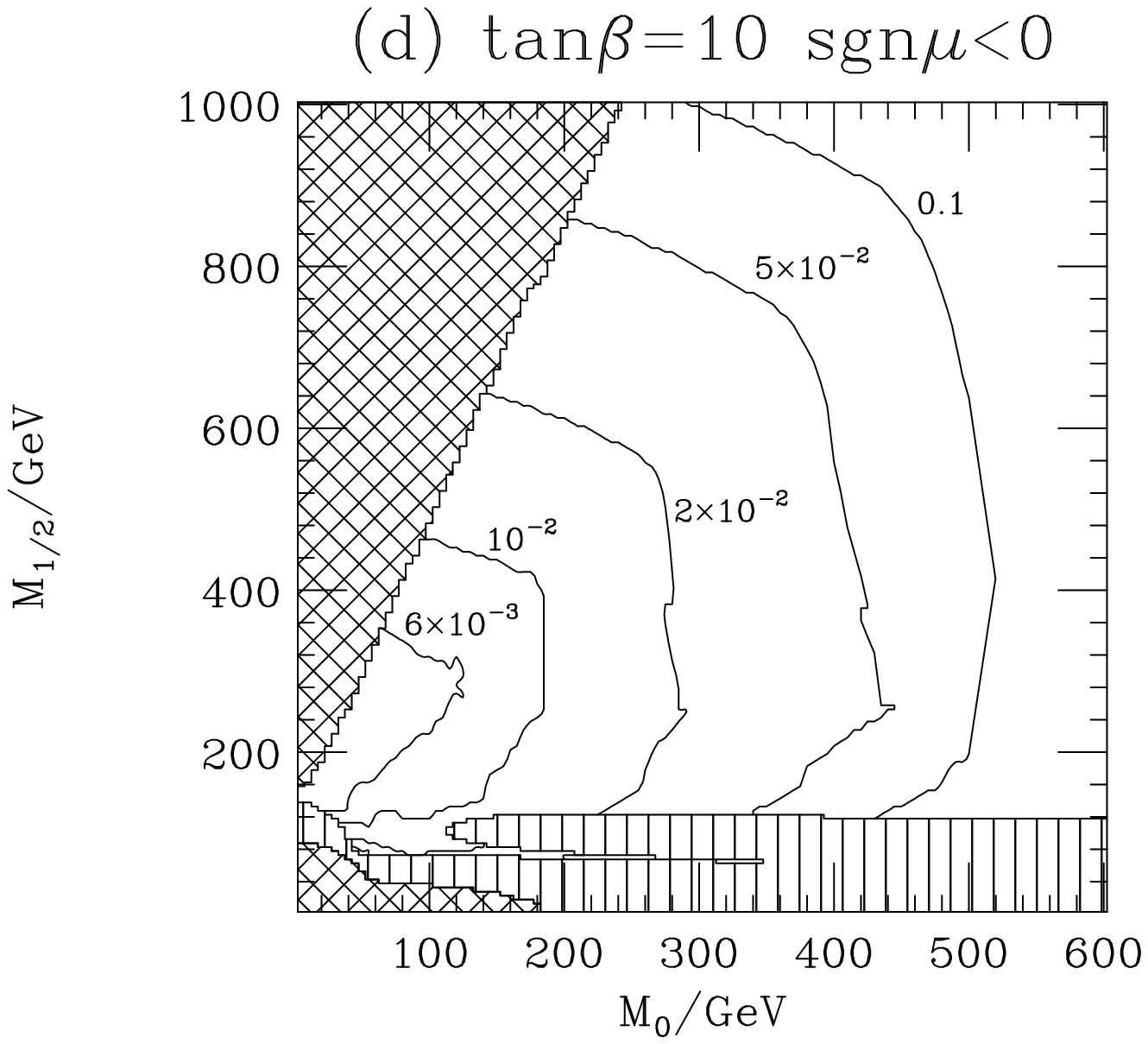}\\
\captionB{Discovery potential for the Tevatron in the $M_0$,
	 $M_{1/2}$ plane, including all the backgrounds, for
	 different values of ${\lam'}_{211}$.}
	{Contours showing the discovery potential of the Tevatron in the
	 $M_0$, $M_{1/2}$ plane for $A_0=0\, \mr{\gev}$  and an integrated
	 luminosity of
  	 $2\  \mr{fb}^{-1}$ for different values of ${\lam'}_{211}$.
	 These contours are a $5\sigma$ excess of the signal above the
	 background. Here, in addition to the cuts on the isolation and $p_T$
  	 of the leptons, the transverse mass and the missing transverse energy
	 described in the text, and a veto on the presence of OSSF leptons
	 we have imposed a cut on the presence of more than two jets. This
	 includes the sparticle pair production background as well as the
	 Standard Model backgrounds. The striped and hatched regions
	 are as described in the caption of Fig.\,\ref{fig:SUSYmass}.}
\label{fig:tevSUSYjetb}
\end{center}
\end{figure}
% End of the Figure %%%%%%%%%%%%%%%%%%%%%%%%%%%%%%%%%%%%%%%%%%%%%%%%%%%%%%%%%%

  The effect of all the cuts on the total background, \ie the Standard Model
  background and the sparticle pair production background is shown in 
  Fig.\,\ref{fig:tevSUSYjet} for different integrated luminosities with 
  ${\lam'}_{211}=10^{-2}$ and in Fig.\,\ref{fig:tevSUSYjetb} for an
  integrated luminosity of $2\  \mr{fb}^{-1}$ with different values 
  of ${\lam'}_{211}$.

  The effect of including the sparticle pair production
  background is to reduce the $5\sigma$ discovery regions. These regions are
  reduced for two reasons: for large $M_{1/2}$ the additional cut removes more
  signal events and hence reduces the statistical significance of the signal; 
  at
  small values of $M_{1/2}$ there is a large background from sparticle pair
  production, relative to the SM background, which
  also reduces the statistical
  significance of the signal. However even for this relatively small value of
  the coupling there are large regions of parameter space in which a signal is
  visible above the background. The ratio of signal to background is still
  larger than one for most of the region where the signal is detectable above
  the background. For $\sgn\mu>0$
  there is only a very
  small region at low $M_{1/2}$ where $S/B$ drops below one and even here
  $S/B>0.5$. However for $\sgn\mu<0$ there are regions of
  low $S/B$ for small values of $M_{1/2}$.
  The discovery range for these \rpv\  processes extends to 
  larger values of 
  $M_{1/2}$ than the $5\sigma$ discovery curve for sparticle pair production
  as only one sparticle is produced which requires a much lower parton--parton
  centre-of-mass energy than sparticle pair production. 

  Again even for small smuon masses with low values of the \rpv\  Yukawa 
  coupling there are regions where a signal of resonant slepton production
  is not visible above the background. However for large couplings the
  signal in these regions is visible above the background. For a coupling
  of ${\lam'}_{211}=0.05$ a smuon mass of $310\,(330)\, \mr{\gev}$  is
  visible above
  the background with 2\,(10)~$\mr{fb}^{-1}$ integrated luminosity,  
  and for a coupling of ${\lam'}_{211}=0.1$ a smuon mass of
  $400\,(430)\, \mr{\gev}$ is visible above the background with
  2\,(10)~$\mr{fb}^{-1}$ integrated luminosity.

%
%  a subsection on the LHC
%
\subsubsection{LHC}
\label{sub:LHC}
 
%%%%%%%%%%%%%%%%%%%%%%%%%%%%%%%%%%%%%%%%%%%%%%%%%%%%%%%%%%%%%%%%%%%%%%%%%%%%%%
%
%  Figure containing the cross sections at the LHC
%
\begin{figure}
\begin{center}
\includegraphics[angle=90,width=0.48\textwidth]{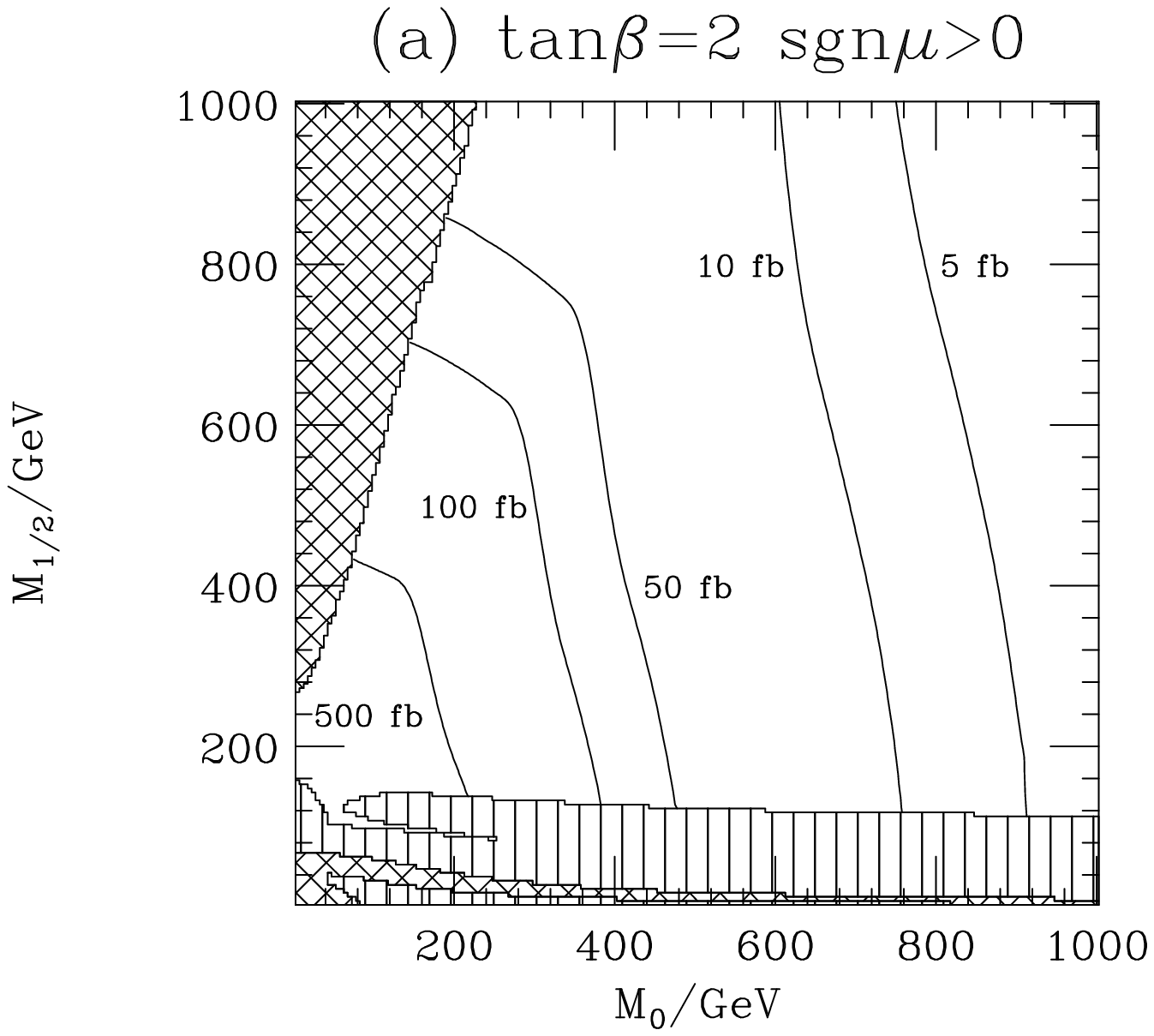}
\hfill
\includegraphics[angle=90,width=0.48\textwidth]{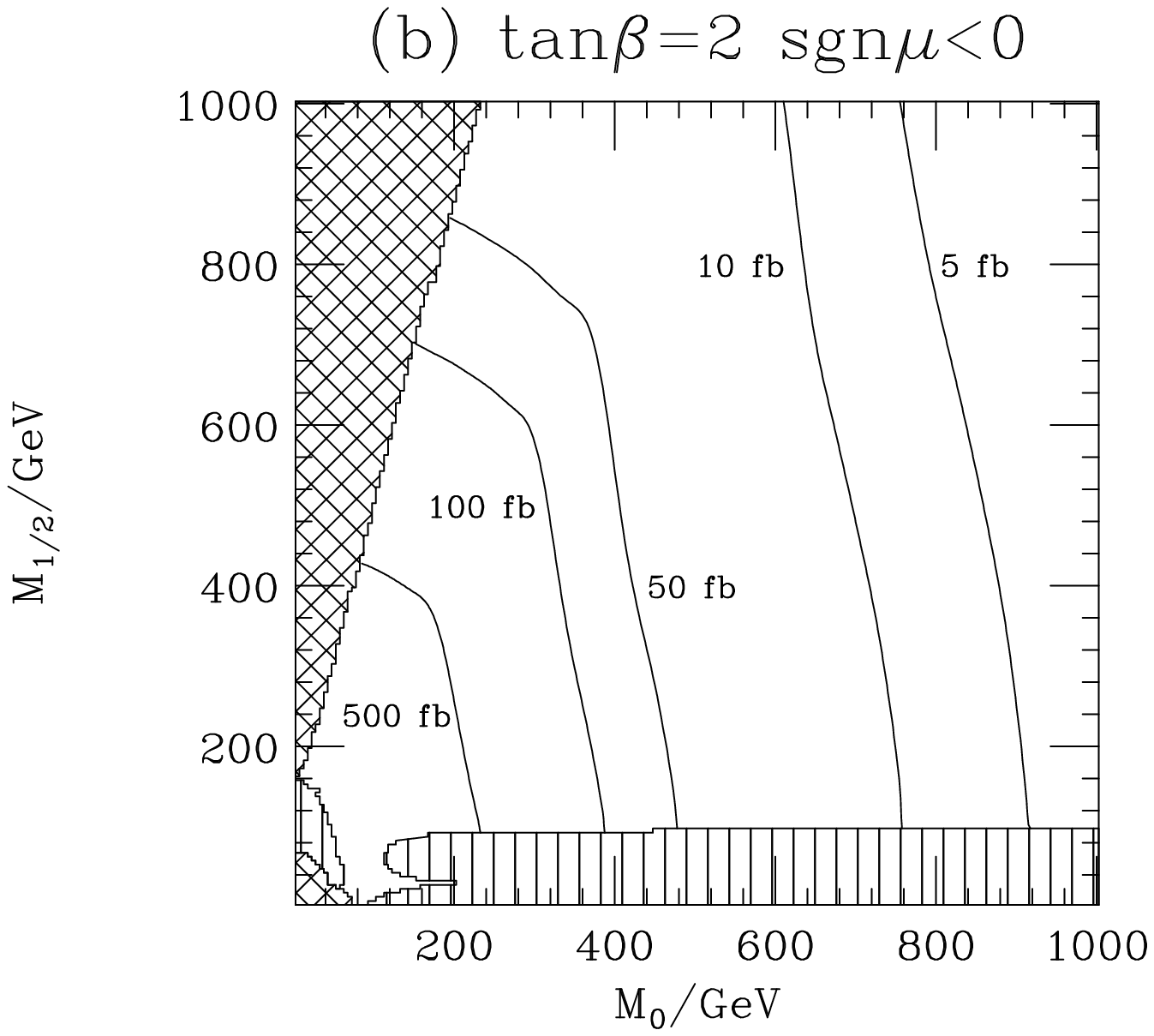}\\
\vskip 15mm
\includegraphics[angle=90,width=0.48\textwidth]{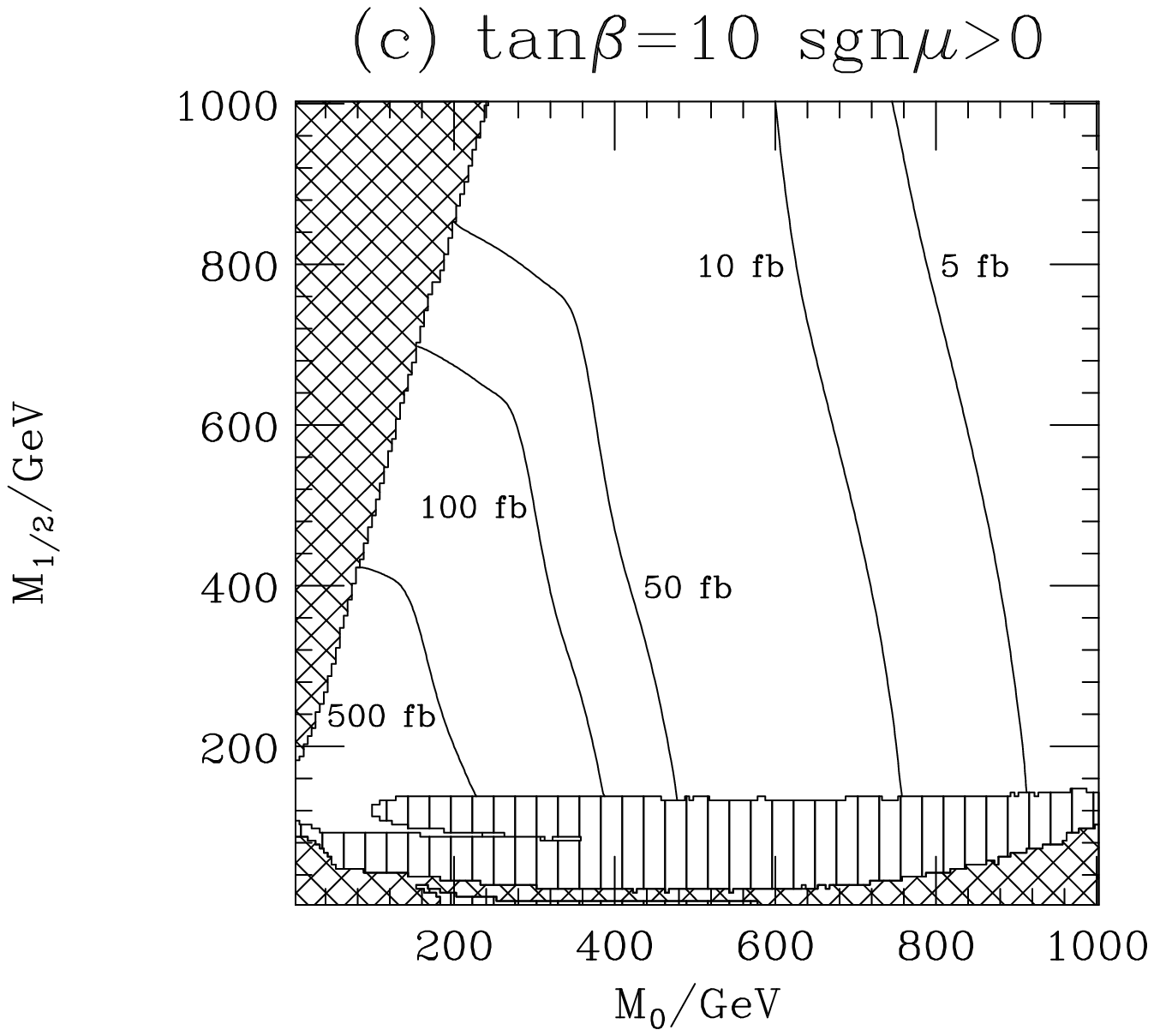}
\hfill
\includegraphics[angle=90,width=0.48\textwidth]{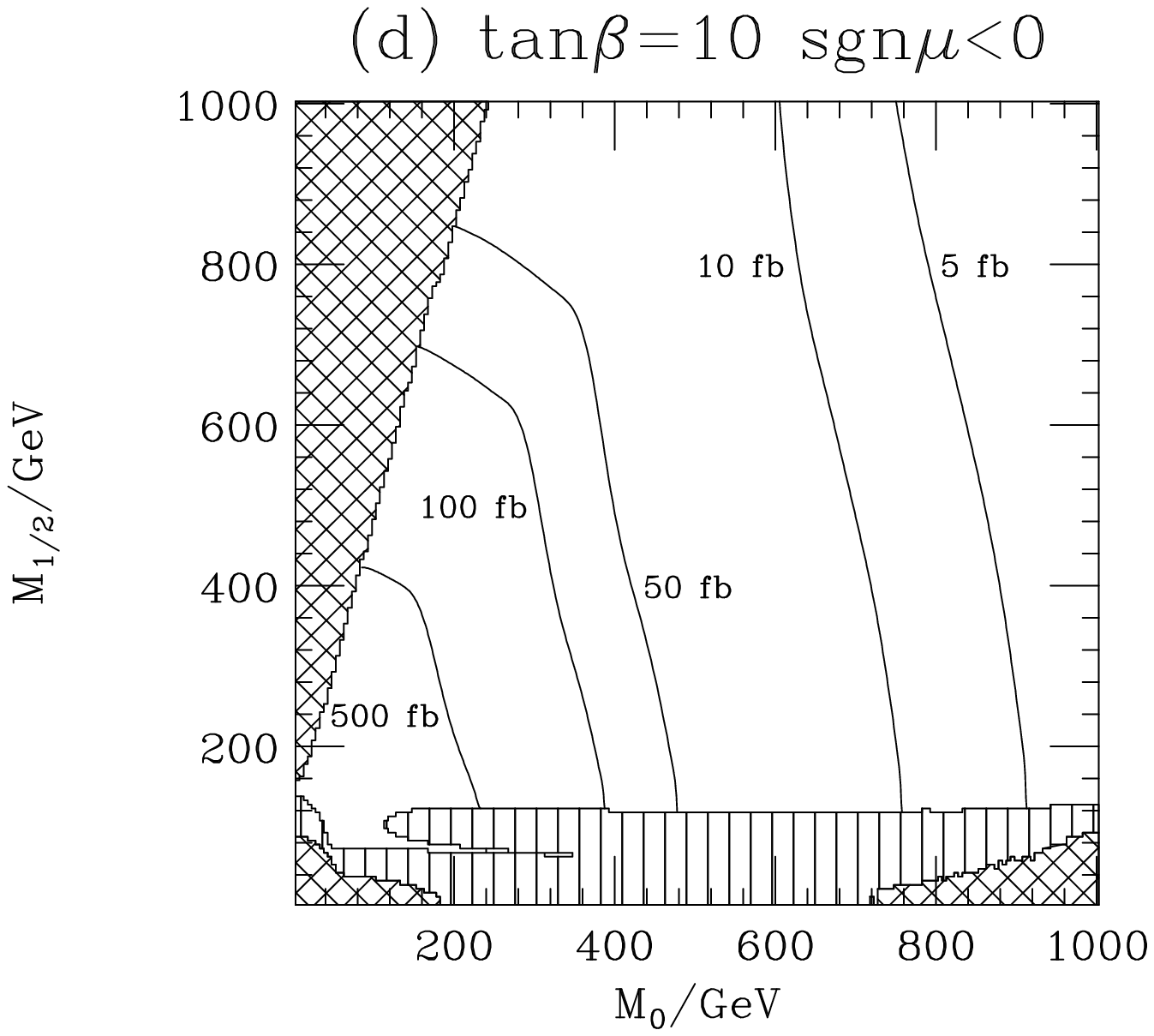}\\
\captionB{Cross section for the production of $\mr{\cht^0\ell}$ at the LHC
	in the $M_0$, $M_{1/2}$ plane.}
	{Contours showing the cross section for the production of a neutralino
	 and a charged lepton at the LHC in the $M_0$, $M_{1/2}$ plane for
       $A_0=0\, \mr{\gev}$  and ${\lam'}_{211}=10^{-2}$ with different values
	 of $\tan\beta$ and $\sgn\mu$. The striped and hatched regions are
	 described in the caption of Fig.\,\ref{fig:SUSYmass}.} 
\label{fig:LHCcross}
\end{center}
\end{figure}
% End of the Figure %%%%%%%%%%%%%%%%%%%%%%%%%%%%%%%%%%%%%%%%%%%%%%%%%%%%%%%%%%

%%%%%%%%%%%%%%%%%%%%%%%%%%%%%%%%%%%%%%%%%%%%%%%%%%%%%%%%%%%%%%%%%%%%%%%%%%%%%%
%
%  Figure containing the cross sections at the LHC
%
\begin{figure}
\begin{center}
\includegraphics[angle=90,width=0.48\textwidth]{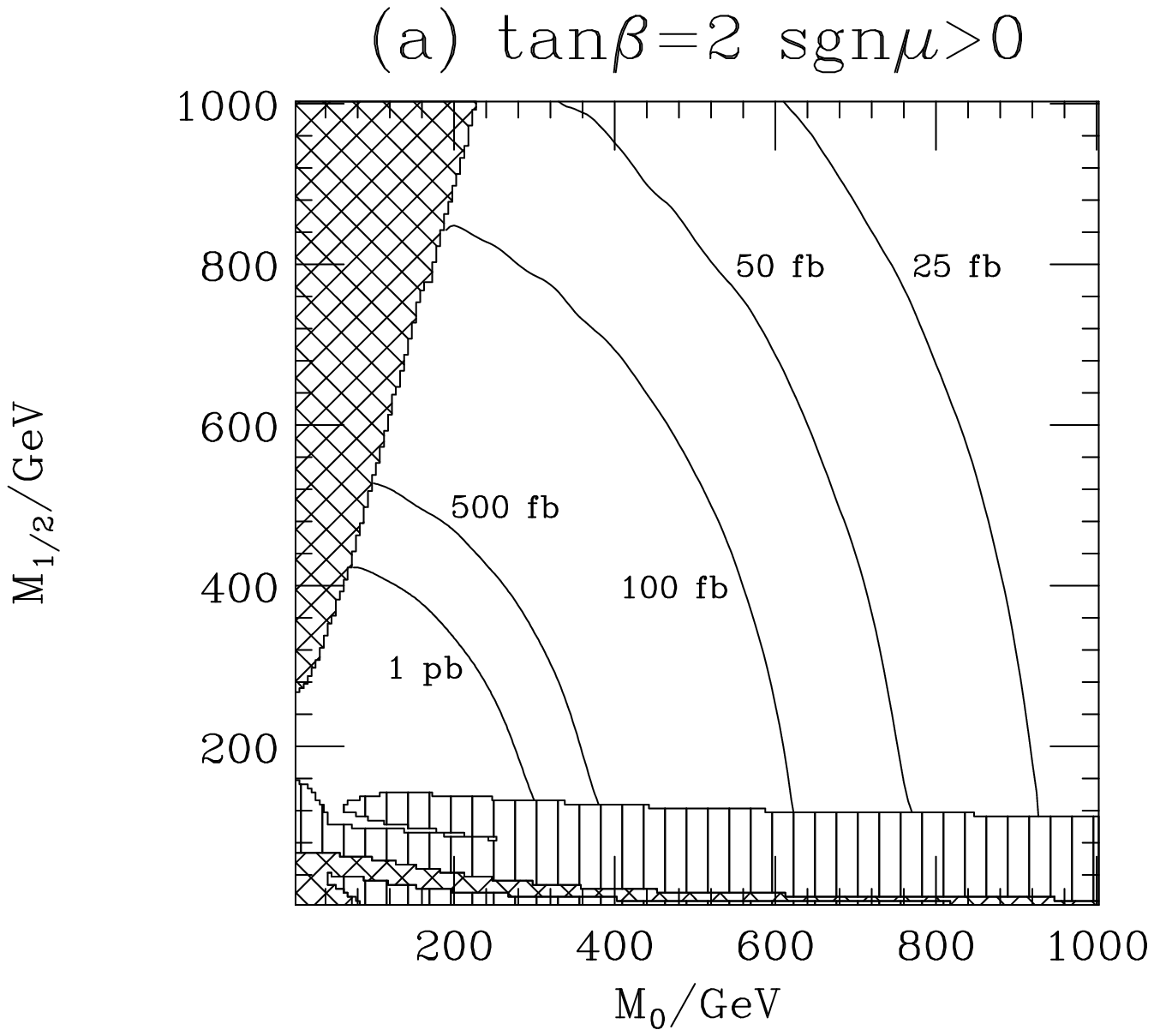}
\hfill
\includegraphics[angle=90,width=0.48\textwidth]{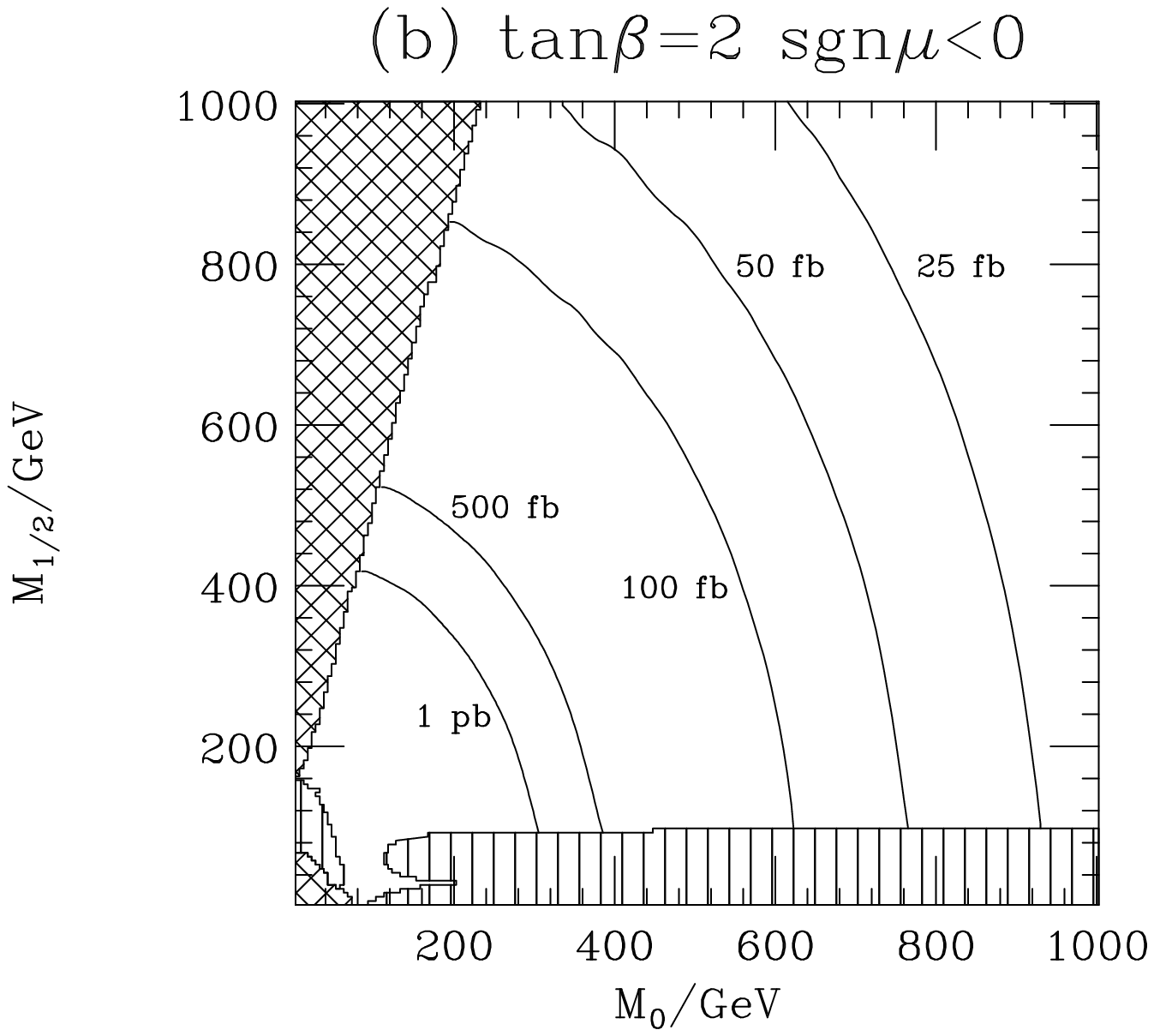}\\
\vskip 15mm
\includegraphics[angle=90,width=0.48\textwidth]{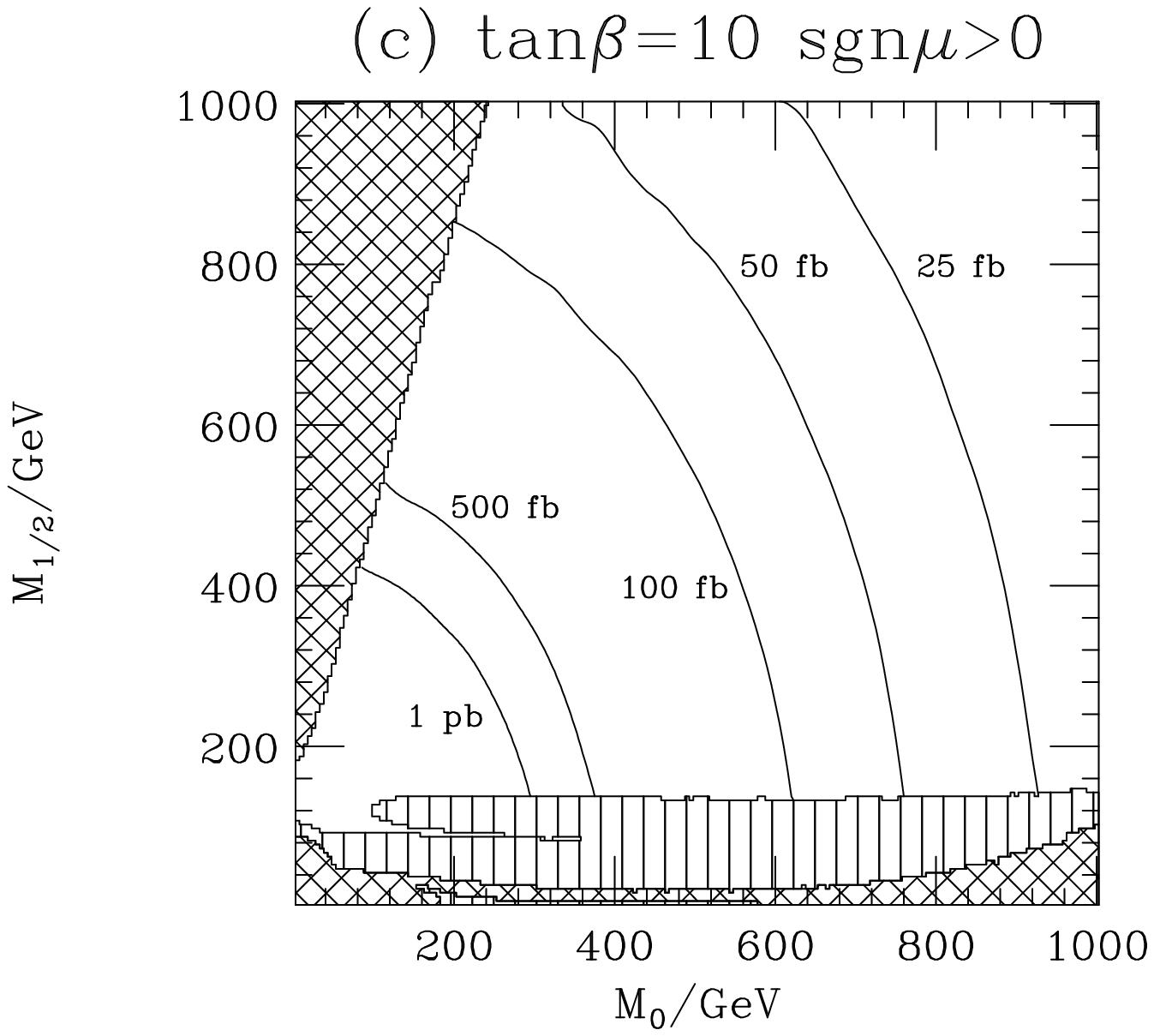}
\hfill
\includegraphics[angle=90,width=0.48\textwidth]{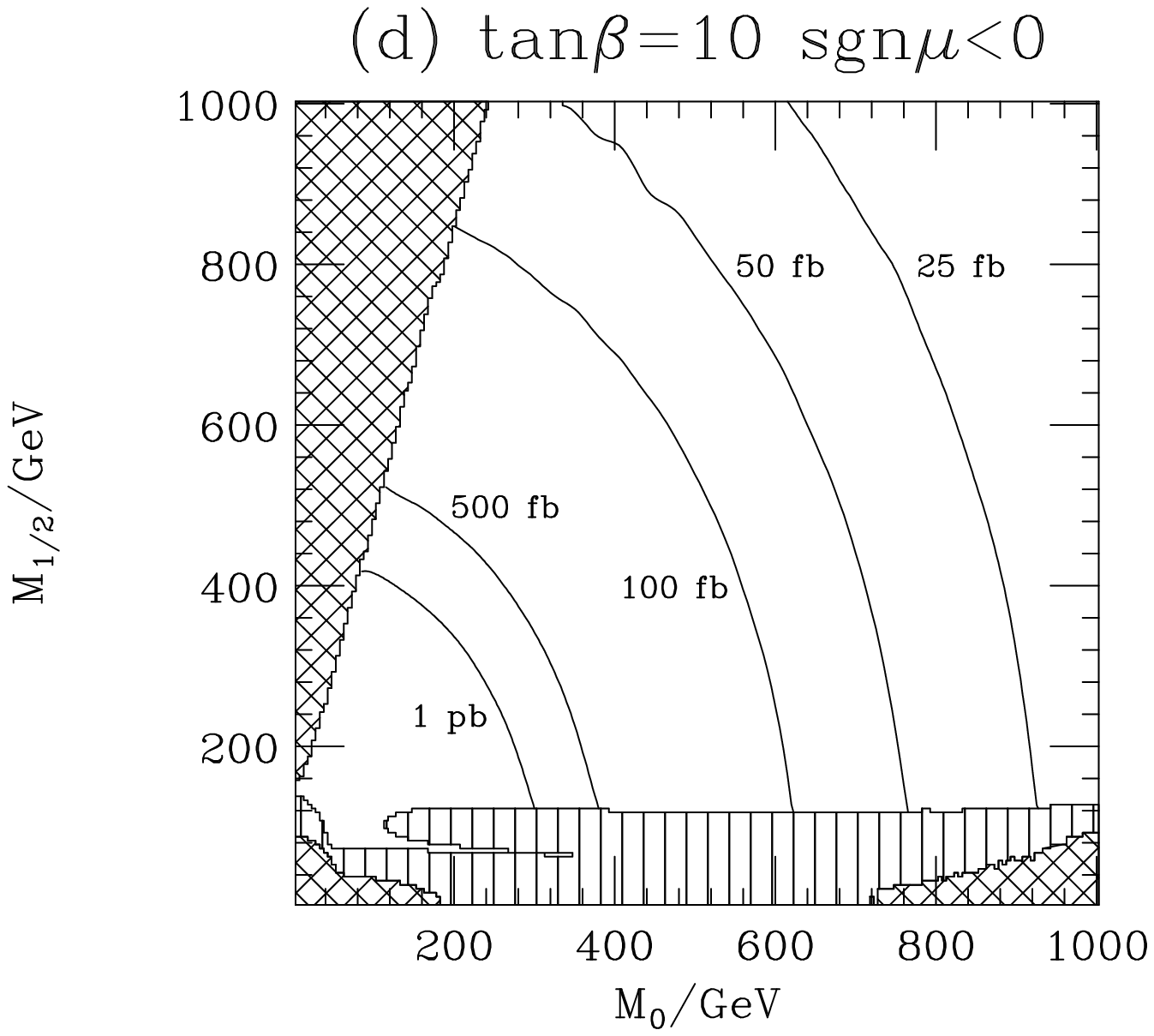}\\
\captionB{Cross section for resonant slepton production followed by a
	 supersymmetric gauge decay at the
	 LHC in the $M_0$, $M_{1/2}$ plane.}
	{Contours showing the cross section for resonant slepton 
	 production followed by a supersymmetric gauge decay at the
	 LHC in the $M_0$, $M_{1/2}$ plane 
	 for \linebreak \mbox{$A_0=0\, \mr{\gev}$}
	  and ${\lam'}_{211}=10^{-2}$ with different values
	 of $\tan\beta$ and $\sgn\mu$. The striped and hatched regions are
	 described in the caption of Fig.\,\ref{fig:SUSYmass}.} 
\label{fig:LHCcross2}
\end{center}
\end{figure}
% End of the Figure %%%%%%%%%%%%%%%%%%%%%%%%%%%%%%%%%%%%%%%%%%%%%%%%%%%%%%%%%%

  The cross section for the production of a charged lepton and a neutralino,
  which is again the dominant production mechanism, at the LHC
  is shown in Fig.\,\ref{fig:LHCcross} in the  $M_0$, $M_{1/2}$ plane 
  with $A_0=0\, \mr{\gev}$ and ${\lam'}_{211}=10^{-2}$ for two different
  values
  of $\tan\beta$ and both values of $\sgn\mu$. The total cross section for
  resonant slepton production followed by a supersymmetric 
  gauge decay is shown in
  Fig.\,\ref{fig:LHCcross2}. As for the Tevatron, the total resonant slepton
  cross section closely follows the slepton mass contours whereas the cross
  section for neutralino--lepton production falls off more quickly at small
  $M_{1/2}$ because the branching ratio for $\mut_L\ra\mu\cht^0_1$
  is reduced due to the production of charginos and the heavier neutralinos.
  We adopted the same procedure 
  described in Section~\ref{sub:tevatron} to
  estimate the acceptance of the cuts
  we imposed. We will again first consider the cuts required to reduce
  the Standard Model backgrounds and then the additional cut used to suppress
  the sparticle pair production background.
   
\begin{figure}
\begin{center}
\includegraphics[angle=90,width=0.48\textwidth]{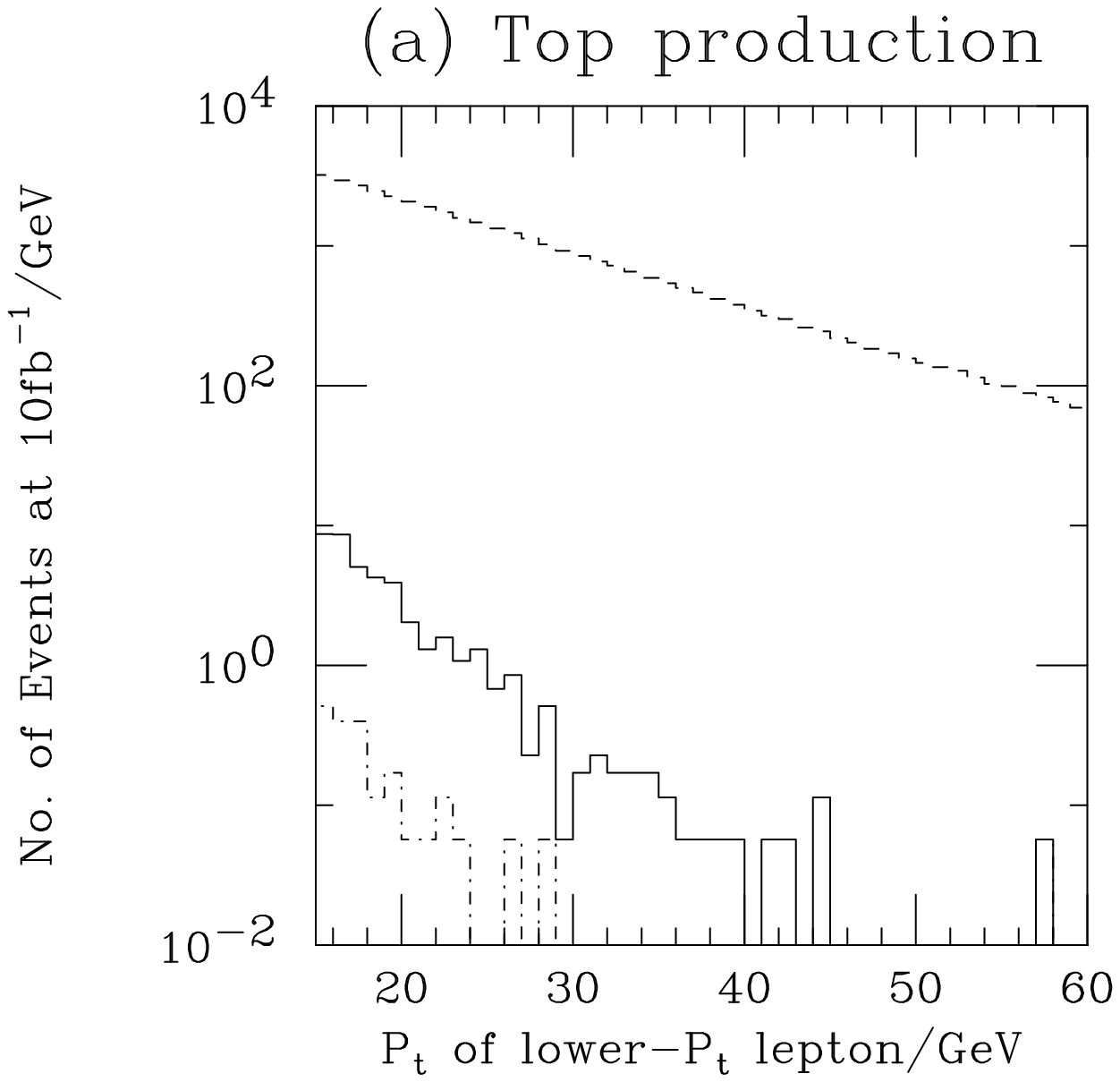}
\hfill
\includegraphics[angle=90,width=0.48\textwidth]{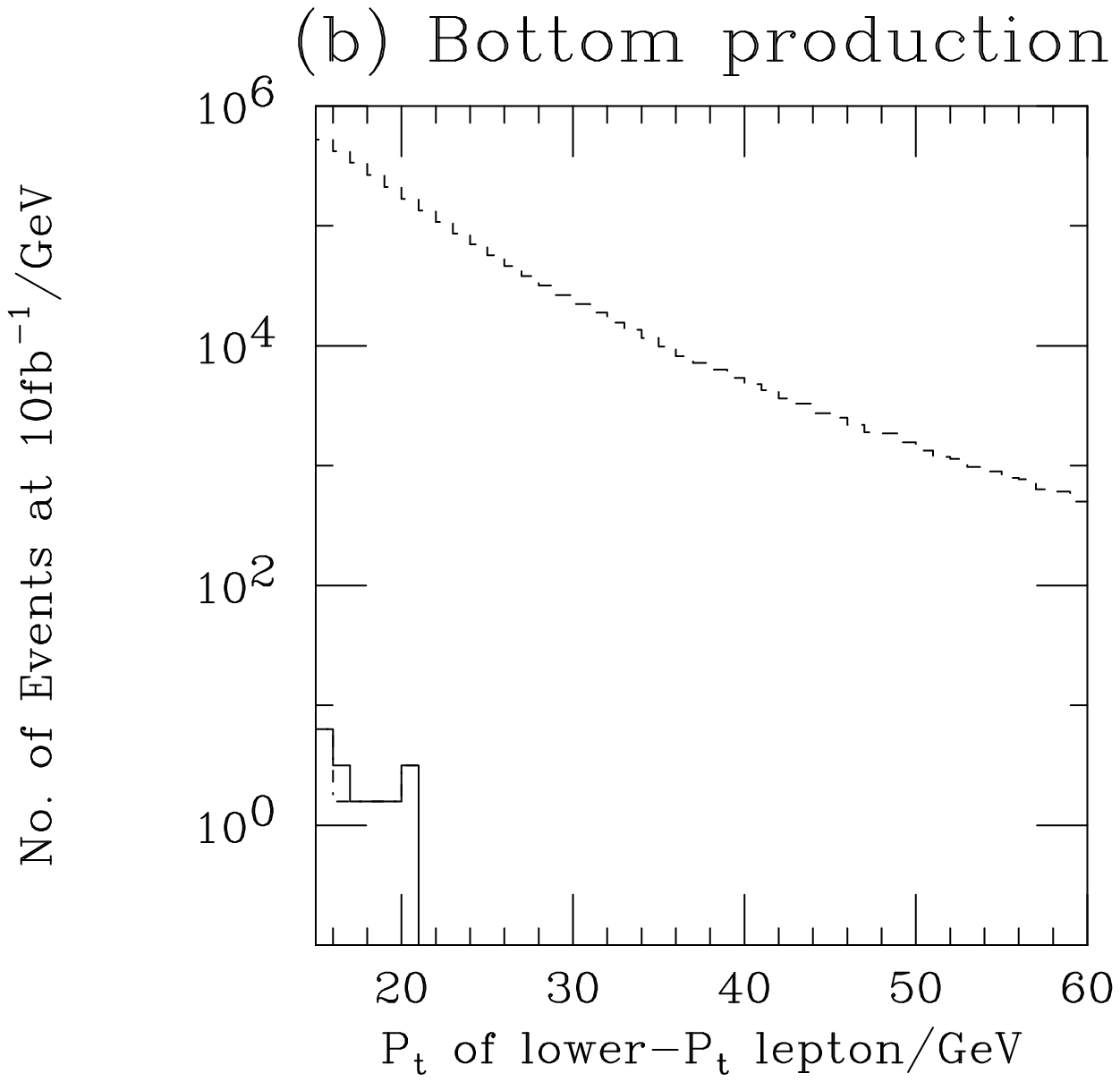}\\
\captionB{Effect of the isolation cuts on the $\mr{t\bar{t}}$ and 
	 $\mr{b\bar{b}}$ backgrounds at the LHC.}
	{Effect of the isolation cuts on the $\mr{t\bar{t}}$ and 
	 $\mr{b\bar{b}}$ backgrounds at 
	the LHC. The dashed line gives the background before any cuts and 
	the solid line shows the effect of the isolation cut described in the
	text. The dot-dash line gives the effect of all the cuts, including
	the cut on the number of jets, for the $\mr{b\bar{b}}$ background 
 	this is almost  indistinguishable from the solid line. As a
	parton-level cut of $40\, \mr{\gev}$  was used in simulating the
	$\mr{b\bar{b}}$ background, the results below $40\, \mr{\gev}$  for
	the lepton $p_T$ do not correspond to the full number of background
	events. The distributions have been normalized
 	to an integrated luminosity of $10\  \mr{fb}^{-1}$.} 
\label{fig:lhcheavyiso}
\end{center}
\end{figure}

%
%  first a subsection on the Standard Model
%
\vskip 5mm
\noindent{\underline{Standard Model Backgrounds}}
\nopagebreak
\vskip 5mm
\nopagebreak
  We applied the following cuts to reduce the Standard Model backgrounds:
\begin{enumerate}

\item A cut requiring all the leptons to be in the central region of the
      detector $|\eta|<2.0$.

\item	A cut on the transverse momentum of the like-sign leptons,
	$p_T^{\mr{lepton}} \geq 40\, \mr{\gev}$.
	This is the lowest cut we could apply given our parton-level cut of
	$p_T^{\mr{parton}}=40\, \mr{\gev}$, for the $\mr{b\bar{b}}$
	background.
	
\item 	An isolation cut on the like-sign leptons so that the
	transverse energy in a cone of radius,
	$\Delta R = \sqrt{\Delta\phi^2+\Delta\eta^2} = 0.4$, about
	the direction of the lepton is less than $5\, \mr{\gev}$.

\item   We reject events with  $60\, \mr{\gev} < M_T < 85\, \mr{\gev}$ 
       ($c.f.$ Eqn.\,\ref{eqn:MTdef}). This cut is
	applied to both of the like-sign leptons.

\item   A veto on the presence of a lepton in the event with the same flavour
        but opposite charge as either of the leptons in the like-sign
	pair if the lepton has  $p_T>10\, \mr{\gev}$  
	and passes the same isolation cut as the like-sign leptons.

\item   A cut on the missing transverse energy,  $\not\!\!E_T<20\, \mr{\gev}$.
\end{enumerate}
  The first two cuts are designed to reduce the background from heavy
  quark, \ie $\mr{b\bar{b}}$ and $\mr{t\bar{t}}$, production which is
  the major source of background before any cuts. However, as can be seen in
  Fig.\,\ref{fig:lhcheavyiso}, after the
  imposition of the $p_T$ and isolation cuts this background is significantly
  reduced. It remains the major source of the error on the background however
  due to the large cross section for $\mr{b\bar{b}}$ production which makes
  it impossible to simulate the full luminosity of the LHC with the resources
  available.

%%%%%%%%%%%%%%%%%%%%%%%%%%%%%%%%%%%%%%%%%%%%%%%%%%%%%%%%%%%%%%%%%%%%%%%%%%%%%%
%
%  Figure containing the effect of the cuts on top and bottom at the LHC
%
% End of the Figure %%%%%%%%%%%%%%%%%%%%%%%%%%%%%%%%%%%%%%%%%%%%%%%%%%%%%%%%%%

%%%%%%%%%%%%%%%%%%%%%%%%%%%%%%%%%%%%%%%%%%%%%%%%%%%%%%%%%%%%%%%%%%%%%%%%%%%%%%
%
%  Figure containing the transverse mass and missing ET at the LHC
%
\begin{figure}
\begin{center}
\includegraphics[angle=90,width=0.48\textwidth]{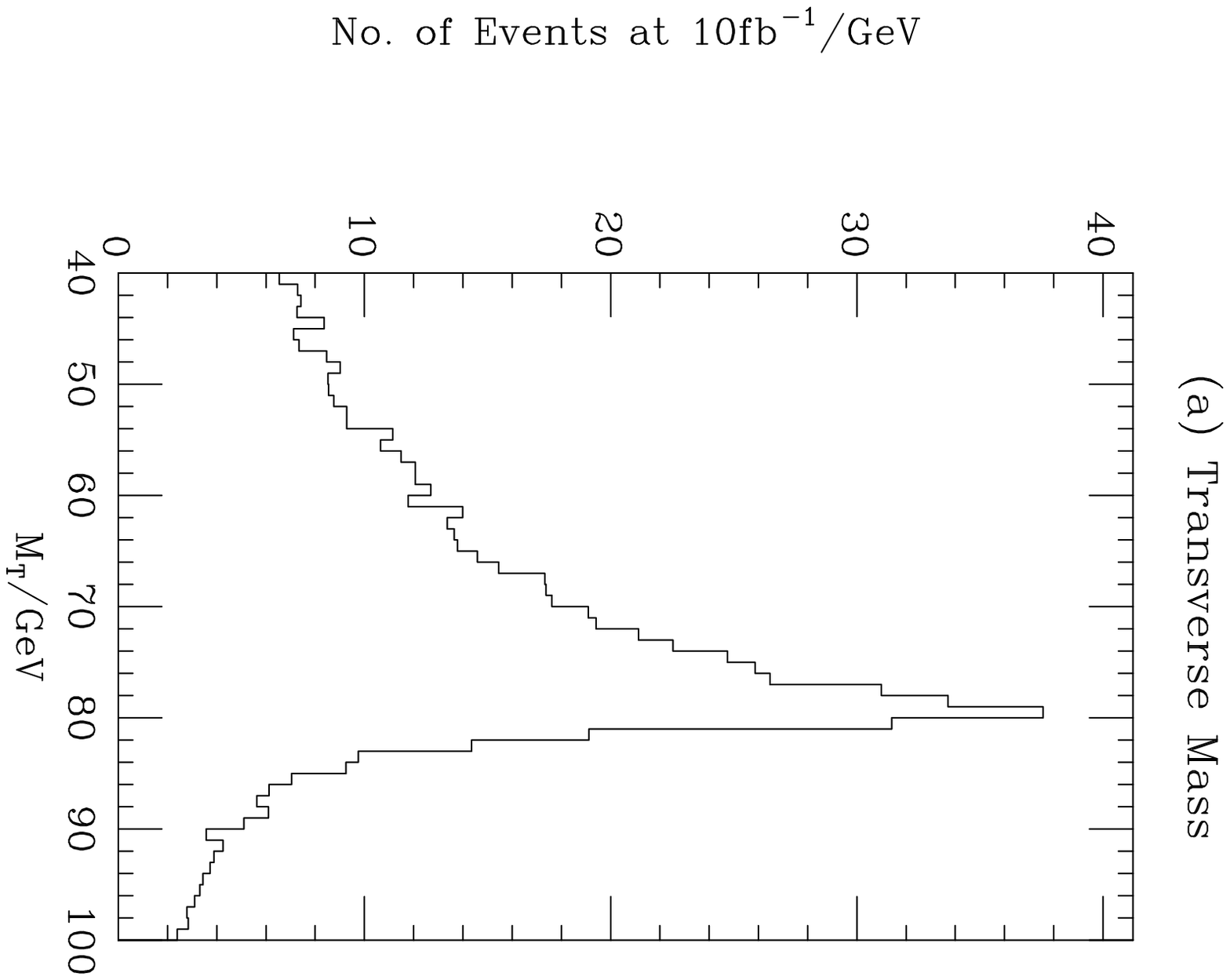}
\hfill
\includegraphics[angle=90,width=0.48\textwidth]{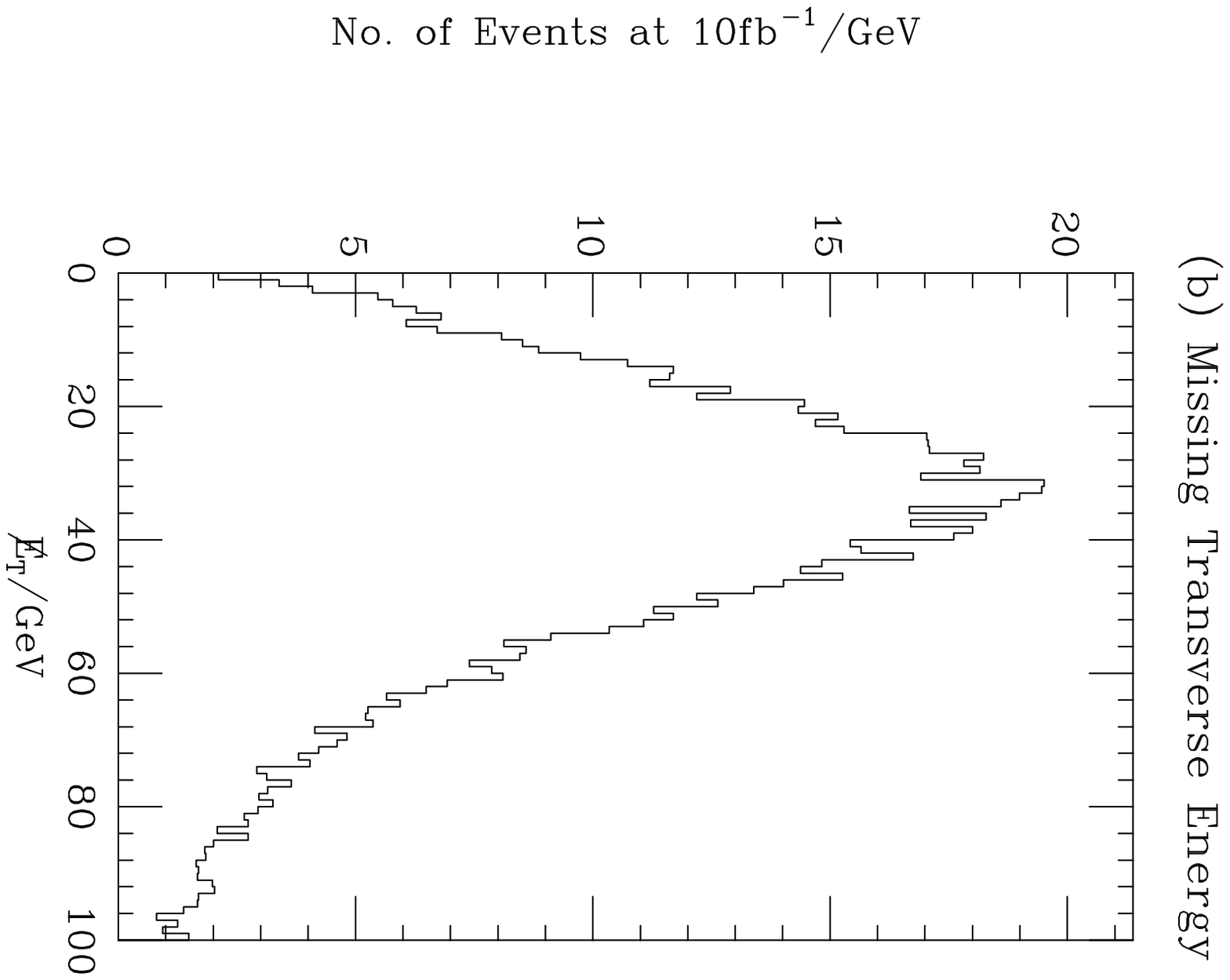}\\
\captionB{$M_T$ and \met\ in WZ events at the LHC.}
	{The transverse mass and missing transverse energy in WZ events
	 at the LHC. The distributions are normalized to
	an integrated luminosity of $10\  \mr{fb}^{-1}$.} 
\label{fig:lhcemiss}
\end{center}
%\end{figure}
%% End of the Figure %%%%%%%%%%%%%%%%%%%%%%%%%%%%%%%%%%%%%%%%%%%%%%%%%%%%%%%%%
%%%%%%%%%%%%%%%%%%%%%%%%%%%%%%%%%%%%%%%%%%%%%%%%%%%%%%%%%%%%%%%%%%%%%%%%%%%%%%
%%
%%  Figure containing the effect of the cuts on WZ and ZZ at the LHC
%%
%\begin{figure}
\begin{center}
\includegraphics[angle=90,width=0.48\textwidth]{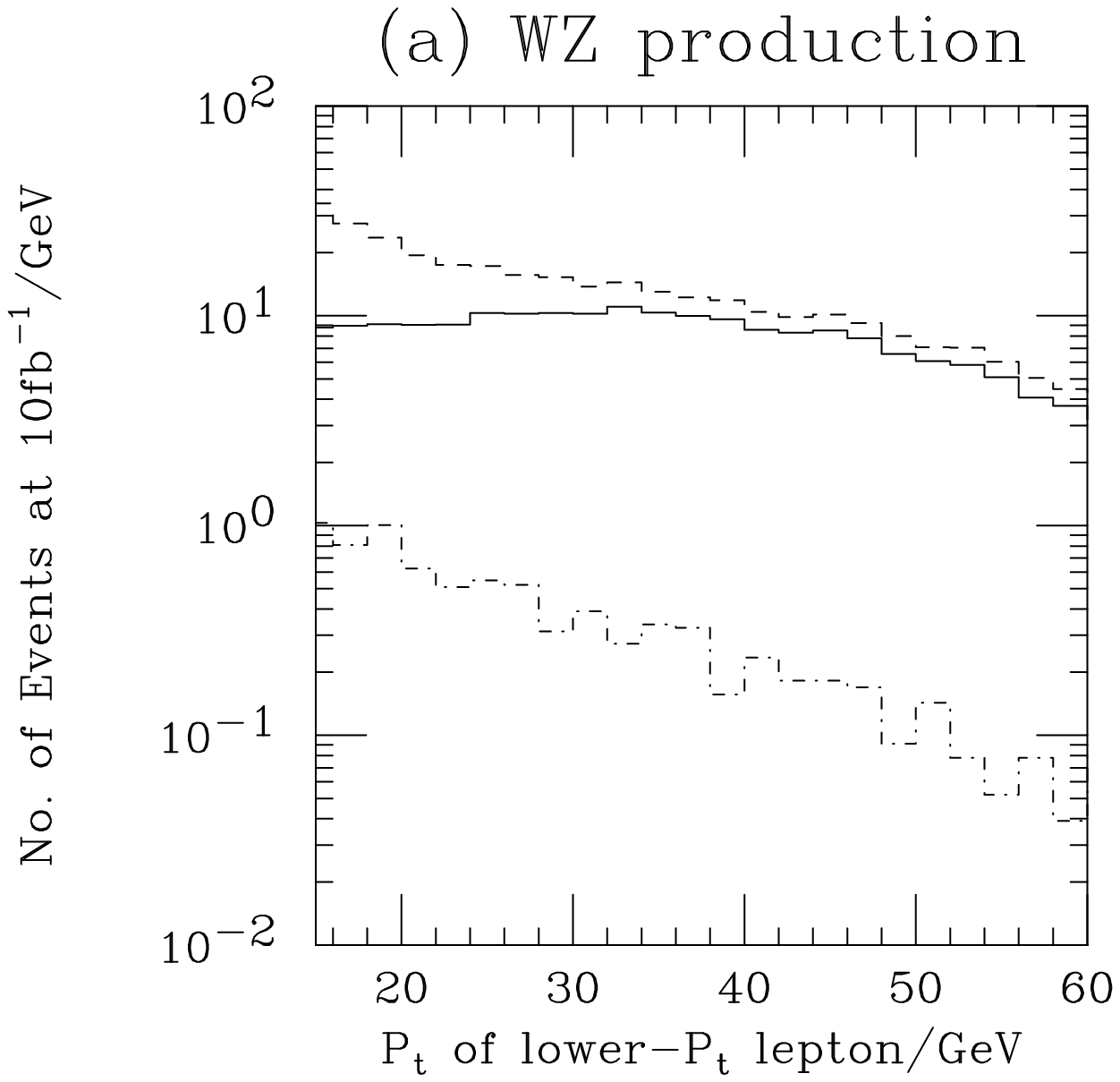}
\hfill
\includegraphics[angle=90,width=0.48\textwidth]{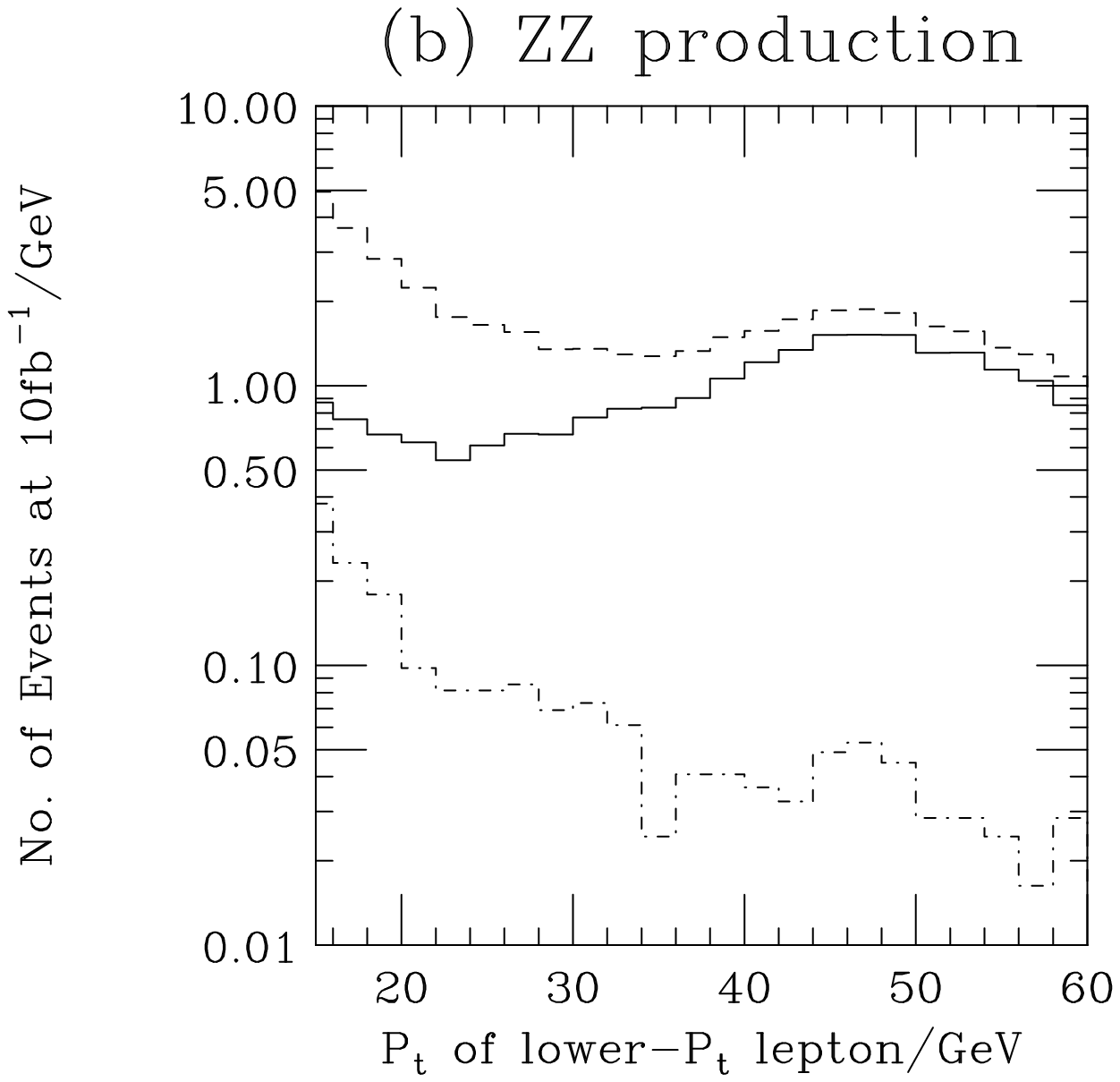}\\
\captionB{Effect of the isolation cuts on the WZ and ZZ backgrounds at 
	the LHC.}
	{Effect of the isolation cuts on the WZ and ZZ backgrounds at 
	the LHC. The dashed line gives the background before any cuts and
	the solid line shows the effect of the isolation cut described in the
	text. The dot-dash line gives the effect of all the cuts, including
	the cut on the number of jets. The distributions are normalized to
	an integrated luminosity of $10\  \mr{fb}^{-1}$.} 
\label{fig:lhcgaugeiso}
\end{center}
\end{figure}
% End of the Figure %%%%%%%%%%%%%%%%%%%%%%%%%%%%%%%%%%%%%%%%%%%%%%%%%%%%%%%%%%

  The remaining cuts reduce the background from gauge boson pair production,
  particularly WZ production, which dominates the Standard Model 
  background after the imposition of the isolation and $p_T$ cuts.
  Fig.\,\ref{fig:lhcemiss}a
  shows that the cut on the transverse mass, \ie removing the region
  $60\, \mr{\gev} < M_T < 85\, \mr{\gev}$  for each of the 
  like sign leptons, will reduce the background
  from WZ production, which is the largest of the gauge boson pair
  production backgrounds. The cut on the missing transverse
  energy, $\not\!\!\!E_T<20\, \mr{\gev}$, 
  will also significantly reduce the background from WZ production, as
  can be seen in Fig.\,\ref{fig:lhcemiss}b. The effect of these cuts is
  shown in Fig.\,\ref{fig:lhcgaugeiso}. Again the simulation of the gauge
  boson pair production backgrounds does not include $\mr{W\gamma}$ 
  production which
  may be an important source of background, but should be significantly
  reduced by the cuts.

%%%%%%%%%%%%%%%%%%%%%%%%%%%%%%%%%%%%%%%%%%%%%%%%%%%%%%%%%%%%%%%%%%%%%%%%%%%%%%
%
%  Table giving the backgrounds for the LHC
%
\begin{table}
\begin{center}
\begin{tabular}{|c|c|c|c|c|}
\hline
	& \multicolumn{4}{c|}{Number of Events} \\
\cline{2-5}
 		& 	     	  &  		    & After isolation,&\\
Background      & After $p_T$ cut & After isolation & $p_T$, $M_T$, \met\
						   cuts & 
		 After all cuts \\
 process        &                 & and $p_T$ cuts  & and OSSF lepton & \\
	        & 		  &		    & veto.  & \\
\hline
WW 		& $3.6\pm0.5$			& $0.0\pm0.06$ 
	        & $0.0\pm0.06$	 		& $0.0\pm0.06$ \\
\hline		                
WZ 		& $239\pm2.5$			& $198.6\pm2.3$
		& $3.8\pm0.3$ 			& $3.8\pm0.3$ \\
\hline				                
ZZ 		& $55.4\pm0.7$			& $45.2\pm0.6$	
		& $1.04\pm0.09$ 		& $1.04\pm0.09$ \\
\hline		                
$\mr{t\bar{t}}$ & $(4.4\pm0.2)\times10^3$ 	& $0.28\pm0.13$
		& $0.06\pm0.06$ 		& $0.06\pm0.06$ \\
\hline		                
$\mr{b\bar{b}}$ & $(4.4\pm0.9)\times10^4$	& $0.0\pm1.6$
		& $0.0\pm1.6$ 			& $0.0\pm1.6$ \\
\hline	                
Single top 	& $36.6\pm1.5$ 		& $0.0\pm0.004$	
		& $0.0\pm0.004$ & $0.0\pm0.004$ \\
\hline		                
\hline		                
Total 	 	& $(4.9\pm0.9)\times10^4$	& $244.1\pm2.9$ 
		& $4.9\pm1.6$ 		& $4.9\pm1.6$ \\
\hline
\end{tabular}
\end{center}
\captionB{Backgrounds to like-sign dilepton production at the LHC.}
	{Backgrounds to like-sign dilepton production at the LHC.
	 The numbers of events
	are based on an integrated luminosity of $10\  \mr{fb}^{-1}$.
	We have calculated an error on the cross section by varying the
	scale between half and twice the hard scale, apart from the
	gauge boson pair cross section where we do not have this
	information and the effect of varying the scale is expected
	to be small
	anyway. The error on the number of events is the error on
	the cross section and the statistical error from the Monte
	Carlo simulation added in quadrature. If no events passed
	the cut the statistical error was taken to be the same
	as if one event had passed the cuts. }
\label{tab:lhcback}
\end{table}
% End of the Table %%%%%%%%%%%%%%%%%%%%%%%%%%%%%%%%%%%%%%%%%%%%%%%%%%%%%%%%%%%

  The effect of all these cuts on the background is shown in 
  Table~\ref{tab:lhcback}.
  This gives a total background after all the cuts of
  $4.9\pm1.6$ events, for 10~$\mr{fb}^{-1}$ integrated luminosity.
  If we take a conservative approach and take a background of
  6.5 events, \ie a $1\sigma$ fluctuation above the central value of our 
  calculation a $5\sigma$ statistical 
  fluctuation would correspond to 16 events, for
  an integrated luminosity of 10~$\mr{fb}^{-1}$. Fig.\,\ref{fig:lhcheavyiso}
  shows that this is a conservative upper bound.

%%%%%%%%%%%%%%%%%%%%%%%%%%%%%%%%%%%%%%%%%%%%%%%%%%%%%%%%%%%%%%%%%%%%%%%%%%%%%%
%
%  Figure containing the signal SM background, no jet cut at the LHC
%
\begin{figure}
\begin{center}
\includegraphics[angle=90,width=0.48\textwidth]{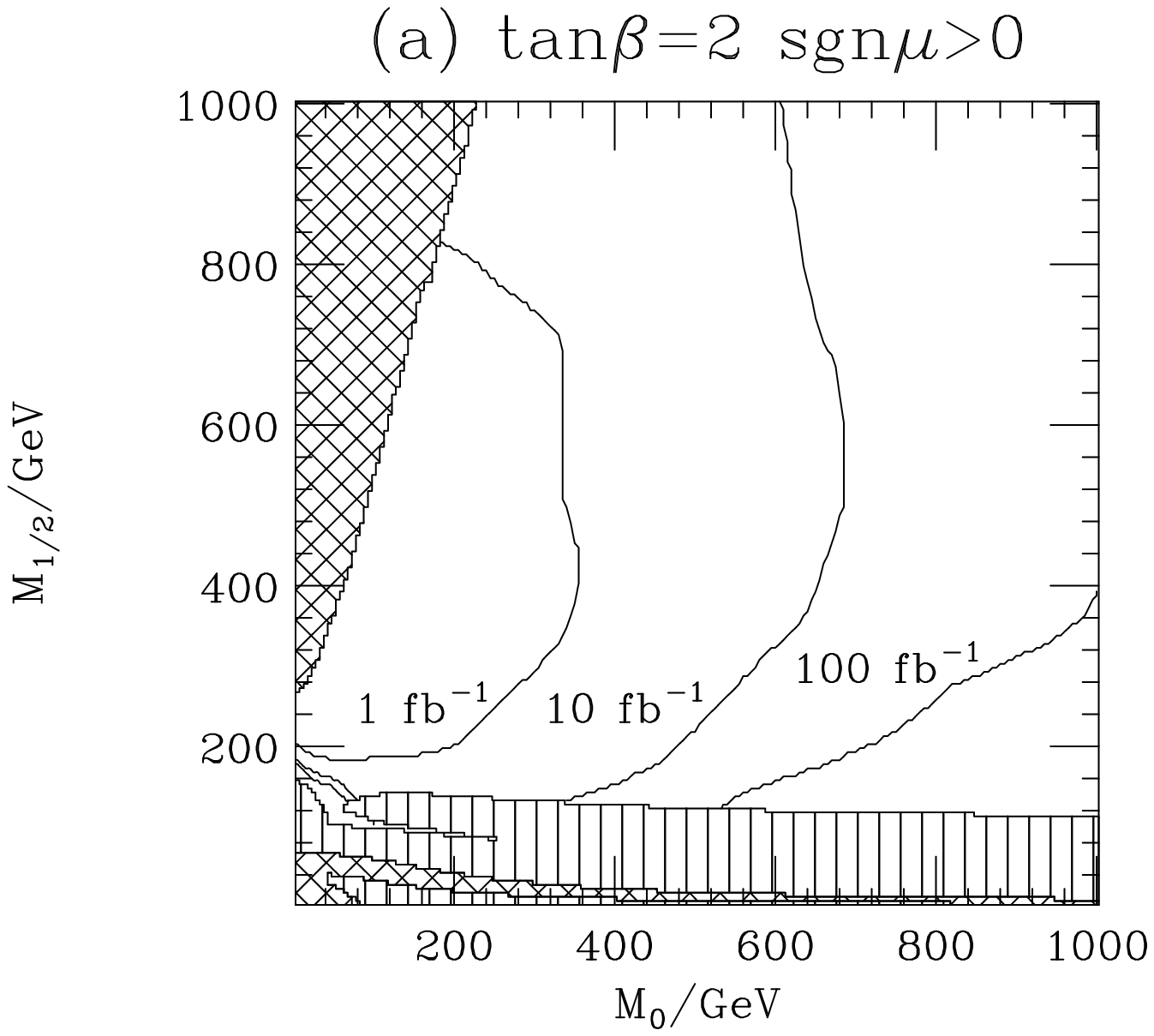}
\hfill
\includegraphics[angle=90,width=0.48\textwidth]{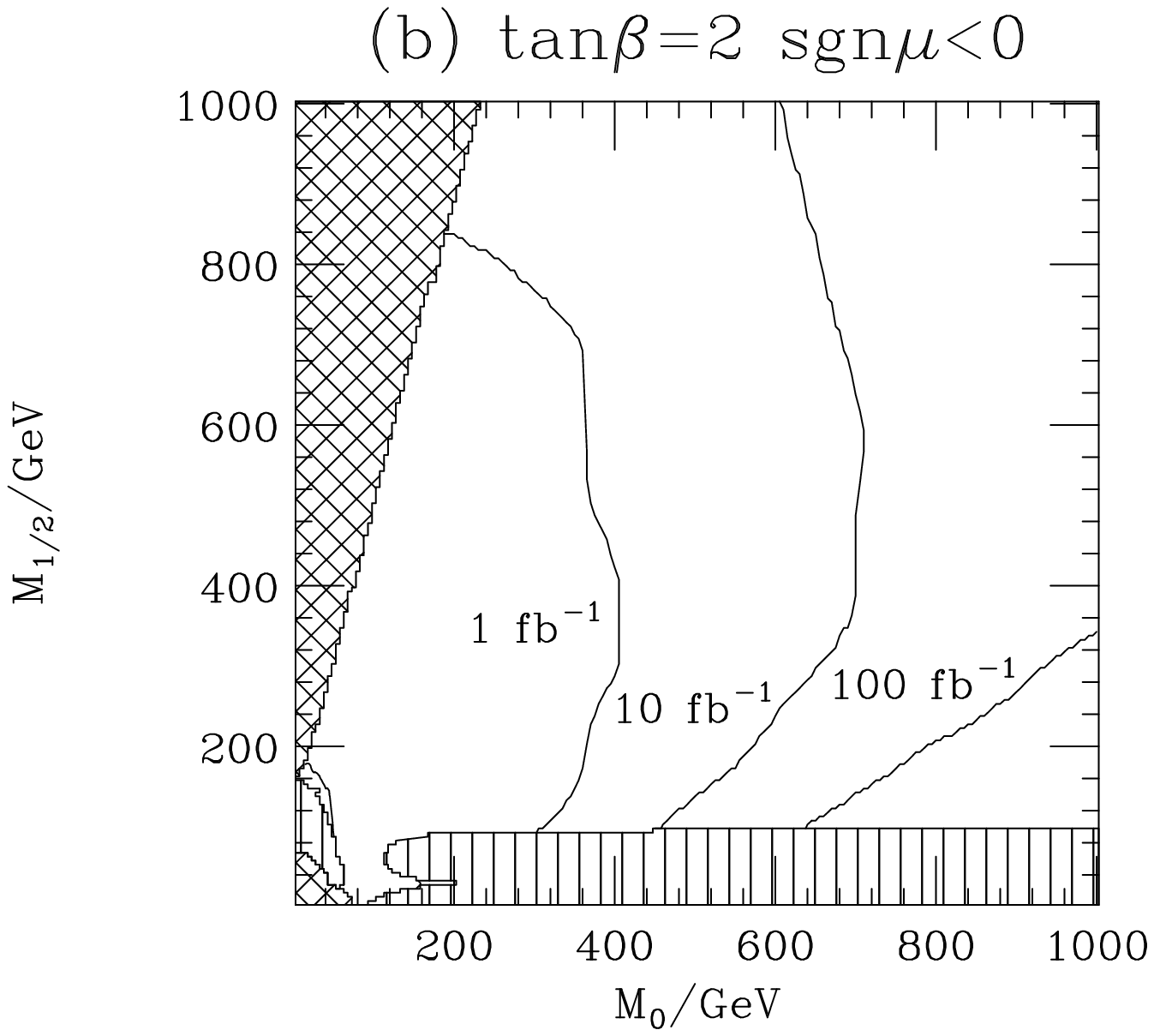}\\
\vskip 15mm
\includegraphics[angle=90,width=0.48\textwidth]{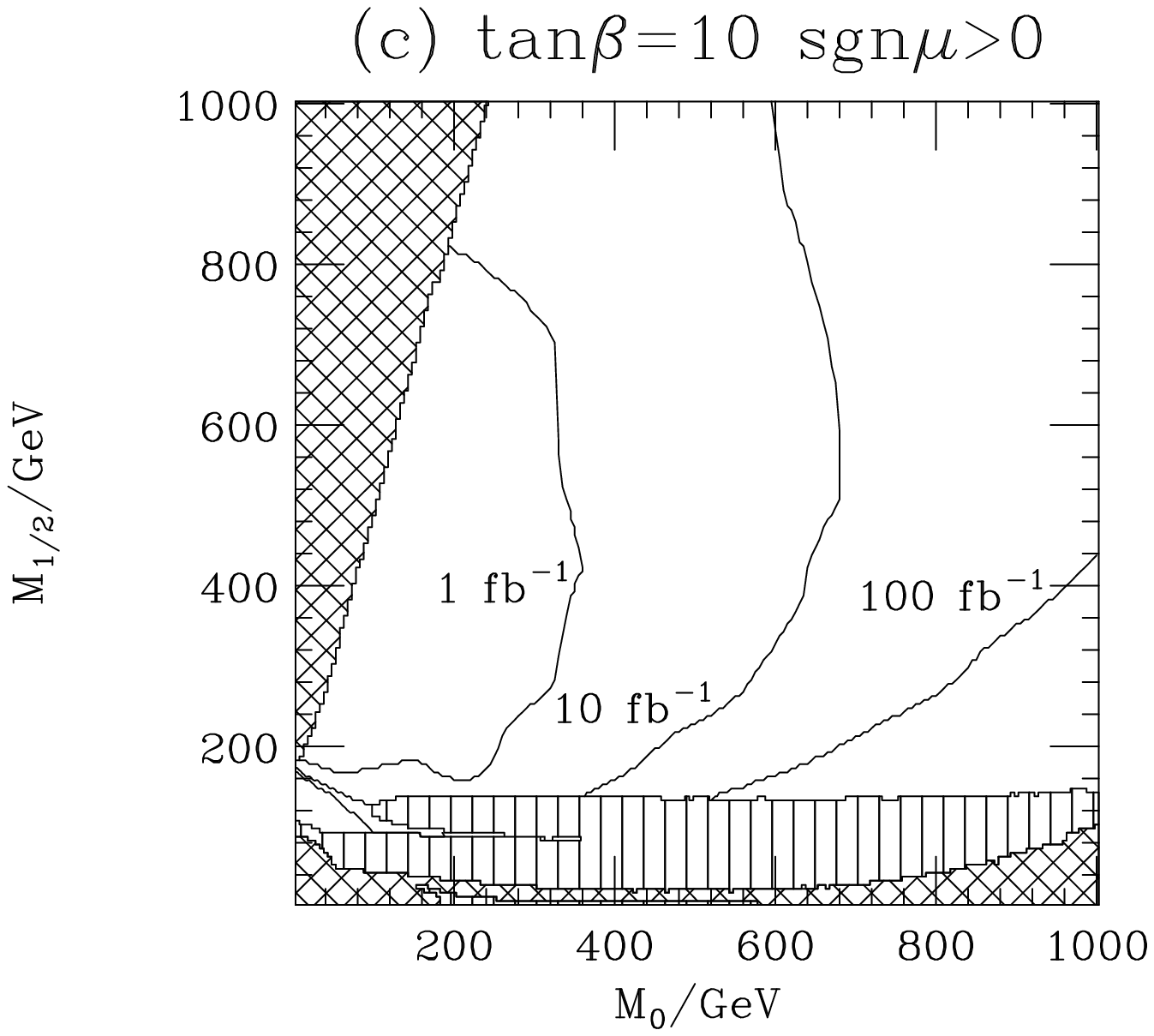}
\hfill
\includegraphics[angle=90,width=0.48\textwidth]{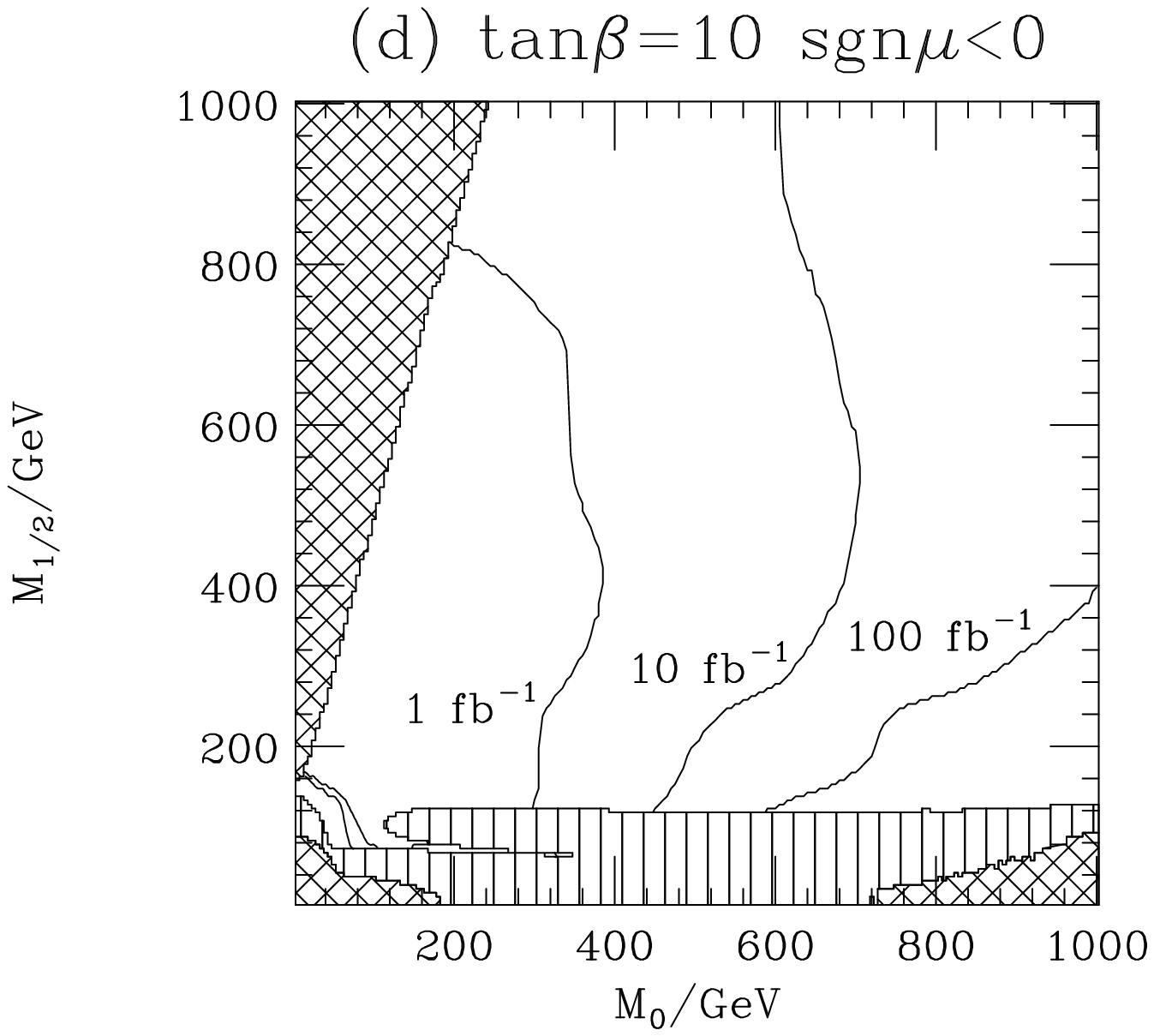}\\
\captionB{Discovery potential at the LHC for the Standard Model backgrounds
	 in the
	$M_0$, $M_{1/2}$ plane with different integrated luminosities.}
	{Contours showing the discovery potential of the LHC in the $M_0$,
	 $M_{1/2}$ plane for ${\lam'}_{211}=10^{-2}$ and $A_0=0\, \mr{\gev}$.
	 These contours are a $5\sigma$ excess of the signal above the
	 background. Here we have imposed cuts on the isolation and $p_T$
  	 of the leptons, the transverse mass and the missing transverse energy
	 described in the text, and a veto on the presence of OSSF leptons.
	 We have only considered the Standard Model background.
	 The striped and hatched regions are
	 described in the caption of Fig.\,\ref{fig:SUSYmass}.} 
\label{fig:lhcSMnojet}
\end{center}
\end{figure}
% End of the Figure %%%%%%%%%%%%%%%%%%%%%%%%%%%%%%%%%%%%%%%%%%%%%%%%%%%%%%%%%%
%%%%%%%%%%%%%%%%%%%%%%%%%%%%%%%%%%%%%%%%%%%%%%%%%%%%%%%%%%%%%%%%%%%%%%%%%%%%%%
%
%  Figure containing the signal SM background, no jet cut at the LHC
%
\begin{figure}
\begin{center}
\includegraphics[angle=90,width=0.48\textwidth]{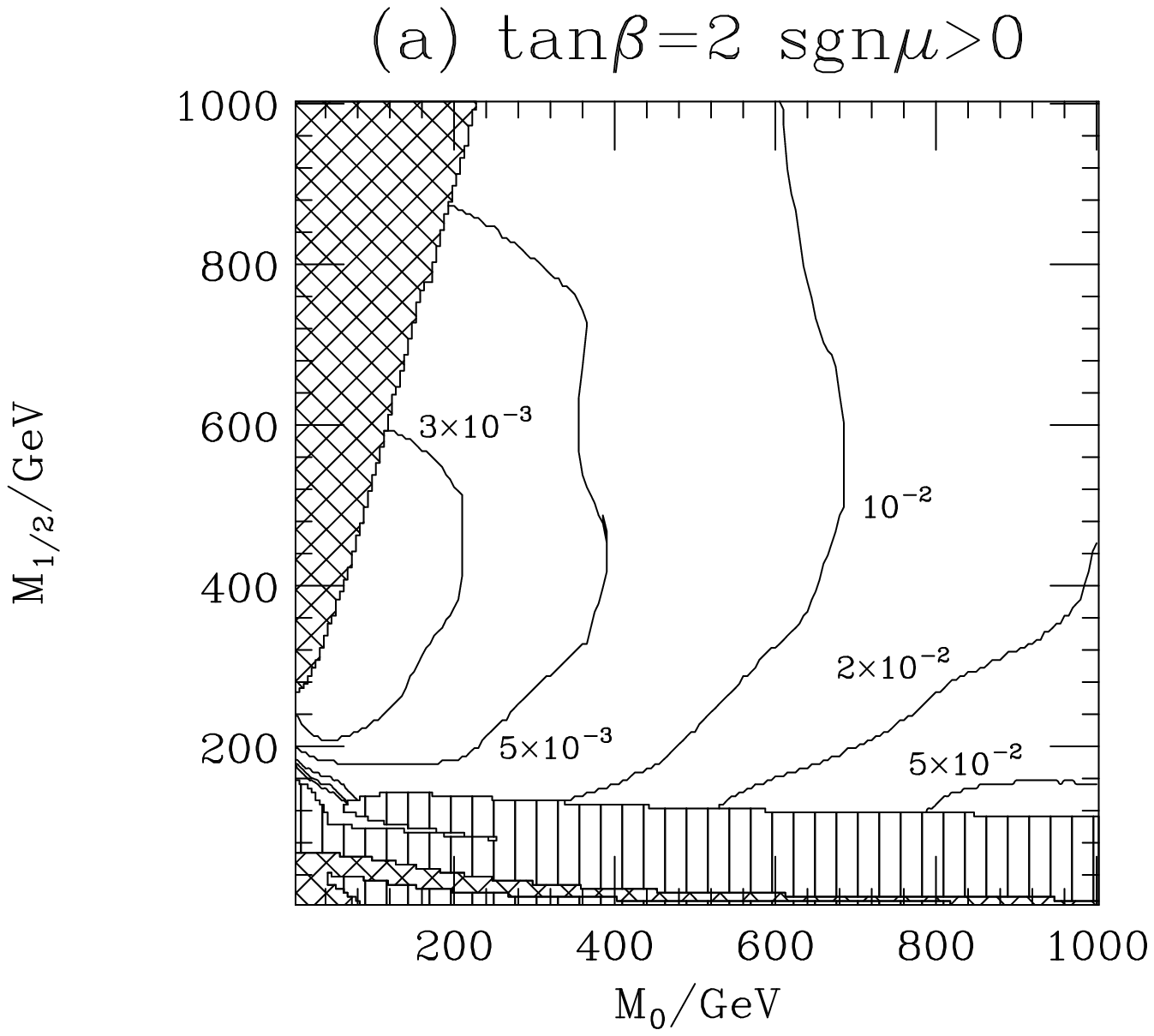}
\hfill						
\includegraphics[angle=90,width=0.48\textwidth]{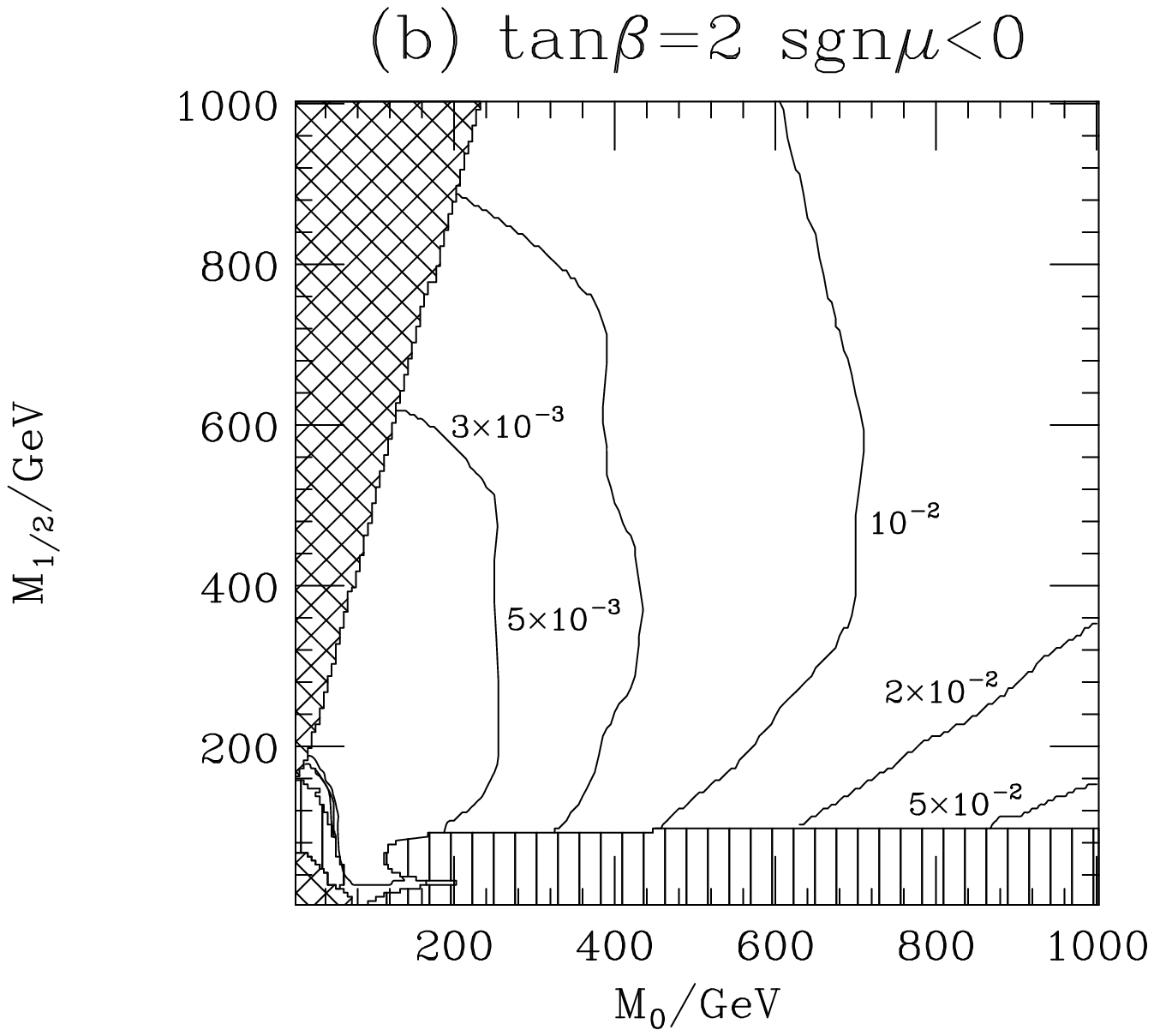}\\
\vskip 15mm					
\includegraphics[angle=90,width=0.48\textwidth]{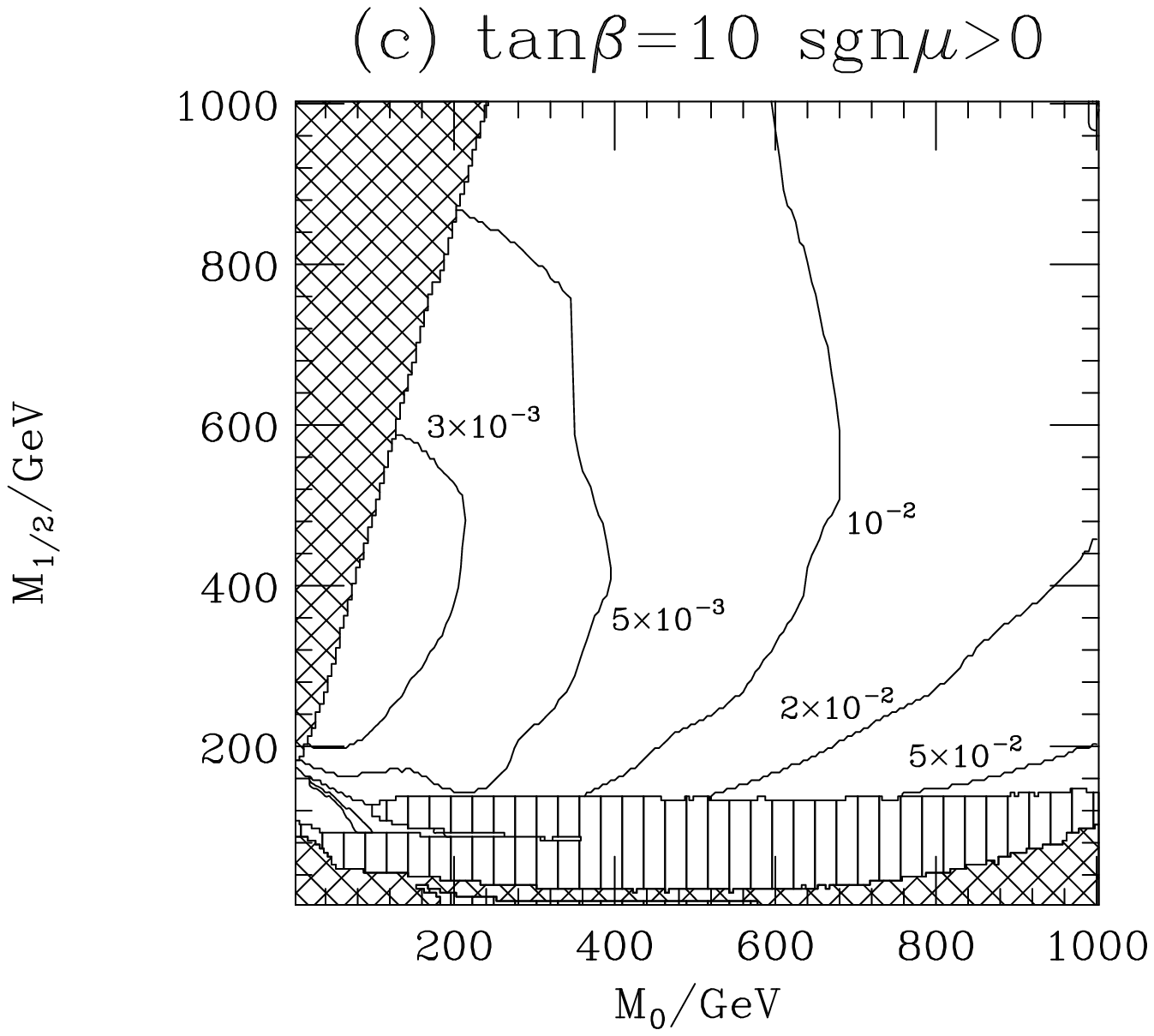}
\hfill						
\includegraphics[angle=90,width=0.48\textwidth]{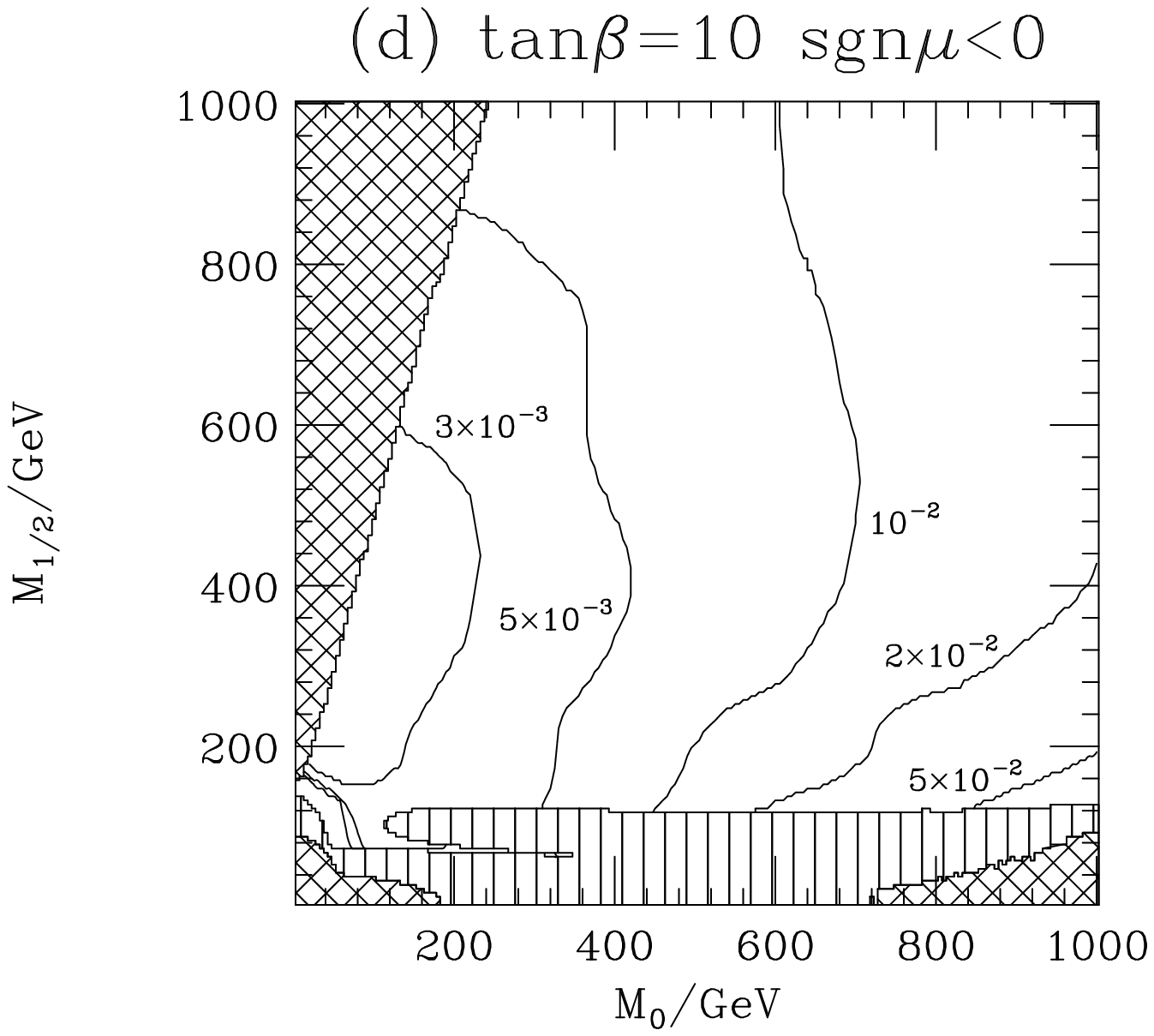}\\
\captionB{Discovery potential at the LHC for the Standard Model backgrounds
	 in the
	$M_0$, $M_{1/2}$ plane with different values of ${\lam'}_{211}$.}
	{Contours showing the discovery potential of the LHC in the $M_0$,
	 $M_{1/2}$ plane for $A_0=0\, \mr{\gev}$  and an integrated 
	luminosity of
	 $10\  \mr{fb}^{-1}$ with different values of ${\lam'}_{211}$.
	 These contours are a $5\sigma$ excess of the signal above the
	 background. Here we have imposed cuts on the isolation and $p_T$
  	 of the leptons, the transverse mass and the missing transverse energy
	 described in the text, and a veto on the presence of OSSF leptons.
	 We have only considered the Standard Model background.
	 The striped and hatched regions are
	 described in the caption of Fig.\,\ref{fig:SUSYmass}.} 
\label{fig:lhcSMnojetb}
\end{center}
\end{figure}
% End of the Figure %%%%%%%%%%%%%%%%%%%%%%%%%%%%%%%%%%%%%%%%%%%%%%%%%%%%%%%%%%

  We adopted the same procedure described in the previous section to 
  obtain the acceptance for the \rpv\  signal given the cuts we have imposed.
  The discovery potential of the
  LHC is shown in Fig.\,\ref{fig:lhcSMnojet}, for ${\lam'}_{211}=10^{-2}$
  with different integrated luminosities, and in Fig.\,\ref{fig:lhcSMnojetb},
  for an integrated luminosity of $10\  \mr{fb}^{-1}$ with different
  values of ${\lam'}_{211}$.
  This is considerably greater
  than the discovery potential of the Tevatron at high $M_0$ and $M_{1/2}$ due
  to the larger centre-of-mass energy of the LHC and hence the larger
  cross sections. In particular, the search potential with one year's running
  at high luminosity, \ie $100\ \mr{fb^{-1}}$, covers large regions of the 
  $M_0$, $M_{1/2}$ plane. At very large values of $M_{1/2}$ this extends
  to regions where the sparticle pair production cross section is
  small due to the high masses of the SUSY particles.

  At small values of $M_0$ and $M_{1/2}$ there are regions of SUGRA parameter
  space which cannot be probed for any couplings due to the cuts we have
  applied.
  However these regions can be excluded by either LEP or the Tevatron and
  we will therefore ignore them in the rest of this analysis. If we
  neglect these regions, the LHC can observe a resonant slepton with
  a mass of up to $510\,(710)\, \mr{\gev}$  for a coupling of
  ${\lam'}_{211}=0.02$ with
  10\,(100)~$\mr{fb}^{-1}$ integrated luminosity and for a coupling
  ${\lam'}_{211}=0.05$ a resonant slepton can be observed with a mass
  of up to $750\,(950)\, \mr{\gev}$ with
  10\,(100)~$\mr{fb}^{-1}$ integrated luminosity. 

  As with the Tevatron analysis, we have neglected the background from
  sparticle pair production. This is reasonable in an initial search for an
  excess of like-sign dilepton pairs over the Standard Model expectation.
  If such an excess were observed it would be necessary to establish
  which process was producing the effect. In the next section, we will
  present the cuts necessary to reduce the background from sparticle pair 
  production and enable a resonant slepton signature to be established
  over all the backgrounds.

\vskip 5mm
\noindent{\underline{SUSY Backgrounds}}
\nopagebreak
\vskip 5mm
\nopagebreak
  The background from sparticle pair production is much more important at the
  LHC than the Tevatron given the much higher cross sections for
  sparticle pair production. The nature of the sparticles produced is also
  different due to the higher energies.
  In the regions of SUGRA parameter space
  where the pair production cross section at the Tevatron is large the
  lightest SUSY particles, \ie the electroweak 
  gauginos, are predominately produced. This
  is because the production of the heavier squarks and gluinos
  is suppressed by
  the higher parton--parton centre-of-mass energies required. However given
  the higher centre-of-mass energy of the LHC the production of the
  coloured sparticles which occurs via the strong interaction dominates the
  cross section. This means that a cut on the number of jets in an event
  will be more effective in reducing the background from sparticle pair
  production. The following cut was applied:
\begin{itemize}
\item 	Vetoing all events when there are more than two jets each with
	$p_T>50\, \mr{\gev}$. 
\end{itemize}

%%%%%%%%%%%%%%%%%%%%%%%%%%%%%%%%%%%%%%%%%%%%%%%%%%%%%%%%%%%%%%%%%%%%%%%%%%%%%%
%
%  Figure containing the signal all backgrounds+jet cut at the LHC
%
\begin{figure}[htp]
\begin{center}
\includegraphics[angle=90,width=0.48\textwidth]{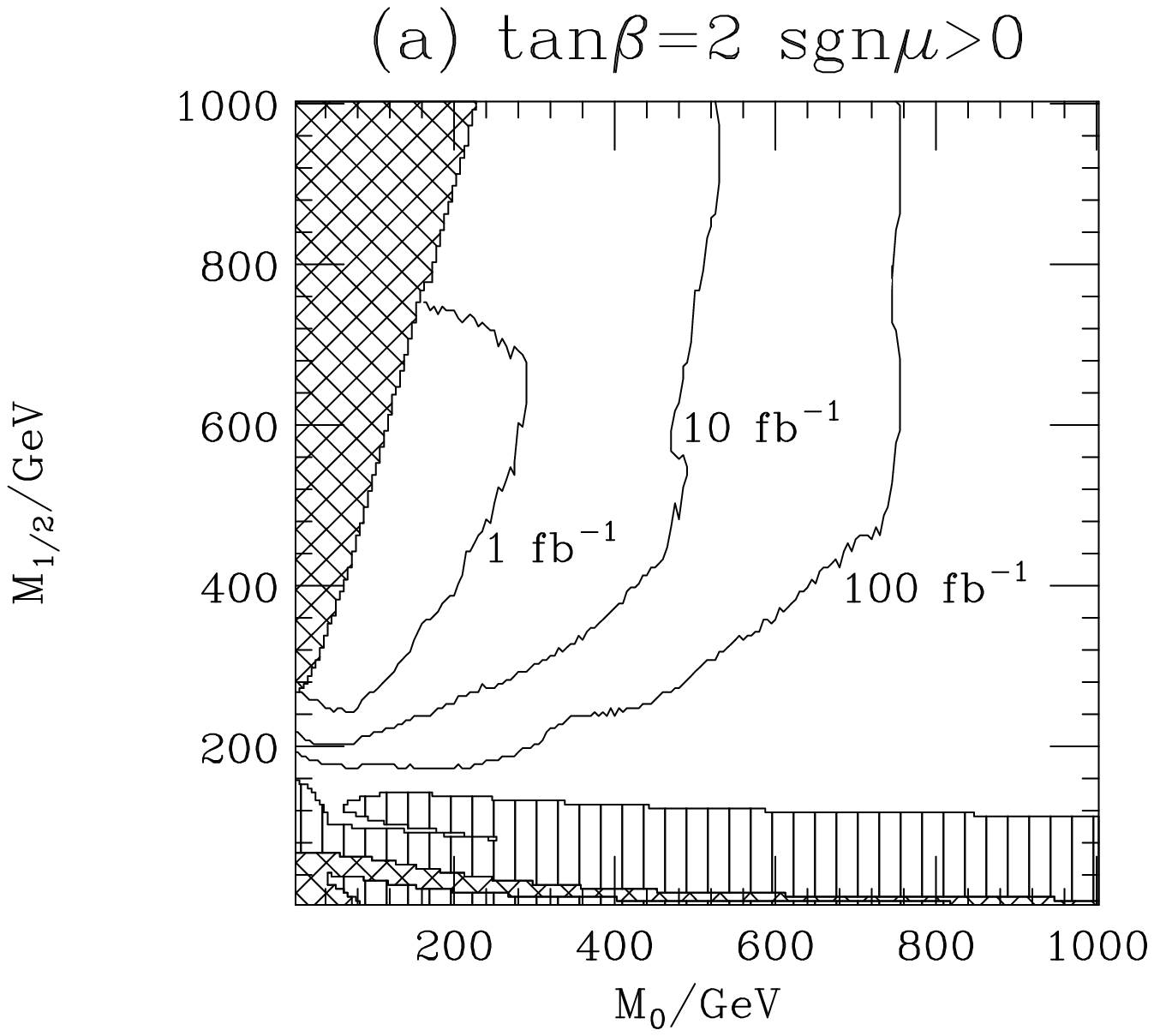}
\hfill
\includegraphics[angle=90,width=0.48\textwidth]{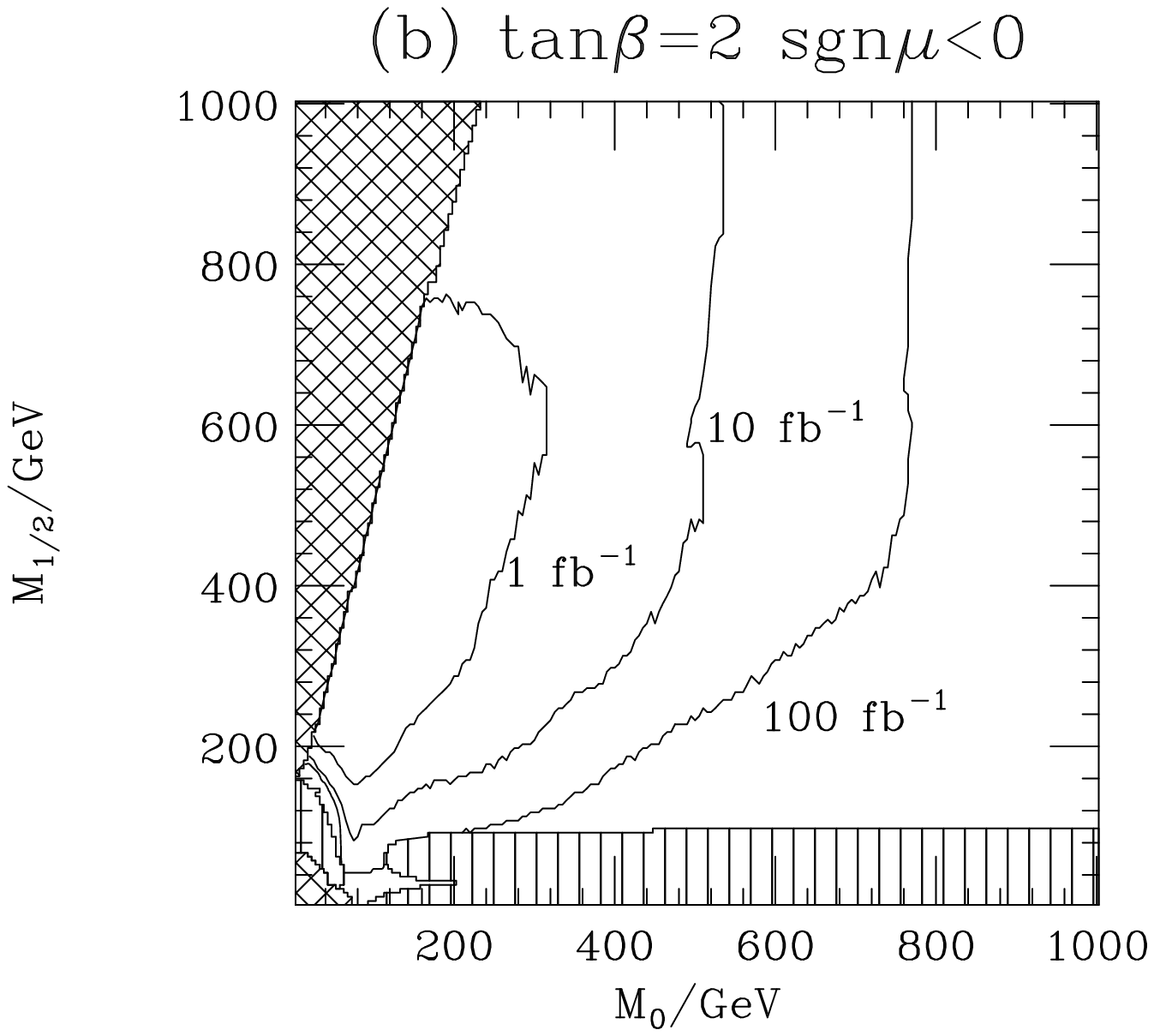}\\
\vskip 15mm
\includegraphics[angle=90,width=0.48\textwidth]{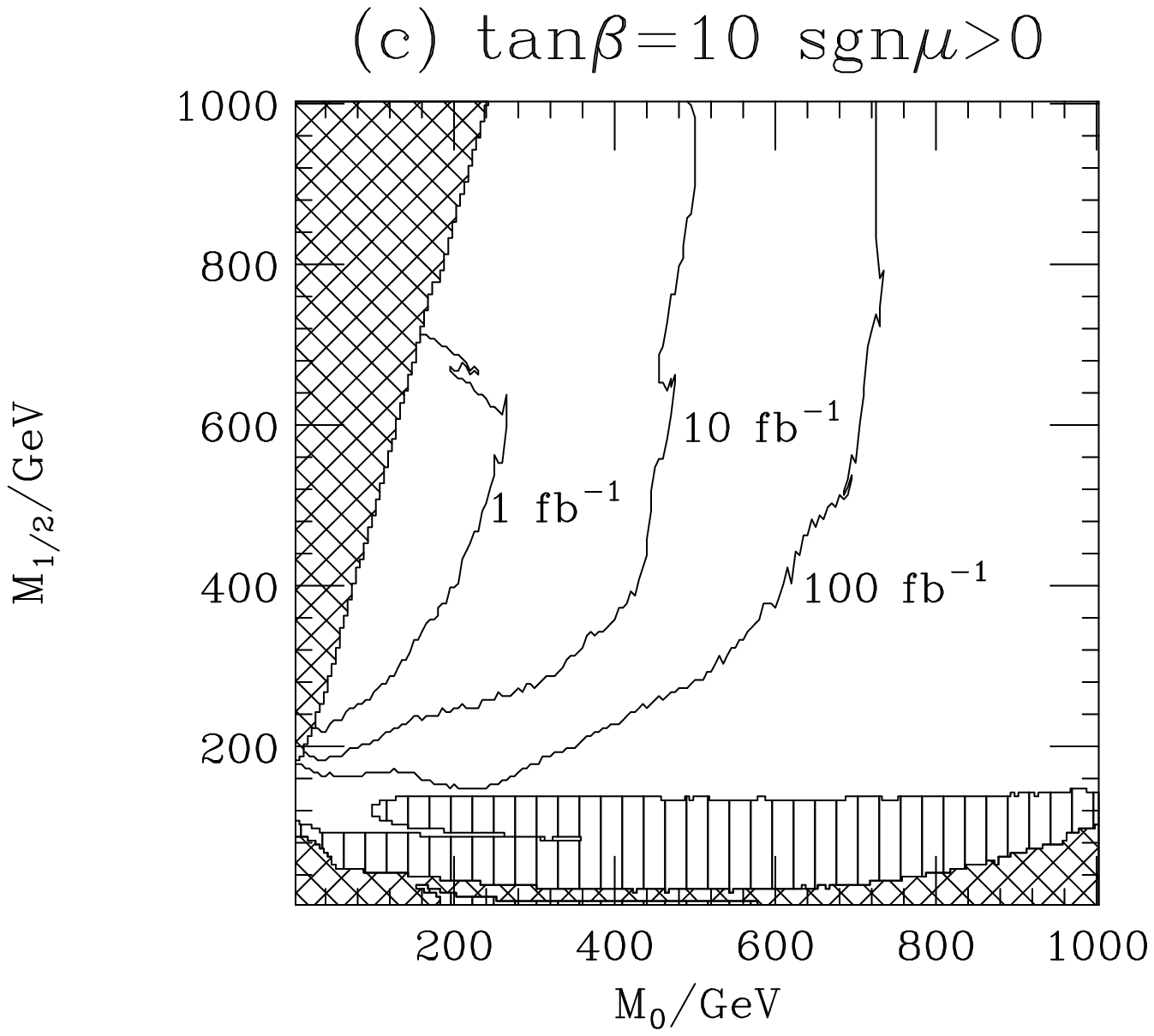}
\hfill
\includegraphics[angle=90,width=0.48\textwidth]{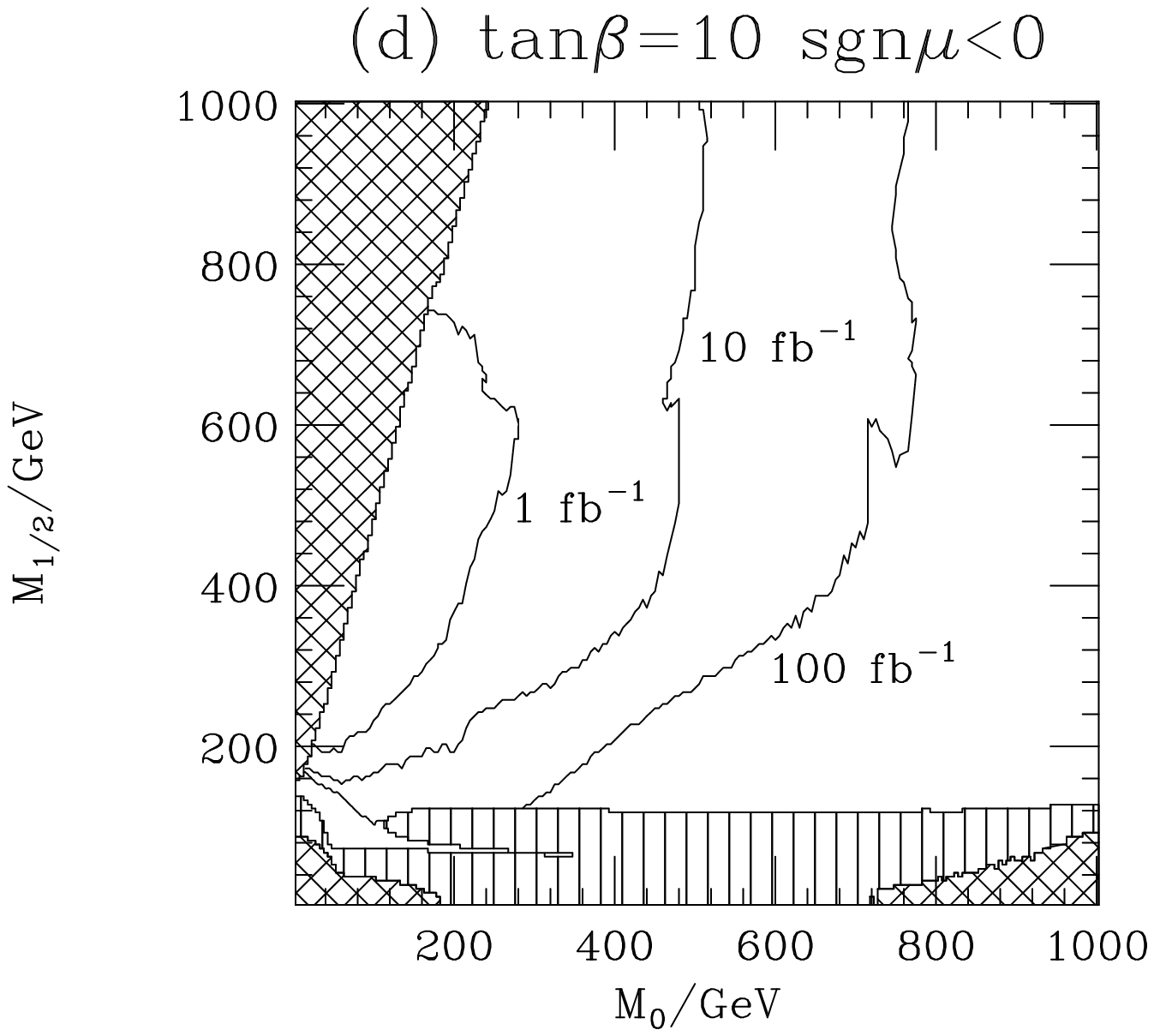}\\
\captionB{Discovery potential at the LHC in the $M_0$, $M_{1/2}$ plane,
	 including 
	all the backgrounds, for different integrated luminosities.}
	{Contours showing the discovery potential of the LHC in the $M_0$,
	 $M_{1/2}$ plane for ${\lam'}_{211}=10^{-2}$ and $A_0=0\, \mr{\gev}$.
	 These contours are a $5\sigma$ excess of the signal above the
	 background. Here, in addition to the cuts on the isolation and $p_T$
  	 of the leptons, the transverse mass and the missing transverse energy
	 described in the text, and a veto on the presence of OSSF leptons,
	 we have imposed a cut on the presence of more than two jets. We
	 have included the sparticle pair production background as well as the
	 Standard Model backgrounds.
	 The striped and hatched regions are
	 described in the caption of Fig.\,\ref{fig:SUSYmass}.} 
\label{fig:lhcSUSYjet}
\end{center}
\end{figure}
% End of the Figure %%%%%%%%%%%%%%%%%%%%%%%%%%%%%%%%%%%%%%%%%%%%%%%%%%%%%%%%%%
%%%%%%%%%%%%%%%%%%%%%%%%%%%%%%%%%%%%%%%%%%%%%%%%%%%%%%%%%%%%%%%%%%%%%%%%%%%%%%
%
%  Figure containing the signal all backgrounds+jet cut at the LHC
%
\begin{figure}[htp]
\begin{center}
\includegraphics[angle=90,width=0.48\textwidth]{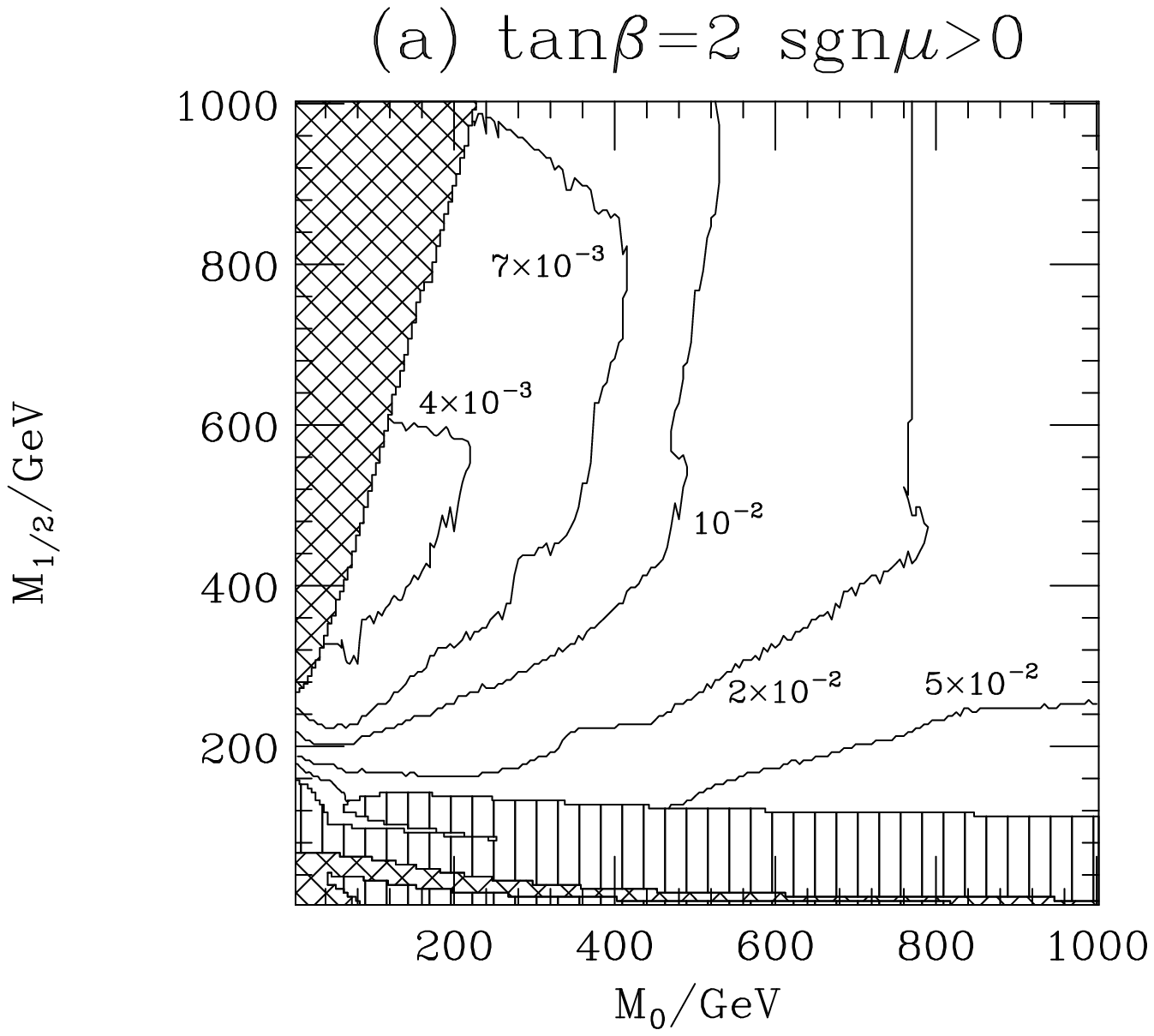}
\hfill						
\includegraphics[angle=90,width=0.48\textwidth]{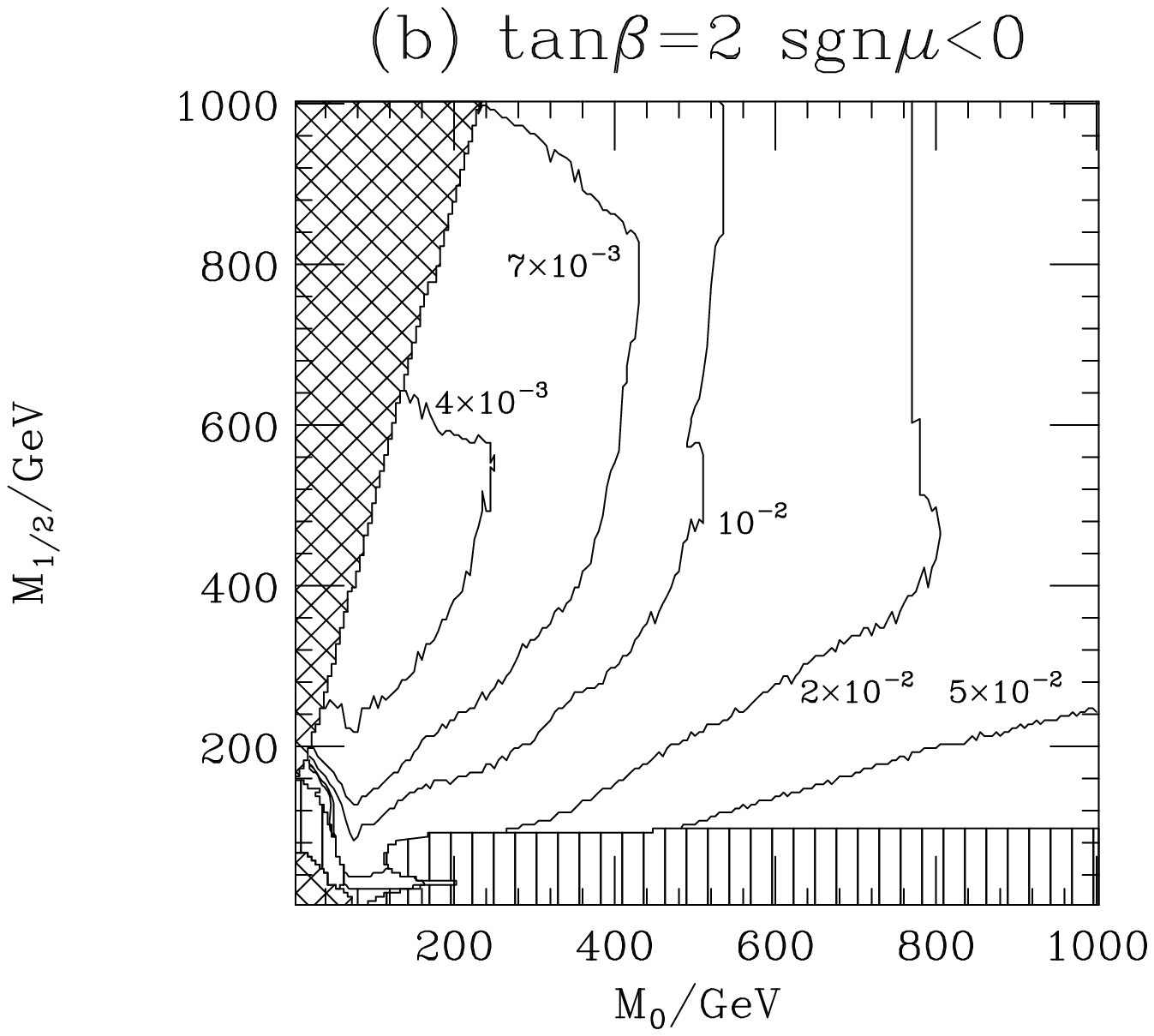}\\
\vskip 15mm					
\includegraphics[angle=90,width=0.48\textwidth]{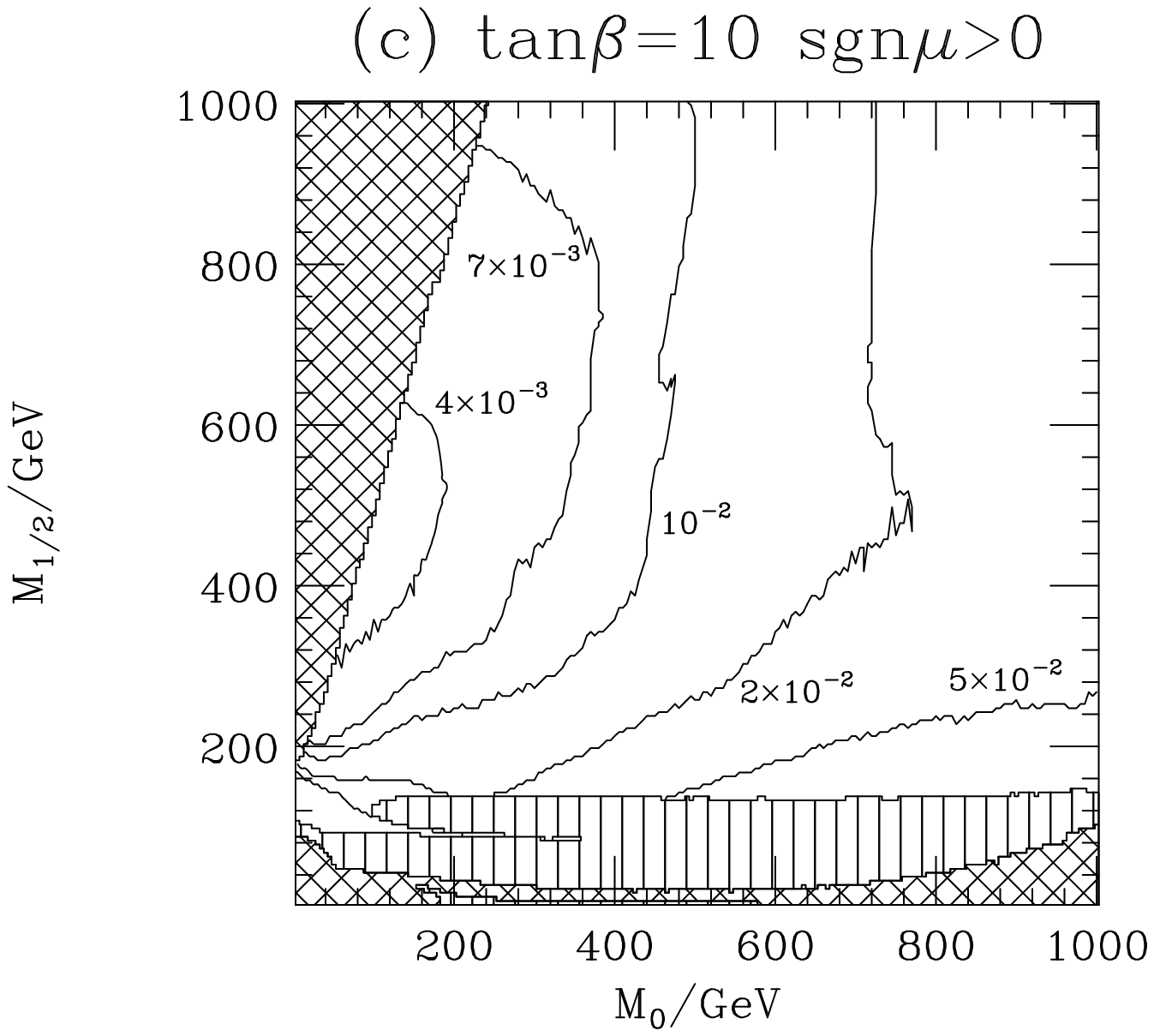}
\hfill						
\includegraphics[angle=90,width=0.48\textwidth]{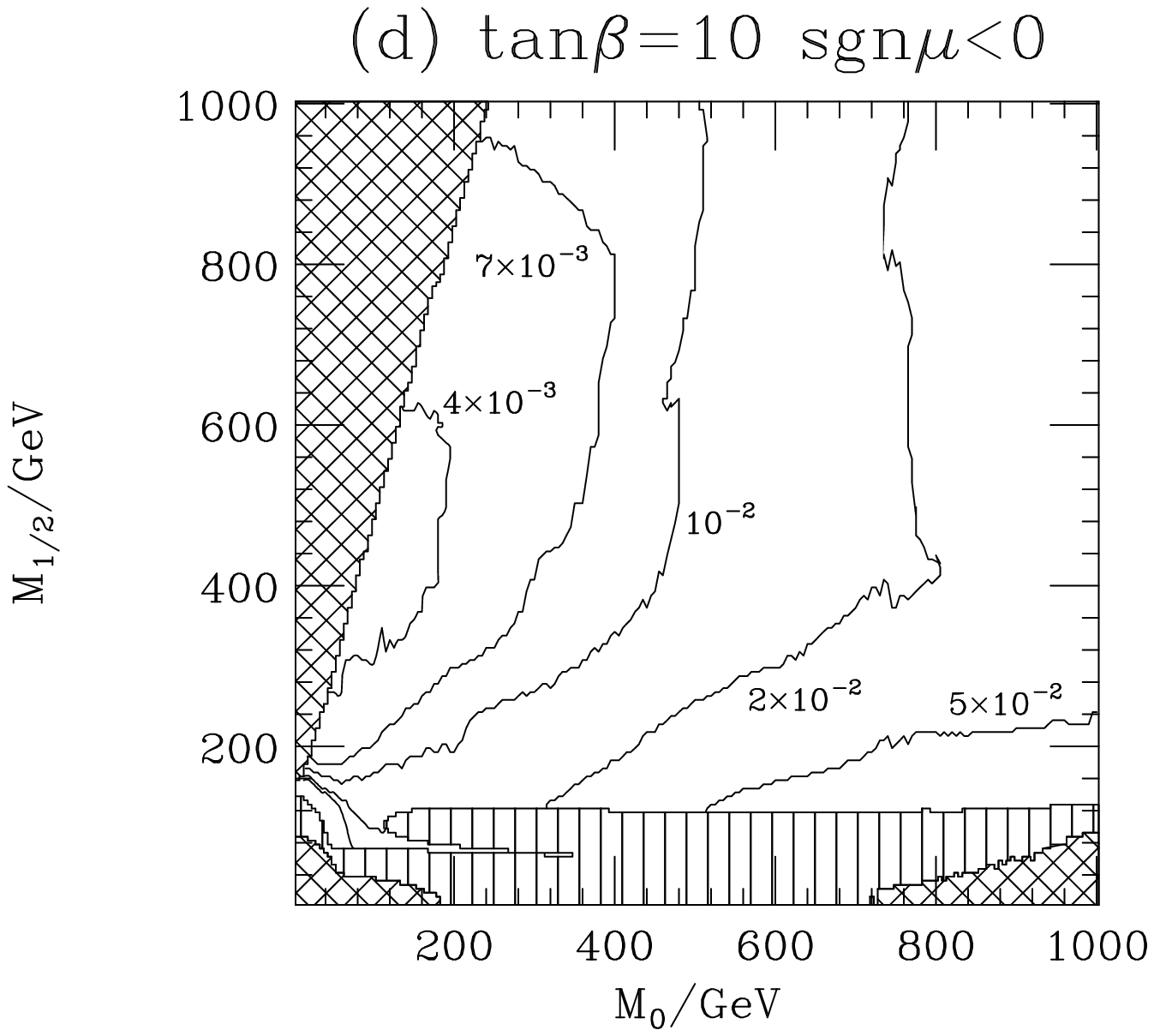}\\
\captionB{Discovery potential at the LHC in the $M_0$, $M_{1/2}$ plane, 
	including 
	all the backgrounds, for different values of ${\lam'}_{211}$.}
	{Contours showing the discovery potential of the LHC in the $M_0$,
	 $M_{1/2}$ plane for $A_0=0\, \mr{\gev}$  and an integrated
	 luminosity of
  	 $10\  \mr{fb}^{-1}$ with different values of ${\lam'}_{211}$.
	 These contours are a $5\sigma$ excess of the signal above the
	 background. Here, in addition to the cuts on the isolation and $p_T$
  	 of the leptons, the transverse mass and the missing transverse energy
	 described in the text, and a veto on the presence of OSSF leptons,
	 we have imposed a cut on the presence of more than two jets. We
	 have included the sparticle pair production background as well as the
	 Standard Model backgrounds.
	 The striped and hatched regions are
	 described in the caption of Fig.\,\ref{fig:SUSYmass}.} 
\label{fig:lhcSUSYjetb}
\end{center}
\end{figure}
% End of the Figure %%%%%%%%%%%%%%%%%%%%%%%%%%%%%%%%%%%%%%%%%%%%%%%%%%%%%%%%%%

  As the sparticle pair production background at the LHC is larger than at
  the Tevatron we needed to simulate more events in order to obtain a reliable
  estimate of the acceptance for this background. This meant that
  with the available resources we were forced to use a coarser scan of
  the $M_0$, $M_{1/2}$ plane. We used a 16 point grid and simulated a 
  different number of events at each point 
  depending on the value of $M_{1/2}$ as
  the sparticle pair production cross section decreases as $M_{1/2}$
  increases. We simulated $10^5$, $10^5$, $10^6$,  and $10^7$ events
  at each of four points in $M_0$ 
  for $M_{1/2}=875\, \mr{\gev}$, $M_{1/2}=625\, \mr{\gev}$,
  $M_{1/2}=375\, \mr{\gev}$ and $M_{1/2}=125\, \mr{\gev}$, respectively.

  Our estimate of the discovery potential of the LHC after this cut, 
  including all the
  backgrounds is given in Fig.\,\ref{fig:lhcSUSYjet}, for
  ${\lam}_{211}=10^{-2}$ with different integrated luminosities, and in
  Fig.\,\ref{fig:lhcSUSYjetb}, for
  an integrated luminosity of $10\  \mr{fb}^{-1}$
  with different values of the \rpv\  Yukawa couplings. As with the
  Tevatron, the discovery potential 
  is reduced in two regions relative to that shown in 
  Figs.\,\ref{fig:lhcSMnojet} and \ref{fig:lhcSMnojetb}.
  The reduction at high $M_{1/2}$ is due to the smaller signal after the
  imposition of the jet cut, whereas the reduction at small $M_{1/2}$ is
  due to the larger background. However there are still large regions of SUGRA
  parameter space in which this process is visible above the background,
  particularly at large $M_{1/2}$ where there is less sensitivity to sparticle
  pair production. Due to the larger backgrounds from sparticle pair
  production there are large regions, at small $M_{1/2}$,
  where the signal is detectable above
  the background although the $S/B$ is small. In general there is a region
  extending around $200\, \mr{\gev}$  in $M_{1/2}$ above the bottom of the
  $5\sigma$ discovery contour for 100~$\mr{fb}^{-1}$ where $S/B<1$.

  If we again neglect the region at small $M_0$ and $M_{1/2}$, which
  cannot be probed for any \rpv\  Yukawa couplings given our cuts, we 
  can obtain a mass reach for the LHC with a given \rpv\  Yukawa coupling.
  Slepton masses of $460\,(600)\, \mr{\gev}$  can be
  discovered with 10\,(100)~$\mr{fb}^{-1}$
  integrated luminosity for a coupling ${\lam'}_{211}=0.05$ and
  slepton masses of $610\,(820)\, \mr{\gev}$  can be observed with
  10\,(100)~$\mr{fb}^{-1}$
  integrated luminosity for a coupling ${\lam'}_{211}=0.1$.

%
%  Subsection on Mass reconstruction
%
\subsection{Mass Reconstruction}

   There are many possible models which lead to an excess of like-sign
   dilepton pairs over the prediction of the Standard Model. Indeed, we
   have seen that within the \rpv\  extension of the MSSM such an excess
   could be due to either sparticle pair production followed by \rpv\  
   decays of the LSPs, or resonant charged slepton production followed
   by a supersymmetric gauge decay of the slepton. The cut
   on the number of jets described above gives one way of discriminating
   between these two scenarios. 

\begin{figure}[htp]
\begin{center}
\includegraphics[angle=90,width=0.48\textwidth]{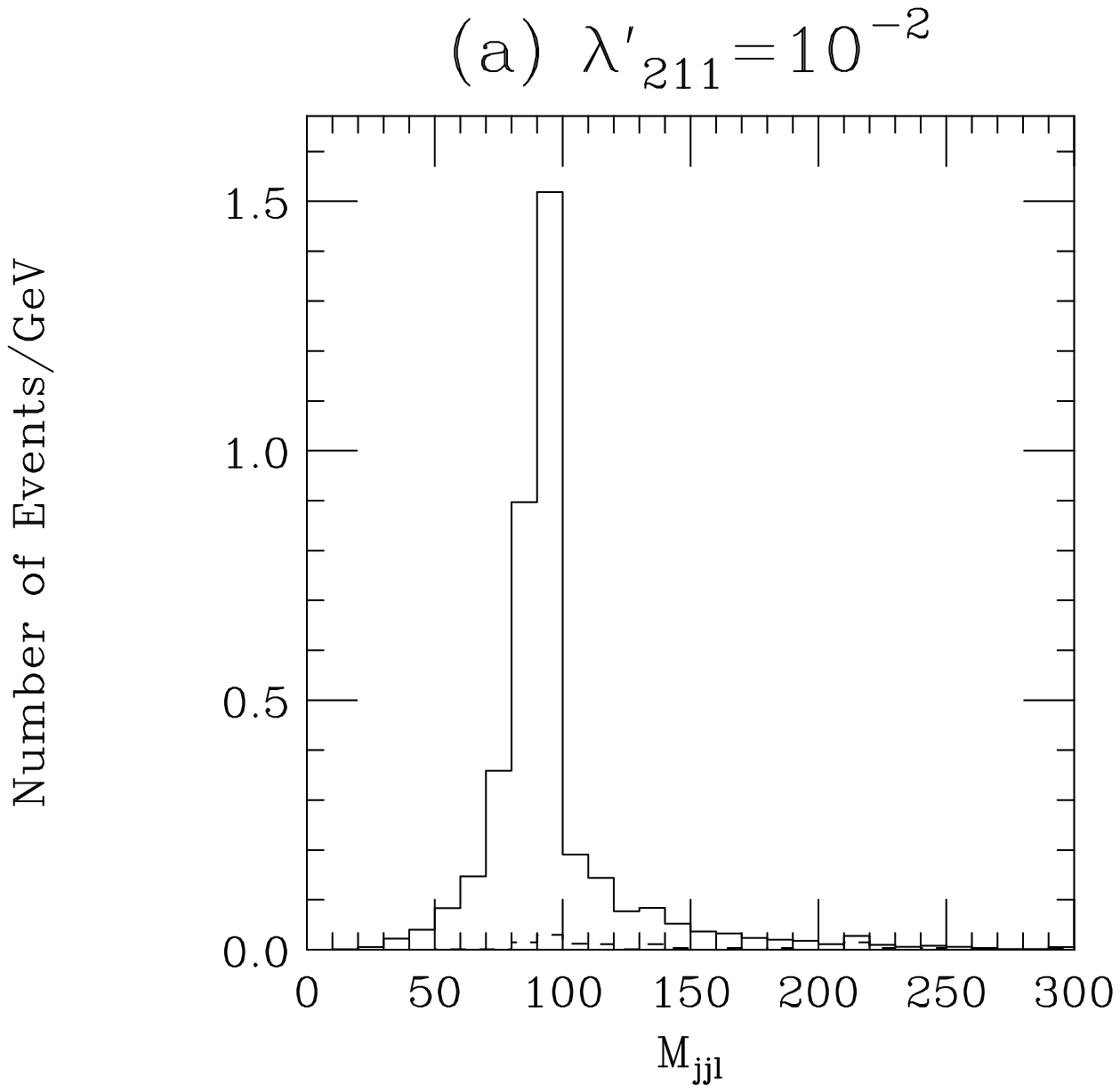}
\hfill
\includegraphics[angle=90,width=0.48\textwidth]{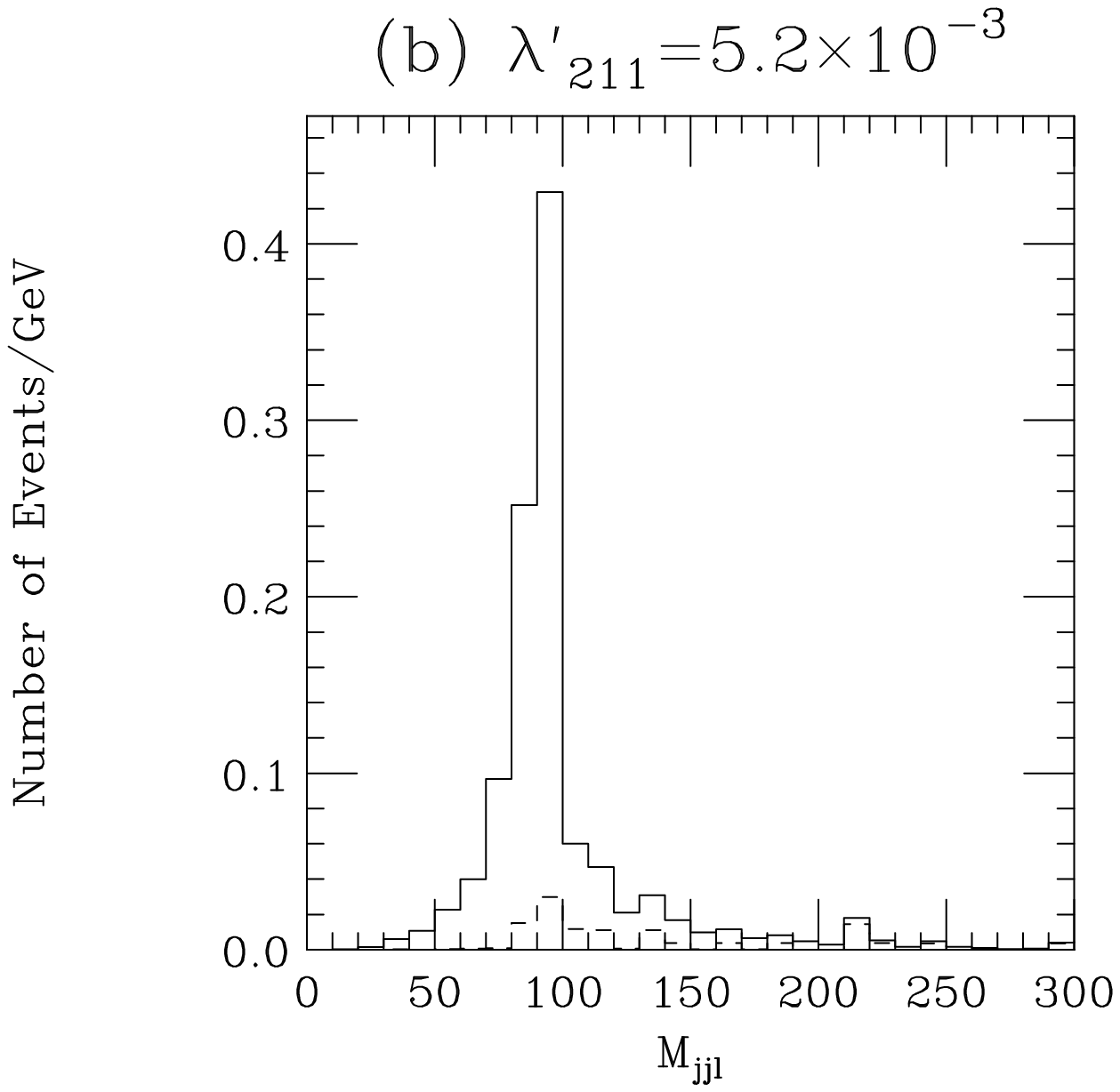}\\
\captionB{Reconstructed  $\mr{\cht^0_1}$  mass at  the Tevatron for 
	$M_0=50\, \mr{\gev}$,  \mbox{$M_{1/2}=250\, \mr{\gev}$},
	$\tan\beta=2$, $\sgn\mu>0$ and $A_0=0\, \mr{\gev}$.}
	{The reconstructed neutralino mass at the Tevatron 
	for $M_0=50\, \mr{\gev}$, \linebreak $M_{1/2}=250\, \mr{\gev}$,
	$\tan\beta=2$, $\sgn\mu>0$ and $A_0=0\, \mr{\gev}$. The value of the
	coupling in
     	(b) is chosen such that after the cuts applied in 
	Section~\ref{sec:results}
	the signal is $5\sigma$ above the background. At this point the 
	lightest neutralino mass is $M_{\cht^0_1}=98.9\, \mr{\gev}$.
	We have normalized the
	distributions to an integrated luminosity of $2\ \mr{fb}^{-1}$.
	The dashed line shows the background and the solid line shows
	the sum of the signal and the background.} 
\label{fig:tevsusyneutmass}
\end{center}
%\end{figure}
% End of the Figure %%%%%%%%%%%%%%%%%%%%%%%%%%%%%%%%%%%%%%%%%%%%%%%%%%%%%%%%%%
%%%%%%%%%%%%%%%%%%%%%%%%%%%%%%%%%%%%%%%%%%%%%%%%%%%%%%%%%%%%%%%%%%%%%%%%%%%%%%
%
%  Figure containing the neutralino masses at the LHC
%
%\begin{figure}
\begin{center}
\includegraphics[angle=90,width=0.48\textwidth]{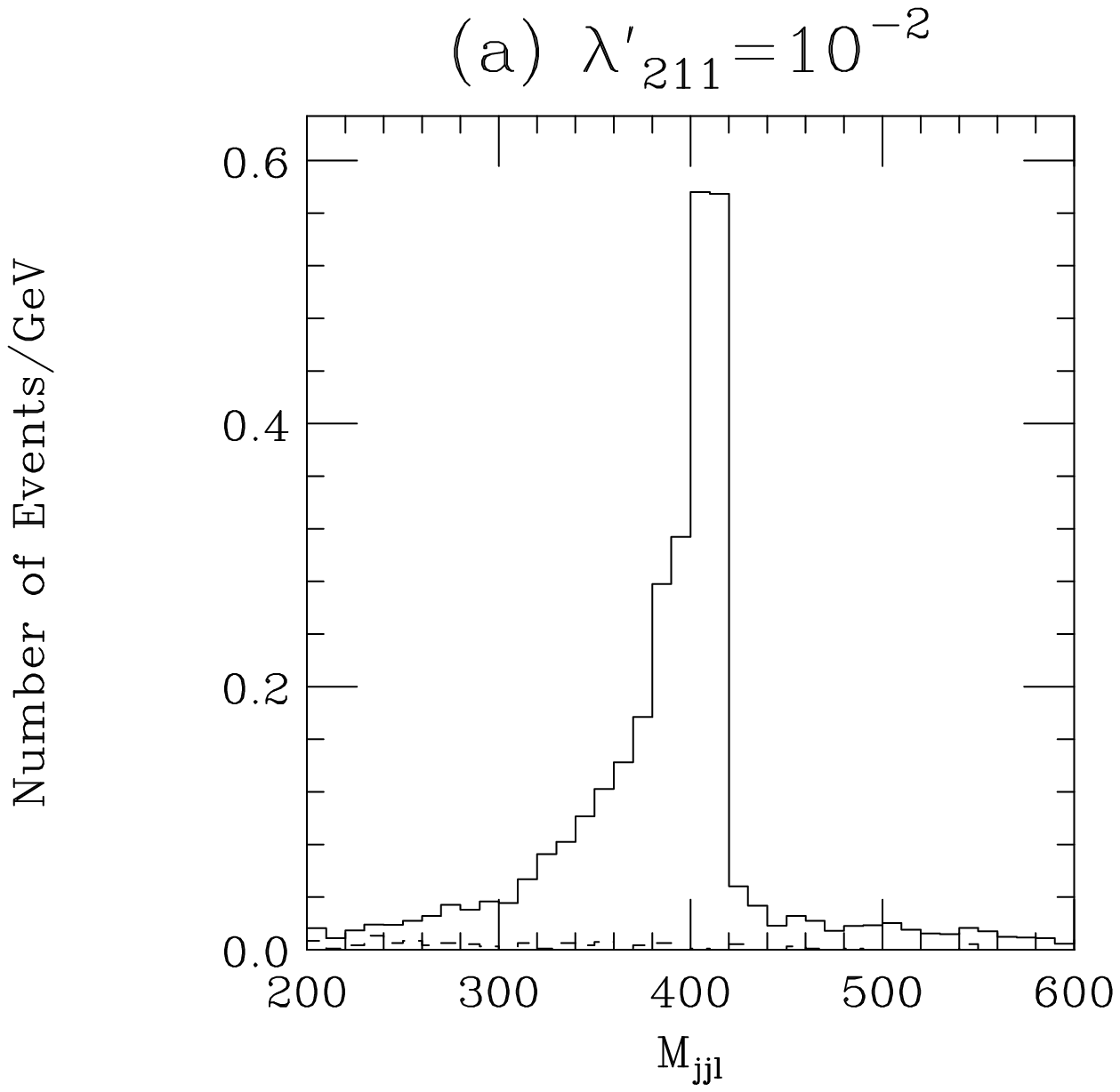}
\hfill
\includegraphics[angle=90,width=0.48\textwidth]{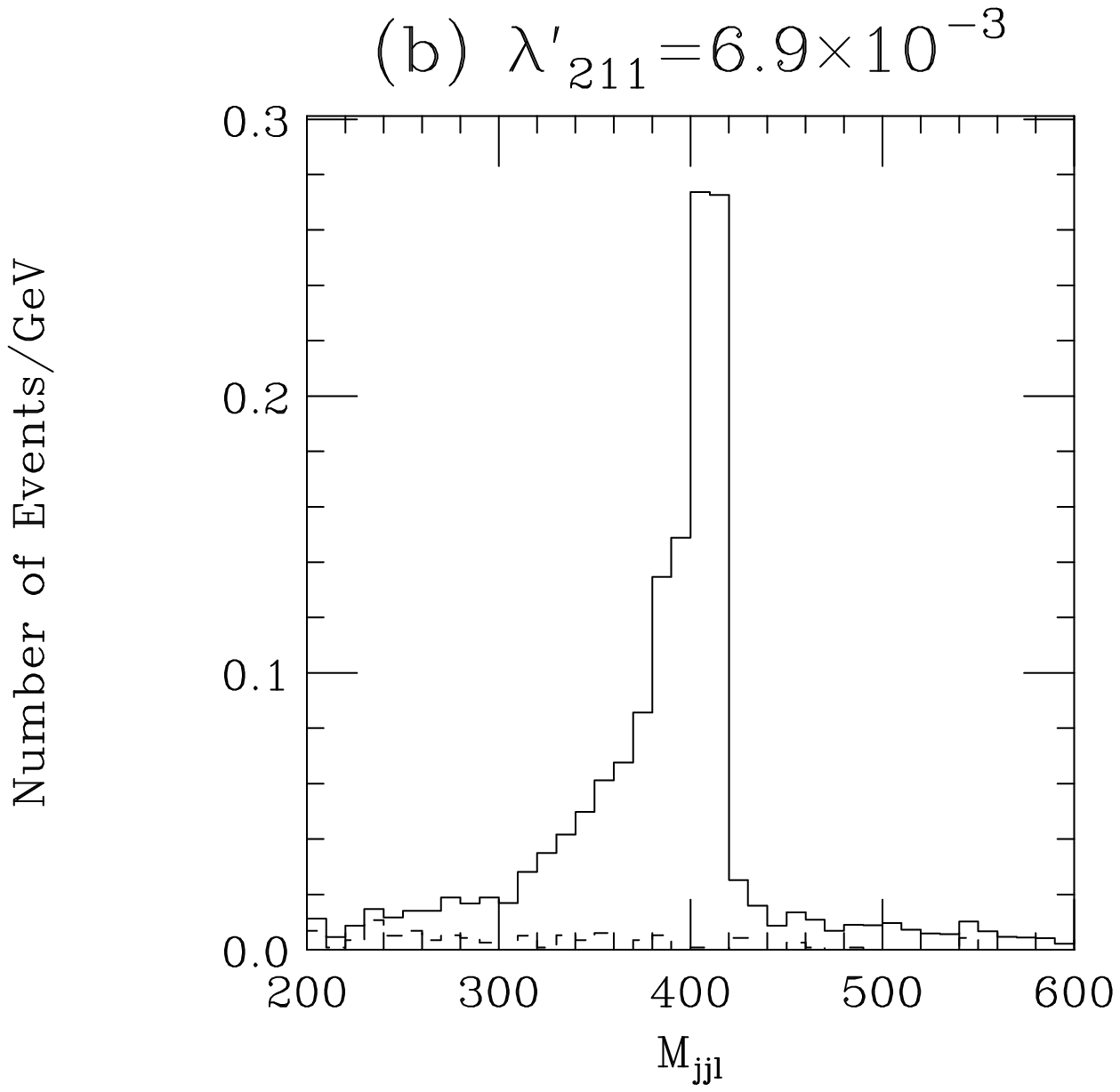}\\
\captionB{Reconstructed $\mr{\cht^0_1}$ mass at the LHC for 
	$M_0=350\, \mr{\gev}$, \mbox{$M_{1/2}=950\, \mr{\gev}$},
	$\tan\beta=10$, $\sgn\mu<0$ and $A_0=0\, \mr{\gev}$.}
	{The reconstructed neutralino mass at the LHC 
	for  $M_0=350\, \mr{\gev}$, \linebreak $M_{1/2}=950\, \mr{\gev}$,
	$\tan\beta=10$, $\sgn\mu<0$ and $A_0=0\, \mr{\gev}$. The value of the
	coupling in (b) is chosen such that after the cuts applied in
	 Section~\ref{sec:results}
	the signal is $5\sigma$ above the background. At this point the
	lightest
	neutralino mass is $M_{\cht^0_1}=418.0\, \mr{\gev}$.
	 We have  normalized the
	distributions to an integrated luminosity of $10\ \mr{fb}^{-1}$.
	The dashed line shows the background and the solid line shows
	the sum of the signal and the background.} 
\label{fig:lhcsusyneutmass}
\end{center}
\end{figure}

   An additional method of distinguishing between these two scenarios 
   is to try to reconstruct the masses of the
   decaying sparticles for resonant slepton production. In principle this
   is
   straightforward. The neutralino decay to a quark--antiquark pair
   and a charged lepton will give two jets (or more after the emission of QCD
   radiation) and a charged lepton. These decay products should be relatively
   close
   together. Therefore to reconstruct the neutralino we took the highest two 
   $p_T$ jets in the event and combined them with the charged lepton which
   was closest
   in $(\eta,\phi)$ space. We only used events in which both jets had 
   $p_T>10\, \mr{\gev}$ 
   in addition to passing all the cuts described in the previous sections,
   \ie both the cuts
   required to suppress the Standard Model and SUSY backgrounds.
   This gives a neutralino candidate. The masses of these candidates are
   shown, for
   a sample point in SUSY parameter space, for both the Tevatron, 
   Fig.\,\ref{fig:tevsusyneutmass}, and the LHC, 
   Fig.\,\ref{fig:lhcsusyneutmass}.
   In both cases, in addition to showing the result for the
   coupling ${\lam'}_{211}=10^{-2}$, we show a coupling for which the
   signal is exactly $5\sigma$ above the background at this point to
   demonstrate what can be seen if the
   signal is only just detectable. Both figures show that the
   reconstructed
   neutralino mass is in good agreement with the simulated value, although
   the situation
   may be worse once detector effects have been included.

%%%%%%%%%%%%%%%%%%%%%%%%%%%%%%%%%%%%%%%%%%%%%%%%%%%%%%%%%%%%%%%%%%%%%%%%%%%%%%
%
%  Figure containing the neutralino masses at the Tevatron
%
% End of the Figure %%%%%%%%%%%%%%%%%%%%%%%%%%%%%%%%%%%%%%%%%%%%%%%%%%%%%%%%%%
%%%%%%%%%%%%%%%%%%%%%%%%%%%%%%%%%%%%%%%%%%%%%%%%%%%%%%%%%%%%%%%%%%%%%%%%%%%%%%
%
%  Figure containing the slepton masses at the Tevatron
%
\begin{figure}
\begin{center}
\includegraphics[angle=90,width=0.48\textwidth]{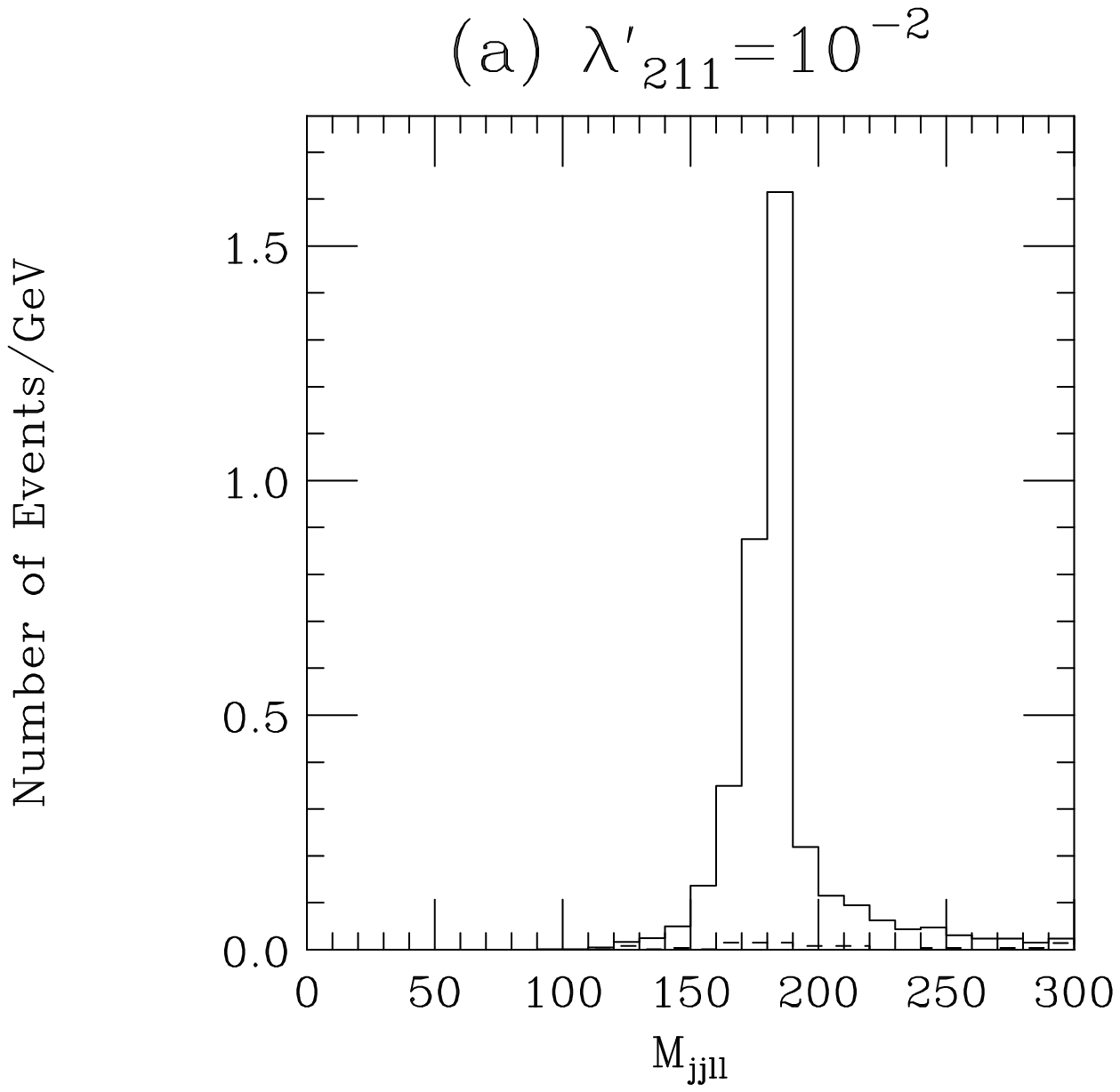}
\hfill
\includegraphics[angle=90,width=0.48\textwidth]{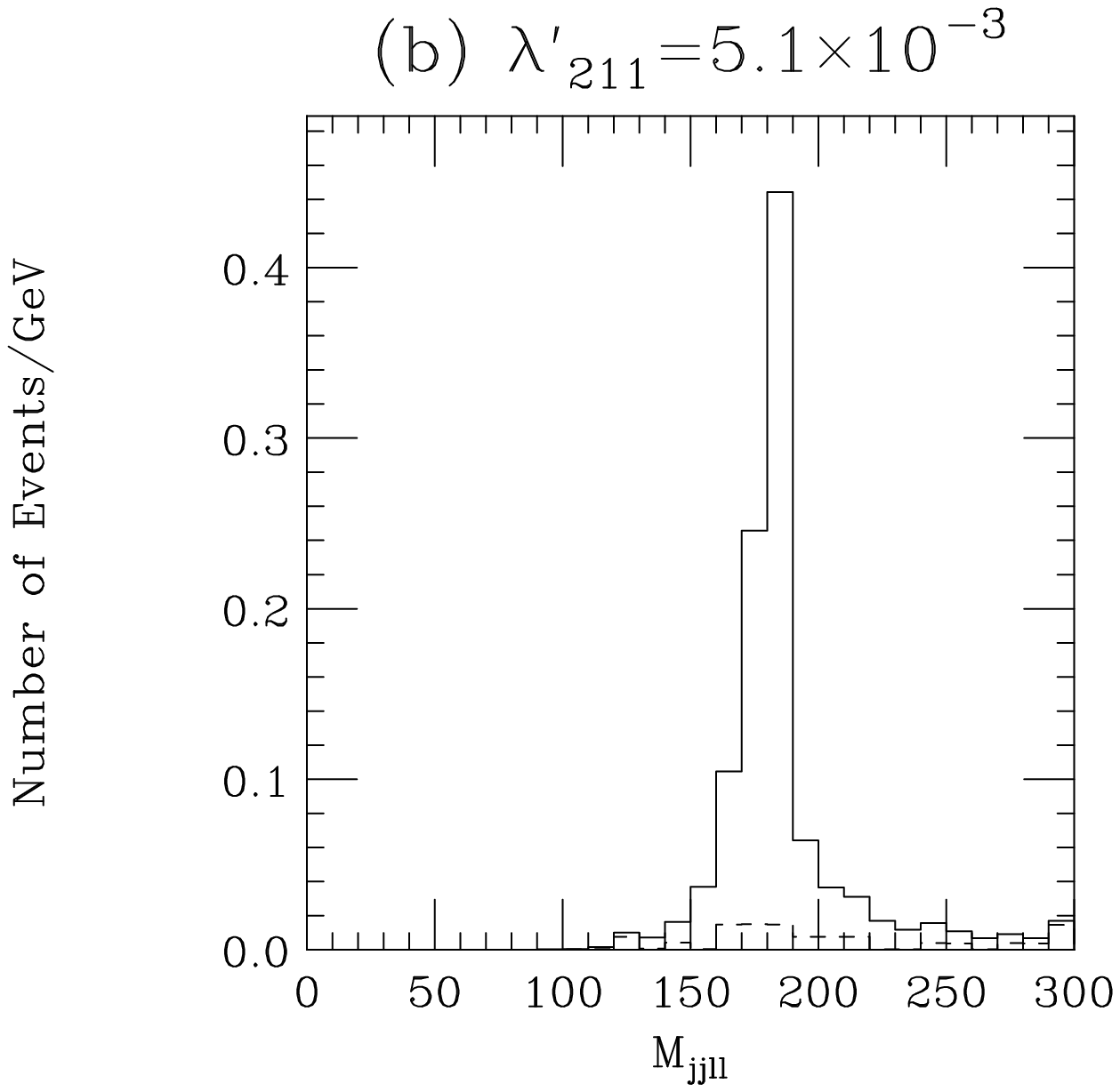}\\
\captionB{Reconstructed $\mut_L$ mass at the Tevatron for 
	 $M_0=50\, \mr{\gev}$, $M_{1/2}=250\, \mr{\gev}$,
	$\tan\beta=2$, $\sgn\mu>0$ and $A_0=0\, \mr{\gev}$.}
	{The reconstructed slepton mass at the Tevatron 
	for  $M_0=50\, \mr{\gev}$, \linebreak $M_{1/2}=250\, \mr{\gev}$,
	$\tan\beta=2$, $\sgn\mu>0$ and $A_0=0\, \mr{\gev}$. The value of the
	coupling in
     	(b) is chosen such that after the cuts applied in
	Section~\ref{sec:results}
	the signal is $5\sigma$ above the background. At this point the
	smuon mass
  	is $M_{\mr{\mut_L}}=189.1\, \mr{\gev}$. We have normalized the
	distributions to an integrated luminosity of $2\ \mr{fb}^{-1}$.
	The dashed line shows the background and the solid line shows
	the sum of the signal and the background.} 
\label{fig:tevsusyslepmass}
\end{center}
%\end{figure}
% End of the Figure %%%%%%%%%%%%%%%%%%%%%%%%%%%%%%%%%%%%%%%%%%%%%%%%%%%%%%%%%%
%%%%%%%%%%%%%%%%%%%%%%%%%%%%%%%%%%%%%%%%%%%%%%%%%%%%%%%%%%%%%%%%%%%%%%%%%%%%%%
%
%  Figure containing the slepton masses at the LHC
%
%\begin{figure}
\begin{center}
\includegraphics[angle=90,width=0.48\textwidth]{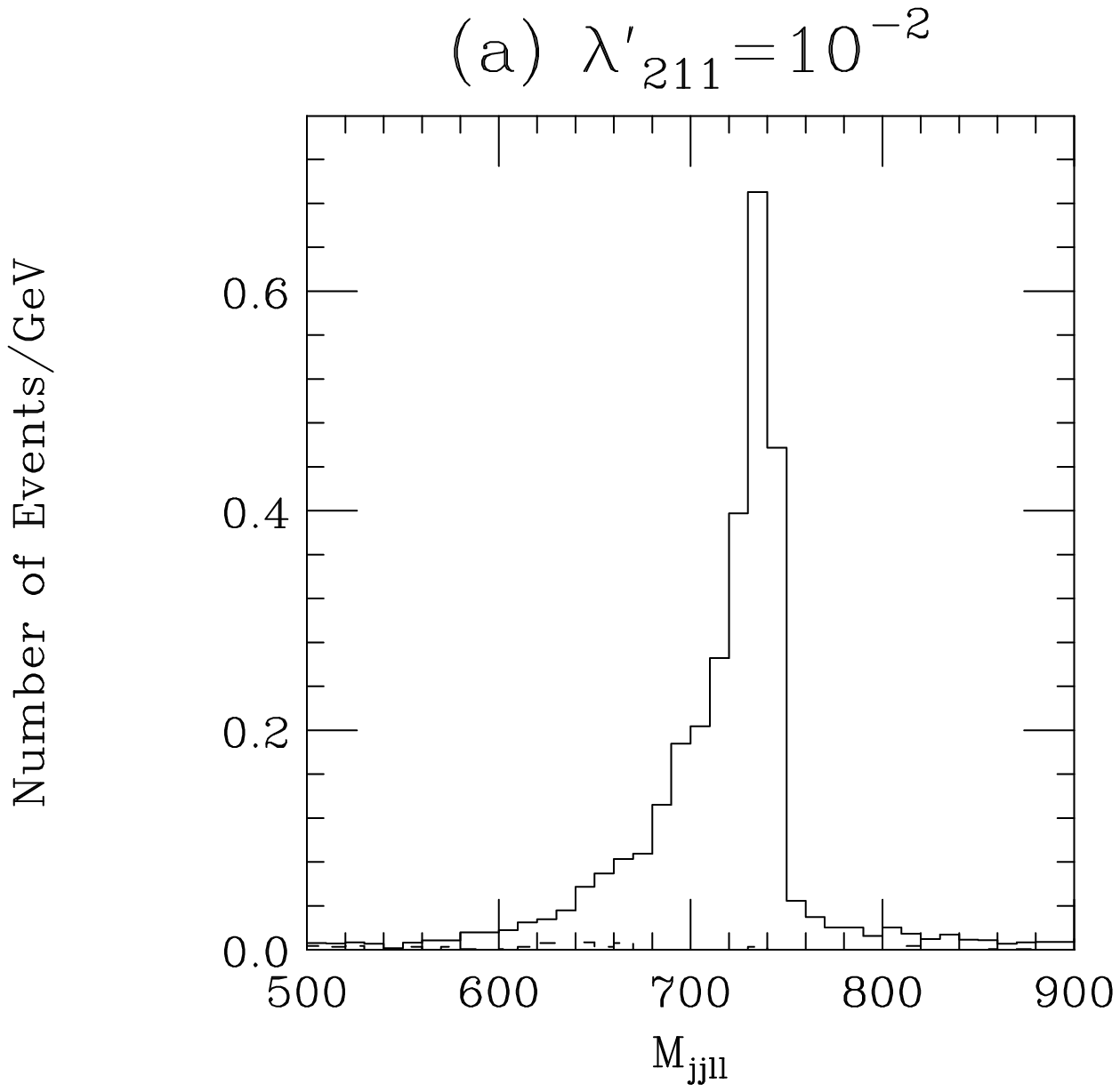}
\hfill
\includegraphics[angle=90,width=0.48\textwidth]{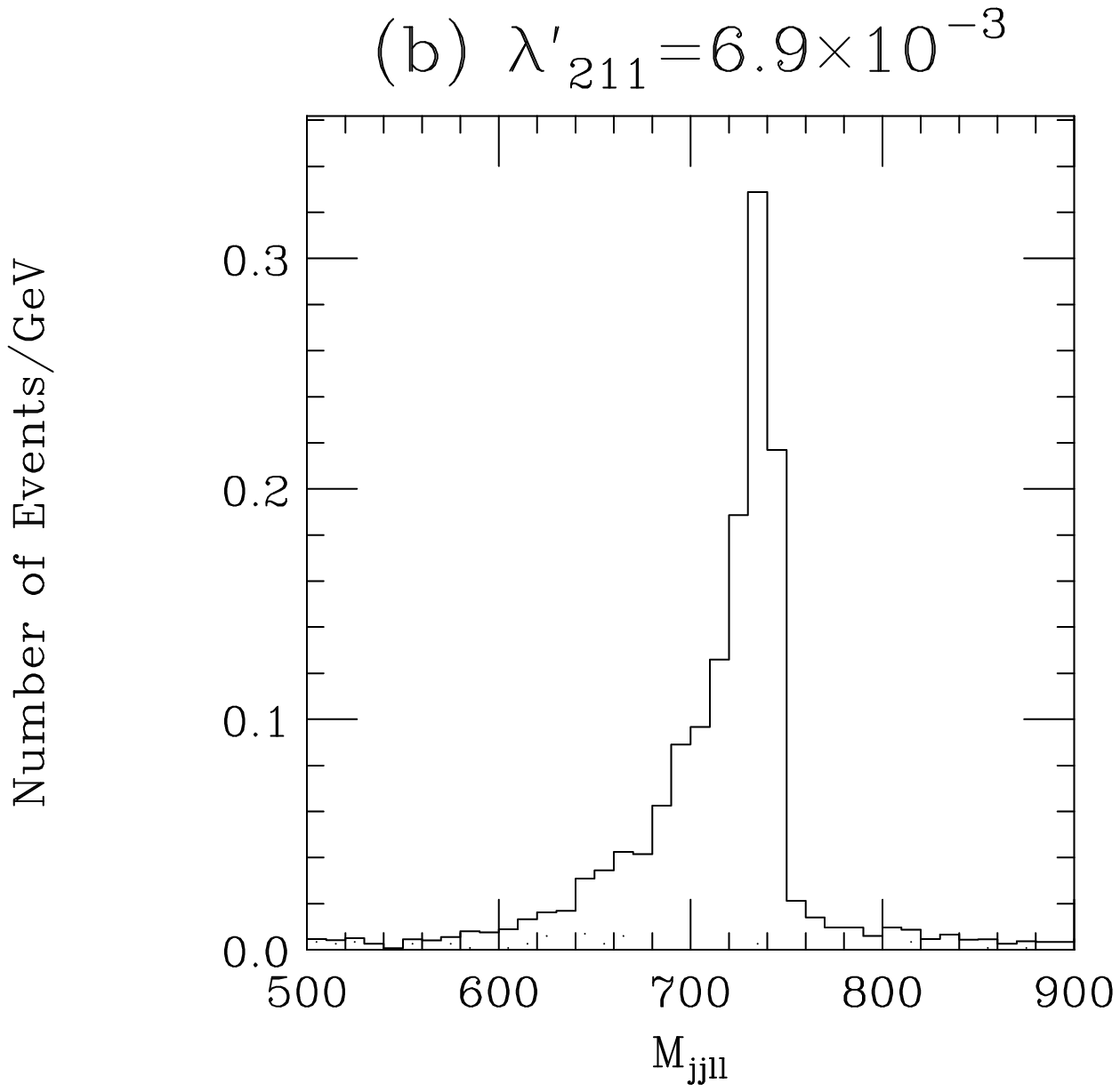}\\
\captionB{Reconstructed $\mut_L$ mass at the LHC for
	 $M_0=350\, \mr{\gev}$, $M_{1/2}=950\, \mr{\gev}$,
	$\tan\beta=10$, $\sgn\mu<0$ and $A_0=0\, \mr{\gev}$.}
	{The reconstructed slepton mass at the LHC 
	for  $M_0=350\, \mr{\gev}$, \linebreak $M_{1/2}=950\, \mr{\gev}$,
	$\tan\beta=10$, $\sgn\mu<0$ and $A_0=0\, \mr{\gev}$. The value of the
	coupling in (b) is chosen such that after the cuts applied in
	 Section~\ref{sec:results}
	the signal is $5\sigma$ above the background. At this point the smuon
	mass is $M_{\mr{\mut_L}}=745.9\, \mr{\gev}$.
	We have normalized the
	distributions to an integrated luminosity of $10\ \mr{fb}^{-1}$.
	The dashed line shows the background and the solid line shows
	the sum of the signal and the background.} 
\label{fig:lhcsusyslepmass}
\end{center}
\end{figure}
% End of the Figure %%%%%%%%%%%%%%%%%%%%%%%%%%%%%%%%%%%%%%%%%%%%%%%%%%%%%%%%%%
    
   We can then combine this neutralino candidate with the remaining lepton in
   the event
   to give a slepton candidate, under the assumption that the like-sign
   leptons were produced in the processes $\mr{\elt^+\ra\ell^+\cht^0_1}$.
   The mass distribution of these slepton
   candidates is shown
   in Fig.\,\ref{fig:tevsusyslepmass} for the Tevatron and 
   Fig.\,\ref{fig:lhcsusyslepmass} for the LHC. Again there is good agreement
   between
   the position of the peak in the distribution and the value of the smuon
   mass used in the simulation.

   The data for both the neutralino and smuon mass reconstructions is
   binned in $10\, \mr{\gev}$  bins. We have used the events in the central
   bin and
   the two bins on either side to  reconstruct the neutralino and smuon
   masses. These reconstructed masses are given in 
   Table~\ref{tab:reconstruct}. For both the points
   we have shown, the reconstructed mass lies between $5\, \mr{\gev}$  and
   $15\, \mr{\gev}$  
   below the simulated sparticle masses. This is due to the loss of
   some of the energy of the jets produced in the neutralino decay from the
   cones used to define the jets. It is common
   to include this effect in the jet energy correction, so this shift would
   probably not be observed in a full experimental simulation.

   The agreement between the results of the simulation and the input values 
   is good provided that the Standard
   Model background is dominant over the background from sparticle pair 
   production and that the lightest neutralino is predominantly produced in
   the smuon decay. This is the case at the points used in 
   Figs.\,\ref{fig:tevsusyneutmass}-\ref{fig:lhcsusyslepmass}. At the
   point \linebreak $M_0=50\, \mr{\gev}$, $M_{1/2}=250\, \mr{\gev}$,
   $\tan\beta=2$, $\sgn\mu>0$ and
   $A_0=0\, \mr{\gev}$ the branching ratio for the decay of the smuon to the
   lightest neutralino is $\mr{BR}(\mr{\mut_L}\ra\cht^0_1\mu^+)=98\%$. 
   Similarly, at the point  $M_0=350\, \mr{\gev}$, $M_{1/2}=950\, \mr{\gev}$,
   $\tan\beta=10$, $\sgn\mu<0$ and $A_0=0\, \mr{\gev}$, the dominant decay
   mode of the smuon is to the lightest neutralino with a branching ratio of 
   $\mr{BR}(\mr{\mut_L}\ra\cht^0_1\mu^+)=99\%$.

%%%%%%%%%%%%%%%%%%%%%%%%%%%%%%%%%%%%%%%%%%%%%%%%%%%%%%%%%%%%%%%%%%%%%%%%%%%%%%
%
%
%
\begin{table}
\begin{center}
\begin{tabular}{|c|c|c|c|c|c|c|c|}
\hline
  Experiment & ${\lam'}_{211}$ & Cuts & Point & 
  \multicolumn{2}{c|}{Neutralino mass/\gev} &
 \multicolumn{2}{c|}{Slepton mass/\gev}\\
\cline{5-8}
  & & &  & Actual & Recon. & Actual & Recon. \\
\hline
 Tevatron & $10^{-2}$          & no & A & 98.9 & 90.3  & 189.1 & 181.6 \\
\hline					                
 Tevatron & $5.2\times10^{-3}$ & no & A & 98.9 & 91.8  & 189.1 & 182.3 \\
\hline					                
 LHC      & $10^{-2}$          & no & B & 418.0 & 404.1 & 745.0 & 734.1 \\
\hline					                
 LHC      & $6.9\times10^{-2}$ & no & B & 418.0 & 405.1 & 745.0 & 732.6 \\
\hline					                
 LHC      & $10^{-2}$          & no & C & 147.6 & 142.7 & 432.0 & 421.3 \\
\hline					                
 LHC      & $10^{-2}$ 	       &yes & C & 147.6 & 143.4 & 432.0 & 423.3 \\
\hline
\end{tabular}
\captionB{Reconstructed neutralino and slepton masses.}
	{Reconstructed neutralino and slepton masses.
	 The following SUGRA points were used in these simulations: point A
	 has 
         $M_0=50\, \mr{\gev}$, $M_{1/2}=250\, \mr{\gev}$, $\tan\beta=2$,
	 $\sgn\mu>0$ and $A_0=0\, \mr{\gev}$; point B has
    	 $M_0=350\, \mr{\gev}$, $M_{1/2}=950\, \mr{\gev}$,     
   	 $\tan\beta=10$, $\sgn\mu<0$ and $A_0=0\, \mr{\gev}$;
	 point C has $M_0=350\, \mr{\gev}$, $M_{1/2}=350\, \mr{\gev}$,
  	 $\tan\beta=10$, $\sgn\mu<0$ and $A_0=0\, \mr{\gev}$. The Tevatron
         and LHC
  	 results are based on an integrated luminosity of $2\  \mr{fb}^{-1}$
	 and $10\  \mr{fb}^{-1}$, respectively.}
\label{tab:reconstruct}
\end{center}
\end{table}
%%%%%%%%%%%%%%%%%%%%%%%%%%%%%%%%%%%%%%%%%%%%%%%%%%%%%%%%%%%%%%%%%%%%%%%%%%%%%%

\begin{figure}
\begin{center}
\includegraphics[angle=90,width=0.48\textwidth]{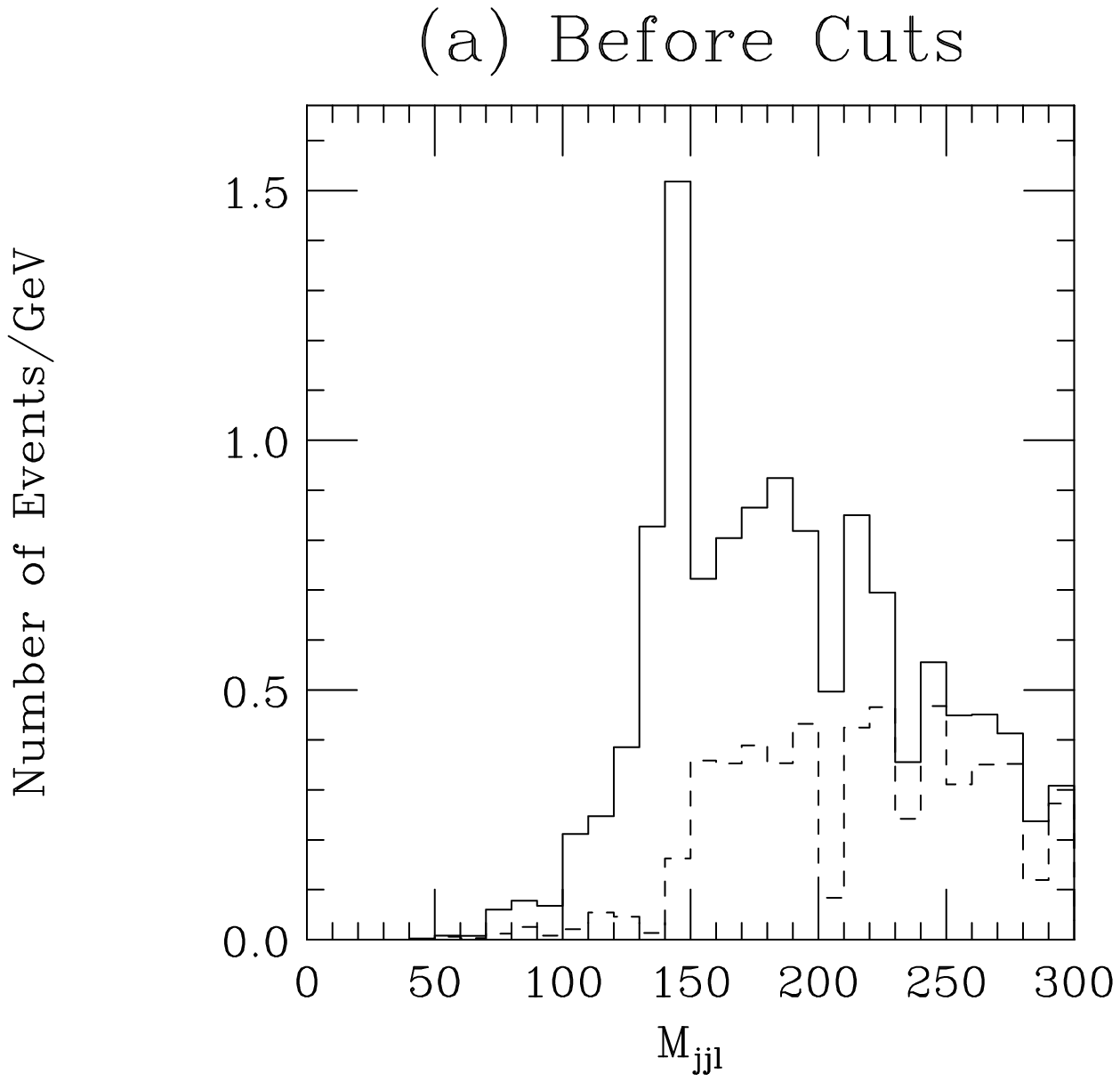}
\hfill
\includegraphics[angle=90,width=0.48\textwidth]{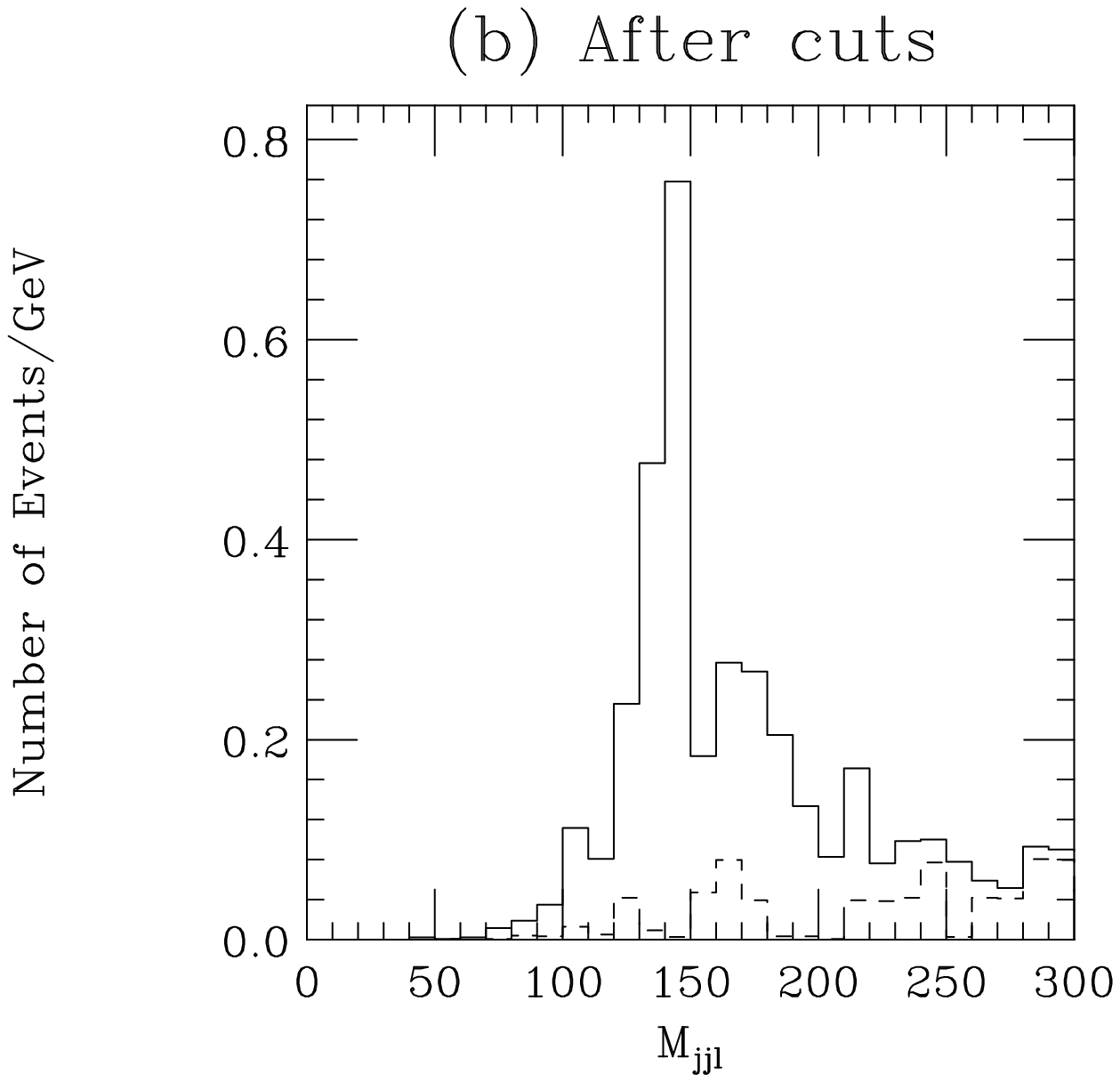}\\
\captionB{Reconstructed $\cht^0_1$ mass at the LHC for 
${\lam'}_{211}=10^{-2}$, $M_0=350\, \mr{\gev}$, $M_{1/2}=350\, \mr{\gev}$,
	$\tan\beta=10$, $\sgn\mu<0$ and $A_0=0\, \mr{\gev}$.}
	{The reconstructed neutralino mass at the LHC 
	for  ${\lam'}_{211}=10^{-2}$, \linebreak
	 $M_0=350\, \mr{\gev}$, $M_{1/2}=350\, \mr{\gev}$,
	$\tan\beta=10$, $\sgn\mu<0$ and $A_0=0\, \mr{\gev}$. At this point the
	lightest neutralino mass is $M_{\cht^0_1}=147.6\, \mr{\gev}$.
	We have normalized the
	distributions to an integrated luminosity of $10\ \mr{fb}^{-1}$. The
	cuts used are described in the text.
	The dashed line shows the background and the solid line shows
	the sum of the signal and the background.} 
\label{fig:lhcsusyneutmasscut}
\end{center}
%\end{figure}
% End of the Figure %%%%%%%%%%%%%%%%%%%%%%%%%%%%%%%%%%%%%%%%%%%%%%%%%%%%%%%%%%
%%%%%%%%%%%%%%%%%%%%%%%%%%%%%%%%%%%%%%%%%%%%%%%%%%%%%%%%%%%%%%%%%%%%%%%%%%%%%%
%
%  Figure containing the slepton masses at the LHC
%
%\begin{figure}
\begin{center}
\includegraphics[angle=90,width=0.48\textwidth]{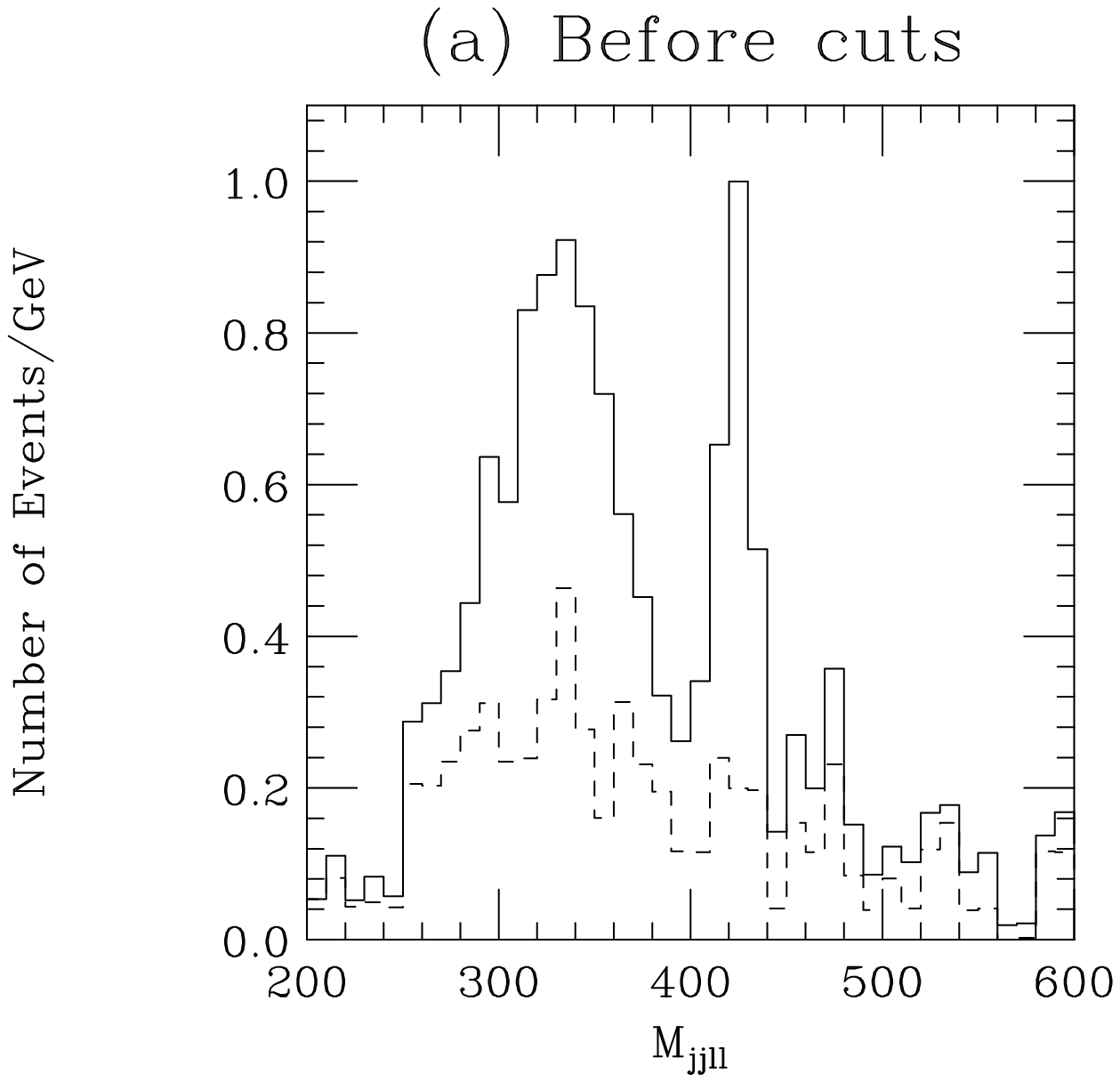}
\hfill
\includegraphics[angle=90,width=0.48\textwidth]{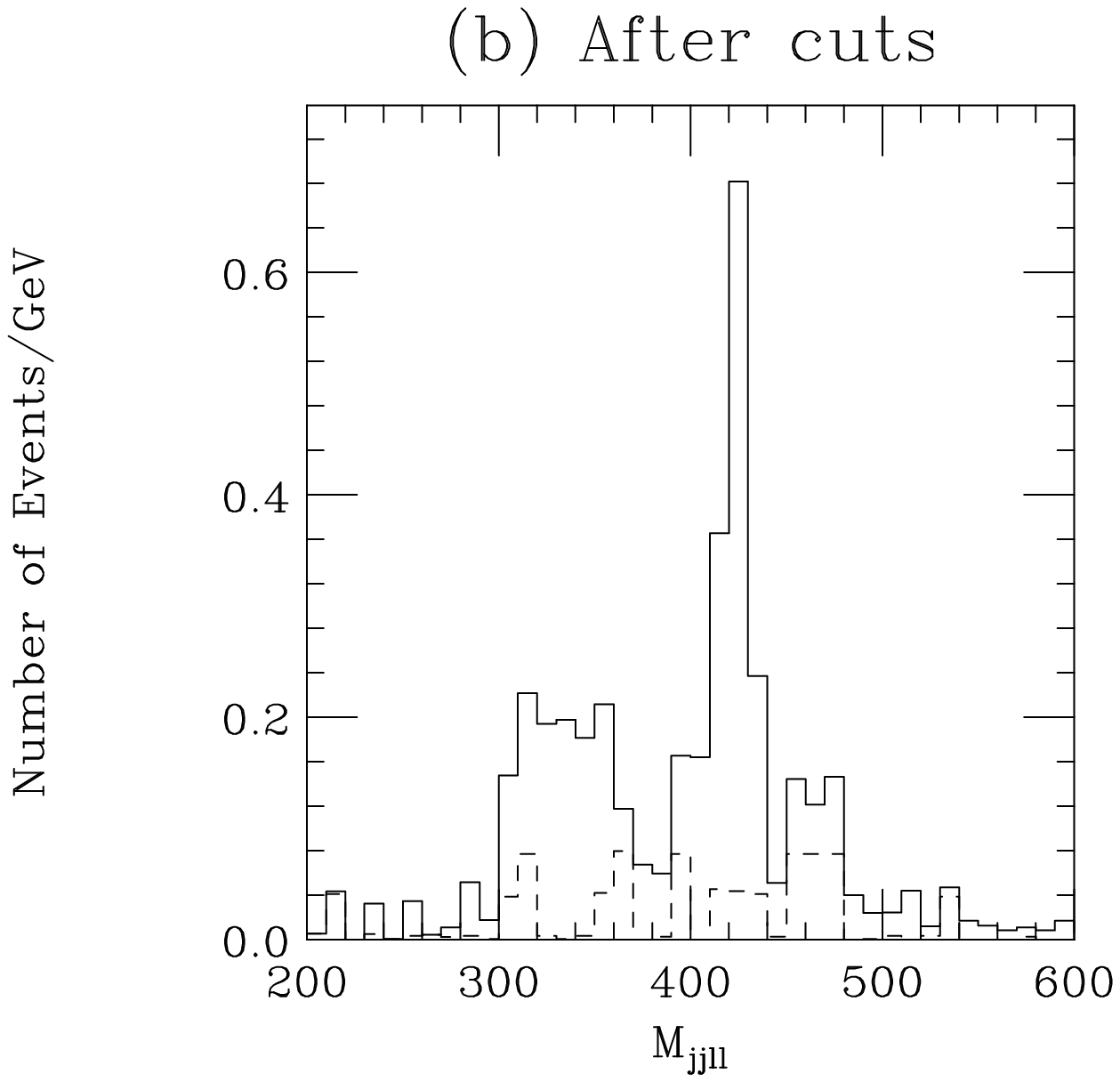}\\
\captionB{Reconstructed $\mut_L$ mass at the LHC for 
	${\lam'}_{211}=10^{-2}$, $M_0=350\, \mr{\gev}$, 
	$M_{1/2}=350\, \mr{\gev}$,
	$\tan\beta=10$, $\sgn\mu<0$ and $A_0=0\, \mr{\gev}$.}
	{The reconstructed slepton mass at the LHC for 
 ${\lam'}_{211}=10^{-2}$, $M_0=350\, \mr{\gev}$, $M_{1/2}=350\, \mr{\gev}$,
$\tan\beta=10$, $\sgn\mu<0$ and $A_0=0\, \mr{\gev}$. At this point the smuon
	mass is $M_{\mr{\mut_L}}=432.0\, \mr{\gev}$. We have normalized the
	distributions to an integrated luminosity of $10\ \mr{fb}^{-1}$. The
	cuts used are described in the text.
	The dashed line shows the background and the solid line shows
	the sum of the signal and the background.} 
\label{fig:lhcsusyslepmasscut}
\end{center}
\end{figure}

  It can however be the case that there is a significant background from
  sparticle pair production and a substantial contribution from the
  production 
  of charginos and heavier neutralinos. This is shown in
  Fig.\,\ref{fig:lhcsusyneutmasscut}a for
  the neutralino mass reconstruction and Fig.\,\ref{fig:lhcsusyslepmasscut}a
  for
  the smuon mass reconstruction.
  Figs.\,\ref{fig:lhcsusyneutmasscut}a and \ref{fig:lhcsusyslepmasscut}a
  show that there is
  a significant background in both distributions. At this point, \ie
  ${\lam'}_{211}=10^{-2}$, $M_0=350\, \mr{\gev}$, $M_{1/2}=350\, \mr{\gev}$,
  $\tan\beta=10$, $\sgn\mu<0$ and $A_0=0\, \mr{\gev}$, the smuon dominantly 
  decays to the lightest chargino, with branching ratio
  $\mr{BR}(\mr{\mut_L\ra\cht^-_1\nu_\mu})=50.9\%$.
  The other important decay modes are to the next-to-lightest
  neutralino, with branching ratio
  $\mr{BR}(\mr{\mut_L}\ra\cht^0_2\mu^+)=28.0\%$, 
  and to the lightest neutralino,
  with branching ratio $\mr{BR}(\mr{\mut_L}\ra\cht^0_1\mu^+)=20.9\%$.

%%%%%%%%%%%%%%%%%%%%%%%%%%%%%%%%%%%%%%%%%%%%%%%%%%%%%%%%%%%%%%%%%%%%%%%%%%%%%%
%
%  Figure containing the neutralino masses at the LHC
%
% End of the Figure %%%%%%%%%%%%%%%%%%%%%%%%%%%%%%%%%%%%%%%%%%%%%%%%%%%%%%%%%%

  In the neutralino distribution there is still a peak at the simulated
  neutralino
  mass, although there is a large tail at higher masses. This tail is mainly
  due to
  the larger sparticle pair production background. The production 
  of charginos and the
  heavier
  neutralinos does not significantly effect this distribution as the heavier
  gauginos will cascade decay to the LSP due to the small \rpv\  coupling.

  In the slepton distribution in addition to the larger background there is
  also a spurious peak in the mass distribution due to the production of the
  lightest chargino and the~$\cht^0_2$.
  As we are not including all of the decay products of the chargino or 
  $\cht^0_2$ in the mass
  reconstruction, the reconstructed slepton mass in signal events where a 
  chargino or 
  heavier neutralino is produced is below the true value.

  We can improve the extraction of both the neutralino and slepton masses by
  imposing some additional cuts. The aim of these cuts is to require that the 
  neutralino candidate and the second lepton are produced back-to-back,
  because in most of the signal events the resonant smuon will only
  have a small transverse momentum due to the initial-state parton shower.
  We therefore 
  require the transverse momenta of the neutralino candidate and the second
  lepton
  to satisfy $|p_T^{\mr{jj\ell_1}}-p_T^{\ell_2}|<20\, \mr{\gev}$, and the
  azimuthal angles to satisfy 
  $||\phi_{\mr{jj\ell_1}}-\phi_{\ell_2}|-180^0|<15^0$.
  $p_T^{\mr{jj\ell_1}}$ is the transverse momentum of the combination of
  the hardest
  two jets in the event and the lepton closest to the jets in $(\eta,\phi)$
  space, \ie the transverse momentum of the neutralino candidate.
  Similarly $\phi_{\mr{jj\ell_1}}$ is the azimuthal
  angle of the combination of the hardest
  two jets in the event and the lepton closest to the jets in $(\eta,\phi)$
  space, \ie the azimuthal angle of the neutralino candidate.

  Figs.\,\ref{fig:lhcsusyneutmasscut}b and 
  \ref{fig:lhcsusyslepmasscut}b show that this significantly reduces the
  background and
  the spurious peak in the slepton mass distribution. At these points it is
  also possible to reconstruct the lightest 
  neutralino, chargino and sneutrino masses using
  the decay chain \mbox{$\mr{\nut\ra\cht^+_1\ell^+}$} followed
  by the decay of the chargino \mbox{$\mr{\cht^+_1\ra\ell^+\nu_\ell\cht^0_1}$}
  and the \rpv\  decay of the lightest neutralino to a lepton and two jets 
  \cite{Moreau:1999bt:Moreau:2000ps:Moreau:2000bs,Abdullin:1999zp}. The
  reconstructed
   neutralino and slepton masses, before and after the imposition of the
   new cuts, are given in Table~\ref{tab:reconstruct}. The same procedure
   as before was used to extract the sparticle masses. There is
   reasonable agreement between the simulated and reconstructed sparticle
   masses  although again the reconstructed values lie between 5 and
   $15\, \mr{\gev}$  
   below the values used in the simulations, due to the loss of energy
   from the cones used to define the jets in the neutralino decay.

%
%  Finally the Conclusions
%
\section{Summary}
\label{sec:conclusions}

  We have considered both the \rpv\  and gauge decay modes of resonant
  sleptons.
  The \rpv\  decay modes can only be used to discover these processes for
  large values of the couplings due to the large QCD background.

  However the like-sign dilepton signature of these processes can be observed 
  above the background for much lower values of the \rpv\  Yukawa couplings.
  We have performed a detailed analysis of the background
  to like-sign dilepton
  production at both Run II of the Tevatron and the LHC. We find a background
  from Standard Model processes of $0.43\pm0.16$ events for $2\ \mr{fb}^{-1}$
  integrated luminosity at the Tevatron and $4.9\pm1.6$ events for 
  $10\  \mr{fb}^{-1}$ integrated luminosity at the LHC, after a set of
  cuts. If we only consider
  this background there are large regions of SUGRA parameter space where
  resonant slepton production followed by a supersymmetric
  gauge decay of the slepton is
  visible above the SM background even for the small values of the \rpv\  
  couplings we considered.

  This is presumably the strategy which would be adopted in any initial
  experimental search, \ie looking for an excess of a given type of event
  over the Standard Model prediction. If such an excess were
  observed it would
  then be necessary to identify which of the many possible models of beyond
  the Standard Model physics was correct.

  In the \rpv\  MSSM
  such an excess of like-sign dileptons can come from two
  possible sources: from sparticle pair production followed by the decay
  of the LSP, and from resonant sparticle production. We have considered the
  background to resonant slepton production from sparticle pair production
  and found that after an additional cut on the number of jets the signal
  from resonant slepton production is visible
  above the combined Standard Model
  and SUSY pair production background for large ranges of SUGRA parameter
  space.

  Finally we have studied the possibility of measuring the mass of the
  resonant slepton and the neutralino into which it decays. Our results
  suggest that this should be possible even if the signal is only just 
  detectable above the background.

  Resonant slepton production offers a potentially interesting channel for
  the discovery of \rpv\  SUSY and can be used, provided the \rpv\  couplings
  are not too small, to discover supersymmetry over a larger range of
  SUSY parameter space than supersymmetric particle pair production due to 
  the larger kinematic reach.

%\chapter{Neutralino}
%\include{neutralino}
%\chapter{Karman Anomaly}
\chapter{KARMEN Anomaly} \label{chap:karmen}

\section[Introduction]{Introduction} \label{sect:karmenintro}

  The KARMEN experiment is designed to search for neutrino
  oscillations by looking for the appearance of $\mr{\nu_e}$ from the
  oscillation $\nu_\mu \ra \nu_{\mr{e}}$ and of $\mr{\bar{\nu}_e}$ from the
  oscillation $\bar{\nu}_\mu \ra \mr{\bar{\nu}_e}$ 
  \cite{Zeitnitz:1994kz:Bodmann:1994py}. While the KARMEN experiment
  has detected no deviations from the Standard Model predictions which are
  consistent with neutrino oscillations, their results do contain an anomaly
  \cite{Armbruster:1995nr}.

  We will first describe the KARMEN experiment and the nature of the
  anomaly followed by a possible explanation of this anomaly in terms of
  R-parity violating supersymmetry. This is followed by a discussion of
  other possible constraints on this model and possible future experimental
  tests of our model.

\begin{figure}
\includegraphics[width=0.9\textwidth]{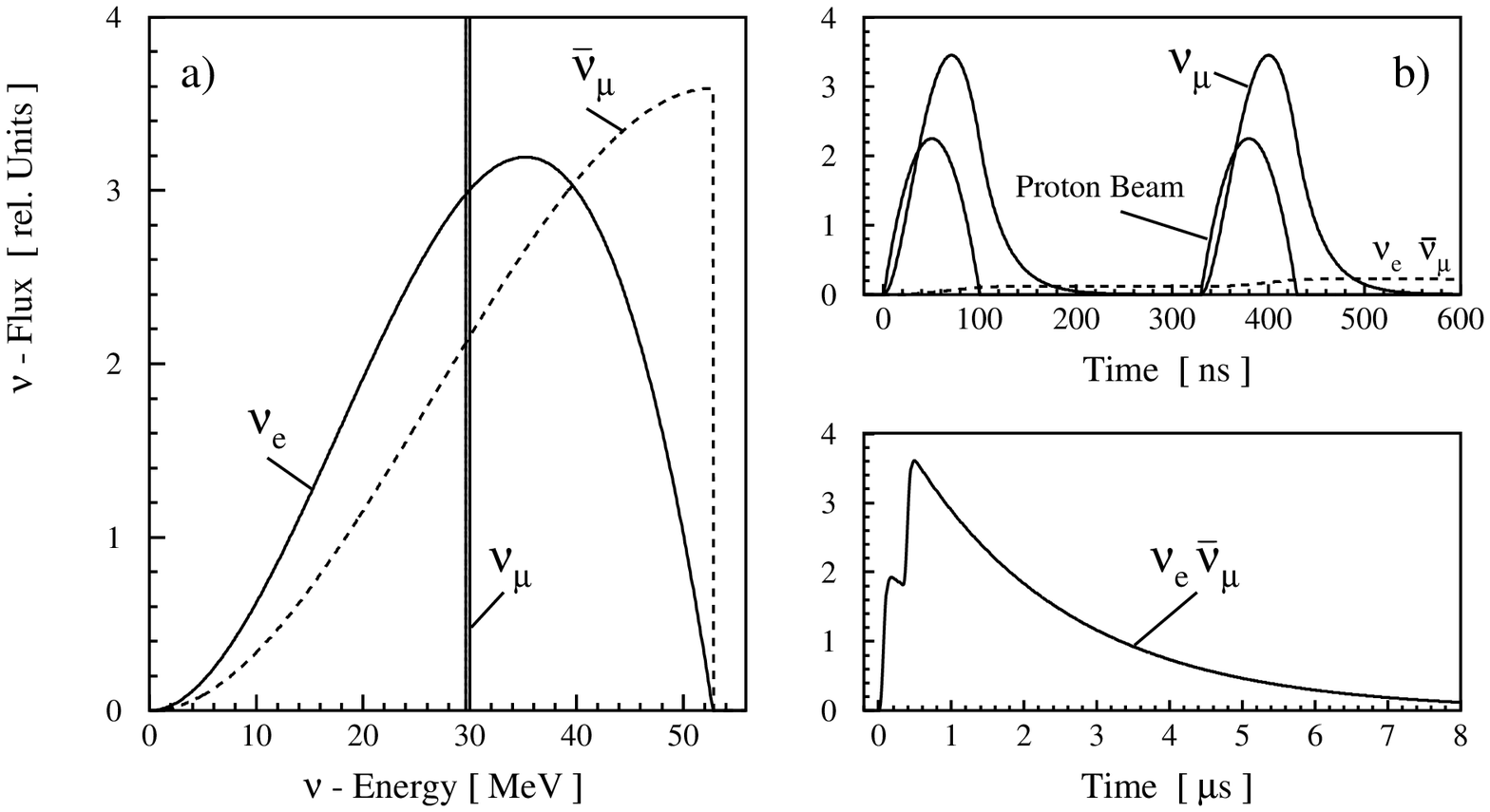}
\captionB{Energy spectrum and time structure of neutrino
 	production in KARMEN.}
	{The energy spectrum (a) and time structure (b) of the neutrinos
 	produced in the KARMEN experiment. In (b) the neutrino production
	is shown on two different time scales for the
	$\mr{\nu_\mu}$ neutrinos, and the $\mr{\bar{\nu}_\mu}$ and
	$\mr{\nu_e}$ neutrinos. (Taken from \cite{Eitel:1997cy}).}
\label{fig:karmenexp}
\end{figure}

  The basic idea of the KARMEN experiment is that a proton beam hits a
  target producing pions. These pions are quickly
  stopped in the target and then decay,
  \mbox{$\mr{\pi}^+\ra\mr{\mu}^+ \mr{\nu_\mu}$}. This gives a source of
  mono-energetic $\nu_\mu$, from the two-body pion decay,
  the spectrum of which is shown in Fig.\,\ref{fig:karmenexp}a.
  The muons then decay,
  $\mr{\mu}^+ \ra \mr{e}^+\mr{\nu_e}\mr{\bar{\nu}_\mu}$, giving equal numbers
  of $\mr{\nu_e}$ and $\mr{\bar{\nu}_\mu}$. The energy spectrum of these
  neutrinos is also shown in Fig.\,\ref{fig:karmenexp}a.

  The experiment uses the ISIS proton beam at the Rutherford Appleton
  Laboratory. This is a pulsed proton beam which has two pulses separated
  by 330~ns. The pulse structure of the proton beam is shown in
  Fig.\,\ref{fig:karmenexp}b.  The pairs of pulses are separated
  by $20$~ms. When the beam
  hits the target pions are promptly produced, followed by $\nu_\mu$ from the
  pion decays due to the short lifetime of the charged pion. This means the
  detector will first detect two pulses of $\mr{\nu_{\mu}}$ during the first
  0.5~$\mu$s, after the proton beam hits the target, followed by the
  $\mr{\nu_e}$ and $\mr{\bar{\nu}_\mu}$ from the muon decays, which occur with
  a lifetime of 2.2~$\mu$s. This gives the expected time distribution for
  neutrino detection by the KARMEN experiment shown in 
  Fig.\,\ref{fig:karmenexp}b.

  The $\mr{\nu_{\mu}}$ can thus be separated from the other
  two types of neutrino by measuring the time at which they are detected
  relative to the time of the beam hitting the target. A full
  description of the experiment can be found in 
  \cite{Zeitnitz:1994kz:Bodmann:1994py,Armbruster:1995nr}. 

%
%  Figure taken from the KARMEN paper
%

  The expectation of the experiment is to see the initial pulse
  followed by an exponential fall-off with a time constant of
  2.2~$\mu$s consistent with the muon decay. However, in addition to
  this signal an additional component is seen which is
  consistent with a Gaussian centered at a time $3.6 \pm
  0.25\  \mu$s after the beam hits the target \cite{Armbruster:1995nr}. 
  This effect was seen in the
  first run of the KARMEN  experiment \cite{Armbruster:1995nr} and has
  since been confirmed by a
  new run with improved shielding to eliminate the background from
  cosmic-rays\cite{Karmennew1,Karmennew2}.\footnote{The KARMEN collaboration
			has recently announced new results
		      	which are discussed in the Addendum.} 

\renewcommand{\arraystretch}{0.5}
  The explanation of this anomaly proposed by the KARMEN collaboration in
  \cite{Armbruster:1995nr} was that a new hypothetical particle, X,
  was produced at the target and deposits energy in the detector when it
  decays. The arrival time, \ie the time of the Gaussian component
  observed by the KARMEN experiment, can be used to estimate the velocity
  of the particle giving
   \mbox{$v_\mr{X}=\left(5.2\pm \begin{array}{c} 2.2 \\  1.4 \\ \end{array}
  \right) \times 10^6\  \mr{ms^{-1}}$}. If we assume that the particle X is
  produced in the decay of the pion, \ie $\pi^+ \ra \mu^+\mr{X}$, its mass is
  $m_\mr{X}=33.9\  \mr{MeV}$ (just below the kinematic limit for
  the decay) and the momentum of the particle in the pion rest frame is
  $p_\mr{X}=0.6\  \mr{MeV}$ \cite{Armbruster:1995nr}. The new particle could
  also be produced by the interaction of the proton beam with the target,
  which we will not consider here. The energy observed in the detector 
  is~$\sim 11-35\, \mr{\mev}$, 
  which is a lot greater than the kinetic energy of the X particle and
  must therefore come from the decay of the X particle,
  if it is produced in the pion decay, $\pi^+ \ra \mu^+\mr{X}$.
  Since the particle passes through over 7~metres of steel, between the
  target and the KARMEN detector, it must also be neutral. A time-of-flight
  likelihood analysis adopting the hypothesis that the anomaly is due
  to a decaying particle has a negative natural log-likelihood ratio of 9,
  \ie  a less than 1~in~$10^4$ chance of being a statistical fluctuation.
\renewcommand{\arraystretch}{1.0}
                                     
  We can  consider various possible candidates for the particle X
  and its decay modes which lead to the observed energy deposit in
  the experiment. There have been a number of different explanations of the
  anomaly \cite{Barger:1995ty,Gninenko:1998ec,Choudhury:1996pj,Lukas:2000fy}:
\begin{itemize}
 \item 	One suggestion in \cite{Armbruster:1995nr} was that X could be
 	the tau neutrino, but as they point out this is excluded by the limit
 	on the tau neutrino mass from ALEPH \cite{Buskulic:1995tj}. It was
        shown in \cite{Barger:1995ty} that an $SU(2)_L$ doublet neutrino was
        excluded. This was reinforced by an improvement \cite{Daum:1995hs} in
        the limit for the branching 
	ratio,\footnote{There have subsequently been two further improvements
			in this bound. The first bound \cite{Formaggio:2000ne}
			only applies for small X lifetimes in the range
			$\sim 10^{-9}-10^{-3}\,\mr{s}$ and is therefore 
 			not relevant for this study. The second 
	                \cite{koglin:2000} is an improvement in the results
		        of \cite{Daum:1995hs} and now excludes branching
		        ratios \mbox{$\mr{BR}(\pi^+\ra\mu^+\mr{X}) 
			< 6.0\times 10^{-10}\  (\mr{95\%}\ \mr{C.L.})$. We
			will use this bound in the rest of this thesis.}}
\begin{equation}
   \mr{BR}(\pi^+\ra\mu^+\mr{X}) < 1.2\times 10^{-8} \ \ \ \ \ \ \ (\mr{95\%}\ 
							\mr{C.L.}),
	   \label{eqn:pionbound}
\end{equation}
	versus the minimum value of $\sim 2\times 10^{-8}$ required in the 
	doublet-neutrino scenario. A sterile neutrino has also been
        considered, however \cite{Barger:1995ty,Govaerts:1996hk} 
	showed that while a sterile neutrino was consistent
	with the current laboratory data, within strict limits on the mixing
        parameters, it may  conflict with the cosmological bounds.

%
%  Another figure taken from a KARMEN paper
%
\begin{figure}
\includegraphics[width=0.9\textwidth]{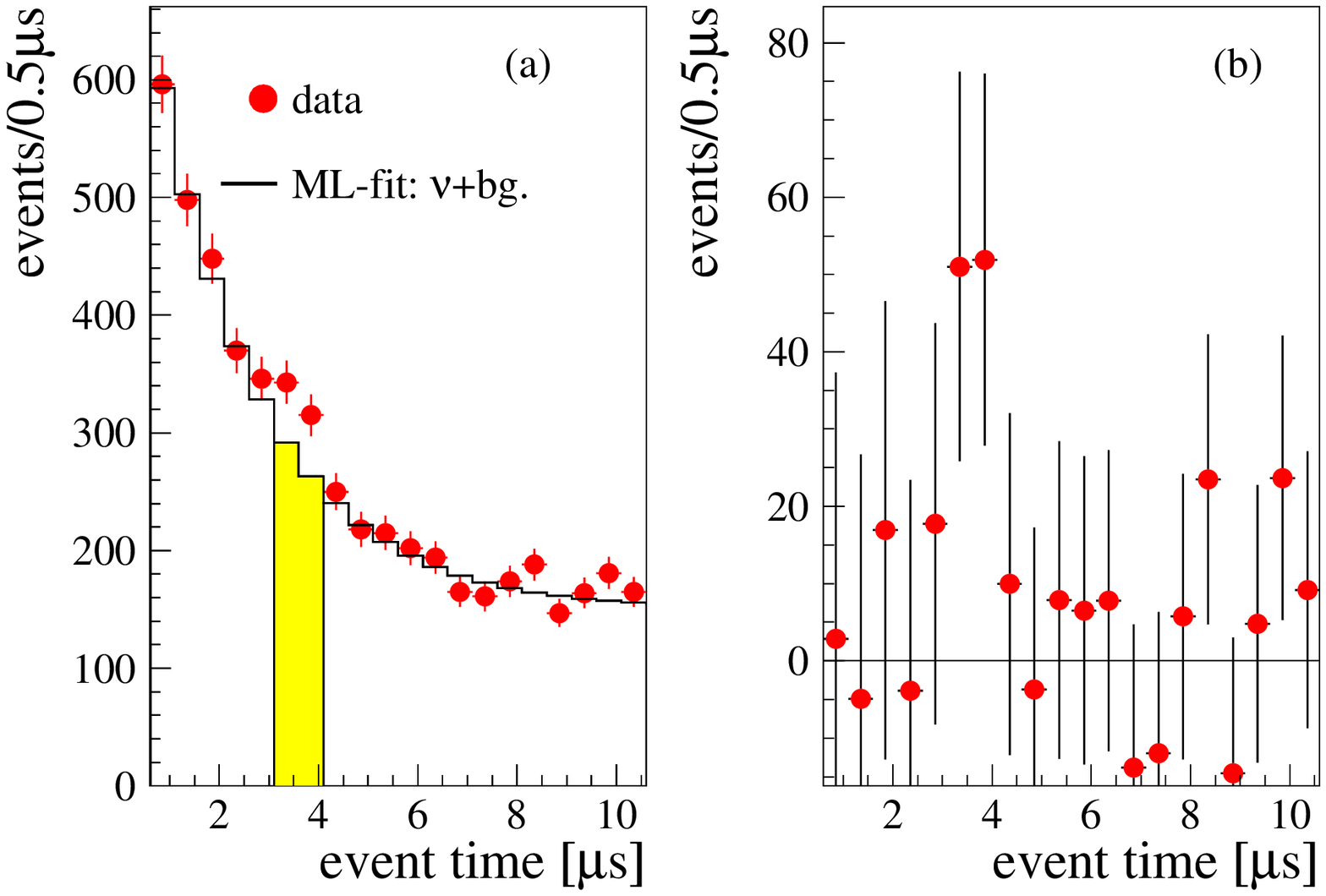}
\captionB{Time distribution of the events observed by the KARMEN experiment.}
	{Time distribution of the events detected by KARMEN 
	 between $0.6-10.6\  \mu\mr{s}$ after the proton beam hits the target.
         The solid line in (a) is the result of a fit to the cosmic ray 
	 background (assumed to be constant in time) and the neutrino-induced
	 events from muon decay. The excess number of events after the
	 events from the cosmic ray background and the neutrinos from muon
	 events have been subtracted is shown in (b). These plots are taken
	 from \cite{Karmennew3}.}
\end{figure}
%
% End of the figure
%
 \item  In \cite{Gninenko:1998ec} the solution proposed was based on an
        anomalous muon decay, $\mu^+\ra \mr{e}^+\mr{X}$, as opposed to the
        pion decay of \cite{Armbruster:1995nr}. X was taken to be a 
        scalar boson with mass $103.9\, \mr{\mev}$
	and kinetic energy $1.7\, \mr{\mev}$.
	This gives too large a value for the energy release in the X decay
	and the required branching ratio is constrained by the 
	bound \linebreak
        \mbox{$\mr{BR}(\mu^+\ra \mr{e}^+\mr{X}) < 
		5.7\times 10^{-4} \ (\mr{90\%}\ \mr{C.L.})$}
	\cite{Bilger:1998rp}. It is therefore necessary to add two more
	scalar bosons to the model into which X can cascade decay to reduce
	the energy deposit in the detector \cite{Gninenko:1998ec}.
	This model is viable but is somewhat contrived.
 
 \item	Another suggestion \cite{Choudhury:1996pj} was that the massive
	particle could be the lightest neutralino in a SUSY model. This
	neutralino then has to decay via an R-parity violating  interaction.
        In \cite{Choudhury:1996pj} the neutralino was assumed to be a photino
	or a zino. The pion decay can  proceed via the \rpv\  operator 
        $L_2 Q_1 \overline{D}_1$,
        \begin{equation}
           \mr{\pi}^+ \longrightarrow \mr{\mu}^+ \mr{\tilde{\gamma}}.
        \end{equation} 
         In \cite{Choudhury:1996pj} only one non-zero \rpv\   operator
	 was considered and hence the neutralino has to decay radiatively 
         \begin{equation}
            \mr{\tilde{\gamma}} \longrightarrow \mr{\gamma}  \mr{\nu_{\mu}},
         \end{equation}
         via the Feynman diagrams shown in Fig.\,\ref{fig:photinoraddecay}. 
         However the new data from the KARMEN experiment appears to suggest
	 that a three-body decay of the new particle, X, is favoured
	 \cite{Karmennew1}. 
%
%  The radiative neutralino Decay
%
\begin{figure}
\vskip 6mm
\begin{center}
\begin{picture}(440,120)
\SetOffset(50,55)
\ArrowLine(0,50)(40,50)
\ArrowLine(70,20)(40,50)
\DashArrowLine(70,80)(70,20){5}
\DashArrowLine(40,50)(70,80){5}
\ArrowLine(70,20)(110,20)
\Photon(70,80)(110,80){5}{5}
\Text(50,30)[r]{$\mr{d}$}
\Text(50,70)[r]{$\mr{\dnt}_R$}
\Text(75,50)[l]{$\mr{\dnt}_R$}
\Text(-2,50)[r]{$\mr{\cht^0_1}$}
\Text(115,20)[l]{$\mr{\nu_\mu}$}
\Text(115,80)[l]{$\mr{\gamma}$}
\SetOffset(270,55)
\ArrowLine(0,50)(40,50)
\ArrowLine(70,20)(70,80)
\DashArrowLine(40,50)(70,20){5}
\ArrowLine(70,80)(40,50)
\ArrowLine(70,20)(110,20)
\Photon(70,80)(110,80){5}{5}
\Text(50,30)[r]{$\mr{\dnt}_R$}
\Text(50,70)[r]{$\mr{d}$}
\Text(75,50)[l]{$\mr{d}$}
\Text(-2,50)[r]{$\mr{\cht^0_1}$}
\Text(115,20)[l]{$\mr{\nu_\mu}$}
\Text(115,80)[l]{$\mr{\gamma}$}
\SetOffset(50,-30)
\ArrowLine(0,50)(40,50)
\ArrowLine(40,50)(70,20)
\DashArrowLine(70,20)(70,80){5}
\DashArrowLine(70,80)(40,50){5}
\ArrowLine(70,20)(110,20)
\Photon(70,80)(110,80){5}{5}
\Text(50,30)[r]{$\mr{d}$}
\Text(50,70)[r]{$\mr{\dnt}_L$}
\Text(75,50)[l]{$\mr{\dnt}_L$}
\Text(-2,50)[r]{$\mr{\cht^0_1}$}
\Text(115,20)[l]{$\mr{\nu_\mu}$}
\Text(115,80)[l]{$\mr{\gamma}$}
\SetOffset(270,-30)
\ArrowLine(0,50)(40,50)
\ArrowLine(70,80)(70,20)
\DashArrowLine(70,20)(40,50){5}
\ArrowLine(40,50)(70,80)
\ArrowLine(70,20)(110,20)
\Photon(70,80)(110,80){5}{5}
\Text(50,30)[r]{$\mr{\dnt}_L$}
\Text(50,70)[r]{$\mr{d}$}
\Text(75,50)[l]{$\mr{d}$}
\Text(-2,50)[r]{$\mr{\cht^0_1}$}
\Text(115,20)[l]{$\mr{\nu_\mu}$}
\Text(115,80)[l]{$\mr{\gamma}$}
\end{picture}
\end{center}
\captionB{Radiative \rpv\  decay of a neutralino.}
	{Radiative \rpv\  decay of a neutralino.}
\label{fig:photinoraddecay}
\end{figure}
% End of figure  

\item A similar model to \cite{Choudhury:1996pj} with a three-body decay for
  the neutralino has been proposed in \cite{Nowakowski:1996dx}. In 
  \cite{Nowakowski:1996dx} the neutralino decay was due to 
  neutralino/neutrino mixing from the bilinear term in the \rpv\  
  superpotential, Eqn.\ref{eqn:Rsuper1}. This explanation of the KARMEN
  anomaly requires the Higgs mixing term in the superpotential $\mu H_1H_2$
  to be unnaturally small, $\mu\leq30\, \mr{\mev}$. Furthermore, this
  scenario implies a \mev\  mass for the tau
  neutrino which is ruled out by cosmological and astrophysical arguments 
  \cite{Sarkar:1996dd,Raffelt:1996wa}.

\item	In any model with a single particle X, produced in the decay
	$\mr{\pi^+\ra\mu^+X}$, there is fine-tuning between the mass of the
	X particle and the difference between the pion and muon masses in
        order to reproduce the experimental results. This fine-tuning is
        approximately one part in $10^4$, \ie 
	$1-M_{\mr{X}}/(m_\pi-m_\mu)=1.8\times10^{-4}$, 
	where $m_\mu$ is the muon mass and $m_\pi$ is the charged pion mass.

 	In \cite{Lukas:2000fy} a brane-word model was proposed which 
	alleviates this problem. In this model the particle X is a sterile
	neutrino which is part of a tower of Kaluza-Klein excitations
	associated with a singlet fermion, with respect to the Standard Model
	gauge group, propagating in $4+d$ dimensions. If the mass splitting
	of the Kaluza-Klein tower is small it is more plausible that one of
	these Kaluza-Klein excitations has the correct mass to explain the
	experimental results than if there is only a single particle, X. The
	problem with this model is that the neighbouring states must not be
	detectable.  
\end{itemize}

  We shall present a model which extends the model of \cite{Choudhury:1996pj}
  by including a three-body decay of the neutralino via trilinear R-parity
  violation. The details of this model are discussed in the next section. We 
  then consider other possible constraints on this model both from  low
  energy experiments,  Section~\ref{sect:karmencoup}, and present day
  collider experiments, Section~\ref{sect:karmenlight}. We also suggest
  possible tests of this model in future experiments.

\section[The Model]{The Model}

  We propose, as in \cite{Choudhury:1996pj}, that X is the lightest
  neutralino with mass $M_{\cht^0_1}=33.9\  \mr{MeV}$ and that this is the
  lightest supersymmetric particle in the model. 
  Since X is effectively stable on collider time-scales
  ($\tau_{\cht^0_1}\gtrsim0.07\,\mr{s}$) our model is experimentally
  very similar to the MSSM. In the GUT-inspired MSSM
  $M_1=\frac53\tan^2\tht_W M_2$,
  where $M_1$ and $M_2$ are the SUSY-breaking masses for the bino and wino
  respectively.\footnote{The gaugino mass terms in the
	 SUSY Lagrangian can be found in  Appendix~\ref{sect:mixing}.}
  Assuming this relation requires that $M_{\cht^0_1}>32.3\, \mr{\gev}$
  from current LEP data
  \cite{Barate:1999rn}. Hence, in order to obtain a very light neutralino in
  the SUSY spectrum we must consider $M_1$ and $M_2$ to be independent
  parameters. A  small value of $M_2$ implies at least one light chargino.
  This can be seen by considering small values of $M_2$ and the chargino mass
  matrix in Eqn.\,\ref{eqn:charginomatrix}. Such a light chargino is
  excluded by experiment. We must therefore consider small values of $M_1$.
  This implies, from the neutralino mass matrix in
  Eqn.\,\ref{eqn:neutralinomatrix}, that
  the lightest neutralino is dominantly bino. We will quantify this in 
  Section~\ref{sect:karmenlight} where we present regions in the 
  \mbox{$(M_1, M_2, \mu, \tan\be)$} parameter space which are consistent with
  all the current experimental limits. These solutions are indeed dominantly
  bino, with a small higgsino component.

  Instead of only considering one non-zero R-parity violating coupling we
  will allow two non-zero couplings. First we allow the coupling
  ${\lam'}_{211}$ to be non-zero so the pion can decay as in
  \cite{Choudhury:1996pj},
\begin{equation}
           \mr{\pi}^+ \longrightarrow \mr{\mu}^+  \mr{\cht^0_1}.
\end{equation}
  The tree-level Feynman diagrams for this decay are shown in
  Fig.\,\ref{fig:piondecay}. In order to allow a three-body decay for the
  lightest neutralino we consider a  non-zero coupling $\lam_{1j1}$, where
  $j$ is 2 or 3. The neutralino can then undergo a three-body decay
\begin{equation}
 \cht^0_1 \longrightarrow \mr{e}^- \nu_j \mr{e}^+.
\end{equation}
  Due to the low mass of the neutralino this is the only kinematically allowed
  three-body decay mode. This decay mode will dominate over the two-body
  radiative decay mode.

%
% Pion Decay to a muon and a neutralino
%
\begin{figure}
\begin{center} 
\begin{picture}(360,60)(0,30)
\SetScale{0.7}
\SetOffset(0,-15)
\ArrowLine(5,76)(60,102)
\ArrowLine(60,102)(5,128)
\DashArrowLine(90,102)(60,102){5}
\ArrowLine(90,102)(145,76)
\ArrowLine(145,128)(90,102)
\Text(80,53)[]{$\mr{\cht^{0}}$}
\Text(80,88)[]{$\mr{\mu^{+}}$}
\Text(25,90)[]{$\mr{\bar{d}}$}
\Text(25,57)[]{$\mr{u}$}
\Text(53,80)[]{$\mr{\mut}_L$}
\Vertex(60,102){1}
\Vertex(90,102){1}
\SetScale{0.7}
\ArrowLine(240,128)(185,128)
\ArrowLine(240,128)(295,128)
\ArrowLine(185,76)(240,76)
\ArrowLine(295,76)(240,76)
\DashArrowLine(240,76)(240,128){5}
\Text(190,98)[]{$\mr{\cht^{0}}$}
\Text(190,45)[]{$\mr{\mu^{+}}$}
\Text(150,98)[]{$\mr{\bar{d}}$}
\Text(150,47)[]{$\mr{u}$}
\Text(160,70)[]{$\mr{\tilde{d}}_R$}
\Vertex(240,128){1}
\Vertex(240,76){1}
\ArrowLine(420,128)(365,128)
\ArrowLine(475,128)(420,128)
\ArrowLine(365,76)(420,76)
\ArrowLine(420,76)(475,76)
\DashArrowLine(420,76)(420,128){5}
\Text(315,45)[]{$\mr{\cht^{0}}$}
\Text(315,98)[]{$\mr{\mu^{+}}$}
\Text(277,98)[]{$\mr{\bar{d}}$}
\Text(277,47)[]{$\mr{u}$}
\Text(287,70)[]{$\mr{\tilde{u}}_{L}$}
\Vertex(420,128){1}
\Vertex(420,76){1}
\end{picture}
\end{center}
\captionB{Pion decay to muon and neutralino.}
	{Pion decay to muon and neutralino.}
\label{fig:piondecay}
\end{figure}
% End of the figure

\begin{figure}
\centering
\includegraphics[angle=90,width=0.8\textwidth]{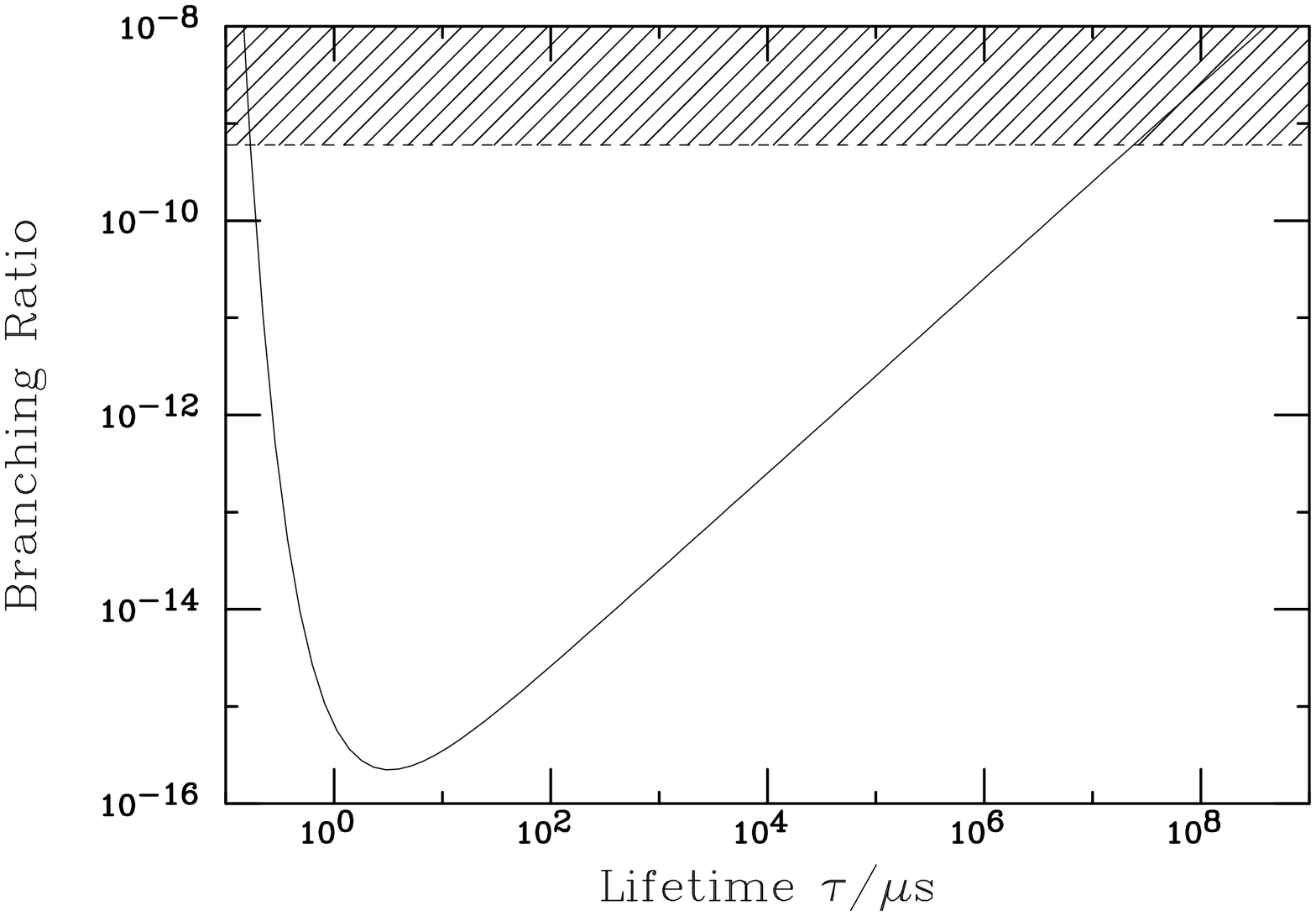}
\captionB{Branching ratio vs. lifetime required to solve the KARMEN anomaly.}
	{Branching ratio for $\mr{\pi^+\ra\mu^+X}$ against lifetime of X
	 required to solve the KARMEN anomaly. The hashed area gives the
	 experimental upper bound, Eqn.\,\ref{eqn:pionbound}. The results of
	 \cite{Formaggio:2000ne} also exclude an additional region down to
	 branching ratios of less than $10^{-12}$ for lifetimes between
	 $\sim 10^{-9}-10^{-3}\,\mr{s}$.}
\label{fig:karmenlife}
\end{figure}

  Fig.\,\ref{fig:karmenlife} shows the values of the branching ratio,
  $\mr{BR(\pi^+\ra\mu^+\cht^0_1)}$, and the neutralino lifetime,
  $\tau_{\cht^0_1}$, which are compatible with the KARMEN data 
  \cite{Armbruster:1995nr}. We can therefore determine the range of couplings
  ${\lam'}_{211}$, ${\lam}_{131}$ (or ${\lam}_{121}$) these solutions
  correspond to in our model by calculating both the pion branching ratio and
  the lifetime of the neutralino. 
  We can understand the shape of this curve, \ie Fig.\,\ref{fig:karmenlife},
  in the following way. The number of events
  detected is the number of neutralinos which decay inside the detector,
  multiplied by the efficiency for detecting such a decay. For one
  neutralino, with lifetime $\tau$, the probability of it decaying inside
  the detector is 
\begin{equation}
  P_{\mr{detected}} =
	 \left[e^{-\frac{t_1}{\tau}}-e^{-\frac{t_2}{\tau}}\right]\!,
\end{equation}
  where $t_1$ is the time the neutralino enters the detector and $t_2$ is the
  time it leaves the detector. The total number of events detected by the
  KARMEN experiment from neutralino decay is
\begin{equation}
  N_{\mr{detected}} = N_{\mr{produced}} \times G \times P_{\mr{detected}},
\end{equation}
  where $N_{\mr{produced}}$ is the number of neutralinos produced and $G$ is a
  geometrical factor describing the fraction of the total solid angle covered
  by the KARMEN experiment multiplied by the probability of detecting a decay
  of the X particle which occurs inside the detector. The number of
  neutralinos produced is 
  $N_{\mr{produced}}=N_\pi\times\mr{BR}(\pi^+\ra\mu^+\cht^0_1)$, where
  $N_\pi$ is the number of pions produced at the target. Hence the branching
  ratio required to solve the KARMEN anomaly is given by
\begin{equation}
  \mr{BR}(\pi^+\ra\mu^+\cht^0_1) = \frac{N_{\mr{detected}}}{N_{\pi} \times G}
	\frac1{\left[e^{-\frac{t_1}{\tau}}-e^{-\frac{t_2}{\tau}}\right]}.
\label{eqn:karmenexp}
\end{equation}
  The ratio $\frac{N_{\mr{detected}}}{N_{\pi}\times G}=1.65\times10^{-17}$.
  The times $t_1$ and $t_2$ at which the particle enters and leaves the
  detector can be calculated from the geometry of the detector and the
  particle's velocity giving
\begin{subequations}
\begin{eqnarray}
          t_1&=2.925\ \mu\mr{s},\\
          t_2&=3.579\ \mu\mr{s}.
\end{eqnarray}
\end{subequations}
  In the limit $\tau\gg t_1,\  t_2$ this reduces to
\begin{equation}
  \mr{BR}(\pi^+\ra\mu^+\cht^0_1) \simeq\frac{N_{\mr{detected}}\times\tau}
	{N_{\pi}
   \times G\,(t_2-t_1)}.
\label{eqn:pionapprox}
\end{equation}
  This gives the linear region in Fig.\,\ref{fig:karmenlife} for
  $\tau>10\ \mu\mr{s}$.

%
% Reproduction of the graph from the KARMEN paper
%
% End of the figure

  The calculation of the anomalous pion decay and the neutralino lifetime are
  presented below. We can use Eqn.\,\ref{eqn:karmenexp} to calculate
  the range of the couplings ${\lam'}_{211}$, ${\lam}_{131}$
  (or~${\lam}_{121}$) which are consistent with the lifetimes and branching
  ratios required to solve the KARMEN anomaly.
%
% Calculation of the RPV Pion Decay rate
%
\subsection[Pion Decay Rate]{Pion Decay Rate}

  We can use chiral perturbation theory to calculate the decay rate of
  the pion via \rpv\  as we would to calculate the Standard Model weak 
  decay rate of the pion. To do this we need to obtain an effective
  Lagrangian for the four-fermion interaction of $\mu^-$, $\mr{\bar{u}}$, 
  $\mr{d}$, and $\chi^0_1$ with the sfermion degrees of freedom integrated
  out, just as we would use the Fermi theory with the W degrees of freedom
  integrated out to perform the calculation of the Standard Model
  decay rate.

  It is easiest to manipulate these Lagrangians in two-component
  notation. Using this notation the relevant terms in the
  fermion-sfermion-neutralino Lagrangian are
\begin{equation}
 \mathcal{L}_{\mr{f\tilde{f}\cht}} = \sqrt{2}
             \left( A_\mu \mut_L \psb_{\cht} \psb_{\mu_L}
                   +A_u \upt_L \psb_{\cht} \psb_{u_L}
                   +A_d \dnt^*_R \psb_{\cht} \psb_{d^c_L} \right)+\mr{h.c.},
\end{equation}
  where the coefficients are given in Table\,\ref{tab:karcp}.
  It should be noted that we have only kept the gaugino pieces
  of the Lagrangian as we are not interested in the higgsino
  case. Similarly, the relevant pieces of the \rpv\  Lagrangian are
\begin{equation}
  \mathcal{L}_{\mathrm{\not \,\!\,R_p}\/\,} = 
         \lam'_{211} \left( \dnt_R \psb_{\mu_L} \psb_{u_L}
                 +\upt^*_L \psb_{\mu_L}\psb_{d_L^c}
                 +\mut^*_L \psb_{u_L} \psb_{d^c_L}  \right)+\mr{h.c.}.
\end{equation}
    We then proceed by integrating out the heavy sfermion degrees of
  freedom to obtain an effective Lagrangian
\begin{equation} 
  \mathcal{L}_{\mr{eff}} = \lam'_{211}\sqrt{2} \left( 
  \frac{A_\mu}{M^2_{\mr{\mut}_L}}\psb_{\cht}\psb_{\mu_L}\psb_{u_L}\psb_{d_L^c}
 +\frac{A_u}{M^2_{\mr{\upt}_L}}\psb_{\cht}\psb_{u_L}\psb_{\mu_L}\psb_{d_L^c}
 +\frac{A_d}{M^2_{\mr{\dnt}_R}}\psb_{\cht}\psb_{d_L^c}\psb_{\mu_L}\psb_{u_L}
    \right)+\mr{h.c.},
  \label{eqn:effective}
\end{equation}
  where $M_{\mr{\mut}_L}$ is the left smuon mass, $M_{\mr{\upt}_L}$ is the 
  left up squark 
  mass and $M_{\mr{\dnt}_R}$ is the right down squark mass.

% Table for the coefficients
\begin{table}
\begin{center}
\begin{tabular}{|c|c|c|c|}
\hline
 Coefficient & General coupling & Pure photino & Pure bino \\
\hline
 $A_\mu$  & $e N'_{l1} + \frac{gN'_{l2}}{\cos\theta_{\rm
  W}}\left(\frac{1}{2}-\sin^2\theta_{\rm W}\right)$
        & $\phantom{-}e \phantom{e_d}$
        & $-\frac{g' Y_{\mu_L}}{2}$\\
\hline
 $A_u$  & $ -e e_u N'_{l1} - \frac{gN'_{l2}}{\cos\theta_{\rm W}}\left(
  \frac{1}{2}-e_u\sin^2\theta_{\rm W}\right)$
        & $ -e e_u $
        & $-\frac{g'Y_{u_L}}{2}$ \\
\hline
 $A_d$  & $ e e_d N'_{l1} - \frac{g e_d \sin^2\theta_{\rm W} 
  N'_{l2}}{\cos\theta_{\rm W}}$
        & $\phantom{-}e e_d$
        & $\phantom{-}\frac{ g'Y_{d_R}}{2}$ \\
\hline
\end{tabular}
\captionB{Coefficients for the fermion-sfermion-neutralino Lagrangian.}
	{Coefficients for the fermion-sfermion-neutralino Lagrangian.
	 The hypercharges of the MSSM fields are given in
	 Table~\ref{tab:superfield}.}
\label{tab:karcp}
\end{center}
\end{table}
%end of table

   To enable us to apply the results of chiral perturbation theory we
  need to rearrange the last two terms in Eqn.\,\ref{eqn:effective}
  using a Fierz transformation. This leads to  Eqn.\,\ref{eqn:Fierz}
  and some tensor-tensor interaction terms which we can neglect as these
  do not contribute to the pion decay,
\begin{equation}
  \mathcal{L}_{\mr{Fierz}} =  \sqrt{2}\lam'_{211}
   \left(\frac{A_\mu}{M^2_{\mr{\mut}_L}}
	-\frac{A_u}{2M^2_{\mr{\upt}_L}}
	-\frac{A_d}{2M^2_{\mr{\dnt}_R}}\right)
   \psb_{\cht}\psb_{\mu_L}\psb_{u_L}\psb_{d_L^c}+\mr{h.c.}.
 \label{eqn:Fierz}
\end{equation}
  This equation can be rewritten in terms of  four-component Dirac spinors
\begin{equation}
  \mathcal{L}_{\mr{Fierz}} =  \sqrt{2}\lam'_{211}
   \left(\frac{A_\mu}{M^2_{\mr{\mut}_L}}
	-\frac{A_u}{2M^2_{\mr{\upt}_L}}
	-\frac{A_d}{2M^2_{\mr{\dnt}_R}}\right)
   \bar{\mu}P_R\cht^0 \bar{u}P_R d+\mr{h.c.},
\end{equation}
 where $P_L=\frac12(1-\ga_5)$ is
 the projection operator for the left-handed component of the fermion field
 and $P_R=\frac12(1+\ga_5)$ is the projection operator for the right-handed
 component of the fermion field. The amplitude for the pion decay can be
 calculated by splitting the amplitude into a pion matrix element and the
 matrix element for the final state,
\begin{equation}
 \mathcal{A} = \frac{i\lam'_{211}}{\sqrt{2}}
 \left(\frac{A_\mu}{M^2_{\mr{\mut}_L}}
      -\frac{A_u}{2M^2_{\mr{\upt}_L}}
      -\frac{A_d}{2M^2_{\mr{\dnt}_R}}\right)
     \bar{u}_\mu(p_1)P_R v_{\cht}(p_2)
 \int d^4x e^{i(p_1+p_2)\cdot x}
           \langle 0 | \bar{u} \gamma_5 d |\pi^-(p_0) \rangle,
\end{equation}
  where  $p_0$ is the pion momentum, $p_1$ is the muon momentum and
  $p_2$ is the neutralino momentum.
  The pion matrix element is given by the standard
  chiral perturbation theory identity
\begin{equation}
  \langle 0 |  j^{\mu 5 a}(x)|\pi^b(p_0)\rangle = -i p_0^\mu f_\pi
  \delta^{ab} e^{-ip_0\cdot x},
\label{eqn:chiral}
\end{equation}
  where $a$ and $b$ are the isospin of the pion, $j^{\mu 5 a}(x)$ is the
  axial-vector current and $f_{\pi}$ is the pion decay constant.
  It should be noted that some authors, including the Particle Data Group
  \cite{Caso:1998tx}, define $f_\pi$ to be $\sqrt{2} f_\pi$ compared to our
  definition. This is because they define Eqn.\,\ref{eqn:chiral} in
  terms of a charged pion basis rather than the isospin basis we have
  used here.

 Using Eqn.\,\ref{eqn:chiral} we can obtain
\begin{equation}
   \langle 0 | \bar{u} \gamma^\mu \gamma_5 d | \pi^-(p_0) \rangle =
   -i\sqrt{2}  p_0^\mu f_\pi e^{-ip_0\cdot x}.
\label{eqn:chiral2}
\end{equation}
  By contracting  Eqn.\,\ref{eqn:chiral2} with the pion four-momentum
  and using the Dirac equation for the up and down quarks we obtain,
\begin{equation}
   \langle 0 | \bar{u} \gamma_5 d | \pi^-(p_0) \rangle = 
   \frac{i\sqrt{2} f_\pi m^2_\pi e^{-ip_0\cdot x}}{(m_u+m_d)},
\end{equation}
  where $m_u$ and $m_d$ are the up and down quark masses respectively.
  Hence the amplitude for this process is given by,
\begin{equation}
 \mathcal{A} =  -\frac{\lam'_{211} f_\pi m^2_\pi }{(m_u+m_d)}
 \left(  \frac{A_\mu}{M^2_{\mr{\mut}_L}}
	-\frac{A_u}{2M^2_{\mr{\upt}_L}}
	-\frac{A_d}{2M^2_{\mr{\dnt}_R}}\right)
 \bar{u}_\mu(p_1)P_R v_{\cht}(p_2) (2\pi)^4 \delta^{(4)}
\left( p_0-p_1-p_2 \right)\!.
\end{equation}
 We can therefore obtain the following partial width:
\begin{equation}
\Gamma(\pi \ra \mu \cht^0_1) =
   \frac{{\lam'}^2_{211} f^2_\pi m^2_\pi p_{\mr{cm}}}{8 \pi (m_u+m_d)^2}
 \left(\frac{A_\mu}{M^2_{\mr{\mut}_L}}
	-\frac{A_u}{2M^2_{\mr{\upt}_L}}
	-\frac{A_d}{2M^2_{\mr{\dnt}_R}}\right)^2
 \left(m^2_\pi -m^2_\mu -m^2_{\cht} \right)\!,
\end{equation}
  where $ p_{\mr{cm}} = \frac{1}{2m_\pi}\sqrt{
                             \left[m^2_\pi -(m_\mu +m_{\cht})^2 \right]
                    \left[m^2_\pi -(m_\mu -m_{\cht} )^2 \right]}$ and
	$m_{\cht}$ is the lightest neutralino mass.

  The partial width for the Standard Model decay mode 
  $\pi^+ \ra \ell^+\nu_\ell$ is given by
\begin{equation}
 \Gamma(\pi^+ \ra \ell^+ \nu_\ell) =
	 \frac{G^2_F f^2_\pi m_\pi  m^2_\ell}{4\pi}
         \left(1-\frac{m^2_\ell}{m^2_\pi}\right)^2\!\!\!.
\end{equation}
\textheight 24cm
  In the Standard Model the dominant pion decay mode is to the muon and
  muon neutrino, and at the
  level of accuracy to which we are working
  we can approximate the total decay rate by the
  partial width for $\mr{\pi^+\ra\mu^+\nu_\mu}$. This gives the branching
%% Footnote saying Subir et al were wrong 
ratio\footnote{This disagrees slightly with the result given in
 	 \cite{Choudhury:1996pj}. 
          We have discussed this with the authors of \cite{Choudhury:1996pj}
          and have agreed on the above result.}
%% End of the footnote
\begin{eqnarray}
  \mr{BR}(\pi^+ \ra \mu^+ \cht^0_1) & = &
			    \frac{{\lam'}^2_{211} m^5_\pi p_{\mr{cm}}}
					{2 G^2_F m^2_\mu (m_u+m_d)^2}
			 \left(\frac{A_\mu}{M^2_{\mr{\mut}_L}}
			-\frac{A_u}{2M^2_{\mr{\upt}_L}}
				-\frac{A_d}{2M^2_{\mr{\dnt}_R}}\right)^2
 			\frac{\left(m^2_\pi -m^2_\mu -m^2_{\cht} \right)}
				{\left(m^2_\pi-m^2_\mu\right)^2} \nonumber \\
   		       & \approx & 1.4 \times 10^{-4} 
			\left(\frac{{\lam'}_{211}}{0.01}\right)^2
			\left(\frac{150 \,  \mr{\gev}}{M_{\ftl}}\right)^4 
			\label{eqn:pionbino}\\
			& < & 6.0\times 10^{-10},\label{eqn:pionlimit}
\end{eqnarray}
  where in Eqn.\,\ref{eqn:pionbino} we have assumed that the sfermion masses
  are degenerate, \ie \linebreak 
  \mbox{$M_{\mr{\mut}_L}=M_{\mr{\upt}_L}=M_{\mr{\dnt}_L}=M_{\ftl}$}, and the
  neutralino is purely bino. In Eqn.\,\ref{eqn:pionlimit} we have quoted the
  experimental bound from Eqn.\,\ref{eqn:pionbound}, which is shown as a
  hashed region in Fig.\,\ref{fig:karmenlife}. This bound can be satisfied by
  either a small \rpv\  coupling or a large sfermion mass. In principle it
  could also be satisfied if there was a fine-tuned cancellation between
  different diagrams for non-degenerate sfermion masses, however we disregard
  this possibility. This bound on the pion branching
  ratio can be translated into an upper bound on ${\lam'}_{211}$,
\begin{equation}
{\lam'}_{211}<
2.1\times10^{-5}\left(\frac{M_{\ftl}}{150\, \mr{\gev}}\right)^2\!\!\!.
\label{eqn:lampion}
\end{equation}

%
% Calculation of the Neutralino lifetime
%

\subsection[Neutralino Lifetime]{Neutralino Lifetime}

{\renewcommand{\arraystretch}{1.5}
\begin{table}
\begin{center}
\begin{tabular}{|c|c|c|c|}
\hline
Coefficient & General Formula & Pure Photino & Pure Bino \\
\hline
 $B_1$ & 
   $-\left( eN'_{l1} 
   +\frac{gN'_{l2}}{\cw}\left[\frac{1}{2}-\ssw\right]\right)$
   & $-e$  &$\phantom{-}\frac{Y_{e_L}}{2}$ \\
\hline
 $B_2$ & $\frac{gN'_{l2}}{2 \cw}$ &\phantom{-} 0 &
$\phantom{-}\frac{Y_{\nu_L}}{2}$ \\
\hline
 $B_3$ &
   $\left(eN'_{l1} -\frac{gN'_{l2}\ssw}{\cw}\right)$ &
 $\phantom{-}e$ & $-\frac{Y_{e_R}}{2}$ \\
\hline
\end{tabular}
\captionB{Coefficients for the neutralino decay.}
	{Coefficients for the neutralino decay.}
\label{tab:LLEneut}
\end{center}
\end{table}}
  We also need to calculate the decay rate for the three-body decay of the
  neutralino produced in the pion decay via the $L_iL_j\overline{E}_k$ term
  in the \rpv\  superpotential, Eqn.\,\ref{eqn:Rsuper1}. We calculate this
  decay rate neglecting
  the momentum flow through the virtual sfermions, as we did in the
  pion decay calculation, and the masses of the final-state particles.
  With these approximations the decay rate is
\begin{eqnarray}
  \Gamma (\cht \ra \ell^+_i \bar{\nu}_j \ell^-_k) &= &
       \frac{M^5_{\cht} \lam^2_{ijk}}{3072 \pi^3} 
    \left( \frac{B^2_1}{M^4_{\elt_{iL}}}  +\frac{B^2_2}{M^4_{\nut_{jL}}} 
           +\frac{B^2_3}{M^4_{\elt_{kR}}}  \right.\nonumber\ \\
 &&\left.
  -\frac{B_1B_2}{M^2_{\elt_{iL}}M^2_{\nut_{jL}}}
     -\frac{B_1B_3}{M^2_{\elt_{iL}}M^2_{\elt_{kR}}}
	-\frac{B_2B_3}{M^2_{\nut_{jL}}M^2_{\elt_{kR}}}
    \right)\!.
\end{eqnarray}
\textheight 23cm
  This is obtained by integrating the matrix element given in 
  Appendix~\ref{sect:decayneut}, Eqn.\,\ref{eqn:LLEneutdecay},
  over the available phase space,
  Eqn.\,\ref{eqn:threebodyphase}. The coefficients are given in
  Table\,\ref{tab:LLEneut}. In the pure bino limit this gives
\begin{equation}
  \Gamma (\cht \ra \ell^+_i \bar{\nu}_j \ell^-_k) = 
          \frac{3\al M^5_{\cht} \lam^2_{ijk}}{1024 \pi^2 \ccw M^4_{\ftl}},
\end{equation}
assuming a common sfermion mass.
  This gives the result for the neutralino lifetime 
\begin{eqnarray}
  \tau_{\mr{bino}}    & = & 1.32 \times 10^{-3}  \frac{1}{\lam^2_{ijk}} 
                       \left(\frac{M_{\ftl}}{150\, \mr{\gev}}\right)^4
                       \left(\frac{33.9\,\mr{MeV}}{M_{\cht}}\right)^5\!\!\!,
\label{eqn:neutlife2}\\
  & <& 23.8\  \mr{s}.
\label{eqn:neutlife}
\end{eqnarray}
  The last inequality comes from using the bound, Eqn.\,\ref{eqn:pionbound},
  on the pion branching ratio and the solutions shown in 
  Fig.\,\ref{fig:karmenlife}. This can be inverted to give a bound on the
  coupling
\begin{equation}
\lam_{131},\  \lam_{121} > 7.45\times 10^{-3}
			\left(\frac{M_{\ftl}}{150\, \mr{\gev}}\right)^2\!\!\!.
\label{eqn:lamlife}
\end{equation}

\subsection[Solutions of the KARMEN Anomaly]{Solutions of the KARMEN Anomaly}

\begin{figure}[t]
\centering
\includegraphics[angle=90,width=0.8\textwidth]{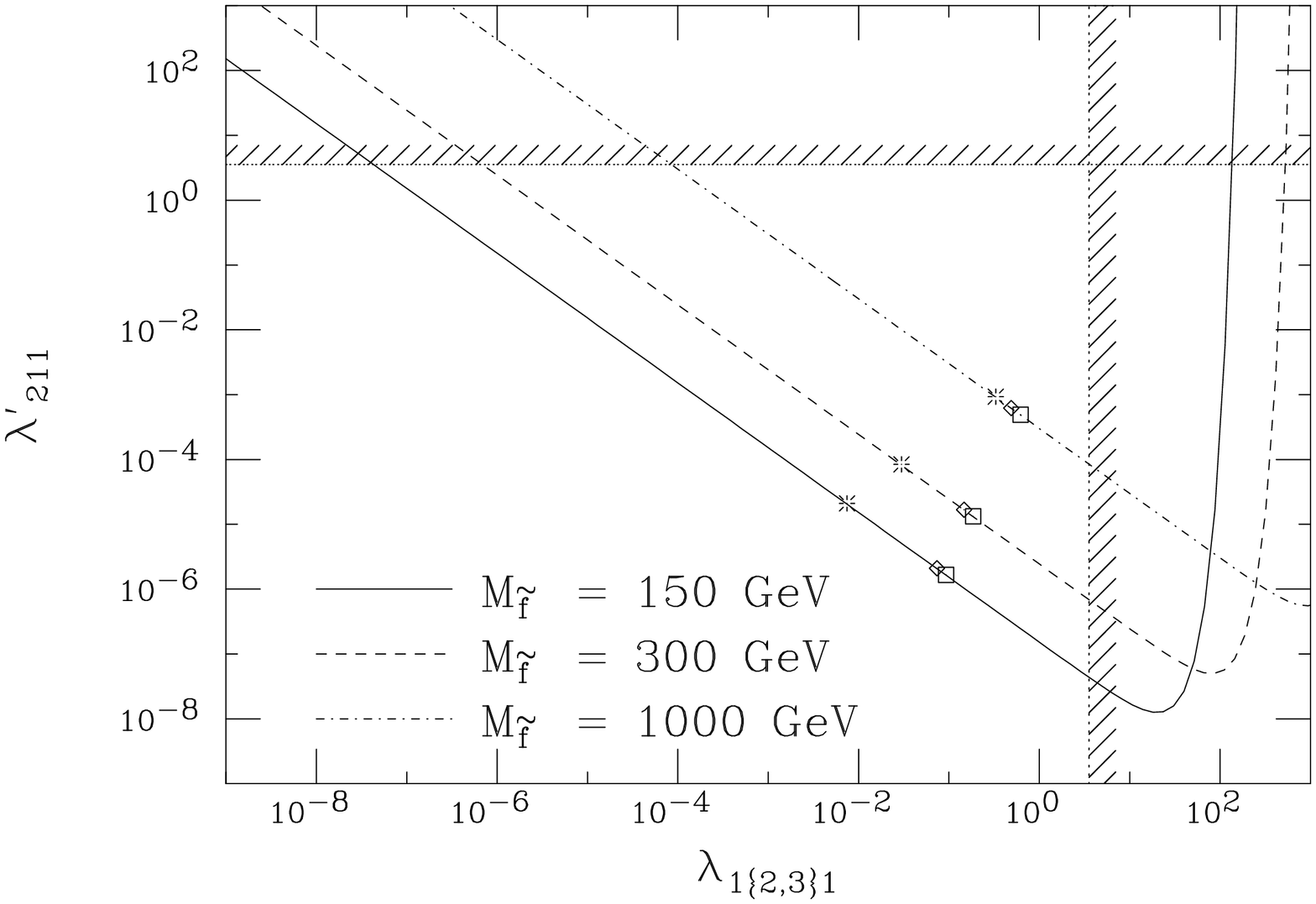}
\captionB{Solutions of the KARMEN anomaly in terms of the \rpv\  couplings.}
	{Solutions to the KARMEN anomaly in terms of the
 R-parity violating couplings $\lambda'_{211}$ and
 $\lambda_{1\{2,3\}1}$, for different (assumed degenerate) sfermion
 masses. The hashed lines indicate upper limits on the couplings from
 perturbativity. The stars and diamonds (squares) give the upper
 limits on the couplings $\lambda'_{211}$ and $\lambda_{121}$
 ($\lambda_ {131}$), respectively. 
 Solutions above and to the left of
 the stars are excluded, as are solutions below and to the right of
 the diamonds (squares).}
\label{fig:couplife}
\end{figure}

  The KARMEN collaboration has produced a graph of the branching ratio
  for $\pi^+ \ra \mu^+ \mr{X}$ against the lifetime of X which is required
  to explain their data, Fig.\,\ref{fig:karmenlife}. In our model, each point
  along the curve in Fig.\,\ref{fig:karmenlife} corresponds to a specific
  anomalous pion branching ratio, Eqn.\,\ref{eqn:pionbino}, and a specific
  neutralino lifetime, Eqn.\,\ref{eqn:neutlife2}. If we assume that the
  scalar fermions are degenerate we can translate these solutions
  into specific values of the couplings ${\lam'}_{211}$ and 
  $\lam_{131}$ (or $\lam_{121}$) for a fixed sfermion mass using
  Eqn.\,\ref{eqn:karmenexp}. This set of solutions in
  \mbox{R-parity} violating SUSY parameter space is shown in
  Fig.\,\ref{fig:couplife} for \mbox{$M_{\ftl}=150\, \mr{\gev}$}  
 (solid line), $M_{\ftl}=300\, \mr{\gev}$  (dashed line)  and 
  $M_{\ftl}=1000\, \mr{\gev}$  (dot-dash line). The hashed lines at
  $\lam,\  \lam'=\sqrt{4\pi}$ give the perturbative limit. For large
  sfermion masses ($>1\, \mr{TeV}$) there is little room for perturbative
  solutions. Solutions above and to the left of the stars are excluded by the
  inequalities in Eqns.\,\ref{eqn:lampion} and \ref{eqn:lamlife}.

%
%   Our solutions to the KARMEN anomaly
%

  Given the perturbative upper bound, $\lam_{ijk}<\sqrt{4\pi}$, and the lower
  bound on the sfermion mass from LEP~2, $M_{\ftl}>100\, \mr{\gev}$, we also
  have a lower limit on the lifetime of the neutralino, 
  $\tau_{\mr{bino}}>2.1\times10^{-5}$s. There is therefore a range of six
  orders of magnitude in the lifetime or three orders of magnitude in the
  coupling for solutions in our model.
%
% Section on the experimental limits on the R-parity violating couplings
%
\section[Limits on the R-parity Violating Couplings]
	{Limits on the R-parity Violating Couplings}
\label{sect:karmencoup}

  There are constraints on the R-parity violating couplings which come from
  low energy experiments. The best bounds at the $2\sigma$-level on the
  couplings we are interested in are  summarized in
  \cite{Allanach:1999ic},
\begin{eqnarray}
{\lam'}_{211} & < & 0.059
	 \left(\frac{M_{\dnt_R}}{100\, \mr{\gev}}\right), \nonumber \\
{\lam}_{121}  & < & 0.049 
	\left(\frac{M_{\elt_R}}{100\, \mr{\gev}}\right) \Rightarrow
	\tau_{\mr{bino}}>0.11\,\mr{s},
\label{eqn:Rppionbounds} \\
{\lam}_{131}  & < & 0.062 
	\left(\frac{M_{\elt_R}}{100\, \mr{\gev}}\right) \Rightarrow
	\tau_{\mr{bino}}>0.07\mr{s}. \nonumber 
\end{eqnarray}
  The bound on ${\lam'}_{211}$ is from the measurement of
  $R_\pi=\Gamma(\pi\ra e\nu)/\Gamma(\pi\ra\mu\nu)$ \cite{Barger:1989rk},
  the bound on $\lam_{121}$ is from charged-current universality
  \cite{Barger:1989rk} and the bound on $\lam_{131}$ is from a measurement of 
  $R_\tau=\Gamma(\tau\ra e\nu\bar{\nu})/\Gamma(\tau\ra\mu\nu\bar{\nu})$ 
  \cite{Barger:1989rk}.

  The bound on ${\lam'}_{211}$ in Eqn.\,\ref{eqn:Rppionbounds} is weaker than
  the constraint on the coupling imposed by the bound on the pion branching
  ratio, Eqn.\,\ref{eqn:pionbound}, and therefore we do not consider it
  further. In Fig.\,\ref{fig:couplife} the bounds on the couplings
  $\lam_{121}$ and $\lam_{131}$ forbid solutions to the right of the diamonds
  and squares, respectively. This leaves a range of solutions of about one
  order of magnitude in ${\lam'}_{211}$ and $\lam_{131}$ (or $\lam_{121}$),
  and corresponds to two orders of magnitude in the pion branching ratio and
  neutralino lifetime. The bounds on the couplings $\lam_{131}$ and
  $\lam_{121}$ give a lower bound on the neutralino lifetime, using
  Eqn.\,\ref{eqn:neutlife2}, which is given in Eqn.\,\ref{eqn:Rppionbounds}. 

  We must also consider bounds on the products of the couplings
  ${\lam'}_{211}\lam_{121}$ and  ${\lam'}_{211}\lam_{131}$. In the first case
  there is an additional contribution to the pion decay
  $\pi^+\ra\tilde{\mu}^+\ra \mr{e^+\nu_e}$, which changes the prediction for
  $R_\pi$. To calculate this decay rate we first need to obtain an effective
  Lagrangian for the four-fermion interaction of the u, $\mr{\bar{d}}$, 
  $\mr{\nu_e}$ and e. This is 
\begin{equation}
   \mathcal{L} = -\frac{{\lam'}_{211}\lam_{121}}{M^2_{\tilde{\mu}_L}}
     		  \bar{d} P_L u \bar{\nu} P_R e,
\end{equation}
 where again the sfermion degrees of freedom have been integrated out. 
 
  We can use chiral perturbation theory to calculate the partial width for
  the decay $\mr{\pi^+\ra e^+ \nu_e}$ including this correction to the
  Standard Model rate. This gives
\begin{equation}
  \Gamma(\pi^+\ra \mr{e^+ \nu_e}) = \frac{G^2_F f^2_\pi m_\pi  m^2_e}{4\pi}
                               \left(1-\frac{m^2_e}{m^2_\pi}\right)^2 \left[1
 					+\frac{m^2_\pi{\lam'}_{211}\lam_{121}}
		         {2\sqrt{2}G_F M^2_{\tilde{\mu}_L}m_e(m_u+m_d)}
				     \right]^2\!\!\!.
\end{equation}
  We can express this as a correction to the ratio $R_\pi$,
\begin{equation}
 R_\pi = R_\pi^{\rm SM} \left[1+\frac{m^2_\pi{\lambda'}_{211}\lambda_{121}}
         {2\sqrt{2}G_{\rm F} M^2_{\tilde{\mu}{\rm L}} m_e (m_u+m_d)}
	 \right]^2\!\!\!.
\end{equation}
  The corresponding Feynman diagram has a different structure from the
  $t$-channel squark exchange which gives the bound on ${\lam'}_{211}$ in
  Eqn.\,\ref{eqn:Rppionbounds}. This leads to a much stricter bound on the
  product of the couplings than on either of the
  couplings individually. At the $2\sigma$ level we obtain
\begin{equation}
 \lambda'_{211}\lambda_{121} < 4.6 \times 10^{-7} \;
 \left(\frac{M_{\mr{\mut}_L}}{100\, \mr{\gev}}\right)^2\!\!\!.
\end{equation}
  Hence in the case of $\lam_{121}$ the maximum sfermion mass which will
  solve the KARMEN anomaly in our model is $400\, \mr{\gev}$,
  using Eqns.\,\ref{eqn:pionapprox},
  \ref{eqn:pionbino} and \ref{eqn:neutlife2}.
  
  The couplings ${\lam'}_{211}$ and $\lam_{131}$ violate muon and tau lepton
  number, respectively. This can therefore lead to the decay 
  $\tau\ra\mu\gamma$. The experimental bound on this decay has recently
  improved \cite{Anastassov:1999fk},
\begin{equation}
 {\rm BR}(\tau \rightarrow \mu \gamma) < 1.0 \times 10^{-6} \quad 
                                         {\rm (90\%~C.L.)}.
\end{equation}
  As the couplings ${\lam'}_{211}\lam_{131}$ only contribute to this decay
  at the two-loop level any bound from this decay is significantly weaker
  than the individual bounds on the couplings, Eqn.\,\ref{eqn:Rppionbounds}.

  There are also severe cosmological bounds on the R-parity violating
  couplings derived from considerations of GUT-scale lepto/baryogenesis in
  the early universe 
  \cite{Campbell:1991fa,Fischler:1991gn,Dreiner:1993vm,Campbell:1992at},
\begin{equation}
\lam,\ \lam',\ \lam'' < 5 \times 10^{-7}\left(\frac{M_{\ftl}}{1\, \mr{TeV}}
	 \right)\!.
\label{eqn:cosmobounds}
\end{equation}
  It was subsequently shown that it is sufficient for just one lepton flavour
  to satisfy this bound \cite{Dreiner:1993vm,Campbell:1992at}. In our model
  both of the couplings must violate this bound. For the case
  $({\lam'}_{211},\lam_{121})$ we must therefore require either all the
  electron or all the tau lepton number violating couplings to satisfy
  Eqn.\,\ref{eqn:cosmobounds}. For the case $({\lam'}_{211},\lam_{131})$ we
  must demand that all the electron number violating couplings satisfy
  Eqn.\,\ref{eqn:cosmobounds}. Another alternative is that
  baryogenesis could occur at the weak scale in which case the bounds in 
  Eqn.\,\ref{eqn:cosmobounds} do not apply.

%  Section on the experimental constraints
%
\section[Experimental Constraints on a Light Neutralino]
	{Experimental Constraints on a Light Neutralino}
\label{sect:karmenlight}

  There are a number of possible experimental constraints on a light
  neutralino which we will summarize here. We then show that there are
  regions of $(M_1,M_2,\mu,\tan\beta)$ parameter space in which all these
  constraints are satisfied for a dominantly bino lightest neutralino with a
  small higgsino component. We consider this scenario to avoid all of the
  constraints. In our model $M_1$ and $M_2$ are not related by the
  supersymmetric GUT relation and we treat them as independent free 
  parameters.

%
%  Bounds from e^+e^- ---> nu nu gamma
%
\subsection[Bounds from $\mr{e^+e^-\ra \nu \bar{\nu} \gamma}$]
	   {Bounds from \boldmath{$\mr{e^+e^-\ra \nu \bar{\nu} \gamma}$}}
  
\begin{figure}[t]
\centering
\includegraphics[angle=90,width=0.6\textwidth]{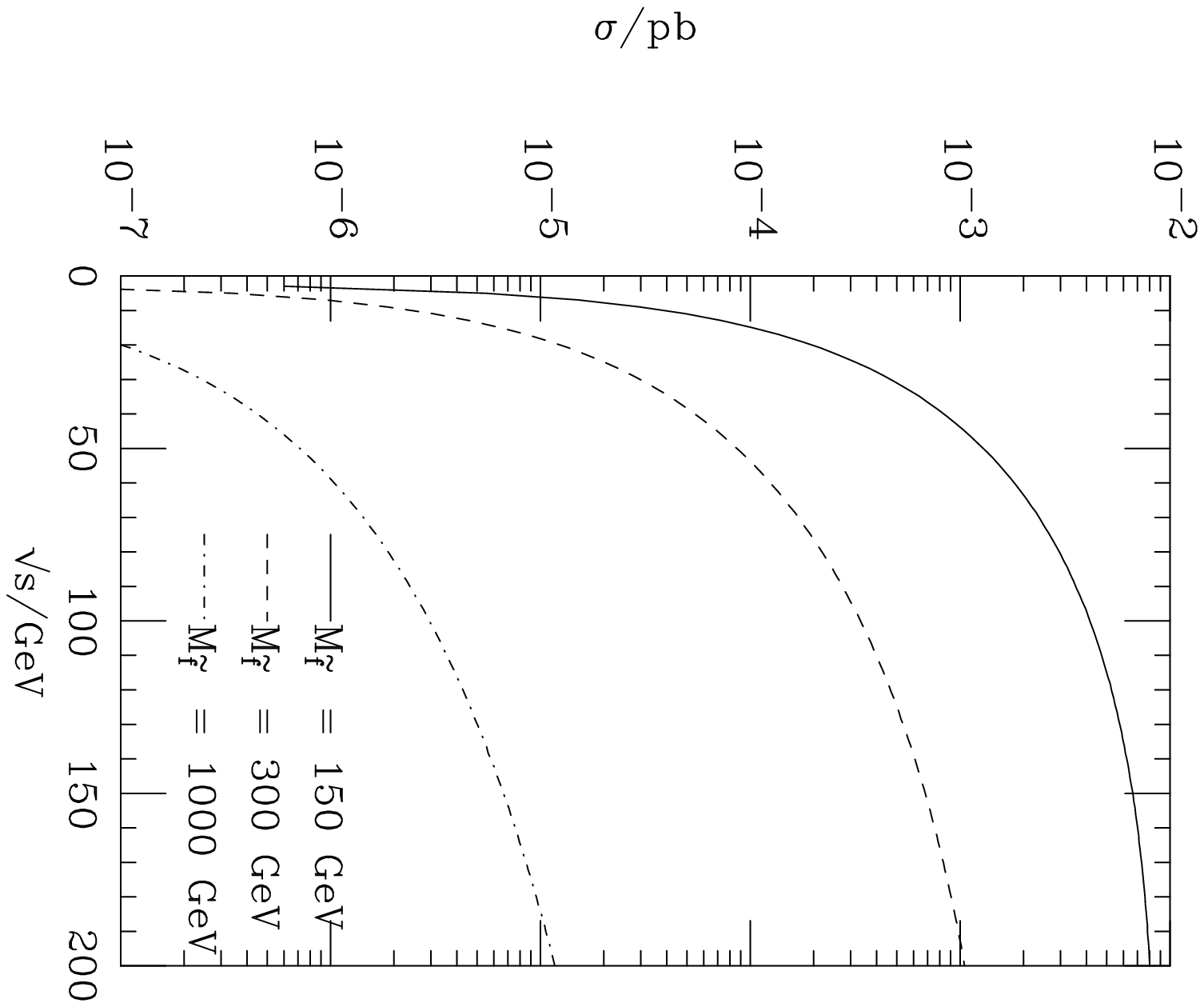}
\captionB{Cross section for neutralino production via 
 	 $\mr{e^+e^-\rightarrow\tilde{\chi}^0_1\tilde{\chi}^ 0_1\gamma}$.}
	{Cross section for the production of a purely bino
 	 neutralino with mass $33.9\, \mr{MeV}$ through
 	 $\mr{e^+e^-\rightarrow\tilde{\chi}^0_1\tilde{\chi}^ 0_1\gamma}$.}
\label{fig:binocross}
\end{figure}

  The Standard Model process $\mr{e^+e^-\ra \nu \bar{\nu} \gamma}$ is
  measured in $\mr{e^+e^-}$ collisions by looking for the presence of a
  photon and missing energy and momentum due to the neutrinos escaping the
  detector \cite{Ma:1978zm,Gaemers:1979fe}. As the lightest neutralino
  lifetime in our model is large enough that the neutralino will escape the
  detector before decaying there will be an additional contribution to this
  process from $\mr{e^+e^-\ra \cht^0_1 \cht^0_1 \gamma}$. The cross section
  for the process $\mr{e^+e^-\ra \tilde{\gamma} \tilde{\gamma} \gamma}$ was
  calculated in \cite{Grassie:1984kq}. We can use this result to obtain the
  cross section for a purely bino LSP by changing the relevant couplings. The
  cross section is shown as a function of the centre-of-mass energy in
  Fig.\,\ref{fig:binocross}. The expected number of events for various
  experiments is given in Table~\ref{tab:gammaevents} assuming a scalar
  fermion mass of $M_{\ftl}=150\, \mr{\gev}$. We used the same cuts on the
  energy and angle of the photon with respect to the beam direction as in 
  \cite{Grassie:1984kq}.

  As can be seen in Table~\ref{tab:gammaevents}, no limits on this process
  can be set by LEP as the expected number of events in much less than one.
  The recent results from
  OPAL give 138 observed events, against the Standard Model expectation of 
  $141.1 \pm 1.1$ events from $\mr{e^+e^-} \ra \nu \bar{\nu} \gamma$
  and the non-physics background of $2.3\pm 1.1$ events.
  There is no evidence for any excess given the statistical error
  of $\pm11.9$ events on the background. As the lightest neutralino in our
  model does not couple to the Z boson, the background at
  $\sqrt{\sh}=M_{\mr{Z}}$ is too large to see a signal.

%
% Photon Cross Section table
%
\begin{table}
\begin{center}
\begin{tabular}{|c|c|c|c|c|}
\hline
 Experiment & Integrated  & Energy 
  & Cross section (fb) & Number of events \\
 & luminosity (pb$^{-1}$) & & & \\
\hline
LEP    & 	6.65	& 130  & 	$5.87$		& 0.04\\
       & 	5.96	& 136  & 	$6.14$		& 0.04\\
       &	9.89	& 161  &  	$7.11$		& 0.07\\
       &	10.28	& 172  &  	$7.44$		& 0.08\\
       &	54.5	& 183  &  	$7.72$		& 0.42 \\
       &	75.0	& 200  & 	$8.05$		& 0.60 \\
\hline
 KEK-B & $1\times10^{5}$&  10.5 & $6.74\times10^{-2}$ 	& 6.7 \\
\hline
 BaBar & $3\times10^4$	& 10.5  & $6.74\times10^{-2}$ 	& 2.0 \\
\hline
 NLC   & $3\times10^5$  & 500   & 6.19                  & 1857 \\
\hline
\end{tabular}
\captionB{Cross sections for the production of
 	$\mr{\tilde{\chi}^0_1\tilde{\chi}^0_1\gamma}$ at $\mr{e^+e^-}$
	 colliders.}
	{Cross sections for the production of
 	$\mr{\tilde{\chi}^0_1\tilde{\chi}^0_1\gamma}$ at $\mr{e^+e^-}$ 
	colliders for
 	(an assumed degenerate) sfermion mass $M_{\tilde f}=150\, \mr{\gev}$
	 and general expectations of the integrated luminosity.}
\label{tab:gammaevents}
\end{center}
\end{table}

  Given the high luminosities at the B-factories KEK-B and BaBar a few events
  may be expected. The Standard Model cross section at this energy is 2.3 fb, 
  corresponding to $230\pm15$ events at KEK-B and $70\pm8$ events at 
  BaBar\footnote{This corresponds to one year of running based on the
	         luminosities given in \cite{Caso:1998tx}.}, where the quoted
		 errors are statistical.
  The statistical uncertainty on the Standard Model rate 
  still exceeds the signal rate and hence we do not expect 
  any sensitivity to a light neutralino.

  At the NLC we expect a substantially higher number of events. The Standard
  Model cross section for the same cuts is about $0.35$~pb for three neutrinos
  \cite{Ambrosani1999:private:Pukhov:1999gg:Ambrosanio:1996it} corresponding
  to $1.1\times10^{5}$ events, with a small relative statistical error of 330
  events. This can therefore provide a test of our model as the expected
  number of events from $\mr{e^+e^-}\ra\cht^0_1\cht^0_1\gamma$ is more than a
  $5\sigma$ fluctuation of the background.
%
%
%  Bounds from the invisible Z width
%
\subsection[Bounds from the Invisible Z Width]
	{Bounds from the Invisible Z Width}

%
%  Figure of the cross sections for neutralino neutralino photon
%
  In our model, as $M_{\cht^0_1}\ll M_\mr{Z}/2$, the decay 
  $\mr{Z^0}\ra \cht^0_1\cht^0_1$ is kinematically accessible. Due to its
  small mass with respect to $M_\mr{Z}$ the neutralino can be considered
  effectively massless like a neutrino. The lifetime of the lightest
  neutralino in our model is such that it will decay outside
  the detector, and therefore the process $\mr{Z^0}\ra \cht^0_1\cht^0_1$
  contributes to the invisible Z width. The current measurement of the
  invisible Z width can be expressed as a number of light neutrino species
  \cite{Caso:1998tx},\footnote{There has been a recent improvement
     \cite{Abbaneo:2000nr} in this result, $N_\nu = 2.9835 \pm 0.0083$. 
	This new result disagrees with the Standard Model at the 
	$2\sigma$-level and hence we must take the
		upper limit to be $3\sigma$ above the central value.}
\begin{equation}
N_\nu = 3.00 \pm 0.08.
\end{equation} 
  We must therefore require that 
  $\Gamma(Z^0\ra\cht^0_1\cht^0_1)< 0.08 \Gamma(Z^0\ra\nu\bar{\nu})$. A purely
  bino neutralino does not couple to the Z at tree level. The dominant
  contribution to the decay $Z^0\ra\cht^0_1\cht^0_1$ comes from the higgsino 
  admixtures of the neutralino, $N_{13}$ and $N_{14}$, in the notation of
  \cite{Haber:1985rc}. This enters with the fourth power in the decay 
  rate $\mr{Z^0}\ra \cht^0_1\cht^0_1$. The ratio of the neutralino to
  neutrino partial widths is
\begin{equation}
\frac{\Gamma(\mr{Z}\ra \cht^0_1 \cht^0_1)}{\Gamma(\mr{Z}\ra \nu\bar{\nu})}
	= \left(|N_{14}|^2-|N_{13}|^2\right)^2\!\!\!,
\end{equation} giving the constraint
\begin{equation}
\left||N_{14}|^2-|N_{13}|^2\right|^{1/2}<0.53\approx \left(0.08
	\right)^{1/4}\!\!\!.
\label{eqn:higgsinobound}
\end{equation} 
  We shall show below that it is straightforward to find regions which
  satisfy this in \linebreak $(M_1,M_2,\mu,\tan\beta)$ parameter
  space.\footnote{As these solutions
  contain $\lesssim 15$\,\% higgsino admixture they also satisfy the bound 
  $\left||N_{14}|^2-|N_{13}|^2\right|^{1/2}<0.30$ from the improved results of
  \cite{Abbaneo:2000nr}.}
%
% Scan of the MSSM parameter space
%
\subsection[Solutions in the MSSM Parameter Space]
	{Solutions in the MSSM Parameter Space}

\begin{figure}
\centering
\includegraphics[angle=90,width=8cm]{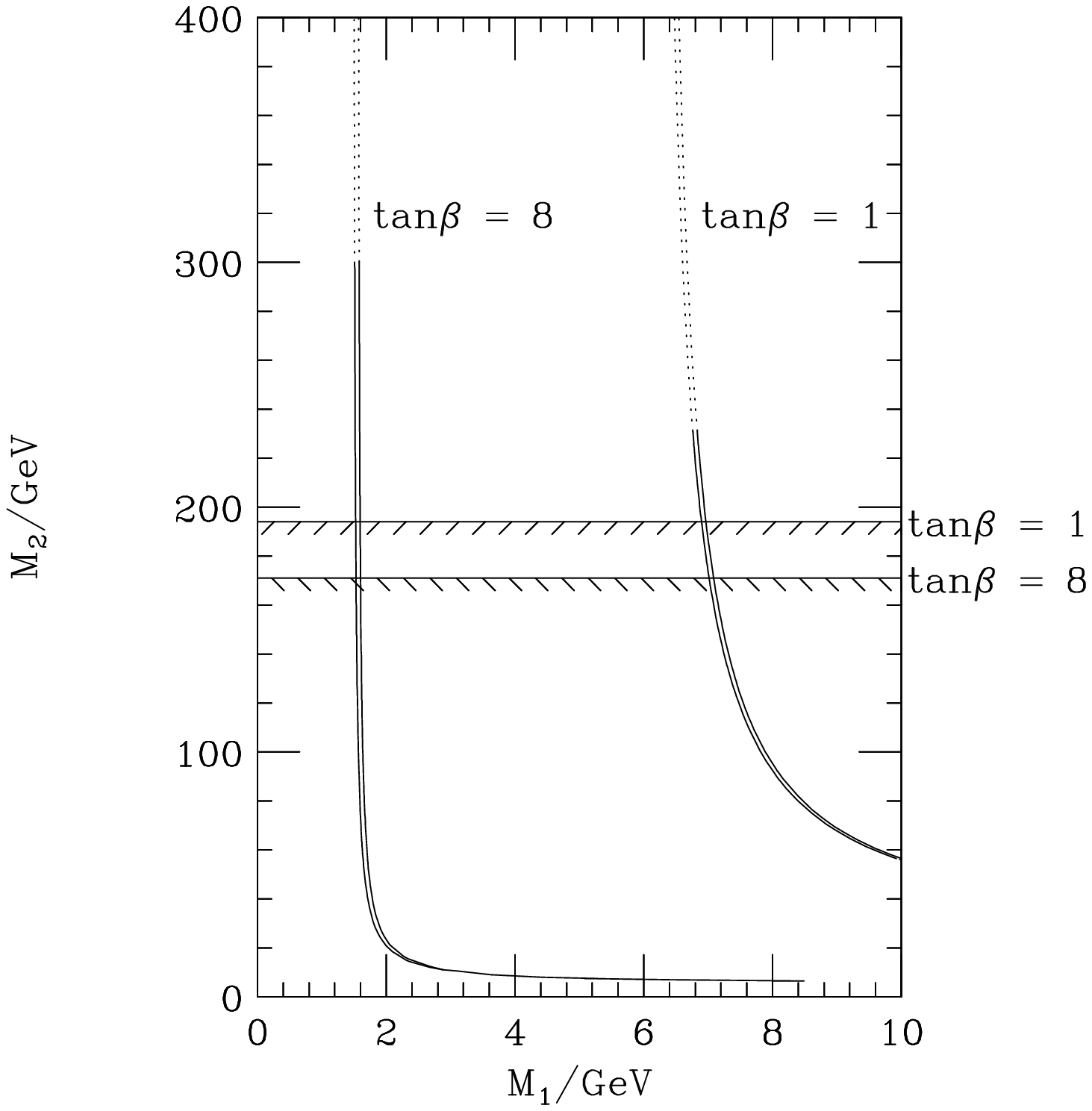}
\captionB{Solutions in $(M_1,M_2,\mu,\tan\beta)$ parameter space.}
	{Solutions in $(M_1,M_2,\mu,\tan\beta)$ parameter space
	 giving a $M_{\cht^0_1}=33.9\, \mr{MeV}$ neutralino for 
	$\mu=300\, \mr{\gev}$ and 2 representative values of $\tan\beta$.
	 The width of the lines is $0.01\, \mr{MeV}$. Below the hashed lines
 	 the chargino mass is less than $150\, \mr{\gev}$. The dotted lines
	 have $\Delta\rho_{\rm SUSY}<10^{-4}$ and
 	 the solid lines have $\Delta\rho_{\rm SUSY}<5 \times 10^{-4}$.}
\label{fig:mssmsols}
\end{figure}

  It is important to establish whether it is possible to have a lightest
  neutralino with \linebreak $M_{\cht^0_1}=33.9\, \mr{\mev}$  within the
  MSSM. We have therefore scanned the MSSM parameter
  space with independent $M_1$, $M_2$, for a neutralino in the mass range
\begin{equation}
33.89\,\mr{\mev} < M_{\cht^0_1} < 33.91\,\mr{\mev}.
\end{equation}
  This gives the neutralino iso-mass curves shown in Fig.\,\ref{fig:mssmsols}
  for $\mu=300\, \mr{\gev}$  and two representative values of $\tan\beta$. 
  We have not been able to find any solutions for $\mu<0\, \mr{\gev}$. Some
  fine-tuning is necessary in order to obtain these solutions. This should 
  not be surprising given the requirement of reproducing the lightest
  neutralino mass needed to solve the KARMEN anomaly. This fine-tuning is a
  few parts in $10^3$ for $\tan\be=1$ and a few parts in $10^2$ for 
  $\tan\beta=8$ \cite{rominino}. The fine-tuning is reduced for larger $M_2$
  and $\mu$ and small $M_1$ because a light neutralino can then be generated
  by the see-saw mechanism. It is also reduced for large values of
  $\tan\beta$ because in the limit  $\beta=\pi/2$ there is a zero mass
  eigenvalue for $M_1\approx0$ \cite{rominino}.
  
  We have checked that the higgsino contribution always satisfies the bound in
  Eqn.\,\ref{eqn:higgsinobound}. In order to avoid an observable
  light chargino we require that $M_{\cht^{\pm}}>150\, \mr{\gev}$, which
  eliminates the region below the hashed lines in Fig.\,\ref{fig:mssmsols} for
  the specified values of $\tan\beta$. The lightest neutralino is dominantly
  bino along the solution curves given in Fig.\,\ref{fig:mssmsols}.
  The next-to-lightest neutralino, $\cht^0_2$, is dominantly wino for 
  $M_2<300\, \mr{\gev}$, but for larger values it is mainly higgsino. For 
  $M_2\gtrsim 110\, \mr{\gev}$, $M_{\mr{\cht^0_2}}\gtrsim 100\, \mr{\gev}$,
  and for 
  $M_2\gtrsim 235\, \mr{\gev}$, $M_{\mr{\cht^0_2}}\gtrsim 200\, \mr{\gev}$.
%
%  Solutions in MSSM parameter space 
%
% End of the figure

%
%  Loop calculations
%
\subsection[Limits from Precision Electroweak Measurements]
	{Limits from Precision Electroweak Measurements}

  We would expect the addition of a new light particle to affect the results
  of the precision electroweak measurements made by the LEP collaborations. We
  can look at this by calculating the contributions in loop diagrams of the
  gauginos. To do this we will need to calculate the gaugino loop diagrams in
  the photon, W and Z propagators, and in the photon--Z mixing. In general we
  can write the results of these loop diagrams as
%
%  Equation for the loop in propagators with diagram
%
\begin{equation}
\begin{picture}(200,20)(0,0)
\SetScale{0.5}
\SetOffset(0,-20)
\Photon(40,50)(80,50){5}{3}
\Photon(120,50)(160,50){5}{3}
\GCirc(100,50){20}{0.7}
\Text(140,25)[c]{\large{$= i\Pi^{\mu\nu}_{\mr{IJ}}(q).$}}
\Text(15,34)[]{$\mu$}
\Text(85,34)[]{$\nu$}
\Text(15,16)[]{I}
\Text(85,16)[]{J}
\end{picture}
\end{equation}
%  End of the equation
  Due to the tensor structure of the diagram this can be written as
\begin{equation}
  \Pi^{\mu\nu}_{\mr{IJ}} = 
	\Pi_{\mr{IJ}}(q^2) g^{\mu\nu}-\Delta(q^2) q^\mu q^\nu.
\end{equation}

  We use the conventions of \cite{Peskin:1992sw,Peskin:1995ev} and, as was
  done there, neglect the terms proportional to $q^\mu q^\nu$ in the W and Z
  propagators. This is valid as we will only be considering processes with
  light fermions as the external particles and these terms are suppressed by
  a factor $m_f^2/M^2_{\mr{Z}}$ with respect to the $g^{\mu\nu}$ terms after
  contraction with the external fermion current, where $m_f$ is the fermion
  mass.
%
%  Loop diagrams with gauginos
%
\begin{figure}[htp]
\begin{center}
\subfigure[Neutralino and chargino loops in the Z propagator.]{
\begin{picture}(440,90)(0,0)
%  Neutralino Loop in the Z
\SetOffset(20,0)
\Photon(40,50)(80,50){5}{5}
\ArrowArcn(100,50)(20,0,180)
\ArrowArcn(100,50)(20,180,360)
\Photon(120,50)(160,50){5}{5}
\Text(30,50)[c]{Z}
\Text(170,50)[c]{Z}
\Text(100,17)[c]{$\mr{\cht^0_i}$}
\Text(100,83)[c]{$\mr{\cht^0_j}$}
% Chargino Loop in the Z
\SetOffset(220,0)
\Photon(40,50)(80,50){5}{5}
\ArrowArcn(100,50)(20,0,180)
\ArrowArcn(100,50)(20,180,360)
\Photon(120,50)(160,50){5}{5}
\Text(30,50)[c]{Z}
\Text(170,50)[c]{Z}
\Text(100,17)[c]{$\mr{\cht^+_i}$}
\Text(100,83)[c]{$\mr{\cht^-_j}$}
\end{picture}}
% Chargino Neutralino loop in the W
\subfigure[Chargino-neutralino loop in the W propagator.]{
\begin{picture}(440,90)(0,0)
\SetOffset(120,0)
\Photon(40,50)(80,50){5}{5}
\ArrowArcn(100,50)(20,0,180)
\ArrowArcn(100,50)(20,180,360)
\Photon(120,50)(160,50){5}{5}
\Text(25,50)[c]{$\mr{W^+}$}
\Text(175,50)[c]{$\mr{W^+}$}
\Text(100,17)[c]{$\mr{\cht^0_i}$}
\Text(100,83)[c]{$\mr{\cht^+_j}$}
\end{picture}}
% Chargino Loop in the Photon
\subfigure[Chargino loop in the photon propagator.]{
\begin{picture}(440,90)(0,0)
\SetOffset(120,0)
\Photon(40,50)(80,50){5}{5}
\ArrowArcn(100,50)(20,0,180)
\ArrowArcn(100,50)(20,180,360)
\Photon(120,50)(160,50){5}{5}
\Text(30,50)[c]{$\mr{\gamma}$}
\Text(170,50)[c]{$\mr{\gamma}$}
\Text(100,17)[c]{$\mr{\cht^+_i}$}
\Text(100,83)[c]{$\mr{\cht^-_i}$}
\end{picture}}
% Chargino Loop in Z photon mixing
\subfigure[Chargino loop in photon--Z mixing.]{
\begin{picture}(440,90)(0,0)
\SetOffset(120,0)
\Photon(40,50)(80,50){5}{5}
\ArrowArcn(100,50)(20,0,180)
\ArrowArcn(100,50)(20,180,360)
\Photon(120,50)(160,50){5}{5}
\Text(30,50)[c]{Z}
\Text(170,50)[c]{$\mr{\gamma}$}
\Text(100,17)[c]{$\mr{\cht^+_i}$}
\Text(100,83)[c]{$\mr{\cht^-_i}$}
\end{picture}}
\end{center}
\captionB{Gaugino loops in the gauge boson propagators.}
	{Gaugino loops in the gauge boson propagators.}
\label{fig:loops}
\end{figure}
% End of the figure  

  We will therefore consider the loops in the propagators shown in 
  Fig.\,\ref{fig:loops}. We have calculated these and obtain
%%%%%%%%%%%%%%%%%%%%%%%%%%%%%%%%%%%%%%%%%%%%%%%%%%%%%%%%%%%%%%%%%%%%%%%%%%%%%
%  SUSY loops in the propagators
{
\renewcommand{\arraystretch}{1.0}
\renewcommand{\jot}{4mm}
\begin{subequations}
\label{eqn:SUSYloops}
\begin{eqnarray}
% Z propagator
 \Pi_{\mr{ZZ}}(q)& = & - \frac{g^2}{16\pi^2\ccw} \sum_{i=1}^4\sum_{j=1}^4
                             \left[\rule{0cm}{1.1cm}
		2\mathcal{R}e\left\{{O''}^L_{ij}{O''}^R_{ij}\right\}
                              M_{\cht^0_i}^2M_{\cht^0_j}^2
                              \left(E-b_0(\cht^0_i\cht^0_jq) \right)
			     \right.  \nonumber \\\nopagebreak
	& &    +\left( |{O''}^L_{ij}|^2+|{O''}^R_{ij}|^2 \right)
			 \left( \rule{0cm}{0.95cm}
              \left[\rule{0cm}{0.85cm}\frac{q^2}{3}-\frac{M_{\cht^0_i}^2}{2}-
				   \frac{M_{\cht^0_j}^2}{2}\right]E
                              -2q^2 b_2(\cht^0_i\cht^0_jq)\right. 
	\nonumber\\ &&\left.\left.
				 +M_{\cht^0_i}^2 b_0(\cht^0_i\cht^0_jq)
                             +(M_{\cht^0_j}^2-M_{\cht^0_i}^2)
                               b_1(\cht^0_i\cht^0_jq)
			\rule{0cm}{0.95cm}\right)\rule{0cm}{1.1cm}
				 \right] 
   				\nonumber\\\nopagebreak
 		& &  - \frac{g^2}{8\pi^2\ccw} \sum_{i=1}^2\sum_{j=1}^2 \left[
		       \rule{0cm}{1.1cm}
			2\mathcal{R}e\left\{{O'}^L_{ij}{O'}^R_{ij}\right\}
                              M_{\cht^+_i}^2M_{\cht^-_j}^2
                              \left(E-b_0(\cht^0_i\cht^0_jq)  \right)\right.
 \nonumber\\ && 
                       +\left( |{O'}^L_{ij}|^2+|{O'}^R_{ij}|^2 \right) 
			        \left( \rule{0cm}{0.95cm}
             \left[\rule{0cm}{0.85cm}\frac{q^2}{3}-\frac{M_{\cht^+_i}^2}{2}
  					-\frac{M_{\cht^-_j}^2}{2}\right]E
	 -2q^2 b_2(\cht^+_i\cht^-_jq)\right.\nonumber\\
                       & &   \left.\left.
			     +M_{\cht^+_i}^2 b_0(\cht^+_i\cht^-_jq)
                             +(M_{\cht^-_j}^2-M_{\cht^+_i}^2)
                               b_1(\cht^+_i\cht^-_jq)\rule{0cm}{0.95cm}\right)
			 \rule{0cm}{1.1cm}\right]\!\!,
% W propagator
\end{eqnarray}
\begin{eqnarray}
\Pi_{\mr{WW}}(q)&=&  -\frac{g^2}{8\pi^2} \sum_{i=1}^4\sum_{j=1}^2 
		       \left[\rule{0cm}{1.1cm}
			2\mathcal{R}e\left\{{O}^L_{ij}{O}^R_{ij}\right\}
                             M_{\cht^0_i}^2M_{\cht^+_j}^2
                     \left(E-b_0(\cht^0_i\cht^+_jq) \right)\right.\nonumber\\
  &&			 	+\left( |{O}^L_{ij}|^2+|{O}^R_{ij}|^2 \right)
			     \left( \rule{0cm}{0.95cm}
             \left[\rule{0cm}{0.85cm}\frac{q^2}{3}-\frac{M_{\cht^0_i}^2}{2}-
				   \frac{M_{\cht^+_j}^2}{2}\right]E
                              -2q^2 b_2(\cht^0_i\cht^+_jq)\right.\nonumber\\
   && \left.\left.
				 +M_{\cht^0_i}^2 b_0(\cht^0_i\cht^+_jq)
                             +(M_{\cht^+_j}^2-M_{\cht^0_i}^2)
                           b_1(\cht^0_i\cht^+_jq) \rule{0cm}{0.95cm}\right)
			\rule{0cm}{1.1cm}	\right]\!\!,\\
%\end{eqnarray}
%\begin{eqnarray}
%  photon propagator
\Pi_{\mr{\gamma\gamma}}(q) & = &  -\frac{e^2}{2\pi^2}\sum_{i=1}^2\left[
	 \frac{q^2}{6}E -q^2 b_2(\cht^+_i\cht^-_iq)\right]\!, \\
% photon Z mixing
\Pi_{\mr{Z \gamma}} (q)& =& \frac{ge}{4\pi^2\cw}\sum_{i=1}^2
				\left({O'}^L_{ij}+{O'}^R_{ij}\right) 
	    \left[ \frac{q^2}{6}E -q^2b_2(\cht^+_i\cht^-_iq)\right]\!. 
\end{eqnarray}
\end{subequations}}
%%%%%%%%%%%%%%%%%%%%%%%%%%%%%%%%%%%%%%%%%%%%%%%%%%%%%%%%%%%%%%%%%%%%%%%%%%%%%
  The coefficients ${O}^{L,R}_{ij}$,
  ${O'}^{L,R}_{ij}$ and ${O''}^{L,R}_{ij}$ in the loop diagrams
  are taken from \cite{Haber:1985rc}. We have used the same notation as 
  \cite{Peskin:1995ev} for the functions in the loop calculations, \ie
{\renewcommand{\jot}{4mm}
%%%%%%%%%%%%%%%%%%%%%%%%%%%%%%%%%%%%%%%%%%%%%%%%%%%%%%%%%%%%%%%%%%%%%%%%%%%%%
%  Definitions of the integrals from P&S
\begin{subequations}
\begin{eqnarray}
% Infinite and scheme dependent pieces
  E                      & = & \frac{2}{\epsilon}-\gamma_E+
			 \log(4\pi)-\log(M^2), \\
% denominator of the propagator
 \Delta(m^2_1,m^2_2,q^2) & = & xm^2_2+(1-x)m^2_1-x(1-x)q^2,
\end{eqnarray} 
\vspace{-8mm}
\begin{eqnarray}
% Integral b0
  b_0(12q)\ \, =& b_0(m^2_1,m^2_2,q^2)  \ \,   = &  \int^1_0 dx 
 		\log \left( \frac{\Delta(m^2_1,m^2_2,q^2)}{M^2}\right)\!, \\
% Integral b1
  b_1(12q)\ \, = & b_1(m^2_1,m^2_2,q^2)  \ \,   = &  \int^1_0 dx x
 	      \log \left( \frac{\Delta(m^2_1,m^2_2,q^2)}{M^2}\right)\!, \\
% Integral b2
  b_2(12q)\ \, = & b_2(m^2_1,m^2_2,q^2)   \ \,  = &  \int^1_0 dx x(1-x)
 	      \log\left(\frac{\Delta(m^2_1,m^2_2,q^2)}{M^2}\right)\!,
\end{eqnarray}
\end{subequations}}
%%%%%%%%%%%%%%%%%%%%%%%%%%%%%%%%%%%%%%%%%%%%%%%%%%%%%%%%%%%%%%%%%%%%%%%%%%%%%%
  where the calculation has been performed in $d=4-\epsilon$ dimensions.

  The effect of new physics from these vacuum polarization diagrams is
  usually parameterized using either the $S$, $T$, and $U$ parameters of 
  \cite{Peskin:1990zt,Marciano:1990dp,Peskin:1992sw} or the $\epsilon_1$,
   $\epsilon_2$, and $\epsilon_3$ parameters of
  \cite{Altarelli:1991zd:Altarelli:1992fk}. The calculation of these
  parameters is based on an expansion in $q^2/M^2_{\mr{new}}$, where $q^2$ is
  the momentum flow through the gauge boson propagator, typically
  $M^2_{\mr{Z}}$ or smaller and $M^2_{\mr{new}}$ is the scale of the new
  physics, which is assumed to be well above~$M^2_{\mr{Z}}$.
  If, however, there are new light particles in the
  spectrum, as is the case in our model, these approximations are
  insufficient and in principle the
  box and vertex corrections must also be calculated \cite{Barbieri:1992qp}.

  This full calculation is beyond the scope of our analysis. There is one
  exception to this which is the ratio of the charged to neutral current
  neutrino--electron/muon scattering events, the $\rho$ parameter. This is
  defined at $q^2=0$ and hence the expansion is trivial. We have calculated
  the correction to the $\rho$ parameter from the full set of chargino and
  neutralino diagrams. The radiative correction to the $\rho$ parameter is
  given by \cite{Caso:1998tx}
\begin{equation}
  \Delta\rho = 
 \frac{\Pi_{\mr{WW}}(0)}{M^2_\mr{W}} - \frac{\Pi_{\mr{ZZ}}(0)}{M^2_\mr{Z}}.
\end{equation}
  The dominant Standard Model contributions to $\Delta\rho$ are from top
  quark and Higgs boson loops. These contributions can be subtracted, 
  assuming $M_{\mr{H}}=M_{\mr{Z}}$, from the
  experimental results giving the contribution from new physics
\begin{equation}
   \Delta\rho_{\mr{NEW}} = (-1.3 \pm 1.2) \times 10^{-3}.
\end{equation}
 This gives the following limit on the contribution to $\Delta\rho$ from
 new physics at the $2\sigma$ level
\begin{equation} 
-3.7\times 10^{-3}< \Delta\rho_{\mr{NEW}} < 1.1\times 10^{-3}.
\end{equation}
  It is interesting to consider the effect on this limit 
  of varying the Standard Model Higgs.
  The following result is for $M_{\mr{H}}=300\, \mr{\gev}$:
\begin{equation}
-4.6\times 10^{-3}< \Delta\rho_{\mr{NEW}} < 0.2\times 10^{-3},
\end{equation}
  although it should be noted that this is 
  a larger value of the Higgs mass than is possible in the MSSM.
  We have calculated the contribution to $\Delta\rho$ from the
  chargino and neutralino loops in
  Fig.\,\ref{fig:loops}, using the results given in Eqn.\,\ref{eqn:SUSYloops}.
  We denote this contribution to the $\rho$ parameter 
  by $\Delta\rho_{\mr{SUSY}}$. We then determine $\Delta\rho_{\mr{SUSY}}$
  along the solution curves given in Fig.\,\ref{fig:mssmsols}. The dotted
  lines indicate solutions for which $\Delta\rho_{\mr{SUSY}}<10^{-4}$, while
  the solid lines indicate solutions for which
  $\Delta\rho_{\mr{SUSY}}<5\times10^{-4}$. This shows that there is no
  conflict with the experimental constraint on the $\rho$ parameter, for
  $M_{\mr{H}}=M_{\mr{Z}}$. There may be some constraints on this
  model for higher values of the Higgs mass although even for Higgs masses
  above those which are possible in the MSSM there are still areas of SUSY
  parameter 
  space in which the lightest neutralino mass can solve the KARMEN anomaly. 

  It is possible to derive bounds on the existence and properties of an
  additional light particle in the mass spectrum on cosmological grounds and 
  from astrophysical processes. The potential bounds on our model are
  discussed in 
  \cite{Choudhury:1999tn}.

%
%  Section on future tests
%
\section[Future Tests]{Future Tests}
\label{sect:futureexp}

\begin{figure}
\centering
\includegraphics[angle=90,width=0.48\textwidth]{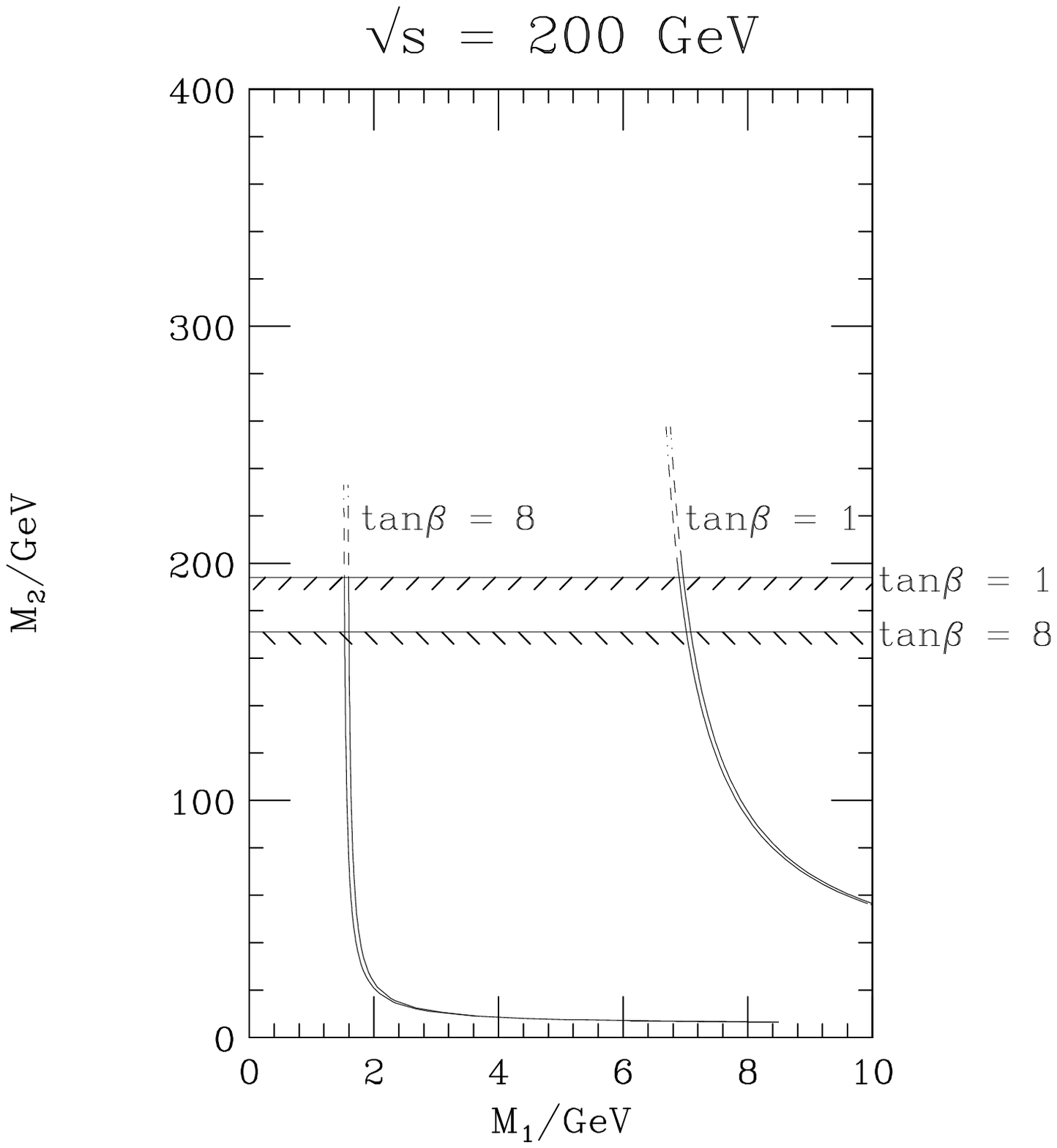}
\hfill
\includegraphics[angle=90,width=0.48\textwidth]{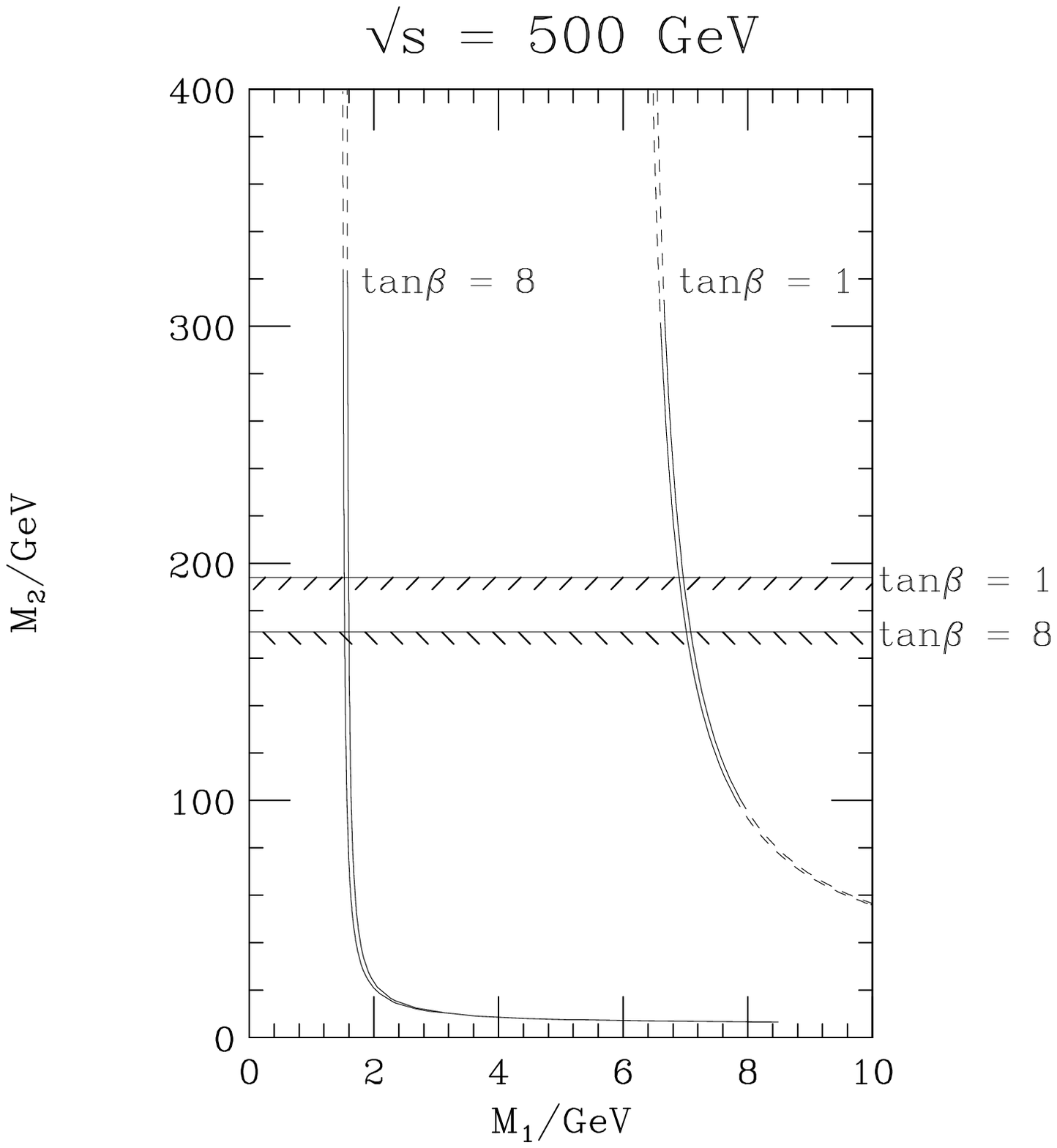}
\captionB{Cross sections for $\mr{e^+e^-\rightarrow
 	{\tilde\chi}^0_2{\tilde\chi}^0_1}$.}
	{Cross sections for $\mr{e^+e^-\rightarrow
 	{\tilde\chi}^0_2{\tilde\chi}^0_1}$ for the solutions in
 	Fig.~\ref{fig:mssmsols}. The solid lines correspond to
 	$0.1\, \mr{pb} < \sigma < 1\, \mr{pb}$,
	the dashed lines to 
	$10\,  \mr{fb} < \sigma < 0.1\, \mr{pb}$, the
 	dotted lines to $1\, \mr{fb}  < \sigma < 10\, \mr{fb}$,
	 and the dot-dashed lines to
 	$\sigma < 1\, \mr{fb}$.}
\label{fig:chi2chi1}
\end{figure}

  Experimentally our model looks very much like the MSSM with non-universal
  gaugino masses and a very light LSP. Hence most future tests of the MSSM
  will also apply to our model. A specific test would be to identify the
  existence of a very light neutralino in, for
  example, neutralino pair production in $\mr{e^+e^-}$ collisions via,
\begin{equation}
   \mr{e}^+\mr{e}^-\longrightarrow \cht^0_2\cht^0_1,
\end{equation}
  followed by a visible decay of the second-to-lightest neutralino. The cross
  section for this process is shown in Fig.\,\ref{fig:chi2chi1} along our
  MSSM solution curves from Fig.\,\ref{fig:mssmsols} for both LEP2
  ($\sqrt{s}=200\, \mr{\gev}$) and the NLC ($\sqrt{s}=500\, \mr{\gev}$).
  This process should be observable provided that it is
  kinematically accessible, \ie $\sqrt{s}>M_{\cht^0_2}$.

%
%  Cross-Sections for chi^0_2 chi^0_1 production in e^+e^-
%

  As the neutralino is sufficiently long lived in our model that it will
  escape the detector before decaying the main difference between our model
  and the MSSM with non-universal gaugino masses is the possibility of
  resonant sparticle production. As we saw in Chapter~\ref{chap:slepton},
  the value of the coupling ${\lam'}_{211}\lesssim10^{-4}$ is too small for
  the observation of resonant slepton production in hadron--hadron collisions.
  However the values of the couplings
  $\lam_{131}$~(or~$\lam_{121}$)$~>10^{-3}$ should allow a test of resonant
  sneutrino production in $\mr{e^+e^-}$ collisions provided
  $M_{\nut}\lesssim\sqrt{s}$. The production of first generation sleptons
  can also be tested by the processes suggested in \cite{Allanach:1997sa}.

\begin{figure}
\centering
\includegraphics[angle=90,width=0.8\textwidth]{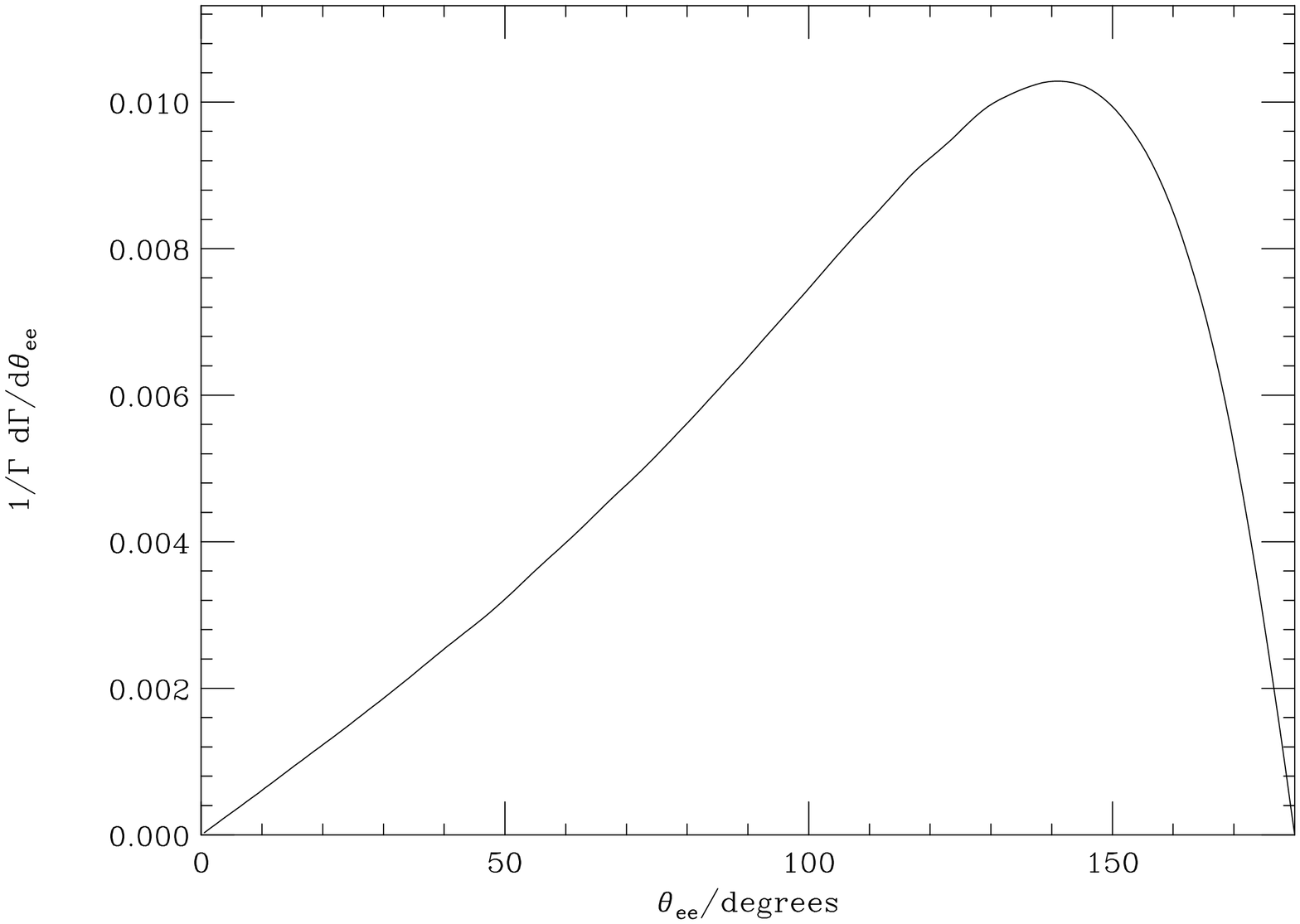}
\captionB{$\frac{d\Gamma}{d\theta_{\mr{ee}}}$ in the KARMEN laboratory frame.}
	{$\frac{d\Gamma}{d\theta_{\mr{ee}}}$ in the KARMEN laboratory frame.}
\label{fig:angle}
\end{figure}

  A further upgrade of the KARMEN detector may allow better resolution of the
  decay of the X particle. In particular there may be tracking so that the
  angular distribution of the decay products can be measured. We therefore
  show the differential decay rate of the neutralino in our model as a
  function of the angle between the electron and positron
  produced in the neutralino decay in the KARMEN laboratory frame,
  Fig.\,\ref{fig:angle}.
  This differs from the corresponding singlet neutrino decay distribution.
\textheight 24cm
\section[Summary]{Summary}

  There have been many reported experimental 
  deviations from the Standard Model in recent years. Indeed most of these
  deviations have had much the same statistical significance as the KARMEN
  time anomaly. However while the statistical significance of most of the
  deviations has reduced with increased statistics, or the effect has been 
  explained away as a systematic effect, this is not the case for the KARMEN
  anomaly. In fact the anomaly persists in the second run of the KARMEN
  experiment, which has a much reduced background \cite{Karmennew1}, with the
  same characteristics as in the initial run of the experiment
  \cite{Armbruster:1995nr}.

  An additional problem in the resolution of this anomaly is that there is no 
  independent experiment with the sensitivity to either confirm or exclude the
  KARMEN result. Although the LSND experiment also studies the decay of pions
  and muons at rest, it lacks the time structure of the beam in the KARMEN
  experiment, which is necessary to extract the anomaly. Since the second run
  of the KARMEN experiment only collects about ten anomaly events per year, a
  definitive resolution of the anomaly will 
  require an upgraded detector with improved tracking capabilities.

  The different phenomenological models which have been proposed to explain
  the \linebreak
  KARMEN results via the production and decay of a new particle are tightly
  constrained. The only viable proposals at present are the production  of a
  singlet neutrino which decays through its large mixing with the tau neutrino
  \cite{Barger:1995ty,Govaerts:1996hk}, the brane-world model of 
  \cite{Lukas:2000fy} or model with a neutralino decaying via \rpv\  
  \cite{Choudhury:1996pj} which we have extended and investigated here.

  It is important to note from our study that a light neutralino is only
  excluded by present day collider experiments if we assume a GUT relation
  between the SUSY-breaking masses for the gauginos.

\textheight 23cm

%\chapter{Conclusions}
%
%    Finally the Conclusions
%
%
\chapter{Conclusions}
\label{chap:conclude}

  Despite the lack of any experimental evidence there are many compelling
  theoretical arguments for the existence of supersymmetry. In this 
  thesis we have argued that if we are to discover whether or not 
  supersymmetry is realized in nature we cannot simply consider the
  simplest possible supersymmetric extension of the Standard Model. 
  Indeed this simplest extension is no better motivated than the R-parity 
  violating  models we have considered. In these models lepton or baryon
  number is violated which leads to very different experimental signatures
  of supersymmetry which must be investigated if we are to explore all 
  possible channels for the discovery of supersymmetry.

  There has however been much less study of these models, particularly
  experimentally, than of the \rp\  conserving MSSM. In hadron--hadron
  collisions studies of these models have been hindered by the lack of a
  Monte Carlo simulation which includes the \rpv\  production processes
  and decays. We have presented the calculations necessary to produce
  a full simulation of these \rpv\  processes including colour coherence
  effects via the angular-ordering procedure. This sees a new type of
  vertex in which three particles in the fundamental representation of
  $SU(3)$ interact at a vertex which violates baryon number.
  The inclusion of these colour coherence
  effects has proven to be important for an accurate simulation of 
  the QCD radiation in Standard Model processes at
  previous experiments and we therefore must include it to give a
  reliable simulation of these \rpv\  processes. 
  The simulation we produced using these results
  has already allowed a full simulation of
  baryon number violating processes at the LHC for the first time 
  \cite{Drage:1999th}.

  Our studies of the colour coherence properties of different resonant
  \rpv\  production processes in hadron--hadron collisions show 
  differences for variables which are sensitive to the colour coherence
  phenomenon. In particular there are differences between resonant slepton
  production, which violates lepton number, and resonant squark
  production, which violates baryon number. We have used this simulation
  to show that the extraction of a signal for resonant slepton production
  via the \rpv\  decay modes of the slepton is much harder than was
  previously believed due to the emission of QCD radiation.
  The differences in the distributions of those variables
  which are sensitive to the presence of colour connections between the
  initial and final states for these processes and QCD jet production
  allow us to improve the extraction of a resonant slepton signal by
  imposing cuts on these variables. However the discovery of resonant
  slepton production via these \rpv\  decay modes is still only possible
  for the largest values of the couplings allowed, given the current
  low-energy limits.

  We therefore studied the discovery potential of the supersymmetric 
  gauge decay modes of resonantly
  produced sleptons. These decay modes give a number of different channels
  for the production of like-sign dilepton pairs, which has a very low
  Standard Model background. This like-sign dilepton signature is
  dominated by the production of a neutralino, which decays via \rpv,
  and a lepton in the supersymmetric gauge decay of a resonant charged
  slepton. We studied the cuts needed to reduce both the Standard 
  Model background and the background from sparticle pair production and
  improve the extraction of a signal. This gives a large discovery 
  potential for resonant slepton production mechanisms, at both the 
  Tevatron and the LHC, and allows us to probe much smaller values of the
  \rpv\  Yukawa couplings than is possible using the \rpv\  decay modes of
  the resonant sleptons. We also showed that if such a signal is discovered
  it should be possible to measure the mass of both the
  resonant slepton and the lightest neutralino.

  Finally, we considered the possibility that we have already seen evidence
  for the existence of \rpv\  in the time anomaly reported by the KARMEN
  collaboration. We have shown that this anomaly could be due to the 
  production via \rpv\  of the lightest neutralino in charged pion decays.
  This neutralino could then decay, via \rpv, inside the KARMEN detector
  giving rise to the observed excess. Surprisingly the existence of this
  light neutralino, decaying via \rpv, is not ruled out by any current 
  experimental data. Hopefully a future run of the KARMEN experiment will 
  both improve the statistical significance of this excess and give us 
  further insights into the physics which is producing it. This model can
  be tested by looking for the production of $\mr{\cht^0_1\cht^0_2}$ at
  LEP, or any future $\mr{e^+e^-}$ collider.

  In conclusion we have produced the first full simulation of \rpv\ 
  violating supersymmetry and have used this simulation to study some 
  possible channels for the discovery of \rpv\  processes. There are many 
  processes which still needed to be examined. Hopefully this work will 
  allow a more detailed study of these processes, particularly by the
  experimental community, in the future. 

\addcontentsline{toc}{chapter}{Addendum}
\chapter*{Addendum}

  While I was in the final stages of preparing this thesis new results
  \cite{Karmennew4} were announced by the KARMEN collaboration at the
  Neutrino~2000 conference. These results show that the results of the
  KARMEN\,2 experiment no longer confirm the anomaly seen in the
  KARMEN\,1 data. However there is still an anomaly for
  the combined KARMEN\,1+2 data set.
  Given this result the planned experiment to investigate the anomaly has
  been cancelled.

\renewcommand{\chaptermark}[1]{\markboth{Appendix
\ \thechapter.\ #1}{}}
\renewcommand{\sectionmark}[1]{\markright{\thesection.\ #1}}
\appendix
%\chapter{Appendix: conventions etc}
%
%  First the appendix on the general SUSY convention we have used
%
\chapter{Feynman Rules and Conventions}
\label{chap:Feynman}
\section[Introduction]{Introduction}	  
  
  There are a number of different conventions which are commonly adopted in
  the literature for the MSSM Lagrangian. While most authors have chosen to
  follow the conventions used in \cite{Haber:1985rc,Gunion:1986yn} other
  conventions, for example \cite{Baer:1987au}, are also commonly used.
  We will present the conventions we have used in this thesis, which
  in general follow those of \cite{Haber:1985rc,Gunion:1986yn}.
  We then present the Feynman rules used in the various decay rate
  and cross-section calculations. 

  We will first present our 
  conventions for the mixing matrices for the charginos, neutralinos and
  sfermions, together with the conversion between the conventions of
  \cite{Haber:1985rc,Gunion:1986yn} and \cite{Baer:1987au}.
  We then give the
  Feynman rules we have used from \cite{Haber:1985rc,Gunion:1986yn} with
  the inclusion of left/right sfermion mixing. In particular we give the
  Feynman rules for the interactions of squarks and sleptons with the
  gauginos, gluinos, electroweak gauge bosons and Higgs bosons of the MSSM.

  We conclude by giving the Feynman rules for the
  R-parity violating superpotential. These Feynman rules are then used in
  Appendix~\ref{chap:decay} 
  to evaluate the decay rates of the sparticles via the different 
  \rpv\  operators and in Appendix~\ref{chap:cross} to calculate the single
  sparticle production cross sections. 
%
%  First the various conventions
%

\section[Mixing]{Mixing}
\label{sect:mixing}
  All of the cross-section and decay rate calculations presented
  in this thesis have been
  implemented in the HERWIG Monte Carlo event generator 
  \cite{HERWIG61}. While internally HERWIG uses the conventions of 
  \cite{Haber:1985rc,Gunion:1986yn,Gunion:1989we}, it relies on ISAJET 
  \cite{Baer:1999sp} for the calculation  of the masses of the sparticles  and
  their \rp\  conserving decay rates. ISAJET also calculates the mixing
  matrices for the electroweak gauginos and scalar fermions. ISAJET uses the
  conventions of \cite{Baer:1987au,Baer:1994xr} and it is  essential that the
  conversion between the two formalisms is correct.

  The major problem arises in the conventions for the various mixing
  matrices, for the charginos and neutralinos, and the left/right mixing of
  the scalar fermions. We will first consider the conversion between the two
  formalisms for the gauginos and then for the left/right sfermion mixing.
  The ISAJET \cite{Baer:1987au} and Haber and Kane \cite{Haber:1985rc}
  conventions for the sign of the $\mu$ and $A$ terms are the 
  same.\footnote{This also agrees with the conventions of the other major
	  SUSY event generators, SPYTHIA \cite{Mrenna:1997hu} and SUSYGEN
	 \cite{Katsanevas:1997fb}.} This corresponds to the following
\begin{subequations}
\begin{eqnarray}
    2m_1  &=& -\mu, \\
    \mu_1 &=& -M_1,\\
    \mu_2 &=& -M_2,
\end{eqnarray}
\end{subequations}
  where $m_1$, $\mu_1$ and $\mu_2$ are in the notation of ISAJET
  \cite{Baer:1987au} and $\mu$, $M_1$ and $M_2$ are in 
  the notation\footnote{In \cite{Haber:1985rc,Gunion:1986yn}, $M'$ and
				$M$ are used instead of $M_1$ and $M_2$,
			        respectively. However the notation
				$M_1$ and $M_2$ has become more common
				and we will use this here.}
  of \cite{Haber:1985rc,Gunion:1986yn}. Here $\mu$ is the mixing parameter
  in the MSSM superpotential, Eqn.\,\ref{eqn:MSSMsuper}, for the two Higgs
  doublets, $M_1$ is the soft SUSY-breaking mass for the bino and $M_2$ is
  the soft SUSY-breaking mass for the winos.
  
  We can compare the Lagrangians of \cite{Baer:1987au} and
  \cite{Haber:1985rc,Gunion:1986yn} to obtain the conversions between them.
  We shall use the conventions of \cite{Haber:1985rc,Gunion:1986yn} 
  for the chargino and neutralino mixing matrices.

%
%  First the chargino mixing
%
\subsection[Charginos]{Charginos}

  In the notation of 
  \cite{Gunion:1986yn}, we define the following two-component spinors for the
  charginos before mixing
\begin{subequations}
\begin{eqnarray}
   \psi^+_j & = & \left( -i\lam^+,\psi^+_{\mr{H_2}} \right)\!,\\
   \psi^-_j & = & \left( -i\lam^-,\psi^-_{\mr{H_1}} \right)\!,
\end{eqnarray}
\end{subequations}
  where $\lam^{+,-}$ are the charged winos, $\psi^-_{\mr{H_1}}$ is the
  charged higgsino associated with the Higgs which gives mass to the
  down-type quarks and $\psi^+_{\mr{H_2}}$ is the charged higgsino
  associated with the Higgs doublet which gives the up-type quarks mass.
  The mass term in the Lagrangian from Eqn.\,A.2 of \cite{Gunion:1986yn} is
\begin{equation}
   \mathcal{L}_{\mr{chargino}} = -\frac12\left(\psi^+, \psi^-\right)
  \left(\begin{array}{cc} 0 & X^T \\ X & 0  \end{array}\right)
	\left(\begin{array}{c} \psi^+\\ \psi^- \end{array}\right) + \mr{h.c.},
\end{equation}
  where the mass matrix is given by
\begin{equation}
X=\left(	\begin{array}{cc} M_2 & \mw\sqrt{2}\sbe \\
			\mw\sqrt{2}\cbe & \mu \end{array}\right)\!.
\label{eqn:charginomatrix}
\end{equation}
  In \cite{Gunion:1986yn} this was diagonalized in two-component notation
  by defining two-component mass eigenstates
\begin{subequations}
\begin{eqnarray}
  \chi^+_i &=& V_{ij}\psi^+_j,\\
  \chi^-_i &=& U_{ij}\psi^-_j,
\end{eqnarray}
\end{subequations}
  where $U$ and $V$ are unitary matrices chosen such that
\begin{equation}
  U^*XV^{-1}= M_D,
\end{equation}
  where $M_D$ is the diagonal chargino mass matrix. The four-component mass
  eigenstates, the charginos, are defined in terms of the two-component
  fields
\begin{equation}
\cht^+_1 = \left(\begin{array}{c}\chi^+_1\\\bar{\chi}^-_1
	\end{array}\right)\!,\ \ \ \ \ \ 
\cht^+_2 = \left(\begin{array}{c}\chi^+_2\\\bar{\chi}^-_2\end{array}\right)\!.
\end{equation}
  We need to express the chargino mass terms in the Lagrangian
  in a four-component notation so we can compare this Lagrangian with the
  conventions of \cite{Baer:1987au}. We define the four-component spinors as
  in \cite{Gunion:1986yn},
\begin{equation}
\widetilde{W} = \left(\begin{array}{c}-i\lam^+\\ i\bar{\lam}^-
 \end{array}\right)\!,\ \ \ \ \
  \ \widetilde{H} = \left(\begin{array}{c} \psi^+_{\mr{H_2}} \\ 
			\psb^-_{\mr{H_1}}\end{array}\right)\!.
\end{equation}
  Using this notation we can express the Lagrangian in four-component notation
\begin{equation}
   \mathcal{L}_{\mr{chargino}} = -\left(\overline{\widetilde{W}},
 \overline{\widetilde{H}}\right)
  \left[  XP_L + X^TP_R\right]
	\left(\begin{array}{c}  \widetilde{W}\\ \widetilde{H} 
\end{array}\right)\!.
\label{eqn:GHchar}
\end{equation}
  This is now in a similar form to the chargino mass term in 
  Eqn.\,2.1\footnote{This Lagrangian is taken from \cite{Tata:1997uf}
  which corrected a sign error in the off-diagonal terms in
  \cite{Baer:1987au}.} of \cite{Baer:1987au}, 
\begin{equation}
   \mathcal{L}_{\mr{chargino}} =-\left(\bar{\lam},\bar{\chi} \right)
\left[M_{\mr{charge}}P_L+M^T_{\mr{charge}}P_R\right]
	\left(\begin{array}{c} \lam\\\chi\end{array}\right)\!,
\end{equation}
 where
\begin{equation}
  M_{\mr{charge}} =
 \left(\begin{array}{cc} \mu_2 & -gv' \\ -gv & 2m_1\end{array}\right)
 =  -\left(\begin{array}{cc} M_2 & \sqrt{2}\mw\cbe \\ \sqrt{2}\mw\sbe &
 \mu\end{array} \right)\!,
\end{equation}
   where $v$ and $v'$ are the vacuum expectation values for the Higgs fields
  which give mass to the up- and down-type quarks, respectively. 
  We now come to the problem in comparing the notations of
  \cite{Gunion:1986yn} and \cite{Baer:1987au}. The wino and higgsino fields
  of \cite{Baer:1987au} are charge conjugates of those used in
  \cite{Gunion:1986yn}, so the following transformation is required
\begin{subequations}
\begin{eqnarray}
 \widetilde{W} &=& \lam^c,\\
 \widetilde{H} &=& \chi^c.
\label{eqn:Cfields}
\end{eqnarray}
\end{subequations}
  This gives the chargino mass matrix in the form
\begin{equation}
   \mathcal{L}_{\mr{chargino}} =
\left(\overline{\widetilde{W}},\overline{\widetilde{H}}\right) 
	\left[M'_{\mr{charge}}P_L+{M'}^T_{\mr{charge}}P_R\right]
\left(\begin{array}{c} \widetilde{W}\\\widetilde{H}\end{array}\right)\!,
\end{equation}
where
\vspace{-1mm}
\begin{equation}
  M'_{\mr{charge}} = 
\left(\begin{array}{cc} -\mu_2 & gv \\ gv' & -2m_1\end{array}\right)
 =  \left(\begin{array}{cc} M_2 & \sqrt{2}\mw\sbe \\ \sqrt{2}\mw\cbe &
 \mu\end{array} \right)\!.
\end{equation}
  This agrees with Eqn.\,\ref{eqn:GHchar}, apart from the overall sign.
  We need to express the fields in terms of the ISAJET mixing matrices.
  The ISAJET mixing matrices are
  \cite{Baer:1987au}, Eqns.\,2.10 and 2.11,
\vspace{-1mm}
\begin{subequations}
\begin{eqnarray}
\left(\begin{array}{c} \widetilde{W}_+ \\
	 \widetilde{W}_-\end{array}\right)_L & = &
\left( \begin{array}{cc}\theta_x\cos\ga_L & -\theta_x\sin\ga_L \\
\phantom{\theta_x}\sin\ga_L & \phantom{-\theta_x}\cos\ga_L \end{array}\right)
\left(\begin{array}{c} \lam \\ \chi \end{array}\right)_L,\\[0.5mm]
\left(\begin{array}{c} (-1)^{\theta_+}\widetilde{W}_+ \\
	               (-1)^{\theta_-}\widetilde{W}_-\end{array}\right)_R &=&
\left( \begin{array}{cc}\theta_y\cos\ga_R & -\theta_y\sin\ga_R \\
\phantom{\theta_y} \sin\ga_R & \phantom{-\theta_y}\cos\ga_R \end{array}\right)
\left(\begin{array}{c} \lam \\ \chi \end{array}\right)_R,
\end{eqnarray}
\end{subequations}
  where the mixing angles $\gamma_L$ and $\gamma_R$, and the sign functions
  $\tht_x$, $\tht_y$, $\tht_+$, and $\tht_-$ are defined in
  \cite{Baer:1987au}.

  We can transform these equations into the notation of Haber and Kane with
  the identification
\vspace{-1mm}
\begin{subequations}
\begin{eqnarray}
	\tilde{\chi}_1 &=& \widetilde{W}_-^c,\\
	\tilde{\chi}_2 &=& \widetilde{W}_+^c.
\end{eqnarray}
\end{subequations}
  This gives
\vspace{-1mm}
\begin{subequations}
\begin{eqnarray}
	P_L\left(\begin{array}{c} \widetilde{W} \\ \widetilde{H}\end{array}
\right) &=& 
  P_L\left(\begin{array}{cc} -\sin\ga_R & -\theta_y\cos\ga_R \\
			     -\cos\ga_R & \theta_y\sin\ga_R 
 \end{array}\right)
  \left(\begin{array}{c} (-1)^{\theta_-}\cht_1 \\
 (-1)^{\theta_+}\cht_2\end{array}
	\right)\!,\\[0.5mm]
	P_R\left(\begin{array}{c} \widetilde{W} \\ 
\widetilde{H}\end{array}\right) &=& 
  P_R\left(\begin{array}{cc} -\sin\ga_L & -\theta_x\cos\ga_L \\
			     -\cos\ga_L & \theta_x\sin\ga_L 
 \end{array}\right)
  \left(\begin{array}{c} \cht_1 \\ \cht_2\end{array}\right)\!.
\end{eqnarray}
\label{eqn:ISAJETcharginomix}
\end{subequations}
  These mixing matrices are now in a form that allows us to compare them
   with the notation of \cite{Gunion:1986yn}, Eqn.\,A.13,
\vspace{-1mm}
\begin{subequations}
\begin{eqnarray}
	P_L\left(\begin{array}{c} \widetilde{W} \\ 
\widetilde{H}\end{array}\right) &=& 
  P_L\left(\begin{array}{cc} V^*_{11} & V^*_{21} \\ 
			     V^*_{12} & V^*_{22}\end{array}\right)
  \left(\begin{array}{c} \cht_1 \\ \cht_2\end{array}\right)\!,\\[0.5mm]
	P_R\left(\begin{array}{c} \widetilde{W} \\ 
\widetilde{H}\end{array}\right) &=& 
  P_R\left(\begin{array}{cc} U_{11} & U_{21} \\ U_{12} & U_{22}
\end{array}\right)
  \left(\begin{array}{c} \cht_1 \\ \cht_2\end{array}\right)\!.
\end{eqnarray}
\label{eqn:HKcharginomix}
\end{subequations}
 By comparing Eqns.\ref{eqn:ISAJETcharginomix} and \ref{eqn:HKcharginomix}
  we can obtain the mixing matrices in the  notation of \cite{Gunion:1986yn} 
\vspace{-1mm}
\begin{subequations}
\begin{eqnarray}
   U &=& \left(\begin{array}{cc} \phantom{\theta_x}-\sin\ga_L & 
\phantom{\theta_x}-\cos\ga_L \\
				-\theta_x\cos\ga_L &
 \phantom{-}\theta_x\sin\ga_L 
\end{array}\right)\!,\\[0.5mm]
   V &=& \left(\begin{array}{cc}
 \phantom{\theta_y}-\sin\ga_R (-1)^{\theta_-} & 
 \phantom{\theta_y}-\cos\ga_R (-1)^{\theta_+} \\
-\theta_y\cos\ga_R (-1)^{\theta_-} & 
\phantom{-}\theta_y\sin\ga_R (-1)^{\theta_+}
\end{array}\right)\!.
\end{eqnarray}
\end{subequations}
  It should be noted that we adopt the opposite sign convention 
  for the chargino masses due to the sign differences in the two Lagrangians.

\subsection[Neutralinos]{Neutralinos}

  We define a two-component fermion field
  for the neutralinos before mixing
\begin{equation}
\psi^0_j = \left(-i\lam',-i\lam^3,\psi^0_{\mr{H_1}},\psi^0_{\mr{H_2}}\right),
\end{equation}
  where $\lam'$ is the bino, $\lam^3$ is the neutral wino,
  $\psi^0_{\mr{H_1}}$ is the higgsino for the Higgs giving mass to the
  down-type quarks and $\psi^0_{\mr{H_2}}$ is the higgsino for the Higgs
  giving mass to the up-type quarks. 
  The Lagrangian for the neutralino masses from Eqn.\,A.18 of
  \cite{Gunion:1986yn} is
\begin{equation}
\mathcal{L}_{\mr{neutralino}} = 
-\frac12\left(\psi^0\right)^TY\psi^0+\mr{h.c.},
\end{equation}
  where
\begin{equation}
\renewcommand{\arraycolsep}{4pt}
 Y = \left(\begin{array}{cccc}
	M_1 & 0	&-\mz\sw\cbe & \phantom{-}\mz\sw\sbe \\
	0  & M_2 & \phantom{-}\mz\cw\cbe & -\mz\cw\sbe \\
	-\mz\sw\cbe & \phantom{-}\mz\cw\cbe & 0 & -\mu \\
	\phantom{-}\mz\sw\sbe & -\mz\cw\sbe & -\mu & 0 \\
     \end{array}\right)\!.
\label{eqn:neutralinomatrix}
\end{equation}
  In \cite{Gunion:1986yn} the Lagrangian was diagonalized in this
  two-component notation to give the mass eigenstates. The diagonalization
  was performed by defining two-component fields
\begin{equation}
\chi^0_i = N_{ij}\psi^0_j, \ \ \ \ \ \ \ \ \ \  i,j=1,\ldots\,\!,4,
\end{equation}
  where $N$ is a unitary matrix satisfying $N^*\, Y\, N^{-1} = N_D$, and
  $N_D$ is the diagonal neutralino mass matrix. The four-component mass 
  eigenstates, the neutralinos, can be defined in terms of the two-component
  fields, \ie
\begin{equation}
    \cht^0_i=\left(\begin{array}{c}\chi^0_i \\
 \bar{\chi}^0_i\end{array}\right)\!.
\end{equation}
  Rather than adopting this approach we need to express the neutralino mass
  terms in four-component notation before performing the diagonalization in
  order to compare this Lagrangian with that of \cite{Baer:1987au}. 
  We use the standard procedure of \cite{Gunion:1986yn} to express this
  in four-component notation by defining four-component Majorana fields
\begin{equation}
  \widetilde{B}   = \left(\begin{array}{c}-i\lam'  \\
				i\bar{\lam'}\end{array}\right)\!,\ \ \ \ \ 
  \widetilde{W}_3 = \left(\begin{array}{c}-i\lam^3 \\
				i\bar{\lam}^3\end{array}\right)\!,\ \ \ \ \ 
  \widetilde{H}_1 = \left(\begin{array}{c}\psi^0_{\mr{H_1}} \\
			\psb^0_{\mr{H_1}}\end{array}\right)\!,\ \ \ \ \ 
  \widetilde{H}_2 = \left(\begin{array}{c}\psi^0_{\mr{H_2}} \\
				\psb^0_{\mr{H_2}}\end{array}\right)\!,
\end{equation}
  where $\widetilde{B}$ is the bino field, $\widetilde{W}_3$ is the neutral
  wino field, $\widetilde{H}_1$ is the field
  for the higgsino associated with the Higgs boson which gives mass to the
   down-type quarks and $\widetilde{H}_2$ is the field for the higgsino
  associated with the Higgs
  boson which gives mass to the up-type quarks.

  This gives the Lagrangian in four-component notation
\begin{equation}
  \mathcal{L}_{\mr{neutralino}} =-\frac12
 \left(\overline{\widetilde{B}},\overline{\widetilde{W}}_3,
\overline{\widetilde{H}}_1,\overline{\widetilde{H}}_2 \right)
\left[Y P_L+Y P_R\right]\left(\begin{array}{c}
 \widetilde{B} \\\widetilde{W}_3\\\widetilde{H}_1\\ \widetilde{H}_2
  \end{array}\right)\!.
\label{eqn:GHneut}
\end{equation}
  We can now compare this with the relevant Lagrangian, in four-component
  notation,
  given in \cite{Baer:1987au}. This Lagrangian is from \cite{Baer:1987au},
  Eqn.\,2.2,\footnote{This Lagrangian is taken from \cite{Tata:1997uf}
  which corrects a sign error in \cite{Baer:1987au}.}
\begin{equation}
  \mathcal{L}_{\mr{neutralino}}=-\frac12
\left( \bar{h}^0,\bar{h}^{'0} , \bar{\lam}_3, \bar{\lam_0} \right)
\left[ M_{\mr{neutral}}P_L+M_{\mr{neutral}}P_R\right]
\left(\begin{array}{c}
 h^0\\ {h'}^0\\ \lam_3\\ \lam_0
\end{array}\right)\!,
\end{equation}
  where $h^0$ is the higgsino partner of the Higgs which gives mass to the
  up-type quarks, ${h'}^0$ is the higgsino partner of the Higgs which gives
  mass to the down-type quarks, $\lam_3$ is the neutral wino,
  $\lam_0$ is the bino and the mass matrix is given by
\begin{equation}
{\renewcommand{\arraycolsep}{10pt}
\renewcommand{\arraystretch}{1.5}
 M_{\mr{neutral}} = \left(\begin{array}{cccc}
  0 	 & -2m_1	 & -\ort gv 	& \phantom{-}\ort g'v \\
  -2m_1  & 0		 & \phantom{-}\ort gv'	& -\ort g'v' \\
 -\ort gv& \phantom{-}\ort g v'	 & \mu_2 	& 0 \\
\phantom{-}\ort g'v& -\ort g'v' 	 & 0 		& \mu_1  
     \end{array}\right)\!.
\label{eqn:ISAJETneutmass}}
\end{equation}
  It is easier to compare this with Eqn.\,\ref{eqn:GHneut} after reordering
  the entries and reexpressing it in terms of $\mz$, $\be$ and $\theta_W$.
  This gives
\begin{equation}
  \mathcal{L}_{\mr{neutralino}}=\frac12
\left( \bar{\lam_0}, \bar{\lam}_3, \bar{h}^{'0},\bar{h}^0  \right)
\left[ M'_{\mr{neutral}}P_L+M'_{\mr{neutral}}P_R\right]
\left(\begin{array}{c}
   \lam_0 \\ \lam_3\\{h'}^0\\h^0
\end{array}\right)\!,
\end{equation}
  where 
\begin{equation}
{\renewcommand{\arraycolsep}{3pt}
 M'_{\mr{neutral}} = \left(\begin{array}{cccc}
	M_1 & 0	&\phantom{-}\mz\sw\cbe & -\mz\sw\sbe \\
	0  & M_2 & -\mz\cw\cbe & \phantom{-}\mz\cw\sbe \\
	\phantom{-}\mz\sw\cbe & -\mz\cw\cbe & 0 & -\mu \\
	-\mz\sw\sbe & \phantom{-}\mz\cw\sbe & -\mu & 0 \\
     \end{array}\right)\!.}
\end{equation}
  This equation then agrees with Eqn.\,\ref{eqn:GHneut} up to an overall sign
  provided we make the following identification
\begin{subequations}
\begin{eqnarray}
  \widetilde{B} &  = &  \lam_0,  \\
  \widetilde{W}_3 &=& \lam_3,\\
  \widetilde{H}_1 &=&-{h'}^0,\\
  \widetilde{H}_2 &=&-h^0.
\end{eqnarray}
\end{subequations}
  The convention from ISAJET for the mixing
  taken from Eqn.\,2.12 of \cite{Baer:1987au} is
\begin{equation}
\left(\begin{array}{c}	\left(-i\ga_5\right)^{\theta_1}\widetilde{Z}_1\\
	        	\left(-i\ga_5\right)^{\theta_2}\widetilde{Z}_2\\
			\left(-i\ga_5\right)^{\theta_3}\widetilde{Z}_3\\
	\left(-i\ga_5\right)^{\theta_4}\widetilde{Z}_4\end{array}\right)
   =  \left(\begin{array}{cccc}
	v_1^{(1)} & v_2^{(1)} & v_3^{(1)} & v_4^{(1)} \\
	v_1^{(2)} & v_2^{(2)} & v_3^{(2)} & v_4^{(2)} \\
	v_1^{(3)} & v_2^{(3)} & v_3^{(3)} & v_4^{(3)} \\
	v_1^{(4)} & v_2^{(4)} & v_3^{(4)} & v_4^{(4)} 
      \end{array}\right)
 \left(\begin{array}{c}h^0\\{h'}^0\\\lam_3\\\lam_0\end{array}\right)\!,
\end{equation}
  where $\widetilde{Z}_i$ are the mass eigenstates obtained by diagonalizing
  the mass matrix given in Eqn.\,\ref{eqn:ISAJETneutmass}, $v_j^{(i)}$ are
  the elements of the mixing matrix and $\tht_i$ is zero (one) if the mass of
  $\widetilde{Z}_i$ is positive (negative).

  After reordering we get
\begin{equation}
\left(\begin{array}{c}	\left(-i\ga_5\right)^{\theta_1}\widetilde{Z}_1\\
	        	\left(-i\ga_5\right)^{\theta_2}\widetilde{Z}_2\\
			\left(-i\ga_5\right)^{\theta_3}\widetilde{Z}_3\\
	\left(-i\ga_5\right)^{\theta_4}\widetilde{Z}_4\end{array}\right)
   =  \left(\begin{array}{cccc}
	v_4^{(1)} & v_3^{(1)} & -v_2^{(1)} & -v_1^{(1)} \\
	v_4^{(2)} & v_3^{(2)} & -v_2^{(2)} & -v_1^{(2)} \\
	v_4^{(3)} & v_3^{(3)} & -v_2^{(3)} & -v_1^{(3)} \\
	v_4^{(4)} & v_3^{(4)} & -v_2^{(4)} & -v_1^{(4)} 
      \end{array}\right)
 \left(\begin{array}{c} \widetilde{B} \\ \widetilde{W} \\
	 \widetilde{H}_1 \\ \widetilde{H}_2
 \end{array}\right)\!.
\end{equation}
  We can therefore obtain the mixing matrix in the notation of
  \cite{Gunion:1986yn} by making the identification
\begin{subequations}
\begin{eqnarray}
	N_{i1}& =& v_4^{(i)}, \\
	N_{i2}& =& v_3^{(i)},\\
	N_{i3}& =& -v_2^{(i)}, \\
	N_{i4}& =& -v_1^{(i)}.
\end{eqnarray}
\end{subequations}
  Again we need to adopt the opposite sign  convention for the neutralino
  masses.

\subsection[Left/Right Sfermion Mixing]{Left/Right Sfermion Mixing}

  In addition to the mixing of the neutralinos and charginos we also need
  to consider the left/right mixing of the sfermions. In general as the 
  off-diagonal terms in the mass matrices are proportional to the fermion
  mass these effects are only important for the third generation sfermions,
  \ie stop, sbottom and stau. As the top mass is much 
  larger than any of the other Standard Model fermion masses these effects
  are particularly important for the top squarks.

 The following mass matrix for the top squarks uses the conventions
 of \cite{Gunion:1986yn}\footnote{ This is the same as \cite{Bartl:1994bu},
 SPYTHIA \cite{Mrenna:1997hu} and SUSYGEN \cite{Katsanevas:1997fb}.}
 and is taken from Eqn.\,4.17 of \cite{Gunion:1986yn},
\begin{equation}
 M^2_{\tilde{t}} = \left(\begin{array}{cc}
 M^2_{\tilde{Q}}+M^2_{\mr{Z}}\cos2\be\left(\frac12-\frac23\ssw\right)+m^2_t &
 m_t\left(A_t-\mu\cot\be\right) \\
 m_t\left(A_t-\mu\cot\be\right) &
 M^2_{\tilde{U}}+\frac23m^2_{\mr{Z}}\cos2\be\ssw+m^2_t
\end{array}\right)\!,
\end{equation}
  where $M_{\tilde{Q}}$ and $M_{\tilde{U}}$ are soft SUSY-breaking masses for
  the left and right top squarks, respectively. $A_t$ is the trilinear soft
  SUSY-breaking term for the interaction of the left and right stop squarks
  with the Higgs boson. This compares with the ISAJET matrix from 
  \cite{Baer:1994xr}
\begin{equation}
 M^2_{\tilde{t}} = \left(\begin{array}{cc}
 M^2_{\tilde{t}_L}+M^2_{\mr{Z}}\cos2\be\left(\frac12-\frac23\ssw\right)
	+m^2_t &
 -m_t\left(A_t-\mu\cot\be\right) \\
 -m_t\left(A_t-\mu\cot\be\right) &
 M^2_{\tilde{t}_R}+\frac23m^2_{\mr{Z}}\cos2\be\ssw+m^2_t
\end{array}\right)\!,
\end{equation}
  where $ M^2_{\tilde{t}_L}=M^2_{\tilde{Q}}$ and
  $M^2_{\tilde{t}_R}=M^2_{\tilde{U}}$. There is a difference in the sign of
  the off-diagonal terms. This means that,
  as the sign convention for the $\mu$ and $A$ terms are the same, there is a
  difference in the relative phases of the two fields in the different
  conventions. Hence we should apply the following change in the ISAJET output
\begin{equation}
\theta_t \longrightarrow -\theta_t,
\end{equation}
 \ie change the sign of the stop mixing angle. The same argument also applies
  to the sbottom and stau mixing angles.

  We adopt the following convention for the sfermion mixing matrices
\begin{equation}
 \left( \begin{array}{c} \qkt_{iL} \\ \qkt_{iR} \end{array}
\right) 
  = \left( \begin{array}{cc}
	 \phantom{-}\cos\theta^i_q & \phantom{-}\sin\theta^i_q \\
                             -\sin\theta^i_q & \phantom{-}\sin\theta^i_q
\end{array}\right)
   \left( \begin{array}{c} \qkt_{i1} \\ \qkt_{i2} \end{array} \right)\!,
\end{equation}
  where  $\qkt_{iL}$ and $\qkt_{iR}$ are the left and right squark fields, for
  the $i$th quark, where $i$ is  u, d, s, c, b, and t. $\qkt_{i1}$ and
  $\qkt_{i2}$ are the squark fields for the mass eigenstates, for the quark
  $i$, and $\theta^i_q$ is the  mixing angle obtained by diagonalizing the
  mass matrix. 
  We denote the mixing matrix above as $Q^i_{\al\be}$ where $i$ is the quark,
  $\be$ is the mass eigenstate and $\al$ is the left/right eigenstate.

  Similarly we denote the lepton mixing as above with the matrix
  $L^i_{\al\be}$ where $i$ is $\mr{e^-}$,
  $\mr{\nu_e}$,  $\mr{\mu^-}$,  $\mr{\nu_\mu}$,
  $\mr{\tau^-}$ and $\mr{\nu_\tau}$,
  $\be$ is the mass eigenstate and $\al$ is the left/right eigenstate. As we
  do not include the right-handed neutrino we will neglect the left/right
  mixing for the sneutrinos.
%
%  Now the Feynman rules for Chargino and Neutralino interactions
%
\section[Gaugino Interactions with the Sfermions]
	{Gaugino Interactions with the Sfermions}
%
%  Figure containing the chargino Feynman rules
%
\begin{figure}[t]
\vskip -4mm
\begin{center} \begin{picture}(180,50)(100,40)
\SetScale{1.0}
\ArrowLine(20,35)(80,35)
\ArrowLine(80,35)(115,70)
\DashArrowLine(115,0)(80,35){5}
\Text(90,57)[]{\large $\mr{u}_i$}
\Text(50,47)[]{\large $\mr{\cht}_j^+$}
\Text(90,12)[]{\large $\mr{\dnt}_{i\al}$}
\Text(300,35)[]{\large 
${\displaystyle-\frac{ig}{2}
\left[\rule{0mm}{6mm} a(\mr{\dnt}_{i\al})(1-\ga_5)+
		 b(\mr{\dnt}_{i\al})(1+\ga_5)\right]} $ }
\Vertex(80,35){1}
\end{picture} \end{center}
\vskip 10mm
\begin{center} \begin{picture}(180,50)(100,40)
\SetScale{1.0}
\ArrowLine(80,35)(20,35)
\ArrowLine(80,35)(115,70)
\DashArrowLine(115,0)(80,35){5}
\Text(90,57)[]{\large $\mr{d}_i$}
\Text(50,47)[]{\large $\mr{\cht}_j^+$}
\Text(90,12)[]{\large $\mr{\upt}_{i\al}$}
\Text(300,35)[]{\large 
${\displaystyle-\frac{ig}{2}
\left[\rule{0mm}{6mm} a(\mr{\upt}_{i\al})(1-\ga_5)+
			b(\mr{\upt}_{i\al})(1+\ga_5)\right] C }$ }
\Vertex(80,35){1}
\end{picture} \end{center}
\vskip 10mm
\captionB{Feynman rules for $\mr{q\qkt\cht^+}$.}
	{Feynman rules for $\mr{q\qkt\cht^+}$.}
\label{fig:chi+q}
%\end{figure}
%\begin{figure}
\vspace{-3mm}
\begin{center} \begin{picture}(180,50)(100,40)
\SetScale{1.0}
\ArrowLine(20,35)(80,35)
\ArrowLine(80,35)(115,70)
\DashArrowLine(115,0)(80,35){5}
\Text(90,57)[]{\large $\mr{\nu}_i$}
\Text(50,47)[]{\large $\mr{\cht}_j^+$}
\Text(90,12)[]{\large $\mr{\elt}_{i\al}$}
\Text(250,35)[]{\large ${\displaystyle-\frac{ig}{2}b(\elt_{i\al})(1+\ga_5)}$ }
\Vertex(80,35){1}
\end{picture} \end{center}
\vskip 10mm
\begin{center} \begin{picture}(180,50)(100,40)
\SetScale{1.0}
\ArrowLine(80,35)(20,35)
\ArrowLine(80,35)(115,70)
\DashArrowLine(115,0)(80,35){5}
\Text(90,57)[]{\large $\mr{\ell}^-_i$}
\Text(50,47)[]{\large $\mr{\cht}_j^+$}
\Text(90,12)[]{\large $\mr{\nut}_{i}$}
\Text(250,35)[]{\large 
${\displaystyle-\frac{ig}{2} \left[\rule{0mm}{6mm}a(\nut_i)(1-\ga_5) 
+b(\nut_i)(1+\ga_5) \right] C}$ }
\Vertex(80,35){1}
\end{picture} \end{center}
\vskip 10mm
\captionB{Feynman rules for $\mr{\ell\elt\cht^+}$.}
	{Feynman rules for $\mr{\ell\elt\cht^+}$.}
	\label{fig:chi+l}
\end{figure}
  The Lagrangians for the interactions of the electroweak gauginos with
  the sfermions and fermions
  are derived in \cite{Gunion:1986yn} without left/right
  mixing of the sfermions. We will therefore take the results for the
  Lagrangians in the left/right sfermion basis and transform them into 
  the mass basis we will use in the decay rate and cross-section calculations.

  We will first consider the interactions of the sfermions and the charginos.
 The relevant Lagrangian, without left/right sfermion mixing, is
\begin{eqnarray}
 \mathcal{L}_{\mr{q\qkt\cht^+}} &= & -g \left[
\bar{u} P_R U_{l1} \cht_l^+ \dnt_L+ 
\bar{d} P_R V_{l1} \chi_l^c \upt_L \right]  \nonumber \\[1mm]
 && 
  +\frac{g m_d}{\rtt \mw \cos \beta} \left[ 
\bar{u} P_R U_{l2} \cht_l^+ \dnt_R +
\bar{d} P_L U^*_{l2} \cht_l^c \upt_L \right] \nonumber \\[1mm]
 && +\frac{g m_u}{\rtt \mw \sin \beta}  \left[ 
\bar{u} P_L V^*_{l2} \cht_l^+ \dnt_L +
\bar{d} P_R V_{l2} \cht_l^c \upt_R \right] +\mr{h.c.}, \label{eqn:chi+2}
\end{eqnarray}
  for one generation of quarks. This is taken from 
  \cite{Gunion:1986yn}, Eqn.~5.3. In Eqn.\,\ref{eqn:chi+2} there is an implied
  summation over the chargino mass eigenstates.
\textheight 24cm
  If we now substitute for the left and right squark eigenstates in terms of
  the mass eigenstates we obtain 
\begin{eqnarray}
 \mathcal{L}_{\mr{q\qkt\cht^+}} & = &
-g \bar{u}_i \left[ \rule{0mm}{7mm}\left( U_{l1} Q^{2i-1}_{1\al}
 -\frac{m_{d_i} U_{l2} Q^{2i-1}_{2\al} }
{\rtt \mw \cos \beta} \right) P_R
- \frac{m_{u_i} V^*_{l2} Q^{2i-1}_{1\al}}{\rtt \mw \sin \beta} P_L 
\right] \cht_l^+ \dnt_{i\al} \nonumber \\
 &&
-g \bar{d}_i \left[ \left( V_{l1} Q^{2i}_{1\al} - \frac{m_{u_i}
  V_{l2} Q^{2i}_{2\al} }
{\rtt \mw \sin \beta}  \right) P_R
- \frac{m_{d_i} U^*_{l2} Q^{2i}_{1\al}}{\rtt \mw \cos \beta} P_L
 \rule{0mm}{7mm} \right]  
\cht_l^c \upt_{i\al} \label{eqn:chi+},
\end{eqnarray}
  where $i$ is the generation of the squark and $\al$ is the mass eigenstate. 
  In Eqn.\,\ref{eqn:chi+}
   there is an implicit summation over the squark and 
   chargino mass eigenstates.
  This leads to the Feynman rules given in Fig.\,\ref{fig:chi+q}, with the
  coefficients given in Table~\ref{tab:chargecp}.

  The Feynman rules for the sleptons, Fig.\,\ref{fig:chi+l}, can
  be obtained by changing the relevant masses and couplings, \ie replacing the
  squark mixing matrices with the slepton mixing matrices and making the
  replacement
\begin{equation}
   e_d\ra -1,\ \ \ \  e_u\ra 0,\ \ \ \  m_d\ra m_e,\ \ \ \  m_u \ra 0.
\label{eqn:replace}
\end{equation}

%
%  Table with the chargino couplings
%
\begin{table}
\renewcommand{\arraystretch}{2.0}
\begin{center}
\begin{tabular}{|l|l|l|l|}
\hline
 Coefficient & Coupling & Coefficient  & Coupling \\
\hline
 $a(\mr{\elt}_{i\al})$ & 0  &
 $b(\mr{\elt}_{i\al})$ & $ U_{l1}L^{2i-1}_{1\al}
                    - \frac{ m_{\ell_i} U_{l2} L^{2i-1}_{2\al}}
                            {\rtt \mw\cos\be}$ \\
\hline
 $a(\mr{\nut}_i)$ & $-\frac{ m_{\ell_i}U^*_{l2}}{\rtt \mw\cos\be}$  &
 $b(\mr{\nut}_i)$ & $V_{l1}$ \\
\hline
 $a(\mr{\dnt}_{i\al})$ &  $-\frac{m_{u_i} V^*_{l2} Q^{2i-1}_{1\al}}
                      {\rtt \mw\sin\be}$ &
 $b(\mr{\dnt}_{i\al})$ & $ U_{l1} Q^{2i-1}_{1\al}
                    -\frac{m_{d_i}U_{l2} Q^{2i-1}_{2\al} }
                            {\rtt \mw\cos\be}$ \\
\hline
 $a(\mr{\upt}_{i\al})$ & $-\frac{m_{d_i} U^*_{l2} Q^{2i}_{1\al}}
                      {\rtt \mw\cos\be}$  &
 $b(\mr{\upt}_{i\al})$ & $ V_{l1}Q^{2i}_{1\al}
                      -\frac{m_{u_i} V_{l2} Q^{2i}_{2\al}}
                      {\rtt \mw\sin\be}$  \\
\hline
\end{tabular}
\captionB{Couplings for the chargino Feynman rules.}
	{Couplings for the chargino Feynman rules.}
\label{tab:chargecp}
\end{center}
\end{table}

  The Lagrangian for the interaction of neutralinos with squarks is 
given in Eqn.~5.5 of \cite{Gunion:1986yn},
\begin{eqnarray}
\mathcal{L}_{\mr{q\qkt\cht^0}} & = & 
	-\rtt\bar{u} \left\{ \frac{g m_u N^*_{l4}}{2\mw\sbe}P_L
  	        +\left[ee_uN'_{l1}+\frac{gN'_{l2}}{\cw}
			\left(\frac{1}{2}-e_u\ssw\right)
\rule{0mm}{7mm}\right]P_R
	    \rule{0mm}{9mm}\right\} \cht^0_l\upt_L \nonumber \\
     & & +\rtt\bar{u}\left[ \rule{0mm}{7mm}
      \left( ee_u{N'}^*_{l1}-\frac{ge_u\ssw {N'}^*_{l2}}{\cw}\right)P_L 
        -\frac{gm_uN_{l4}}{2\mw\sbe}P_R \right]
\rule{0mm}{7mm}\cht^0_l\upt_R \nonumber\\[1mm]
     & &-\rtt\bar{d}  \left\{ \frac{g m_d N^*_{l3}}{2\mw\cbe}P_L
  	        +\left[ee_dN'_{l1}-\frac{gN'_{l2}}{\cw}
			\left(\frac{1}{2}+e_d\ssw\right)
\rule{0mm}{7mm}\right]P_R
	   \rule{0mm}{9mm} \right\} \cht^0_l\dnt_L \nonumber\\
     & & +\rtt\bar{d}\left[ 
      \left( ee_d{N'}^*_{l1}-\frac{ge_d\ssw {N'}^*_{l2}}{\cw}\right)P_L 
        -\frac{gm_dN_{l3}}{2\mw\cbe}P_R \rule{0mm}{7mm}\right]\cht^0_l\dnt_R
\nonumber\\ 
&& +\mr{h.c.},
\end{eqnarray}
\textheight 23cm
  for one generation of squarks. Here there is an implied summation over the
  neutralino mass eigenstates.

  This Lagrangian can be expressed in terms of the squark mass eigenstates,
\begin{eqnarray}
\mathcal{L}_{\mr{q\qkt\cht^0}} & = & 
	\rtt\bar{u}_i \left\{\rule{0mm}{9mm}
 -\left[\left\{ee_uN'_{l1}+\frac{gN'_{l2}}{\cw}
	\left(\frac{1}{2}-e_u\ssw\right) \rule{0mm}{6mm}\right\}Q^{2i}_{1\al}
		+\frac{gm_{u_i}N_{l4}}{2\mw\sbe}Q^{2i}_{2\al}
	 \rule{0mm}{7mm}	 \right]P_R
		  \right. \nonumber\\
 &&     \left.  +\left[-\frac{g m_{u_i} N^*_{l4}}{2\mw\sbe}Q^{2i}_{1\al}
	                  +\left( ee_u{N'}^*_{l1}-
			\frac{ge_u\ssw {N'}^*_{l2}}{\cw}\right)Q^{2i}_{2\al}
	 \rule{0mm}{7mm}	 \right]P_L	
\rule{0mm}{9mm}	\right\}\cht^0_l\upt_{i\al} \nonumber \\
  && +\rtt\bar{d}_i \left\{\rule{0mm}{9mm}
		 \left[-\frac{g m_{d_i} N^*_{l3}}{2\mw\cbe}Q^{2i-1}_{1\al}
	                  +\left( ee_d{N'}^*_{l1}-
			\frac{ge_d\ssw {N'}^*_{l2}}{\cw}\right)Q^{2i-1}_{2\al}
	 \rule{0mm}{7mm}	 \right]P_L \right. \nonumber\\
 &&  \left.        	 -\left[\left\{ee_dN'_{l1}-\frac{gN'_{l2}}{\cw}
		\left(\frac{1}{2}+e_d\ssw\right)\right\}Q^{2i-1}_{1\al}
		+\frac{gm_{d_i}N_{l3}}{2\mw\cbe}Q^{2i-1}_{2\al}
	 \rule{0mm}{7mm}	 \right]P_R
	\rule{0mm}{9mm}\right\}\cht^0_l\dnt_{i\al} \nonumber\\
 && +\mr{h.c.},
\label{eqn:neutL}
\end{eqnarray}
  where $i$ is the generation of the squark (assuming no mixing between
  the generations)
  and $\al$ is its mass eigenstate. Again there is an implied summation
  over the squark and neutralino mass eigenstates.
%
%  Figures containing the neutralino Feynman rules
%
\begin{figure}
\vskip -10mm
\begin{center} \begin{picture}(180,50)(100,40)
\SetScale{1.0}
\ArrowLine(20,35)(80,35)
\ArrowLine(80,35)(115,70)
\DashArrowLine(115,0)(80,35){5}
\Vertex(80,35){1}
\Text(90,57)[]{\large $\mr{u}_i$}
\Text(50,47)[]{\large $\mr{\cht}_j^0$}
\Text(90,12)[]{\large $\mr{\upt}_{i\al}$}
\Text(300,35)[]{\large 
${\displaystyle \frac{-i}{\rtt}\left[ a(\mr{\upt}_{i\al})(1-\ga_5)
		+b(\mr{\upt}_{i\al})(1+\ga_5)\rule{0mm}{5.1mm}\right]}$}
\end{picture} \end{center}
\vskip 10mm
\begin{center} \begin{picture}(180,50)(100,40)
\SetScale{1.0}
\ArrowLine(20,35)(80,35)
\ArrowLine(80,35)(115,70)
\DashArrowLine(115,0)(80,35){5}
\Text(90,57)[]{\large $\mr{d}_i$}
\Text(50,47)[]{\large $\mr{\cht}_j^0$}
\Text(90,12)[]{\large $\mr{\dnt_{i\al}}$}
\Vertex(80,35){1}
\Text(300,35)[]{\large 
${\displaystyle\frac{-i}{\rtt}\left[ a(\mr{\dnt}_{i\al})(1-\ga_5)
	+b(\mr{\dnt}_{i\al})(1+\ga_5)\rule{0mm}{5.1mm}\right]}$}
\end{picture} \end{center}
\vskip 10mm
\captionB{Feynman rules for $\mr{q\qkt\cht^0}$.}
	{Feynman rules for $\mr{q\qkt\cht^0}$.}
\label{fig:chi0q}
%\end{figure}
%\begin{figure}
\begin{center} \begin{picture}(180,50)(100,40)
\SetScale{1.0}
\ArrowLine(20,35)(80,35)
\ArrowLine(80,35)(115,70)
\DashArrowLine(115,0)(80,35){5}
\Vertex(80,35){1}
\Text(90,57)[]{\large $\mr{\nu}_i$}
\Text(50,47)[]{\large $\mr{\cht}_j^0$}
\Text(90,12)[]{\large $\mr{\nut_i}$}
\Text(300,35)[]{\large 
${\displaystyle\frac{-i}{\rtt}b(\mr{\nut}_i)(1+\ga_5)}$}
\end{picture} \end{center}
\vskip 10mm
\begin{center} \begin{picture}(180,50)(100,40)
\SetScale{1.0}
\ArrowLine(20,35)(80,35)
\ArrowLine(80,35)(115,70)
\DashArrowLine(115,0)(80,35){5}
\Text(90,57)[]{\large $\mr{\ell}_i$}
\Text(50,47)[]{\large $\mr{\cht}_j^0$}
\Text(90,12)[]{\large $\mr{\elt}_{i\al}$}
\Text(300,35)[]{\large 
${\displaystyle\frac{-i}{\rtt}\left[ a(\mr{\elt}_{i\al})(1-\ga_5)
	+b(\mr{\elt}_{i\al})(1+\ga_5)\rule{0mm}{5.1mm}\right]}$}
\Vertex(80,35){1}
\end{picture} \end{center}
\vskip 10mm
\captionB{Feynman rules for $\mr{\ell\elt\cht^0}$.}
	{Feynman rules for $\mr{\ell\elt\cht^0}$.}\label{fig:chi0l}
%\end{figure}
%\begin{figure}[htp]
\begin{center} \begin{picture}(180,50)(100,40)
\SetScale{1.0}
\ArrowLine(20,35)(80,35)
\ArrowLine(80,35)(115,70)
\DashArrowLine(115,0)(80,35){5}
\Vertex(80,35){1}
\Text(90,57)[]{\large $\mr{u}^{c_1}_i$}
\Text(50,47)[]{\large $\mr{\glt}^a$}
\Text(90,10)[]{\large $\mr{\upt}^{c_2}_{i\al}$}
\Text(280,35)[]{\large 
${\displaystyle\frac{-ig_s{\bf t}^a_{c_1c_2}}{\rtt}\left[ a(\mr{\upt}_{i\al})
(1-\ga_5)+b(\mr{\upt}_{i\al})(1+\ga_5)\rule{0mm}{5.1mm}\right]}$}
\end{picture} \end{center}
\vskip 10mm
\begin{center} \begin{picture}(180,50)(100,40)
\SetScale{1.0}
\ArrowLine(20,35)(80,35)
\ArrowLine(80,35)(115,70)
\DashArrowLine(115,0)(80,35){5}
\Text(90,57)[]{\large $\mr{d}^{c_1}_i$}
\Text(50,47)[]{\large $\mr{\glt}^a$}
\Text(90,10)[]{\large $\mr{\dnt}^{c_2}_{i\al}$}
\Vertex(80,35){1}
\Text(280,35)[]{\large 
${\displaystyle\frac{-ig_s{\bf t}^a_{c_1c_2}}{\rtt}\left[ a(\mr{\dnt}_{i\al})
(1-\ga_5)+b(\mr{\dnt}_{i\al})(1+\ga_5)\rule{0mm}{5.1mm}\right]}$}
\end{picture} \end{center}
\vskip 10mm
\captionB{Feynman rules for $\mr{q\qkt\glt}$.}
	{Feynman rules for $\mr{q\qkt\glt}$. The colours of the gluino, 
	quark and squark are $a$, $c_1$ and $c_2$, respectively.}
\label{fig:gluinoq}
\end{figure}

%
%  Table with the neutralino couplings
%
\begin{table}
\renewcommand{\arraystretch}{2.0}
\begin{center}
\begin{tabular}{|c|c|}
\hline
 Coefficient & Coupling\\
\hline
 $a(\mr{\dnt}_{i\al})$ & $\frac{g m_{d_i} N^*_{l3}}{2\mw\cbe}Q^{2i-1}_{1\al}
                        -Q^{2i-1}_{2\al}\left( e e_d {N'}^*_{l1} 
                        -\frac{g e_d \ssw {N'}^*_{l2}}{\cw}\right)$ \\
\hline
 $b(\mr{\dnt}_{i\al})$     & $\frac{g m_{d_i} N_{l3}}{2\mw\cbe}Q^{2i-1}_{2\al}
                         +Q^{2i-1}_{1\al}\left( e e_d {N'}_{l1} 
                      -\frac{g {N'}_{l2}\left(\frac{1}{2}+ e_d \ssw\right) 
                        }{\cw} \right)$\\
\hline
 $a(\mr{\upt}_{i\al})$ & $\frac{g m_{u_i} N^*_{l4}}{2\mw \sbe}Q^{2i}_{1\al}
                   -Q^{2i}_{2\al} \left( e e_u {N'}^*_{l1} 
                    -\frac{g e_u \ssw {N'}^*_{l2}}{\cw}\right)$ \\
\hline
 $b(\mr{\upt}_{i\al})$ & $\frac{g m_{u_i} N_{l4}}{2\mw \sbe}Q^{2i}_{2\al}
                    +Q^{2i}_{1\al}\left( e e_u {N'}_{l1} 
                    +\frac{g {N'}_{l2}\left(\frac{1}{2}- e_u \ssw\right) 
                     }{\cw} \right)$ \\
\hline
 $a(\mr{\elt}_{i\al})$ & 
	$\frac{gm_{\ell_i} N^*_{l3}}{2\mw \cbe}L^{2i-1}_{1\al}
                  +L^{2i-1}_{2\al}\left( e  {N'}^*_{l1} 
                  -\frac{g  \ssw {N'}^*_{l2}}{\cw}\right)$\\
\hline
 $b(\mr{\elt}_{i\al})$     & 
	$\frac{gm_{\ell_i} N_{l3}}{2\mw \cbe}L^{2i-1}_{2\al}
                        -L^{2i-1}_{1\al}\left( e  {N'}_{l1} 
                       +\frac{ g{N'}_{l2} \left(\frac{1}{2}- \ssw\right) 
                         }{\cw} \right) $  \\
\hline
 $a(\mr{\nut}_i)$   & 0   \\
\hline
 $b(\mr{\nut}_i)$        &  $\frac{g{N'}_{l2}}{2\cw}$ \\
\hline
\end{tabular}
\captionB{Couplings for the neutralino Feynman rules.}
	{Couplings for the neutralino Feynman rules.}
\label{tab:neutcp}
\end{center}
\end{table}
\textheight 24cm
  We can obtain the Feynman rules from Eqn.\,\ref{eqn:neutL}, which
  gives the Feynman rules shown in Fig.\,\ref{fig:chi0q}, where the couplings
  are given in Table~\ref{tab:neutcp}.
  The Feynman rules for the interactions of leptons and
  sleptons with the neutralinos, Fig.\,\ref{fig:chi0l}, can be obtained by
  taking the Feynman rules for the squarks and replacing the relevant masses
  and mixings, as in Eqn.\,\ref{eqn:replace}.

%
%  Lets put the Gluino Feynman Rules in too
%
\section[Gluino Interactions with the Squarks]
	{Gluino Interactions with the Squarks}

  The Lagrangian for the interaction of the gluino and the squarks in given
  in Eqn.\,C.89 of \cite{Haber:1985rc},
\begin{equation}
 \mathcal{L}_{\mr{q\qkt\glt}} = -\rtt g_s{\bf t}^a_{c_1c_2}
		\left[ \bar{\glt}_aP_Lq^{c_2}\qkt^{c_1*}_L
			+\bar{q}^{c_1} P_R \glt_a \qkt^{c_2}_L
                        -\bar{\glt}_aP_Rq^{c_2}\qkt^{c_1*}_R
			-\bar{q}^{c_1} P_L \glt_a \qkt^{c_2}_R
		\right],
\end{equation}
where for simplicity we have only considered one flavour of quark. Again we
can replace the left/right eigenstates with the mass eigenstates giving
\begin{equation}
 \mathcal{L}_{\mr{q\qkt\glt}} = -\rtt g_s{\bf t}^a_{c_1c_2}
		\left[ \bar{\glt}_aP_Lq^{c_2} Q^{i}_{1\al}\qkt^{c_1*}_{i\al}
		+\bar{q}^{c_1} P_R \glt_a Q^{i}_{1\al}\qkt^{c_2}_{i\al}
                -\bar{\glt}_aP_Rq^{c_2} Q^i_{2\al}\qkt^{c_1*}_{i\al}
		-\bar{q}^{c_1} P_L \glt_a Q^i_{2\al}\qkt^{c_2}_{i\al}
		\right]
\end{equation}
The Feynman rules for this process
are given in Fig.\,\ref{fig:gluinoq}, and the relevant couplings are in 
Table~\ref{tab:gluinocp}.

\begin{table}
\renewcommand{\arraystretch}{2.0}
\begin{center}
\renewcommand{\tabcolsep}{5mm}
\begin{tabular}{|c|c|c|c|}
\hline
 Coefficient & Coupling & Coefficient  & Coupling\\
\hline
 $a(\mr{\upt}_{i\al})$ &  $-Q^{2i}_{2\al}$ &
 $b(\mr{\upt}_{i\al})$ & $Q^{2i}_{1\al}$ \\
\hline
 $a(\mr{\dnt}_{i\al})$ & $-Q^{2i-1}_{2\al}$ &
 $b(\mr{\dnt}_{i\al})$ & $Q^{2i-1}_{1\al}$\\
\hline
\end{tabular}
\captionB{Couplings for the gluino Feynman rules.}
	{Couplings for the gluino Feynman rules.}
\label{tab:gluinocp}
\end{center}
\end{table}

%
%  Now the Feynman Rules for gauge bosons and sfermions after left/right
%  mixing
%
\section[Gauge Boson Interactions with the Sfermions and Fermions]
	{Gauge Boson Interactions with the Sfermions and Fermions}

  We also need the Feynman rules for the interactions of squark--antisquark
  pairs with the gauge bosons of the MSSM. The relevant 
  Lagrangian is given in Eqn.\,C.66 of \cite{Haber:1985rc},
\begin{eqnarray}
  \mathcal{L}_{\mr{\qkt\qkt V}} &=& 
   \frac{-ig}{\rtt}\left[ \rule{0mm}{7mm}W^+_\mu\upt^*_L\dbmu\dnt_L 
			+W^-_\mu\dnt^*_L\dbmu\upt_L \right]
	 -\frac{ig}{\cw}Z_\mu\left[  \rule{0mm}{7mm}
   \left(\frac{1}{2}-e_u\ssw\right)\upt^*_L\dbmu\upt_L\right.
\nonumber\\[1.0mm]
 && \left. -\left(\frac{1}{2}+e_d\ssw\right)\dnt^*_L\dbmu\dnt_L
   -e_u\ssw \upt^*_R\dbmu\upt_R -e_d\ssw \dnt^*_R\dbmu\dnt_R \rule{0mm}{7mm}
 \right]\nonumber \\[1.0mm]
 &&  -ieA_\mu\left[ \rule{0mm}{7mm}
   	  e_u\left( \upt^*_L\dbmu\upt_L+\upt^*_R\dbmu\upt_R\right)
  	 +e_d\left( \dnt^*_L\dbmu\dnt_L+\dnt^*_R\dbmu\dnt_R\right)\right]\!.
\end{eqnarray}
\textheight 24cm
  This can be expressed in terms of the mass eigenstates giving,
\begin{eqnarray}
  \mathcal{L}_{\mr{\qkt\qkt V}} &=& 
 -\frac{ig}{\rtt}\left[  
          Q^{2i}_{1\al}Q^{2i-1}_{1\be}W^+_\mu\upt^*_{i\al}\dbmu\dnt_{i\be} 
         +Q^{2i-1}_{1\al}Q^{2i}_{1\be}W^-_\mu\dnt^*_{i\al}\dbmu\upt_{i\be}
\right] \nonumber \\
 && -\frac{ig}{\cw}Z_\mu\left[ -e_u\ssw\upt^*_{i\al}\dbmu\upt_{i\al}
			       -e_d\ssw\dnt^*_{i\al}\dbmu\dnt_{i\al}
+\frac{1}{2}Q^{2i}_{1\al}Q^{2i}_{1\be}\upt^*_{i\al}\dbmu\upt_{i\be}\right.
\nonumber \\
 && \left.
-\frac{1}{2}Q^{2i-1}_{1\al}Q^{2i-1}_{1\be}\dnt^*_{i\al}\dbmu\dnt_{i\be}\right]
 -ieA_\mu\left[
 e_u\upt^*_{i\al}\dbmu\upt_{i\al}+e_d\dnt^*_{i\al}\dbmu\dnt_{i\al}\right]\!,
\end{eqnarray}
  where there is now an implied summation over the squark mass eigenstates.
  Hence we obtain the Feynman rules in Fig.\,\ref{fig:Zsquarks}, 
  where the couplings for the $Z\mr{\qkt\qkt^*}$ vertex are given in
  Table~\ref{tab:Zcp}. The Feynman rules for
  the interactions of the sleptons with the gauge bosons can be obtained by
  replacing the relevant couplings and are given in Fig.\,\ref{fig:Zsleptons}.
  The couplings for these processes are also given in Table~\ref{tab:Zcp}.
\begin{table}
\begin{center}
\renewcommand{\arraystretch}{1.4}
\begin{tabular}{|c|c|c|c|}
\hline
\multicolumn{4}{|c|}{Squark couplings} \\
\hline
 $ Z^{\al\be}_{u_i}$ & $\frac{1}{2}\left(-Q^{2i}_{1\al}Q^{2i}_{1\be}
			+2e_u\ssw\delta_{\al\be}\right)$ &
 $ Z^{\al\be}_{d_i}$ & $\frac{1}{2}\left(Q^{2i-1}_{1\al}Q^{2i-1}_{1\be}
			+2e_d\ssw\delta_{\al\be}\right)$ \\
\hline
\multicolumn{4}{|c|}{Slepton couplings} \\
\hline
 $ Z^{\al\be}_{\nu_i}$ & $-\frac{1}{2}\delta_{\al=1,\be=1}$ &
 $ Z^{\al\be}_{\ell_i}$ & $\frac{1}{2}\left(L^{2i-1}_{1\al}L^{2i-1}_{1\be}
			-2\ssw\delta_{\al\be}\right)$ \\
\hline
\multicolumn{4}{|c|}{Quark couplings} \\
\hline
 $Z_{u_L}$ & $-\frac{1}{4}\left(1-2e_u\ssw\right)$ & 
 $Z_{d_L}$ & $ \frac{1}{4}\left(1+2e_d\ssw\right)$ \\ 
 $Z_{u_R}$ & $ \frac{1}{2}e_u\ssw$ & 
 $Z_{d_R}$ & $\frac{1}{2}e_d\ssw$\\ 
\hline
\multicolumn{4}{|c|}{Lepton couplings} \\
\hline
 $Z_{\nu_L}$ & $-\frac{1}{4}$ & 
 $Z_{\ell_L}$ & $ \frac{1}{4}\left(1-2\ssw\right)$ \\ 
 $Z_{\nu_R}$ & $0$ & 
 $Z_{\ell_R}$ & $-\frac{1}{2}\ssw$\\ 
\hline
\end{tabular}
\end{center}
\captionB{Couplings of squarks, sleptons, quarks and leptons to the
	 $\mr{Z^0}$.}
	{Couplings of squarks, sleptons, quarks and leptons to the
	 $\mr{Z^0}$.}
\label{tab:Zcp}
\end{table}
  For completeness, we include the Lagrangian for the interactions of the
  quarks 
  with the gauge bosons which is given in Eqn.\,C.61 of \cite{Haber:1985rc}, 
\begin{eqnarray}
  \mathcal{L}_{\mr{qqV}} &=&
	\frac{-g}{\rtt}\left[\rule{0mm}{4mm}W^+_\mu\bar{u}\ga^\mu P_L d
			+W^-_\mu\bar{d}\ga^\mu P_L u \right]
-eA_\mu\left(e_u \bar{u}\ga^\mu u + e_d \bar{d}\ga^\mu d  \right)
\nonumber\\
&&-\frac{g}{\cw}Z_\mu\left\{\rule{0mm}{9mm}\bar{u}\ga^\mu\left[\rule{0mm}{7mm}
	\left(\frac{1}{2}-e_u\ssw\right)P_L-e_u\ssw P_R\right]u\right.
\nonumber\\
 		&&  \left. -\bar{d}\ga^\mu\left[\rule{0mm}{7mm}
	 	\left(\frac{1}{2}+e_d\ssw\right)P_L+e_d\ssw P_R\right]d
		\rule{0mm}{9mm}\right\}\!.
\label{eqn:LfermionZ}
\end{eqnarray}
  The relevant Feynman rules for both the quarks and the leptons, which can be
  obtained by replacing the relevant couplings, are given in
  Fig.\,\ref{fig:Zquarks}.
  We have neglected the effect of quark mixing in these Feynman rules.
\textheight 23cm

\begin{figure}
\begin{center} \begin{picture}(180,50)(100,40)
\SetScale{1.0}
\Photon(20,35)(80,35){5}{5}
\DashArrowLine(115,70)(80,35){5}
\DashArrowLine(80,35)(115,0){5}
\Text(90,55)[]{\large $p$}
\Text(90,12)[]{\large $p'$}
\Text(120,70)[l]{\large $\mr{\upt}_{i\al}$}
\Text(15,35)[r]{\large $\mr{\ga}$}
\Text(120,0)[l]{\large $\mr{\upt}_{i\al}$}
\Text(250,35)[]{\large$-iee_u(p+p')^\mu$}
\Vertex(80,35){1}
\end{picture} \end{center}
\vskip 20mm
\begin{center} \begin{picture}(180,50)(100,40)
\SetScale{1.0}
\Photon(20,35)(80,35){5}{5}
\DashArrowLine(115,70)(80,35){5}
\DashArrowLine(80,35)(115,0){5}
\Text(90,55)[]{\large $p$}
\Text(90,12)[]{\large $p'$}
\Text(120,70)[l]{\large $\mr{\dnt}_{i\al}$}
\Text(15,35)[r]{\large $\mr{\ga}$}
\Text(120,0)[l]{\large $\mr{\dnt}_{i\al}$}
\Text(250,35)[]{\large$-iee_d(p+p')^\mu$}
\Vertex(80,35){1}
\end{picture} \end{center}
\vskip 20mm
\begin{center} \begin{picture}(180,50)(100,40)
\SetScale{1.0}
\Photon(20,35)(80,35){5}{5}
\DashArrowLine(115,70)(80,35){5}
\DashArrowLine(80,35)(115,0){5}
\Text(90,55)[]{\large $p$}
\Text(90,12)[]{\large $p'$}
\Text(120,70)[l]{\large $\mr{\upt}_{i\al}$}
\Text(15,35)[r]{\large $\mr{W^-}$}
\Text(120,0)[l]{\large $\mr{\dnt}_{i\be}$}
\Text(250,35)[]{\large 
${\displaystyle\frac{-ig}{\rtt}Q^{2i}_{1\al}Q^{2i-1}_{1\be}(p+p')^\mu}$}
\Vertex(80,35){1}
\end{picture} \end{center}
\vskip 20mm
\begin{center} \begin{picture}(180,50)(100,40)
\SetScale{1.0}
\Photon(20,35)(80,35){5}{5}
\DashArrowLine(115,70)(80,35){5}
\DashArrowLine(80,35)(115,0){5}
\Text(90,55)[]{\large $p$}
\Text(90,12)[]{\large $p'$}
\Text(120,70)[l]{\large $\mr{\upt}_{i\al}$}
\Text(15,35)[r]{\large $\mr{Z^0}$}
\Text(120,0)[l]{\large $\mr{\upt}_{i\be}$}
\Text(250,35)[]{\large
${\displaystyle\frac{ig}{\cw}Z^{\al\be}_{u_i}(p+p')^\mu}$}
\Vertex(80,35){1}
\end{picture} \end{center}
\vskip 20mm
\begin{center} \begin{picture}(180,50)(100,40)
\SetScale{1.0}
\Photon(20,35)(80,35){5}{5}
\DashArrowLine(115,70)(80,35){5}
\DashArrowLine(80,35)(115,0){5}
\Text(90,55)[]{\large $p$}
\Text(90,12)[]{\large $p'$}
\Text(120,70)[l]{\large $\mr{\dnt}_{i\al}$}
\Text(15,35)[r]{\large $\mr{Z^0}$}
\Text(120,0)[l]{\large $\mr{\dnt}_{i\be}$}
\Text(250,35)[]{\large
${\displaystyle\frac{ig}{\cw}Z^{\al\be}_{d_i}(p+p')^\mu}$}
\Vertex(80,35){1}
\end{picture} \end{center}
\vskip 20mm
\captionB{Feynman rules for the interactions of the squarks and the gauge
 	 bosons.}
{Feynman rules for the interactions of the squarks and the gauge bosons.
 The couplings of the W and Z to the squarks are given in Table~\ref{tab:Zcp}.
 The momenta of the outgoing squarks, $p$ and $p'$, should be taken in the
 direction of the arrows.}
\label{fig:Zsquarks}
\end{figure}

%
%  Feynman rules for the interactions of the gauge bosons and sleptons
%
\begin{figure}[p!]
\begin{center} \begin{picture}(180,50)(100,40)
\SetScale{1.0}
\Photon(20,35)(80,35){5}{5}
\DashArrowLine(115,70)(80,35){5}
\DashArrowLine(80,35)(115,0){5}
\Text(90,55)[]{\large $p$}
\Text(90,12)[]{\large $p'$}
\Text(120,70)[l]{\large $\elt_{i\al}$}
\Text(15,35)[r]{\large $\ga$}
\Text(120,0)[l]{\large $\elt_{i\al}$}
\Text(250,35)[]{\large$ie(p+p')^\mu$}
\Vertex(80,35){1}
\end{picture} \end{center}
\vskip 20mm
\begin{center} \begin{picture}(180,50)(100,40)
\SetScale{1.0}
\Photon(20,35)(80,35){5}{5}
\DashArrowLine(115,70)(80,35){5}
\DashArrowLine(80,35)(115,0){5}
\Text(90,55)[]{\large $p$}
\Text(90,12)[]{\large $p'$}
\Text(120,70)[l]{\large $\nut_i$}
\Text(15,35)[r]{\large $\mr{W^-}$}
\Text(120,0)[l]{\large $\elt_{i\al}$}
\Text(250,35)[]{\large
${\displaystyle\frac{-ig}{\rtt}L^{2i-1}_{1\al}(p+p')^\mu}$}
\Vertex(80,35){1}
\end{picture} \end{center}
\vskip 20mm
\begin{center} \begin{picture}(180,50)(100,40)
\SetScale{1.0}
\Photon(20,35)(80,35){5}{5}
\DashArrowLine(115,70)(80,35){5}
\DashArrowLine(80,35)(115,0){5}
\Text(90,55)[]{\large $p$}
\Text(90,12)[]{\large $p'$}
\Text(120,70)[l]{\large $\nut_{i\al}$}
\Text(15,35)[r]{\large $\mr{Z^0}$}
\Text(120,0)[l]{\large $\nut_{i\be}$}
\Text(250,35)[]{\large
${\displaystyle\frac{ig}{\cw}Z^{\al\be}_{\nu_i}(p+p')^\mu}$}
\Vertex(80,35){1}
\end{picture} \end{center}
\vskip 20mm
\begin{center} \begin{picture}(180,50)(100,40)
\SetScale{1.0}
\Photon(20,35)(80,35){5}{5}
\DashArrowLine(115,70)(80,35){5}
\DashArrowLine(80,35)(115,0){5}
\Text(90,55)[]{\large $p$}
\Text(90,12)[]{\large $p'$}
\Text(120,70)[l]{\large $\elt_{i\al}$}
\Text(15,35)[r]{\large $\mr{Z^0}$}
\Text(120,0)[l]{\large $\elt_{i\be}$}
\Text(250,35)[]{\large
${\displaystyle\frac{ig}{\cw}Z^{\al\be}_{e_i}(p+p')^\mu}$}
\Vertex(80,35){1}
\end{picture} \end{center}
\vskip 30mm
\captionB{Feynman rules for the interactions of the sleptons and the gauge
	 bosons.}
{Feynman rules for the interactions of the sleptons and the gauge bosons.
 The couplings of the sleptons to the gauge bosons are given in 
 Table~\ref{tab:Zcp} and the momenta of the outgoing sleptons, $p$ and $p'$,
 should be taken in the directions of the arrows.}
\label{fig:Zsleptons}
\end{figure}
%
% Feynman rules for the interactions of gauge bosons and the quarks
%
\begin{figure}
\vskip -15mm
\begin{center} \begin{picture}(180,45)(100,40)
\SetScale{1.0}
\Photon(14,24.5)(56,24.5){4}{4}
\ArrowLine(80.5,49)(56,24.5)
\ArrowLine(56,24.5)(80.5,0)
\Text(85,49)[l]{\large $\mr{\bar{u}}$}
\Text(9,24.5)[r]{\large $\mr{\ga}$}
\Text(85,0)[l]{\large $\mr{u}$}
\Text(250,24.5)[]{\large${\displaystyle-iee_u\ga^\mu}$}
\Vertex(56,24.5){1}
\end{picture} \end{center}
\vskip 6mm
\begin{center} \begin{picture}(180,45)(100,40)
\SetScale{1.0}
\Photon(14,24.5)(56,24.5){4}{4}
\ArrowLine(80.5,49)(56,24.5)
\ArrowLine(56,24.5)(80.5,0)
\Text(85,49)[l]{\large $\mr{\bar{d}}$}
\Text(9,24.5)[r]{\large $\mr{\ga}$}
\Text(85,0)[l]{\large $\mr{d}$}
\Text(250,24.5)[]{\large${\displaystyle-iee_d\ga^\mu}$}
\Vertex(56,24.5){1}
\end{picture} \end{center}
\vskip 6mm
\begin{center} \begin{picture}(180,45)(100,40)
\SetScale{1.0}
\Photon(14,24.5)(56,24.5){4}{4}
\ArrowLine(80.5,49)(56,24.5)
\ArrowLine(56,24.5)(80.5,0)
\Text(85,49)[l]{\large $\mr{\ell^+}$}
\Text(9,24.5)[r]{\large $\mr{\ga}$}
\Text(85,0)[l]{\large $\mr{\ell^-}$}
\Text(250,24.5)[]{\large${\displaystyle ie\ga^\mu}$}
\Vertex(56,24.5){1}
\end{picture} \end{center}
\vskip 6mm
\begin{center} \begin{picture}(180,45)(100,40)
\SetScale{1.0}
\Photon(14,24.5)(56,24.5){4}{4}
\ArrowLine(80.5,49)(56,24.5)
\ArrowLine(56,24.5)(80.5,0)
\Text(85,49)[l]{\large $\mr{\bar{u}}$}
\Text(9,24.5)[r]{\large $\mr{W^-}$}
\Text(85,0)[l]{\large $\mr{d}$}
\Text(250,24.5)[]{\large${\displaystyle\frac{-ig}{2\rtt}\ga^\mu(1-\ga_5)}$}
\Vertex(56,24.5){1}
\end{picture} \end{center}
\vskip 6mm
\begin{center} \begin{picture}(180,45)(100,40)
\SetScale{1.0}
\Photon(14,24.5)(56,24.5){4}{4}
\ArrowLine(80.5,49)(56,24.5)
\ArrowLine(56,24.5)(80.5,0)
\Text(85,49)[l]{\large $\mr{\bar{\nu}}$}
\Text(9,24.5)[r]{\large $\mr{W^-}$}
\Text(85,0)[l]{\large $\mr{\ell^-}$}
\Text(250,24.5)[]{\large${\displaystyle\frac{-ig}{2\rtt}\ga^\mu(1-\ga_5)}$}
\Vertex(56,24.5){1}
\end{picture} \end{center}
\vskip 6mm
\begin{center} \begin{picture}(180,45)(100,40)
\SetScale{1.0}
\Photon(14,24.5)(56,24.5){4}{4}
\ArrowLine(80.5,49)(56,24.5)
\ArrowLine(56,24.5)(80.5,0)
\Text(85,49)[l]{\large $\mr{\bar{u}}$}
\Text(9,24.5)[r]{\large $\mr{Z^0}$}
\Text(85,0)[l]{\large $\mr{u}$}
\Text(250,24.5)[]{\large
${\displaystyle\frac{ig}{\cw}\ga^\mu\left[ Z_{u_L}(1-\ga_5) +
			 Z_{u_R}(1+\ga_5)\right]}$}
\Vertex(56,24.5){1}
\end{picture} \end{center}
\vskip 6mm
\begin{center} \begin{picture}(180,45)(100,40)
\SetScale{1.0}
\Photon(14,24.5)(56,24.5){4}{4}
\ArrowLine(80.5,49)(56,24.5)
\ArrowLine(56,24.5)(80.5,0)
\Text(85,49)[l]{\large $\mr{\bar{d}}$}
\Text(9,24.5)[r]{\large $\mr{Z^0}$}
\Text(85,0)[l]{\large $\mr{d}$}
\Text(250,24.5)[]{\large
${\displaystyle\frac{ig}{\cw}\ga^\mu\left[ Z_{d_L}(1-\ga_5) +
			 Z_{d_R}(1+\ga_5)\right]}$}
\Vertex(56,24.5){1}
\end{picture} \end{center}
\vskip 6mm
\begin{center} \begin{picture}(180,45)(100,40)
\SetScale{1.0}
\Photon(14,24.5)(56,24.5){4}{4}
\ArrowLine(80.5,49)(56,24.5)
\ArrowLine(56,24.5)(80.5,0)
\Text(85,49)[l]{\large $\mr{\bar{\nu}}$}
\Text(9,24.5)[r]{\large $\mr{Z^0}$}
\Text(85,0)[l]{\large $\mr{\nu}$}
\Text(250,24.5)[]{\large
${\displaystyle\frac{ig}{\cw}\ga^\mu Z_{\nu_L}(1-\ga_5)}$}
\Vertex(56,24.5){1}
\end{picture} \end{center}
\vskip 6mm
\begin{center} \begin{picture}(180,45)(100,40)
\SetScale{1.0}
\Photon(14,24.5)(56,24.5){4}{4}
\ArrowLine(80.5,49)(56,24.5)
\ArrowLine(56,24.5)(80.5,0)
\Text(85,49)[l]{\large $\mr{\ell^+}$}
\Text(9,24.5)[r]{\large $\mr{Z^0}$}
\Text(85,0)[l]{\large $\mr{\ell^-}$}
\Text(250,24.5)[]{\large
${\displaystyle\frac{ig}{\cw}\ga^\mu\left[ Z_{\ell_L}(1-\ga_5) +
			 Z_{\ell_R}(1+\ga_5)\right]}$}
\Vertex(56,24.5){1}
\end{picture} \end{center}
\vskip 12mm
\captionB{Feynman rules for the interactions of the quarks and the gauge
	 bosons.}
	{Feynman rules for the interactions of the quarks and the gauge
	 bosons.}
\label{fig:Zquarks}
\end{figure}
%
%  Now the Feynman rules for Higgs bosons
%
\section[Higgs Boson Interactions with the Sfermions and Fermions]
	{Higgs Boson Interactions with the Sfermions and Fermions}

  The Lagrangian for the interaction of the Higgs bosons of the MSSM with the
  sfermions is given in \cite{Gunion:1986yn}. This Lagrangian is given below
  without left/right sfermion mixing,\footnote{We have neglected the	
  			terms arising from the presence of
			an additional Higgs singlet.}
  this is taken \nopagebreak from Eqn.\,4.19 of \cite{Gunion:1986yn},        
\begin{equation}
\begin{split}
\mathcal{L}_{\mr{H\qkt\qkt^*}} 
&= -\frac{g\mz}{\cw}\left[\rule{0mm}{4mm}H^0_1\cos(\al+\be)
	-H^0_2\sin(\al+\be)\right]
\\[1.5mm]
&\left[ \rule{0mm}{7mm}\left(\frac{1}{2}-e_u\ssw\right)\upt^*_L\upt_L
				    +e_u\ssw \upt^*_R\upt_R \right. \\[1.5mm]
&	\left. -\left(\frac{1}{2}+e_d\ssw\right)\dnt^*_L\dnt_L
		    +e_d\ssw \dnt^*_R\dnt_R \rule{0mm}{7mm}\right] \\[1.5mm]
 & -\frac{gm^2_d}{\mw\cbe}\left(\dnt^*_L\dnt_L+\dnt^*_R\dnt_R\right)
  	\left(H^0_1\cos\al-H^0_2\sin\al\right) \\[1.5mm]
 & -\frac{gm^2_u}{\mw\sbe}\left(\upt^*_L\upt_L+\upt^*_R\upt_R\right)
  	\left(H^0_1\sin\al+H^0_2\cos\al\right)\\[1.5mm]
 & -\frac{gm_d}{2\mw\cbe}\left(\dnt^*_R\dnt_L+\dnt^*_L\dnt_R\right)
    \left[\rule{0mm}{4mm}\left(A_d\cos\al-\mu\sin\al\right)H^0_1
	-\left(\mu\cos\al+A_d\sin\al\right)H^0_2\right]\\[1.5mm]
 & -\frac{gm_u}{2\mw\sbe}\left(\upt^*_R\upt_L+\upt^*_L\upt_R\right)
    \left[\rule{0mm}{4mm}\left(A_u\sin\al-\mu\cos\al\right)H^0_1
	+\left(\mu\sin\al+A_u\cos\al\right)H^0_2\right]\\[1.5mm]
 & -\frac{igm_d}{2\mw}\left(A_d\tan\beta+\mu\right)
     \left(\dnt^*_R\dnt_L-\dnt^*_L\dnt_R\right)H^0_3\\[1.5mm]
 &-\frac{igm_u}{2\mw}\left(A_u\cot\beta+\mu\right)
     \left(\upt^*_R\upt_L-\upt^*_L\upt_R\right)H^0_3\\[1.5mm]
 &+\frac{g}{\rtt \mw}\left(m^2_d\tan\be+m^2_u\cot\be-\mw^2\sin2\be \right)
    \left(H^+\upt^*_L\dnt_L+H^-\dnt^*_L\upt_L\right)\\[1.5mm]
 &+\frac{gm_um_d\left(\cot\be+\tan\be\right)}{\rtt \mw}
    \left(H^+\upt^*_R\dnt_R+H^-\dnt^*_R\upt_R\right)\\[1.5mm]
 &+\frac{gm_d}{\rtt \mw}\left(A_d\tan\be+\mu\right)
    \left(H^+\upt^*_L\dnt_R+H^-\dnt^*_R\upt_L\right)\\[1.5mm]
 &+\frac{gm_u}{\rtt \mw}\left(A_d\tan\be+\mu\right)
    \left(H^+\upt^*_R\dnt_L+H^-\dnt^*_L\upt_R\right)\!,
\end{split}
\end{equation}
  where $\al$ is the mixing angle for the CP-even Higgs bosons,
  $H^2_0$ is the lighter CP-even Higgs boson field, $H^1_0$ is the heavier
  CP-even Higgs boson field, $H^3_0$ is the pseudo-scalar Higgs boson field
  and all the other terms have been  defined previously. As before, this
  Lagrangian is only for one generation of squarks.

%
%  Higgs Feynman rules
%  
\begin{figure}
\vspace{-8mm}
\begin{center} \begin{picture}(180,50)(100,40)
\SetScale{1.0}
\DashLine(20,35)(80,35){5}
\DashLine(80,35)(115,70){5}
\DashLine(115,0)(80,35){5}
\Vertex(80,35){1}
\Text(120,70)[l]{\large $\mr{\upt}^*_{i\be}$}
\Text(50,45)[]{\large $\mr{h_0}$}
\Text(120,0)[l]{\large $\mr{\upt}_{i\al}$}
\Text(300,35)[]{\large $igH^1_{\upt_{i\al}\upt_{i\be}}$}
\end{picture} \end{center}
\vskip 10mm
\begin{center} \begin{picture}(180,50)(100,40)
\SetScale{1.0}
\DashLine(20,35)(80,35){5}
\DashLine(80,35)(115,70){5}
\DashLine(115,0)(80,35){5}
\Vertex(80,35){1}
\Text(120,70)[l]{\large $\mr{\dnt}^*_{i\be}$}
\Text(50,45)[]{\large $\mr{h_0}$}
\Text(120,0)[l]{\large $\mr{\dnt}_{i\al}$}
\Text(300,35)[]{\large $igH^1_{\dnt_{i\al}\dnt_{i\be}}$}
\end{picture} \end{center}
\vskip 10mm
\begin{center} \begin{picture}(180,50)(100,40)
\SetScale{1.0}
\DashLine(20,35)(80,35){5}
\DashLine(80,35)(115,70){5}
\DashLine(115,0)(80,35){5}
\Vertex(80,35){1}
\Text(120,70)[l]{\large $\mr{\upt}^*_{i\be}$}
\Text(50,45)[]{\large $\mr{H_0}$}
\Text(120,0)[l]{\large $\mr{\upt}_{i\al}$}
\Text(300,35)[]{\large $igH^2_{\upt_{i\al}\upt_{i\be}}$}
\end{picture} \end{center}
\vskip 10mm
\begin{center} \begin{picture}(180,50)(100,40)
\SetScale{1.0}
\DashLine(20,35)(80,35){5}
\DashLine(80,35)(115,70){5}
\DashLine(115,0)(80,35){5}
\Vertex(80,35){1}
\Text(120,70)[l]{\large $\mr{\dnt}^*_{i\be}$}
\Text(50,45)[]{\large $\mr{H_0}$}
\Text(120,0)[l]{\large $\mr{\dnt}_{i\al}$}
\Text(300,35)[]{\large $igH^2_{\dnt_{i\al}\dnt_{i\be}}$}
\end{picture} \end{center}
\vskip 10mm
\begin{center} \begin{picture}(180,50)(100,40)
\SetScale{1.0}
\DashArrowLine(20,35)(80,35){5}
\DashArrowLine(80,35)(115,70){5}
\DashArrowLine(115,0)(80,35){5}
\Vertex(80,35){1}
\Text(120,70)[l]{\large $\mr{\upt}_{i\be}$}
\Text(50,45)[]{\large $\mr{A_0}$}
\Text(120,0)[l]{\large $\mr{\upt}^*_{i\al}$}
\Text(300,35)[]{\large $gH^3_{\upt_{i\al}\upt_{i\be}}$}
\end{picture} \end{center}
\vskip 10mm
\begin{center} \begin{picture}(180,50)(100,40)
\SetScale{1.0}
\DashArrowLine(20,35)(80,35){5}
\DashArrowLine(80,35)(115,70){5}
\DashArrowLine(115,0)(80,35){5}
\Vertex(80,35){1}
\Text(120,70)[l]{\large $\mr{\dnt}_{i\be}$}
\Text(50,45)[]{\large $\mr{A_0}$}
\Text(120,0)[l]{\large $\mr{\dnt}^*_{i\al}$}
\Text(300,35)[]{\large $gH^3_{\dnt_{i\al}\dnt_{i\be}}$}
\end{picture} \end{center}
\vskip 10mm
\begin{center} \begin{picture}(180,50)(100,40)
\SetScale{1.0}
\DashArrowLine(20,35)(80,35){5}
\DashArrowLine(80,35)(115,70){5}
\DashArrowLine(115,0)(80,35){5}
\Vertex(80,35){1}
\Text(120,70)[l]{\large $\mr{\upt}_{i\al}$}
\Text(50,45)[]{\large $\mr{H^+}$}
\Text(120,0)[l]{\large $\mr{\dnt}^*_{i\be}$}
\Text(300,35)[]{\large $igH^c_{\upt_{i\al}\dnt_{i\be}}$}
\end{picture} \end{center}
\vskip 10mm
\captionB{Feynman rules for the interactions of the Higgs bosons and the
	 squarks.}
	{Feynman rules for the interactions of the Higgs bosons and the
	 squarks. The sign of the $\mr{A_0}\mr{\qkt}_{i\al}\mr{\qkt}_{i\be}$
	 vertex changes under a change of sign of the squark momenta. The 
	 couplings of the Higgs bosons to the squarks are given in
	 Table.~\ref{tab:UDDhiggs}.}
\label{fig:Higgssquark}
\end{figure}\begin{table}[htp]
\renewcommand{\arraystretch}{1.5}
\begin{center}
\begin{tabular}{|l|l|} \hline
 Coefficient & Coupling  \\
\hline
  $H^1_{\dnt_{i\al}\dnt_{i\be}}$ &
  $ -\frac{\mz\sin(\al+\be)}{\cw} \left[ 
    Q^{2i-1}_{1\al}Q^{2i-1}_{1\be}( \frac{1}{2}+e_d\ssw )
    -e_d \ssw Q^{2i-1}_{2\al}Q^{2i-1}_{2\be} \right]$ \\
 & $ +\frac{m_{d_i}^2\sa}{\mw \cbe}\left[
     Q^{2i-1}_{1\al}Q^{2i-1}_{1\be}+ Q^{2i-1}_{2\al}Q^{2i-1}_{2\be} 
 \right] $ \\
 & $ +\frac{m_{d_i}}{2\mw \cbe}\left( A_{d_i}\sa+\mu\ca\right)
    \left[ Q^{2i-1}_{2\al}Q^{2i-1}_{1\be}+
Q^{2i-1}_{1\al}Q^{2i-1}_{2\be}
\right] $  \\
\hline
  $H^1_{\upt_{i\al}\upt_{i\be}}$ &
  $ \frac{\mz\sin(\al+\be)}{\cw} \left[ 
    Q^{2i}_{1\al}Q^{2i}_{1\be}( \frac{1}{2}-e_u\ssw )
    +e_u \ssw Q^{2i}_{2\al}Q^{2i}_{2\be} \right]$ \\
 & $ -\frac{m_{u_i}^2\ca}{\mw \sbe}\left[
     Q^{2i}_{1\al}Q^{2i}_{1\be}+ Q^{2i}_{2\al}Q^{2i}_{2\be}  \right] $ \\
 & $ -\frac{m_{u_i}}{2\mw \sbe}\left(A_{u_i}\ca +\mu\sa\right)
    \left[ Q^{2i}_{2\al}Q^{2i}_{1\be}+ Q^{2i}_{1\al}Q^{2i}_{2\be}\right] $  \\
\hline
  $H^2_{\dnt_{i\al}\dnt_{i\be}}$ &
  $ \frac{\mz\cos(\al+\be)}{\cw} \left[ 
    Q^{2i-1}_{1\al}Q^{2i-1}_{1\be}( \frac{1}{2}+e_d\ssw )
    -e_d\ssw Q^{2i-1}_{2\al}Q^{2i-1}_{2\be} \right]$ \\
 & $ -\frac{m_{d_i}^2\ca}{\mw \cbe}\left[
     Q^{2i-1}_{1\al}Q^{2i-1}_{1\be}+ Q^{2i-1}_{2\al}Q^{2i-1}_
{2\be} \right] $ \\
 & $ +\frac{m_{d_i}}{2\mw \cbe}\left( \mu\sa-A_{d_i}\ca\right)
    \left[ Q^{2i-1}_{2\al}Q^{2i-1}_{1\be}+
Q^{2i-1}_{1\al}Q^{2i-1}_{2\be} \right] $  \\
\hline
  $H^2_{\upt_{i\al}\upt_{i\be}}$ &
  $ -\frac{\mz\cos(\al+\be)}{\cw} \left[ 
    Q^{2i}_{1\al}Q^{2i}_{1\be}( \frac{1}{2}-e_u\ssw )
    +e_u \ssw Q^{2i}_{2\al}Q^{2i}_{2\be} \right]$ \\
 & $ -\frac{m_{u_i}^2\sa}{\mw \sbe}\left[
     Q^{2i}_{1\al}Q^{2i}_{1\be}+ Q^{2i}_{2\al}Q^{2i}_{2\be}  \right] $ \\
 & $ -\frac{m_{u_i}}{2\mw \sbe}\left( A_{u_i}\sa-\mu\ca\right)
    \left[ Q^{2i}_{2\al}Q^{2i}_{1\be}+Q^{2i}_{1\al}Q^{2i}_{2\be}\right] $  \\
\hline
  $H^3_{\dnt_{i\al}\dnt_{i\be}}$ &
  $ \de_{\al\neq\be}\frac{m_{d_i}}{2\mw}\left(A_{d_i}\tan\beta+\mu\right) $\\
\hline
  $H^3_{\upt_{i\al}\upt_{i\be}}$ &
  $ \de_{\al\neq\be}\frac{m_{u_i}}{2\mw}\left(A_{u_i}\cot\beta+\mu\right) $\\
\hline
  $H^c_{\upt_{i\al}\dnt_{i\be}}$ &
   $\frac{1}{\rtt \mw}\left[Q^{2i}_{1\al}Q^{2i-1}_{1\be} \left(
      m_{d_i}^2\tan\beta +m_{u_i}^2\cot\beta-\mw^2\sin2\beta\right)\right.$\\
 & $+Q^{2i}_{2\al}Q^{2i-1}_{2\be}m_{u_i}m_{d_i}\left(\cot
\beta+\tan\beta\right)$ \\
 & $ \left.+Q^{2i}_{1\al}Q^{2i-1}_{2\be}m_{d_i} \left(A_{d_i}\tan
\beta +\mu \right)
     +Q^{2i}_{2\al}Q^{2i-1}_{1\be}m_{u_i} \left( A_{u_i}\cot\beta+\mu 
\right) \right]$\\
\hline
\end{tabular}
\captionB{Higgs couplings to the squarks.}{Higgs couplings to the squarks.}
\label{tab:UDDhiggs}
\end{center}
%\end{table}
%\begin{table}[htp]
\begin{center}
\begin{tabular}{|l|l|}
\hline
 Coefficient & Coupling  \\
\hline
  $H^1_{\elt_{i\al}\elt_{i\be}}$ &
  $ -\frac{\mz\sin(\al+\be)}{\cw} \left[ 
    L^{2i-1}_{1\al}L^{2i-1}_{1\be}( \frac{1}{2}-\ssw )
    +\ssw L^{2i-1}_{2\al}L^{2i-1}_{2\be} \right]$ \\
 & $ +\frac{m_{\ell_i}^2\sa}{\mw \cbe}\left[
     L^{2i-1}_{1\al}L^{2i-1}_{1\be}+ L^{2i-1}_{2\al}L^{2i-1}_{2\be}
 \right] $ \\
 & $ +\frac{m_{\ell_i}}{2\mw \cbe}\left( \mu\ca+A_{e_i}\sa\right)
    \left[ L^{2i-1}_{2\al}L^{2i-1}_{1\be}+ L^{2i-1}_{1\al}L^{2i-1}_{2\be}
 \right] $  \\
\hline
  $H^1_{\nut_i\nut_i}$ &
  $ \frac{\mz\sin(\al+\be)}{2\cw}$ \\
\hline
  $H^2_{\elt_{i\al}\elt_{i\be}}$ &
  $ \frac{\mz\cos(\al+\be)}{\cw} \left[ 
    L^{2i-1}_{1\al}L^{2i-1}_{1\be}( \frac{1}{2}-\ssw )
    +\ssw L^{2i-1}_{2\al}L^{2i-1}_{2\be} \right]$ \\
 & $ -\frac{m_{\ell_i}^2\ca}{\mw \cbe}\left[
     L^{2i-1}_{1\al}L^{2i-1}_{1\be}+ L^{2i-1}_{2\al}L^{2i-1}_{2\be}
 \right] $ \\
 & $ +\frac{m_{\ell_i}}{2\mw \cbe}\left(\mu\sa- A_{e_i}\ca\right)
    \left[ L^{2i-1}_{2\al}L^{2i-1}_{1\be}+ L^{2i-1}_{1\al}L^{2i-1}_{2\be}
 \right] $  \\
\hline
  $H^2_{\nut_i\nut_i}$ &
  $ -\frac{\mz\cos(\al+\be)}{2\cw}$ \\
\hline
  $H^3_{\elt_{i\al}\elt_{i\be}}$ &
  $ \de_{\al\neq\be}\frac{m_{\ell_i}}{2\mw}
	\left(A_{e_i}\tan\beta+\mu\right)$  \\
\hline
  $H^c_{\nut\elt_{i\al}}$ &
   $\frac{1}{\rtt \mw}\left[
	L^{2i-1}_{1\al}\left(m_{\ell_i}^2\tan\beta-\mw^2\sin2\beta\right)
   	+L^{2i-1}_{2\al}m_{\ell_i}
	\left(A_{e_i}\tan\beta +\mu\right)\right]$\\
\hline
\end{tabular}
\captionB{Higgs couplings to the sleptons.}{Higgs couplings to the sleptons.} 
\label{tab:LQDhiggs}
\end{center}
\end{table}\begin{figure}
\begin{center} \begin{picture}(180,50)(100,40)
\SetScale{1.0}
\DashLine(20,35)(80,35){5}
\DashLine(80,35)(115,70){5}
\DashLine(115,0)(80,35){5}
\Vertex(80,35){1}
\Text(120,70)[l]{\large $\mr{\nut}^*_{i}$}
\Text(50,45)[]{\large $\mr{h_0}$}
\Text(120,0)[l]{\large $\mr{\nut}_{i}$}
\Text(300,35)[]{\large $igH^1_{\nut_{i}\nut_{i}}$}
\end{picture} \end{center}
\vskip 10mm
\begin{center} \begin{picture}(180,50)(100,40)
\SetScale{1.0}
\DashLine(20,35)(80,35){5}
\DashLine(80,35)(115,70){5}
\DashLine(115,0)(80,35){5}
\Vertex(80,35){1} 
\Text(120,70)[l]{\large $\mr{\elt}^*_{i\be}$}
\Text(50,45)[]{\large $\mr{h_0}$}
\Text(120,0)[l]{\large $\mr{\elt}_{i\al}$}
\Text(300,35)[]{\large $igH^1_{\elt_{i\al}\elt_{i\be}}$}
\end{picture} \end{center}
\vskip 10mm
\begin{center} \begin{picture}(180,50)(100,40)
\SetScale{1.0}
\DashLine(20,35)(80,35){5}
\DashLine(80,35)(115,70){5}
\DashLine(115,0)(80,35){5}
\Vertex(80,35){1}
\Text(120,70)[l]{\large $\mr{\nut}^*_{i\be}$}
\Text(50,45)[]{\large $\mr{H_0}$}
\Text(120,0)[l]{\large $\mr{\nut}_{i\al}$}
\Text(300,35)[]{\large $igH^2_{\nut_{i\al}\nut_{i\be}}$}
\end{picture} \end{center}
\vskip 10mm
\begin{center} \begin{picture}(180,50)(100,40)
\SetScale{1.0}
\DashLine(20,35)(80,35){5}
\DashLine(80,35)(115,70){5}
\DashLine(115,0)(80,35){5}
\Vertex(80,35){1}
\Text(120,70)[l]{\large $\mr{\elt}^*_{i\be}$}
\Text(50,45)[]{\large $\mr{H_0}$}
\Text(120,0)[l]{\large $\mr{\elt}_{i\al}$}
\Text(300,35)[]{\large $igH^2_{\elt_{i\al}\elt_{i\be}}$}
\end{picture} \end{center}
\vskip 10mm
\begin{center} \begin{picture}(180,50)(100,40)
\SetScale{1.0}
\DashArrowLine(20,35)(80,35){5}
\DashArrowLine(80,35)(115,70){5}
\DashArrowLine(115,0)(80,35){5}
\Vertex(80,35){1}
\Text(120,70)[l]{\large $\mr{\elt}_{i\be}$}
\Text(50,45)[]{\large $\mr{A_0}$}
\Text(120,0)[l]{\large $\mr{\elt}^*_{i\al}$}
\Text(300,35)[]{\large $gH^3_{\elt_{i\al}\elt_{i\be}}$}
\end{picture} \end{center}
\vskip 10mm
\begin{center} \begin{picture}(180,50)(100,40)
\SetScale{1.0}
\DashArrowLine(20,35)(80,35){5}
\DashArrowLine(80,35)(115,70){5}
\DashArrowLine(115,0)(80,35){5}
\Vertex(80,35){1}
\Text(120,70)[l]{\large $\mr{\nut}_{i\al}$}
\Text(50,45)[]{\large $\mr{H^+}$}
\Text(120,0)[l]{\large $\mr{\elt}^*_{i\be}$}
\Text(300,35)[]{\large $igH^c_{\nut_{i\al}\elt_{i\be}}$}
\end{picture} \end{center}
\vskip 10mm
\captionB{Feynman rules for the interactions of the Higgs bosons and the
	 sleptons.}
	{Feynman rules for the interactions of the Higgs bosons and the
	 sleptons. The sign of the 
	$\mr{A_0}\tilde{\ell}_{i\al}\tilde{\ell}_{i\be}$ vertex changes under
	 a change of sign of the slepton momenta. The couplings of the Higgs
	 bosons to the sleptons are given in Table.~\ref{tab:LQDhiggs}.}
\label{fig:Higgsslepton}
\end{figure}\begin{figure}
\begin{center} \begin{picture}(180,50)(100,40)
\SetScale{1.0}
\DashLine(20,35)(80,35){5}
\ArrowLine(80,35)(115,70)
\ArrowLine(115,0)(80,35)
\Vertex(80,35){1}
\Text(120,70)[l]{\large $\mr{u}_i$}
\Text(50,45)[]{\large $\mr{h_0}$}
\Text(120,0)[l]{\large $\mr{\bar{u}}_i$}
\Text(300,35)[]{\large $-igU^1_i$}
\end{picture} \end{center}
\vskip 10mm
\begin{center} \begin{picture}(180,50)(100,40)
\SetScale{1.0}
\DashLine(20,35)(80,35){5}
\ArrowLine(80,35)(115,70)
\ArrowLine(115,0)(80,35)
\Vertex(80,35){1}
\Text(120,70)[l]{\large $\mr{d}_i$}
\Text(50,45)[]{\large $\mr{h_0}$}
\Text(120,0)[l]{\large $\mr{\bar{d}}_i$}
\Text(300,35)[]{\large $-igD^1_i$}
\end{picture} \end{center}
\vskip 10mm
\begin{center} \begin{picture}(180,50)(100,40)
\SetScale{1.0}
\DashLine(20,35)(80,35){5}
\ArrowLine(80,35)(115,70)
\ArrowLine(115,0)(80,35)
\Vertex(80,35){1}
\Text(120,70)[l]{\large $\mr{u}_i$}
\Text(50,45)[]{\large $\mr{H_0}$}
\Text(120,0)[l]{\large $\mr{\bar{u}}_i$}
\Text(300,35)[]{\large $-igU^2_i$}
\end{picture} \end{center}
\vskip 10mm
\begin{center} \begin{picture}(180,50)(100,40)
\SetScale{1.0}
\DashLine(20,35)(80,35){5}
\ArrowLine(80,35)(115,70)
\ArrowLine(115,0)(80,35)
\Vertex(80,35){1}
\Text(120,70)[l]{\large $\mr{d}_i$}
\Text(50,45)[]{\large $\mr{H_0}$}
\Text(120,0)[l]{\large $\mr{\bar{d}}_i$}
\Text(300,35)[]{\large $-igD^2_i$}
\end{picture} \end{center}
\vskip 10mm
\begin{center} \begin{picture}(180,50)(100,40)
\SetScale{1.0}
\DashLine(20,35)(80,35){5}
\ArrowLine(80,35)(115,70)
\ArrowLine(115,0)(80,35)
\Vertex(80,35){1}
\Text(120,70)[l]{\large $\mr{u}_i$}
\Text(50,45)[]{\large $\mr{A_0}$}
\Text(120,0)[l]{\large $\mr{\bar{u}}_i$}
\Text(300,35)[]{\large $-gU^3_i\ga_5$}
\end{picture} \end{center}
\vskip 10mm
\begin{center} \begin{picture}(180,50)(100,40)
\SetScale{1.0}
\DashLine(20,35)(80,35){5}
\ArrowLine(80,35)(115,70)
\ArrowLine(115,0)(80,35)
\Vertex(80,35){1}
\Text(120,70)[l]{\large $\mr{d}_i$}
\Text(50,45)[]{\large $\mr{A_0}$}
\Text(120,0)[l]{\large $\mr{\bar{d}}_i$}
\Text(300,35)[]{\large $-gD^3_i\ga_5$}
\end{picture} \end{center}
\vskip 10mm
\begin{center} \begin{picture}(180,50)(100,40)
\SetScale{1.0}
\DashArrowLine(20,35)(80,35){5}
\ArrowLine(80,35)(115,70)
\ArrowLine(115,0)(80,35)
\Vertex(80,35){1}
\Text(120,70)[l]{\large $\mr{u}_i$}
\Text(50,45)[]{\large $\mr{H^+}$}
\Text(120,0)[l]{\large $\mr{\bar{d}}_i$}
\Text(300,35)[]{\large $ig\left[D^c_i\left(1+\ga_5\right)
			+U^c_i\left(1-\ga_5\right)\right]$}
\end{picture} \end{center}
\vskip 10mm
\captionB{Feynman rules for the interactions of the Higgs bosons and the
	 quarks.}
	{Feynman rules for the interactions of the Higgs bosons and the
	 quarks. The couplings of the Higgs bosons to the quarks are given in 
	 Table.~\ref{tab:higgsqk}.}
\label{fig:Higgsquark}
\end{figure}

  We can express this Lagrangian in terms of the squark mass eigenstates
giving
\begin{eqnarray}
\mathcal{L}_{\mr{H\qkt\qkt^*}}
 &=& -\frac{g\mz}{\cw}\left[H^0_1\cos(\al+\be)-H^0_2\sin(\al+\be)\right]
\nonumber\\[1.5mm]
&& \left[\rule{0mm}{9mm}\left\{\rule{0mm}{7mm}
\left(\frac{1}{2}-e_u\ssw\right)Q^{2i}_{1\al}Q^{2i}_{1\be}
+e_u\ssw Q^{2i}_{2\al}Q^{2i}_{2\be}\right\}\upt^*_{i\al}\upt_{i\be}\right.  
\nonumber\\[1.5mm]
 &&        \left.\rule{0mm}{9mm}-\left\{\rule{0mm}{7mm}
\left(\frac{1}{2}+e_d\ssw\right)Q^{2i-1}_{1\al}Q^{2i-1}_{1\be}
-e_d\ssw Q^{2i-1}_{2\al}Q^{2i-1}_{2\be} \right\}\dnt^*_{i\al}\dnt_{i\be}
\right]\nonumber\\[1.5mm]
 && -\frac{gm^2_d}{\mw\cbe}\left(Q^{2i-1}_{1\al}Q^{2i-1}_{1\be}
		+Q^{2i-1}_{2\al}Q^{2i-1}_{2\be} \right)
\left(H^0_1\cos\al-H^0_2\sin\al\right)\dnt^*_{i\al}\dnt_{i\be} 
\nonumber\\[1.5mm]
 && -\frac{gm^2_u}{\mw\sbe}\left(Q^{2i}_{1\al}Q^{2i}_{1\be}
			+Q^{2i}_{2\al}Q^{2i}_{2\be} \right)
  	\left(H^0_1\sin\al+H^0_2\cos\al\right)\upt^*_{i\al}\upt_{i\be}
\nonumber\\[1.5mm]
 && -\frac{gm_d}{2\mw\cbe}\left(Q^{2i-1}_{2\al}Q^{2i-1}_{1\be}+
 Q^{2i-1}_{1\al}Q^{2i-1}_{2\be}\right) \nonumber\\[1.5mm]
 &&   \left[\rule{0mm}{4mm}\left(A_{d_i}\cos\al-\mu\sin\al\right)H^0_1
-\left(\mu\cos\al+A_{d_i}\sin\al\right)H^0_2\right]\dnt^*_{i\al}\dnt_{i\be}
\nonumber\\[1.5mm]
 && -\frac{gm_u}{2\mw\sbe}\left(Q^{2i}_{2\al}Q^{2i}_{1\be}+
 Q^{2i}_{1\al}Q^{2i}_{2\be}\right)\nonumber\\[1.5mm]
 &&   \left[\rule{0mm}{7mm}\left(A_{u_i}\sin\al-\mu\cos\al\right)H^0_1
+\left(\mu\sin\al+A_{u_i}\cos\al\right)H^0_2\right]\upt^*_{i\al}\upt_{i\al}
\nonumber\\[1.5mm]
 && -\frac{igm_d}{2\mw}\left(A_{d_i}\tan\beta+\mu\right)
     \left(Q^{2i-1}_{2\al}Q^{2i-1}_{1\be}-Q^{2i-1}_{1\al}Q^{2i-1}_{2\be}
 \right)H^0_3
    \dnt^*_{i\al}\dnt_{i\be}\nonumber\\[1.5mm]
 && -\frac{igm_u}{2\mw}\left(A_{u_i}\cot\beta+\mu\right)
     \left(Q^{2i}_{2\al}Q^{2i}_{1\be}-Q^{2i}_{1\al}Q^{2i}_{2\be}  \right)H^0_3
 \upt^*_{i\al}\upt_{i\be}\nonumber\\[1.5mm]
 &&+\frac{g}{\rtt \mw}\left(m^2_d\tan\be+m^2_u\cot\be-\mw^2\sin2\be
 \right)\nonumber\\[1.5mm]
 &&   \left(H^+Q^{2i}_{1\al}Q^{2i-1}_{1\be}\upt^*_{i\al}\dnt_{i\be}
  	 +H^-Q^{2i-1}_{1\al}Q^{2i}_{1\be}\dnt^*_{i\al}\upt_{i\be}\right)
\nonumber\\[1.5mm]
 &&+\frac{gm_um_d\left(\cot\be+\tan\be\right)}{\rtt \mw}
    \left(H^+Q^{2i}_{2\al}Q^{2i-1}_{2\be}\upt^*_{i\al}\dnt_{i\be}
	 +H^-Q^{2i-1}_{2\al}Q^{2i}_{2\be}\dnt^*_{i\al}\upt_{i\be}\right)
\nonumber\\[1.5mm]
 &&+\frac{gm_d}{\rtt \mw}\left(A_{d_i}\tan\be+\mu\right)
    \left(H^+Q^{2i}_{1\al}Q^{2i-1}_{2\be}\upt^*_{i\al}\dnt_{i\be}
	 +H^-Q^{2i-1}_{2\al}Q^{2i}_{1\be}\dnt^*_{i\al}\upt_{i\be}\right)
\nonumber\\[1.5mm]
 &&+\frac{gm_u}{\rtt \mw}\left(A_{d_i}\tan\be+\mu\right)
    \left(H^+Q^{2i}_{2\al}Q^{2i-1}_{1\be}\upt^*_{i\al}\dnt_{i\be}
         +H^-Q^{2i-1}_{1\al}Q^{2i}_{2\be}\dnt^*_{i\al}\upt_{i\be}\right)\!,
\nonumber\\
\end{eqnarray}
  where there is an implied summation over the squark mass eigenstates.
  The $\al$ in the sines and cosines is the mixing angle for the CP-even Higgs
  bosons, whereas that appearing in the subscripts is the mass eigenstate of
  the squark. We can derive the Feynman rules for the interactions of the
  Higgs bosons with the squarks, Fig.\,\ref{fig:Higgssquark}, from this
  Lagrangian. The relevant couplings are given in Table~\ref{tab:UDDhiggs}.
  In the Feynman rules we have used the notation $\mr{h_0}$ for the lighter
  scalar Higgs, $\mr{H_0}$ for the heavier scalar Higgs
  and $\mr{A_0}$ for the pseudo-scalar Higgs. These 
  correspond to $\mr{H^2_0}$, $\mr{H^1_0}$
  and $\mr{H^3_0}$ respectively in the notation used above.
  The Feynman rules for the interactions of the sleptons can be derived by
  replacing the relevant couplings in the above Lagrangian. This gives the
  Feynman rules shown in Fig.\,\ref{fig:Higgsslepton}, with the couplings
  given in Table~\ref{tab:LQDhiggs}.

  Again for completeness we will include the Feynman rules for the interaction
  of the Higgs bosons of the MSSM with the fermions. The relevant Lagrangian
  from Eqn.\,4.10 of \cite{Gunion:1986yn} is,
\begin{eqnarray}
\mathcal{L}_{\mr{Hq\bar{q}}} &=& 
 -\frac{gm_u}{2\mw\sbe}
\left[\rule{0mm}{4mm}\bar{u}u\left(H^0_1\sin\al+H^0_2\cos\al\right)
				-i\bar{u}\ga_5uH^0_3\cbe\right]\nonumber \\
 &&-\frac{gm_d}{2\mw\cbe}
\left[\rule{0mm}{4mm}\bar{d}d\left(H^0_1\cos\al-H^0_2\sin\al\right)
				-i\bar{d}\ga_5dH^0_3\sbe\right]\\
 &&+\frac{g}{2\rtt \mw}\left
\{\rule{0mm}{5.5mm} H^+\bar{u}\left[
\rule{0mm}{4mm}\left(m_d\tan\be+m_u\cot\be\right)
+\left(m_d\tan\be-m_u\cot\be\right)\ga_5\right]d +\mr{h.c.}\right\}\!.
\nonumber
\end{eqnarray}
  This gives the Feynman rules shown in Fig.\,\ref{fig:Higgsquark}, using the
  couplings given in Table~\ref{tab:higgsqk}.

\begin{table}
\begin{center}
\renewcommand{\arraystretch}{2.5}
\begin{tabular}{|c|c|c|c|}
\hline
 ${\displaystyle U^1_i}$ & ${\displaystyle  \frac{m_{u_i}\ca}{2\mw \sbe}}$ &
 ${\displaystyle D^1_i}$ & ${\displaystyle -\frac{m_{d_i}\sa}{2\mw\cbe}}$ \\
 ${\displaystyle U^2_i}$ & ${\displaystyle  \frac{m_{u_i}\sa}{2\mw \sbe}}$ & 
 ${\displaystyle D^2_i}$ & ${\displaystyle  \frac{m_{d_i}\ca}{2\mw\cbe}}$ \\
 ${\displaystyle U^3_i}$ & ${\displaystyle  \frac{m_{u_i}\cot\be}{2\mw}}$ &
 ${\displaystyle D^3_i}$ & ${\displaystyle  \frac{m_{d_i}\tan\be}{2\mw}}$ \\
 ${\displaystyle U^c_i}$ & ${\displaystyle 
			 \frac{m_{u_i}\cot\be}{2\rtt \mw}}$ &
 ${\displaystyle D^c_i}$ & ${\displaystyle 
			 \frac{m_{d_i}\tan\be}{2\rtt \mw}}$ \\
\hline
\end{tabular}
\end{center}
\captionB{Higgs couplings to the quarks.}
	{Higgs couplings to the quarks.}
\label{tab:higgsqk}
\end{table}

%
%  Now the Feynman Rules for R-parity violation
%
\section[R-parity Violating Feynman Rules]
	{R-parity Violating Feynman Rules}

  Here we present the Feynman rules for R-parity violating supersymmetry
  with arbitrary left/right sfermion mixing.
  In general these terms in the superpotential, Eqn.\,\ref{eqn:Rsuper1},
  give  rise to a number of
  interactions, the Yukawa-type coupling of two fermions and a sfermion, and 
  scalar-scalar interaction terms. As we are only interested in the
  Yukawa-type terms we will only consider the Feynman rules which couple two
  fermions and a sfermion. We derive the Feynman rules from the
  superpotential including mixing between the left and right  
  sfermions. Due to our definition of the couplings in the superpotential,
  Eqn.\,\ref{eqn:Rsuper1}, there are differences in our Feynman rules and
  hence cross sections and decay rates when compared with
\cite{Baltz:1998gd}.

  We follow the standard procedure defined in \cite{Haber:1985rc} to
obtain the Lagrangian from the superpotential. We will consider the
superpotential term by term. The first term in the
\rpv\  superpotential, Eqn.\,\ref{eqn:Rsuper1}, \ie the LLE term, gives the
Lagrangian
\begin{equation}
 {\cal L}_{\mr{LLE}}  = - \lambda_{ijk} \left( 
 \elt_{kR}^* \bar{\nu}^c_i  P_{L} \ell_j   
 +\elt_{jL} \bar{\ell}_k  P_{L} \nu_i   
 -\nut_j \bar{\ell}_{k}  P_{L} \ell_i
  +\mr{h.c.} \right)\!,
\end{equation}
  where we have used the antisymmetries of the coupling $\lam_{ijk}$, \ie
  $\lam_{ijk}=-\lam_{jik}$. This Lagrangian therefore only applies for $i>j$. 

  After the inclusion of left/right sfermion mixing this gives the Lagrangian
\begin{equation}
 {\cal L}_{\mr{LLE}}  = - \lambda_{ijk} \left( 
 L^{2k-1}_{2\al}\elt_{k\al}^* \bar{\nu}^c_i  P_{L} \ell_j   
 +L^{2j-1}_{1\al}\elt_{j\al} \bar{\ell}_k  P_{L} \nu_i   
 -\nut_j \bar{\ell}_k  P_{L} \ell_i
  +\mr{h.c.} \right)\!,
\end{equation} 
 where there is an implied summation over the slepton mass eigenstates.
 This Lagrangian gives the Feynman rules shown in Fig.\,\ref{fig:LLE}.

\begin{figure}
\begin{center} \begin{picture}(180,40)(100,40)
\SetScale{1.0}
\DashArrowLine(20,35)(80,35){5}
\ArrowLine(80,35)(115,70)
\ArrowLine(115,0)(80,35)
\Text(90,58)[]{\large $\mr{\ell}_k$}
\Text(50,47)[]{\large $\mr{\elt}_{j\al}$}
\Text(90,12)[]{\large  $\mr{\nu}_i$}
\Text(250,35)[]{\large 
${\displaystyle-\frac{i\lambda_{ijk}L^{2j-1}_{1\al}}{2}(1-\ga_5)}$ }
\Vertex(80,35){1}
\end{picture} \end{center}
\vskip 12mm
\begin{center} \begin{picture}(180,40)(100,40)
\SetScale{1.0}
\DashArrowLine(20,35)(80,35){5}
\ArrowLine(80,35)(115,70)
\ArrowLine(115,0)(80,35)
\Text(90,58)[]{\large $\ell_k$}
\Text(50,47)[]{\large $\nut_j$}
\Text(90,12)[]{\large  $\ell_i$}
\Text(250,35)[]{\large ${\displaystyle\frac{i\lambda_{ijk}}{2}(1-\ga_5)}$ }
\Vertex(80,35){1}
\end{picture} \end{center}
\vskip 12mm
\begin{center} \begin{picture}(180,40)(100,40)
\SetScale{1.0}
\DashArrowLine(20,35)(80,35){5}
\ArrowLine(80,35)(115,70)
\ArrowLine(80,35)(115,0)
\Text(90,58)[]{\large $\ell_j$}
\Text(50,47)[]{\large $\elt_{k\al}$}
\Text(90,12)[]{\large  ${\nu}_i$}
\Text(250,35)[]{\large 
${\displaystyle-\frac{i\lambda_{ijk}L^{2k-1}_{2\al}}{2}(1+\ga_5)C}$}
\Vertex(80,35){1}
\end{picture} \end{center}
\vskip 12mm
\captionB{Feynman rules for LLE.}
	{Feynman rules for LLE. The standard rules for charge
	conjugation matrices apply. In this case 
	the $\nu_i$ should be transposed in the
	last diagram.}
\label{fig:LLE}
\end{figure}

\begin{figure}
\vspace{-5mm}
\begin{center} \begin{picture}(180,50)(100,40)
\SetScale{1.0}
\DashArrowLine(20,35)(80,35){5}
\ArrowLine(80,35)(115,70)
\ArrowLine(115,0)(80,35)
\Text(90,60)[]{\large $\mr{d}_k$}
\Text(50,47)[]{\large $\mr{\dnt}_{j\al}$}
\Text(90,12)[]{\large  $\nu_i$}
\Text(250,35)[]{\large ${\displaystyle
-\frac{i\lambda_{ijk}'Q^{2j-1}_{1\al}}{2}(1-\ga_5)}$ }
\Vertex(80,35){1}
\end{picture} \end{center}
\vskip 10mm
\begin{center} \begin{picture}(180,50)(100,40)
\SetScale{1.0}
\DashArrowLine(20,35)(80,35){5}
\ArrowLine(80,35)(115,70)
\ArrowLine(115,0)(80,35)
\Text(90,60)[]{\large $\mr{d}_k$}
\Text(50,47)[]{\large $\mr{\upt}_{j\al}$}
\Text(90,12)[]{\large  $\ell^{-}_i$}
\Text(250,35)[]{\large ${\displaystyle
\frac{i\lambda_{ijk}'Q^{2j}_{1\al}}{2}(1-\ga_5)}$ }
\Vertex(80,35){1}
\end{picture} \end{center}
\vskip 10mm
\begin{center} \begin{picture}(180,50)(100,40)
\SetScale{1.0}
\DashArrowLine(20,35)(80,35){5}
\ArrowLine(80,35)(115,70)
\ArrowLine(115,0)(80,35)
\Text(90,60)[]{\large $\mr{d}_k$}
\Text(50,47)[]{\large $\nut_i$}
\Text(90,12)[]{\large  $\mr{d}_j$}
\Text(250,35)[]{\large ${\displaystyle
-\frac{i\lambda_{ijk}'}{2}(1-\ga_5)}$ }
\Vertex(80,35){1}
\end{picture} \end{center}
\vskip 10mm
\begin{center} \begin{picture}(180,50)(100,40)
\SetScale{1.0}
\DashArrowLine(20,35)(80,35){5}
\ArrowLine(80,35)(115,70)
\ArrowLine(115,0)(80,35)
\Text(90,60)[]{\large $\mr{d}_k$}
\Text(50,47)[]{\large $\elt_{i\al}$}
\Text(90,12)[]{\large  $\mr{u}_j$}
\Text(250,35)[]{\large ${\displaystyle
\frac{i\lambda_{ijk}'L^{2i-1}_{1\al}}{2}(1-\ga_5)}$ }
\Vertex(80,35){1}
\end{picture} \end{center}
\vskip 10mm
\begin{center} \begin{picture}(180,50)(100,40)
\SetScale{1.0}
\DashArrowLine(20,35)(80,35){5}
\ArrowLine(80,35)(115,70)
\ArrowLine(80,35)(115,0)
\Text(90,60)[]{\large $\mr{d}_j$}
\Text(50,47)[]{\large $\mr{\dnt}_{k\al}$}
\Text(90,12)[]{\large  $\nu_i$}
\Text(250,35)[]{\large ${\displaystyle
-\frac{i\lambda_{ijk}'Q^{2k-1}_{2\al}}{2}(1+\ga_5)C}$ }
\Vertex(80,35){1}
\end{picture} \end{center}
\vskip 10mm
\begin{center} \begin{picture}(180,50)(100,40)
\SetScale{1.0}
\DashArrowLine(20,35)(80,35){5}
\ArrowLine(80,35)(115,70)
\ArrowLine(80,35)(115,0)
\Text(90,60)[]{\large $\mr{u}_j$}
\Text(50,47)[]{\large $\mr{\dnt}_{k\al}$}
\Text(90,12)[]{\large  $\ell^-_i$}
\Text(250,35)[]{\large ${\displaystyle
\frac{i\lambda_{ijk}'Q^{2k-1}_{2\al}}{2}(1+\ga_5)C}$ }
\Vertex(80,35){1}
\end{picture} \end{center}
\vskip 10mm
\captionB{Feynman rules for LQD.}
	{Feynman rules for LQD. The standard rules for charge
	conjugation matrices apply. In this case 
	the lepton in the last two rules should be
	transposed.}
\label{fig:LQD}
\end{figure}
  We can apply the same procedure to obtain the Lagrangian for the second term
  in Eqn.\,\ref{eqn:Rsuper1}, \ie the LQD term,
 \begin{eqnarray}
 {\cal L}_{\mr{LQD}}  &=& - \lambda_{ijk}' 
  \left( 
 \dnt_{kR}^{*}  \bar{\nu}^{c}_i  P_{L} d_j    
 -\dnt_{kR}^{*} \bar{\ell}^c_i    P_{L} u_{j} 
 +\dnt_{jL}     \bar{d}_{k}        P_{L} \nu_i\right. \nonumber  \\  
 && \left.-\upt_{jL}     \bar{d}_k        P_{L} \ell_{i}
  +\nut_{i}       \bar{d}_k        P_{L} d_{j}
  -\elt_{iL}     \bar{d}_k        P_{L} u_{j}  +\mr{h.c.} 
 \right)\!,
 \end{eqnarray}
  where again there is an implied summation over the sfermion mass
  eigenstates.
  We can include the left/right sfermion mixing in the Lagrangian, giving
 \begin{eqnarray}
 {\cal L}_{\mr{LQD}}  &=& - \lambda_{ijk}' 
  \left( 
 Q^{2k-1}_{2\al}\dnt_{k\al}^*  \bar{\nu}^c_i  P_{L} d_j    
 -Q^{2k-1}_{2\al}\dnt_{k\al}^* \bar{\ell}^c_i    P_{L} u_j 
 +Q^{2j-1}_{1\al}\dnt_{j\al}  \bar{d}_k P_{L} \nu_i\right. \nonumber  \\  
 && \left.-Q^{2j}_{1\al}\upt_{j\al}     \bar{d}_{k}        P_{L} \ell_{i}
  +\nut_{i}       \bar{d}_{k}        P_{L} d_{j}
  -L^{2i-1}_{1\al}\elt_{i\al}     \bar{d}_{k}        P_{L} u_{j}  +\mr{h.c.} 
 \right)\!. 
 \end{eqnarray}
  This gives the Feynman rules in  Fig.\,\ref{fig:LQD}.

  The last term in the superpotential, Eqn.\,\ref{eqn:Rsuper1},
  gives the Lagrangian
 \begin{equation}
 {\cal L}_{\mr{UDD}}  = - \lambda_{ijk}''\varepsilon_{c_1c_2c_3} 
  \left( 
 \upt_{iR}^{*c_1}  \bar{d}_k^{c_3}  P_{L} {d^c}_{j}^{c_2}    
 +\dnt_{kR}^{*c_3} \bar{u}_{i}^{c_1}  P_{L} {d^c}_{j}^{c_2}
  +\mr{h.c.} \right)\!,
 \end{equation}
  where $c_1$, $c_2$ and $c_3$ are the colour indices.

  We can include the left/right mixing in this equation, giving
 \begin{equation}
 {\cal L}_{\mr{UDD}}  = - \lambda_{ijk}''\varepsilon_{c_1c_2c_3} 
  \left( 
 Q^{2i}_{2\al}\upt_{i\al}^{*c_1} \bar{d}_{k}^{c_3}  P_{L} {d^c}_{j}^{c_2}    
 +Q^{2k-1}_{2\al}\dnt_{k\al}^{*c_3} \bar{u}_{i}^{c_1}  P_{L} {d^c}_{j}^{c_2}
  +\mr{h.c.} \right)\!.
 \end{equation}
 The Feynman rules from this Lagrangian are shown in 
Fig.\,\ref{fig:UDD}.

\begin{figure}
\vspace{-5mm}
\begin{center} \begin{picture}(180,50)(100,40)
\SetScale{1.0}
\DashArrowLine(20,35)(80,35){5}
\ArrowLine(115,70)(80,35)
\ArrowLine(115,0)(80,35)
\Text(90,60)[]{\large $\mr{d}_k^{c_3}$}
\Text(50,47)[]{\large $\mr{\upt}_{i\al}^{c_1}$}
\Text(90,7)[]{\large  $\mr{d}_j^{c_2}$}
\Text(250,35)[]{\large 
${\displaystyle\frac{i\lambda_{ijk}''Q^{2i}_{2\al}
\varepsilon_{c_1c_2c_3}}{2}C^\dagger(1+\ga_5)}$ }
\Vertex(80,35){1}
\end{picture} \end{center}
\vskip 10mm
\begin{center} \begin{picture}(180,50)(100,40)
\SetScale{1.0}
\DashArrowLine(20,35)(80,35){5}
\ArrowLine(115,70)(80,35)
\ArrowLine(115,0)(80,35)
\Text(90,60)[]{\large $\mr{u}_i^{c_1}$}
\Text(50,47)[]{\large $\mr{\dnt}_{k\al}^{c_3}$}
\Text(90,7)[]{\large  $\mr{d}_j^{c_2}$}
\Text(250,35)[]{\large
${\displaystyle\frac{i\lambda_{ijk}''Q^{2k-1}_{2\al}
\varepsilon_{c_1c_2c_3}}{2}C^\dagger
(1+\ga_5)}$ }
\Vertex(80,35){1}
\end{picture} \end{center}
\vskip 10mm
\captionB{Feynman rules for UDD.}
	{Feynman rules for UDD. The standard rules for charge conjugation
	 matrices apply. In this case the $\mr{d}_j$ should be transposed.}
\label{fig:UDD}
\end{figure}

%%%%%%%%%%%%%%%%%%%%%%%%%%%%%%%%%%%%%%%%%%%%%%%%%%%%%%%%%%%%%%%%%%%%%%%%%%%%%%

%
%  Now the Decay Rate Calculations
%
\chapter{Decay Rate Calculations}
\label{chap:decay}
\section[Introduction]{Introduction}

In this appendix we present
the matrix elements for the decays of the sfermions, charginos, neutralinos
and gluinos via \rpv. The next appendix gives the matrix elements for the
\rpv\  production cross
sections, most of which can simply be obtained by crossing the various
decay matrix elements. Throughout we allow for more than one \rpv\ 
coupling to be non-zero.

In order to simplify the notation for the matrix elements, for both the
decay rates presented in this appendix and the cross sections in
Appendix~\ref{chap:cross}, we introduce
the following functions
\vspace{-4mm}
\begin{subequations}
\begin{eqnarray}
{\displaystyle R(\tilde{a},m_{bc}^2)} &\equiv&
{\displaystyle\frac{1}{(m_{bc}^2-M_{\tilde{a}}^2)^2 +\Gamma_{\tilde{a}}^2 
M_{\tilde{a}}^2},}\\[1.5mm]
{\displaystyle
S(\tilde{a},\tilde{b},m_{cd}^2,m_{ef}^2)} &\equiv&{\displaystyle
 R(\tilde{a},m_{cd}^2) R(\tilde{b},m_{ef}^2) \!
 \left[(m_{cd}^2-M_{\tilde{a}}^2)(m_{ef}^2-M_{\tilde{b}}^2) + 
   \Gamma_{\tilde{a}} \Gamma_{\tilde{b}} M_{\tilde{a}}
  M_{\tilde{b}}\right]\!,\,\,\,\,\,\,\,\,\,\,\,\,\,\,\,\,\,\,\,\,\,}
\end{eqnarray}
\end{subequations}
where $m_{bc}^2 = (p_b+p_c)^2$, and $M_{\tilde a}$ and $\Gamma_{\tilde a}$
are the mass and the width of the sfermion ${\tilde a}$,
respectively. 
The various terms in the matrix elements can be more easily
expressed in terms of 
\vspace{-3mm}
\begin{subequations}
\label{eqn:kindef}
\begin{eqnarray} 
{\displaystyle \Psi(\tilde{a},1,2,3)} & 
{\displaystyle \equiv }&
{\displaystyle R(\tilde{a},m^2_{12}) \left(m^2_{12}-m^2_1-m^2_2\right)}
 \nonumber \\[0.5mm]
&&{\displaystyle  \left[\rule{0cm}{0.5cm}
\left(a^2(\tilde{a})+b^2(\tilde{a})\right)
\left(M^2_0+m^2_3-m^2_{12}\right)+4a(\tilde{a})b(\tilde{a})m_3 M_0
\right]\!,}\\[0.7mm]
{\displaystyle \Upsilon(\tilde{a},1,2,3) }&
{\displaystyle \equiv}&
{\displaystyle S(\tilde{a}_1,\tilde{a}_2,m_{12}^2,m_{12}^2)
		\left(m^2_{12}-m^2_1-m^2_2\right)}
\nonumber \\[0.5mm] &&
{\displaystyle
\left[\rule{0cm}{0.5cm}\left(a(\tilde{a}_1)
a(\tilde{a}_2)+b(\tilde{a}_1)b(\tilde{a}_2)\right)
\left(M^2_0+m^2_3-m^2_{12}\right)\right.}\nonumber\\[0.5mm]
&&{\displaystyle \left. +2\left(a(\tilde{a}_1)b(\tilde{a}_2) 
+a(\tilde{a}_2)b(\tilde{a}_1)\right) m_3 M_0\rule{0cm}{0.5cm} \right]\!,}
\\[0.7mm]
{\displaystyle \Phi(\tilde{a},\tilde{b},1,2,3)}& \equiv &
{\displaystyle S(\tilde{a},\tilde{b},m_{12}^2,m_{23}^2)\!
\left[\rule{0cm}{0.5cm}m_1m_3a(\tilde{a})a(\tilde{b})
\left(m^2_{12}+m^2_{23}-m^2_1-m^2_3\right) \right.}\nonumber \\[0.5mm] &&
{\displaystyle +m_1M_0b(\tilde{a})a(\tilde{b})\left(m^2_{23}-m^2_2-m^2_3
\right)}
\nonumber \\[0.5mm] &&{\displaystyle 
+m_3M_0a(\tilde{a})b(\tilde{b})\left(m^2_{12}-m^2_1-m^2_2\right)}
\nonumber \\[0.5mm] &&{\displaystyle  \left.  +b(\tilde{a})b(\tilde{b})
\left(m^2_{12}m^2_{23}-m^2_1m^2_3-M^2_0m^2_2\right)
\rule{0cm}{0.5cm}\right]\!,}
\end{eqnarray}
\end{subequations}
where $\tilde{a}_1$ and $\tilde{a}_2$ are the mass eigenstates of the
relevant SUSY particle. The functions $a$ and $b$ are
gaugino-sfermion-fermion coupling constants and are given in the
following tables in Appendix~\ref{chap:Feynman}:
Table\,\ref{tab:chargecp} for the charginos,
Table\,\ref{tab:neutcp} for the neutralinos and
Table\,\ref{tab:gluinocp} for the gluino.  The couplings are defined
such that $a(\tilde{c}^*)=b(\tilde{c})$ and $b(\tilde{c}^*)=a(\tilde
{c})$.  In all the above expressions $M_0$ is the mass of the decaying
sparticle and 1, 2, 3, are the decay products. As we will consider the
decays for arbitrary couplings we will use the indices $i,j,k=1,2,3$ to
represent the generations of the particles and the indices $\al,\be=1,2$
to represent the mass eigenstates of the sfermions. We have not included
the right-handed neutrino and therefore we will neglect the left/right
mixing for the sneutrinos.

\section[Sfermions]{Sfermions}

Here we present the matrix elements for the two-body sfermion decays
including left/right mixing. In general the spin- and colour-averaged
matrix elements have the form
\begin{equation}
  |\overline{\me}(a \rightarrow b,c)|^2 =  C^a_{bc} (M_a^2-m_b^2-m_c^2), 
\end{equation}
where $C^a_{bc}$ is the product of the colour factor and the coupling for the
process. These factors are tabulated for the various sfermion decays
in Table\,\ref{tab:scalarcp} where $N_c$ denotes the number of colours.

\renewcommand{\arraystretch}{1.5}
\begin{table}
\begin{center}
\begin{tabular}{|l|l|l|}
\hline
Operator &  Process & Product of the colour factor and coupling $C^a_{bc}$ \\
\hline
 LLE     & $\mr{\elt}_{j\al}^- \longrightarrow \bar{\nu}_i \ell^-_k$ &
            $|\lam_{ijk}|^2 |L^{2j-1}_{1\al}|^2$ \\
\hline
 LLE     & $\elt_{k\al}^- \longrightarrow \nu_i \ell^-_j$ &
            $|\lam_{ijk}|^2 |L^{2k-1}_{2\al}|^2$ \\
\hline
 LLE     & $\nut_j \longrightarrow \ell^+_i \ell^-_k$ &
             $|\lam_{ijk}|^2$ \\
\hline
 LQD     & $\elt^-_{i\al} \longrightarrow \mr{\bar{u}}_j \mr{d}_k$ &
             $N_c|\lam_{ijk}'|^2 |L^{2i-1}_{1\al}|^2$ \\
\hline
 LQD     & $\nut_i \longrightarrow \mr{\bar{d}}_j \mr{d}_k$ & 
            $N_c|\lam_{ijk}'|^2$  \\
\hline
 LQD     & $\mr{\dnt}_{j\al} \longrightarrow \bar{\nu}_i \mr{d}_k$ &
            $|\lam_{ijk}'|^2 |Q^{2j-1}_{1\al}|^2$ \\
\hline
 LQD     &  $\mr{\upt}_{j\al} \longrightarrow \ell_i^+ \mr{d}_k$ &
            $|\lam_{ijk}'|^2 |Q^{2j}_{1\al}|^2$ \\
\hline
 LQD     &  $\mr{\dnt}_{k\al} \longrightarrow \nu_i \mr{d}_j$ &
            $|\lam_{ijk}'|^2 |Q^{2k-1}_{2\al}|^2$ \\
\hline
 LQD     &  $\mr{\dnt}_{k\al} \longrightarrow \ell_i^- \mr{u}_j$ &
            $|\lam_{ijk}'|^2 |Q^{2k-1}_{2\al}|^2$ \\
\hline
 UDD     &  $\mr{\upt}_{i\al} \longrightarrow \mr{\bar{d}}_j \mr{\bar{d}}_k$ &
            $(N_c-1)! |\lam_{ijk}''|^2 |Q^{2i}_{2\al}|^2$ \\
\hline
 UDD     &  $\mr{\dnt}_{k\al} \longrightarrow \mr{\bar{u}}_i \mr{\bar{d}}_j$ &
            $(N_c-1)! |\lam_{ijk}''|^2 |Q^{2k-1}_{2\al}|^2$ \\
\hline
\end{tabular}
\captionB{Coefficients for the sfermion \rpv\  decays.}
	{Coefficients for the sfermion \rpv\  decays.}
\label{tab:scalarcp}
\end{center}
\end{table}

  In all these terms the Roman indices represent the generation of the
particle and the Greek indices the mass eigenstate of the sfermions
when there is mixing. The decay rate can be obtained by
integrating over the two body phase space. This gives
\begin{equation}
  \Gamma(a \rightarrow b,c) = \frac{|\overline{\me}(a 
\rightarrow b,c)|^2 p_{\mr{cm}}}  {8 \pi M_a^2},
\end{equation}
where $p_{\mr{cm}}$ is the final-state momentum in the rest frame of the
decaying particle,
\begin{equation}
   p_{\mr{cm}}^2 = \frac{1}{4M_a^2}
           \left[M_a^2-(m_b+m_c)^2\right]
           \left[M_a^2-(m_b-m_c)^2\right]\!. \nonumber
\end{equation}

\section[Charginos]{Charginos}
  
Most of the chargino \rpv\  decay rates have already been calculated
\cite{Dreiner:1996dd} in the case of no left/right sfermion mixing for the
first two operators in the \rpv\  superpotential. We recalculate these
rates with left/right sfermion mixing. First we consider the LLE decays of the
chargino. There are three possible decay modes:
\begin{enumerate}
\item $\mr{\cht}^+_l \longrightarrow \bar{\nu}_i \ell^+_j \nu_k$;
\item $\mr{\cht}^+_l \longrightarrow \nu_i \nu_j \ell^+_k$;
\item $\mr{\cht}^+_l \longrightarrow \ell^+_i \ell^+_j \ell^-_k$.
\end{enumerate}
The Feynman diagrams for these decays are shown in
Fig.\,\ref{fig:LLEchar}.
\begin{figure}
\begin{center} 
\begin{picture}(360,80)(0,0)
\SetScale{0.7}
\ArrowLine(5,78)(60,78)
\ArrowLine(60,78)(105,105)
\ArrowLine(129,26)(84,53)
\ArrowLine(129,80)(84,53)
\DashArrowLine(84,53)(60,78){5}
\Text(25,63)[]{$\mr{\cht}^+_l$}
\Text(55,72)[]{$\mr{\nu}_k$}
\Text(75,18)[]{$\mr{\ell}^+_j$}
\Text(75,56)[]{$\mr{\bar{\nu}}_i$}
\Text(45,40)[]{$\mr{\elt}_{k\al}$}
\Vertex(60,78){1}
\Vertex(84,53){1}
\ArrowLine(185,78)(240,78)
\ArrowLine(240,78)(285,105)
\ArrowLine(264,53)(309,26)
\ArrowLine(309,80)(264,53)
\DashArrowLine(264,53)(240,78){5}
\Text(150,63)[]{$\mr{\cht}^+_l$}
\Text(200,56)[]{$\mr{\ell}^+_k$}
\Text(200,18)[]{$\mr{\nu}_j$}
\Text(180,70)[]{$\mr{\nu}_i$}
\Text(170,40)[]{$\mr{\elt}_{i\al}$}
\Vertex(240,78){1}
\Vertex(264,53){1}
\ArrowLine(365,78)(420,78)
\ArrowLine(420,78)(465,105)
\ArrowLine(444,53)(489,26)
\ArrowLine(489,80)(444,53)
\DashArrowLine(444,53)(420,78){5}
\Text(277,63)[]{$\mr{\cht}^+_l$}
\Text(330,18)[]{$\mr{\nu}_i$}
\Text(310,72)[]{$\mr{\nu}_j$}
\Text(330,58)[]{$\mr{\ell}^+_k$}
\Text(300,40)[]{$\mr{\elt}_{j\al}$}
\Vertex(420,78){1}
\Vertex(444,53){1}
\end{picture} 
\begin{picture}(360,80)(0,0)
\SetScale{0.7}
\SetOffset(-50,0)
\ArrowLine(185,78)(240,78)
\ArrowLine(285,105)(240,78)
\ArrowLine(264,53)(309,26)
\ArrowLine(309,80)(264,53)
\DashArrowLine(240,78)(264,53){5}
\Text(150,63)[]{$\mr{\cht}^+_l$}
\Text(200,56)[]{$\mr{\ell}^+_j$}
\Text(200,18)[]{$\mr{\ell}^-_k$}
\Text(180,72)[]{$\mr{\ell}^+_i$}
\Text(170,40)[]{$\mr{\nut}_i$}
\Vertex(240,78){1}
\Vertex(264,53){1}
\ArrowLine(365,78)(420,78)
\ArrowLine(465,105)(420,78)
\ArrowLine(444,53)(489,26)
\ArrowLine(489,80)(444,53)
\DashArrowLine(420,78)(444,53){5}
\Text(277,63)[]{$\mr{\cht}^+_l$}
\Text(330,18)[]{$\mr{\ell}^-_k$}
\Text(310,74)[]{$\mr{\ell}^+_j$}
\Text(330,56)[]{$\mr{\ell}^+_i$}
\Text(300,40)[]{$\mr{\nut}_j$}
\Vertex(420,78){1}
\Vertex(444,53){1}
\end{picture}
\vskip -10mm
\end{center}
\captionB{LLE decays of the $\mr{{\tilde\chi}^+}$.}
	{LLE decays of the $\mr{{\tilde\chi}^+}$.The index $l=1,2$ gives the
	 mass eigenstate of the chargino, the index $\al=1,2$ gives the mass
	 eigenstate of the slepton and the indices $i,j,k=1,2,3$ give the
	 generations of the particles.}
\label{fig:LLEchar}
\end{figure}
  The spin-averaged matrix elements are given by:
\begin{eqnarray}
\lefteqn{|\overline{\me}(\cht^+_l \rightarrow \bar{\nu}_i
\ell^+_j\nu_k)|^2 
= } & \nonumber \\
%Amp Square piece
 && \frac{g^2 \lam_{ijk}^2}{2} \left[
  \alsm |L^{2k-1}_{2\al}|^2 \Psi(\elt^*_{k\al},\nu_i,\ell_j,\nu_k)
%Light/Heavy Piece
 +2L^{2k-1}_{21} L^{2k-1}_{22} 
   \Upsilon(\elt^*_k,\nu_i,\ell_j,\nu_k) \right]\!;
\end{eqnarray}
\begin{eqnarray}
|\overline{\me}(\cht^+_l \rightarrow \nu_i \nu_j \ell^+_k)|^2
  & =
& \frac{g^2 \lam_{ijk}^2}{2} \left[ 
  \alsm |L^{2i-1}_{1\al}|^2 \Psi(\elt_{i\al},\nu_j,\ell_k,\nu_i)\right.
 \nonumber \\
%Amp Square pieces
 && 
 +\alsm |L^{2j-1}_{1\al}|^2
  \Psi(\elt_{j\al},\nu_i,\ell_k,\nu_j)
\nonumber \\
%light/heavy pieces
 &&  2L^{2i-1}_{11} L^{2i-1}_{12}\Upsilon(\elt_i,\nu_j,\ell_k,\nu_i)
   +2L^{2j-1}_{11} L^{2j-1}_{12}\Upsilon(\elt_j,\nu_i,\ell_k,\nu_j) 
\nonumber \\
% true interference term
 &&  \left.+\alsm\besm2  L^{2i-1}_{1\al}L^{2j-1}_{1\be} 
       \Phi(\elt_{j\be},\elt_{i\al},\nu_i,\ell_k,\nu_j) \right]\!;\\
%\end{eqnarray}
%\vskip -5mm
%\begin{eqnarray}
|\overline{\me}(\cht^+_l \rightarrow \ell^+_i \ell^+_j \ell^-_k)|^2& = &
{\displaystyle \frac{g^2 \lam_{ijk}^2}{2} \left[ \rule{0mm}{0.8cm}
%Amp Square pieces 
 \Psi(\nut_i,\ell_j,\ell_k,\ell_i)
+ \Psi(\nut_j,\ell_i,\ell_k,\ell_j)\right.} \nonumber \\
% interference piece
 &&\displaystyle{\left. +2 \Phi(\nut_j,\nut_i,\ell_i,\ell_k,\ell_j)
\rule{0mm}{0.8cm} \right]\!.}
\end{eqnarray}
We go beyond the results of \cite{Dreiner:1996dd} to include the decay
$ \cht^+ \longrightarrow \bar{\nu}_i \ell^+_j \nu_k$. 

We now consider the LQD decays of the chargino. There are four
possible decay modes:
\begin{enumerate}
\item $\mr{\cht}^+_l \longrightarrow \mr{\bar{\nu}}_i \mr{\bar{d}}_j\mr{u}_k$;
\item $\mr{\cht}^+_l \longrightarrow \mr{\ell}^+_i \mr{\bar{u}}_j \mr{u}_k$;
\item $\mr{\cht}^+_l \longrightarrow \mr{\ell}^+_i \mr{\bar{d}}_j \mr{d}_k$;
\item $\mr{\cht}^+_l \longrightarrow \mr{\nu}_i \mr{u}_j \mr{\bar{d}}_k$.
\end{enumerate}

\begin{figure}
\begin{center} 
\begin{picture}(360,80)(0,0)
\SetScale{0.7}
\ArrowLine(5,78)(60,78)
\ArrowLine(60,78)(105,105)
\ArrowLine(129,26)(84,53)
\ArrowLine(129,80)(84,53)
\DashArrowLine(84,53)(60,78){5}
\Text(25,63)[]{$\mr{\cht}^+_l$}
\Text(55,72)[]{$\mr{u}_k$}
\Text(75,20)[]{$\mr{\bar{\nu}}_i$}
\Text(75,58)[]{$\mr{\bar{d}}_j$}
\Text(45,40)[]{$\mr{\dnt}_{k\al}$}
\Vertex(60,78){1}
\Vertex(84,53){1}
\ArrowLine(185,78)(240,78)
\ArrowLine(285,105)(240,78)
\ArrowLine(264,53)(309,26)
\ArrowLine(309,80)(264,53)
\DashArrowLine(240,78)(264,53){5}
\Text(150,63)[]{$\mr{\cht}^+_l$}
\Text(200,58)[]{$\mr{\bar{d}}_j$}
\Text(200,20)[]{$\mr{d}_k$}
\Text(180,72)[]{$\mr{\ell}^+_i$}
\Text(170,40)[]{$\mr{\nut}_i$}
\Vertex(240,78){1}
\Vertex(264,53){1}
\ArrowLine(365,78)(420,78)
\ArrowLine(465,105)(420,78)
\ArrowLine(489,26)(444,53)
\ArrowLine(444,53)(489,80)
\DashArrowLine(420,78)(444,53){5}
\Text(277,63)[]{$\mr{\cht}^+_l$}
\Text(330,16)[]{$\mr{\ell}^+_i$}
\Text(310,74)[]{$\mr{\bar{d}}_j$}
\Text(330,58)[]{$\mr{d}_k$}
\Text(300,40)[]{$\mr{\upt}_{j\al}$}
\Vertex(420,78){1}
\Vertex(444,53){1}
\end{picture} 
\begin{picture}(360,80)(0,0)
\SetScale{0.7}
\ArrowLine(5,78)(60,78)
\ArrowLine(60,78)(105,105)
\ArrowLine(129,26)(84,53)
\ArrowLine(129,80)(84,53)
\DashArrowLine(84,53)(60,78){5}
\Text(25,63)[]{$\mr{\cht}^+_l$}
\Text(55,72)[]{$\mr{u}_k$}
\Text(75,20)[]{$\mr{\bar{u}}_j$}
\Text(75,58)[]{$\mr{\ell}^+_i$}
\Text(45,40)[]{$\mr{\dnt}_{k\al}$}
\Vertex(60,78){1}
\Vertex(84,53){1}
\ArrowLine(185,78)(240,78)
\ArrowLine(240,78)(285,105)
\ArrowLine(264,53)(309,26)
\ArrowLine(309,80)(264,53)
\DashArrowLine(264,53)(240,78){5}
\Text(150,63)[]{$\mr{\cht}^+_l$}
\Text(200,58)[]{$\mr{\bar{d}}_k$}
\Text(200,20)[]{$\mr{u}_j$}
\Text(180,72)[]{$\mr{\nu}_i$}
\Text(170,40)[]{$\mr{\elt}_{i\al}$}
\Vertex(240,78){1}
\Vertex(264,53){1}
\ArrowLine(365,78)(420,78)
\ArrowLine(420,78)(465,105)
\ArrowLine(444,53)(489,26)
\ArrowLine(489,80)(444,53)
\DashArrowLine(444,53)(420,78){5}
\Text(277,63)[]{$\mr{\cht}^+_l$}
\Text(330,17)[]{$\mr{\nu}_i$}
\Text(310,75)[]{$\mr{u}_j$}
\Text(330,58)[]{$\mr{\bar{d}}_k$}
\Text(300,40)[]{$\mr{\dnt}_{j\al}$}
\Vertex(420,78){1}
\Vertex(444,53){1}
\end{picture}
\vskip -10mm
\end{center}
\captionB{LQD decays of the $\mr{{\tilde\chi}^+}$.}
	{LQD decays of the $\mr{{\tilde\chi}^+}$.The index $l=1,2$ gives the
	 mass eigenstate of the chargino, the index $\al=1,2$ gives the mass
	 eigenstate of the sfermion and the indices $i,j,k=1,2,3$ give the
	 generations of the particles.}
\label{fig:LQDchar}
\end{figure}
  The Feynman diagrams for these decays are shown in Fig.\,\ref{fig:LQDchar}.
  The spin- and colour-averaged matrix elements are given by
\begin{eqnarray}
\lefteqn{|\overline{\me}(\cht^+_l \rightarrow \bar{\nu}_i \mr{\bar{d}}_j
\mr{u}_k)|^2 =} & \nonumber \\
%Amp Square piece
 && \frac{g^2 {\lam'}_{ijk}^2 N_c}{2} \left[ 
  \alsm |Q^{2k-1}_{2\al}|^2 \Psi(\mr{\dnt}^*_{k\al},\nu_i,\mr{d}_j,\mr{u}_k)
%Light/Heavy Piece
 +2 Q^{2k-1}_{21}Q^{2k-1}_{22}
 \Upsilon(\mr{\dnt}^*_k,\nu_i,\mr{d}_j,\mr{u}_k) \right]\!;\,\,\,\,\,\\[3mm]
%\end{eqnarray}
%\begin{eqnarray}
\lefteqn{|\overline{\me}(\cht^+_l \rightarrow \ell^+_i \mr{\bar{u}}_j
\mr{u}_k)|^2 = } & \nonumber \\
%Amp Square piece
 &&\frac{g^2{\lam'}_{ijk}^2N_c }{2}  \left[ 
 \alsm |Q^{2k-1}_{2\al}|^2\Psi(\mr{\dnt}^*_{k\al},\ell_i,\mr{u}_j,\mr{u}_k)
%Light/Heavy Piece
 +2 Q^{2k-1}_{21}Q^{2k-1}_{22}
 \Upsilon(\mr{\dnt}^*_k,\ell_i,\mr{u}_j,\mr{u}_k)\right]\!;\,\,\,\,\,\,\,\,\,
\\[3mm]
%\end{eqnarray}
%\begin{eqnarray}
\lefteqn{|\overline{\me}(\cht^+_l \rightarrow \ell^+_i \mr{\bar{d}}_j
\mr{d}_k)|^2 =} 
& \nonumber \\
%Amp Square pieces
 && \frac{g^2 {\lam'}_{ijk}^2 N_c}{2}  \left[
  \Psi(\nut_i,\mr{d}_j,\mr{d}_k,\ell_i)
 +\alsm|Q^{2j}_{1\al}|^2  \Psi(\mr{\upt}_{j\al},\ell_i,\mr{d}_k,\mr{d}_j)
\right. \nonumber \\
%light/heavy piece
 && \left. +2 Q^{2j}_{11} Q^{2j}_{12}
 \Upsilon(\mr{\upt}_j,\ell_i,\mr{d}_k,\mr{d}_j) 
% true interference term
  +2\alsm  Q^{2j}_{1\al}
       \Phi(\mr{\upt}_{j\al},\nut_i,\ell_i,\mr{d}_k,\mr{d}_j) 
\right]\!;\\[2mm]
% next ME
\lefteqn{|\overline{\me}(\cht^+_l \rightarrow \nu_i \mr{u}_j
 \mr{\bar{d}}_k)|^2
       =} 
& \nonumber \\
%Amp Square Pieces
 && \frac{g^2 {\lam'}_{ijk}^2 N_c}{2} \left[
  \alsm|L^{2i-1}_{1\al}|^2 \Psi(\mr{\elt}_{i\al},\mr{u}_j,\mr{d}_k,\nu_i)
 +\alsm|Q^{2j-1}_{1\al}|^2 \Psi(\mr{\dnt}_{j\al},\nu_i,\mr{d}_k,\mr{u}_j)
\right. \nonumber \\
% light/heavy pieces
 &&  +2 L^{2i-1}_{11}L^{2i-1}_{12}\Upsilon(\elt_i,\mr{u}_j,\mr{d}_k,\nu_i) 
   +2 Q^{2j-1}_{11} Q^{2j-1}_{12}
 \Upsilon(\mr{\dnt}_j,\nu_i,\mr{d}_k,\mr{u}_j) \nonumber \\
% true interference term 
 && \left. +2\alsm\besm L^{2i-1}_{1\al}Q^{2j-1}_{1\be}
       \Phi(\mr{\dnt}_{j\be},\elt_{i\al},\nu_i,\mr{d}_k,\mr{u}_j) \right]\!.
\end{eqnarray}

We now come to the baryon number violating decays. We do not assume
that there is only one non-zero \rpv\  coupling. This means that
more than one coupling contributes to these decays. It may seem that
this will only matter in the case where more than one $\lam''$
coupling is taken to be non-zero, however there can be more than one
diagram even with only one coupling non-zero, \eg $\lam''_{112}$ will
give two diagrams for each of the decay modes. In this case one of
these diagrams is obtained from the other simply by crossing the
identical fermions in the final state.

%\pagebreak
There are two possible decay modes:
\begin{enumerate}
\item $ \cht^+_l \longrightarrow \mr{u}_i \mr{u}_j \mr{d}_k $;
\item $ \cht^+_l \longrightarrow \mr{\bar{d}}_i \mr{\bar{d}}_j\mr{\bar{d}}_k$.
\end{enumerate}
The Feynman diagrams for these decays are shown in
Fig.\,\ref{fig:UDDchar}. The spin- and colour-averaged matrix elements
for these processes with left/right sfermion mixing are given by:
\begin{eqnarray}
\lefteqn{|\overline{\me}(\cht^+_l \rightarrow
 \mr{u}_i \mr{u}_j \mr{d}_k)|^2 = } & 
\nonumber \\
%Amp Square Terms
 &&\frac{g^2 N_c!}{2(1+\delta_{ij})} \left[ 
{\lam''}_{jik}^2 \alsm|Q^{2i-1}_{2\al}|^2
\Psi(\mr{\dnt}^*_{i\al},\mr{u}_j,\mr{d}_k,\mr{u}_i)
+{\lam''}_{ijk}^2\alsm|Q^{2j-1}_{2\al}|^2
\Psi(\mr{\dnt}^*_{j\al},\mr{u}_i,\mr{d}_k,\mr{u}_j)
\right.\nonumber \\
%Light/Heavy terms
 &&+2{\lam''}_{jik}^2Q^{2i-1}_{21}Q^{2i-1}_{22}
\Upsilon(\mr{\dnt}^*_i,\mr{u}_j,\mr{d}_k,\mr{u}_i)
  +2{\lam''}_{ijk}^2Q^{2j-1}_{21}Q^{2j-1}_{22}
\Upsilon(\mr{\dnt}^*_j,\mr{u}_i,\mr{d}_k,\mr{u}_j) 
\nonumber \\
% true interference term
 && \left.  + 2{\lam''}_{ijk} {\lam''}_{jik}\alsm\besm Q^{2i-1}_{2\al}
Q^{2j-1}_{2\be}
 \Phi(\mr{\dnt}^*_{j\be},\mr{\dnt}^*_{i\al},\mr{u}_i,\mr{d}_k,\mr{u}_j) 
\right]\!;\\[2mm] 
%\end{eqnarray}
%\begin{eqnarray}
\lefteqn{|\overline{\me}(\cht^+_l \rightarrow \mr{\bar{d}}_i \mr{\bar{d}}_j 
\mr{\bar{d}}_k)|^2 =}& \nonumber \\
%Amp Square pieces
 && \frac{g^2 N_c!}{2(1+\delta_{ij}+\delta_{jk}+\delta_{ik})} \left[ 
 {\lam''}_{ijk}^2
 \alsm|Q^{2i}_{2\al}|^2\Psi(\mr{\upt}^*_{i\al},\mr{d}_j,\mr{d}_k,\mr{d}_i)
\right.\nonumber \\
 &&
 +{\lam''}_{jki}^2\alsm|Q^{2j}_{2\al}|^2
\Psi(\mr{\upt}^*_{j\al},\mr{d}_i,\mr{d}_k,\mr{d}_j)
\nonumber \\
 &&+{\lam''}_{kij}^2\alsm|Q^{2k}_{2\al}|^2
\Psi(\mr{\upt}^*_{k\al},\mr{d}_i,\mr{d}_j,\mr{d}_k)
%Light\Heavy pieces
+2
{\lam''}_{ijk}^2Q^{2i}_{21}Q^{2i}_{22}
\Upsilon(\mr{\upt}^*_i,\mr{d}_j,\mr{d}_k,\mr{d}_i)
\nonumber \\
 &&+2 {\lam''}_{jki}^2Q^{2j}_{21}Q^{2j}_{22}
\Upsilon(\mr{\upt}^*_j,\mr{d}_i,\mr{d}_k,\mr{d}_j)
 +2{\lam''}_{kij}^2Q^{2k}_{21}Q^{2k}_{22}
\Upsilon(\mr{\upt}^*_{k\al},\mr{d}_i,\mr{d}_j,\mr{d}_k)
\nonumber \\
% true interference terms
 && -2{\lam''}_{ijk}{\lam''}_{jki} \alsm\besm Q^{2i}_{2\al}Q^{2j}_{2\be}
      \Phi(\mr{\upt}^*_{j\be},\mr{\upt}^*_{i\al},\mr{d}_i,\mr{d}_k,\mr{d}_j) 
\nonumber \\
 &&  - 2{\lam''}_{ijk}{\lam''}_{kij}\alsm\besm Q^{2i}_{2\al}Q^{2k}_{2\be}
       \Phi(\mr{\upt}^*_{k\be},\mr{\upt}^*_{i\al},\mr{d}_i,\mr{d}_j,\mr{d}_k) 
\nonumber \\
 && \left. - 2{\lam''}_{jki}{\lam''}_{kij}
	\alsm\besm Q^{2j}_{2\al}Q^{2k}_{2\be}
      \Phi(\mr{\upt}^*_{k\be},\mr{\upt}^*_{j\al},\mr{d}_j,\mr{d}_i,\mr{d}_k)
\right]\!.
\end{eqnarray}

  The coefficients in the chargino matrix elements are given in 
  Table\,\ref{tab:chargecp}.
  When the chargino mass matrix is diagonalized it is possible to get
  negative eigenvalues in which case the physical field is
  $\ga_5\chi$ rather than $\chi$.  This
  means that the coefficients  $a(\nut_i)$, $a(\mr{\upt}_{i\al})$, and
  $a(\mr{\dnt}_{i\al})$ change sign if the chargino mass is negative.

  The partial widths can be obtained from these matrix elements by
integrating over any two of $m^2_{12}$, $m^2_{23}$ and $m^2_{13}$. The
partial width is given by \cite{Caso:1998tx}
\begin{equation}
 \Gamma(0 \ra 1,2,3) = \frac1{(2\pi)^3}\frac1{32M^3_0}
  \int^{\left(m^2_{12}\right)_{\mr{max}}}_{\left(m^2_{12}\right)_{\mr{min}}}
                            dm^2_{12}
 \int^{\left(m^2_{23}\right)_{\mr{max}}}_{\left(m^2_{23}\right)_{\mr{min}}}
                            dm^2_{23}\,
 				|\overline{\me}|^2,
\label{eqn:threebodyphase}
\end{equation}
where 
\begin{itemize} 
  \item $\left(m^2_{12}\right)_{\mr{max}}=(M_0-m_3)^2$, 
  \item $\left(m^2_{12}\right)_{\mr{min}}=(m_1+m_2)^2$,
  \item $\left(m^2_{23}\right)_{\mr{max}}=(E^*_2+E^*_3)^2
             -\left(\sqrt{{E^*_2}^2-m^2_2}-\sqrt{{E^*_3}^2-m^2_3}\right)^2$,
\textheight 24cm
  \item $\left(m^2_{23}\right)_{\mr{min}}=(E^*_2+E^*_3)^2
             -\left(\sqrt{{E^*_2}^2-m^2_2}+\sqrt{{E^*_3}^2-m^2_3}\right)^2$,
  \item $E^*_2=\left(m^2_{12}-m^2_1+m^2_2\right)/2m_{12}$ and
            $E^*_3=\left(M^2_0-m^2_{12}-m^2_3\right)/2m_{12}$
            are the energies of particles 2 and 3 in the $m_{12}$ rest frame.
\end{itemize}
  These double integrals are complicated and in general, for massive
  final-state particles, cannot be performed analytically. However the
  kinematic functions defined in Eqn.\,\ref{eqn:kindef} only depend on two
  of the possible kinematic variables. This is because there are only two
  independent variables with the third given by momentum 
  conservation,\linebreak
  \ie $m^2_{12}+m^2_{13}+m^2_{23}=M^2_0+m_1^2+m_2^2+m_3^2$. We can therefore
  split the matrix element up into terms containing these functions and
  integrate each term over different kinematic variables. This allows us to
  perform the first integral for each of these terms analytically leaving the
  second integral to be performed numerically. Thus we obtain a number of
  one-dimensional integrals which must be performed numerically rather than
  one two-dimensional integral. It is more efficient to calculate these
  one-dimensional integrals rather than the two-dimensional integral. 

\section[Neutralinos]{Neutralinos}
\label{sect:decayneut}

The total three-body decay rate of a photino was first computed in
\cite{Dawson:1985vr} in the limit where the sfermion is much heavier
than the decaying photino and assuming massless final-state particles. In
\cite{Butterworth:1993tc} the general photino matrix element squared
was given, allowing for the computation of final-state
distributions. In \cite{Dreiner:1994tj,Baltz:1998gd} this was extended
to the general case of a neutralino. In \cite{Baltz:1998gd} arbitrary
sfermion mixing was included as well.  We have recalculated the rates
with only left/right sfermion mixing (neglecting inter-generational
sfermion mixing.)  We 
use a different convention for both the \rpv\  superpotential and the
MSSM Lagrangian, which is more appropriate for implementation in
HERWIG. The LLE, LQD and UDD decay modes are shown in
Figs.\,\ref{fig:LLEneut},~\ref{fig:LQDneut} and \ref{fig:UDDneut},
 respectively.

\begin{figure}
\begin{center} 
\begin{picture}(360,80)(0,0)
\SetScale{0.7}
\ArrowLine(5,78)(60,78)
\ArrowLine(105,105)(60,78)
\ArrowLine(84,53)(129,26)
\ArrowLine(129,80)(84,53)
\DashArrowLine(60,78)(84,53){5}
\Text(25,63)[]{$\mr{\cht}^{0}_l$}
\Text(55,73)[]{$\mr{\bar{\nu}}_i$}
\Text(75,18)[]{$\mr{\ell}^-_k$}
\Text(75,56)[]{$\mr{\ell}^+_j$}
\Text(45,40)[]{$\mr{\nut}_i$}
\Vertex(60,78){1}
\Vertex(84,53){1}
\ArrowLine(185,78)(240,78)
\ArrowLine(285,105)(240,78)
\ArrowLine(264,53)(309,26)
\ArrowLine(309,80)(264,53)
\DashArrowLine(240,78)(264,53){5}
\Text(150,63)[]{$\mr{\cht}^{0}_l$}
\Text(200,56)[]{$\mr{\bar{\nu}}_i$}
\Text(200,18)[]{$\mr{\ell}^-_k$}
\Text(180,70)[]{$\mr{\ell}^+_j$}
\Text(170,40)[]{$\mr{\elt}_{j\al}$}
\Vertex(240,78){1}
\Vertex(264,53){1}
\ArrowLine(365,78)(420,78)
\ArrowLine(420,78)(465,105)
\ArrowLine(489,26)(444,53)
\ArrowLine(489,80)(444,53)
\DashArrowLine(444,53)(420,78){5}
\Text(277,63)[]{$\mr{\cht}^{0}_l$}
\Text(330,15)[]{$\mr{\ell}^+_j$}
\Text(310,76)[]{$\mr{\ell}^-_k$}
\Text(330,56)[]{$\mr{\bar{\nu}}_i$}
\Text(300,40)[]{$\mr{\elt}_{k\al}$}
\Vertex(420,78){1}
\Vertex(444,53){1}
\end{picture}
\vspace{-10mm}
\end{center}
\captionB{LLE decays of the $\mr{{\tilde\chi}^0}$.}
	{LLE decays of the $\mr{{\tilde\chi}^0}$. The index $l=1,\ldots\,\!,4$
	 gives the mass eigenstate of the neutralino, the index $\al=1,2$
	 gives the mass eigenstate of the slepton and the indices
	 \mbox{$i,j,k=1,2,3$} give the generations of the particles.}
\label{fig:LLEneut}
\end{figure}

There are four decay modes:
\begin{enumerate}
\item $\mr{\cht}^0_l\longrightarrow
	\mr{\bar{\nu}}_i \mr{\ell}^+_j \mr{\ell}^-_k$;
\item $\mr{\cht}^0_l\longrightarrow
	\mr{\bar{\nu}}_i \mr{\bar{d}}_j\mr{d}_k$;
\item $\mr{\cht}^0_l\longrightarrow
	\mr{\ell}^+_i \mr{\bar{u}}_j \mr{d}_k$;
\item $\mr{\cht}^0_l\longrightarrow
	\mr{\bar{u}}_i \mr{\bar{d}}_j \mr{\bar{d}}_k$,
\end{enumerate}
as well as their charge conjugates, since the neutralino is a
Majorana fermion.
\textheight 24cm
The spin- and colour-averaged matrix elements are given 
below\footnote{We have a slight disagreement with \cite{Baltz:1998gd}
concerning the sign of the width of the sfermions. This is numerically
insignificant since when the sfermion is on-shell HERWIG treats this
as a two-body decay. The authors of \cite{Baltz:1998gd} agree with our
signs. We thank Paolo Gondolo for discussions of this point.}
% LLE decay rate
\begin{eqnarray}
\lefteqn{|\overline{\me}(\cht^0_l \rightarrow \bar{\nu}_i \ell_j^
+\ell_k^-)|^2 =}  &
   \nonumber \\
%Amp Square pieces
   &&   {\lam}_{ijk}^2 \left[
       \Psi(\nut_i,\ell_j,\ell_k,\nu_i)
     +\alsm|L^{2j-1}_{1\al}|^2\Psi(\elt_{j\al},\nu_i,\ell_k,\ell_j) 
\right. \nonumber \\ 
 &&
+\alsm|L^{2k-1}_{2\al}|^2\Psi(\elt^*_{k\al},\nu_i,\ell_j,\ell_k) 
\nonumber \\
% light\heavy pieces
 &&  +2 L^{2j-1}_{11}L^{2j-1}_{12}\Upsilon(\elt_j,\nu_i,\ell_k,\ell_j) 
   +2 L^{2k-1}_{21}L^{2k-1}_{22}\Upsilon(\elt^*_k,\nu_i,\ell_j,\ell_k)
\nonumber \\
% true interference bits
  && -\alsm2L^{2j-1}_{1\al}\Phi(\elt_{j\al},\nut_i,\nu_i,\ell_k,\ell_j)
   - \alsm 2L^{2k-1}_{2\al}\Phi
(\elt^*_{k\al},\nut_i,\nu_i,\ell_j,\ell_k)  
\nonumber \\
  && \left.
  - \alsm\besm 2L^{2j-1}_{1\al}L^{2k-1}_{2\be}
                        \Phi(\elt^*_{k\be},\elt_{j\al},\ell_j,\nu_i,\ell_k) 
    \right]\!, \label{eqn:LLEneutdecay}\\
%\end{eqnarray}
% First LQD neutralino decay rate
%\begin{eqnarray}
\lefteqn{|\overline{\me}(\cht^0_l \rightarrow \bar{\nu}_i \mr{\bar{d}}_j
\mr{d}_k)|^2 =} & \nonumber \\
%Amp square pieces
 && {\lam'}_{ijk}^2 N_c  \left[ 
   \Psi(\nut_i,\mr{d}_j,\mr{d}_k,\nu_i)
  +\alsm|Q^{2j-1}_{1\al}|^2\Psi(\mr{\dnt}_{j\al},\nu_i,\mr{d}_k,\mr{d}_j)
 \right. \nonumber \\
  && +\alsm|Q^{2k-1}_{2\al}|^2
\Psi(\mr{\dnt}^*_{k\al},\nu_i,\mr{d}_j,\mr{d}_k) \nonumber \\
% light/heavy pieces
 && +2Q^{2j-1}_{11}Q^{2j-1}_{12}\Upsilon(\mr{\dnt}_j,\nu_i,\mr{d}_k,\mr{d}_j) 
   +2Q^{2k-1}_{21}Q^{2k-1}_{22}
\Upsilon(\mr{\dnt}^*_k,\nu_i,\mr{d}_j,\mr{d}_k)\nonumber \\
% true interference bits
 && - \alsm 2Q^{2j-1}_{1\al}
\Phi(\mr{\dnt}_{j\al},\nut_i,\nu_i,\mr{d}_k,\mr{d}_j)
   - \alsm 2Q^{2k-1}_{2\al}
\Phi (\mr{\dnt}^*_{k\al},\nut_i,\nu_i,\mr{d}_j,\mr{d}_k)\nonumber \\
 && \left.- \alsm\besm2Q^{2j-1}_{1\al}Q^{2k-1}_{2\be}
      \Phi(\mr{\dnt}^*_{k\be},\mr{\dnt}_{j\al},\mr{d}_j,\nu_i,\mr{d}_k) 
    \right]\!,\\
%\end{eqnarray}
% Second LQD neutralino decay rate
%\begin{eqnarray}
\lefteqn{|\overline{\me}(\cht^0_l 
\rightarrow \ell^+_i \bar{u}_j d_k)|^2 =} & \nonumber \\
%Amp Square pieces
 && {\lam'}_{ijk}^2 N_c  \left[ 
  \alsm|L^{2i-1}_{1\al}|^2 \Psi(\elt_{i\al},\mr{u}_j,\mr{d}_k,\ell_i)
  +\alsm|Q^{2j}_{1\al}|^2 
\Psi(\mr{\upt}_{j\al},\ell_i,\mr{d}_k,\mr{u}_j) \right. \nonumber \\
 && +\alsm|Q^{2k-1}_{2\al}|^2
 \Psi(\mr{\dnt}^*_{k\al},\ell_i,\mr{u}_j,\mr{d}_k) 
% light/heavy pieces
    +2 L^{2i-1}_{11}L^{2i-1}_{12} 
\Upsilon(\elt_i,\mr{u}_j,\mr{d}_k,\ell_i) \nonumber \\
 &&  +2 Q^{2j}_{11}Q^{2j}_{12} \Upsilon(\mr{\upt}_j,\ell_i,\mr{d}_k,\mr{u}_j) 
    +2 Q^{2k-1}_{21}Q^{2k-1}_{22}
\Upsilon(\mr{\dnt}^*_k,\ell_i,\mr{u}_j,\mr{d}_k) \nonumber \\
% true interference bits 
 &&  -\alsm\besm2L^{2i-1}_{1\al}Q^{2j}_{1\be}
  \Phi(\mr{\upt}_{j\be},\elt_{i\al},\ell_i,\mr{d}_k,\mr{u}_j) \nonumber \\
 && - \alsm\besm2L^{2i-1}_{1\al}Q^{2k-1}_{2\be}
  \Phi(\mr{\dnt}^*_{k\be},\elt_{i\al},\ell_i,\mr{u}_j,\mr{d}_k)\nonumber \\
 &&  \left.- \alsm\besm 2Q^{2j}_{1\al}Q^{2k-1}_{2\be}
  \Phi(\mr{\dnt}^*_{k\be},\mr{\upt}_{j\al},\mr{u}_j,\ell_i,\mr{d}_k) 
    \right]\!,
\end{eqnarray}
\textheight 23cm
% UDD decay rate
\begin{eqnarray}
\lefteqn{|\overline{\me}
(\cht^0_l \rightarrow \bar{u}_i \bar{d}_j \bar{d}_k)|^2 =}
 & \nonumber \\
%Amp Square pieces
 && {\lam''}_{ijk}^2 N_c!  \left[  
  \alsm|Q^{2i}_{2\al}|^2\Psi(\mr{\upt}^*_{i\al},\mr{d}_j,\mr{d}_k,\mr{u}_i)
 +\alsm|Q^{2j-1}_{2\al}|^2
\Psi(\mr{\dnt}^*_{j\al},\mr{u}_i,\mr{d}_k,\mr{d}_j) \right. \nonumber \\
 && +\alsm|Q^{2k-1}_{2\al}|^2
\Psi(\mr{\dnt}^*_{k\al},\mr{u}_i,\mr{d}_j,\mr{d}_k)
% light/heavy pieces
    + 2 Q^{2i}_{21}Q^{2i}_{22}
\Upsilon(\mr{\upt}^*_i,\mr{d}_j,\mr{d}_k,\mr{u}_i) \nonumber \\
  && +2 Q^{2j-1}_{21}Q^{2j-1}_{22}  
\Upsilon(\mr{\dnt}^*_j,\mr{u}_i,\mr{d}_k,\mr{d}_j)
    +2 Q^{2k-1}_{21}Q^{2k-1}_{22} 
\Upsilon(\mr{\dnt}^*_k,\mr{u}_i,\mr{d}_j,\mr{d}_k) \nonumber \\
% true interference bits
 && - \alsm\besm 2Q^{2i}_{2\al}Q^{2j-1}_{2\be}
\Phi(\mr{\dnt}^*_{j\be},\mr{\upt}^*_{i\al},\mr{u}_i,\mr{d}_k,\mr{d}_j)
\nonumber \\
 &&- \alsm\besm 2Q^{2i}_{2\al}Q^{2k-1}_{2\be}
\Phi(\mr{\dnt}^*_{k\be},\mr{\upt}^*_{i\al},\mr{u}_i,\mr{d}_j,\mr{d}_k)
\nonumber \\ 
 &&  \left.- \alsm\besm 2Q^{2j-1}_{2\al}Q^{2k-1}_{2\be}
\Phi(\mr{\dnt}^*_{k\be},\mr{\dnt}^*_{j\al},\mr{d}_j,\mr{u}_i,\mr{d}_k) 
    \right]\!.
\end{eqnarray}

% LQD Neutralino Decay Feynman diagrams
\begin{figure}
\begin{center} 
\begin{picture}(360,80)(0,0)
\SetScale{0.7}
\ArrowLine(5,78)(60,78)
\ArrowLine(105,105)(60,78)
\ArrowLine(84,53)(129,26)
\ArrowLine(129,80)(84,53)
\DashArrowLine(60,78)(84,53){5}
\Text(25,63)[]{$\mr{\cht}^{0}_l$}
\Text(55,72)[]{$\mr{\bar{\nu}}_i$}
\Text(75,20)[]{$\mr{d}_k$}
\Text(75,56)[]{$\mr{\bar{d}}_j$}
\Text(45,40)[]{$\mr{\nut}_{i\al}$}
\Vertex(60,78){1}
\Vertex(84,53){1}
\ArrowLine(185,78)(240,78)
\ArrowLine(285,105)(240,78)
\ArrowLine(264,53)(309,26)
\ArrowLine(309,80)(264,53)
\DashArrowLine(240,78)(264,53){5}
\Text(150,63)[]{$\mr{\cht}^{0}_l$}
\Text(200,54)[]{$\mr{\bar{\nu}}_i$}
\Text(200,20)[]{$\mr{d}_k$}
\Text(180,72)[]{$\mr{\bar{d}}_j$}
\Text(170,40)[]{$\mr{\dnt}_{j\al}$}
\Vertex(240,78){1}
\Vertex(264,53){1}
\ArrowLine(365,78)(420,78)
\ArrowLine(420,78)(465,105)
\ArrowLine(489,26)(444,53)
\ArrowLine(489,80)(444,53)
\DashArrowLine(444,53)(420,78){5}
\Text(277,63)[]{$\mr{\cht}^{0}_l$}
\Text(330,18)[]{$\mr{\bar{\nu}}_i$}
\Text(310,74)[]{$\mr{d}_k$}
\Text(330,58)[]{$\mr{\bar{d}}_j$}
\Text(300,40)[]{$\mr{\dnt}_{k\al}$}
\Vertex(420,78){1}
\Vertex(444,53){1}
\end{picture} 
\begin{picture}(360,80)(0,0)
\SetScale{0.7}
\ArrowLine(5,78)(60,78)
\ArrowLine(105,105)(60,78)
\ArrowLine(84,53)(129,26)
\ArrowLine(129,80)(84,53)
\DashArrowLine(60,78)(84,53){5}
\Text(25,63)[]{$\mr{\cht}^{0}_l$}
\Text(55,72)[]{$\mr{\ell}^{+}_i$}
\Text(75,20)[]{$\mr{d}_k$}
\Text(75,56)[]{$\mr{\bar{u}}_j$}
\Text(45,40)[]{$\mr{\elt}_{i\al}$}
\Vertex(60,78){1}
\Vertex(84,53){1}
\ArrowLine(185,78)(240,78)
\ArrowLine(285,105)(240,78)
\ArrowLine(264,53)(309,26)
\ArrowLine(309,80)(264,53)
\DashArrowLine(240,78)(264,53){5}
\Text(150,63)[]{$\mr{\cht}^{0}_l$}
\Text(200,56)[]{$\mr{\ell}^{+}_i$}
\Text(200,20)[]{$\mr{d}_k$}
\Text(180,72)[]{$\mr{\bar{u}}_j$}
\Text(170,40)[]{$\mr{\upt}_{j\al}$}
\Vertex(240,78){1}
\Vertex(264,53){1}
\ArrowLine(365,78)(420,78)
\ArrowLine(420,78)(465,105)
\ArrowLine(489,26)(444,53)
\ArrowLine(489,80)(444,53)
\DashArrowLine(444,53)(420,78){5}
\Text(277,63)[]{$\mr{\cht}^{0}_l$}
\Text(330,17)[]{$\mr{\ell}^{+}_i$}
\Text(310,75)[]{$\mr{d}_k$}
\Text(330,56)[]{$\mr{\bar{u}}_j$}
\Text(300,40)[]{$\mr{\dnt}_{k\al}$}
\Vertex(420,78){1}
\Vertex(444,53){1}
\end{picture}
\vskip -10mm
\end{center}
\captionB{LQD decays of the $\mr{{\tilde\chi}^0}$.}
	{LQD decays of the $\mr{{\tilde\chi}^0}$.The index $l=1,\ldots\,\!,4$
         gives the mass eigenstate of the neutralino, the index $\al=1,2$
         gives the mass eigenstate of the sfermion and the indices 
         \mbox{$i,j,k=1,2,3$} give the generations of the particles.}
\label{fig:LQDneut}
\end{figure}
% End of the figure
  
  The relevant coefficients are given in Table\,\ref{tab:neutcp}. 
  Again, when the neutralino mass matrix is diagonalized, negative
  eigenvalues can be obtained and the fields must be rotated.
  This changes the sign of some
  of the coefficients in Table\,\ref{tab:neutcp}: the coefficients
  $a(\tilde{c})$ change sign, and hence the coefficients
  $b(\tilde{c}^*)$ also change sign.
  The partial widths can be obtained by
  integrating the matrix elements in the same way as for the
  chargino decays.
  
\section[Gluinos]{Gluinos}

These decay rates are calculated here with left/right mixing. There
are three possible decay modes, two via the LQD operator and one via
the UDD operator:
\begin{enumerate}
\item $\mr{\glt} \longrightarrow \bar{\nu}_i \mr{\bar{d}}_j \mr{d}_k$;
\item $\mr{\glt} \longrightarrow \ell^+_i \mr{\bar{u}}_j \mr{d}_k$;
\item $\mr{\glt} \longrightarrow \mr{u}_i \mr{d}_j \mr{d}_k $.
\end{enumerate}

\begin{figure}
\begin{center} 
\begin{picture}(360,80)(0,0)
\SetScale{0.7}
\SetOffset(-50,0)
\ArrowLine(185,78)(240,78)
\ArrowLine(285,105)(240,78)
\ArrowLine(264,53)(309,26)
\ArrowLine(309,80)(264,53)
\DashArrowLine(240,78)(264,53){5}
\Text(150,63)[]{$\mr{\glt}$}
\Text(200,56)[]{$\mr{\bar{\nu}}_i$}
\Text(200,18)[]{$\mr{d}_k$}
\Text(180,74)[]{$\mr{\bar{d}}_j$}
\Text(170,40)[]{$\mr{\dnt}_{j\al}$}
\Vertex(240,78){1}
\Vertex(264,53){1}
\ArrowLine(365,78)(420,78)
\ArrowLine(420,78)(465,105)
\ArrowLine(489,26)(444,53)
\ArrowLine(489,80)(444,53)
\DashArrowLine(444,53)(420,78){5}
\Text(277,63)[]{$\mr{\glt}$}
\Text(330,18)[]{$\mr{\bar{\nu}}_i$}
\Text(310,74)[]{$\mr{d}_k$}
\Text(330,56)[]{$\mr{\bar{d}}_j$}
\Text(300,40)[]{$\mr{\mr{\dnt}}_{k\al}$}
\Vertex(420,78){1}
\Vertex(444,53){1}
\end{picture}
\begin{picture}(360,80)(0,0)
\SetScale{0.7}
\SetOffset(-50,0)
\ArrowLine(185,78)(240,78)
\ArrowLine(285,105)(240,78)
\ArrowLine(264,53)(309,26)
\ArrowLine(309,80)(264,53)
\DashArrowLine(240,78)(264,53){5}
\Text(150,63)[]{$\mr{\glt}$}
\Text(200,56)[]{$\mr{\ell}^+_i$}
\Text(200,18)[]{$\mr{d}_k$}
\Text(180,74)[]{$\mr{\bar{u}}_j$}
\Text(170,40)[]{$\mr{\upt}_{j\al}$}
\Vertex(240,78){1}
\Vertex(264,53){1}
\ArrowLine(365,78)(420,78)
\ArrowLine(420,78)(465,105)
\ArrowLine(489,26)(444,53)
\ArrowLine(489,80)(444,53)
\DashArrowLine(444,53)(420,78){5}
\Text(277,63)[]{$\mr{\glt}$}
\Text(330,18)[]{$\mr{\ell}^+_i$}
\Text(310,74)[]{$\mr{d}_k$}
\Text(330,56)[]{$\mr{\bar{u}}_j$}
\Text(300,40)[]{$\mr{\dnt}_{k\al}$}
\Vertex(420,78){1}
\Vertex(444,53){1}
\end{picture}
\vskip -10mm
\end{center}
\captionB{LQD decays of the $\mr{\glt}$.}
	{LQD decays of the $\mr{\glt}$. The index $\al=1,2$ gives the mass
	 eigenstate of the sfermion and the indices $i,j,k=1,2,3$ give the
	 generations of the particles.}
\label{fig:LQDgluino}
\end{figure}
Since the gluino is a Majorana fermion the charge conjugate decay
modes are also possible.  The Feynman diagrams for these processes
are shown in Fig.\,\ref{fig:LQDgluino} and Fig.\,\ref{fig:UDDgluino},
respectively.  The spin- and colour-averaged matrix elements with
left/right sfermion mixing are given by
% First LQD gluino decay rate
\begin{eqnarray}
\lefteqn{|\overline{\me}(\mr{\glt} \rightarrow \bar{\nu}_i \mr{\bar{d}}_j
\mr{d}_k)|^2 =} & \nonumber \\[1.5mm]
%Amp Square terms
 && \frac{ {\lam'}_{ijk}^2 g_s^2}{2} \left[ 
 \alsm|Q^{2j-1}_{1\al}|^2\Psi(\mr{\dnt}_{j\al},\nu_i,\mr{d}_k,\mr{d}_j) 
  +\alsm|Q^{2k-1}_{2\al}|^2\Psi(\mr{\dnt}^*_{k\al},\nu_i,\mr{d}_j,\mr{d}_k)
 \right. \nonumber \\[1.5mm]
% light/heavy pieces
 && +2Q^{2j-1}_{11}Q^{2j-1}_{12}\Upsilon(\mr{\dnt}_j,\nu_i,\mr{d}_k,\mr{d}_j) 
   +2Q^{2k-1}_{21}Q^{2k-1}_{22}
\Upsilon(\mr{\dnt}^*_k,\nu_i,\mr{d}_j,\mr{d}_k)\nonumber \\[1.5mm]
% true interference bits
 && \left.- \alsm\besm2Q^{2j-1}_{1\al}Q^{2k-1}_{2\be}
       \Phi(\mr{\dnt}^*_{k\be},\mr{\dnt}_{j\al},\mr{d}_j,\nu_i,\mr{d}_k) 
    \right]\!,
\\[3mm]
%\end{eqnarray}
% Second LQD gluino decay rate
%\begin{eqnarray}
\lefteqn{|\overline{\me}(\mr{\glt} \rightarrow \ell^+_i \mr{\bar{u}}_j
\mr{d}_k)|^2 =} & \nonumber \\[1.5mm]
 %Amp Square
 &&\frac{{\lam'}_{ijk}^2 g_s^2}{2}  \left[ 
  \alsm|Q^{2j}_{1\al}|^2 \Psi(\mr{\upt}_{j\al},\ell_i,\mr{d}_k,\mr{u}_j) 
  +2 Q^{2j}_{11}Q^{2j}_{12} 
\Upsilon(\mr{\upt}_j,\ell_i,\mr{d}_k,\mr{u}_j) \right.
\nonumber \\[1.5mm] 
% light/heavy pieces
 &&   +\alsm|Q^{2k-1}_{2\al}|^2 
\Psi(\mr{\dnt}^*_{k\al},\ell_i,\mr{u}_j,\mr{d}_k)
    +2 Q^{2k-1}_{21}Q^{2k-1}_{22}
\Upsilon(\mr{\dnt}^*_k,\ell_i,\mr{u}_j,\mr{d}_k) \nonumber \\[1.5mm]
% true interference bits 
 &&  \left.- \alsm\besm 2Q^{2j}_{1\al}Q^{2k-1}_{2\be}
\Phi(\mr{\dnt}^*_{k\be},\mr{\upt}_{j\al},\mr{u}_j,\ell_i,\mr{d}_k) 
    \right]\!,
\end{eqnarray}
% UDD decay rate
\begin{eqnarray}
\lefteqn{|\overline{\me}(\mr{\glt} \rightarrow \mr{\bar{u}}_i \mr{\bar{d}}_j 
\mr{\bar{d}}_k)|^2 = }  & \nonumber \\
%Amp Square Terms
 &&
 \frac{{\lam''}_{ijk}^2 (N_c-1)!}{2}  \left[ \alsm|Q^{2i}_{2\al}|^2\Psi(
\mr{\upt}^*_{i\al},\mr{d}_j,\mr{d}_k,\mr{u}_i)
 +2 Q^{2i}_{21}Q^{2i}_{22}
\Upsilon(\mr{\upt}^*_i,\mr{d}_j,\mr{d}_k,\mr{u}_i) \right. \nonumber \\
 && +\alsm|Q^{2j-1}_{2\al}|^2
\Psi(\mr{\dnt}^*_{j\al},\mr{u}_i,\mr{d}_k,\mr{d}_j)
% light/heavy pieces
+2 Q^{2j-1}_{21}Q^{2j-1}_{22}  
\Upsilon(\mr{\dnt}^*_j,\mr{u}_i,\mr{d}_k,\mr{d}_j)
     \nonumber \\
  && +\alsm|Q^{2k-1}_{2\al}|^2
\Psi(\mr{\dnt}^*_{k\al},\mr{u}_i,\mr{d}_j,\mr{d}_k) 
    +2 Q^{2k-1}_{21}Q^{2k-1}_{22} 
\Upsilon(\mr{\dnt}^*_k,\mr{u}_i,\mr{d}_j,\mr{d}_k) \nonumber \\
% true interference bits
 && + \frac{1}{N_c-1}\alsm\besm 2 Q^{2i}_{2\al}Q^{2j-1}_{2\be}
  \Phi(\mr{\dnt}^*_{j\be},\mr{\upt}^*_{i\al},\mr{u}_i,\mr{d}_k,\mr{d}_j)
\nonumber \\
 && +\frac{1}{N_c-1}\alsm\besm2 Q^{2i}_{2\al}Q^{2k-1}_{2\be}
   \Phi(\mr{\dnt}^*_{k\be},\mr{\upt}^*_{i\al},\mr{u}_i,\mr{d}_j,\mr{d}_k)
\nonumber \\ 
 &&  +\left. \frac{1}{N_c-1}\alsm\besm2Q^{2j-1}_{2\al}Q^{2k-1}_{2\be}
    \Phi(\mr{\dnt}^*_{k\be},\mr{\dnt}^*_{j\al},\mr{d}_j,\mr{u}_i,\mr{d}_k) 
    \right]\!.
\end{eqnarray}
The coefficients for these matrix elements are given in
Table\,\ref{tab:gluinocp}. As the gluino mass is not obtained by
diagonalizing a mass matrix it cannot be negative. The partial widths can be
obtained
integrating the matrix elements in the same way as for the chargino decays.

%
% Hard Processes
%
\chapter{Cross-section Calculations}
\label{chap:cross}

\section[Introduction]{Introduction}

  Here we present all the matrix elements for $2\ra2$ scattering processes
  via single sparticle exchange. We
  disregard those possibilities where the sfermion resonance
   is not kinematically probed, \eg
  \begin{equation}
   \mr{d}_j+\mr{\bar{d}}_k\ra {\tilde\nu}_i\ra{\tilde\nu}_i+\mr{Z^0}.
  \end{equation}

  All of the single neutralino, chargino and gluino production cross
  sections can be obtained by crossing from the decay matrix elements we
  have already presented in Appendix~\ref{chap:decay}. This crossing
  will lead to the invariants $m^2_{12}$, $m^2_{23}$, and $m^2_{13}$
  being replaced by the usual invariants $s$, $t$ and $u$. There is also an
  overall sign change due to exchanging fermions between the initial and
  final states. Furthermore it should be remembered that the decay matrix
  elements have been averaged over the spin and colour of the
  initial particle. 

  The cross sections for the remaining processes are
  presented below. In all cases the formulae have been averaged over the
  initial spins and colours. The initial-state masses have all been set
  to zero, except where they appear in a coupling constant. In $t$- and
  $u$-channel fermion propagators the fermion masses have been neglected
  as well.

%
%LQD processes
%
\section[LQD Processes]{LQD Processes}

%
%  LQD Weak processes
%
\subsection[Resonant Slepton Production followed by Weak Decay]
	   {Resonant Slepton Production  followed by Weak Decay}

  There are three processes which can occur via the production of a
resonant slepton followed by a weak decay of this slepton. These are:
\begin{enumerate}
\item $\mr{d}_j \mr{\bar{d}}_k \longrightarrow \elt^*_i \mr{W^-}$;
\item $\mr{u}_j \mr{\bar{d}}_k \longrightarrow \nut^*_i \mr{W^+}$;
\item $\mr{u}_j \mr{\bar{d}}_k \longrightarrow \tilde{\tau}_1^* \mr{Z^0}$.
\end{enumerate}
  We have not included processes where the
  resonance is not accessible, \eg
  $\mr{u}_j \mr{\bar{d}}_k \longrightarrow \tilde{\tau}_2^* \mr{Z^0}$.
  The matrix elements are given below:
\begin{eqnarray}
% sneutrino to charged slepton and W
\displaystyle{\lefteqn{|\overline{\me}(\mr{d}_j \mr{\bar{d}}_k \ra
\elt^*_{i\al} \mr{W^-})|^2 = }} & \nonumber\\
 &&\displaystyle{ 
\frac{g^2{\lam'}_{ijk}^2 |L^{2i-1}_{1\al}|^2}{2 \mw^2 N_c}\left[
\rule{0mm}{0.8cm}
	\sh^2 p^2_{\mr{cm}} R(\nut_i,\sh) \right.
	+\frac{1}{4\uh^2} \left(2\mw^2(\uh\tha-M^2_{\elt_{i\al}} \mw^2)
						+\uh^2\sh\right) \nonumber}\\ 
 &&\displaystyle{ 
\left.-\frac{\sh\left(\sh-M^2_{\nut_i}\right)R(\nut_i,\sh)}{2\uh}
	\left(\mw^2(2M^2_{\elt_{i\al}}-\uh)+\uh(\sh-M^2_{\elt_{i\al}})
\right)\rule{0mm}{0.8cm}\right]\!;}\\[3mm]
% charged slepton to sneutrino and W
\lefteqn{|\overline{\me}(\mr{u}_j \mr{\bar{d}}_k
   \rightarrow \nut^*_i \mr{W^+})|^2
 =} & 
 \nonumber\\
&&\displaystyle{ \frac{g^2{\lam'}_{ijk}^2}{2 \mw^2 N_c}\left[\rule{0mm}{0.9cm}
      \alsm  |L^{2i-1}_{1\al}|^4 \sh^2 p^2_{\mr{cm}} R(\elt_{i\al},\sh) 
     +2|L^{2i-1}_{11}|^2 |L^{2i-1}_{12}|^2 \sh^2 p^2_{\mr{cm}} S(\elt_{i1},
\elt_{i2},\sh,\sh)
     \right. \nonumber} \\
&&\displaystyle{
+\frac{1}{4\uh^2} \left(2\mw^2(\uh\tha-M^2_{\nut_i}\mw^2) +\uh^2\sh\right) 
  \nonumber} \\ 
&& \displaystyle{\left.  -\alsm\frac{|L^{2i-1}_{1\al}|^2
	\sh\left(\sh-M^2_{\elt_{i\al}}\right) R(\elt_{i\al},\sh)}{2\uh} 
  \left(\mw^2(2M^2_{\nut_i}-\uh)+\uh(\sh-M^2_{\nut_i})\right)
 \rule{0mm}{0.9cm}  \right]\!;} \\[3mm]
% stau_2 to stau_1 and Z 
\displaystyle{\lefteqn{|\overline{\me}(\mr{u}_j \mr{\bar{d}}_k
 \rightarrow  \elt^*_{i1}
  \mr{Z^0})|^2 =}} &
  \nonumber\\
  &&\displaystyle{\frac{g^2{\lam'}_{ijk}^2}{N_c \mz^2\cos^2 \theta_W} \left[
    \alsm |L^{2i-1}_{1\al}|^2 |Z^{\al1}_{\ell_i}|^2 \sh^2 p^2_{\mr{cm}} 
R(\elt_{i\al},\sh) \right.}\nonumber \\
 &&\displaystyle{ +\frac{|L^{2i-1}_{11}|^2 Z_{u_L}^2}{\uh^2}
	\left(2\mz^2(\uh\tha-M^2_{\elt_{i1}}\mz^2) +\uh^2\sh\right)}
\nonumber \\
 &&\displaystyle{ +\frac{|L^{2i-1}_{11}|^2 Z_{d_R}^2}{\tha^2}
	\left(2\mz^2(\uh\tha-M^2_{\elt_{i1}}\mz^2) +\tha^2\sh\right)
+2L^{2i-1}_{11}L^{2i-1}_{12}Z^{11}_{\ell_i}Z^{21}_{\ell_i} \sh^2 p^2_{\mr{cm}}
    S(\elt_{i1},\elt_{i2},\sh,\sh)\nonumber} \\
  && \displaystyle{
+\alsm \frac{L^{2i-1}_{1\al} L^{2i-1}_{11}Z^{\al1}_{\ell_i} Z_{u_L}
        \sh \left(\sh-M^2_{\elt_{i\al}}\right)R(\elt_{i\al},\sh) }{\uh}
	\left( \mz^2(2M^2_{\elt_{i1}}-\uh) +\uh
  (s-M^2_{\elt_{i1}})\right) \nonumber} \\
  &&\displaystyle{
 -\alsm \frac{L^{2i-1}_{1\al} L^{2i-1}_{11}Z^{\al1}_{\ell_i} Z_{d_R}
        \sh  \left(\sh-M^2_{\elt_{i\al}}\right)R(\elt_{i\al},\sh)}{\tha}
	\left( \mz^2(2M^2_{\elt_{i1}}-\tha) +\tha(s-M^2_{\elt_{i1}})
\right)\nonumber}\\
  &&\displaystyle{\left. +\frac{2 |L^{2i-1}_{11}|^2  Z_{u_L} Z_{d_R}}{\uh\tha}
	\left( 2\mz^2(M^2_{\elt_{i1}}-\tha)(M^2_{\elt_{i1}}-\uh) 
-\sh\tha\uh \right)
\rule{0mm}{0.8cm}	\right]\!,}
\end{eqnarray}
where in all the above equations
\begin{equation}
   p_{\mr{cm}}^2 = \frac{1}{4\sh}
           \left[\sh-(m_1+m_2)^2\right]
           \left[\sh-(m_1-m_2)^2\right]\!, \nonumber
\end{equation}
and $m_1$, $m_2$ are the masses of the final-state particles. The couplings
for these processes can be found in Table\,\ref{tab:Zcp}.

  In the Monte Carlo simulation all of the cross-section integrals are
  performed using the multi-channel Monte Carlo integration technique
  described in  Chapter~\ref{chap:monte}.

%
% LQD R slash processes
%
\subsection[Resonant Slepton  Production followed by \rpv\  Decay]
       	   {Resonant Slepton  Production followed by \boldmath{\rpv}\  Decay}
  There are four processes in which we can produce a resonant slepton
via \rpv\   which then decays back to Standard Model particles via a \rpv\   
decay. These are:
\begin{enumerate}
\item $\mr{d}_j \mr{\bar{d}}_k \longrightarrow \mr{d}_l \mr{\bar{d}}_m$;
\item $\mr{u}_j \mr{\bar{d}}_k \longrightarrow \mr{u}_l \mr{\bar{d}}_m$;
\item $\mr{d}_j \mr{\bar{d}}_k \longrightarrow \ell^-_l \ell^+_m$;
\item $\mr{u}_j \mr{\bar{d}}_k \longrightarrow \nu_l \ell^+_m$.
\end{enumerate}
  The first two of these processes only require non-zero LQD couplings
whereas the second two require both non-zero LQD and LLE couplings.
 The matrix elements are presented below for an arbitrary number of
non-zero \rpv\  couplings:
% Resonant sneutrino to quarks
\begin{eqnarray}
|\overline{\me}(\mr{d}_j \mr{\bar{d}}_k \rightarrow
 \mr{d}_l \mr{\bar{d}}_m)|^2 &= &
 \frac{1}{4}\displaystyle\sum_{i,n=1,3}
 {\lam'}_{ijk}{\lam'}_{ilm}{\lam'}_{njk}{\lam'}_{nlm} 
S(\nut_i,\nut_n,\sh,\sh)
   \sh\left(\sh-m^2_{\mr{d}_l}-m^2_{\mr{d}_m} \right) \nonumber \\
&& +\frac{1}{4}\displaystyle\sum_{i,n=1,3} 
  {\lam'}_{ijl}{\lam'}_{ikm}{\lam'}_{njl}{\lam'}_{nkm}
  \frac{(m^2_{\mr{d}_l}-\hat{t})(m^2_{\mr{d}_m}-\hat{t})}
{(\hat{t}-M^2_{\nut_i})(\hat{t}-M^2_{\nut_n})}; \\[2mm]
&& \nonumber \\
% Resonant slepton to quarks
|\overline{\me}(\mr{u}_j \mr{\bar{d}}_k \rightarrow \mr{u}_l
 \mr{\bar{d}}_m)|^2& = &
 \frac{1}{4}\displaystyle \sum_{\al,\be=1,2}\displaystyle\sum_{i,n=1,3}
 {\lam'}_{ijk}{\lam'}_{ilm}{\lam'}_{njk}{\lam'}_{nlm}
 |L^{2i-1}_{1\al}|^2|L^{2n-1}_{1\be}|^2 \nonumber \\
&&  
  S(\elt_{i\al},\elt_{n\be},\sh,\sh)
   \sh\left(\sh-m^2_{\mr{u}_l}-m^2_{\mr{d}_m} \right)\!; \\[2mm]
&& \nonumber \\
% Resonant sneutrino to leptons
|\overline{\me}(\mr{d}_j \mr{\bar{d}}_k \rightarrow \ell^-_l \ell^+_m)|^2 &= &
 \frac{1}{4N_c}\displaystyle\sum_{i,n=1,3}
 {\lam'}_{ijk}{\lam'}_{njk}{\lam}_{ilm}{\lam}_{nlm} 
S(\nut_i,\nut_n,\sh,\sh)
   \sh\left(\sh-m^2_{\ell_m}-m^2_{\ell_l} \right)\!; \nonumber\\&&\\
&& \nonumber \\
% Resonant slepton to leptons
|\overline{\me}(\mr{u}_j \bar{d}_k \rightarrow \nu_l \ell^+_m)|^2 &= &
 \frac{1}{4N_c}\displaystyle \sum_{\al,\be=1,2}\displaystyle\sum_{i,n=1,3}
 {\lam'}_{ijk}{\lam'}_{njk}{\lam}_{ilm}{\lam}_{nlm}
 |L^{2i-1}_{1\al}|^2|L^{2n-1}_{1\be}|^2  \nonumber \\
& & 
  S(\elt_{i\al},\elt_{n\be},\sh,\sh)
   \sh\left(\sh-m^2_{\ell_m} \right)\!.
\end{eqnarray}

%
% LQD Higgs Processes
% 
\subsection[Resonant Slepton Production followed by Higgs Decay]
	{Resonant Slepton Production followed by Higgs Decay}

  There are a number of processes which can occur via the production of a
  resonant slepton followed by the decay of the resonant slepton
  to either a neutral or charged
  Higgs:
\begin{enumerate}
\item $\mr{d}_j \mr{\bar{d}}_k \longrightarrow \elt^*_{i\alpha} \mr{H^-}$;
\item $\mr{u}_j \mr{\bar{d}}_k \longrightarrow \nut^*_i \mr{H^+}$;
\item $\mr{u}_j \mr{\bar{d}}_k \longrightarrow
 \elt^*_{i\beta} \mr{h_0/H_0/A_0}$.
\end{enumerate}
  As we only include processes where there is a possibility of a
  resonant production mechanism, the process
  $\mr{d}_j \mr{\bar{d}}_k \longrightarrow \nut^*_i \mr{h_0/H_0/A_0}$
  is not included. For the same reason we also have not included the processes
  $\mr{u}_j \mr{\bar{d}}_k \longrightarrow \elt^*_{iL} \mr{h_0/H_0/A_0}$
  for the first two slepton
  generations. This is because HERWIG does not include left/right sfermion
  mixing for the first two generation sleptons and
  the initial state only couples to the left-handed slepton. The process 
  $\mr{u}_j \mr{\bar{d}}_k \longrightarrow \elt^*_{i2} \mr{h_0/H_0/A_0}$ is
  also not included for the third generation ($i=3$) as there is no
  accessible resonance.

  Since the
  matrix elements have the same form for all the neutral Higgs processes
  we use the notation $\mr{H}^l_0$ where $l$=1,2,3 is $\mr{ h_0}$,
  $\mr{H_0}$ and $\mr{A_0}$, respectively. The matrix elements for these
  processes are given below:
\begin{eqnarray}
% sneutrino to charged slepton and H
{\displaystyle \lefteqn{|\overline{\me}(\mr{d}_j \mr{\bar{d}}_k
 \rightarrow  \elt^*_{i\al} 
\mr{H^-})|^2 =}}\nonumber\\
 &{\displaystyle \frac{g^2 {\lam'}_{ijk}^2}{4N_c}\left[
	 |H^c_{\nut\elt_{i\al}}|^2 \sh  R(\nut_i,\sh)
	+\frac{4|L^{2i-1}_{1\al}|^2|D^c_j|^2}{\uh^2}
	 \left(\uh\tha-M^2_{ \elt_{i\be}}M^2_{\mr{H^-}}\right)
\rule{0mm}{0.8cm}\right]\!;} \\
 & \nonumber \\
% charged slepton to sneutrino and H
{\displaystyle\lefteqn{|\overline{\me}(\mr{u}_j \mr{\bar{d}}_k 
  \rightarrow \nut^*_i \mr{H^+})|^2
=}} \nonumber\\
\nopagebreak[4]
 &{\displaystyle \frac{g^2{\lam'}^2_{ijk}}{4 N_c}\left[\rule{0mm}{0.8cm}
	\alsm |L^{2i-1}_{i\al}|^2 |H^c_{\nut\elt_{i\al}}|^2
		\sh  R(\elt_{i\al},\sh) 
	+ 2  L^{2i-1}_{i1}L^{2i-1}_{i2} H^c_{\nu\elt_{i1}}H^c_{\nu\elt_{i2}}
		\sh  S(\elt_{i1},\elt_{i2},\sh,\sh) 
\right.} \nonumber \\
 &{\displaystyle \left.
 +\frac{4|U^c_j|^2}{\uh^2}\left(\uh\tha-M^2_{\nut_i}M^2_{\mr{H^+}}
 \right)\rule{0mm}{0.8cm}\right]\!;} \\
 & \nonumber \\
% stau_2 to stau_1 and H 
{\displaystyle \lefteqn{|\overline{\me}(\mr{u}_j \bar{d}_k 
\rightarrow \elt^*_{i\be}
\mr{H}^l_0)|^2 =}}
\nonumber\\
 & {\displaystyle \frac{g^2 {\lam'}_{ijk}^2}{4N_c}\left[\rule{0mm}{0.8cm}
	\alsm  |L^{2i-1}_{i\al}|^2  |H^l_{\elt_{i\al}\elt_{i\be}}|^2
		\sh  R(\elt_{i\al},\sh)
	+2  L^{2i-1}_{i1}L^{2i-1}_{i2}
		 H^l_{\elt_{i1}\elt_{i\be}}H^l_{\elt_{i2}\elt_{i\be}}
		\sh S(\elt_{i1},\elt_{i2},\sh,\sh) 
 \right.}\nonumber \\
 & {\displaystyle \left. +\frac{|L^{2i-1}_{1\be}|^2|D^l_j|^2}{\uh^2}
		\left(\uh\tha-M^2_{ \elt_{i\be}}M^2_{\mr{H}^l_0}\right)
 	  +\frac{|L^{2i-1}_{1\be}|^2|D^l_k|^2}{\tha^2}
\left(\uh\tha-M^2_{ \elt_{i\be}}M^2_{\mr{H}^l_0}\right)
\rule{0mm}{0.8cm} \right]\!.}
\end{eqnarray}
The couplings involved in the various processes can be found in
Tables\,\ref{tab:LQDhiggs} and \ref{tab:higgsqk}.

\section[UDD Processes]{UDD Processes}
\subsection[Resonant Squark Production   followed by Weak Decay]
	{Resonant Squark  Production followed by Weak Decay}

   There are four processes which can occur via the production of a
resonant squark followed by a weak decay of this squark:
\begin{enumerate}
\item $\mr{d}_j \mr{d}_k \longrightarrow \mr{\dnt}^*_{i\be} \mr{W^-}$;
\item $\mr{u}_i \mr{d}_j \longrightarrow \mr{\upt}^*_{k\be} \mr{W^+}$;
\item $\mr{d}_j \mr{d}_k \longrightarrow \mr{\tilde{t}}^*_1 \mr{Z^0}$;
\item $\mr{u}_i \mr{d}_j \longrightarrow \mr{\tilde{b}}^*_1 \mr{Z^0}$.
\end{enumerate}
  Again we do not include processes where the resonance is not
  accessible, \ie 
  $\mr{d}_j \mr{d}_k \longrightarrow \mr{\tilde{t}}^*_2 \mr{Z^0}$ and
  $\mr{u}_i \mr{d}_j \longrightarrow \mr{\tilde{b}}^*_2 \mr{Z^0}$.
  The matrix elements for these processes are given by
% sup to sdown and W
\begin{eqnarray}
{\displaystyle
\lefteqn{|\overline{\me}(\mr{d}_j \mr{d}_k \rightarrow
		 \mr{\dnt}^*_{i\be} \mr{W^-})|^2 =}
\nonumber }\\[1.5mm]
&&
{\displaystyle  
\frac{g^2 {\lam''}_{ijk}^2 (N_c-1)!|Q^{2i-1}_{1\be}|^2\sh^2 p^2_{\mr{cm}}}
   {2 N_c \mw^2}\left[\rule{0mm}{0.8cm}
 \alsm |Q^{2i}_{2\al}|^2 |Q^{2i}_{1\al}|^2
  R(\mr{\upt}_{i\al},\sh) \right.}
    \\[1.5mm]
&& {\displaystyle \left.+2 Q^{2i}_{21}Q^{2i}_{22}Q^{2i}_{11}Q^{2i}_{12}
  S(\mr{\upt}_{i1},\mr{\upt}_{i2},\sh,\sh)\rule{0mm}{0.8cm} \right]\!,}
\nonumber\\[1.5mm]
% sdown to sup and W
{\displaystyle\lefteqn{|\overline{\me}(\mr{u}_i \mr{d}_j \rightarrow
		 \mr{\upt}^*_{k\be} \mr{W^+})|^2  = }\nonumber}\\
&&
 {\displaystyle
  \frac{g^2 {\lam''}_{ijk}^2 (N_c-1)!\sh^2 p^2_{\mr{cm}}|Q^{2i}_{1\be}|^2}
  {2 N_c \mw^2}
 \left[\rule{0mm}{0.8cm}
\alsm|Q^{2i-1}_{2\al}|^2 |Q^{2i-1}_{1\al}|^2 R(\mr{\dnt}_{k\al},\sh)
 \right.}\nonumber\\[1.5mm]
 &&{\displaystyle 
  \left.+2Q^{2i-1}_{21}Q^{2i-1}_{22}Q^{2i-1}_{11}Q^{2i-1}_{12}
 S (\mr{\dnt}_{k1},\mr{\dnt}_{k2},\sh,\sh)\rule{0mm}{0.8cm}
  \right]\!, }\\[3mm]
% stop_2 to stop_1 and Z 
{\displaystyle\lefteqn{|\overline{\me}(\mr{d}_j \mr{d}_k \rightarrow
 \mr{\upt}^*_{i1} \mr{Z^0})|^2 =} 
\nonumber }\\[1.5mm]
 &&{\displaystyle 
 \frac{g^2 {\lam''}_{ijk}^2 (N_c-1)!}{N_c \mz^2\cos^2\theta_W}\left[
\rule{0mm}{0.8cm}
	\alsm |Q^{2i}_{2\al}|^2|Z^{\al1}_{u_i}|^2 \sh^2 p^2_{\mr{cm}} 
R(\mr{\upt}_{i\al},\sh)
  \right.  \nonumber }\\[1.5mm]
 &&{\displaystyle
 	+2 Q^{2i}_{21}Q^{2i}_{22}Z^{11}_{u_i}Z^{21}_{u_i}\sh^2 p^2_{\mr{cm}}
S(\mr{\upt}_{i1},\mr{\upt}_{i2},\sh,\sh)
 +\frac{|Q^{2i}_{21}|^2Z_{d_R}^2}{\uh^2}
	\left(2\mz^2(\uh\tha-M^2_{\mr{\upt}_{i1}}\mz^2) +\uh^2\sh\right)
\nonumber }\\[1.5mm]
 &&{\displaystyle   +\frac{|Q^{2i}_{21}|^2Z^2_{d_R}}{\tha^2}
	\left(2\mz^2(\uh\tha-M^2_{\mr{\upt}_{i1}}\mz^2) +\tha^2\sh\right)
\nonumber }\\[1.5mm]
 &&{\displaystyle  
	-\frac{2|Q^{2i}_{21}|^2Z_{d_R}^2}{\uh\tha}
  	 \left(2\mz^2(M^2_{\mr{\upt}_{i1}}-\uh)(M^2_{\mr{\upt}_{i1}}-\tha)-
\sh\tha\uh\right)}\nonumber\\[1.5mm]
 &&{\displaystyle
  +\alsm\frac{Q^{2i}_{2\al}Q^{2i}_{21}Z^{\al1}_{u_i}Z_{d_R}}{\uh}
	 \sh(\sh-M^2_{\mr{\upt}_{i\al}})R(\mr{\upt}_{i\al},\sh)
\left(\mz^2(2M^2_{\mr{\upt}_{i1}}-\uh) +\uh(\sh-M^2_{\mr{\upt}_{i1}})\right)
 \nonumber }\\[1.5mm]
 &&{\displaystyle   \left.
  +\alsm\frac{Q^{2i}_{2\al}Q^{2i}_{21}Z^{\al1}_{u_i}Z_{d_R}}{\tha}
	 \sh(\sh-M^2_{\mr{\upt}_{i\al}})R(\mr{\upt}_{i\al},\sh)
\left(\mz^2(2M^2_{\mr{\upt}_{i1}}-\tha) +\tha(\sh-M^2_{\mr{\upt}_{i1}})\right)
 \rule{0mm}{0.8cm}\right]\!,  }\nonumber\\
\end{eqnarray}
\begin{eqnarray}
% bottom_2 to bottom_1 and Z 
{\displaystyle \lefteqn{|\overline{\me}(\mr{u}_i \mr{d}_k 
\rightarrow \mr{\dnt}^*_{j1} \mr{Z^0})|^2 =} 
\nonumber }\\
 &&{\displaystyle  
\frac{g^2 {\lam''}_{ijk}^2 (N_c-1)!}{N_c \mz^2 \cos^2\theta_W}\left[
  \rule{0mm}{0.8cm}
\alsm|Q^{2j-1}_{2\al}|^2|Z^{\al1}_{d_j}|^2  \sh^2 p^2_{\mr{cm}} 
R(\mr{\dnt}_{j\al},\sh)\right. \nonumber}\\ 
 &&{\displaystyle 
      +2Q^{2j-1}_{21}Q^{2j-1}_{22}Z^{11}_{d_j}Z^{21}_{d_j}\sh^2 p^2_{\mr{cm}}
	 S(\mr{\dnt}_{j1},\mr{\dnt}_{j2},\sh^2,\sh^2)
\nonumber }\\
 &&{\displaystyle 
	 +\frac{|Q^{2j-1}_{21}|^2Z_{u_R}^2}{\uh^2}
	\left(2\mz^2(\uh\tha-M^2_{\mr{\dnt}_{j1}}\mz^2)+\uh^2\sh\right)
+\frac{|Q^{2j-1}_{21}|^2Z_{d_R}^2}{\tha^2}
\left(2\mz^2(\uh\tha-M^2_{\mr{\dnt}_{j1}}\mz^2)+\tha^2\sh\right)\nonumber}\\
&&{\displaystyle
	-\frac{2|Q^{2j-1}_{21}|^2Z_{u_R}Z_{d_R}}{\uh\tha}
	\left(2\mz^2(M^2_{\mr{\dnt}_{j1}}-\uh)(M^2_{\mr{\dnt}_{j1}}-\tha)-
\sh\tha\uh\right) 
\nonumber }\\
 &&{\displaystyle  
+\alsm\frac{Q^{2j-1}_{2\al}Q^{2j-1}_{21}Z^{\al1}_{d_j}Z_{u_R}}{\uh}
	\sh(\sh-M^2_{\mr{\dnt}_{j\al}})R(\mr{\dnt}_{j\al},\sh)
\left(\mz^2(2M^2_{\mr{\dnt}_{j1}}-\uh) +\uh(\sh-M^2_{\mr{\dnt}_{j1}})\right)
\nonumber }\\
 &&{\displaystyle  \left. 
+\alsm\frac{Q^{2j-1}_{2\al}Q^{2j-1}_{21}Z^{\al1}_{d_j}Z_{d_R}}{\tha}
	\sh(\sh-M^2_{\mr{\dnt}_{j\al}})R(\mr{\dnt}_{j\al},\sh)
\left(\mz^2(2M^2_{\mr{\dnt}_{j1}}-\tha) +\tha(\sh-M^2_{\mr{\dnt}_{j1}})\right)
\rule{0mm}{0.8cm}\right]\!,}\nonumber\\
\end{eqnarray}
  The coefficients for these processes can be found in Table\,\ref{tab:Zcp}.

\subsection[Resonant Squark  Production followed by \rpv\  Decay]
	{Resonant Squark  Production followed by \boldmath{\rpv}\  Decay}

  There are two processes in which a resonant squark is produced via
the \bv\  term in the superpotential where these squarks then decay to
Standard Model particles:
\begin{enumerate}
\item $\mr{d}_j \mr{d}_k \longrightarrow  \mr{d}_l \mr{d}_m$;
\item $\mr{u}_i \mr{d}_j \longrightarrow  \mr{u}_l \mr{d}_m$.
\end{enumerate}
  The matrix elements are given by
\begin{eqnarray}
% sup to quarks
{\displaystyle|\overline{\me}(\mr{d}_j \mr{d}_k \rightarrow
 \mr{d}_l \mr{d}_m)|^2} &=
 &{\displaystyle \frac{(N_c-1)!^2}{4N_c}
 \sum_{\al,\be=1,2}
\sum_{i,n=1,3}
  {\lam''}_{ijk}{\lam''}_{ilm}{\lam''}_{njk}{\lam''}_{nlm}
 |Q^{2i}_{2\al}|^2|Q^{2n}_{2\be}|^2 \nonumber} \\
&&{\displaystyle  S(\mr{\upt}_{i\al},\mr{\upt}_{n\be},\sh,\sh)
   \sh\left(\sh-m^2_{\mr{d}_l}-m^2_{\mr{d}_m} \right)\!,} \\[3mm]
% sdown to quarks 
{\displaystyle |\overline{\me}(\mr{u}_i \mr{d}_j \rightarrow 
\mr{u}_l \mr{d}_m)|^2}& = 
 &{\displaystyle  \frac{(N_c-1)^2}{4N_c}\sum_{\al,\be=1,2}
\sum_{k,n=1,3}
  {\lam''}_{ijk}{\lam''}_{lmk}{\lam''}_{ijn}{\lam''}_{lmn}
 |Q^{2k-1}_{2\al}|^2|Q^{2n-1}_{2\be}|^2
  }\nonumber\\
&&{\displaystyle  S(\mr{\dnt}_{i\al},\mr{\dnt}_{n\be},\sh,\sh)
   \sh\left(\sh-m^2_{\mr{u}_l}-m^2_{\mr{d}_m} \right)\!.  }
\end{eqnarray}

\subsection[Resonant Squark Production followed by Higgs Decay]
	{Resonant Squark Production followed by Higgs Decay}
  
There are a number of processes which occur via the production of a
resonant squark which subsequently decays to either a neutral or
charged Higgs. Again we only consider those processes for which a
resonance is possible, \ie we neglect the processes \linebreak
 $\mr{d}_j \mr{d}_k \longrightarrow \mr{\upt}^*_{iR} \mr{h_0/H_0/A_0}$ and
 $\mr{u}_i \mr{d}_j \longrightarrow \dnt^*_{iR}      \mr{h_0/H_0/A_0}$
  for the first two generations and the processes 
 $\mr{d}_j \mr{d}_k \longrightarrow \tilde{t}^*_{i2} \mr{h_0/H_0/A_0}$ and
 $\mr{u}_i \mr{d}_j \longrightarrow \tilde{b}^*_{i2} \mr{h_0/H_0/A_0}$
for the third generation, where we consider left/right sfermion
mixing as these processes cannot occur via a resonant diagram:
\begin{enumerate}
\item $\mr{d}_j \mr{d}_k \longrightarrow \mr{\dnt}^*_{i\be} \mr{H^-}$;
\item $\mr{u}_i \mr{d}_j \longrightarrow \mr{\upt}^*_{k\be} \mr{H^+}$;
\item $\mr{d}_j \mr{d}_k \longrightarrow \mr{\upt}^*_{i1} \mr{h_0/H_0/A_0}$;
\item $\mr{u}_i \mr{d}_j \longrightarrow \mr{\dnt}^*_{i1}  \mr{h_0/H_0/A_0}$.
\end{enumerate}
 Due to our
notation for the squark mixing matrices in the case of no left/right sfermion
mixing the right squark is denoted as the second mass eigenstate.
The matrix elements for these processes are given below:
\begin{eqnarray}
% sup to sdown and H^-
{\displaystyle \lefteqn{|\overline{\me}(\mr{d}_j \mr{d}_k \rightarrow
 \mr{\dnt}^*_{i\be} \mr{H^-})|^2 
= } \nonumber} \\[1mm]
 &&{\displaystyle \frac{g^2 (N_c-1)!}{4N_c} \left[ \rule{0mm}{0.8cm}
	\alsm  {\lam''}_{ijk}^2|Q^{2i}_{2\al}|^2|H^c_{\upt_{i\al}\dnt_
{i\be}}|^2 
		\sh R(\mr{\upt}_{i\al},\sh) \right.}\nonumber\\[1mm] &&
{\displaystyle + 2{\lam''}_{ijk}^2 Q^{2i}_{21}Q^{2i}_{22}
		H^c_{\upt_{i1}\dnt_{i\be}}H^c_{\upt_{i2}\dnt_{i\be}}
\sh S(\mr{\upt}_{i1},\mr{\upt}_{i2},\sh,\sh)
+\frac{4 {\lam''}_{jik}^2|U^c_j|^2|Q^{2i-1}_{2\be}|^2}{\uh^2}
		\left(\uh\tha-M^2_{\mr{\dnt}_{i\be}}M^2_{\mr{H^-}}\right)
 \nonumber} \\[1mm]
 &&{\displaystyle \left.
  +\frac{4 {\lam''}_{kij}^2|U^c_k|^2|Q^{2i-1}_{2\be}|^2}{\tha^2}
		\left(\uh\tha-M^2_{\mr{\dnt}_{i\be}}M^2_{\mr{H^-}}\right)
 \rule{0mm}{0.8cm}\right]\!;}
\\[3mm]
%\end{eqnarray}
%\begin{eqnarray}
% sdown to sup and H^+
{\displaystyle\lefteqn{|\overline{\me}(\mr{u}_i \mr{d}_j 
\rightarrow \mr{\upt}^*_{k\be} \mr{H^+})|^2 = } \nonumber} \\[1mm]
 && {\displaystyle\frac{g^2(N_c-1)!}{4N_c} \left[ \rule{0mm}{0.8cm}
\alsm {\lam''}_{ijk}^2|Q^{2k-1}_{2\al}|^2H^c_{\upt_{k\be}\dnt_{k\al}}|^2
		\sh  R(\mr{\dnt}_{k\al},\sh)\right.\nonumber}\\[1mm] &&
{\displaystyle	+2{\lam''}_{ijk}^2 Q^{2k-1}_{21} Q^{2k-1}_{22}
          	H^c_{\upt_{k\be}\dnt_{k1}}H^c_{\upt_{k\be}\dnt_{k2}}
            	\sh S(\mr{\dnt}_{k1},\mr{\dnt}_{k2},\sh,\sh)}
 \nonumber \\[1mm]
 &&{\displaystyle \left.
+\frac{4 {\lam''}_{kij}^2|D^c_i|^2|Q^{2k}_{2\be}|^2}{\uh^2}
		\left(\uh\tha-M^2_{\mr{\dnt}_{k\al}}M^2_{\mr{H^+}}\right)
\rule{0mm}{0.8cm}\right]\!;} \\[3mm]
% stop_2 to stop_1 and h_0 
{\displaystyle\lefteqn{|\overline{\me}(\mr{d}_j \mr{d}_k 
\rightarrow \mr{\upt}^*_{i1} \mr{H}^l_0)|^2 = }}
\nonumber \\[1mm]
 &&{\displaystyle \frac{g^2{\lam''}_{ijk}^2(N_c-1)!}{4N_c}
\left[\rule{0mm}{0.8cm}
	\alsm|Q^{2i}_{2\al}|^2|H^l_{\upt_{i\al}\upt_{i1}}|^2 \sh 
 R(\mr{\upt}_{i\al},\sh)\right.}\nonumber\\[1mm] &&{\displaystyle
	+2Q^{2i}_{21}Q^{2i}_{22}H^l_{\upt_{i1}\upt_{i1}}H^l_
{\upt_{i2}\upt_{i1}}
 	 \sh S(\mr{\upt}_{i1},\mr{\upt}_{i2},\sh,\sh) }
\nonumber \\[1mm]
 && {\displaystyle\left. +\frac{|Q^{2i}_{21}|^2|D^l_j|^2}{\tha^2}
		\left(\uh\tha-M^2_{\mr{\upt}_{i1}}M_{\mr{H^l_0}}^2\right)
   +\frac{|Q^{2i}_{21}|^2|D^l_k|^2}{\uh^2}
		\left(\uh\tha-M^2_{\mr{\upt}_{i1}}M_{\mr{H^l_0}}^2\right)
\rule{0mm}{0.8cm}\right]\!;}
% \\[3mm]
\end{eqnarray}
\begin{eqnarray}
% bottom_2 to bottom_1 and h_0 
{\displaystyle \lefteqn{|\overline{\me}(\mr{u}_i \mr{d}_k \rightarrow
 \mr{\dnt}^*_{j1} \mr{H}^l_0)|^2 = 
}} \nonumber \\
 &&{\displaystyle \frac{g^2{\lam''}_{ijk}^2(N_c-1)!}{4 N_c}
\left[\rule{0mm}{0.8cm}
	\alsm |Q^{2j-1}_{2\al}|^2|H^l_{\dnt_{j\al}\dnt_{j1}}|^2 
\sh R(\mr{\dnt}_{j\al},\sh)\right.}\nonumber\\ &&
{\displaystyle+2Q^{2j-1}_{21}Q^{2j-1}_{22}
H^l_{\dnt_{j1}\dnt_{j1}}H^l_{\dnt_{j2}\dnt_{j1}}
	 \sh S(\mr{\dnt}_{j1},\mr{\dnt}_{j2},\sh,\sh)}
\nonumber \\
 &&{\displaystyle \left. +\frac{|Q^{2j-1}_{21}|^2|U^l_i|^2}{\tha^2}
		\left(\uh\tha-M^2_{\mr{\dnt}_{j1}}M_{\mr{H^l_0}}^2\right)
	  +\frac{|Q^{2j-1}_{21}|^2|D^l_k|^2}{\uh^2}
\left(\uh\tha-M^2_{\mr{\dnt}_{j1}}M_{\mr{H^l_0}}^2\right)
\rule{0mm}{0.8cm}\right]\!.}
\end{eqnarray}
  The coefficients for the various processes can be found in 
  Tables\, \ref{tab:UDDhiggs} and \ref{tab:higgsqk}.

\addcontentsline{toc}{chapter}{Bibliography}
%\bibliography{thesis3}

\printindex
\end{document}